\newcommand{\bfr}{ {\bf r}} 
\newcommand{\bfrp}{ {\bf r'}} 
\newcommand{\bfrpp}{ {\bf r''}} 
\newcommand{\vrp}{ {\bf r'}} 
\newcommand{\vx}{ {\bf x}}   
\newcommand{\vxp}{ {\bf x'}} 
\newcommand{\vxpp}{ {\bf x''}} 
\newcommand{\vk}{ {\bf k}} 

\newcommand{\bracket}[3]{\ensuremath{\langle #1 | #2 | #3 \rangle}}
\newcommand{\ket}[1]{\ensuremath{| #1 \rangle}}
\newcommand{\bra}[1]{\ensuremath{\langle #1 |}}
\newcommand{\refeq}[1]{(\ref{#1})} 

\newenvironment{rcases}
  {\left.\begin{aligned}}
  {\end{aligned}\right\rbrace}

\documentclass[reprint,amsmath,amssymb,aps,rmp,showpacs,showkeys,floatfix,longbibliography]{revtex4-1}

\usepackage[english]{babel}
\usepackage{placeins}
\usepackage{comment}
\usepackage{graphicx}
\usepackage{rotating}
\usepackage{amsmath,amssymb}
\usepackage{amsthm}
\usepackage{bm}
\usepackage{color}
\usepackage{natbib}
\usepackage{longtable}
\usepackage{dcolumn}
\usepackage[usenames,dvipsnames]{xcolor}
\usepackage{epsfig}
\usepackage{braket}
\usepackage{dsfont}

\definecolor{aaltoOrange}{RGB}{255,121,0}%

\DeclareMathOperator{\sgn}{sgn}

\usepackage{feynmp}

\usepackage{ifpdf}
\ifpdf
\fi

\makeatletter
\def\endfmffile{%
  \fmfcmd{\p@rcent\space the end.^^J%
          end.^^J%
          endinput;}%
  \if@fmfio
    \immediate\closeout\@outfmf
  \fi
  \IfFileExists{\thefmffile.mp}{\immediate\write18{mpost \thefmffile}}{}
  \let\thefmffile\relax
}
\makeatother

\usepackage{array}
\newcolumntype{x}[1]{>{\centering\let\newline\\\arraybackslash\hspace{0pt}}m{#1}}
\newcolumntype{z}[1]{>{\raggedright\let\newline\\\arraybackslash\hspace{0pt}}m{#1}}
\newcolumntype{y}[1]{>{\let\newline\\\arraybackslash\hspace{6pt}}m{#1}}
\usepackage{colortbl}
\makeatletter
\newcommand{\thickhline}{%
    \noalign {\ifnum 0=`}\fi \hrule height 1pt
    \futurelet \reserved@a \@xhline
}
\newcolumntype{[}{@{\vrule width 1pt\hspace{0pt}}}
\newcolumntype{]}{@{\hspace{0pt}\vrule width 1pt}}
\newcolumntype{!}{@{\vrule width 1pt}}
\makeatother

\begin{document}
\widetext

\title{The $\bm{GW}$ compendium: A practical guide to theoretical photoemission spectroscopy}

\author{Dorothea Golze}
\affiliation{Department of Applied Physics, Aalto University School of Science, 00076-Aalto, Finland}
\author{Marc Dvorak}
\affiliation{Department of Applied Physics, Aalto University School of Science, 00076-Aalto, Finland}
\author{Patrick Rinke}
\affiliation{Department of Applied Physics, Aalto University School of Science, 00076-Aalto, Finland}

\date{\today}

\begin{abstract}
The $GW$ approximation in electronic structure theory has become a widespread tool for predicting electronic excitations in chemical compounds and materials. In the realm of theoretical spectroscopy, the $GW$ method provides access to charged excitations as measured in direct or inverse photoemission spectroscopy. The number of $GW$ calculations in the past two decades has exploded with increased computing power and modern codes. The success of $GW$ can be attributed to many factors: favorable scaling with respect to system size, a formal interpretation for charged excitation energies, the importance of dynamical screening in real systems, and its practical combination with other theories. In this review, we provide an overview of these formal and practical considerations. We expand, in detail, on the choices presented to the scientist performing $GW$ calculations for the first time. We also give an introduction to the many-body theory behind $GW$, a review of modern applications like molecules and surfaces, and a perspective on methods which go beyond conventional $GW$ calculations. This review addresses chemists, physicists and material scientists with an interest in theoretical spectroscopy. It is intended for newcomers to $GW$ calculations but can also serve as an alternative perspective for experts and an up-to-date source of computational techniques.
\end{abstract}

\maketitle

\tableofcontents
\section{Introduction}
	\label{sec:intro}
		Electronic structure theory derives from the fundamental laws of quantum mechanics and describes the behavior of electrons -- the glue that shapes all matter. To understand the properties of matter and the behavior of molecules, the quantum mechanical laws must be solved numerically because a pen and paper solution is not possible. In this context, Hedin's $GW$ method \cite{Hedin:1965} has become the \emph{de facto} standard for electronic structure properties as  measured by direct and inverse photoemission experiments, such as quasiparticle band structures and molecular excitations.

Electronic structure theory covers the quantum mechanical spectrum of computational materials science and quantum chemistry. The fundamental aim of computational science is to derive understanding entirely from the basic laws of physics, i.e. quantum mechanical first principles, and increasingly also to make predictions of new properties or new materials and new molecules for specific tasks. The exponential increase in available computer power and new methodological developments are two major factors in the growing impact of this field for practical applications to real systems. As a result of these advances, computational science is establishing itself as a viable complement to the purely experimental and theoretical sciences. 

Hedin published the $GW$ method in 1965, in the same time period as the foundational density-functional theory (DFT) papers \cite{Hohenberg/Kohn:1964,Kohn/Sham:1965}. While DFT has shaped the realm of first principles computational science like no other method today, $GW$'s fame took a little longer to develop\footnote{The theory directly comparable to DFT is the Luttinger-Ward formalism, not the $GW$ approximation. Here, we focus on the chronological development of practical electronic structure calculations with Kohn-Sham DFT or $GW$.}. DFT's success has been facilitated by the computational efficiency of the local-density \cite{Kohn/Sham:1965} or generalized gradient approximations \cite{Becke:1988a, Lee/Yang/Parr:1992, Perdew/Burke/Ernzerhof:1996} (LDA and GGA) of the exchange-correlation functional that make DFT applicable to polyatomic systems containing up to several thousand atoms. $GW$, however, only saw its first applications to realistic materials 20 years after its inception \cite{Hybertsen/Louie:1985,Godby/etal:1986}, due to its much higher computational expense. Soon after it was realized that $GW$ can overcome some of the most notorious deficiencies of common density functionals such as the self-interaction error, the absence of long-range polarization effects and the Kohn-Sham band-gap problem.

The $GW$ approach is now an integral part of electronic structure theory and readily available in major electronic structure codes. It is taught at summer schools along side DFT and other electronic structure methods. This review is intended as a tutorial that complements showcases of $GW$'s achievements with a practical guide through the theory, its implementation and actual use. For $GW$ novices, the review offers a gentle introduction to the $GW$ concept and its application areas. Regular $GW$ users can consult this review as a handbook in their day-to-day use of the $GW$ method. For seasoned $GW$ users and $GW$ experts it might serve as a reference for key applications and some of the subtler points of the $GW$ framework.

In this review we take a more practical approach towards the $GW$ method. We will recap the basic theory starting from theoretical spectroscopic view point as a probe of the electronic structure. Aiming at $GW$ practitioners, we will illustrate how the $GW$ approach emerges from the theoretical spectroscopy framework as a practical scheme for electronic structure calculations. A more in-depth discussion of  the theoretical foundations of many-body theory can be found in textbooks, e.g., \cite{Gross/Runge:MPT,Fetter/Walecka,Szabo/Ostlund:1989,martin_reining_ceperley_2016,bechstedt_2014}, while the $GW$ theory itself is covered in excellent early reviews \cite{Hedin/Lundqvist:GW,Aryasetiawan/Gunnarsson:1998,Hedin:1999}. $GW$ began to flourish at the beginning of the 21st century and two reviews succinctly summarized the state of the field until then \cite{Aulbur/Jonsson/Wilkins:2000,Onida/Reining/Rubio:2002}. Our review bridges the ensuing gap of almost twenty years, after which only several more specialized reviews addressed different aspects and applications of $GW$ calculations \cite{Rinke/pssb,Giantomassi/etal:2011,Ping/etal:2013,Faber/etal:2014,Bruneval/Gatti:2014,Marom:2017,Gerosa/etal:2018,Kang/etal:2019} and complements a recent review \cite{Reining2017}.

A considerable part of our review is devoted to the presentation of different $GW$ implementations. We will discuss the practical considerations that $GW$ users have to make when they decide on a particular $GW$ implementation or code for their work. Moreover, we will guide the reader through computational decisions that might affect the accuracy of their $GW$ calculations and illustrate them with concrete examples from the $GW$ literature. An important aspect in this regard is the issue of self-consistency in $GW$, which we cover in detail. 

Although the $GW$ method might be best known for its success in predicting the band gaps of solids, we will present its diversity and discuss a range of different quantities that can be computed with the $GW$ method. Since no method is perfect, we will conclude with a critical outlook on the challenges faced by the $GW$ method and discuss ways to go beyond $GW$.

We conclude this introduction with a quote from H. J. Monkhorst, who  wrote in a laudation in 2005: \emph{It is therefore my conviction that, rather sooner then later, we will see a resurgence of the precise many-body approach to solid-state theory as we envisioned. Almost assuredly the $GW$ method will be the tool of choice.} \cite{Monkhorst:2005}. In 2019, we can say that Monkhorst was right.

\section{Theoretical spectroscopy} 
	\label{sec:TS} 
		\subsection{Direct and inverse photoemission spectroscopy}
\label{sec:direct_inv_spec}

Spectroscopic measurements are an important component in the characterization of materials.  Any spectroscopic technique perturbs the system under investigation and promotes it into an excited state. Experimentally, the challenge then lies in the correct interpretation of the system's response. From a theoretical point of view, however, the challenge is to find (or develop) a suitable, accurate and, most of all, computationally tractable approach to describe the response of the system. Experimental and theoretical spectroscopy are often complementary and, when combined, they are a powerful approach.\par
Many-body perturbation theory (MBPT) provides a rigorous and systematic quantum mechanical framework to describe the spectral properties of a system that connects central quantities like the Green's function, the self-energy, and the dielectric function with each other. The poles of the single-particle Green's function, the central object in MBPT, correspond to the electron addition and removal energies probed in direct and inverse photoemission, which is explained in detail at the end of Section~\ref{sec:direct_inv_spec}. In contrast, information about neutral excitations probed in optical or energy loss spectra can be extracted from the dielectric function. In this review, we will not address optical properties or other neutral excitations and instead focus on the single-particle Green's function and its connection to direct and inverse photoemission spectroscopy.\par
\begin{figure} 
    \includegraphics[width=\columnwidth]{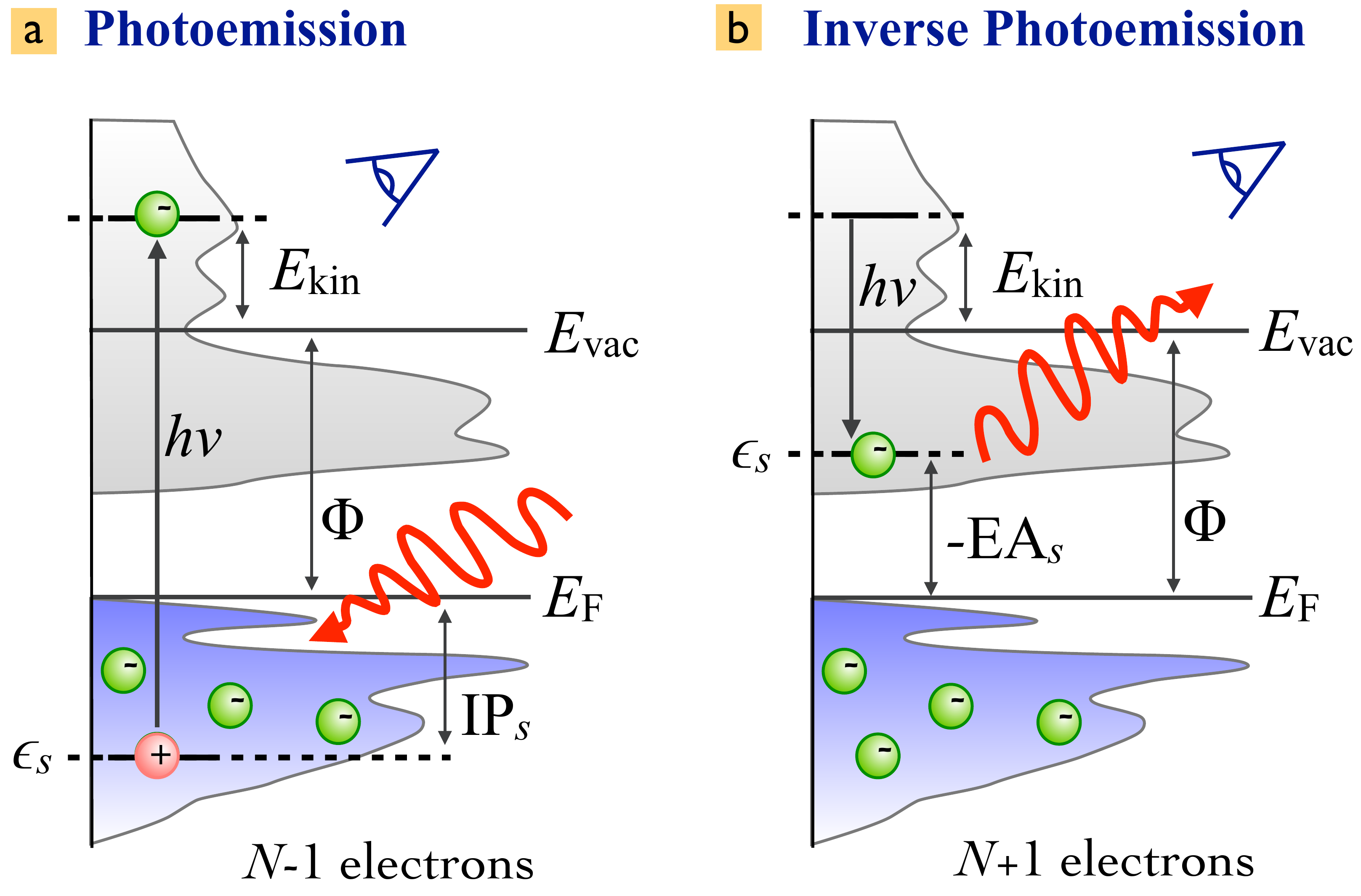}
	\caption{{\small \label{fig:(I)PES}
        	 Schematic of the photoemission (PES) and inverse photoemission
		 (IPES) process. In PES (a) an electron is excited by an
		 incoming photon from a previously
		 occupied valence state (lower shaded region) into the 
		 continuum (gray shaded region, starting above the vacuum level
		 $E_{\text{vac}}$). In IPES (b) an injected electron with kinetic
		 energy $E_{\text {kin}}$ undergoes a radiative transition into
		 an unoccupied state (upper shaded region) thus 
		 emitting a photon in the process. Figure adapted from \cite{Rinke/etal:2005}.
		 }}
\end{figure}
In photo-electron spectroscopy (PES) \cite{Himpsel:1983,Plummer/Eberhardt:1982,ARPES:1992}, electrons are ejected from a sample upon irradiation with visible or ultraviolet light (UPS) or with X-rays (XPS), as sketched in Figure~\ref{fig:(I)PES}(a). 
 The energy of the bound state $\epsilon_s$ can be reconstructed from the photon energy $h\nu$, the work function $\Phi$ and the kinetic energy $E_{\text{kin}}$ of the photoelectrons that reach the detector\footnote{Throughout this article the energy zero is chosen to be the top of the valence bands for extended systems and the vacuum level for finite systems.}\par
\begin{equation}
\label{Eq:def:Ei}
  \text{IP}_s=-\epsilon_s= h \nu - E_{\rm kin} - \Phi, \quad  \textrm{for} \quad \epsilon_s < E_{\rm F}  .
\end{equation}
The ionization potential $\text{IP}_s$ is defined as the energy that is required to remove an electron from the bound initial state $s$ of the neutral sample, where the energy of the state is below the Fermi level ($E_{\rm F}$). It is always a positive number and related to $\epsilon_s$ as shown in Equation~\eqref{Eq:def:Ei}. \par
By inverting  the photoemission process, as schematically shown in  Figure~\ref{fig:(I)PES}(b), the unoccupied states can be probed. This technique is commonly referred to as inverse photoemission spectroscopy
(IPES) or bremsstrahlung isochromat spectroscopy (BIS) \cite{Dose:1985,Smith:1988,IPES:1992}. In IPES, an incident electron with energy $E_{\text{kin}}$ is scattered in the sample emitting \emph{bremsstrahlung}. Eventually it will undergo a radiative transition into a lower-lying unoccupied state, emitting a photon that carries the transition energy $h\nu$. The energy of the final, unoccupied state $\epsilon_s$ can be deduced from the measured photon energy according to
\begin{equation}
\label{Eq:def:Ef}
  -\text{EA}_s=\epsilon_s = E_{\rm kin} - h \nu + \Phi, \quad\textrm{for}\quad \epsilon_s \geq E_\mathrm{F}
\end{equation}
EA denotes the electron affinity, which we define as the energy needed to detach an electron from a negatively charged species and which is thus the negative of $\epsilon_s$. EA can be a positive or negative number. It is negative when the additional electron is in an unbound state, and positive when the electron is bound.\par 
The experimental observable in photoemission spectroscopy is the photocurrent, which is the probability of emitting an electron with the kinetic energy $E_{\rm kin}$ within a certain time interval. It is related to the intrinsic spectral function $A(\bfr,\bfrp,\omega)$ of the electronic system, given by the imaginary part of the single-particle Green's function\footnote{Atomic units $4\pi\epsilon_0=h=e=m_e=1$, where $e$ and $m_e$ are the charge and mass of an electron, respectively, will be used in the remainder of this article.}:
\cite{Onida/Reining/Rubio:2002,Almbladh/Hedin:1983}
\begin{equation}
\label{Eq:def_A}
  A(\bfr,\bfrp,\omega)=\frac{1}{\pi}{\rm Im} \: G(\bfr,\bfrp,\omega)\sgn(E_{\rm F}-\omega),
\end{equation}
where $\omega$ denotes an energy (frequency). The single-particle Green's function, $G(\bfr,\bfrp,\omega)$, is the probability amplitude that a particle created or destroyed at $\bfr$ is correlated with the adjoint process at $\bfrp$ $-$ it will be discussed in detail later. The actual dependence of the photocurrent on the spectral function is quite complicated because the coupling to the exciting light and electron loss processes in the sample, as well as surface effects, have to be taken into account. To our knowledge, no comprehensive theory yet exists for this relation and we therefore proceed with the discussion of the spectral function and will return to the photocurrent later.\par
The energies $\epsilon_s$ in Equations~\eqref{Eq:def:Ei} and \eqref{Eq:def:Ef} are the removal and addition energies of the photoelectron, respectively, and we refer to the transition amplitudes from the $N$ to the $N\pm 1$-body states as $\psi_s(\bfr)$ (see also Section \ref{sec:G}): 
\begin{eqnarray}
\label{Eq:eng_ex_h}
  \left.
  \begin{array}{lcl}
    \epsilon_s&=&E(N)-E(N-1,s)\\
    \psi_s(\bfr)&=&\bracket{N-1,s}{\hat{\psi}(\bfr)}{N}
  \end{array}
  \quad \right\} 
  \textrm{for} \quad \epsilon_s < E_{\rm F}\\
  \label{Eq:def_es}
  \left.
  \begin{array}{lcl}
    \epsilon_s&=&E(N+1,s)-E(N) \\
    \psi_s(\bfr)&=&\bracket{N}{\hat{\psi}(\bfr)}{N+1,s}
  \end{array}
  \quad \right\} 
  \textrm{for} \quad \epsilon_s \geq E_{\rm F}
\end{eqnarray}
The states $\ket{N,s}$ are many-body eigenstates (wave functions in real space) of the $N$-electron Schr\"odinger equation $\hat{H}\ket{N,s}=E(N,s)\ket{N,s}$, $\hat{H}$ is the many-body Hamiltonian and  $E(N,s)=\bracket{N,s}{\hat{H}}{N,s}$ is the corresponding total energy. The field operator $\hat{\psi}(\bfr)$ annihilates an electron at point $\mathbf{r}$ from the many-body states $\ket{N}$ or $\ket{N+1}$. The representation given in Equations~\eqref{Eq:eng_ex_h} and \eqref{Eq:def_es} is particularly insightful because it allows a direct interpretation of $\epsilon_s$ as the photoexcitation energy from the $N$-particle ground state with total energy $E(N)$ into an excited state $s$ of the ($N$-1)-particle system with total energy $E(N-1,s)$ upon removal of an electron in the photoemission process. Similarly, the addition energy that is released in the radiative transition in inverse photoemission is given by the total energy difference of the  excited ($N$+1)-particle system and the ground state.\par
To build a practical  scheme for calculating the energies in Equations~\eqref{Eq:eng_ex_h} and \eqref{Eq:def_es}  we introduce the definition of the single-particle Green's function\footnote{We consider only the zero temperature $G$ and assume $\mu = E_\mathrm{F}$.}
\begin{equation}
\label{def:G}
  G(\bfr,\sigma, t,\bfrp,\sigma' t')=-i\bracket{N}{\hat{T}\{\hat{\psi}(\bfr,\sigma,t)
                   \hat{\psi}^\dagger(\bfrp,\sigma',t') \}}{N}
\end{equation}
where $\hat{T}$ is the time ordering operator for the times $t$ and $t'$ and $\sigma$ the spin. $\hat{T}$ arranges the field operators so that the earlier time is to the right and acts on the ground state $\ket{N}$ first. $G$ allows for \textit{both} time orderings: $t > t'$ or $t' > t$. This definition of the Green's function is particularly insightful because it illustrates the process of adding and removing electrons from the system, as done in photoemession spectroscopy. Assuming the time-ordering is as shown in Equation~\eqref{def:G}, 
$\hat{\psi}^\dagger(\bfrp,\sigma',t')$ will create an electron with spin $\sigma'$ at time $t'$ in point $\vrp$. This electron will then propagate through the system, until it is annihilated by $\hat{\psi}(\bfr,\sigma,t)$ at a later time $t$ in position $\bfr$. The Green's function is therefore also often called a \emph{propagator}. We will return to this propagator picture of $G$ in later sections of this review.\par
To make contact with Equations~\eqref{Eq:eng_ex_h} and \eqref{Eq:def_es}, we need to Fourier transform the Green's function from the time to the energy axis. For a time-independent Hamiltonian this then produces the spectral or Lehman representation of $G$ \cite{Gross/Runge:MPT,Fetter/Walecka}
\begin{align}
  G(\bfr,\bfrp,\omega)= & \lim_{\eta \rightarrow 0^+}   \sum_s   \psi_s(\bfr) \psi_s^*(\bfrp) \times \nonumber \\
                  &  \times \left[
                  \frac{\Theta(\epsilon_s-E_{\rm F})}{\omega-(\epsilon_s - i\eta)} +
		  \frac{\Theta(E_{\rm F}-\epsilon_s)}{\omega-(\epsilon_s + i\eta)}  \right] 
\label{G:Leh}
\end{align}
where we have assumed a spin paired system and summed over the spin quantum number shown in Equation~\eqref{def:G}. The two terms in brackets are for the two time orderings in $G$. $\Theta$ is the Heaviside step function, which is zero for negative arguments and one for positive arguments\footnote{For the remainder of the review, $\eta$ is always assumed to be a positive infinitesimal.}. It kills any processes that do not obey the correct time ordering, as determined by the created/annihilated particle's energy relative to $E_{\rm F}$. This representation illustrates that the many-body excitations of the system that are associated with the removal or addition of an electron are given by the poles of the single-particle Green's function. The diagonal spectral function
\begin{align}
  A(\bfr,\bfr,\omega)&=\frac{1}{\pi}{\rm Im} \: G(\bfr,\bfr,\omega) \sgn(E_{\rm F}-\omega)\\
                      &=\sum_s
		       \psi_s(\bfr)\psi_s^*(\bfr)
		       \delta(\omega-\epsilon_s) 
\label{Eq:A_NI}
\end{align}
then assumes the intuitive form of a (many-body) density of states.

\subsection{The quasiparticle concept}
\label{sec:qp_concept}

\begin{figure} 
        \includegraphics[width=0.99\linewidth]{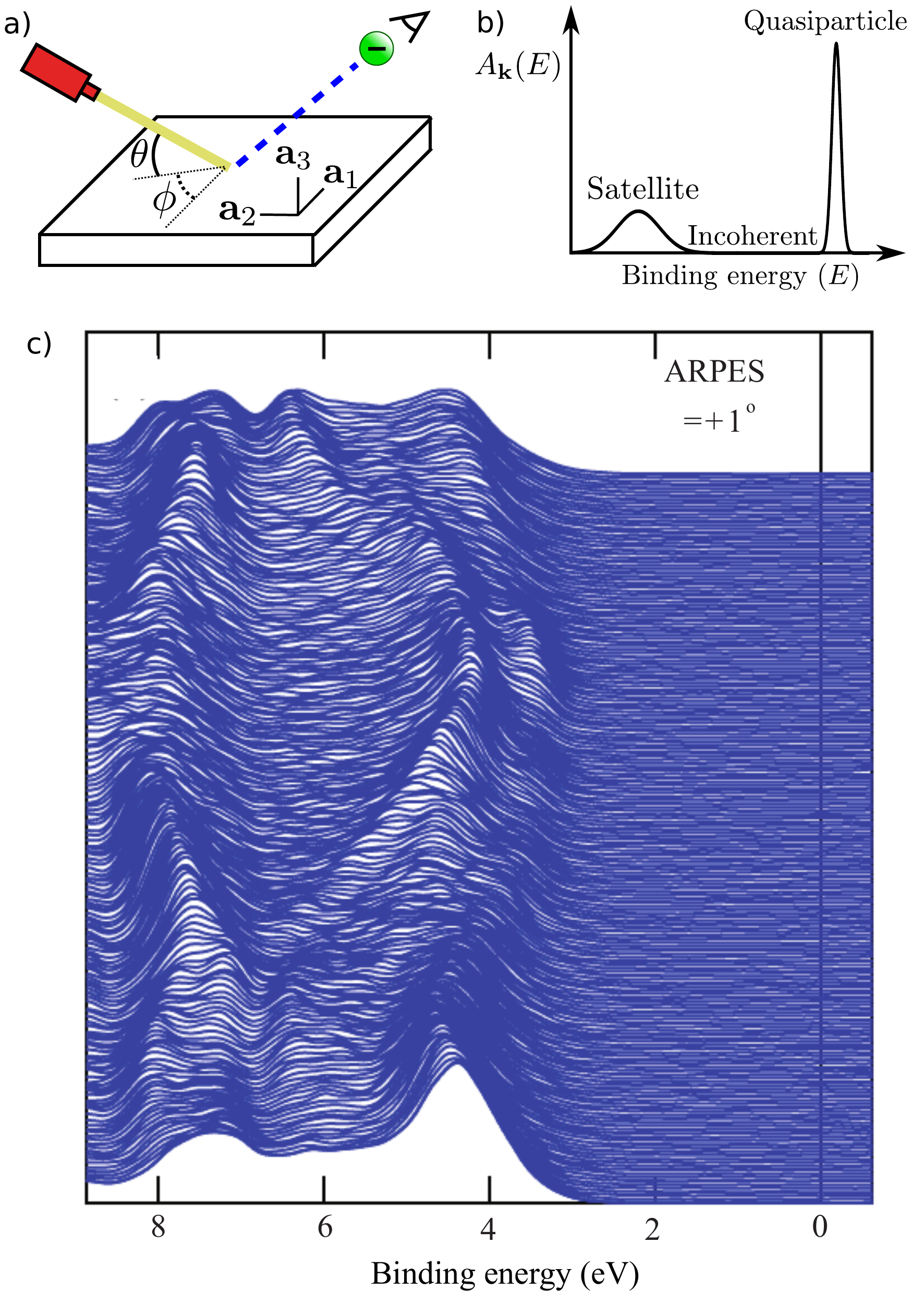}
	\caption{{\small \label{fig:PES_spec_BS}
        	 (a) Schematic representation of an ARPES experiment. By varying the angles $\theta$ and $\phi$ with respect to the crystallographic axes ($\mathbf{a}_i$), the measured spectrum is direction, or $\mathbf{k}$, dependent. In practice, the detector angle is usually varied with respect to a fixed beam. (b) A typical spectral function features a sharp peak attributed to the quasiparticle, an incoherent background, and satellites away from the single particle peak. (c) ARPES data of the upper valance bands of ZnO \cite{Kobayashi:2009,Kobayashi:privcom}. The corresponding $G_0W_0$ band structure of ZnO is shown in Figure~\ref{fig:zno_bands}.}}
\end{figure}
In periodic solids, the crystal has special crystallographic directions so that spectra are direction dependent, with the direction indexed by a wave vector $\mathbf{k}$. By varying the direction of the incident beam relative to the crystallographic axes ($\mathbf{a}_i$), one can map the $\mathbf{k}$ dependent PES spectra, as shown in Figure~\ref{fig:PES_spec_BS}(a). This technique is called angle resolved photoemission spectroscopy (ARPES). Figure~\ref{fig:PES_spec_BS}(c) shows data from a typical ARPES measurement and Figure~\ref{fig:PES_spec_BS}(b) a schematic of the spectral function at a single $\vk$-point. The spectra usually exhibit distinct peaks that are attributed to particle-like states but have a finite width. There can also be additional, broader peaks away from the main peak called satellites. However, the spectral function in Equation~\eqref{Eq:A_NI} contains only Dirac-delta functions which appear as infinitely sharp peaks. The broadening of the spectral function comes from the sum of many delta functions close in energy, which merge to form a peak of finite width. If the contributing delta functions are closely packed around one energy, the peak is attributed to a \textit{quasiparticle} \cite{Landau:1980}.\par
To further motivate the association of \emph{quasiparticles} with particle-like excitations it is insightful to consider non-interacting electrons. In that case, the spectral function consists of a series of delta peaks
\begin{equation}
  A_{ss'}(\omega)=\bracket{\psi_s}{A(\omega)}{\psi_{s'}}=
                   \delta_{ss'}\delta(\omega-\epsilon_s),  
 \label{eq:adelta}
\end{equation} 
each of which corresponds to the excitation of a non-interacting particle, see Appendix~\ref{sec:notation} for the integral notation used in Equation~\eqref{eq:adelta}. The many-body states $\ket{N}$ and $\ket{N\pm 1}$ become single Slater determinants so that the exact excited states are characterized by a single creation or  annihilation operator acting on the ground state. The excitation energies $\epsilon_s$ and the wave functions $\psi_s(\bfr)$ are thus the eigenvalues and eigenfunctions of the single-particle Hamiltonian.\par
When the electron-electron (or electron-ion) interaction is turned on, the exact eigenstates $\ket{N,s}$ are no longer single Slater determinants. As a consequence, the matrix elements of the spectral function $A_{ss'}(\omega)$ will contain contributions from many non-vanishing transition amplitudes.  If these contributions merge into a clearly identifiable peak that appears to be derived from a single delta-peak broadened by the  electron-electron interaction, this structure can be interpreted as a single-particle like excitation -- the \emph{quasiparticle}. The broadening of the quasiparticle peak in the spectral function is associated with the lifetime $\tau$ of the excitation due to electron-electron scattering, whereas the area underneath the peak is interpreted as the renormalisation $Z$ of the quasiparticle.  This renormalisation factor quantifies the reduction in spectral weight due to the electron-electron interaction compared to an independent electron, though the total spectral weight is conserved. We can combine these various arguments and say that the quasiparticle peak for state $s$ will exhibit the following shape: 
\begin{equation}
\label{eq:simplified_SF}
  A_{ss}(\omega)\approx \frac{1}{\pi} \left| \frac{Z_s}{\omega-(\epsilon_s+i\Gamma)} \right|
  \quad.
\end{equation}
In contrast to the exact energies of the many-body states,  which are poles of the Green's function on the real axis, the quasiparticle poles reside in the complex plane and are no longer eigenvalues of the $N$-body Hamiltonian. The real part of this complex energy is associated with the energy of the quasiparticle excitation and the imaginary part with its  inverse lifetime $\Gamma=2/\tau$. \par
\begin{figure} 
        \includegraphics[width=0.99\linewidth]{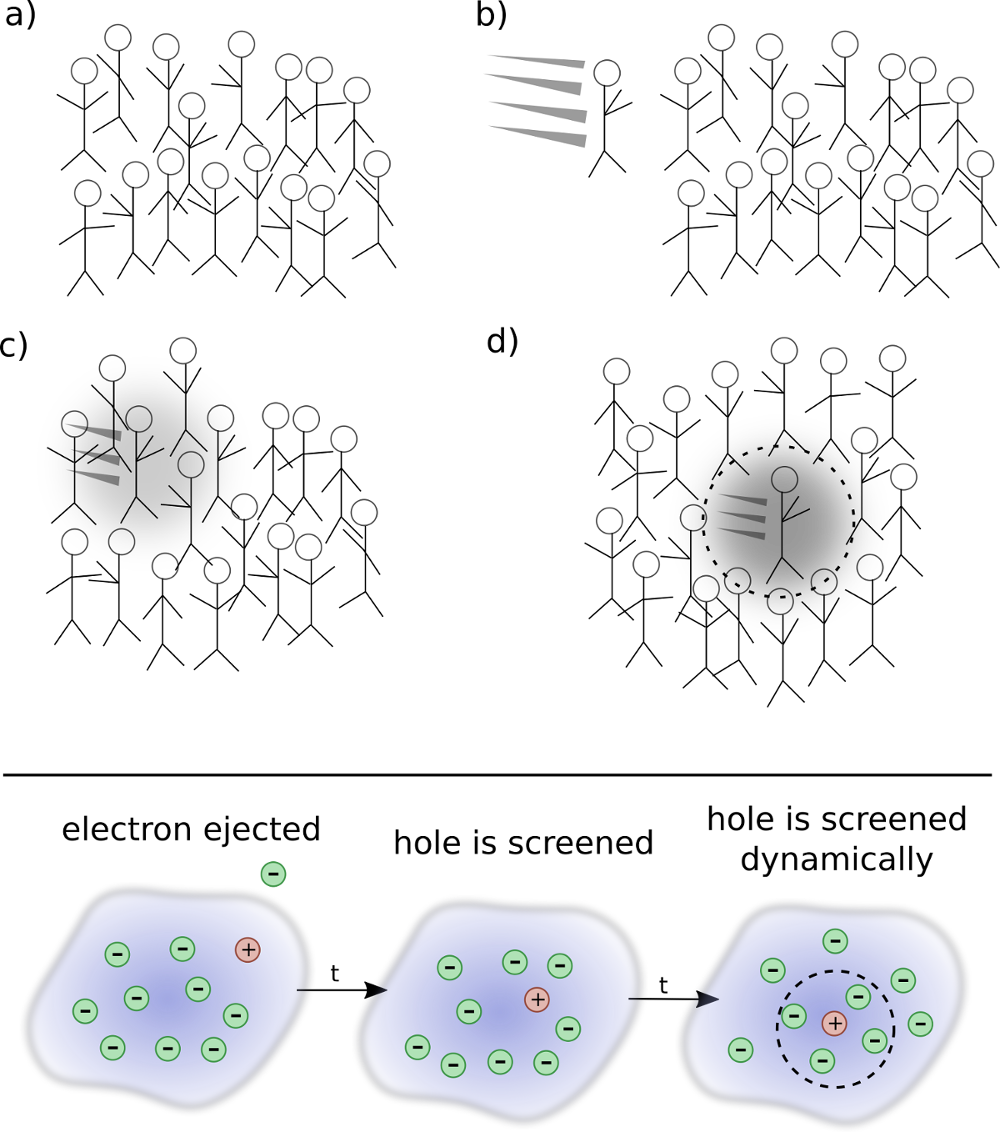}
        \caption{{\small \label{fig:quasi-person}
        	 Top: Depiction of the quasiparticle concept. (a) A crowd of people is analogous to the electronic ground state. A new person (that represents an additional electron) enters the crowd in (b). The new person begins to interact with other people who, in turn, interact back with the new person in (c) and form a polarisation cloud. An effective, or renormalized, object, the quasi-person, moves through the crowd in (d). Even though it is an interacting system, the many-person state in (d) can still be connected to, or identified by, the single person added to the crowd. This connection allows us to identify the quasi-person.
        	 Bottom: Schematic representation of photoemission spectroscopy.}}
\end{figure}

To develop a more intuitive understanding of quasiparticles, it is insightful to adopt a real-space picture. The quasiparticle concept can be explained by analogy with a crowd of people, as shown in Figure~\ref{fig:quasi-person}. Picture a group of people, such as at a concert or festival, all crowded into the same area. Not wanting to get too close to each other to preserve their own space, people in the crowd interact with each other. If one person gets too close to another, their mutual repulsion eventually takes over and separates them again. The exact description of the crowd requires the location of each individual person at all times. This is a very difficult task because of the constant interactions, or repulsions, between individual people. This collection of people and their occasional fluctuations are grouped together and labeled the ground state.\par
A new person arrives and pushes their way into the crowd. We can think of this new person as the electron in inverse photoemission that is injected into the system. The new person enters in a specific direction with a certain energy. As they enter the group, they repeatedly interact with other people as they continue their trajectory, as shown in Figure \ref{fig:quasi-person}(c). These repeated interactions repel people in their immediate vicinity and form a small halo of free space around the incoming person. People seem to move out of their way on their journey, forming a polarisation cloud created by the absence of other people around them. The intruder's motion and their polarisation cloud can be taken together to form a new composite object, a quasi-person, which appears as a slowed-down version of the newcomer.  From far away, one does not need to describe the precise motion of all $N+1$ people in the group, but only the motion of this quasi-person propagating through the crowd.\par
By analogy, a quasiparticle can then be considered a combination of an additional electron or hole in the system that interacts with its surrounding polarisation cloud. The situation corresponding to photemission spectroscopy is depicted in the bottom panel of Figure \ref{fig:quasi-person}. As time increases, the bare hole left by removal of the interaction is screened. The quasiparticle therefore embodies an electron state with the perturbation of its own surrounding. The feedback via interactions of the particle with surrounding electrons is termed the \textit{self-energy}. Over time, the propagating quasiparticle can decay into many different elementary excitations, giving it a finite lifetime. Essential quasiparticle properties are dispersion, lifetime, weight, and satellite spectrum. The latter arises from the collective excitations in the medium. 

\subsection{Comparison to experimental spectra}
\label{sec:exp_spec} 
We have now identified quasiparticles as one possible source for peaks in experimental photoemission spectra. Before we introduce the $GW$ approximation as  a tractable computational approach for calculating quasiparticle energies, we will first address the photocurrent, which is the quantity measured in direct photoemission experiments. Then we will briefly discuss the reconstruction of the band structure information, as well as other sources of peaks in spectrum.\par
Establishing rigorous links between the spectral function and the photocurrent is still a challenge for theory \cite{Hedin:1999,Lee/Gunnarsson/Hedin:1999,Minar2011,Minar2013}. The photocurrent $J_{\mathbf{k}}(h\nu)$ is the probability per unit time of emitting a photoelectron with momentum $\mathbf{k}$ and energy $E_{\rm kin,\mathbf{k}}$ due to an incident photon with the energy $h\nu$. The spectral function defined in Equation~\eqref{Eq:def_A} describes the removal of an electron from the sample, but does not include intermediate steps on the way to the detector where the electron loses energy. Therefore, it does not correspond to $J_{\mathbf{k}}(h\nu)$. However, the spectral function can be related to the photocurrent by using the sudden approximation \cite{Hedin:1999,Hedin2002} assuming that the ejected photoelectron is immediately decoupled from the sample.  $J_{\mathbf{k}}(h\nu)$ is then given by \cite{Hedin:1999}
\begin{equation}
    J_{\mathbf{k}}(h\nu) =\ \sum_{s,s'} \Delta_{\mathbf{k}s}A_{ss'}(E_{\rm kin,\mathbf{k}}-h\nu) \Delta_{s'\mathbf{k}} \quad,
\end{equation}
where $A_{ss'}(\omega)=\bracket{\psi_s}{A(\omega)}{\psi_{s'}}$ are matrix elements of the spectral function defined in Equation~\eqref{Eq:def_A}. $\Delta_{\mathbf{k},s}$ are matrix elements of the dipole operator which describe the coupling to photons. The dipole matrix elements capture the promotion of the electron to a highly excited state (often assumed to be a plane wave), i.e., they describe the transition between the initial and the final electron state. 
 In this final state, the electron travels to the detector. On the way, it crosses the surface of the sample, which adds a further perturbation to its path and its energy. The transition matrix elements affect the amplitudes of the spectrum and add selection rules that give rise to the suppression or enhancement of certain peaks. In practice, one compares only matrix elements of the spectral function to the experiment disregarding the effects of the dipole matrix. Furthermore, it is often assumed that only the diagonal elements of the spectral function are dominating.\par
\begin{figure} 
    \includegraphics[width=0.99\columnwidth]{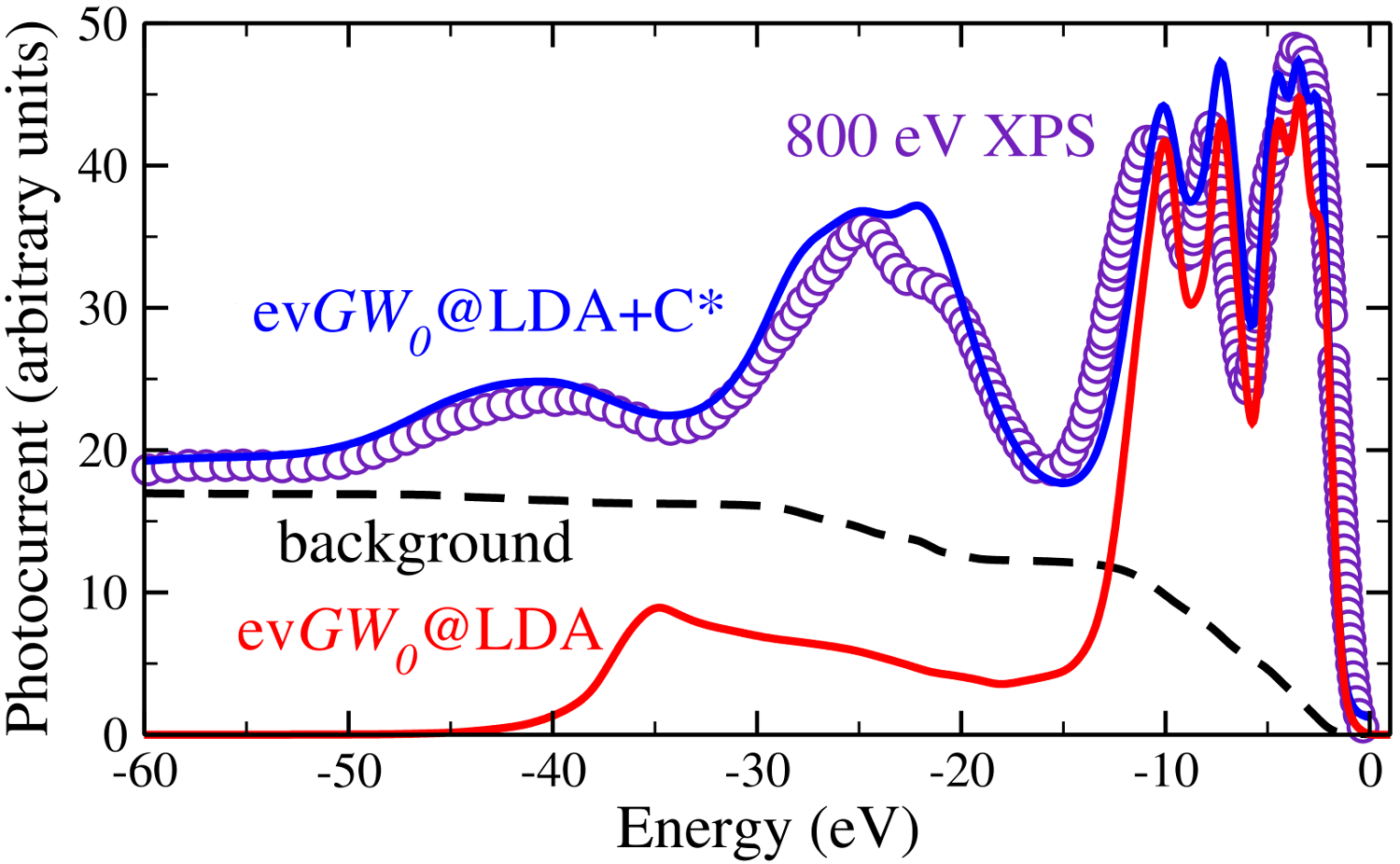}
	\caption{{\small \label{fig:Si_XPS_GW_C}
        	 X-ray photoemission spectrum with 800 eV incident energy compared to two calculated spectra. The red line shows the ev$GW_0$@LDA spectrum (see Section \ref{sec:scgw}), whereas the blue spectrum contains additional vertex corrections in form of a cumulant expansion (see Section \ref{sec:beyond}). The ev$GW_0$@LDA+C$^*$ spectrum contains the addition of the Shirley background (shown by the black dashed line) and loss effects of the outgoing photoelectron. Figure adapted from \cite{Guzzo/etal:2011}.  }}
\end{figure}
For the reconstruction of the band structure, e.g. with ARPES, the comparison between theory and experiment is hampered from the experimental side. In ARPES studies of crystalline materials, the emitted photons or electrons inevitably have to pass the surface of the crystal to reach the detector. Therefore, information about their transverse momentum $k_\bot$ is lost. This is because the crystal's translational symmetry is broken at the surface and only the in-plane momentum $k_{\parallel}$ is conserved. To reconstruct the three-dimensional band structure of the solid from experimental data, assumptions are often made about the dispersion of the final states \cite{Himpsel:1983,Plummer/Eberhardt:1982,Dose:1985,Smith:1988,Hora/Scheffler:1984}.  {\it Ab initio} calculations as described in this article can aid in the assignment of the measured peaks. Either way, some layer of interpretation between theoretical and experimental band structures is required.\par
Apart from quasiparticle excitations, a typical photoemission experiment  provides a rich variety of additional information.  In core-electron emission for instance, inelastic losses or multi-electron excitations such as shake-ups and  shake-offs lead to satellites in the spectrum. Satellites can also appear in the valence region. The outgoing photoelectron or the hole left behind can, for example, excite other quasiparticles like plasmons, phonons or magnons. This gives rise to additional peaks, the so-called plasmon or magnon satellites or phonon side bands, that are typically separated from the quasiparticle peak by multiples of the plasmon, magnon or phonon energy. The broad peak in Figure~\ref{fig:Si_XPS_GW_C}, which shows integrated spectra and therefore has no $\mathbf{k}$ dependence, near -40 eV is an example of a satellite. Satellites are collective effects that are not described within the quasiparticle picture.

\section{Hedin's $\bm{GW}$ equations}
	\label{sec:GW}
\label{sec:Hedin_GW}
Having introduced the general Green's function framework and quasiparticle concept, we are prepared to consider the concrete formalism for $GW$. $GW$ is an approximation to an exact set of coupled integro-differential equations called Hedin's equations \cite{Hedin:1965}, the full derivation of which can be found in Appendix \ref{sec:MB}.

We build up Hedin's equations from a perturbation theory perspective. We can conveniently represent the perturbation expansion for $G$ that we introduced in the time domain in Equation~\eqref{def:G} and in the energy domain in Equation~\eqref{G:Leh} with the Feynman diagram technique. Feynman diagrams are a pictorial way of representing many-body and Green's function theory. We cover the necessary basics in this section and refer the interested reader to an excellent book on Feynman diagrams \cite{Mattuck:1992}.

\begin{figure} 
       \includegraphics[width=0.7\columnwidth]{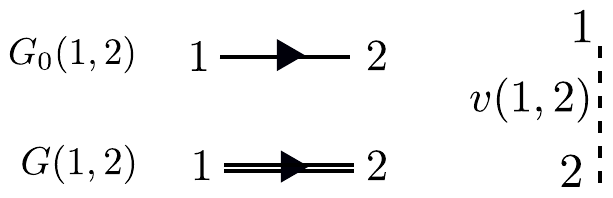}
	\caption{{\small \label{fig:feynman_parts}
        	 The most basic pieces of diagrammatic perturbation theory are $G_0$ and $v$. From these, all other quantities can be built. The interaction $v(1,2)$ is instantaneous. Therefore, the dashed line is perpendicular to the time axis. The arrows in $G_0$ and $G$ point in only one direction, but both time orderings are included.}}
\end{figure}

The perturbation expansion begins with the noninteracting Green's function, denoted $G_0$. $G_0$ is the probability amplitude for a noninteracting particle to propagate from one spacetime point to another. In the diagrammatic technique, $G_0$ is represented by a solid line with an arrow. The ends of $G_0$ indicate spacetime points. The generic notation $1=(\mathbf{r}_1,t_1,\sigma_1)$ refers to the spatial coordinate $\mathbf{r}_1$, time $t_1$, and spin variable $\sigma_1$. $G_0$, shown in Figure \ref{fig:feynman_parts}, is one of the basic building blocks for the perturbation expansion.

$G$ is the probability amplitude for the \textit{interacting} system that a particle creation at 2 is correlated with a particle annihilation at 1. The rules of quantum mechanics dictate that we must sum over all possible paths for the particle to move from $2$ to $1$ $-$ this generates the exact $G$, which is represented by a bold, or double, line with an arrow in the diagram language. Every possible process between $1$ and $2$ contributes a different amplitude to the total $G$. The different processes which connect $1$ to $2$ depend on interactions with other particles in the system at the times between $t_1$ and $t_2$. Without these interactions, the problem is already solved with $G_0$.

Recall from the definition of $G$ in Equation~\eqref{def:G} that $G$ contains \textit{two} time orderings. The second time ordering implies that the annihilation process may come before the creation. Remember that the field operators in Equation~\eqref{def:G} act on the interacting ground state. In the ground state, there is some charge for the annihilation operator $\hat{\psi}(\mathbf{r})$ to ``act" on, even without any preceding creation process, so that the reverse time ordering in $G$ makes sense. Feynman diagrams do not explicitly show both time orderings in $G$, but it is important to remember that $G$ and $G_0$ lines implicitly contain both time orderings.

\begin{figure} 
       \includegraphics[width=0.99\columnwidth]{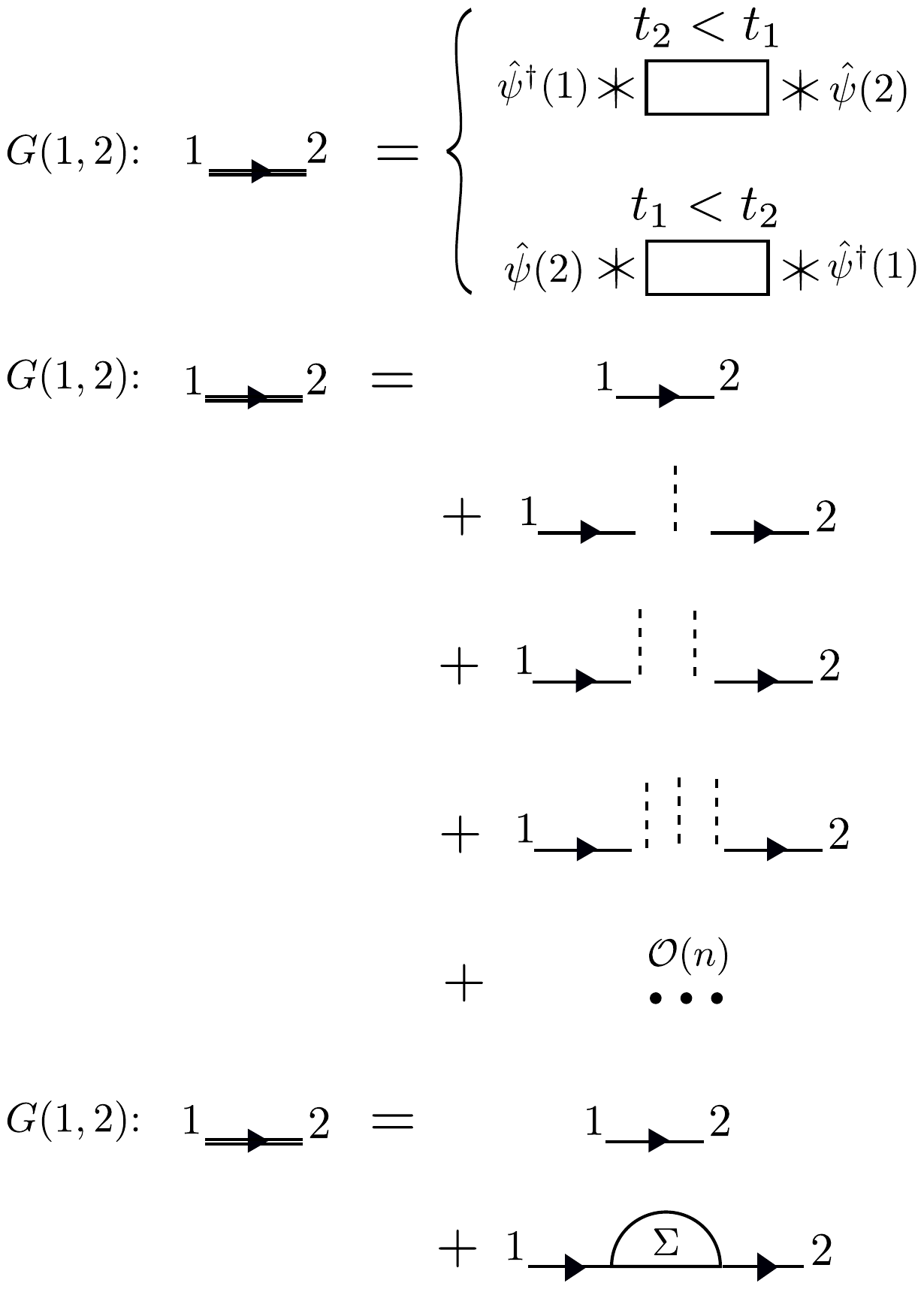}
	\caption{{\small \label{fig:self_energy}
        	  The exact $G$ contains amplitudes from all possible paths between $1$ and $2$. Amplitudes from all of these paths are represented by the rectangle placed between the field operators, the action of which is represented by $*$ symbols. These terms can be calculated order-by-order with perturbation theory. At a given order $n$, we must connect $n$ interaction lines at internal times in all possible $-$ and allowed $-$ ways. Concrete examples of diagrams are in Figure \ref{fig:diagrams}. All terms of the topology which can be inserted between two $G_0$ lines form the reducible self-energy. 
        	  }}
\end{figure}

\begin{figure} 
       \includegraphics[width=0.99\columnwidth]{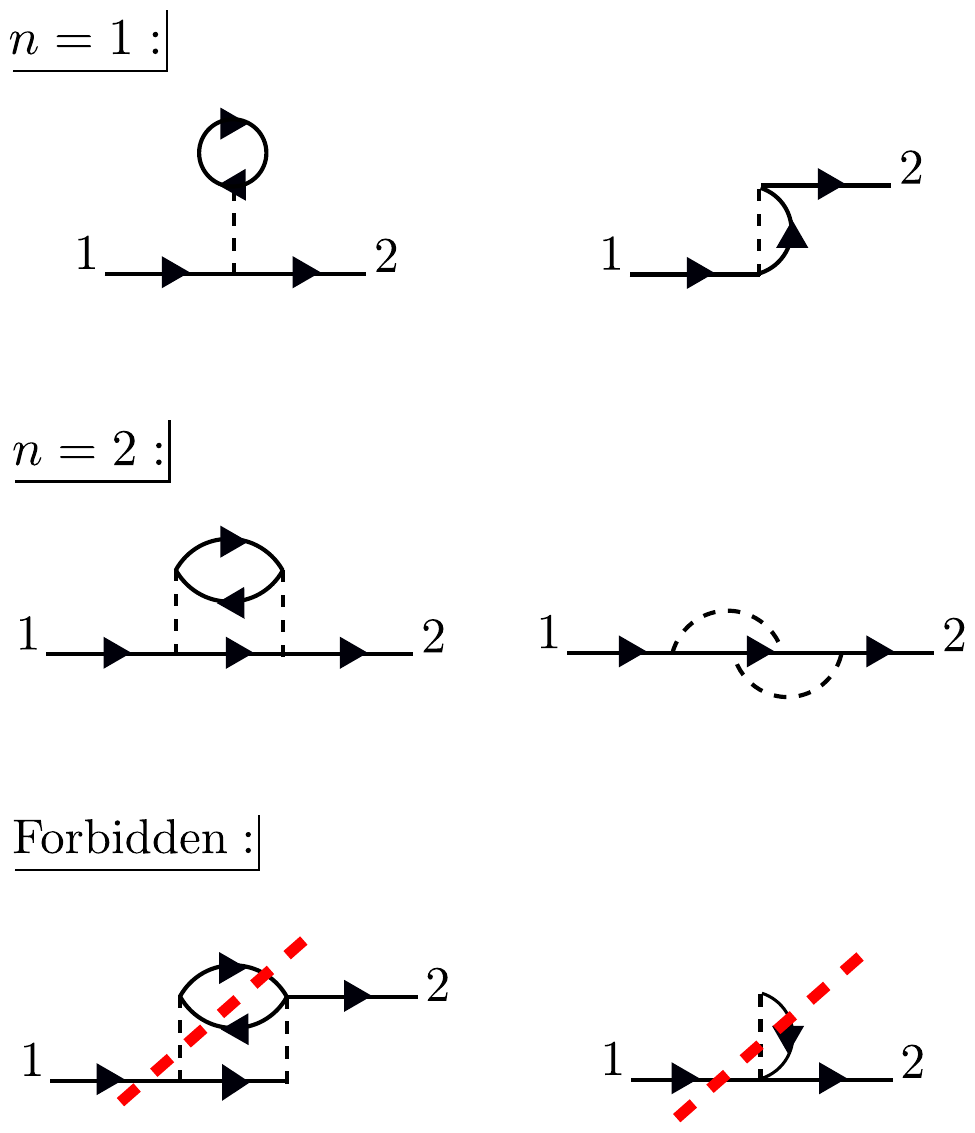}
	\caption{{\small \label{fig:diagrams}
        	 At first order, $n=1$, there are only two possible self-energy diagrams. These are the diagrams of the Hartree-Fock approximation, the direct electrostatic interaction (left) and exchange (right). Two possible $n=2$ diagrams are also shown (there are others). The bottom diagrams are forbidden because they do not have two $G_0$ lines at each end of the interaction lines. When drawing the diagrams, a certain degree of flexibility is allowed and they must be interpreted carefully. For example, the curved interaction lines above must still be treated as instantaneous in a calculation.}}
\end{figure}

The times between $t_1$ and $t_2$ are called internal times. We can add up all the processes contributing to $G$ in a certain order depending on the number of times the particle interacts with other particles in the system. These interactions occur only at internal times, and the number of internal interactions is the order of the diagram. To efficiently represent all these internal interactions, we use a dashed line to represent the interaction in the diagram. At a given order $n$, we construct all possible processes, or diagrams, which connect $n$ interaction lines with $G_0$ lines at $1$ and $2$. We connect all of the dashed lines appearing at internal times with additional $G_0$ lines. There is a very specific set of rules for how these arrangements can be done. Wick's theorem defines how to \textit{contract} these pieces \cite{Fetter/Walecka}. A simple principle is enough to demonstrate the idea, however. Because the Coulomb interaction is a two-body operator, each dashed line must have two $G_0$ lines at each end. To compute the exact $G$, the expansion must be taken to infinite order, $n \rightarrow \infty$, adding up all possible processes along the way. The process of building up all diagrams in the perturbation expansion is shown in Figure \ref{fig:self_energy}. A few example diagrams, as well as a couple of forbidden diagrams, are shown in Figure \ref{fig:diagrams}.

To go further with our analysis, we must dissect the internal structure of the diagrams and separate it into pieces. By considering the possible topologies of internal parts allowed by the contraction rules, we can group the parts into different categories. Here, the ``topology'' of the piece is determined by the number of $G_0$ lines and interactions at two different times, without considering the internal structure between the two chosen times. Those parts which have two $G_0$ lines sticking out are called a self-energy diagram. The full self-energy ($\Sigma$) can be inserted between two $G_0$ lines to form $G$ (one must also include the separate $G_0$ term). Topologies which connect two $G_0$ legs to an interaction line are labeled a vertex. Summing over all pieces with this topology creates the full vertex ($\Gamma$), which depends on three spacetime points. Finally, the diagram parts which end in two dashed lines sum up to the effective, or screened, interaction ($W$).

\begin{figure} 
       \includegraphics[width=0.79\columnwidth]{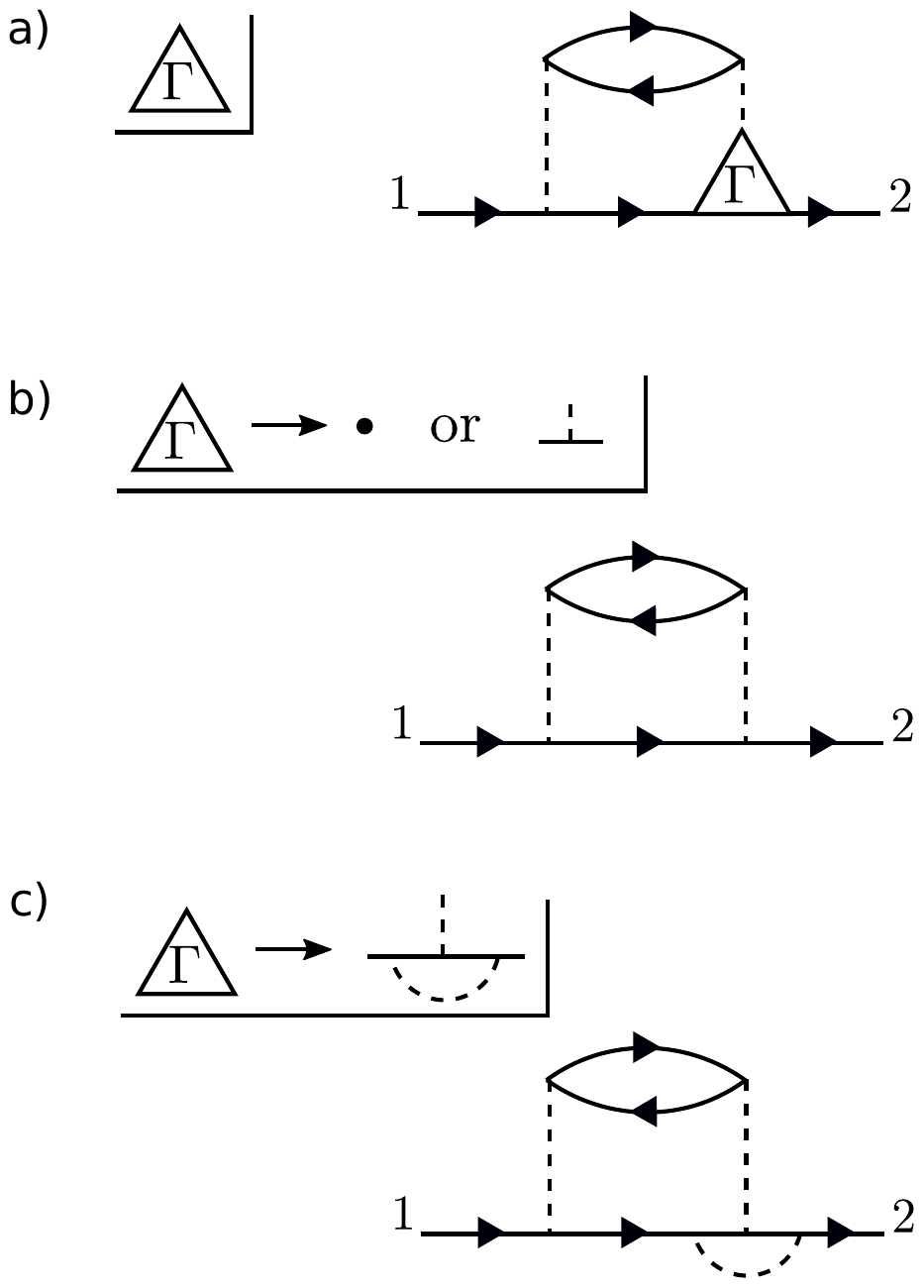}
	\caption{{\small \label{fig:vertex}
        	 The exact vertex $\Gamma$, shown in (a), can be replaced with approximations to simplify the calculation. The approximation in (b) is referred to as a ``single spacetime point'' because the vertex has no internal structure. In contrast, the vertex in (c) has internal structure. The diagram shown here is only an example to demonstrate the role of $\Gamma$ and does not correspond to the exact self-energy or the $GW$ self-energy. }}
\end{figure}

Conceptually, the vertex is the most difficult to understand. Figure~\ref{fig:vertex} demonstrates the effect of the vertex in a specific example. The diagram shown in Figure~\ref{fig:vertex}(a) is meant to contain the exact vertex, $\Gamma$. $\Gamma$ has three corners and can be inserted where two $G_0$ lines meet an interaction line. By simply letting these three pieces meet without any internal structure, we replace $\Gamma$ with a single spacetime point, as shown in Figure~\ref{fig:vertex}(b). Alternatively, we could allow the vertex to include the curved interaction line shown in Figure~\ref{fig:vertex}(c). In that case, the vertex has internal structure.

Hedin's equations can be interpreted as the self-consistent formulation of these topologically distinct building blocks. While Hedin followed a formal and systematic derivation, a heuristic motivation is to group all diagrams of a certain topology together and replace them with a single dressed, or renormalized, object with the same topology. A critical aspect of this replacement is their energy dependence. By replacing many diagrams of perturbation theory with a single object of the same shape, we reduce the number of objects to be computed. However, the information apparently missing due to the reduction in objects is encoded in the energy dependence of the dressed quantity. The final result, Hedin's equations (Appendix \ref{sec:MB}), are a compact and self-contained set of five integro-differential equations. Despite the reduction in the number of objects to be treated compared to the perturbation expansion, the functional differential equations coupling these pieces are extremely difficult to solve exactly.

Hedin recognized this difficulty and suggested the $GW$ approximation. As mentioned above, the vertex is the building block which has a single interaction connected to two $G_0$ lines. Unlike the other building blocks, the vertex depends on three spacetime points instead of two, making it the most difficult to compute. To simplify the theory, Hedin suggested replacing $\Gamma$ with a single spacetime point. In Hedin's equations, the exact self-energy is $\Sigma = i GW\Gamma$. With the replacement $\Gamma(1,2,3) = \delta(1,2)\delta(1,3)$, Hedin's approximation gives $\Sigma = i GW$, hence the name of the $GW$ approximation. In this approximation, Hedin's equations are
\begin{align}
\label{eq:dyson}
  G(1,2)&= G_0(1,2)   \\
  & +\int G_0(1,3)   \Sigma(3,4)  G(4,2)d(3,4)  \nonumber  \\
   \label{eq:gamma_GW}
  \Gamma(1,2,3)&=\delta(1,2)\delta(1,3) \\
\label{eq:P_GW}
  \chi_0(1,2)&=-i G(1,2)G(2,1) \\ 
\label{eq:W_GW}
  W(1,2)&=v(1,2)+\! \! \int \!  \! v(1,3)\chi_0(3,4)W(4,2)d(3,4)  \\ 
\label{eq:Sigma_GW}
  \Sigma(1,2)&=i G(1,2)W(1^+,2) \, 
\end{align}
where the Hartree potential is included in the solution for $G_0$\footnote{$1^+$ means the time $t_1$ evaluated at an infinitesimally later time. Such time infinitesimals appear in order to define the time-ordering for quantities meant to be evaluated in the instantaneous limit. The ground state density, for example, is time independent but must be written as $n(1) = -iG(1,1^+)$ so that the time-ordering makes sense.}. 

\begin{figure} 
       \includegraphics[width=\columnwidth]{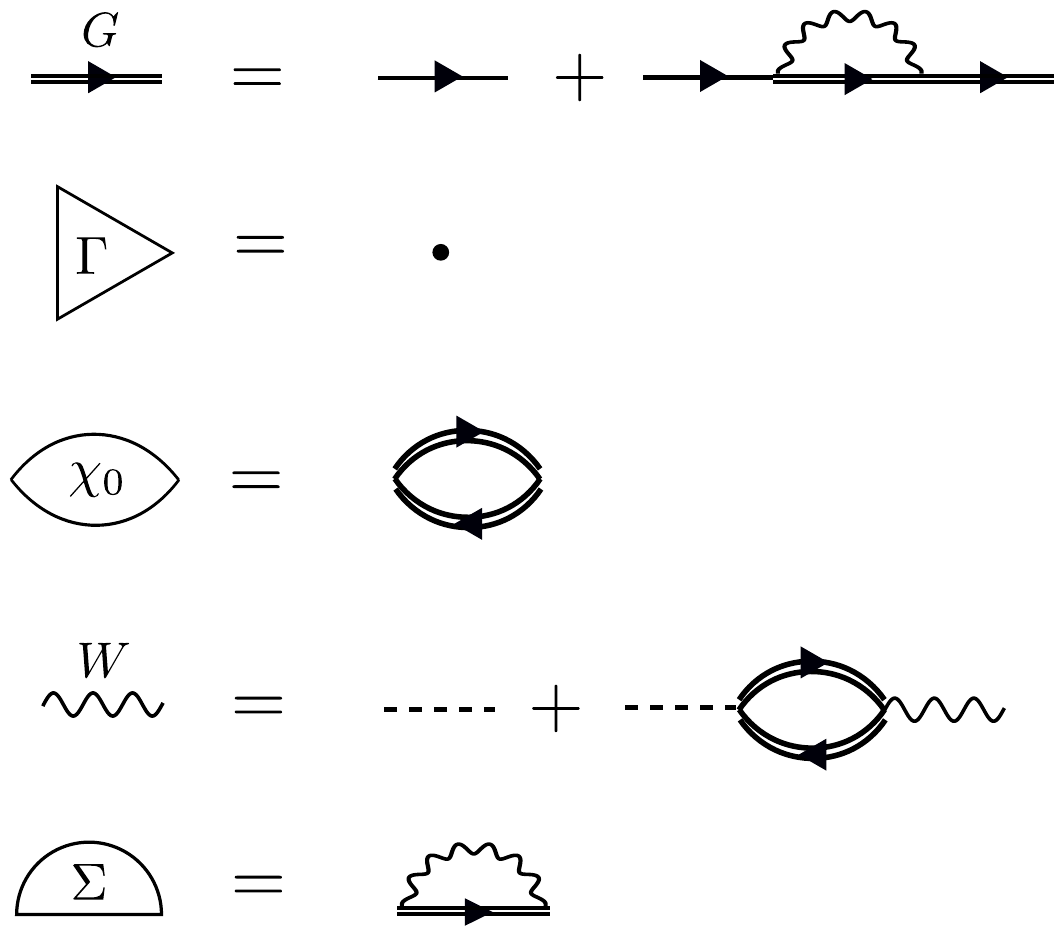}
	\caption{{\small \label{fig:gw_diagram}
        	 Diagrammatic representation of Equations \eqref{eq:dyson}-\eqref{eq:Sigma_GW}. The $GW$ approximation reduces the self-energy to a product of $G$ with $W$. The first equation (Dyson's equation) has a $G$ line on the left- and right-hand sides. This equation can be iterated, inserting $G_0 + G_0 \Sigma G$ in place of each $G$ on the RHS, forming the Dyson series. The same iterative procedure for $W$ forms its own Dyson series.}}
\end{figure}

These are the $GW$ equations, which are translated into the diagram language in Figure~\ref{fig:gw_diagram}. There is one final important point regarding the \textit{reducibility} of the quantities in Hedin's equations. $\Sigma$, $\chi_0$, and $\Gamma$ in Hedin's equations are all irreducible, which means that they cannot be broken into smaller pieces with the same topology. To generate the full, or reducible, quantity from its irreducible part, the irreducible component is iterated in a series similar to the perturbation expansion for $G$. Series of this type are commonly called Dyson series, and Dyson's equation refers to the equation for $G$ shown in Equation~\eqref{eq:dyson} ($W$ in Equation~\eqref{eq:W_GW} also obeys a Dyson series). Dyson's equation is of great importance in many-body physics, and we return to it in later sections in the context of self-consistent $GW$. It is common in the literature to use the same symbol for both reducible and irreducible components with the same topology, especially when discussing the self-energy. Almost always, the symbol refers to the quantity as it appears in Hedin's equations. When discussing or calculating the self-energy, this implies that one is interested in the irreducible self-energy.

In Hedin's equations, the screened Coulomb interaction $W$ plays a central role. Screening is based on the simple idea that charges in the system rearrange themselves to minimize their interaction. In polarizable materials, screening is significant, and the effective interaction is noticeably weaker than the bare one. $W$ is also dependent on frequency or a time difference. The frequency dependence of $W$ is critical to both the physics and the numerical implementation of $GW$. Even though the bare interaction is instantaneous, $W$ is time difference dependent because it is built from repeated bare interactions at different times. This series of bare interactions to form $W$ can be built by iterating the fourth line in Figure~\ref{fig:gw_diagram}. The underlying $G_0$ lines which connect the bare interactions in the $W$ expansion are themselves dependent on a time difference, so that if we vary the initial or final times the entire expansion changes magnitude. This series of repeated bare interactions connected by $G_0$ is the microscopic mechanism for the quasiparticle screening concept developed in Section~\ref{sec:qp_concept}. The frequency dependence of $W$ is what allows the system to relax and screen the quasiparticle. The $GW$ self-energy is similar to the bare exchange in Hartree-Fock theory, which can be written as the product of $G$ with $v$. Given the similarity between the $GW$ self-energy and bare exchange, $GW$ can be thought of as a dynamically screened version of Hartree-Fock.

The $GW$ equations should still be solved self-consistently since all four quantities are coupled to each other. As with other nonlinear equations, including the equations of mean-field theories like Kohn-Sham DFT or Hartree-Fock theory, the $GW$ equations can be solved by iteration. In principle, the prescription is clear. Start from a given $G_0$ and iterate Equations \eqref{eq:dyson}-\eqref{eq:Sigma_GW} to self-consistency (sc$GW$). However, remarkably few fully self-consistent solutions of the $GW$ equations have been performed in the last 50 years. The first calculations for the homogeneous electron gas (HEG) were reported at the turn of the previous century \cite{Holm/vonBarth:1998,Holm:1999,Garcia-Gonzalez/Godby:2001} and reported worse agreement with experiment on quasiparticle band widths and satellite structure compared to non-self-consistent calculations. They were quickly followed by calculations for real solids, like silicon and sodium \cite{Schoene/Eguiluz:1998,Ku/Eguiluz:2002}. Self-consistency was then dropped for several years because of its high computational expense and the success of non-self-consistent approximations. More recent sc$GW$ studies for atoms \cite{Delaney/Garcia-Gonzalez/Rubio/Rinke/Godby:2004,Stan/etal:2006,Stan/etal:2009}, molecules \cite{Rostgaard/Jacobsen/Thygesen:2010,Caruso/etal:2012,Marom/etal:2012,Caruso/etal:2013_H2,Caruso/etal:2013_tech}, conventional solids \cite{Kutepov/etal:2009,Kresse/etal:2018} and actinides \cite{Kutepov/etal:2012} have been reported. In practice, non-self-consistent calculations are much more common, and even self-consistent $GW$ calculations come in different types. sc$GW$ is discussed in more detail in Section \ref{sec:scgw}.


\section{The $\bm{G_0W_0}$ approach: Concept and implementation}
	\label{sec:g0w0inpractice}
\subsection{The $\bm{G_0W_0}$ equations}
\label{subsec:g0w0equations}

The lowest rung in the hierarchy of $GW$ approximations is the widely used $G_0W_0$ approach. Starting from a mean-field Green's function, $G_0W_0$ calculations correspond to the first iteration of Hedin's equations.
We denote the self-energy of such single-shot perturbation calculations $\Sigma_{0}$. Since we always refer to the single-shot self-energy in   Section~\ref{sec:g0w0inpractice}, we drop the label. Furthermore, we define the single-particle Hamiltonians 
\begin{eqnarray}
\hat{h}^0 &=& -\frac{1}{2}\nabla^2 + v_{\mathrm{ext}} \label{eq:h_0}   \\
\hat{h} &=&-\frac{1}{2}\nabla^2 + v_{\mathrm{ext}} + v_{\mathrm{H}} = \hat{h}^0 + v_{\mathrm{H}}  \label{h_hartreee} \\
\hat{h}^{\mathrm{MF}} &=& -\frac{1}{2}\nabla^2 + v_{\mathrm{ext}} + v_{\mathrm{H}} +  v^{\mathrm{MF}}_{\sigma} = \hat{h} + v^{\mathrm{MF}}_{\sigma} \label{eq:h_MF}
\end{eqnarray}
where $v_{\mathrm{ext}}$ is the external potential, $v_{\mathrm{H}}$ is the Hartree potential, and $v^{\mathrm{MF}}_{\sigma}$ is the mean-field (MF) exchange-correlation potential. The spin channel is denoted by $\sigma$. Possible mean-field Hamiltonians are the Kohn-Sham (KS) or Hartree-Fock (HF) Hamiltonians.\par
From Dyson's equation for $G$, one can derive an effective single-particle eigenvalue problem referred to as the quasiparticle (QP) equation. The solutions of the QP equation are then given by
\begin{align}
   \label{Eq:qp_eq}
    \hat{h}^{\mathrm{MF}} (\bfr)\psi_{s\sigma}(\bfr) - & \int d\bfrp v^{\rm MF}_{\sigma}(\bfr,\bfrp)\psi_{s\sigma}(\bfrp) +   \\
   &\int \! d\bfrp \Sigma_\sigma(\bfr,\bfrp,\epsilon_{s\sigma}) \psi_{s\sigma}(\bfrp) = 
      \epsilon_{s\sigma} \psi_{s\sigma}(\bfr)\nonumber \; .
\end{align}
The self-energy is calculated with a $G_0$ chosen to match the initial mean-field calculation based on $\hat{h}^{\mathrm{MF}}$. The solution of  Equation~{\eqref{Eq:qp_eq}} provides the QP energies $\{\epsilon_{s\sigma}\}$ and wave functions $\{\psi_{s\sigma}\}$.\par
Most commonly, the QP wave functions are approximated with the eigenfunctions $\{\phi_{s\sigma}^0\}$ of the mean-field Hamiltonian. Projecting each side of Equation~\eqref{Eq:qp_eq} onto 
$\phi_{s\sigma}^0$ yields a set of QP equations 
\begin{equation}
\label{Eq:qpe}
  \epsilon_{s\sigma}=\epsilon_{s\sigma}^0+\bracket{\phi_{s\sigma}^0}{\Sigma_\sigma\left(\epsilon_{s\sigma}\right)-v^{\rm MF}_{\sigma}}{\phi_{s\sigma}^0},
\end{equation}
where $\{\epsilon_{s\sigma}^0\}$ are the eigenvalues of $\hat{h}^{\mathrm{MF}}$. Solving 
Equation~\eqref{Eq:qpe}, the QP energy $\epsilon_{s\sigma}$ is obtained by correcting the mean-field eigenvalue $\epsilon_{s\sigma}^0$. \par
To solve Equation~\eqref{Eq:qpe}, we have to calculate the $G_0W_0$ self-energy $\Sigma_{\sigma}$,
\begin{equation}
\label{Eq:S=G0W0_rsp}
\begin{split}
 \Sigma_{\sigma}(\mathbf{r},\mathbf{r}',\omega)=& \\
\frac{i}{2\pi}\int & d\omega'  e^{i\omega'\eta} G_0^{\sigma}(\mathbf{r},\mathbf{r}',\omega+\omega')W_0(\mathbf{r},\mathbf{r}',\omega')
 \end{split}
\end{equation}
where $\omega$ is the frequency at which the self-energy is computed. Equation~\eqref{Eq:S=G0W0_rsp} is the frequency space version of Equation~\eqref{eq:Sigma_GW} for the $GW$ self-energy. The Green's function $G_0^{\sigma}$ stems from the aforementioned mean-field Hamiltonian and is given by
\begin{equation}
 G_0^{\sigma}(\mathbf{r},\mathbf{r}',\omega) = \sum_m\frac{\phi_{m\sigma}^0(\mathbf{r})\phi_{m\sigma}^{0*}(\mathbf{r}')}{\omega-\epsilon_{m\sigma}^0-i\eta 
\sgn(E_{\rm F}-\epsilon_{m\sigma}^0)}.
 \label{eq:greensfkt}
\end{equation}
 $W_0$ in Equation~\eqref{Eq:S=G0W0_rsp} is the screened Coulomb interaction in the random-phase approximation (RPA) 
\begin{equation}
 W_0(\mathbf{r},\mathbf{r}',\omega) = \int  d\mathbf{r}''\varepsilon^{-1}(\mathbf{r},\mathbf{r}'',\omega)v(\mathbf{r''},\mathbf{r}'),
 \label{eq:W0}
\end{equation}
with the bare Coulomb interaction $v(\mathbf{r},\mathbf{r'})=1/|\mathbf{r}-\mathbf{r}'|$ and the dynamical dielectric function $\varepsilon$. The latter is given by
\begin{equation}
 \varepsilon(\mathbf{r},\mathbf{r}',\omega) = \delta(\mathbf{r},\mathbf{r}') - \int d\mathbf{r}'' 
v(\mathbf{r},\mathbf{r}'')\chi_0(\mathbf{r}'',\mathbf{r}',\omega) .
  \label{eq:epsilon}
\end{equation}
In $G_0W_0$, the irreducible polarizability $\chi_0$, 
%
\begin{equation}
 \begin{split}
  \chi_0&(\mathbf{r},\mathbf{r}',\omega) = \\
   &-\frac{i}{2\pi}\sum_{\sigma}\int d\omega' G_0^{\sigma}(\bfr,\bfrp,\omega+\omega')G_0^{\sigma}(\bfrp,\bfr,\omega'),
 \end{split}
\end{equation}
simplifies to the Adler-Wiser expression \cite{Adler:1962,Wiser:1963}
%
\begin{equation}
 \begin{split}
\chi_0(\mathbf{r},\mathbf{r}',\omega) =& \\
  \sum_{\sigma}\sum_i^{\mathrm{occ}}&\sum_a^{\mathrm{virt}} 
\left\{\frac{\phi_{i\sigma}^{0*}(\mathbf{r})\phi_{a\sigma}^{0}(\mathbf{r})\phi_{a\sigma}^{0*}(\mathbf{r}')\phi_{i\sigma}^0(\mathbf{r}')}{\omega-(\epsilon_{a\sigma}^0-\epsilon_{i\sigma}^0)+i\eta}
\right.\\[0.5em]
&\qquad\left.-\frac{\phi_{i\sigma}^{0}(\mathbf{r})\phi_{a\sigma}^{0*}(\mathbf{r})\phi_{a\sigma}^{0}(\mathbf{r}')\phi_{i\sigma}^{0*}(\mathbf{r}')}{\omega+(\epsilon_{a\sigma}^0-\epsilon_{i\sigma}
^0)-i\eta 
}\right\} ,
  \label{eq:chi0}
 \end{split}
\end{equation}
where the index $i$ denotes an occupied and $a$ an unoccupied (also called virtual) single-particle orbital.\par
For numerical convenience as well as insight into the underlying physics, the $G_0W_0$ self-energy is often split into a correlation part $\Sigma^c_{\sigma}$,
\begin{equation}
  \Sigma_\sigma^c (\bfr,\bfrp,\omega)= \frac{i}{2\pi}\int d\omega' 
                           G_0^\sigma(\bfr,\bfrp,\omega+\omega')  
                         W_0^c(\bfr,\bfrp,\omega') \, ,
\label{eq:sigma_c}
\end{equation}
where $W^c_0$ is defined as
\begin{equation}
W^c_0(\bfr,\bfrp,\omega) = W_0(\bfr,\bfrp,\omega)-v(\bfr,\bfrp)\,,
\label{eq:wc}
\end{equation}
 and an exchange part
\begin{align}
   \Sigma_\sigma^x (\bfr,\bfrp) &= \frac{i}{2\pi}\int d\omega' 
                           e^{i\omega'\eta} G_0^\sigma(\bfr,\bfrp,\omega+\omega')  
                         v(\bfr,\bfrp) \label{eq:sigma_x_full}\\[5pt]
   &=-\sum^{\mathrm{occ}}_i \phi_{i\sigma}^0(\mathbf{r})\phi_{i\sigma}^{0*}(\mathbf{r'})v(\mathbf{r},\mathbf{r}') \label{eq:sigma_x}.
 \end{align}
Note that the exponential factor in Equation~\eqref{eq:sigma_x_full} is necessary to close the integration contour, whereas $W^c_0(\bfr,\bfrp,\omega)$ falls of quickly with increasing frequency and we can take the zero limit of $\eta$ before integrating.  
For a derivation of Equation~\eqref{eq:sigma_x} see Ref.~\cite{vanSetten/etal:2013}. We introduce the following notation for the $(s,s)$-diagonal matrix elements of the 
self-energy, 
\begin{equation}
 \Sigma_{s\sigma}(\omega) = \bracket{\phi_{s\sigma}^0}{\Sigma_{\sigma}(\omega)}{\phi_{s\sigma}^0}.
\end{equation}
The same notation is also used for matrix elements of the mean-field potential $v^{\rm MF}_{s\sigma} = \bracket{\phi_{s\sigma}^0}{v^{\rm MF}_{\sigma}}{\phi_{s\sigma}^0}$.\footnote{For simplicity, we will drop spin variables in the following parts of Section~\ref{sec:g0w0inpractice}.}\par
In the literature, $G_0$ is often referred to as the ``non-interacting'' Green's function. However, this is technically only correct if $G_0$ is constructed from an initial calculation based on $\hat{h}^0$. This is the definition of $G_0$ in formal many-body theory. However, often times in the theoretical literature, the Hartree potential is included in the $G_0$ solution and excluded from the self-energy. This is the case of starting the calculation from $\hat{h}$ in Equation~\eqref{h_hartreee}. For $G_0W_0$ in practice, we usually start from $\hat{h}^{\mathrm{MF}}$, which implies that we start from a mean-field Green's function rather than a non-interacting one. Conceptually, such a mean-field $G_0$ is closer to the interacting $G$ than the true $G_0$. This is precisely why the mean-field $G_0$ serves as such a useful starting point for $GW$ calculations -- it is closer to a self-consistent solution for $G$ than a true non-interacting $G_0$ is. When consulting literature references, keep in mind that $GW$ calculations most likely refer to a mean-field $G_0$.

\begin{figure} 
    \includegraphics[width=0.99\linewidth]{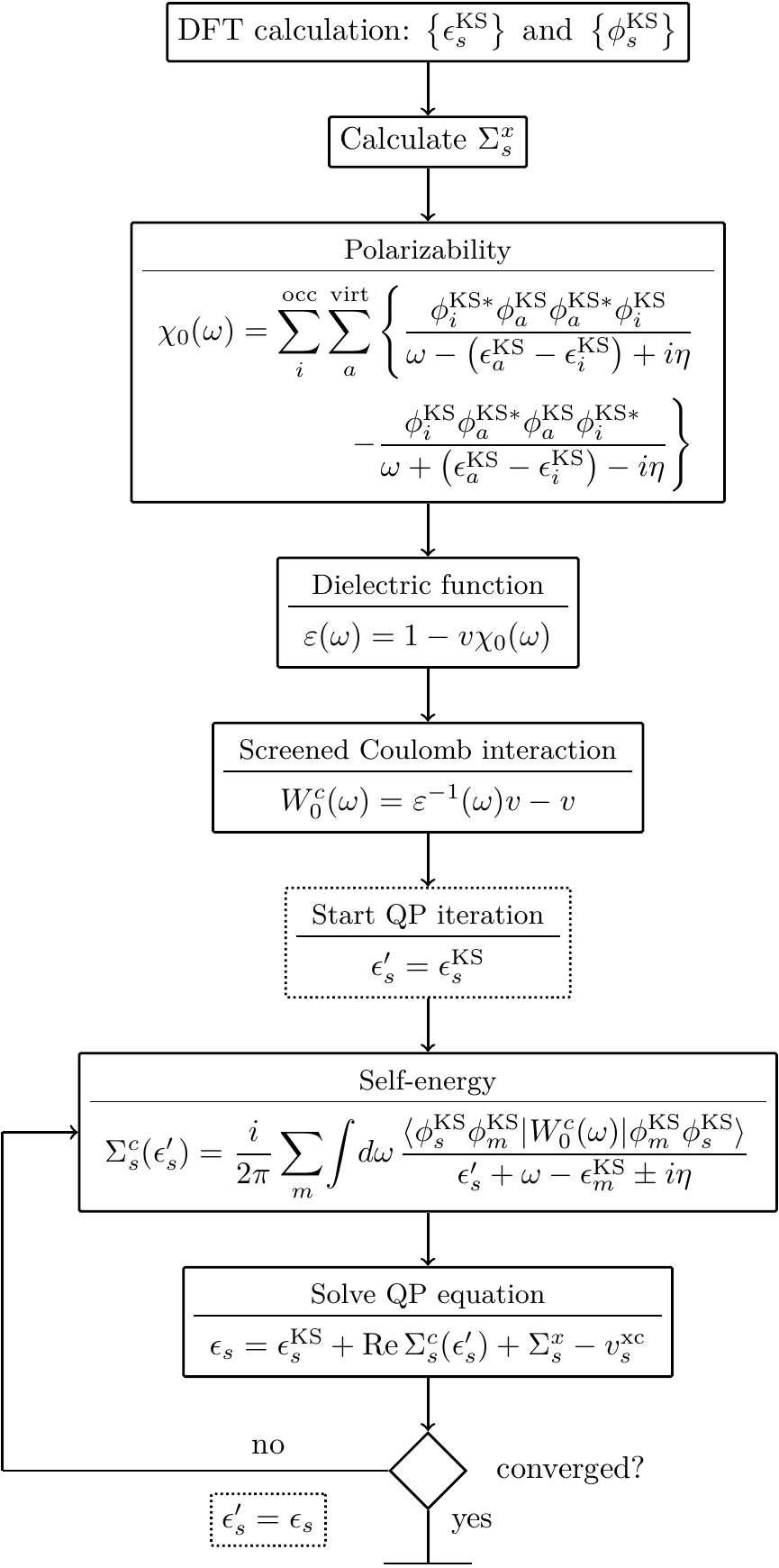}
	\caption{{\small Flowchart for a $G_0W_0$ calculation starting from a KS-DFT calculation. The KS energies $\{\epsilon_s^{\mathrm{KS}}\}$ and orbitals $\{\phi_s^{\mathrm{KS}}\}$ are used as 
input for the $G_0W_0$ calculation. For the full expressions of $\chi_0$, $\varepsilon$ and $W_0^c$ see Equations~\eqref{eq:W0}-\eqref{eq:chi0} and \eqref{eq:wc}. The spin 
has been omitted for simplicity. \label{fig:g0w0procedure}
        }}
\end{figure}
%

\subsection{Procedure}
\label{subsec:g0w0procedure}

 $G_0W_0$ calculations are usually performed on top of KS-DFT or HF calculations. A flowchart for a typical $G_0W_0$ calculation starting from a KS-DFT Hamiltonian is shown in Figure~\ref{fig:g0w0procedure}. Note that details of the flowchart depend on the treatment of the frequency dependence discussed in Section~\ref{subsec:g0w0freq}. Figure~\ref{fig:g0w0procedure} starts with the KS energies $\{\epsilon_s^{\rm KS}\}$, KS orbitals $\{\phi_s^{\rm KS}\}$ and the exchange-correlation potential $v^{\rm 
xc}$ from a DFT calculation. The exchange part of the self-energy $\Sigma_s^x$ is directly computed from the DFT orbitals. For the correlation term $\Sigma_s^c$, the frequency integral over $G_0$ and $W_0$ must be computed, see Equation~\eqref{eq:sigma_c}. If the integral is evaluated numerically, $W_0$ is computed for a set of frequencies $\{\omega\}$. The procedure to obtain $W_0$ is as follows: First, the irreducible polarizability $\chi_0$ (Equation~\eqref{eq:chi0}) is computed with the KS energies and orbitals. Second, $\chi_0$ is used to calculate the dielectric function $\varepsilon$ (Equation~\eqref{eq:epsilon}). From the inverse of $\varepsilon$ and the bare Coulomb interaction $v$, we finally obtain the correlation part of the screened Coulomb interaction, see Equations~\eqref{eq:W0} and \eqref{eq:wc}.\par
Since the QP energies appear on both sides of Equation~\eqref{Eq:qpe}, an iterative procedure is required. More precisely, the correlation term of the self-energy depends on $\epsilon_s$ and must be updated at each step. Note that only $G_0$ is a function of the QP energy, while $W_0^c$ depends solely on the frequencies of the integration grid. Therefore, $W_0^c$ can be pre-computed before entering the QP cycle, as displayed in Figure~\ref{fig:g0w0procedure}. \par
\begin{figure} 
    \includegraphics[width=0.99\linewidth]{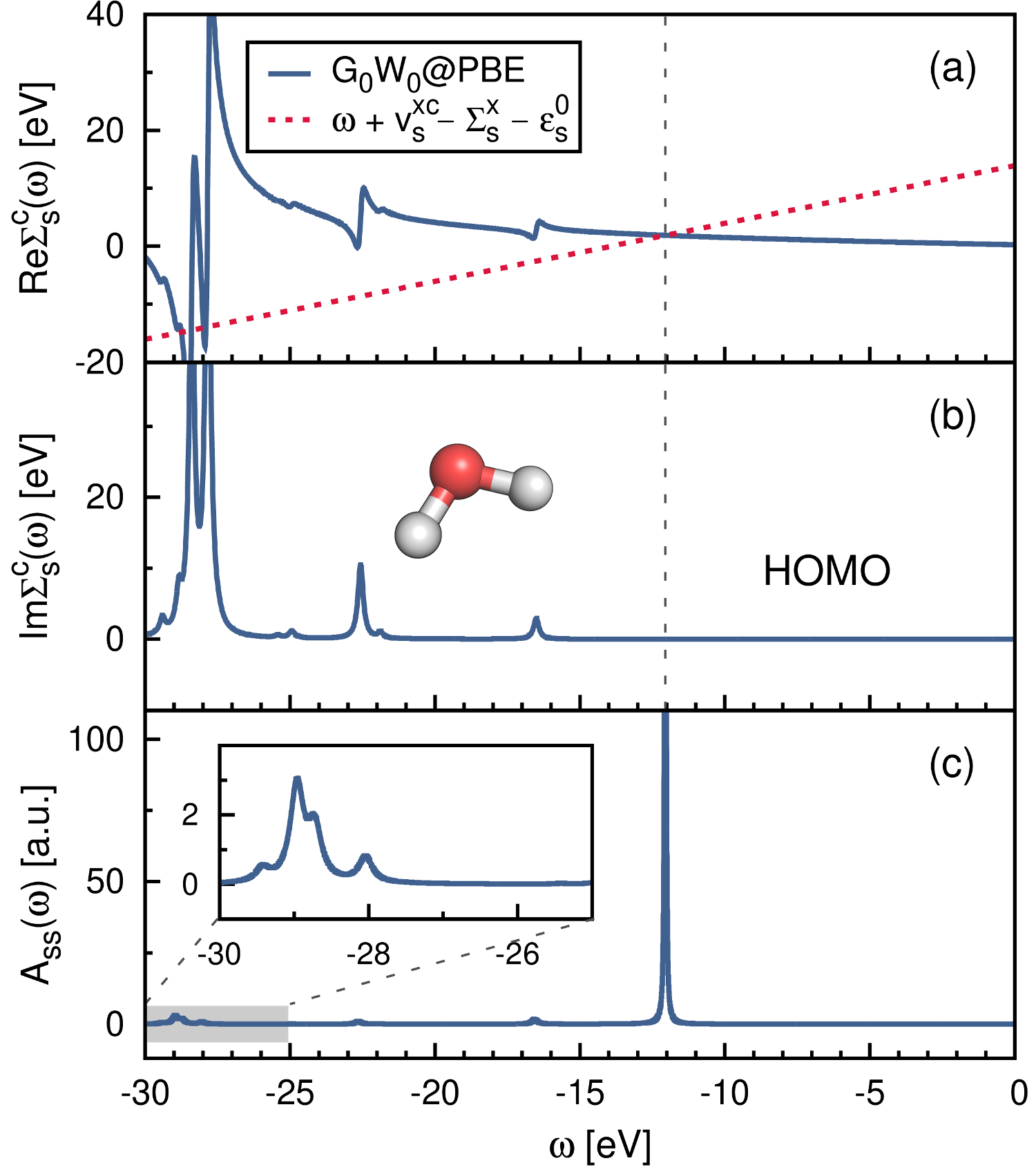}
	\caption{{\small (a) Real and (b) imaginary part of the self-energy $\Sigma^c(\omega)$. Displayed is the diagonal matrix element $\Sigma^c_s=\bracket{s}{\Sigma^c(\omega)}{s}$ for the HOMO 
of the water molecule. The gray-dashed line at $\approx -12.0$~eV indicates the QP solution $\epsilon_s$. (c) Spectral function $A_{ss}(\omega)$ computed from Equation~\eqref{eq:As}. The PBE functional is used as starting in combination with the cc-pV4Z basis set. Further computational details are given in Appendix~\ref{app:computational_details}.}\label{fig:sigma_homo}
        }
\end{figure}

 The correlation self-energy $\Sigma_s^c$ is a complex quantity. However, the imaginary part of $\Sigma_s^c$ is generally small for frequencies around the QP energy\footnote{While the imaginary part might be small, it is nonetheless important as its inverse is proportional to the lifetime of the state.}, see Figures~\ref{fig:sigma_homo}(a) and (b), where $\Sigma^c_s(\omega)$ is plotted for the highest occupied molecular orbital (HOMO) of the water molecule. To solve Equation~\eqref{Eq:qpe}, often only the real part of $\Sigma_s^c$ is used, which simplifies the matrix algebra to real operations and reduces the computational cost.\par
\begin{figure} 
    \includegraphics[width=0.99\linewidth]{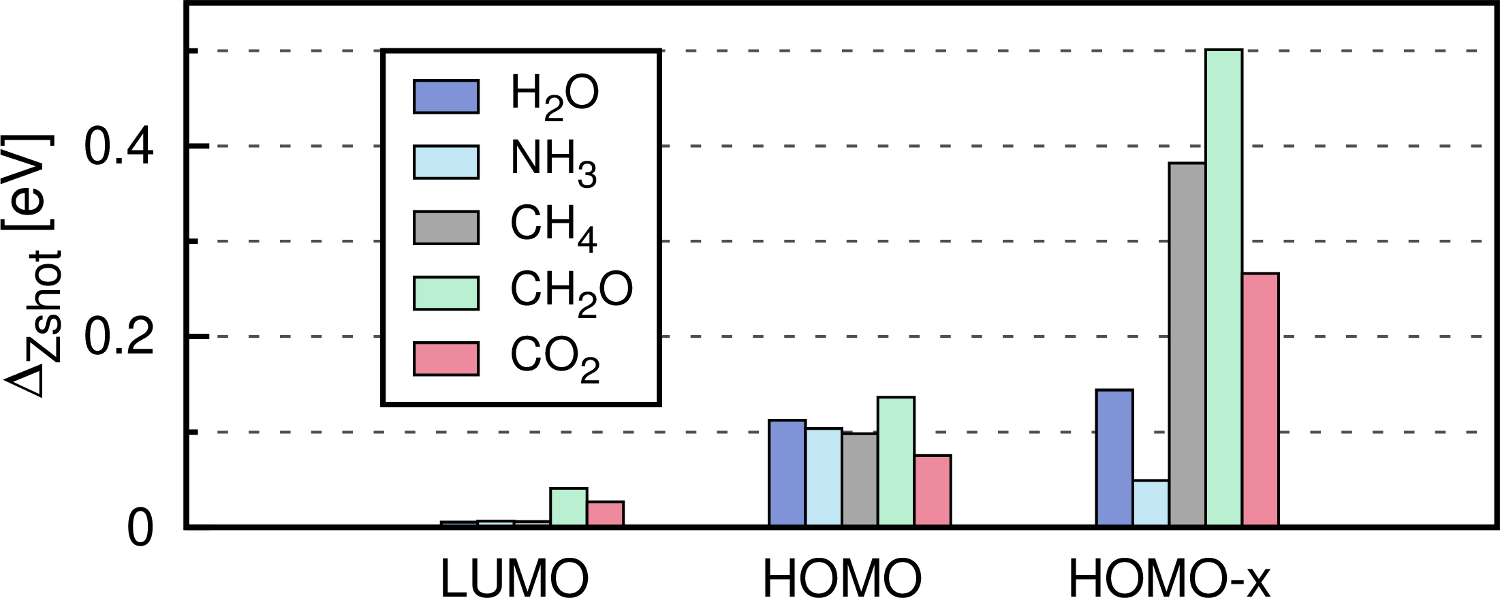}
	\caption{{\small  Error introduced by linearizing the QP equation, $\Delta_{\rm Zshot} = |\epsilon_s^{\rm iter}-\epsilon_s^{\rm Zshot}|$, where  $\epsilon_s^{\rm iter}$ has been obtained from the iterative procedure and $\epsilon_s^{\rm Zshot}$ from Equation~\eqref{Eq:qpe_lin}. ``HOMO-x" indicates deeper valence states. The PBE functional is used as starting point in combination with the cc-pV4Z basis set. Further computational details are given in Appendix~\ref{app:computational_details}. \label{fig:zshot}
        }}
\end{figure}

 A common technique to avoid the re-calculation of the self-energy at each iteration step of the QP cycle is the linearization of Equation~\eqref{Eq:qpe} 
\cite{Giantomassi/etal:2011,Liu2016,Wilhelm2016}. Assuming that the difference between QP and mean-field energies is relatively small, the matrix elements $\Sigma_s^c$ can be Taylor expanded to 
first-order around $\epsilon_s^0$:
 \begin{align}
\label{Eq:qpe_lin}
  \epsilon_s&=\epsilon_s^0+Z_s
              \bracket{\phi_s^0}{ \Sigma(\epsilon_s^0)-v^{\rm MF}}{\phi_s^0}\\
\label{Eq:Z_s}
	      Z_{s}&=\left[1-\frac{d}{d\omega}  \bracket{\phi_s^0}{ \Sigma(\omega)}{\phi_s^0}_{\omega=\epsilon_s^0}\right]^{-1} \, .
\end{align}
The self-energy matrix elements are now only required at the mean-field energies $\epsilon_s^0$. $Z_{s}$ is known as the QP renormalization factor, because it measures how much spectral weight the QP peak carries (see also Equation~\eqref{eq:simplified_SF} in Section~\ref{sec:qp_concept}). The QP solution (main peak) is characterized by large  $Z_s$ values, which lie around 0.7 to 0.8 for simple insulators, semiconductors and metals \cite{Aulbur/Jonsson/Wilkins:2000,Laasner2014} and around 0.9 for the molecules in Figure~\ref{fig:zshot}. Small $Z_s$ values indicate satellite features. \par
The linearization error depends on the state $s$. The deviation from the full iterative solution usually is in the range of 0.1~eV for the HOMO, as shown in Figure~\ref{fig:zshot} for a set of small molecules.  The Taylor expansion of the QP equation becomes less and less accurate for larger binding energies because the absolute distance between DFT eigenvalues and QP energies increases (i.e. the $G_0W_0$ correction increases). For the deeper valence states,  the linearization error is already as large as 0.5~eV (see Figure~\ref{fig:zshot}).\par
Another alternative to iterating the QP equation is to find a graphical solution. As shown in Figure~\ref{fig:sigma_homo}(a), the real part of the self-energy matrix elements is computed and plotted on a fine grid of real frequencies $\{\omega\}$ around the expected solution. All intersections of the straight line $\omega+v_s^{\rm{xc}}-\Sigma_s^x-\epsilon_s^0$ with $\operatorname{Re}\Sigma^c_s(\omega)$ are then solutions of Equation~\eqref{Eq:qpe}. The intersection with the largest spectral weight $Z_s$ is the QP solution and is characterized by a small slope of $\operatorname{Re}\Sigma^c_s$.\par
Another way to calculate the QP excitations is to compute the diagonal elements of the spectral function, $A_{ss}(\omega)$, for a set of frequencies as shown in Figure~\ref{fig:sigma_homo}(c). This is the most accurate procedure to obtain QP energies among the methods discussed here. $A_{ss}(\omega)$ is computed from the complex self-energy ($\Sigma^c_s=\operatorname{Re}\Sigma^c_s+i\operatorname{Im}\Sigma^c_s$),
\begin{align}
 A_{ss}(\omega) = {}& \frac{1}{\pi}\operatorname{Im}G_{ss}(\omega)\sgn(E_{\rm F}-\omega)\\
             = {}&\frac{1}{\pi} \operatorname{Im}\left[(\omega-\epsilon_s^0-(\Sigma^c_s(\omega)+\Sigma^x_s-v_s^{\rm xc}))^{-1}\right]\label{eq:As}\\ 
             {}&\times\sgn(E_{\rm F}-\omega)\nonumber,
\end{align} 
where we employed Equation~\eqref{Eq:def_A}, the Dyson equation, $G=G_0+G_0\Sigma G$, and used only the diagonal matrix elements of $\Sigma$. Figure~\ref{fig:sigma_homo}(c) confirms that the solution at around $\approx-12.0$~eV is the main solution. The spectral weight of the other solutions, e.g, the satellite peaks in the frequency range -30 to -25~eV, is indeed very small.  \par
The aforementioned iterative procedure is computationally far more efficient than the graphical solution or the calculation of $A_{ss}(\omega)$. The number of required QP cycles $N_{\rm QP}$ typically ranges between 5 to 15 and the self-energy only has to be computed for  $N_{\rm QP}$ many frequencies. However, the spectral function takes also the imaginary part of the self-energy into account. This is essential for the accurate computation of 
 satellite features in the $GW$ spectrum \cite{Reining2017,Zhou/etal:2015}. Satellites  fall usually in a region where $\operatorname{Re}\Sigma^c_s$ has poles, as demonstrated in Figure~\ref{fig:sigma_homo}(a). In these regions, the imaginary part 
$\operatorname{Im}\Sigma^c_s$ exhibits complementary peaks and is non-zero (Kramers-Kronig relation), see Figure~\ref{fig:sigma_homo}(b). Note that the graphical solution indicates the expected range of the satellite peaks, but does not predict their positions accurately because the imaginary part is omitted.\par

\subsection{Frequency treatment}
\label{subsec:g0w0freq}
The frequency integration in Equation~\eqref{Eq:S=G0W0_rsp} is one of the major difficulties in a $G_0W_0$ calculation since both functions that are integrated, $G_0$ and $W_0$, have 
poles infinitesimally  above and below the real frequency axis.  In principle, a numerical integration of Equation~\eqref{Eq:S=G0W0_rsp} is possible, but potentially unstable since the integrand 
needs to be evaluated in regions in which it is ill-behaved. However, a toolbox of approximate and exact alternatives is available. The most frequently used methods are summarized in the following.
%
\subsubsection{Plasmon-pole models}
\label{subsec:plasmon-pole}
The simplest way to calculate the frequency integral is to approximate the frequency dependence of the dielectric function $\varepsilon$ and thus the screened Coulomb interaction $W_0$ by a plasmon pole model (PPM) \cite{Hybertsen/Louie:1986}. The PPM approximation takes advantage of the fact that $\varepsilon^{-1}$
is usually dominated by a pole at the plasma frequency $\omega_p$ \cite{Hybertsen/Louie:1986}. This pole corresponds to a collective charge-neutral excitation (a plasmon) in the material. Assuming that only one plasmon branch is excited, the shape of $\varepsilon$ can be modeled by a single-pole function 
\begin{equation}
 \operatorname{Re}\varepsilon^{-1}(\omega) \approx 1 + \frac{\Omega^2}{\omega^2-\tilde{\omega}^2},
 \label{eq:plasmon_pole}
\end{equation}
where $\Omega$ and $\tilde{\omega}$ are two parameters in the model, whose squares are proportional to $\omega_p^2$, see Ref.~\cite{Giantomassi/etal:2011}. 
$\varepsilon$, $\Omega$ and $\tilde{\omega}$ are matrices typically expressed in a plane wave basis because PPMs are mostly used for periodic systems. Note that Equation~\eqref{eq:plasmon_pole} holds for each matrix element and that we take the square of the matrix elements in Equation~\eqref{eq:plasmon_pole} and not the square of the matrix itself. Using a model function for $\varepsilon^{-1}$, the expression for $W_0$ is greatly simplified resulting in an analytic expression for the self-energy, see Ref.~\cite{Deslippe/etal:2012}.\par
The two parameters, $\Omega$ and $\tilde{\omega}$, can be determined in several ways leading to different flavors of the PPM approximation \cite{Giantomassi/etal:2011}. The most common PPMs are the Hybertsen-Louie (HL) \cite{Hybertsen/Louie:1986} and the Godby-Needs (GN) \cite{Godby1989} model. The parameters in the HL model are obtained by requiring that the PPM reproduces the value of $\varepsilon^{-1}$ in the static limit ($\omega =0$) and that the so-called $f$-sum rule is fulfilled. The $f$-sum rule  is a generalized frequency sum rule relating the imaginary part of $\varepsilon^{-1}$ to $\omega_p$ and the electron density in reciprocal space \cite{Johnson1974}. In the HL model, the low and high real frequency limits are exact and $\varepsilon$ has to be calculated explicitly only at $\omega = 0$. The parameters of the GN PPM are determined by calculating $\varepsilon$ at $\omega = 0$ and an imaginary frequency point $i\omega_p'$, where $\omega_p'$ is typically chosen to be close to the plasma frequency $\omega_p$. The latter corresponds to the energy of the plasmon peak in the electron energy loss spectra (EELS) and can be obtained from experiment. Alternatively, $\omega_p'$ follows from the average electronic density $\rho_0$ per volume, $w_p'=\sqrt{4\pi\rho_0}$ \cite{Giantomassi/etal:2011}.\par
A comparison between different PPMs, namely the HL, GN, Linden-Horsch \cite{Linden1988}, and Engel-Farid \cite{Engel1993} model, can be found in Refs.~\cite{Larson2013,Stankovski/etal:2011}. There it was shown that the GN model best reproduces the inverse dielectric function and the corresponding QP energies of reference calculations with an exact full-frequency treatment, such as the contour deformation discussed in Section~\ref{subsubsec:CD}. However, even the accuracy of the GN model decreases further away from the Fermi energy, i.e. for low-lying occupied and high-lying unoccupied states \cite{Laasner2014,Cazzaniga:2012}.\par
While PPMs made the first $G_0W_0$ calculations tractable 
\cite{Hybertsen/Louie:1985,Hybertsen/Louie:1986}, full frequency calculations are now the norm, because the effects of the plasmon-pole approximation on the overall accuracy of the calculation are often hard to judge \cite{Stankovski/etal:2011,Miglio/etal:2012}. Moreover, the imaginary part of the self-energy becomes non-zero only at the plasmon poles, which implies that QP lifetimes cannot  properly be calculated with PPMs, see Equation~\eqref{eq:lifetime} and Section~\ref{sec:lifetimes}.
However, PPMs are still used in large scale $G_0W_0$ calculations 
\cite{Deslippe/etal:2012}, for example for  solids \cite{Jain/Chelikowsky/Louie:2011,Reyes-Lillo2016}, surfaces \cite{Loeser/etal:2012}, 2D materials \cite{Dvorak2015, 
Qiu2016,Drueppel2018}, graphene 
nano ribbons \cite{Talirz2017,Wang2016}  or polymers \cite{Hogan/etal:2013,Lueder2016}.\par
The application of PPMs to molecules is conceptually less straightforward, because the dominant charge neutral excitations in molecules are not necessarily collective. This raises the question of how to define the plasma frequency $\omega_p$ of a molecule.  Nevertheless, PPMs have also been used in benchmark studies for molecules, where mean absolute deviations of 0.5~eV from accurate frequency integration methods were reported \cite{Setten2015}.\par
The plasmon-pole model can be extended to an arbitrary number of poles, as proposed by Rehr and coworkers \cite{Soininen2003,Soininen2005,Kas2007}. If many frequencies are required to determine the parameters in the model, the computational cost for the evaluation of $\Sigma$ is not necessarily reduced compared to full-frequency methods. However, multi-pole models are also well-defined for finite systems since the existence of a distinct plasmon peak is no longer an inherent assumption of the model \cite{Kas2007}.

\subsubsection{Contour deformation}
\label{subsubsec:CD}

\begin{figure} 
    \includegraphics[width=0.95\linewidth]{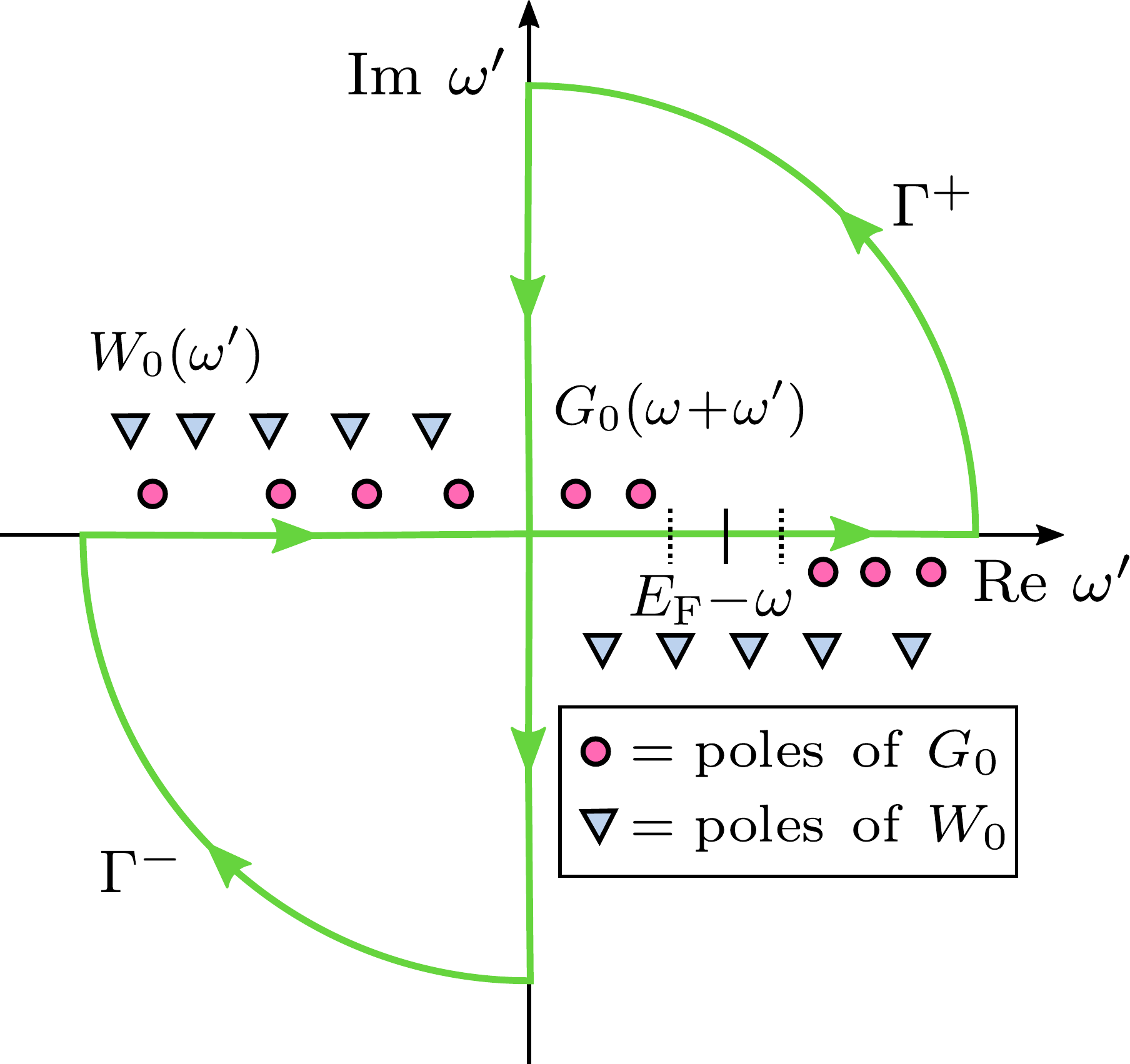}
	\caption{{\small  \label{fig:contours}
      Contour deformation technique: Integration paths in the complex plane to evaluate $\Sigma^c(\omega)$. $\Gamma^+$ and $\Gamma^{-}$ are the integration contours, which are chosen such that the 
poles of $G_0$, but not the poles of $W_0$ are enclosed. $\Gamma^+$ encircles the upper right and $\Gamma^-$ the lower left part of the complex plane. $\omega'$ denotes frequencies of the integration grid and $\omega$ the frequency at which $\Sigma^c$ is calculated. }}
\end{figure}
The contour deformation (CD) approach is a widely used, full-frequency integration technique for the calculation of $\Sigma^c(\omega)$ 
\cite{Godby/Schlueter/Sham:1988,Lebegue2003,Gonze2009,Govoni/etal:2015,Golze2018,Kotani/etal_2:2007,Blase/Attaccalite/Olevano:2011}. In the CD approach, the real-frequency integration is carried out by using the contour integral, see Figure~\ref{fig:contours}. 
By extending the integrand to the complex plane, the numerically unstable integration along the real-frequency axis, where the poles of $G_0$ and $W_0$ are located, is avoided.\par
The integral along the contours shown in Figure~\ref{fig:contours} has four terms: an integral along the real (Re) and the imaginary axis (Im) and along the arcs.
\begin{equation}
 \begin{split}
    \frac{i}{2\pi}&\oint d\omega' G_0(\omega+\omega')W_0(\omega') \\ &= \int_{\rm Re}\cdots~ + \int_{\rm Im} \cdots~+ \int_{\rm arc\,\Gamma^+} \cdots~+ \int_{\rm arc\,\Gamma^-}\cdots
 \end{split}
 \label{eq:contourconcept}
\end{equation}
%
The contour integral is evaluated by taking the contours to infinity, which implies that the radius of the arcs is infinite. For infinitely large $\omega'$, $G_0(\omega+\omega')W_0(\omega')$ vanishes and the integral along the arcs of contours $\Gamma^+$ and $\Gamma^-$ is zero. Therefore, we can compute the real-frequency integral by subtracting the imaginary-frequency integral from the contour integral.  After rearranging Equation~\eqref{eq:contourconcept} and using Equation~\eqref{Eq:S=G0W0_rsp}, we obtain
\begin{equation}
 \begin{split}
  \Sigma(\mathbf{r},\mathbf{r}',\omega) ={} & \frac{i}{2\pi}\oint d\omega' G_0(\mathbf{r},\mathbf{r}',\omega+\omega') 
W_0(\mathbf{r},\mathbf{r}',\omega') \\
                   -&\frac{1}{2\pi} \int_{-\infty}^{\infty} d\omega' G_0(\mathbf{r},\mathbf{r}',\omega+i\omega') W_0(\mathbf{r},\mathbf{r}',i\omega'),
 \end{split}
 \label{eq:Sigma_path_integral}
\end{equation}
where the second term is the integral along the imaginary axis.\par
The contours $\Gamma^+$ and $\Gamma^{-}$ are chosen such that only the poles of $G_0$ fall into $D_{\Gamma^{+}}$ and $D_{\Gamma^{-}}$, which denote the subsets of the complex plane encircled by $\Gamma^{+}$ and $\Gamma^{-}$, respectively. The location of the poles of $G_0(\omega+\omega')$ depends on the frequency $\omega$ at which the self-energy is computed. Recalling  Equation~\eqref{eq:greensfkt}, the poles of $G_0$ lie at the complex frequencies 
 \begin{equation}
  \omega_{m}' = \epsilon_{m}^0 -\omega + i\eta~\mathrm{sgn}(E_{\rm F}-\epsilon_{m}^0).
 \label{eq:poles_G}
\end{equation}
For $\omega < E_{\rm F}$, these poles can enter only $D_{\Gamma^{+}}$ and must arise from occupied states. Our example in Figure~\ref{fig:contours} displays a case were $\omega < 
\epsilon_{\rm ({HOMO} - 1)}$. Two poles, namely $[\epsilon_{(\rm HOMO)}-\omega]$ and $[\epsilon_{(\rm {HOMO}-1)}-\omega]$, fall into $D_{\Gamma^{+}}$. For an even smaller $\omega$, more poles from deeper occupied states will shift into $D_{\Gamma^{+}}$. Conversely, for $\omega > E_{\rm{F}}$, the poles from the unoccupied states will 
enter $D_{\Gamma^{-}}$.\par
We can now calculate the residues of the poles that are in $D_{\Gamma^{+}}$ or $D_{\Gamma^{-}}$. Employing the residue theorem, the contour integral is then replaced by a sum over these residues:
\begin{equation}
 \begin{split}
        \frac{i}{2\pi}&\oint d\omega' G_0(\omega+\omega')W_0(\omega') \\   = {}& -\sum_{\omega_m^{'}\in D_{\Gamma^+}} 
\text{Res}\left\{G_0(\omega+\omega')W_0(\omega'),\omega_m^{'}\right\} \\
                   {}&   +\sum_{\omega_m^{'}\in D_{\Gamma^-}} \text{Res}\left\{G_0(\omega+\omega')W_0(\omega'),\omega_m^{'}\right\}.
 \end{split}
\end{equation}
The integral along the imaginary frequency axis is smooth \cite{GW_space-time_method:1998,Giantomassi/etal:2011} and the integration is performed numerically. The size of the frequency grid for the numerical integration needs to be carefully converged. For more details and a derivation of the final CD equations see, e.g., \cite{Golze2018} or \cite{Govoni/etal:2015}.

\subsubsection{Analytic Continuation}
\label{subsec:ac}
\begin{figure} 
    \includegraphics[width=0.99\linewidth]{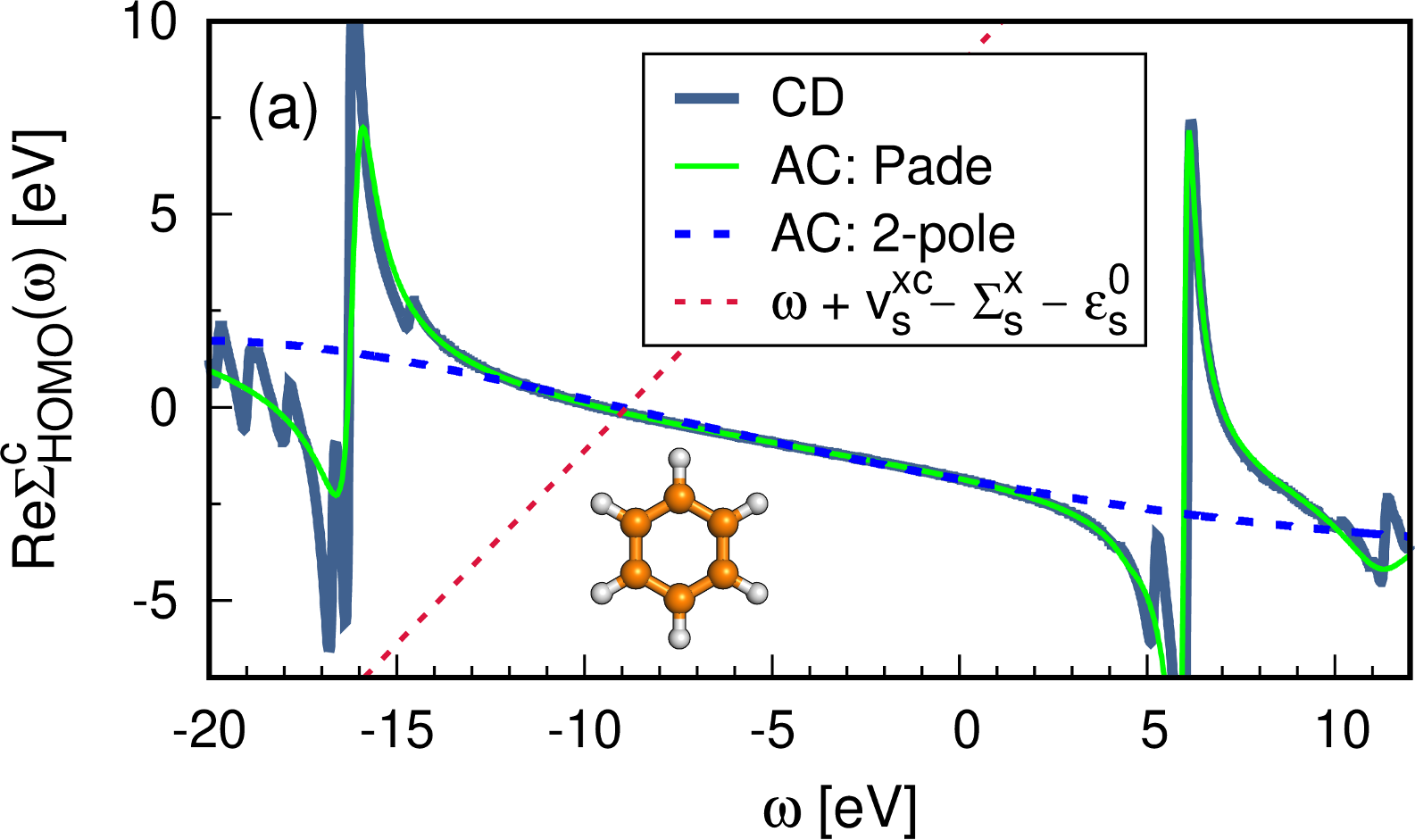}
	\caption{{\small \label{fig:freq_treat}
	$G_0W_0$@PBE self-energy matrix elements for the HOMO of benzene obtained with different frequency integration techniques: contour 
deformation (CD) and analytic continuation (AC) using the Pad\'{e} model with 128 parameters and the 2-pole model. See Appendix~\ref{app:computational_details} for further computational details.   }}
\end{figure}
%
Analytic continuation (AC) from the imaginary to the real frequency axis is another method in our toolbox that enables an integration over the full-frequency range. The AC technique exploits the fact that the integral of the self-energy along the imaginary frequency axis,  
\begin{equation}
 \begin{split}
 \Sigma^c(\mathbf{r},\mathbf{r}', i\omega) = -\frac{1}{2\pi} \int_{-\infty}^{\infty} d\omega' G_0&(\mathbf{r},\mathbf{r}',i\omega+i\omega')\\
&\times W_0(\mathbf{r},\mathbf{r}',i\omega').
 \end{split}
\end{equation}
is smooth and easy to evaluate, unlike the integral along the real-frequency axis. However, the QP energies and spectral functions are measured for real frequencies. To return from the imaginary to the real frequency axis, the procedure is as follows: The self-energy is first calculated for a set of imaginary frequencies $\{i\omega\}$ and then continued to the real-frequency axis by fitting the matrix elements $\Sigma^c_s(i\omega)$ to a multipole model. A common analytic form is, e.g., the so-called 2-pole-model \cite{Rojas/Godby/Needs:1995,GW_space-time_method:1998}
\begin{equation}
  \Sigma^c_s(i\omega) \approx \sum_{j=1}^2 \frac{a_{s,j}}{i\omega+b_{s,j}} + a_{s,0}.
  \label{Eq:two-pole}
\end{equation}
which has been widely used for $G_0W_0$ calculations of materials \cite{GW_space-time_method:1998,Friedrich2010,Pham/etal:2013} and molecules \cite{KeSanhuang:2011,Setten2015,Wilhelm2016}. The unknown complex coefficient $a_{s,j}$ and $b_{s,j}$ are determined by a nonlinear least-squares fit. From the identity theorem of complex analysis we know that the analytic form of a complex differentiable function on the real and imaginary axis are identical. Therefore, we can finally calculate the self-energy in the real-frequency domain by replacing $i\omega$ with $\omega$ in Equation~\eqref{Eq:two-pole}.\par
An alternative multipole model function is the popular Pad\'{e} approximant \cite{Liu2016,Setten2015,Wilhelm2018}, which is more flexible but contains more parameters. In the Pad\'{e} approximation, the complex fitting coefficients are not obtained by a nonlinear least-squares fit, but recursively from the matrix elements $\Sigma^c_s$ and the imaginary frequencies $\{i\omega\}$ \cite{Vidberg1977}. \par
In Figure~\ref{fig:freq_treat} we compare the real self-energy matrix elements $\operatorname{Re}\Sigma^c_{s}$ obtained from the AC approach to an implementation on the real-frequency axis such as the CD method. The Pad\'{e} approximation reproduces the self-energy exactly in the frequency range around the QP energy of the HOMO. The deviation in the HOMO-QP energy is smaller than 10$^{-4}$~eV with respect to the CD results. By using a Pad\'{e} approximant with a large number of parameters, even some features of the pole structure at higher and lower frequencies are reproduced, as shown in Figure~\ref{fig:freq_treat}. The 2-pole model, on the contrary, is significantly less accurate and yields an error of around 0.1~eV in the first ionization potential. For a comprehensive comparison between the Pad\'{e} and 2-pole model see Ref.~\cite{Setten2015}.\par
The reliability of the AC approach is limited to valence excitations, because the self-energy structure of deeper states shows poles closer to the QP solution. Our recent work \cite{Golze2018} showed that the AC technique fails drastically to describe the complicated features of the self-energy for core states resulting in errors of 10-20~eV for the core-level binding energies. Furthermore, satellite features are difficult to obtain. As discussed in Section~\ref{subsec:g0w0procedure}, satellites lie in regions, in which  $\operatorname{Re}\Sigma^c_s$ has poles. As evident from Figure~\ref{fig:freq_treat}, these poles can only partly be reproduced by the AC.\par
The convergence parameters for the AC approach are the number of frequency points $\{i\omega\}$, for which the self-energy is computed, and the size of the frequency grid for the numerical integration over $\omega'$. In practice, the same grid is often employed for $\{i\omega\}$ and $\{i\omega'\}$ \cite{KeSanhuang:2011, Wilhelm2016}.

%
\subsubsection{Fully Analytic Approach}
\label{subsec:fully_analytic}
The integral in Equation~\eqref{Eq:S=G0W0_rsp} can be carried out fully analytically. In this case, the Adler-Wiser sum-over-states representation of the polarizability introduced in Section~\ref{subsec:g0w0equations} is not used. Instead we start from the reducible 
polarizability $\chi(\omega)$. In the spectral representation,  $\chi(\omega)$ is given as sum of its poles $n$ in the complex 
plane
\begin{equation}
 \begin{split}
 \chi(\bfr,&\bfrp,\omega)  = \\&\sum_n\rho_n(\bfr)\rho_n^*(\bfrp)\left( \frac{1}{\omega+i\eta-\Omega_n}-\frac{1}{\omega-i\eta+\Omega_n}\right).
 \end{split}
 \label{eq:chi_irreducible}
\end{equation}
The pole positions $\Omega_n$ correspond to charge neutral excitation energies and $\rho_n(\bfr)$ denotes transition densities. Equation~\eqref{eq:chi_irreducible} would be exact for the exact  $\Omega_n$ and $\rho_n(\bfr)$. Both quantities are obtained by solving 
a conventional eigenvalue problem. The equations that are solved are identical to the Casida equations \cite{Casida1995}, except that for $G_0W_0$ the exchange-correlation kernel is omitted (otherwise it would be time-dependent density-functional theory). \par
The reducible polarizability $\chi(\omega)$ can then be expanded in terms of $\chi_0$ in a Dyson series
\begin{equation}
 \chi(\omega) =  \chi_0(\omega) + \chi_0(\omega)v\chi(\omega).
\end{equation}
and we can thus rewrite $W_0$ given in Equation~\eqref{eq:W0} in terms of $\chi(\omega)$ 
\begin{equation}
 \begin{split}
    W_0(\bfr,\bfrp,\omega) &= v(\bfr,\bfrp) + \\&\int d\mathbf{r}''d\mathbf{r}'''v(\bfr,\mathbf{r}'')\chi(\mathbf{r}'',\mathbf{r}''',\omega)v(\mathbf{r}''',\bfrp).
 \end{split}
\end{equation}
Inserting Equation~\eqref{eq:chi_irreducible} yields a pole expansion for $W_0$. The self-energy integral can then be solved analytically and we obtain a 
closed expression for $\Sigma^c_s(\omega)$:
\begin{equation}
    \Sigma_s^c(\omega) = \sum_m \sum_{n} \frac{\bracket{\phi_s^0\phi_m^0}{P_n}{\phi_m^0\phi_s^0}}{\omega -\epsilon_m^0 + (\Omega_n-i\eta)\sgn(E_{\mathrm{F}} -\epsilon_m^0)},
    \label{eq:sigma_pole}
\end{equation}
where $P_n(\bfr,\bfrp) = \rho_n(\bfr)\rho_n^*(\bfrp)$.
More precisely, $\Sigma^c_s(\omega)$ also becomes a sum over the poles $\Omega_n$. Equation~\eqref{eq:sigma_pole} is therefore similar to the PPM approximation, except that we sum over the exact poles of $W_0$ and not over the poles of a $W_0$-model function. A detailed description of the fully-analytic frequency treatment can be found in Ref.~\cite{vanSetten/etal:2013}. 
Equivalent expressions are also given in Hedin's review article from 1999 \cite{Hedin:1999} and were applied in Refs. \cite{Tiago/Chelikowsky:2006,Bruneval:2012,Bruneval2016}.

\subsubsection{Comparison of accuracy and computational cost}
\label{subsec:freq_comparison}

In the previous sections three full-frequency integration techniques have been introduced: the CD, the AC and the fully-analytic approach. The CD and fully analytic method compute the self-energy directly for real frequencies. By design, the fully analytic approach is in principle the most exact one since it is parameter-free, except for the dependence on the basis set and the broadening parameter $\eta$. However, the same accuracy can already be achieved with the CD using a moderately-sized numerical integration grid for the imaginary frequency term \cite{Golze2018}. In the AC approach the self-energy is calculated on the imaginary frequency axis, which is fairly featureless. The accuracy of the AC approach depends on the features of the self-energy on the real axis and on the flexibility of the model function, which continues the self-energy to real frequencies. 

Generally, QP energies of valence states are well reproduced \cite{Setten2015}, while the AC is likely to fail for deeper states as discussed in Section~\ref{subsec:ac}. In the PPM approximation, the self-energy integral is simplified by introducing a model function for $W_0$. The accuracy is therefore determined by the chosen model function and generally difficult to estimate.\par
The fully-analytic approach is the computationally most expensive method in our toolbox since solving the eigenvalue problem to obtain the poles of $\chi$ is an $\mathcal{O}(N^6)$ step, where $N$ defines the size of the system. The scaling of the CD and AC approach is generally lower, but depends on the details of the implementation. The CD method requires more computational resources than the AC methods due to the additional sum over the residues of the poles of $G_0$. The overhead is relatively small for QP energies of valence states, but increases for deeper states due to the steady increase of the number of residues. The PPM is computationally the most efficient method, because the dielectric function $\varepsilon$ used to compute $W_0$ has to be calculated only at a few frequency points to determine the parameters of the PPM. 

\subsection{Basis sets}
\label{sec:basis_sets}
In any $GW$ implementation, the QP wave functions $\psi_s$ and also the mean-field orbitals $\{\phi^0_{s}\}$ are expanded in a set of normalized basis functions 
$\{\varphi_j\}$. Since in $G_0W_0$ the QP wave functions are approximated by the KS-DFT or HF ones, we expand in practice only $\phi^0_{s}$,
\begin{equation}
  \phi^0_{s}(\bfr) = \sum_jc_{sj}\varphi_j(\bfr),
 \label{eq:expansion}
\end{equation}
 where $c_{sj}$ are the expansion coefficients that have to be determined. Performing the $G_0W_0$ calculation in a basis transforms the expression 
for $W_0$, $\varepsilon$, and $\chi_0$ into matrix equations suitable for implementation in computer codes. %

The basis set choice is often guided by the type of system under investigation. In the following we will introduce the most common basis sets with brief comments on their suitability.

\subsubsection{Plane waves}
\label{sec:planewaves}
 For periodic systems, the energy spacing between discrete enery levels can vanish, in which case the single-particle eigenvalues form bands.  According to Bloch's theorem \cite{Bloch1929}, the single-particle states can be written as Bloch waves
\begin{equation}
 \label{eq:bloch}
 \phi_{n \mathbf{k}}^{0}(\bfr)=\operatorname{e}^{i\mathbf{k}\cdot \bfr} u_{n \mathbf{k}}(\bfr),
\end{equation}
where $\mathbf{k}$ is a wave vector in the first Brillouin zone and $n$ is the band index. The index $s$ that we had used in Equation~\eqref{eq:expansion} and throughout to label states  now becomes the compound index $n\mathbf{k}$. The functions $u_{n \mathbf{k} }(\bfr)$ have the periodicity of the lattice  and can be expanded in plane waves $\{\varphi_{\mathbf{G}}\}$,
\begin{align}
u_{n \mathbf{k} }(\bfr) & =  \sum_{\mathbf{G}}c_{n \mathbf{k}}(\mathbf{G}) \varphi_{\mathbf{G}}(\bfr)\\[5pt]
 \varphi_{\mathbf{G}}(\bfr) &= \frac{1}{\sqrt{\Omega}}\operatorname{e}^{i\mathbf{G}\cdot\bfr}
 \end{align}
where $\Omega$ is the volume of the periodic cell and $\mathbf{G}$ is a reciprocal lattice vector. Reciprocal lattice vectors $\mathbf{G}$ are given by 
$\mathbf{G}\cdot\mathbf{t}=2\pi n$, where $n$ is a positive integer and $\mathbf{t}$ is a translation vector of the unit cell. $G^2$ is directly proportional to the kinetic energy $E$ of a free 
electron. The size of the basis set is characterized by the largest $\mathbf{G}$ 
vector and is usually given in terms of the energy $E$ that corresponds to the largest reciprocal lattice vector, $E =  G_{\mathrm{max}}^2 / 2$. All $\mathbf{G}$ vectors with equal or smaller energies are included in the basis set. 

The first $GW$ calculations~\cite{Hybertsen/Louie:1985,Hybertsen/Louie:1986,Godby/etal:1986} were performed for solids with plane wave basis sets. Also today plane waves are common in state-of-the-art $GW$ implementations, see Table \ref{Tab:GWcodes} for a list of $GW$ codes. The real-space representation of $G_0$, $W_0$, $\varepsilon$, and $\chi_0$  given in Equations~\eqref{eq:greensfkt}-\eqref{eq:chi0} can be easily transformed into a basis of plane waves by Fourier transforms. For expressions of these quantities in plane waves see, e.g., Ref.~\cite{Hueser/Olsen/Thygesen:2013}.\par
Plane wave basis sets are suitable for describing the slowly varying electron density in the valence region, where only the valence orbitals  are non-zero. However, the valence wave functions tend to oscillate rapidly close to the nuclei due to orthogonality constraint with respect to the core orbitals. Representing these oscillations requires a large number of plane waves. Plane waves are therefore used in combination with pseudopotentials or the projector-augmented-wave methods \cite{Bloechl1994} to approximate the effect of the core electrons. We will introduce the pseudopotential concept in Section~\ref{subsec:pseudopotentials} and return to plane waves in the context of the projector augmented wave scheme in Section~\ref{subsec:paw}.

\begin{table}[t]
\renewcommand*{\arraystretch}{1.4}
\centering
\caption{ Selection of $G_0W_0$ codes and large program packages with $G_0W_0$ implementations and corresponding basis sets.}
\label{tab:team}
\begin{tabular}{[y{2.7cm}!x{2.6cm}!x{3.1cm}]}
\thickhline
\bfseries Code\cellcolor{blue!10} &
\bfseries Basis set \cellcolor{blue!10} &
\bfseries Ref. \cellcolor{blue!10}\\\thickhline
Berkeley$GW$     & Plane waves         & \cite{Deslippe/etal:2012}  \\\thickhline
Yambo           & Plane waves         & \cite{Marini/etal:2009}  \\\thickhline
WEST            & Plane waves         & \cite{Govoni/etal:2015}  \\\thickhline
SaX             & Plane waves         & \cite{MartinSamos/Bussi:2009}  \\\thickhline
Sternheimer$GW$ & Plane waves         & \cite{Giustino2010,Schlipf2019} \\\thickhline
ABINIT          & Plane waves (PAW)   & \cite{Gonze2009} \\\thickhline
VASP            & Plane waves (PAW)   & \cite{Shishkin06,Liu2016}  \\\thickhline
GPAW            & Plane waves (PAW)   & \cite{Hueser/Olsen/Thygesen:2013} \\\thickhline
Fiesta          & Gaussian            & \cite{Blase/Attaccalite/Olevano:2011} \\\thickhline
Turbomole       & Gaussian            & \cite{vanSetten/etal:2013} \\\thickhline
CP2K            & Gaussian            & \cite{Wilhelm2016,Wilhelm2018} \\\thickhline
MOLGW           & Gaussian            & \cite{Bruneval2016} \\\thickhline
FHI-aims        & NAO                 & \cite{Xinguo/implem_full_author_list,Golze2018} \\\thickhline
exciting        & FLAPW               & \cite{Exciting:2014}  \\\thickhline
SPEX            & FLAPW               & \cite{Friedrich2010}\\\thickhline
FHI-gap         & FLAPW & \cite{Jiang/etal:2013} \\\thickhline
Tombo           & Augmented           & \cite{Ono/etal:2015} \\ \thickhline
Questaal        & LMTO               & \cite{Methfessel2000,questaal} \\  \thickhline
\end{tabular}
\label{Tab:GWcodes}
\end{table}

\subsubsection{Localized basis sets}
While plane waves are mostly used for periodic systems, they can in principle also be used for finite systems by placing, e.g., the molecule in a sufficiently large unit cell to avoid spurious 
interactions with the neighboring cells. However, large unit cells require a very large plane wave basis set and are therefore computationally expensive. Molecular systems can be more efficiently 
described by atom-centered localized basis sets. 

The most common basis functions of this type are Gaussian basis sets
\begin{equation}
 \varphi_{\alpha,l,m}(\mathbf{r}) = N_l r^lY_{l,m}(\theta,\phi)e^{-\alpha r^2},
\end{equation}
where $N_l$ is a normalization constant and $Y_{l,m}(\theta,\phi)$ are spherical harmonic functions given in spherical coordinates ($r$, $\theta$, $\phi$). A Gaussian type orbital is characterized by the exponent $\alpha$ and the angular and 
magnetic quantum numbers $l$ and $m$, which are dictated by the basis set selection. The design of Gaussian basis sets requires careful optimization regarding the number of functions, their respective angular momentum and exponents $\alpha$. In quantum chemistry, Gaussian basis sets are widely used and ample experience exists in designing suitable basis sets for correlated methods such as coupled cluster theory. These Gaussian basis sets can then also be used in $GW$ calculations.\par
Another type of localized basis functions used in $GW$ calculations are numeric atom-centered 
orbitals (NAOs),
\begin{equation}
 \varphi_{l,m,\mu}(\bfr) = N_l \frac{u_{\mu}(r)}{r} Y_{l,m}(\theta,\phi)
 \label{eq:NAO}
\end{equation}
where $u_{\mu}(r)$ are radial functions that are not restricted to any particular shape. The radial part of NAOs is tabulated on dense grids and is fully flexible. Gaussian radial functions can be considered as special types of this general NAO form.\par
Slater type functions, which posses an exponential decay at long range and a cusp at the position of the nuclei, have been also used in $GW$ calculations \cite{Stan/etal:2009}. However, this basis set type is less common.\par
Local basis functions, in particular NAOs that derive from atomic orbitals, are well suited to describe rapid oscillations of wave functions near the nucleus. They are therefore the obvious choice  for QP calculations  of core and semi-core states.
%
\subsubsection{Augmented basis sets}
\label{subsec:augmentedbas}
Augmented plane waves (APW) are another basis set type that includes the rapidly varying oscillations near the nuclei. APW methods use the so-called muffin tin approximation, which is a physically motivated approximation to the shape of the potential in solid state systems \cite{MartinBook,Slater1937}. The shape of the potential resembles a muffin tin:  it is peaked at the nuclei and predominantly spherical close to it, while it is flat in between. Therefore, real space is partitioned into non-overlapping (muffin-tin) spheres $\Omega_{\text{MT},a}$ centered around each nuclei $a$ and interstitial regions $\Omega_{\text{I}}$ between these spheres. The valence wave functions are then expanded in localized NAO-like functions (Equation~\eqref{eq:NAO}) inside the spheres and plane waves in the interstitial regions.\par
By construction, the APW basis sets produce wave functions with a discontinuity in the first derivative at the muffin-tin boundaries. The linear APW (LAPW) was proposed to guarantee that the solution in the muffin-tin matches continuously and differentiably onto the plane wave part in the interstitial region \cite{Andersen1975}. With this extension, the explicit form of the LAPW basis functions is
\begin{align*}
&\varphi_{\mathbf{G}}(\bfr) \\
&=\begin{cases}
 \Omega^{-1}\operatorname{e}^{i\mathbf{G}\cdot\bfr}&\bfr\in\Omega_{\text{I}}\\
 \sum_{lm}(A_{lm}^a u_l^a(r) + B_{lm}^a\dot{u}_l^a(r))Y_{lm}(\theta,\phi) & \bfr\in\Omega_{\text{MT,a}}
\end{cases}
\end{align*}
where $u_l^a(r)$ and its  derivative $\dot{u}_l^a(r)$ are radial functions centered at the atom $a$. The coefficients $A_{lm}$ and $B_{lm}$ are determined such that continuity in value and derivative of the basis functions at the muffin-tin boundaries is ensured. 

LAPW basis sets can be extended by additional local orbitals, LAPW+lo, that are completely localized in the muffin-tin spheres and go to zero at the boundaries. Inclusion of such local orbitals significantly improves the variational freedom, e.g., the description of $d$ and $f$ electrons \cite{Singh1991}. It has furthermore been shown that these local orbitals are particularly important for the unoccupied state convergence in $GW$ calculations \cite{Friedrich2011,Jiang/Blaha:2016,Jiang:2018}.\par
A general form of the LAPW method are full-potential LAPW (FLAPW) methods that make no approximations on the shape of the potential \cite{Wimmer1981} and which are nowadays standard in LAPW codes. Recently a number of FLAPW $GW$ codes have emerged~\cite{Friedrich/etal:2006,Friedrich2010,Friedrich/etal:2012,Exciting:2014,Jiang/etal:2013}. \par
Linear muffin-tin orbital (LMTO) schemes are very similar to LAPW basis sets, except that the basis functions in the interstitial region are not plane waves \cite{Andersen1975}, but for example smooth Hankel functions \cite{Methfessel2000}. \par
%
In these augmented basis sets it is straightforward to include core and semicore states in the  Green's function $G_0$ (Equations~\eqref{eq:greensfkt}) and the polarizability $\chi_0$ (Equations~\eqref{eq:chi0}) and therefore in the self-energy. This, in principle, improves the description of QP excitations of valence states and band gaps, even though it has been found that the difference to carefully adjusted plane wave-based projector augmented-wave (PAW) calculations (see Section~\ref{subsec:paw}) is typically less than 100~meV \cite{Nabok2016}. However, the same study reported  larger differences for deep-lying and very localized $d$ and $f$ states \cite{Nabok2016}. Core excitations are in principle also accessible with FLAPW basis sets. However, these have not been thoroughly investigated yet.\par
For local and semi-local DFT functionals, the (F)LAPW basis sets have become the ultimate accuracy reference, closely followed by NAOs \cite{Lejaeghere/etal:2014,Lejaeghere/etal:2016}. For $G_0W_0$, first steps in systematically benchmarking solids were made only recently \cite{Setten2017}. For molecules, $G_0W_0$ benchmark calculations emerged during the last years and we will discuss them in Section~\ref{subsec:gw100}. The jury is therefore still out on which basis set is most accurate for solids.

\subsubsection{Pseudopotentials}
\label{subsec:pseudopotentials}
$GW$ calculations can be grouped in two categories: those that take all electrons of the system into consideration and those that partition into valence and core electrons. In this latter case, only the valence electrons enter the $GW$ (and the preceding DFT) calculation explicitly, whereas the effect of the core electrons is taken into account only indirectly, for example through a pseudopotential. Such core-valence partitioning is motivated by the observation that deep core states are relatively inert and do not contribute to chemical bonding. The advantage of using a partitioning scheme is that the electron number in the $GW$ calculation is reduced, which decreases the computational cost. An obvious disadvantage is that the core electrons may have an effect on the valence electrons, which will be difficult to include appropriately in the $GW$ calculation and then may lead to incorrect results.\par
Pseudopotentials have been the default way to partition electrons \cite{MartinBook,HutterBook}. In a pseudopotential, the core electrons are removed and the Coulomb potential of the nucleus and the inner-shell electrons is replaced by a smooth effective potential that acts on the valence electrons. The potentials are generated from calculations of isolated atoms. They are constructed such that the wave function of the valence electrons match those of an all-electron calculation outside the core region or outside a chosen radius around the nuclei. Inside the core region, the functions are smooth and nodeless. Additional norm-conservation criteria \cite{Hamann1979,Bachelet1982}, which preserve the orthonormality condition for the pseudo wave function, are usually applied \cite{Troullier/Martins:1991,fhi98PP,Goedecker1996}. The resulting potential is finite at the origin of the atom and shallow. Pseudopotentials are mostly used in combination with plane waves since the smooth and shallow potentials greatly reduce the required plane wave cutoff and make plane wave $GW$ calculations with these basis sets feasible. In addition, pseudopotentials  have been used for $GW$ calculations with localized functions to reduce the basis set size \cite{Blase/Attaccalite/Olevano:2011,Wilhelm2016}.\par
The majority of pseudopotential development took place in DFT \cite{HutterBook}. Optimizing the parameters in the pseudopotential to ensure transferability is a complex task and requires thorough testing \cite{Shirley/Martin/Bachelet/Ceperley:1990,Goedecker1996}. Transferability means that one and the same pseudopotential should be adequate for an atom in different chemical environments. Similar to localized
basis sets, the parameters of pseudopotentials are precomputed, similar to localized basis sets, and then tabulated for download in libraries like the Pseudo-Dojo \cite{VanSetten/etal:2018,Garcia/etal:2018}.\par
In $GW$, the consistency between pseudopotential and all-electron calculations will almost inevitably be violated. To be fully consistent, the DFT pseudopotentials would have to be cast aside and $GW$ pseudopotentials be used. However, no such $GW$ pseudopotentials have been developed until now, due to the complexity of the $GW$ self-energy, which does not lend itself easily to pseudoization. Even if we had $GW$ pseudopotentials, they would then have to first be used in the preceding DFT calculation, in which they would break the DFT core-valence consistency. Unless we perform fully self-consistent $GW$ calculations, we are stuck with an inconsistency dilemma.\par
\begin{figure*}[t!] 
    \includegraphics[width=0.99\linewidth]{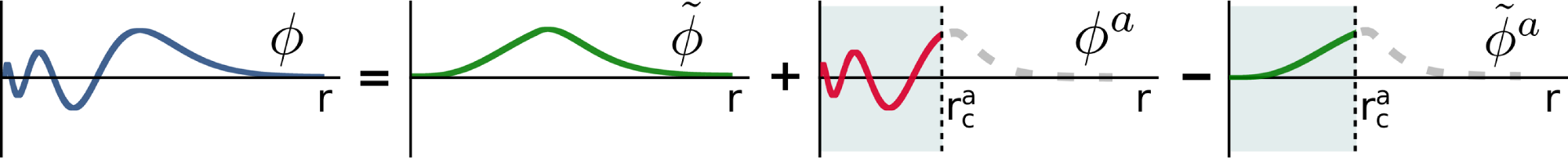}
	\caption{{\small \label{fig:paw}
	Schematic representation of the projector augmented wave (PAW) scheme. The all-electron wave function $\phi$ is constructed from the smooth auxiliary function $\tilde{\phi}$ and corrections from the hard and smooth atom-centered auxiliary wave functions $\phi^a$ and $\tilde{\phi^a}$, respectively. }}
\end{figure*}
Early efforts towards $GW$ pseudopotentials introduced core polarization effects into DFT pseudopotentials \cite{Shirley/Martin:1993,Lee/Needs:2003}. By extending the $GW$ formalism to include core contributions in the dielectric screening and the self-energy, such core-polarization potentials have also been tried successfully in the $GW$ method \cite{Shirley/Zhu/Louie:1997}. However, developments in this direction did not continue. The default procedure today for plane wave $GW$ codes is to use only well tested DFT pseudopotentials for the required elements \cite{Govoni2018}. Care has to be taken that the scattering states (i.e. the unoccupied states) of the pseudopotential are described well and do not introduce ghost states \cite{Gonze/Kaeckell/Scheffler:1990}. If no good pseudopotentials are available, it is recommended  to either generate customized pseudopotentials, use the PAW method or employ genuine all-electron basis sets. Pseudopotential approaches have to be employed with care in particular for materials with localized $d$ and $f$ electrons. Specific issues in $GW$ calculation of these materials are discussed in Section~\ref{sec:solids}.\par

\subsubsection{Projector augmented-wave method (PAW)}
\label{subsec:paw}
The PAW method is commonly used in plane wave $G_0W_0$ implementations, see Table~\ref{Tab:GWcodes}. It enables computational feasibility and ensures transferability between different chemical environments. The PAW method has been derived by Bl\"ochl combining ideas from the pseudopotential method and LAPW basis sets \cite{Bloechl1994}. The idea is to express the KS all-electron wave function $\phi_s^0$ for state $s$ in terms of a smooth auxiliary function $\tilde{\phi}_s^0$ and correction terms, which restore the oscillating behavior in the core region. Note that for Bloch waves, the label $s$ contains the $\mathbf{k}$ and band index $n$. \par
To construct $\tilde{\phi}_s^0$, we define a linear transformation $\mathcal{\hat{T}}$ which establishes a connection between $\phi_s^0$ and $\tilde{\phi}_s^0$,
\begin{align}
    \ket{ \phi_s^0} &= \mathcal{\hat{T}}\ket{\tilde{\phi}_s^0}.
\end{align}
Since the all-electron wave function is already smooth at a certain distance from the nuclei, we partition the space similarly to LAPW schemes: in atom-specific augmentation regions $\Omega_a$ around the nuclei, where $a$ is the atom index, and an interstitial region $\Omega_{\mathrm{I}}$. The augmentation regions are characterized by the cutoff radii $r_c^a$, which should be chosen such that the augmentation spheres do not overlap.
Outside the augmentation regions, $\tilde{\phi}_s^0$ should be identical to the all-electron wave function. $\mathcal{\hat{T}}$ should thus modify  $\phi_s^0$ only in $\Omega_{\mathrm{a}}$ and we define
\begin{equation}
    \mathcal{\hat{T}} = 1 + \sum_a\mathcal{\hat{T}}^a
\end{equation}
where the atom-centered transformation, $\hat{\mathcal{T}}^a$, acts only within
$\Omega_{\mathrm{a}}$.\par
The transformation operator is derived by introducing atom-centered functions as in LAPW, which is described in detail in Refs.~\cite{MartinBook,Rostgaard2009}. The all-electron wave function can then be rewritten as 
\begin{equation}
    \phi^0_s(\mathbf{r}) = \tilde{ \phi^0_s}(\mathbf{r}) + \sum_a(\phi^{a}_s(\mathbf{r})-\tilde{\phi}^{a}_s(\mathbf{r})),
    \label{eq:paw_phi}
\end{equation}
where the atom-centered hard and smooth auxiliary wave functions are denoted by $\phi^a_s$ and $\tilde{\phi}^a_s$, respectively. ``Hard'' refers to rapidly varying functions in the core region. The concept of the PAW scheme is  visualized in a simplified way in Figure~\ref{fig:paw}. By adding $\phi^a$ to $\tilde{\phi}_s^0$  we obtain the oscillating behavior in the core region, but we have to subtract the smooth function $\tilde{\phi}^a$ to cancel the contribution of $\tilde{\phi}_s^0$ in $\Omega_a$. That implies that the following conditions must hold
\begin{equation*}
\begin{rcases}
 \phi_s^0(\bfr) &= \tilde{\phi}_s^0(\bfr)\\
 \phi_s^a(\bfr) &= \tilde{ \phi}_s^a(\bfr)
\end{rcases}
\bfr\in\Omega_{\mathrm{I}}
\quad
\begin{rcases}
 \phi_s^0(\bfr) &= \phi_s^a(\bfr)\\
 \tilde{\phi}_s^0(\bfr) &= \tilde{ \phi}_s^a(\bfr)
\end{rcases}
\bfr\in\Omega_{\mathrm{a}}
\end{equation*}
\par
The atom-centered auxiliary wave functions can be expanded in a finite set of local basis functions $\{\varphi_j^a\}$ and $\{\tilde{\varphi}_j^a\}$ and a set of projector functions $\tilde{p}_j^a$, where `$\sim$' indicates again smooth functions. These expansions are given by
\begin{align}
\phi_s^a(\bfr) &= \sum_j\varphi_j^a(\bfr)\Braket{\tilde{p}_j^a|\tilde{\phi}_s^0}\\
\tilde{\phi}_s^a(\bfr) &= \sum_j\tilde{\varphi}_j^a(\bfr)\Braket{\tilde{p}_j^a|\tilde{\phi}^0_s}.
\end{align}
The variational object in a PAW calculation is $\tilde{\phi}_s^0$. The latter is expanded using, e.g., a plane wave basis set, for which a low energy cutoff can be used due to its smoothness. The local basis sets and projector functions needed to compute the second and third terms in Equation~\eqref{eq:paw_phi} are tabulated for each element of the periodic table. For specific choices of these basis sets see, e.g., Refs.~\cite{Kresse1999,Rostgaard2009}. Details regarding the practical implementation within a plane wave code are given in Ref.~\cite{Kresse1999} and for real-space grid codes in Ref.~\cite{Mortensen2005,Enkovaara2010}.\par
$GW$ calculations within the PAW schemes employ usually the frozen core approximation \cite{Shishkin06,Liu2016}. The core states are localized at the atoms and confined in $\Omega_a$. In the frozen core approximation we assume that the KS core states are identical to the atomic core states $\alpha$, i.e., $\phi_s^0=\varphi_{\alpha}^a$. In this approximation, the decomposition given in Equation~\eqref{eq:paw_phi} is not used for the core states. However, the effect of the core on the valence states is correctly described.\par
The accuracy of the expansion in Equation~\eqref{eq:paw_phi} depends on the completeness of the set of localized basis  and projector functions ($\{\varphi_j^a\}$, $\{\tilde{\varphi}_j^a\}$ and $\{\tilde{p}_j^a\}$). Achieving completeness is easy for occupied states, but practically impossible if $s$ corresponds to a high-energy empty state, which has been discussed by \onlinecite{Klimes/Kaltak/Kresse:2014}. 
However, for a $GW$ calculation, many of these high-energy empty states need to be included, which is explained in more detail in Section~\ref{subsec:basis_convergence}. These incompleteness issues lead to a violation of the norm-conservation for the unoccupied states, which can be the source of substantial errors, in particular for elements with $d$ and $f$ electrons. This error can be avoided using norm-conserving instead of the standard PAW potentials for $GW$ calculations \cite{Klimes/Kaltak/Kresse:2014}.

\subsection{Basis set convergence}
\label{subsec:basis_convergence}
\begin{figure} 
    \includegraphics[width=0.99\linewidth]{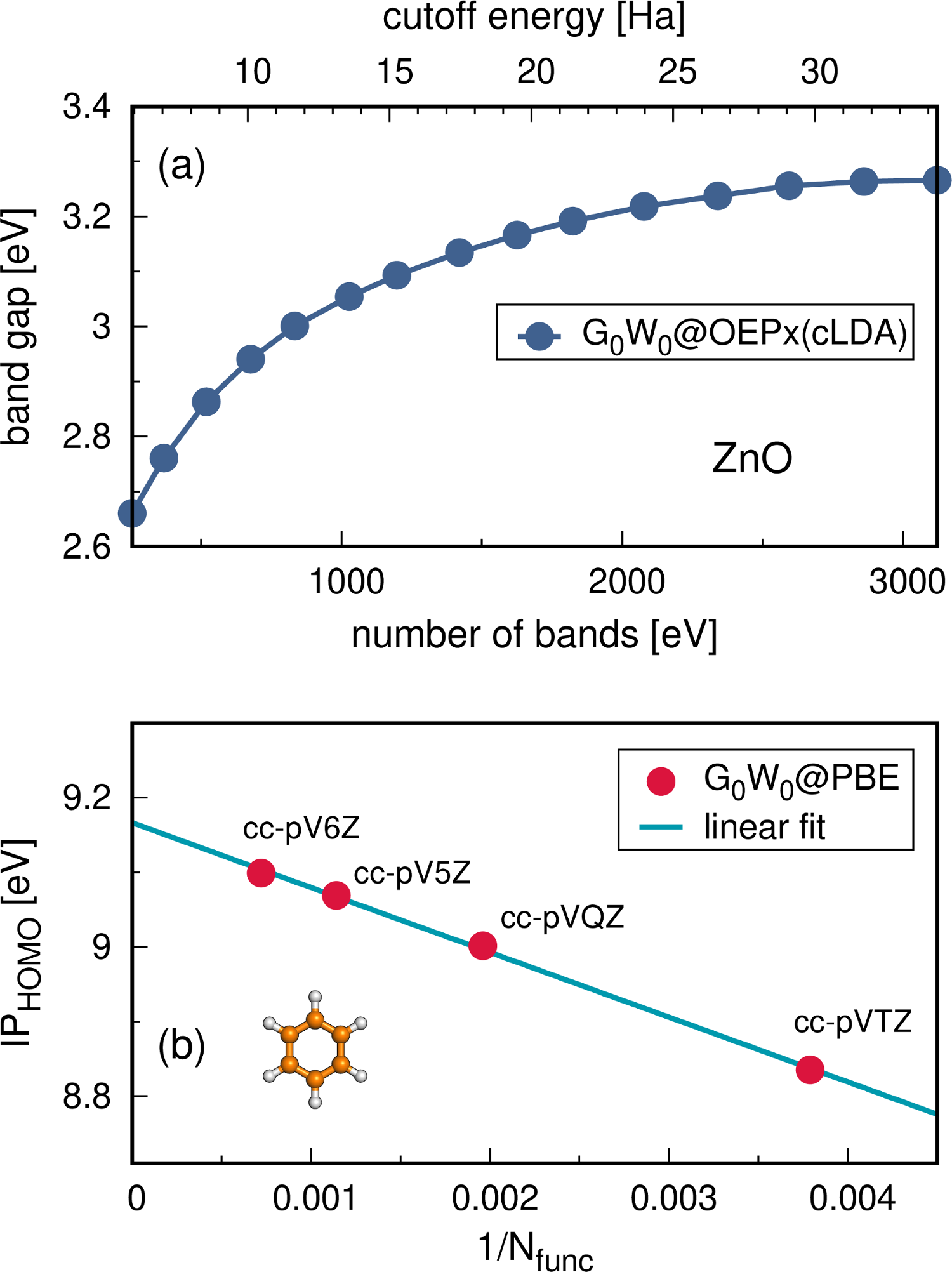}
	\caption{{\small \label{fig:basis_convergence}
	Basis set convergence for $G_0W_0$ calculations. (a) Convergence for a plane wave basis set. Bandgap of wurtzite ZnO dependent on the number of bands and on the corresponding cutoff energy 
(data from SI of Ref. \cite{Yan/etal:2012}). (b) Convergence and extrapolation procedure for a localized basis set. Ionization potential (IP) for the HOMO of benzene plotted with respect to the inverse of the number of basis functions $N_{\mathrm{func}}$ using the cc-pV$n$Z basis set series. Further computational details are given in Appendix~\ref{app:computational_details}.}}
\end{figure}

The first criteria for a reliable $G_0W_0$ calculation is that the underlying DFT or HF calculation is converged. This convergence has to be checked for all basis set types.
The second convergence criteria is the size of the basis set in the $G_0W_0$ calculation itself. In quantum chemistry it is well established that correlated electronic structure methods converge slowly with respect to the number of basis functions \cite{Kendall1992,Kutzelnigg1992,Klopper1999}. The same has been also observed for $G_0W_0$ \cite{Shih2010, Friedrich2011,Yan/etal:2012,Bruneval:2012,Bruneval/Marques:2013,vanSetten/etal:2013,Jacquemin2015,Setten2015,Bruneval2016,Wilhelm2016}. It has been demonstrated that the convergence rate of $G_0W_0$ is comparable to other correlated methods such as second-order M{\o}ller-Plesset perturbation theory (MP2) and the coupled cluster singles, doubles and perturbative triples [CCSD(T)] method \cite{Bruneval/Marques:2013}.\par 
Converging $G_0W_0$ excitations within a plane wave basis set is straightforward since the cutoff energy  can be continuously increased, see Figure~\ref{fig:basis_convergence}(a). In addition, extrapolation schemes to the complete basis set limit have been reported to reduce the computational cost \cite{Klimes/Kaltak/Kresse:2014,Maggio2017b,Govoni2018}.
Conversely, for localized basis sets only a limited number of basis set sizes is available and a steady increase in size as for plane waves is not possible. Therefore, extrapolation techniques must always be used to obtain converged $G_0W_0$ energies. This is displayed in Figure~\ref{fig:basis_convergence}(b), where the first IP of benzene is plotted with respect to the inverse of the basis set size. Shown are results for the Dunning basis set family cc-pV$n$Z ($n$=3-6) \cite{Dunning1989,Wilson1996}, which was designed to smoothly reach the complete basis set limit. Increasing values of $n$ indicate increasingly large basis sets. The convergence is smooth, but very slow, as shown in Figure~\ref{fig:basis_convergence}(b).\par
The extrapolation to an infinite number of basis functions is performed by a linear regression with respect to the inverse of the total number of basis functions. The extrapolated value of 9.17~eV in Figure~\ref{fig:basis_convergence} is 0.07~eV larger than the IP obtained at the cc-pV6Z level showing that even the largest basis set cannot converge the $G_0W_0$ energies completely. This linear fitting procedure is a well-established scheme to extrapolate $G_0W_0$ energies and has been tested in large benchmark studies 
\cite{Setten2015}.  Alternatively, linear regression has also been performed with respect to $C_n^{-3}$, where $C_n$ is the cardinal number of the basis set, i.e., 3 for cc-pVTZ, 4 for cc-pVQZ , 5 for cc-pV5Z and 6 for cc-pV6Z. Extrapolation with respect to $C_n^{-3}$ is well-established for correlated methods \cite{Helgaker1997}. The inverse of the basis set number corresponds roughly to $C_n^{-3}$. The average difference between the two extrapolation schemes for $G_0W_0$ energies is indeed very small with 0.04~eV \cite{Setten2015}. \par
A common misconception in the plane wave community is that the number of unoccupied states that enter a $G_0W_0$ calculation is a separate convergence parameter. The number of unoccupied states that can be resolved with a given basis set typically grows with the size of that basis set, i.e., the Hilbert space of that basis grows. This implies that more empty states  enter the sums in the Green's function (Equation~\eqref{eq:greensfkt}) and the polarizability (Equation~\eqref{eq:chi0}). Since it is computationally expensive to generate a large number of unoccupied states in the preceding plane wave DFT or HF calculation, plane wave $G_0W_0$ practitioners reduced the number of unoccupied states that enter the $GW$ calculation in order to save computational time \cite{Stankovski/etal:2011,Setten2017}. Localized basis sets on the other hand are significantly smaller and typically all virtual states are computed even in DFT-only calculations. Figure~\ref{fig:basis_convergence}(a) gives an impression of the scale for the plane wave case. It displays the convergence of the band gap of wurtzite ZnO with respect to the number of bands (states) \cite{Yan/etal:2012}. The convergence rate of ZnO is particularly slow \cite{Stankovski/etal:2011,Friedrich2011} compared to other semiconductors, e.g., silicon \cite{Friedrich/etal:2006}. The band gap finally converges at around 30 Ha. At this point almost 3000 bands have been included in the $G_0W_0$ calculation.\par
While it might seem appealing to reduce the number of required unoccupied states to less than 3000, Figure~\ref{fig:basis_convergence}(a) illustrates that a reduction is not possible due to the slow convergence. Since the number of resolvable, unoccupied states is coupled to the plane wave cutoff \cite{Setten2017,Stankovski/etal:2011,Gao2016}, one should always include the maximum number of bands in the $G_0W_0$ calculation for a given plane wave cutoff. Such a procedure also greatly simplifies the convergence study since only one and not two parameters need to be converged.
%
\subsection{Elimination of unoccupied state summation}
\label{subsec:sternheimergw}
\begin{figure} 
    \includegraphics[width=0.99\linewidth]{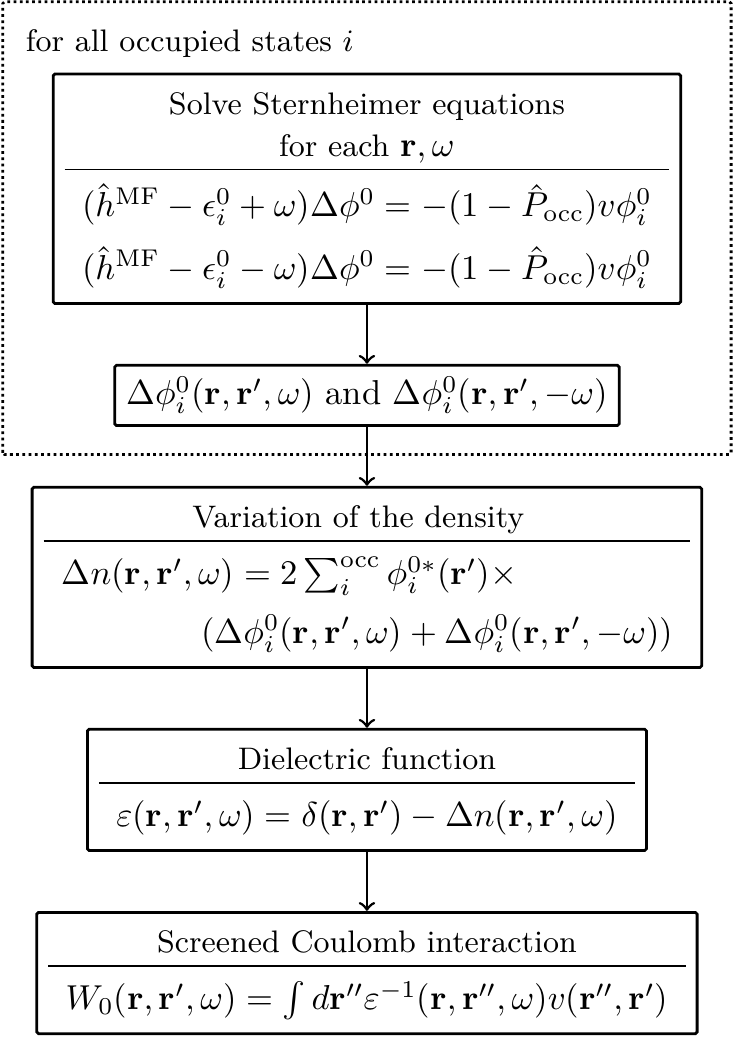}
	\caption{{\small \label{fig:sternheimer}
Non-self-consistent Sternheimer approach for obtaining $W_0$ without empty states. $\Delta \phi^0$ and $v$ have a parametric dependence on the real space point $\bfr$ and the frequency $\omega$.}}
\end{figure}

The complications around the virtual state convergence raised in the previous section for plane wave basis sets can be bypassed completely by eliminating the explicit summation over unoccupied states in the Green's function $G_0$ (Equation~\eqref{eq:greensfkt}) and the polarizability $\chi_0$ (Equation~\eqref{eq:chi0}) \cite{Berger2010,Giustino2010,Umari/Stenuit/Baroni:2010,Wilson2008,Wilson2009,Reining1997}. A practical method for building a perturbation theory without explicit reliance on unoccupied states was pioneered in the context of density functional perturbation theory (DFPT) \cite{Baroni1987,Gonze1995,Gonze1997}. Here we will briefly review how the DFPT concept can be transferred to $G_0W_0$. For a general introduction to the DFPT formalism see \onlinecite{Baroni/RevModPhys:2001}.\par
The central object in DFPT is the response function that measures the response of a system to a perturbation $\Delta V$. In the $G_0W_0$ context, we are interested in the response to the introduction of an additional charge to the system at point $\bfr$. The additional charge perturbs the charge density of the system. The response function mediates the charge density change and the perturbation. We now wish to calculate the response function without invoking the sum over states expression introduced in Equation~\eqref{eq:chi0}. \par

We start with the change in the charge density $\Delta n(\bfr,\bfrp,\omega)$ given for a spin-unpolarized system by \cite{Giustino2010}
\begin{equation}
 \begin{split}
    \Delta n (\bfr,\bfrp,\omega) = 2 \sum_{i}^{\rm occ} &\phi_i^{0*}(\bfrp)\times\\ &(\Delta\phi_i^0(\bfr,\bfrp,\omega) + \Delta\phi_i^0(\bfr,\bfrp,-\omega)).
 \end{split}
\end{equation}
Here $\Delta\phi_i^0(\bfr,\bfrp,\pm\omega)$ is the frequency-dependent variation of the occupied mean-field state $i$. Instead of expanding $\Delta\phi_i^0(\bfr,\bfrp,\pm\omega)$ in the basis of unperturbed mean-field states $\phi_i^0(\bfr)$ it is calculated directly with the Sternheimer equation \cite{Baroni1987,Giustino2010}
\begin{equation}
 (\hat{h}^{\mathrm{MF}} - \epsilon_i^0\pm \omega)\Delta\phi_i^0(\bfr,\bfrp,\pm\omega) = -(1-\hat{P}_{\rm occ}) \Delta V(\bfr,\bfrp) \phi_i^0(\bfrp).
 \label{eq:sternheimer}
\end{equation}
$\hat{P}_{\rm {occ}}$ is a projection operator on the occupied states, $\hat{P}_{\rm {occ}} = \sum_i^{\rm occ}\ket{\phi_i^0}\bra{\phi_i^0}$. \par
Sternheimer $G_0W_0$ formalisms differ in their choice of $\Delta V$. 
There are two possible choices for $\Delta V$. 
The first one is to set the perturbation to the bare Coulomb interaction $v(\bfr,\bfrp)$ \cite{Reining1997}
\begin{equation}
 \Delta V (\bfr,\bfrp) = v(\bfr,\bfrp).
\end{equation}
This choice is known as non-self-consistent Sternheimer $GW$. 
The Sternheimer $G_0W_0$ formalism is shown in Figure~\ref{fig:sternheimer}. Quantities that depend on $\bfrp$ are expanded in a basis $\{\varphi_k(\bfrp)\}$, see Equation\eqref{eq:expansion}, and both sides of Equation~\eqref{eq:sternheimer} are projected onto $\varphi_l(\bfrp)$. This leads to a linear set of equations with a parametric dependence on $\bfr$ and $\pm\omega$. Solving the Sternheimer equation for each real-space grid point $\bfr$ and the frequencies $\pm\omega$ yields $\Delta \phi_i^0(\bfr,\bfrp,\pm\omega$) for the occupied state $i$. From the latter we can evaluate the induced charge density $\Delta n$. The dielectric function given in Equation~\eqref{eq:epsilon} can be rewritten in terms of $\Delta n$ \cite{Reining1997,Lambert/Giustino:2013}
\begin{equation}
    \varepsilon(\bfr,\bfrp,\omega) = \delta(\bfr,\bfrp) - \Delta n(\bfr,\bfrp,\omega).
\end{equation}
$W_0$ is then calculated from the inverse of $\varepsilon$ according to Equation~\eqref{eq:W0} as usual.\par
The second choice for $\Delta V$ is to set it to the screened Coulomb interaction
\begin{equation}
    \Delta V(\bfr,\bfrp,\omega)=W_0(\bfr,\bfrp,\omega),
\end{equation}
leading to the self-consistent Sternheimer $GW$ formalism introduced by \onlinecite{Giustino2010}. In this scheme, the (self-consistent) induced charge density $\Delta n^{\rm {SC}}$ generates a potential $\Delta V_{\rm{scr}}$, which screens the bare Coulomb potential $v$ due to the perturbative charge in the system. From $\Delta V_{\rm{scr}}$ we can directly calculate $W_0$ as
\begin{align}
    \Delta V_{\rm{scr}} (\bfr,\bfrp,\omega) &= \int d\mathbf{r}''\Delta n^{\rm {SC}} (\bfr,\mathbf{r}'',\omega) v(\mathbf{r}''\bfrp)\\
    W_0(\bfr,\bfrp,\omega) &= v(\bfr,\bfrp) +  \Delta V_{\rm{scr}}(\bfr,\bfrp,\omega).\label{eq:W_sternheimer_sc}
\end{align}
It can be shown that Equation~\eqref{eq:W_sternheimer_sc} is equivalent to Equation~\eqref{eq:W0} \cite{Giustino2010}. Since $W_0$ appears on the right-hand side of Equation~\eqref{eq:sternheimer}, it must  be solved self-consistently. In the first step, $W_0$ is initialized with $v$. From the solutions of the Sternheimer equation, we calculate $\Delta n^{\rm {SC}}$, $\Delta V_{\rm{scr}}$ and finally $W_0$.\par
Both schemes yield $W_0$, but the non-self-consistent approach requires fewer steps. However, the dimensions of the dielectric matrix increase with system size and its inversion might become a bottleneck for large systems, in particular when using plane wave basis sets. In this case the self-consistent scheme might be more efficient.\par
The two schemes discussed so far address the elimination of empty states in $W_0$. Removing the sum over virtual states in $G_0$ is also possible by using a similar strategy as for $W_0$, see \onlinecite{Giustino2010} for a detailed description. Once $G_0$ and $W_0$ have been obtained, the self-energy is composed as usual and the frequency integration is performed with the methods described in Section~\ref{subsec:g0w0freq}. Sternheimer approaches have been implemented  for plasmon pole models \cite{Reining1997}, the analytic continuation \cite{Giustino2010} and contour deformation \cite{Govoni/etal:2015}. \par
The Sternheimer $GW$ approach is primarily used in plane wave implementations \cite{Lambert/Giustino:2013,Reining1997,Govoni/etal:2015,Nguyen2012,Pham/etal:2013}.  We are only aware of one  non-plane wave implementation using mixed representations of real space and localized basis sets \cite{Huebener2012a,Huebener2012b}.
As discussed in Section~\ref{subsec:basis_convergence}, converging a $G_0W_0$ calculation with plane waves requires a very large number of empty states. The calculation of all empty states in the preceding DFT or HF calculation is computationally expensive and can easily become a computational and storage bottleneck. In the Sternheimer approach, the preceding DFT step is significantly simplified since only occupied states have to be calculated. For localized basis sets, no such benefit is found in DFPT or Sternheimer since the number of virtual states is typically not that large and rarely a bottleneck \cite{Shang/etal:2018}.\par
Sternheimer $G_0W_0$ saves not only computational time in the preceding mean-field calculation, but also by not having to carry out the sums over states in $G_0$ and $\chi_0$. However, it concomitantly loses time in the Sternheimer iterations. To our knowledge, a detailed comparison of the computational cost to conventional $G_0W_0$ implementations has not been reported yet. To speed up Sternheimer $G_0W_0$, projection techniques for representing the dielectric matrix in an optimal, smaller basis \cite{Nguyen2012,Pham/etal:2013,Govoni/etal:2015,Wilson2008,Wilson2009} and Lanczos-chain algorithms that efficiently obtain the Sternheimer solution over a broad frequency range \cite{Umari/Stenuit/Baroni:2010} have been developed. Furthermore, all the Sternheimer equations, that need to be solved for each $\bfr$ and $\omega$, are independent from each other facilitating massively parallel implementation \cite{Govoni/etal:2015,Schlipf2019}. \par
The Sternheimer approach does not reduce the basis set size, i.e., the plane wave cutoff or equivalently the size of the real-space grid, nor does it change the formal scaling of $G_0W_0$ with respect to system size. However, it is an elegant way to facilitate easier convergence, since the temptation of converging the number of virtual states separately is removed.\par
A modified Sternheimer ansatz has been developed for FLAPW basis sets which accounts for response contributions outside the Hilbert space spanned by the basis set \cite{Betzinger2012,Betzinger2013,Betzinger2015}. 
This modified approach thus allows the basis set size to be decreased, unlike the classical Sternheimer technique. The explicit summation over unoccupied states is not completely removed, but the number of empty states needed for convergence is strongly reduced.

\subsection{Starting point dependence and optimization}
\label{subsec:g0w0start}
The results of a $G_0W_0$ calculation depend on the wave functions $\{\phi_s^0\}$ and the energies $\epsilon_s^0$ that are used as input for the Green's function ($G_0$) and the screened Coulomb interaction 
($W_0$). The single-particle wave functions and energies are determined by the choice of the single-particle mean-field Hamiltonian $\hat{h}^{\mathrm{MF}}$, e.g., by the chosen DFT functional. To denote this dependence, we will introduce the notation $G_0W_0$@$starting\: point$.

\subsubsection{How severe is the dependence on the reference state?}
\begin{figure} 
        \includegraphics[width=0.99\linewidth]{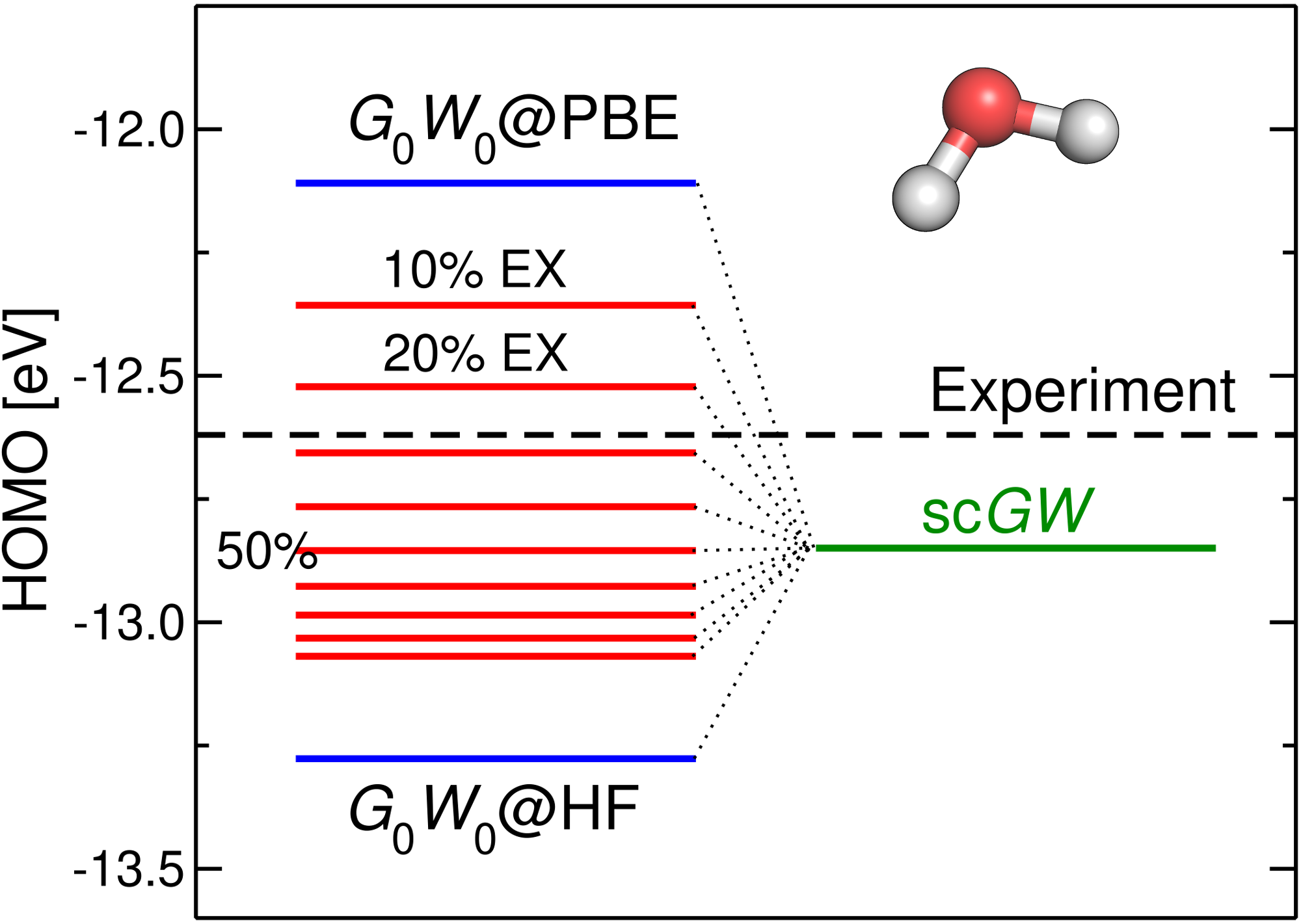}
	\caption{{\small \label{fig:spd_H2O}
        	 Starting point dependence of $G_0W_0$: the left side shows the $G_0W_0$ HOMO energy of the water molecule for hybrid functional starting points with different amounts of exact 
             exchange. The HOMO energy in self-consistent $GW$ (sc$GW$) is shown on the right. The dashed line marks the experimental value of 12.62~eV \cite{Page1988,NIST}. All $GW$ values are extrapolated to the exact basis set limit using the cc-pV$n$Z ($n$=3-5) basis sets. Further computational details are given in Appendix~\ref{app:computational_details}. }}
\end{figure}

%
Until the early 2000s, the majority of all $G_0W_0$ calculations were based on DFT 
calculations using the local-density approximation (LDA) or a generalized gradient approximation (GGA) \cite{Aulbur/Jonsson/Wilkins:2000,Aryasetiawan/Gunnarsson:1998,Onida/Reining/Rubio:2002}. With the advent of exact-exchange based DFT functionals in the solid state community and the proliferation of $G_0W_0$ codes that are based on quantum chemical codes, a more diverse range of starting 
points became available. It was soon realized that $G_0W_0$ can exhibit a pronounced starting-point dependence for semiconductors
\cite{Rinke/etal:2005,Fuchs/etal:2007}. In the last years, the starting point dependence has also been intensively discussed for molecules 
\cite{Koerzdoerfer/Marom:2012,Marom/etal:2012,Bruneval/Marques:2013,Gallandi2015,Gallandi2016,Caruso2016}.\par
Figure~\ref{fig:spd_H2O} illustrates the starting-point dependence for the HOMO of the water molecule. For the underlying DFT calculations, the PBE-based hybrid (PBEh) functional family \cite{PBE0_1,PBE0_2,PBE0_3} was scanned. The PBEh functional family is characterized by an adjustable fraction $\alpha$ of HF exchange. The exchange-correlation energy $E_{\mathrm{xc}}$ is therefore $\alpha$-dependent and given by
\begin{equation}
 E_{\mathrm{xc}} = \alpha E_x^{\mathrm{EX}} + (1-\alpha)E_{x}^{\mathrm{PBE}} + E_{c}^{\mathrm{PBE}}, \quad \alpha\in[0,1],
\end{equation}
where $E_x^{\mathrm{EX}}$ denotes the HF exchange energy. $E_{x}^{\mathrm{PBE}}$ and $E_{c}^{\mathrm{PBE}}$ are the PBE exchange and correlation energy, respectively. To illustrate the starting point dependence in $G_0W_0$, the mixing parameter $\alpha$ in PBEh was varied from 0 to 1 and then a subsequent $G_0W_0$ calculation was performed. The resulting $G_0W_0$ HOMO energies shown in Figure~\ref{fig:spd_H2O} span a range of more than 1~eV. Although a 1~eV spread appears large, it is much smaller than the range of the corresponding mean-field energies $\epsilon^0_{\rm{HOMO}}$ that decrease from -7~eV to -14~eV with increasing $\alpha$. \par
The strong dependence of $G_0W_0$ on the starting point can be largely attributed to over- and under-screening. From the Adler-Wiser expression for $\chi_0$ (Equation~\eqref{eq:chi0}) it can be deduced that the screening strength in $G_0W_0$ is inversely proportional to the eigenvalue gap of the starting point. Since HF typically overestimates gaps, it will under-screen. PBE, on the other hand, underestimates gaps and therefore over-screens. 

Another source of error in the KS orbital energies is the spurious self-interaction term \cite{Perdew/Zunger:1981}. The one-electron self-interaction error (SIE) arises from an incomplete cancellation of the unphysical electrostatic Hartree energy  of an electron with itself by the exchange-correlation term. The SIE is more pronounced for localized than delocalized orbitals \cite{Koerzdoerfer2009}. This can be intuitively understood: an electron in a localized orbital has a stronger self-interaction because its wave function is more confined. As a result, the localized orbitals are destabilized with respect to more delocalized orbitals. This can lead to a wrong ordering of the orbital energies in the underlying DFT calculations, which carries over to the $GW$ spectrum. The SIE can be mitigated by a larger amount of exact exchange, which also restores the correct ordering for the QP energies \cite{Marom2011,Marom/etal:2012,Koerzdoerfer/Marom:2012}.\par
The $G_0W_0$ starting point dependence generally lies in the range of 1.0~eV for HOMO energies of molecules \cite{Marom/etal:2012}, but increases for deeper states. For solids, the spread can exceed 2~eV for the band gap \cite{Fuchs/etal:2007}. This beckons for a judicious choice of the starting point in $G_0W_0$ calculations or an elimination of the starting point dependence. The dependence on the preceding mean-field calculation can be eliminated or reduced by employing some form of self-consistency as discussed in Section~\ref{sec:scgw} or, as proposed only very recently, by replacing $G_0$ by a renormalized singles Green's function \cite{Jin2019}. In this section we focus on the optimal choice of the starting point. The PBEh family of DFT functionals is convenient for this porpose, since one parameter (the amount of exact exchange $\alpha$) governs the behavior of the starting point. Several schemes have been developed to find the optimal $\alpha$ value within the PBEh functional family \cite{Koerzdoerfer/Marom:2012,Pinheiro2015,Dauth2016,Bois2017,Atalla2013}. We summarize them in the following.

\subsubsection{Consistent starting point scheme}
\begin{figure} 
        \includegraphics[width=0.99\linewidth]{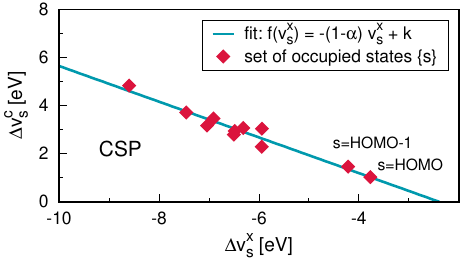}
	\caption{{\small \label{fig:csp}
        	 CSP scheme representative for a small molecule. $\Delta v_s^c$ (Equation~\eqref{eq:vcpbe}) is plotted with respect to  $\Delta v_s^x$ (Equation~\eqref{eq:vxpbe}) for a set of occupied states $s$. The HOMO and HOMO-1 states are indicated.  The new $\alpha$ value is obtained from the slope of the straight line fitted through the red symbols. Data retrieved from Ref.~\cite{Koerzdoerfer/Marom:2012}.}}
\end{figure}
%
K\"orzd\"orfer and Marom developed a consistent starting point (CSP) scheme that seeks a PBEh reference state (i.e. starting point) that best resembles the $G_0W_0$ spectrum. Splitting both the $G_0W_0$ 
self-energy and the starting hybrid functional into their respective exchange and correlation parts (see Equations \eqref{eq:sigma_x} and \eqref{eq:sigma_c})  allows us to rewrite the QP equation (Equation 
\eqref{Eq:qpe}) as follows \cite{Koerzdoerfer/Marom:2012,Marom:2017}
\begin{align}
 \epsilon_{s} ={}& \epsilon_s^0 +(1-\alpha) \Delta v_{s}^\mathrm{x}+ \Delta v_{s}^\mathrm{c} \label{eq:KoerzMaromalpha}\\
 \Delta v_{s}^\mathrm{x} :={}& \bracket{\phi_{s}^0}{ \Sigma^{\mathrm{x}}-v_\mathrm{x}^\mathrm{PBE}}{\phi_{s}^0} \label{eq:vxpbe}\\
 \Delta v_{s}^\mathrm{c} :={}& \bracket{\phi_{s}^0}{ \Sigma^{\mathrm{c}}-v_\mathrm{c}^\mathrm{PBE}}{\phi_{s}^0} . \label{eq:vcpbe}
\end{align}
$v_\mathrm{x}^\mathrm{PBE}$ and $v_\mathrm{c}^\mathrm{PBE}$ are the exchange and correlation part of the PBE exchange-correlation potential, respectively. The optimal $\alpha$ is determined so that the shift between $G_0W_0$ and PBEh for the occupied states is approximately a constant $k$
\begin{align}
\Delta v_{s}^\mathrm{c} +(1-\alpha) \Delta v_{s}^\text{x}\approx k,\quad s \in \text{occ}.
\label{eq:csp_condition}
\end{align}
If Equation~\eqref{eq:csp_condition} is satisfied, the positions of the PBEh orbital energies relative to each other are as close as possible to the $G_0W_0$ energies. In this case, the QP correction amounts to a rigid shift of the PBEh spectrum. The value of $\alpha$ for which Equation~\eqref{eq:csp_condition} is satisfied yields the optimal starting point in the CSP scheme. If the PBEh and the $G_0W_0$@PBEh spectrum matched perfectly, the constant $k$ would be zero. However, in general it is not possible to find a starting point whose spectrum matches the $G_0W_0$ spectrum exactly.\par
For a given guess of $\alpha$,  $\Delta v_{s}^\mathrm{x}$ and $\Delta v_{s}^\mathrm{c}$ are calculated according to Equations~\eqref{eq:vxpbe} and \eqref{eq:vcpbe}. $\Delta 
v_{s}^\mathrm{c}$ is plotted as a function of $\Delta v_{s}^\mathrm{x}$ for a set of occupied states $s$ as data points, see Figure~\ref{fig:csp}. A straight line fit determines a new $\alpha$ which is used to calculate new DFT eigenvalues and orbitals from PBEh($\alpha_{\textnormal{new}})$. From the new eigenvalues and orbitals, a new self-energy is calculated and Equations~\eqref{eq:vxpbe} and \eqref{eq:vcpbe} are reassessed. This procedure is continued until the $\alpha$ of the linear fit equals the initial $\alpha$. Then the optimal $\alpha$ has been found. \par
By construction, 
the PBEh($\alpha$) eigenvalues are now, up to the rigid shift $k$, consistent with the $G_0W_0$ spectrum. Typical CSP $\alpha$ values
lie in the range of $0.25-0.30$.  The CSP scheme has been tested on several organic molecules that are used in organic electronics and yields good agreement with photoemission spectra in all cases \cite{Koerzdoerfer/Marom:2012,Koerzdoerfer/Marom:2012_2,Marom:2017}.

\subsubsection{Deviation from the straight line  scheme}
\begin{figure} 
        \includegraphics[width=0.99\linewidth]{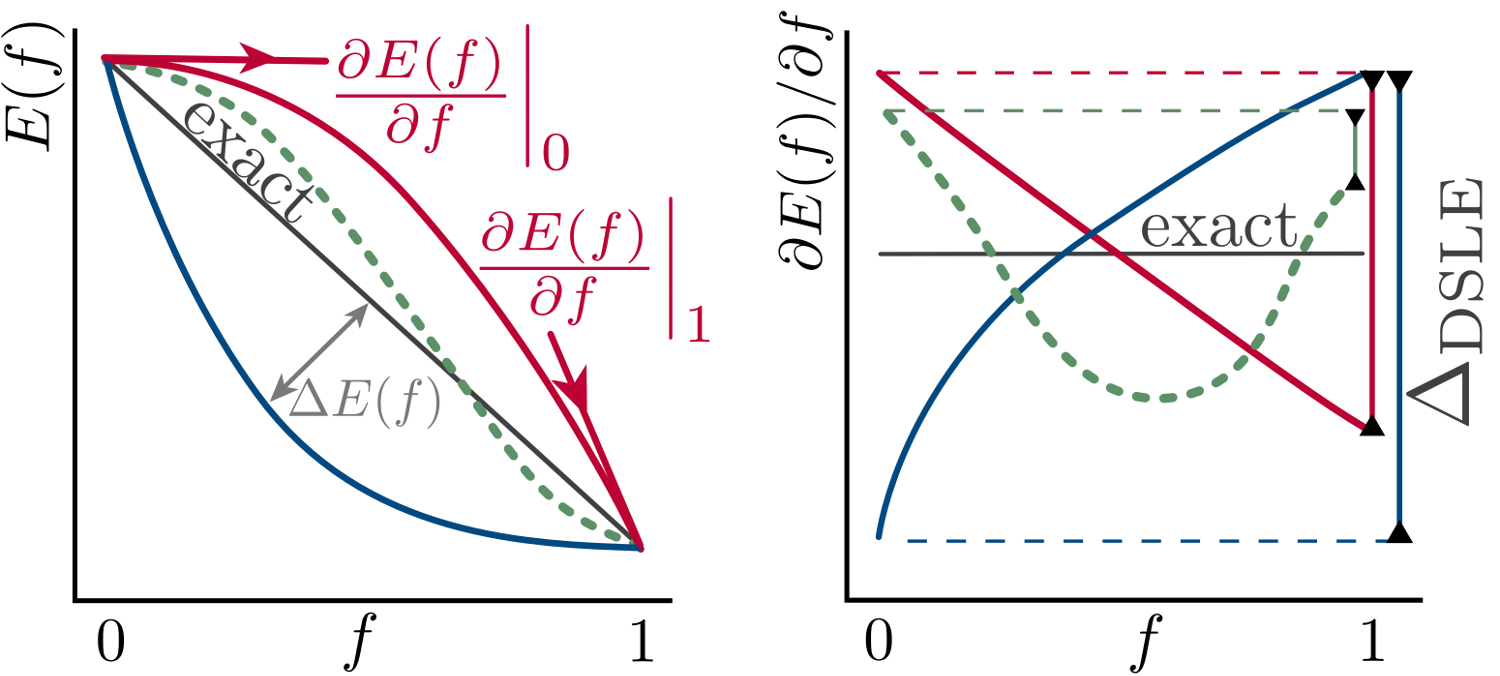}
	\caption{{\small \label{fig:DSLE}
        	 Schematic representation of the straight line condition for total energies $E$ (left) and derivatives $\partial E/ \partial f$ (right). $f$ is the occupation number. The DSLE is shown for three different cases: convex (blue), concave (red) and mixed curvature (green). Figure reprinted from Ref.~\cite{Dauth2016}.}}
\end{figure}

A physically more rigorously motivated optimization scheme is based on the deviation from the straight line error (DSLE) \cite{Dauth2016}. In 1982, Perdew and co-workers showed that the total energy $E$ of any many-electron system should change linearly when varying the electron number continuously from $N$ to $N-1$ electrons \cite{Perdew/etal:1982},
\begin{equation}
 E(f) = (1-f) E(N-1) + fE(N) \qquad f\in[0,1].
\end{equation}
The function $E(f)$ is a piecewise linear function of the occupation number $f$, with cusps at every integer value of $f$, see Figure~\ref{fig:DSLE}. Standard DFT functionals, however, violate this piecewise linearity condition and yield energies that deviate from the straight line at fractional occupation numbers $f$ \cite{Mori-Sanchez2006,Ruzsinszky2006,Kraisler2013}. The straight-line condition applies not only to DFT, but to any total energy method (we will address $GW$ total energies in Section~\ref{sec:gs}). Therefore, we can use the DSLE to find an optimal starting point for $G_0W_0$.  \par
The slope of the total energy as a function of occupation gives the Kohn-Sham eigenvalue or, in the $GW$ case, the quasiparticle excitation energy for a given electron number,
\begin{align}
    \frac{\partial E}{\partial f}\bigg|_{f=1}= E(N) - E(N-1) ={}& \epsilon_{\rm HOMO,N}\label{eq:dsleslopef0}\\
    \frac{\partial E}{\partial f}\bigg|_{f=0}= E(N) - E(N-1) ={}& \epsilon_{\rm LUMO, N-1}\label{eq:dsleslopef1},
\end{align}
where $\epsilon_{\rm HOMO,N}$ is the QP energy of the HOMO for the neutral system ($N$). $\epsilon_{\rm LUMO, N-1}$ denotes the QP energy of the lowest unoccupied orbital (LUMO) for the charged system ($N-1$).
It is evident from Equation~\eqref{eq:dsleslopef0} and \eqref{eq:dsleslopef1} that the slopes must be identical for $f=0$ and $f=1$ in an exact theory. In other words, a necessary condition for piecewise linearity is that the energy for removing an electron from the neutral system equals the energy for adding an electron to the positively charge system, i.e., the IP for the neutral systems and the electron affinity (EA) of the charged system should be equal. The difference  
\begin{align}
 \Delta_{\text{DSLE}} & = \text{EA}_{\text{LUMO}}(N-1) -\text{IP}_{\text{HOMO}}(N) \\
 &=- \epsilon_{\text{LUMO,N-1}}+\epsilon_{\text{HOMO,N}}\label{eq:DSLE}
\end{align}
should thus be zero and if it is not zero it quantifies the deviation from the straight line error $\Delta_{\text{DSLE}}$.\par
The idea is now to find a PBEh starting point for which $G_0W_0$@PBEh minimizes $\Delta_{\text{DLSE}}$. The optimal $\alpha$ value for PBEh can be found by the following procedure: We select a set of PBEh functionals with $\alpha$ values between 0 and 1. Then two $G_0W_0$ calculations are performed for each starting point. One for the neutral system that yields $\epsilon_{\text{HOMO,N}}$ and a second one for the cation to obtain  $\epsilon_{\text{LUMO,N-1}}$. Following Equation~\eqref{eq:DSLE}, we calculate the difference between these two energies to estimate the deviation from the straight line condition. The PBEh($\alpha$) functional that yields the smallest $\Delta_{\text{DSLE}}$ will be the optimal starting point.\par
This DSLE scheme has been tested for small molecular systems, where it has been shown that the optimal $\alpha$ values are distributed around $0.35-0.40$ for the first IP \cite{Dauth2016,Caruso2016}. The reported deviation from the CCSD(T) reference is smaller than 0.2~eV \cite{Caruso2016}. The drawback of the DSLE scheme is that the optimization is restricted to the HOMO. For other states, the straight line condition could still be formulated, but the corresponding $G_0W_0$ calculations could not be performed because the electron occupation function would no longer correspond to an equilibrium distribution (see Section~\ref{subsec:non-equilibrium} for $GW$ calculations out of equilibrium). If we removed an electron from a lower lying occupied state, the sums over occupied and virtual orbitals in the polarizability $\chi_0$ would no longer be rigorously defined, i.e. the energy differences in the denominator in Equation~\eqref{eq:chi0} would exhibit the wrong sign. 

%
\subsubsection{IP-theorem schemes}
Several other schemes were developed that are, in spirit, similar to the CSP and DSLE optimization approaches, but are explicitly based on the ionization potential (IP) theorem. The latter states that in exact DFT the negative of the KS orbital can be strictly assigned to the first ionization potential $\text{IP}_{\text{HOMO}}$ \cite{Almbladh/Barth:1985,Levy/Perdew/Sahni:1984} 
\begin{equation}
  \text{IP}_{\text{HOMO}} = -\epsilon_{\text{HOMO}}^{\text{KS}}.
\end{equation}
This statement is not true for any other KS state and not valid for approximate DFT functionals. Atalla \textit{et al.} proposed a scheme that exploits the IP-theorem and minimizes the $G_0W_0$ correction for the HOMO level with respect to the amount of exact exchange $\alpha$ in a PBEh starting point \cite{Atalla2013},
\begin{align}
\Delta v_{\text{HOMO}}^\mathrm{c} +(1-\alpha) \Delta v_{\text{HOMO}}^\text{x} = 0.
\label{eq:IP_condition}
\end{align}

This approach is similar to the CSP scheme in  Equation~\eqref{eq:csp_condition}. The difference is that we find the PBEh functional whose  HOMO energy matches that of $G_0W_0$@PBEh for the same $\alpha$, whereas CSP looks for the closest spectral match between PBEh and $G_0W_0$@PBEh. HOMO excitations obtained from this IP-theorem-tuned scheme 
agree reasonably well with CCSD(T) reference data \cite{Bois2017}, but are not expected to reproduce the whole excitation spectrum properly \cite{Atalla2013}. They generally yield large $\alpha$s (around 0.8) and produce HOMOs and HOMO-LUMO gaps that are too large (i.e. under-screened).

Finding a PBEh($\alpha$) functional that fulfills the IP-theorem by enforcing consistency with the $G_0W_0$ spectrum is one option. An alternative approach is to IP tune the hybrid functionals themselves \cite{Bois2017} by minimizing 
\begin{equation}
 \Delta_{\text{IP}} = \left|\epsilon_{\text{HOMO}}^{\text{KS}}(\alpha)-(E(N,\alpha)-E(N-1,\alpha))\right| 
\end{equation}
with respect to $\alpha$. These IP-tuned hybrids already give accurate KS-HOMO energies. Recent benchmark studies for molecular systems showed that $G_0W_0$ corrections on top of $\Delta_{\text{IP}}$-tuned functionals provide spectral properties in good agreement with experiment for the whole excitation spectrum \cite{Refaely-Abramson2012,Egger2014,Gallandi2015,Gallandi2016,Knight2016,Bois2017}. In particular, EAs are well reproduced with mean-average deviation (MADs) $< 0.2$~eV from  CCSD(T) references, while the MAD reported for IP$_{\text{HOMO}}$ is 0.1~eV \cite{Knight2016}.


%
\subsection{Computational complexity and cost}
\label{subsec:scaling}

Of all the $GW$ schemes described in this review, $G_0W_0$ is the computationally most efficient one. Only the diagonal elements of the self-energy are needed and the Green's function that enters is 
always $G_0$. The fully interacting $G$, on the contrary, depends on the full self-energy and can only be obtained by iterating Dyson's equation $G=G_0+G_0\Sigma G$.\par
The computational complexity of $G_0W_0$ depends on the frequency integration method and design of the algorithm. The most accurate integration technique, the fully-analytic approach, requires the solution of the full Casida equations, which is an $\mathcal{O}(N^6)$ step with respect to the system size $N$, see Section~\ref{subsec:freq_comparison}. In the canonical implementation, the computational cost is reduced to $\mathcal{O}(N^4)$. Different implementations with $N^4$ complexity have been developed employing a variety of numerical techniques specific for the respective basis set \cite{Shishkin06,Blase/Attaccalite/Olevano:2011,Deslippe/etal:2012,Xinguo/implem_full_author_list,Govoni/etal:2015,Wilhelm2016}. For example, the $\mathcal{O}(N^4)$ algorithm proposed by Ren $\textit{et al.}$ employs localized basis functions and the AC method \cite{Xinguo/implem_full_author_list}. The computational and memory costs for the four-center electron repulsion integrals (4c-ERIs) are reduced by refactoring the latter in two- and three-center Coulomb integrals using a resolution-of-the-identity (RI) approach with a so-called Coulomb metric \cite{Vahtras1993}. The accuracy of this algorithm has been validated for valence excitations and EAs by comparing to the fully analytic approach \cite{Setten2015}. However, for core states it was recently shown that AC fails and that CD is required \cite{Golze2018}. The computational complexity of CD remains unchanged for valence states, but increases to $\mathcal{O}(N^5)$ for the deep states.\par
The $\mathcal{O}(N^4)$ scaling and overall cost of canonical $GW$ implementations restricts the tractable system size and prohibits the study of many systems that are relevant in the chemistry and physics community, such as solid-liquid interfaces, molecules in solution, complex alloys, nanostructures or hybrid interfaces, that require large simulation cells with hundreds to thousands of atoms. To make $G_0W_0$ calculations feasible for larger systems, the scaling and computational complexity have been scrutinized. Developments have proceeded in two directions: 1) reducing the prefactor, i.e. the overall computational cost at $\mathcal{O}(N^4)$ scaling, or 2) reducing the scaling.\par
\begin{figure} 
        \includegraphics[width=0.99\linewidth]{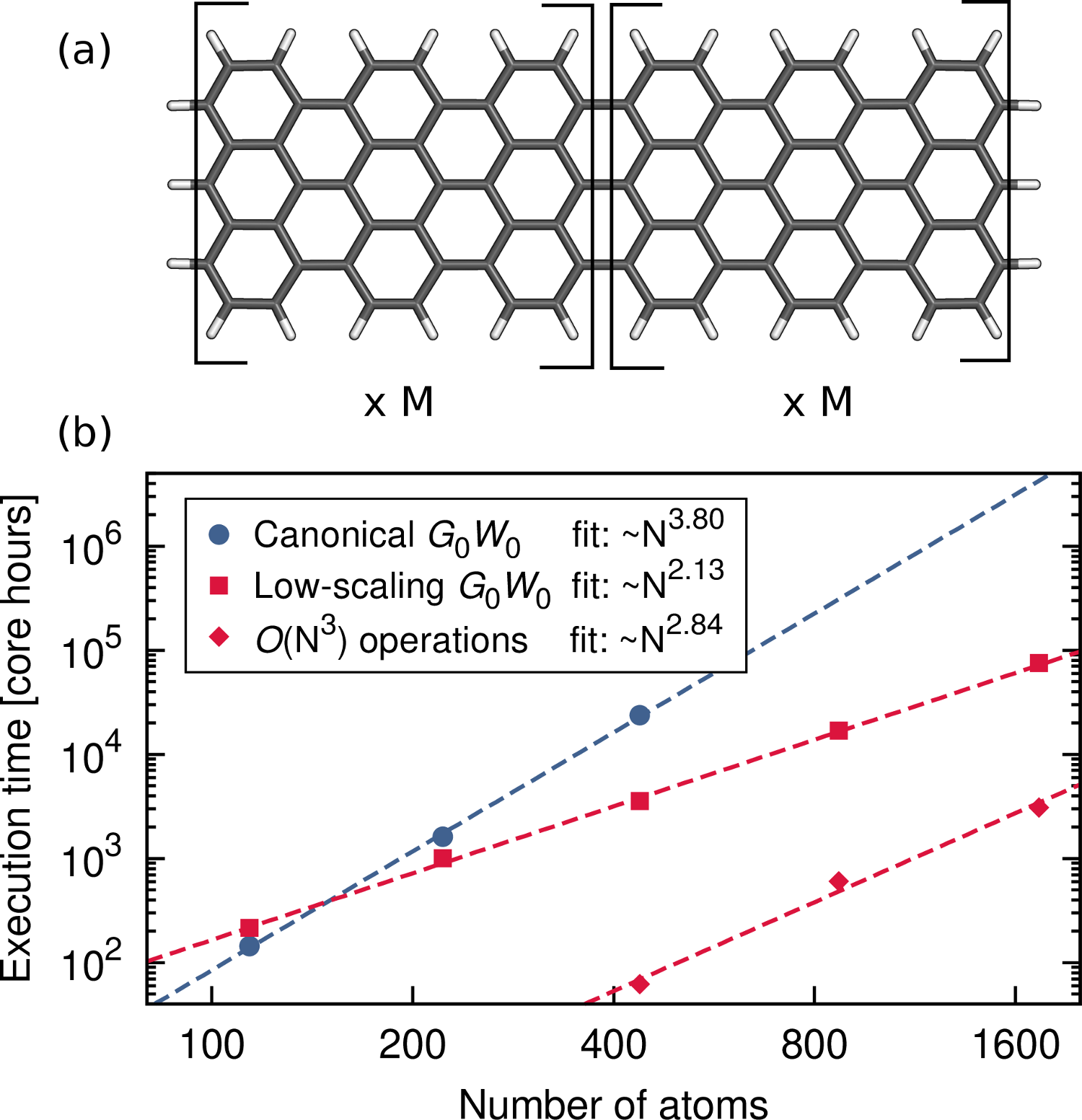}
	\caption{{\small \label{fig:scaling}
        	Scaling of state-of-the-art $G_0W_0$ implementations with respect to system size using graphene nanoribbons as a benchmark system. (a) Smallest graphene nanoribbon unit with 114 atoms. 
(b) Comparison  of the scaling of the canonical $G_0W_0$ \cite{Wilhelm2016} and the low-scaling implementation \cite{Wilhelm2018}. The latter requires operations of at most $\mathcal{O}(N^3)$ complexity (red diamonds). Dashed lines represent least-square fits of exponent and prefactor. Data retrieved from Ref.~\cite{Wilhelm2018}. Both algorithms are implemented in the CP2K program package. 
    }}
\end{figure}
The prefactor has been reduced by low-rank approximations of $\chi_0$, which map $\chi_0$ onto a smaller basis \cite{Wilson2008,Wilson2009,Govoni/etal:2015,DelBen2019,Umari/Stenuit/Baroni:2009}. Another approach is the elimination of the sum over empty orbitals in $\chi_0$ and in $G_0$ \cite{Govoni/etal:2015,Umari/Stenuit/Baroni:2010,Lambert/Giustino:2013,Giustino2010,Pham/etal:2013} by solving the Sternheimer equation \cite{Sternheimer1954}, which we discussed in Section~\ref{subsec:sternheimergw}. Others developed techniques to reduce the number of unoccupied states \cite{Bruneval2008,Bruneval2016a}.
%
The prefactor can also be controlled by choosing an optimal basis set for the respective system under study. In the last years, several algorithms for localized basis sets have been developed \cite{Blase/Attaccalite/Olevano:2011,KeSanhuang:2011,Xinguo/implem_full_author_list,vanSetten/etal:2013,Wilhelm2016,Bruneval2016}. These basis sets are generally smaller than traditional plane wave basis sets and considerably more efficient for molecules. However, the development of reliable $G_0W_0$ algorithm for periodic systems based on localized basis sets is still underway \cite{Wilhelm2017}.\par
The reduction of the exponent to $\mathcal{O}(N^3)$ complexity has been addressed in different ways. Foerster \textit{et al.} developed a cubic-scaling $G_0W_0$ algorithm using Gaussian basis sets and exploiting locality in the electronic structure, albeit with a high prefactor \cite{Foerster/etal:2011}. Recently, two cubic-scaling algorithms have been devised \cite{Liu2016,Wilhelm2018}, which are both variants of the $\mathcal{O}(N^3)$ $GW$ space time method \cite{Rojas/Godby/Needs:1995}. The key step of these algorithms is the computation of the irreducible polarizability in imaginary time, $\chi_0(it) = -iG_0(it)G_0(-it)$ and the subsequent transformation to imaginary frequencies $i\omega$. The time-ordered non-interacting Green's function in imaginary time is given by
\begin{align}
\begin{split}
G_0(\bfr,\bfrp,it)= \left\{ 
\begin{array}{ll}
  i\sum\limits_i^\text{occ} \phi_i^0(\bfr)\phi_i^{0*}(\bfrp)\exp(-\epsilon_i^0t)\,, &  t <0\,,
\\[0.5em]
 - i\sum\limits_a^\text{virt} \phi_a^0(\bfr)\phi_a^{0*}(\bfrp)\exp(-\epsilon_a^0t)\,, &  t >0\,.
\end{array}
\right.
\end{split}
\label{Greens_time}
\end{align}
Inserting $G_0(it)$, the summation over occupied and virtual states is now decoupled in $\chi_0(it)$ and can be performed separately, which is fundamental for the reduction to $\mathcal{O}(N^3)$ complexity. \par
Liu \textit{et al.} based their cubic-scaling algorithm on a plane wave basis set in combination with a PAW scheme and reported a linear-scaling with the number of $\mathbf{k}$ points used to sample the Brillouin zone \cite{Liu2016}. In combination, this paves the way for $GW$ calculations of large periodic systems. Wilhelm \textit{et al.} employed a Gaussian basis set and exploited sparse matrix algebra by using an overlap metric for the RI approximation (RI-SVS) to refactor the 4c-ERIs \cite{Vahtras1993}. The step with the largest prefactor, the computation of $\chi_0$, is reduced from $\mathcal{O}(N^4)$ to $\mathcal{O}(N^2)$ in this algorithm, while the other operations scale with $N^3$. A comparison between the $\mathcal{O}(N^4)$ algorithm developed by Ren \textit{et al. } \cite{Xinguo/implem_full_author_list} and the low-scaling algorithm is shown in Figure~\ref{fig:scaling} for graphene nanoribbons. The canonical algorithm is restricted to system sizes of less than 500 atoms, while systems with more than 1600 atoms can be addressed with the low-scaling implementation. These are some of the largest $G_0W_0$ calculations with high accuracy and full-frequency integration reported so far. The mean absolute deviation with respect to the canonical reference implementation in FHI-aims \cite{Xinguo/implem_full_author_list} is less than 35~meV for the $GW$100 test set \cite{Wilhelm2018}, which is discussed more in detail in Section~\ref{subsec:gw100}.\par
An actual linear scaling algorithm was devised within the framework of stochastic $GW$\cite{Neuhauser/etal:2014} and applied to silicon clusters with 1000 atoms. However, the verification of its general reliability is still the subject of ongoing research \cite{Vlcek2017}.

\subsection{Practical guidelines}
In summary, the following points should be considered when conducting $G_0W_0$ calculations:
\begin{enumerate}
    \item \textit{Frequency integration technique}\\
          A sufficiently accurate method for the frequency integration of Equation\eqref{Eq:S=G0W0_rsp} has to be chosen. The required precision depends on the systems and in particular on the states of interest, see Section~\ref{subsec:g0w0freq}.
    \item \textit{Basis set choice}\\
          The decision should be guided by the system of interest. Localized basis sets are generally more efficient for finite systems, while plane wave $G_0W_0$ codes are currently superior for extended systems (see detailed discussion in Section~\ref{sec:basis_sets}).
    \item \textit{Basis set convergence}\\
          $GW$ calculations have to be carefully converged with respect to basis set size. Extrapolation procedures to the complete basis set limits might be required as demonstrated in Section \ref{subsec:basis_convergence}.
    \item \textit{Starting point}\\
          The QP energies depend strongly on the functional of the preceding DFT calculation as shown in Section~\ref{subsec:g0w0start}. While for solid state systems GGA functionals are often suitable starting points (see also Section~\ref{sec:solids}), hybrid functionals perform better for molecules. A judicious choice of the starting point is necessary.
   \item \textit{Convergence of technical parameters}\\
          $G_0W_0$ calculations usually require the convergence of a few additional parameters, which are strongly implementation dependent. Such a parameter is, e.g., the size of the integration grid for the imaginary frequency terms in the CD and AC approach. $G_0W_0$ practitioners should always carefully check their $GW$ code to ensue the robustness and convergence of all available settings and parameters
\end{enumerate}

The $G_0W_0$ approximation provides computationally efficient access to the whole QP spectrum. Despite these appealing features $G_0W_0$ has certain drawbacks. The most severe is the dependence on the starting point discussed in Section~\ref{subsec:g0w0start}. Furthermore, the ground state energy and density cannot be computed. In Sections~\ref{sec:scgw} and \ref{sec:gs}, we will show how these drawbacks can be overcome.

\section{Beyond $\bm{G_0W_0}$: self-consistency schemes}
    \label{sec:scgw}
    
\begin{figure}
	\includegraphics[width=0.97\linewidth]{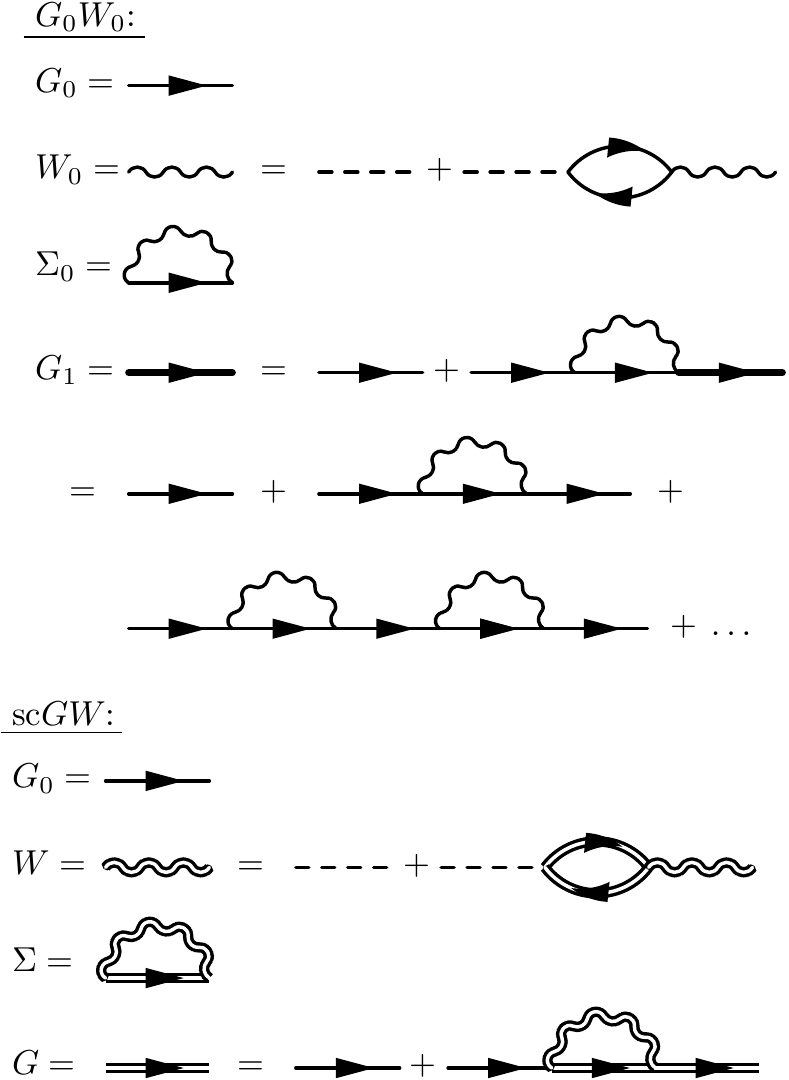}
	\caption{ \label{fig:diag_scGW}
	        	 $G_0W_0$ and sc$GW$ in terms of Feynman diagrams. In $G_0W_0$, the irreducible self-energy is constructed from $G_0$ and $W_0$. The Green's function updated with the lowest order self-energy, $G_1$ (shown as the bold Green's function line), contains an infinite series of $\Sigma$ insertions. In fully self-consistent $GW$, the starting point dependence is removed and all quantities in the diagrammatic expansion are fully dressed. Here, we assume a true $G_0$ starting point instead of a mean-field $G_0$ so that subtraction of $v^{\mathrm{MF}}$ is not necessary to include in the diagrams.}
\end{figure}

\subsection{Fully self-consistent $\bm{GW}$}
	  \label{sec:scGW}

To go beyond $G_0 W_0$, one must include some level of self-consistency in Hedin's $GW$ equations. The conceptually purest approach to $GW$ is to perform full self-consistency in the $GW$ equations, denoted as sc$GW$. It is also the most expensive. As introduced in Section~\ref{sec:GW}, all four quantities are iterated until self-consistency in the Green's function is achieved. Until now, self-consistent $GW$ is the rarest form of $GW$ for reasons of computational expense and conceptual controversy (see below), although that is slowly changing. 

The first sc$GW$ calculation was performed for the homogeneous electron gas (HEG) by \onlinecite{Holm/vonBarth:1998}, after the same authors had previously applied partial self-consistency (sc$GW_0$) \cite{Holm/vonBarth:1996}. Later studies were extended to the 2D HEG \cite{Garcia-Gonzalez/Godby:2001}. sc$GW$ deteriorates the spectral properties of the HEG compared to $G_0W_0$. This deterioration manifests itself in a quasiparticle bandwidth that is larger than the free electron one and a broad and featureless satellite spectrum. Both results contradict experimental evidence for alkali metals which are HEG-like \cite{Holm/vonBarth:1996}. Also, band gaps of simple semiconductors are greatly overestimated by sc$GW$ \cite{Schoene/Eguiluz:1998,Kresse/etal:2018}. sc$GW$ calculations for atoms \cite{Stan/etal:2009}  and molecules \cite{Rostgaard/Jacobsen/Thygesen:2010,Caruso/etal:2012,Marom/etal:2012,Caruso/etal:2013_H2,Caruso/etal:2013_tech} show improvements over $G_0W_0$ for the first ionization energies and transport properties \cite{Strange/etal:2011} of finite systems. With regard to the whole spectrum, however, sc$GW$ is usually outperformed by $G_0W_0$ with a judicious starting-point choice \cite{Marom/etal:2012,Caruso/etal:2013_tech,Knight2016}.

sc$GW$ is computationally more demanding than $G_0W_0$ because the full Green's function must be stored and calculated \cite{Caruso/etal:2013_tech}, increasing memory and computation requirements. In $G_0W_0$, the full Green's function is only required in $\mathcal{O}(N^3)$ schemes (see Section~\ref{subsec:scaling}). Other implementations make use of the fact that intermediate quantities can be expressed in terms of the mean-field wave functions and eigenvalues, which reduces the computational complexity (see Section~\ref{subsec:g0w0equations}). Furthermore, iterations of the $GW$ equations for sc$GW$ are expensive. $\chi_0$, $W$, and $\Sigma$ must be computed at each iteration, with a potentially high computational time for even a single evaluation of $\Sigma$.

Conceptually, the additional self-consistency in the Green's function adds more reducible diagrams compared to $G_0W_0$, as Figure~\ref{fig:diag_scGW} illustrates. In $G_0W_0$, the bare Coulomb interaction is screened by a series of sequentially interacting electron-hole pairs, or ``bubbles.'' In sc$GW$, this structure is preserved, but the bubbles are now composed of interacting Green's function lines instead of non-interacting $G_0$ lines. This effect is a general feature of iterating Green's function diagrams. By iterating diagrams for a given approximation, initial $G_0$ lines at internal times become interacting $G$ lines. Already after the first iteration of the cycle ($G_1$ in Figure~\ref{fig:diag_scGW}), the Green's function contains sequential self-energy insertions.

Let us look more closely at how this occurs. Recall from Equation~\eqref{eq:dyson} that Dyson's equation is
\begin{align}
  G(1,2)&= G_0(1,2) \\ 
  &+\int G_0(1,3)   \Sigma(3,4)  G(4,2)d(3,4) \nonumber .  
\end{align}
The first guess at the full $G$, labeled $G_1$, would then be
\begin{align}
  \label{eq:g1}
  G_1(1,2)&= G_0(1,2) \\  
   &+\int G_0(1,3)  \Sigma_0(3,4)  G_1(4,2)d(3,4) \nonumber ,  
\end{align}
where we have inserted $G_1(4,2)$ in place of $G(4,2)$ on the right-hand-side (RHS). $\Sigma_0$ is the first estimate to the self-energy, evaluated with $G_0$ wherever $G$ lines enter the self-energy diagram. $G_1$ appears on both sides of Equation~\eqref{eq:g1} $-$ by replacing $G_1(4,2)$ on the RHS with the \textit{entire} RHS, one can generate a reducible diagram for $G_1$ that is $\mathcal{O}(\Sigma_0^2)$.
At this point, the series for $G_1$ contains three parts: $G_0$, a term of order $\mathcal{O}(\Sigma_0)$, and a term of order $\mathcal{O}(\Sigma_0^2)$. These contributions to $G_1$ are shown in  Figure~\ref{fig:diag_scGW}. 
By further iterating $G_1$ on the RHS, one can generate all reducible diagrams which contribute to $G_1$. Despite the infinite number of reducible diagrams generated by this prescription, $G_1$ is still computed only with the $G_0W_0$ self-energy because we have not updated $\Sigma_0$. This example also demonstrates why it is conceptually much simpler to work only with the irreducible self-energy and avoid this infinite, reducible series. Indeed, iterating Equation~\eqref{eq:g1} to find $G_1$ \textit{while keeping $\Sigma_0$ fixed} is equivalent to generating the entire reducible series for $G_1$.

\begin{figure}
	\includegraphics[width=0.96\linewidth]{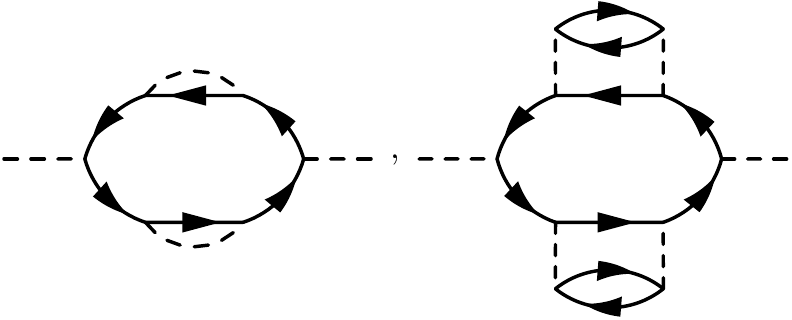}
	\caption{ \label{fig:diag_scGW_HO}
	        	 Two of the reducible diagrams of the screened Coulomb interaction in sc$GW$ that are not present in $G_0W_0$.}
\end{figure}

In sc$GW$, the $G_0W_0$ calculation of $\Sigma_0$ to build $G_1$ is only the first step. Next, we update $\Sigma_0$ to $\Sigma_1$ by inserting $G_1$ into the self-energy diagram. The diagram contains one obvious $G$ line ($\Sigma = iGW$), but contains more that are hidden in the polarisability entering $W$. By updating the polarisability with $G_1$ in place of $G_0$, the diagrams contained in $G_1$ enter the screened Coulomb interaction. Just as before, we can generate all reducible diagrams contributing to the updated Green's function ($G_2$) by iterating the Dyson series
\begin{align}
  \label{eq:g2}
  G_2(1,2)&= G_0(1,2)  \\ 
   & +\int G_0(1,3) \Sigma_1(3,4) G_2(4,2)d(3,4)  \nonumber 
\end{align}
for a fixed $\Sigma_1$. Continuing to update $\Sigma$ and iterate $G$ introduces more and more reducible diagrams. The solution is self-consistent when $G$ entering $\Sigma$ is the same as $G$ from iterating Dyson's equation.

\begin{figure*} 
    \includegraphics[width=\linewidth]{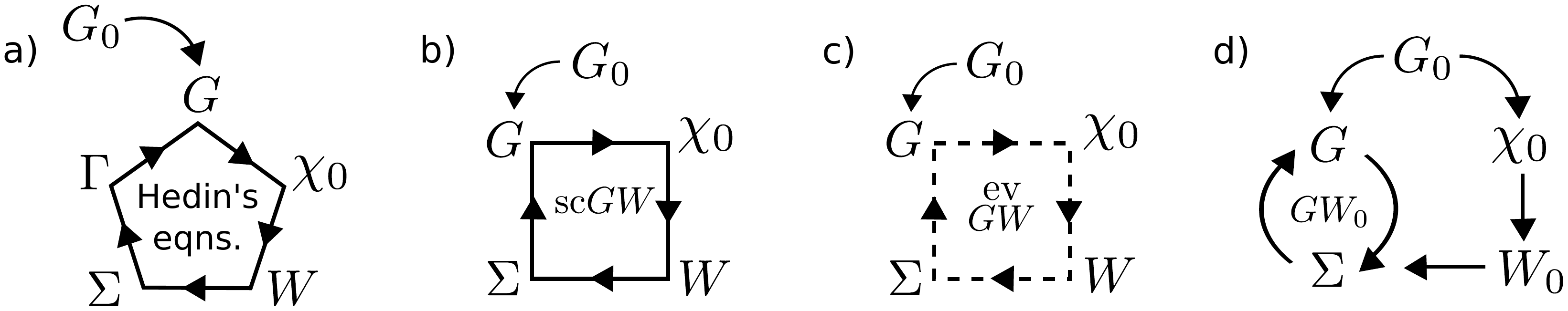}
	\caption{{\small \label{fig:Hedin_loops}
        	 Schematic of Hedin's full set of equations (a) and Hedin's $GW$ approximation (b-d). In a), all five quantities are iterated to self-consistency. In b), self-consistent $GW$ (sc$GW$), $\Gamma$ is set to a single spacetime point and the remaining four quantities are determined self-consistently. Eigenvalue self-consistent $GW$ shown in c), ev$GW$, updates only the quasiparticle energies while leaving the wave functions unchanged. In the sc$GW_0$ or ev$GW_0$ procedures shown in d), one iterates $G$ to self-consistency in Dyson's equation but does not update $\chi_0$ or $W$.   	 
                	 }}
\end{figure*}

In real sc$GW$ calculations, the procedure is slightly different. $G$ and $\Sigma$ are updated together instead of iterating to find $G_{i+1}$ at a fixed $\Sigma_i$. After the first iteration of Equation~\eqref{eq:g1}, the updated $-$ but not yet self-consistent $-$ $G_1$ is inserted into $\Sigma$. This way, $\Sigma$ is updated at each iteration along with $G$. The combined iterations are much more efficient because $G$ and $\Sigma$ converge together. Bear in mind that the efficient method of updating $G$ and $\Sigma$ at each step does not form the same easy-to-interpret series for $G_1$ as in Figure~\ref{fig:diag_scGW}. Note that even after one iteration to find $G$ and $\Sigma$, we would already go beyond $G_0W_0$.

Based on Figure~\ref{fig:diag_scGW}, the fully dressed Green's function and screened Coulomb interaction in sc$GW$ can be interpreted as double renormalisations of $G_0$ and $W_0$ through the two Dyson's equations in the $GW$ equations. However, the third Dyson equation that the vertex function (Equation~\eqref{eq:gamma}) would introduce is missing from the $GW$ equations. The absence of the vertex function has important consequences. Figure~\ref{fig:diag_scGW_HO} shows two of the reducible diagrams that enter $W$ in sc$GW$, but that are not present in $W_0$. The diagram on the left shows the polarisation bubble with the insertion of one interaction, or scattering event, in each arch. It is part of a series of sequential scattering events and builds additional interactions into the screened Coulomb interaction. The other diagram, however, is problematic. After the creation of the first electron-hole pair, both the electron and the hole interact with a new electron-hole pair. The two new electron-hole pairs are composed of the same Green's function lines as the initial electron-hole pair, even though the initial pair still exists. Therefore, the two later pairs do not account for the fact that the initial electron-hole pair has already been created $-$ they should somehow omit the pair already created. The electron (or hole) thus interacts or correlates with itself. The problems that have been identified\footnote{It has also been pointed out that the exact $W$ from RPA does not satisfy the f-sum rule.} for sc$GW$ can be attributed to diagrams like the one on the RHS of Figure~\ref{fig:diag_scGW_HO} \cite{Romaniello/Guyot/Reining:2009}.

We would also like to briefly comment on partial self-consistency in the Green's function. The above discussion points to the screened Coulomb interaction as the major source of imbalance between self-consistency and the missing vertex corrections. This imbalance can be partially fixed by keeping $W$ fixed at the $W_0$ level and iterating only the Green's function to self-consistency (sc$GW_0$). This is shown schematically in Figure~\ref{fig:Hedin_loops}. This approximation is partially motivated by a ``best $G$, best $W$'' philosophy that can improve agreement with experiment in certain situations.

Before leaving the discussion of self-consistent $GW$, we introduce one of the most technical and modern aspects of Green's function theory being researched: the existence of multiple solutions for $G$ from a single Dyson equation \cite{tandetzky_prb_92}. This issue has been studied in detail for the zero-dimensional one-point model (OPM) \cite{tarantino_prb_96,Berger_2014,Lani_2012} and has been produced numerically \cite{kozik_prl_114,prl_gunnarsson_119,parcollet_prb_97}. In the analytic OPM, there exist two interacting $G$ which can be mapped from the same $G_0$ \cite{Stan_2015,Rossi_2015}. One of these solutions is the physical $G$ for all values of interaction strength. Here, the physical solution is characterized by a smooth connection to $G_0$, unlike the unphysical solution for $G$ which diverges at zero interaction strength. The reverse map, from $G$ to $G_0$, has two solutions for $G_0$ which must be disentangled at a certain interaction strength. At this point, the physical $G_0$ \textit{switches} between the two solutions, so that solving the problem for all interaction strengths requires switching solution methods at this point. Otherwise, one would obtain a physical $G_0$ for some interaction strengths and an unphysical $G_0$ for others. In the OPM, this is now understood. However, it is not well understood if or how this same phenomenon emerges in more realistic systems.

\subsection{Eigenvalue self-consistency and level alignment}
\label{sec:ev-scGW}

There are a few strategies to reduce the expense of sc$GW$ while still including more physics than $G_0W_0$. The simplest form of performing approximate self-consistency in $GW$ is to iterate in the eigenvalues (ev$GW$).  After completion of the $G_0W_0$ loop, the real parts of the quasiparticle energies obtained from Equation~\eqref{Eq:qpe} or \eqref{Eq:qpe_lin} are reinserted into the non-interacting Green's function $G_0$ (Equation~\eqref{eq:greensfkt}) in place of the starting eigenvalues. Through $G_0$, the change in eigenvalues permeates through $W_0$ to the self-energy and eventually to the quasiparticle energies (ev$GW$). After iterating until the input quantities equal the output, the equations are self-consistent in the eigenvalues. 

Eigenvalue self-consistency was already proposed in the first $G_0W_0$ calculation for real materials \cite{Hybertsen/Louie:1986} and has since been applied frequently (see e.g. \onlinecite{Shishkin/Kresse:2006} for a more in-depth analysis). However, since only the real part of the quasiparticle energies is used and the wave functions are not updated self-consistently, the starting point dependence cannot be eliminated entirely. For example, a study for azabenzenes demonstrates that although the starting point dependence is reduced from 1.4~eV in $G_0W_0$, it cannot be lowered beyond 0.4~eV \cite{Marom/etal:2012}. 

For molecular systems, it has been shown that eigenvalue self-consistency improves the HOMO-LUMO gaps, which are then in good agreement with experiment \cite{Blase/Attaccalite/Olevano:2011,Wilhelm2016}.
However, examining the entire spectrum reveals that ev$GW$ does not lead to consistent improvements over $G_0W_0$. The ev$GW$ spectra are overly stretched with respect to the experimental spectra, such that large deviations (on the order of 1 eV) from experiment occur for lower lying states. Moreover, for most systems, the orbital ordering deviates from experimental observations \cite{Marom/etal:2012}. This is in line with observations for semiconductors and insulators, that find band gaps to be considerably overestimated in ev$GW$ \cite{Shishkin/Kresse:2006}. The reason for this overestimation in solids lies in the fact that the insertion of the quasiparticle energies into the screened Coulomb interaction leads to an underscreening, which should be compensated by the missing vertex corrections, as discussed previously. 

Just as sc$GW_0$ ameliorates problems in self-consistent $GW$, one can perform eigenvalue self-consistency only in $G$ to circumvent underscreening errors. Iterating the eigenvalues in only $G$ produces an ev$GW_0$ scheme that gives band gaps in good agreement with experiment \cite{Shishkin/Kresse:2006}. However, for open shell systems, it was observed that eigenvalue self-consistency in $G$ strongly affects the calculated multiplet splittings \cite{Lischner/etal:2012}. 

\begin{figure} 
    \includegraphics[width=0.99\linewidth]{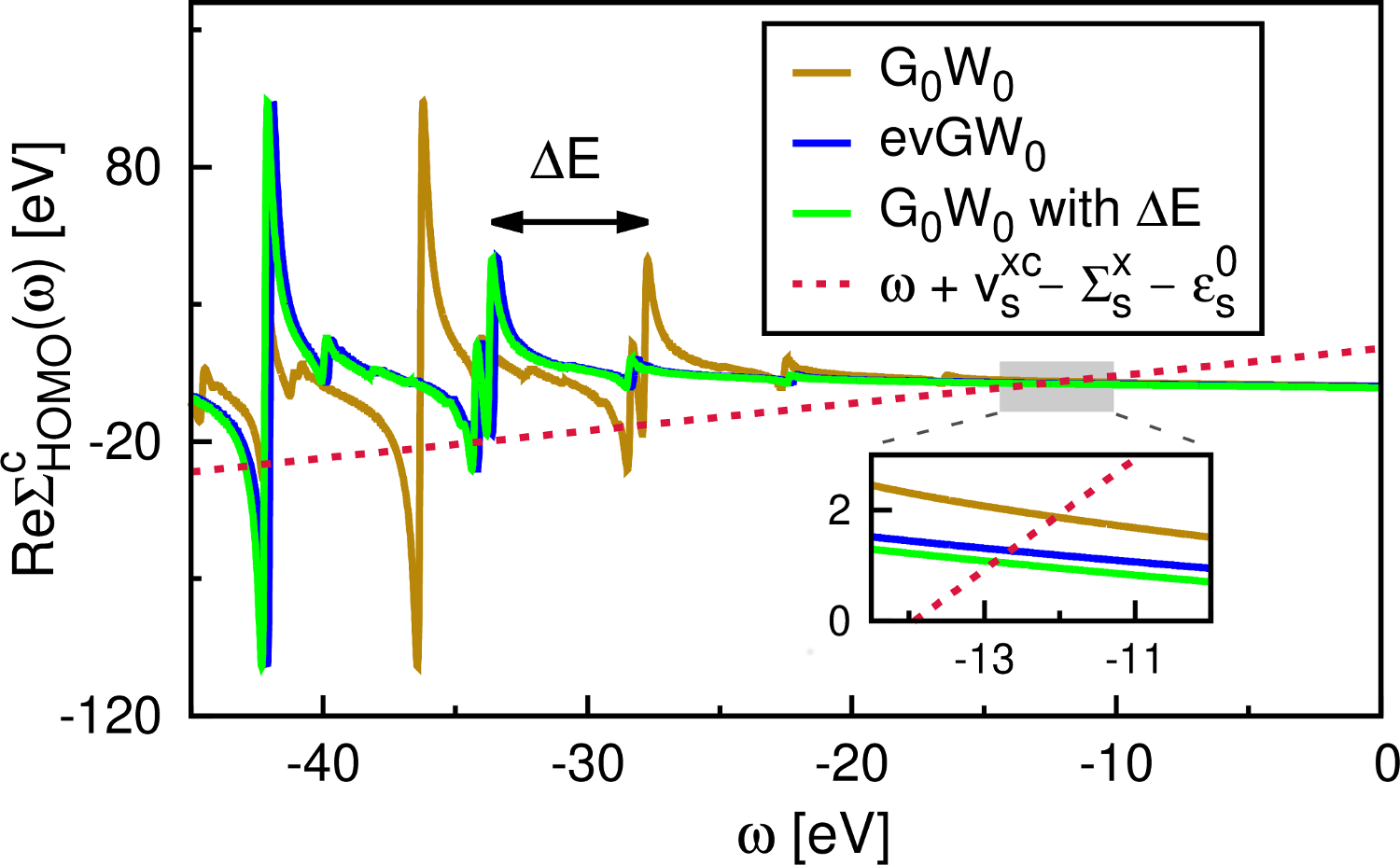}
	\caption{{\small \label{fig:sigma_evgw0_deltaE}
	Self-energy matrix elements for the HOMO of a single water molecule obtained with $G_0W_0$, ev$GW_0$ and level-aligned $G_0W_0$. In all three cases PBE is used as starting point. The inlet shows the graphical solutions of the QP equation. See Appendix~\ref{app:computational_details} for further computational details.   }}
\end{figure} 

The effect of eigenvalue self-consistency in $G$ on the self-energy is demonstrated in Figure~\ref{fig:sigma_evgw0_deltaE} for an ev$GW_0$ calculation. Compared to $G_0W_0$, the structure of the self-energy is almost identical, but shifted to lower energies. The Green's function in the eigenvalue self-consistent $GW_0$ scheme is given by
\begin{equation}
 G_{\textnormal{ev}}^{\sigma }(\mathbf{r},\mathbf{r}',\omega) =  \sum_m\frac{\phi_{m\sigma}^0(\mathbf{r})\phi_{m\sigma}^{0*}(\mathbf{r}')}{\omega-\epsilon_{m\sigma} - i\eta \sgn(E_{\rm F}-\epsilon_{m\sigma})} \label{eq:evgreensfkt}
\end{equation}
with
\begin{align}
   \epsilon_{m\sigma}&=\epsilon_{m\sigma}^0+\Delta \epsilon_{m\sigma}\\
  \Delta\epsilon_{m\sigma}&=\Sigma_{m\sigma}\left(\epsilon_{m\sigma}\right)-v^{\rm MF}_{m\sigma}.
\end{align}
Inserting the $GW$ corrections $\Delta\epsilon_{m\sigma}$ into the Green's function results in a shift of the poles in the self-energy, see Equation~\eqref{eq:sigma_pole}. On the real axis, the poles of $\Sigma_s^c$ are located at $\omega_{i\sigma}^{n} = \epsilon_{i\sigma}^0 + \Delta\epsilon_{i\sigma} -\Omega_{n\sigma}$ and $\omega_{a\sigma}^{n} = \epsilon_{a\sigma}^0 + \Delta\epsilon_{a\sigma} +\Omega_{n\sigma}$, where $i$ indicates again occupied and $a$ virtual states. Starting from a GGA functional, the correction $\Delta \epsilon_{m\sigma}$ is negative for occupied and positive for virtual states. Compared to a $G_0W_0$ scheme, the poles $\omega_{i\sigma}^n$ are now located at lower and the poles $\omega_{a\sigma}^n$ at higher frequencies.\par
A simplified version of ev$GW_0$ was originally suggested by Hedin~\cite{Hedin:1965} and has been revisited a few times by others~\cite{Hedin:1999,Pollehn1998,martin_reining_ceperley_2016}. Instead of using an individual shift $\Delta \epsilon_{m\sigma}$ for each state $m$, a global shift $\Delta E$ is employed:
\begin{align}
 G_{\Delta E}^{\sigma }(\mathbf{r},\mathbf{r}',&\omega) =\nonumber\\  \sum_m&\frac{\phi_{m\sigma}^0(\mathbf{r})\phi_{m\sigma}^{0*}(\mathbf{r}')}{\omega-(\epsilon_{m\sigma}^0+\Delta E_{\sigma})- i\eta \sgn(E_{\rm F}-\epsilon_{m\sigma}^0)},
 \label{eq:deltagreensfkt}
\end{align}
where $G^{\sigma}_0(\omega-\Delta E_{\sigma}) = G_{\Delta E}^{\sigma }(\omega) $. The QP equation (Equation~\ref{Eq:qpe}) then transforms into
\begin{equation}
  \epsilon_{s\sigma}=\epsilon_{s\sigma}^0+\Sigma_{s\sigma}\left(\epsilon_{s\sigma}-\Delta E_{\sigma}\right)-v^{\rm MF}_{s\sigma}.
  \label{Eq:qpe_delta}
\end{equation}
For metals, the shift $\Delta E$ is chosen in such a way that the $G_0W_0$ Fermi energy aligns with that of the starting point calculation, i.e. with the Fermi level of $G_0$. For systems with a energy gap, the highest occupied state is aligned, i.e., the valence band maximum for solids or the HOMO for finite systems. The latter is motivated by ``DFT Koopman's theorem", which states that only the KS energy of the HOMO can be rigorously assigned to the ionization potential when starting from an exact DFT functional~\cite{Almbladh/Barth:1985,Levy/Perdew/Sahni:1984}. In that case $\Delta E$ would be zero.

The shift can be determined by demanding self-consistency for the highest occupied state
\begin{equation}
    \epsilon_{\rm HOMO,\sigma} = \epsilon_{\rm HOMO,\sigma}^0 + \Delta E_{\sigma}.
    \label{eq:Hedinshiftcodition}
\end{equation}
Inserting Equation~\eqref{eq:Hedinshiftcodition} into Equation~\eqref{Eq:qpe_delta} yields the explicit expression 
\begin{equation}
  \Delta E_{\sigma} = \Sigma_{\rm HOMO,\sigma}\left(\epsilon_{\rm{HOMO},\sigma}^0\right)-v^{\rm MF}_{\rm HOMO,\sigma}.
\end{equation}
Adjusting the energy scale of $G_0^{\sigma}$ by $\Delta E$ translates to a rigid shift of the self-energy as shown in Figure~\ref{fig:sigma_evgw0_deltaE}. The results are very similar to ev$GW_0$ in the frequency range where the quasiparticle solution is expected. 

The $\Delta E$ scheme is less frequently used for the calculation of quasiparticle energies than eigenvalue self-consistent schemes. However, it has been shown that it substantially improves satellite spectra \cite{Pollehn1998}. The same holds for the ev$GW_0$ scheme, which has been employed to calculate satellite spectra of VO$_2$ \cite{Gatti2015} and bulk sodium \cite{Zhou/etal:2015}. 

\subsection{Self-consistency via a new ground state}
\label{sec:QSGW}

Building on the idea of iterating in the quasiparticle energies, one can go one step further and also incorporate wave function changes. An elegant way to achieve this is to find the variationally best mean-field potential to a given self-energy \cite{Godby/etal:1986,Godby/etal:1987,Casida:1995,Schilfgaarde/Kotani/Faleev:2006,Kotani/etal:2007}. This mean-field potential can then be used to generate new eigenvalues and eigenfunctions to construct a new $G_0$ for the next iteration of the $GW$ cycle. 

If the new potential is local, this iteration can be formalized exactly in the optimized effective potential (OEP) framework \cite{Casida:1995,Kuemmel/Kronik:2008}, which is equivalent to the Sham-Schl\"uter equation \cite{Godby/etal:1986,Godby/etal:1987}. The OEP framework and the Sham-Schl\"uter equation only guarantee that the density generated by the local potential matches the $GW$ density. The eigenvalue spectrum of the local potential will not be the same as the $GW$ spectrum. For explicitly non-local potentials, no formally exact match between the $GW$ self-energy and the potential has been found because the self-energy is non-local and frequency dependent, while the constructed potential is non-local but static. 

An approximate non-local potential can be found by introducing the $GW$ Hamiltonian $\hat{h}^{GW}(\omega)=\hat{h}^0+v_{\mathrm{H}}+\Sigma^{GW}(\omega)$.
The mean-field Hamiltonian $\hat{h}^{\text{MF}}$ that best reproduces the effects of $\Sigma^{GW}$ is defined as $\hat{h}^{\mathrm{MF}} = \hat{h}^0 + v_{\mathrm{H}} + v^{\mathrm{MF}}$, see also Equations~\eqref{eq:h_0}-\eqref{eq:h_MF} for the definitions of the Hamiltonians. $v^{\mathrm{MF}}$ can then be obtained by minimizing $|| \hat{h}^{GW}-\hat{h}^{\text{MF}}||$ \cite{Schilfgaarde/Kotani/Faleev:2006,Kotani/etal:2007}. An approximate minimization finally yields an analytic expression for the (static and Hermitian) mean-field potential \cite{Faleev/Schilfgaarde/Kotani:2004,Schilfgaarde/Kotani/Faleev:2006,Kotani/etal:2007}
\begin{equation}
   v^\text{MF}_{ij}=\frac{1}{2}  \left[ \left[ \mathrm{Re} \, \Sigma(\epsilon_i) \right]_{ij}+ \left[ \mathrm{Re} \, \Sigma (\epsilon_j) \right]_{ij} \right] 
\end{equation}
where ``$\mathrm{Re}$" signifies here the Hermitian part of $\Sigma(\epsilon_k)$ 
\begin{equation}
\left[ \mathrm{Re} \, \Sigma(\epsilon_k) \right]_{ij} = \frac{1}{2} \left[ \Sigma(\epsilon_k) + \Sigma(\epsilon_k)^{\dagger} \right]_{ij} .
\end{equation}
The quasiparticle energies for the Green's function $G$ are then given by the self-consistent $G_0$ that follows from $v^\text{MF}_{ij}$ \cite{Schilfgaarde/Kotani/Faleev:2006}. Satellites or the incoherent part of the spectra function are not captured by this approximation. This is why the scheme is commonly referred to as quasiparticle self-consistent $GW$ (QS$GW$).
Reports of a starting point dependence for metal oxides \cite{Liao/Carter:2011,Isseroff/Carter:2012} have not yet been reproduced by other groups with a different implementation. In general, the QS$GW$ scheme converges to a unique solution.

An alternative definition for a non-local mean-field potential is given by the static Coulomb hole plus screened exchange (COHSEX) approximation to $GW$ \cite{Hedin:1965,Hedin/Lundqvist:GW}:
\begin{equation}
\label{eq:sigma_COHSEX}
	v_\sigma^\text{MF,COHSEX}(\bfr,\bfrp)=\Sigma^{\rm COH}_{\sigma} (\bfr,\bfrp)+\Sigma^{\rm SEX}_{\sigma} (\bfr,\bfrp).
\end{equation}
The screened exchange (SEX) term is defined in analogy to the exact-exchange self-energy in Equation~\eqref{eq:sigma_x} but with the statically screened Coulomb interaction instead of the bare one
\begin{equation}
  \Sigma^{\rm SEX}_\sigma (\bfr,\bfrp) 
                            = -\sum_i^{\rm occ} \phi_{\sigma i}(\bfr) \phi_{\sigma i}^{*}(\bfrp) W(\bfr,\bfrp,\omega=0)  \, ,
\label{eq:sigma_sex_2}
\end{equation}
where $\phi_{\sigma i}$ are eigenfunctions of the COHSEX mean-field Hamiltonian. 
The static Coulomb hole (COH) term, on the other hand, becomes local in space
\begin{equation}
\label{eq:sigma_coh}
  \Sigma^{\rm COH}_\sigma (\bfr,\bfrp)=  \delta(\bfr-\bfrp)\left[W(\bfr,\bfrp,\omega=0) - v(\bfr,\bfrp)\right] \, .
\end{equation}
The statically screened Coulomb interaction $W(\bfr,\bfrp,\omega=0)$, which enters in Equations~\eqref{eq:sigma_sex_2} and \eqref{eq:sigma_coh}, is obtained by inserting the COHSEX  eigevalues and eigenfunctions in Equations~\eqref{eq:W0}-\eqref{eq:chi0} for $\displaystyle{\omega=0}$. Like in QSGW, $v_\sigma^\text{MF,COHSEX}(\bfr,\bfrp)$ produces new eigenvalues and eigenfunctions, which yield a new self-energy. The COHSEX equation can then be iterated until self-consistency is achieved. 

COHSEX can also serve as an improved starting point compared to KS-DFT for a perturbative $G_0W_0$ calculation. After completing a COHSEX calculation, one can use the self-consistent COHSEX eigenvalues and wave functions for a perturbative $G_0W_0$ calculation with the full, dynamical $W$. In the case of VO$_2$ \cite{Gatti/etal:2007}, $G_0W_0$@LDA fails to open the band gap. On the other hand, $G_0W_0$@COHSEX opens a band gap, in agreement with experiment. The improved COHSEX starting point is especially important for materials with localized electrons \cite{Aguilera/etal:2011}. When comparing to benchmark coupled cluster data on organic molecules, the COHSEX starting point decreases the mean absolute error of $G_0W_0$ compared to $G_0W_0$@PBE \cite{Knight2016}. In Ge under pressure, $G_0W_0$@COHSEX predicts a direct gap at the $\Gamma$ point while $G_0W_0$@LDA predicts band overlap \cite{Jain/etal:2014}.

Self-consistency in $GW$ is a topic that is still being researched. While the results from true self-consistent $GW$ are usually in worse agreement with experiment than $G_0W_0$ \cite{Shishkin/Kresse:2006,Schoene/Eguiluz:1998,Kresse/etal:2018}, studies of self-consistency are still necessary to advance the field. We can learn about shortcomings of the theory or assess challenging materials. Most importantly, self-consistent $GW$ implementations are a necessary foundation to go \textit{beyond} $GW$ in the future, as discussed in Section~\ref{sec:beyond}.

\section{Solids}
	\label{sec:solids}
		
Solids were the first testbed of $GW$ theory in real materials. The basic quantity to compute in solids is the band structure. Unlike in molecules with single-particle states given by molecular orbitals, single particles in solids occupy Bloch waves, defined in Equation~\eqref{eq:bloch} and indexed by their wave vector $\mathbf{k}$. Diagonalizing the Hamiltonian at each $\mathbf{k}$ gives its own set of single-particle eigenvalues. One can conveniently visualize the eigenvalues at different $\mathbf{k}$-points by varying $\mathbf{k}$ continuously along some path, placed on the x-axis, and plotting the eigenvalues on the y-axis. Eigenvalues change continuously with $\mathbf{k}$, forming separate bands of states. The collection of single-particle states in bands is similar to the grouping of different combinations of bonding and anti-bonding states in molecules or polymers. Quasiparticle Hamiltonians are also $\mathbf{k}$-dependent, and energies at all $\mathbf{k}$ form a band structure which can be compared to angle resolved PES and IPES spectra as the incident momentum is varied (see Figure~\ref{fig:PES_spec_BS}).

\begin{figure} 
        \includegraphics[width=0.99\linewidth]{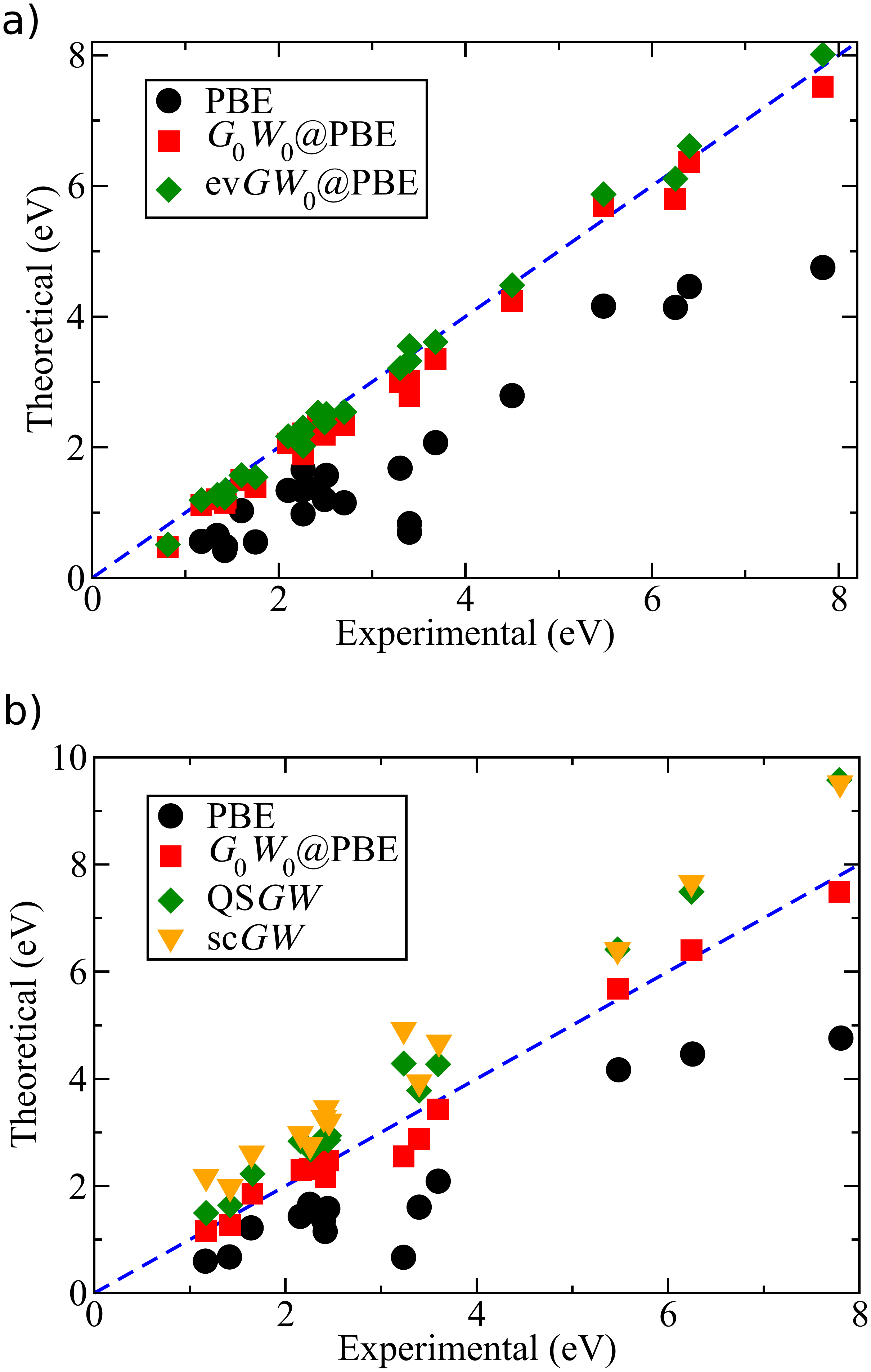}
	\caption{{\small \label{fig:gw_gaps} 
        	 (a) Band gaps of semiconductors and insulators computed with PBE, $G_0W_0$, and ev$GW_0$ in the all-electron, linearized augmented plane wave (LAPW) framework. Data taken from  \onlinecite{Jiang/Blaha:2016}. (b) Band gaps of semiconductors and insulators computed with PBE, $G_0W_0$, QS$GW$, and sc$GW$ in the projector-augmented-wave (PAW) framework. Data taken from  \onlinecite{Kresse/etal:2018}.}}
\end{figure}

Computing quasiparticle band structures with $GW$ gives the quasiparticle band gap, analogous to the HOMO-LUMO gap in molecules. The first $G_0W_0$ calculations for real materials \cite{Hybertsen/Louie:1985,Hybertsen/Louie:1986,Strinati/etal:1980,Strinati/etal:1982} focused on semiconductors (Si, Ge) and insulators (diamond, LiCl). $G_0W_0$ calculations in semiconductors and insulators give a uniform improvement in band gap over estimates with either Hartree-Fock or Kohn-Sham eigenvalues, as shown in Figure~\ref{fig:gw_gaps}(a). This improvement was the first major success of the $GW$ theory. 

The success of $GW$ applied to semiconductors continued with other studies \cite{Godby/etal:1986,Godby/etal:1987,Godby/etal:1987b,Godby/Schlueter/Sham:1988,Blase/etal:1995}. Many common semiconductors lack semicore states and are well described by pseudopotentials (see Section~\ref{subsec:pseudopotentials}). These factors reduce the computational complexity for $GW$ calculations and made early, realistic $GW$ calculations of semiconductors feasible. Screening in simple semiconductors can also be approximated by model dielectric functions like  plasmon-pole models (see Sec.~\ref{subsec:plasmon-pole}), eliminating the need for a numerical evaluation of the self-energy integral. 

It was later realized that semicore $d$-electrons in semiconductors such as GaN, ZnO, ZnS, ZnSe or CdS \cite{Rohlfing/Krueger/Pollmann:1995_CdS,Rohlfing/Krueger/Pollmann:1997_semicore,Rohlfing/Krueger/Pollmann:1998} and metals such as Cu \cite{Marini/etal:2002} and Au \cite{Rangel/etal:2012} have a strong influence  on $GW$ calculations. Due to the strong overlap of the atomic $d$ functions with the atomic $s$ and $p$ functions in the same shell, the exchange self-energy is very sensitive to the inclusion of semicore states. If only the semicore $d$ states are explicitly included as valence state in the $G_0W_0$ calculation, while the $s$ and $p$ states in the same shell are frozen in the core of a pseudopotential, the subsequent $G_0W_0$ calculation will produce an incorrect band gap  \cite{Rohlfing/Krueger/Pollmann:1995_CdS}. This problem can be solved by explicitly including the entire shell as valence in the $G_0W_0$ calculation \cite{Rohlfing/Krueger/Pollmann:1995_CdS, Luo/Louie:2002,Tiago/Ismail-Beigi/Louie:2003,Fleszar/Hanke:2005}, by using exact-exchange pseudopotentials and exact-exchange starting points \cite{Rinke/etal:2005,Qteish/etal:2006,Rinke/pssb} or by all-electron calculations \cite{Friedrich/etal:2006,Shishkin06,Friedrich2010,Exciting:2014,Jiang/Blaha:2016}.

\subsection{Band gaps}
\label{sec:gaps}

Figure~\ref{fig:gw_gaps}(a) shows the quasiparticle band gap computed with $G_0W_0$ and ev$GW_0$ with a modern all-electron LAPW code \cite{Jiang/Blaha:2016} for several different semiconductors and insulators. Perfect agreement between theory and experiment would place all data points on the dashed blue line. Generally, Kohn-Sham eigenvalues based on a multiplicative (local or semi-local) exchange-correlation potential (here PBE) underestimate the band gap and Hartree-Fock eigenvalues overestimate (not shown in Figure~\ref{fig:gw_gaps}). Eigenvalue self-consistency (ev$GW_0$) improves the agreement with experiment even further than $G_0W_0$, when starting from a local or semi-local DFT calculation. 

Figure~\ref{fig:gw_gaps}(b) compares band gaps computed with different self-consistency schemes \cite{Kresse/etal:2018} for a different set of semiconductors and insulators than in panel (a). $G_0W_0$@PBE again provides good agreement with experiment (i.e. the red squares are close to the diagonal). Van Schilfgaarde's QSGW scheme \cite{Schilfgaarde/Kotani/Faleev:2006} and fully self-consistent $GW$ calculations consistently overestimate band gaps.

With the predictive accuracy of $G_0W_0$ band gaps validated, we provide a few examples in which $GW$ calculations helped to resolve band gap controversies. One case is InN. In the early 2000s, alloys of GaN and InN were revolutionizing light-emitting diode (LED) technology. However, the band gap of InN was believed to be almost 2~eV \cite{Butcher/Tansley:2005}, which would have severely limited the usefulness of InGaN alloys to tune the emission of LEDs. Through $G_0W_0$ calculations and more refined experiments, the real value of the InN band gap was found to be 0.7~eV \cite{Bechstedt/Furthmueller:2002,Furthmueller/etal:2005,Rinke/etal:2006}, paving the way for the LEDs we know today. Another example is hybrid perovskites that have triggered a new boom in the emergent photovoltaic materials field. The prototypical material is methylammonium lead triiodide (CH$_3$NH$_3$PbI$_3$ or in short MAPI). Unusually, local or semi-local DFT calculations already predict a band gap in good agreement with measurements, which had caused initial confusion in the field. However, when spin-orbit effects, which are particularly strong in this materials class, are incorporated in the DFT calculations, the band gap becomes significantly underestimated again. $G_0W_0$ and QSGW calculations that include spin-orbit effects then predict the correct band gap \cite{Umari/Mosconi/DeAngelis:2014,Brivio/etal:2014}. 

Finally, we consider high pressure physics. At high pressures ($\sim 100$ GPa), many materials experience band gap closure and transition from an insulator to a metal. There can also be many competing structural phases, each with their own metallization pressure, that are difficult to disentangle in experiments. $GW$ is an excellent tool to theoretically predict the metallization pressure for different structural phases and help interpret experimental results \cite{Tse/etal:2008,Ramzan/etal:2010,Jin/etal:2016,Yong:2017,Khairallah/etal:2008}. Solid hydrogen is a noteworthy example of metallization at high pressure, first predicted in 1935 \cite{Wigner/etal:1935}. The metallic hydrogen puzzle is an exceptionally difficult one that is still not fully understood, but $GW$ calculations help corroborate experimental measurements and support the existence of certain structural phases \cite{Lebegue/etal:2012,Mcminis/etal:2015,Dvorak/etal:2014}.

\subsection{Band structures and band parameters}
\label{sec:bandstructure}

\begin{figure} 
        \includegraphics[width=0.99\linewidth]{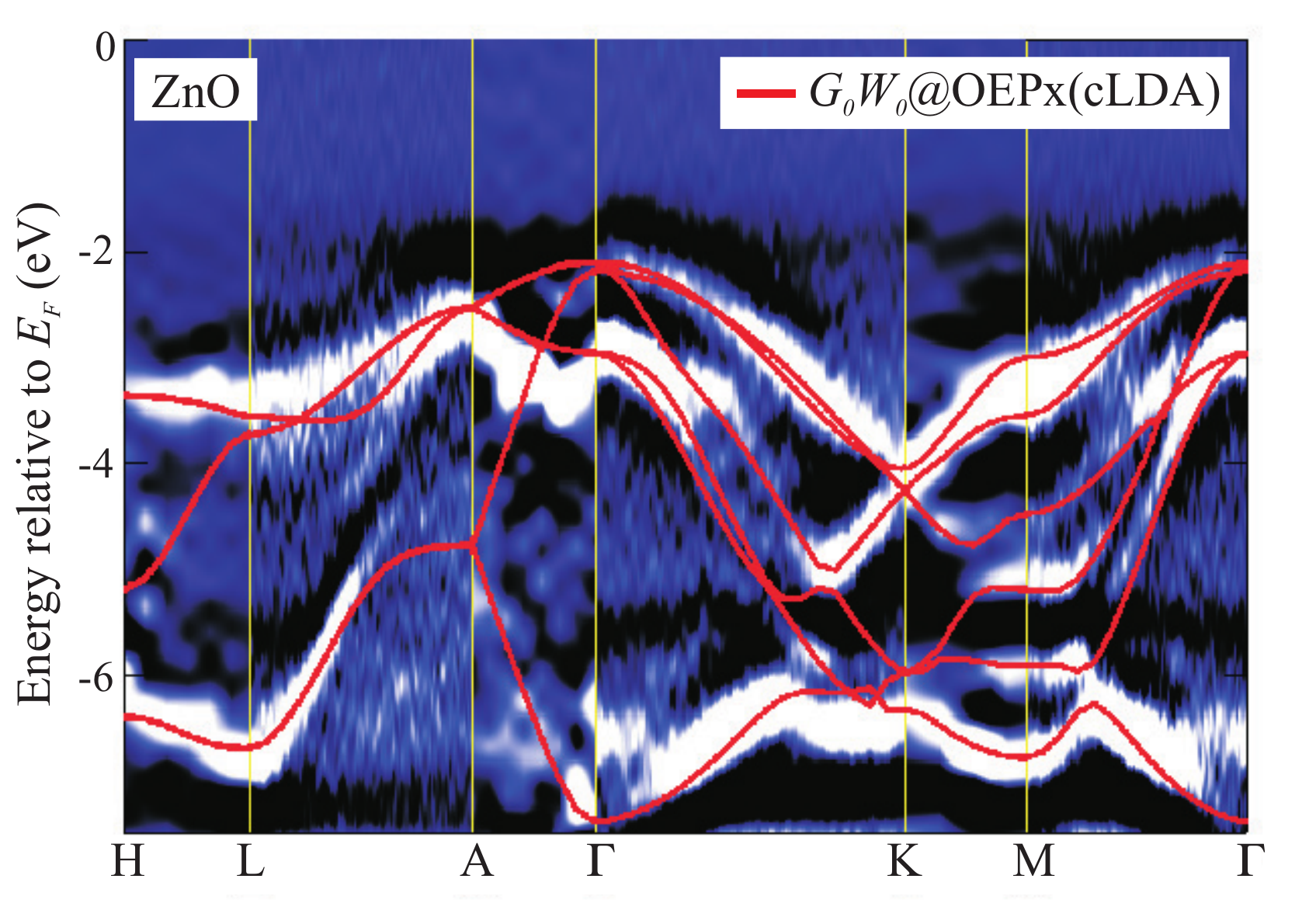}
	\caption{{\small \label{fig:zno_bands}
              $G_0W_0$ band structure of ZnO superimposed on experimental ARPES data \cite{Yan/etal:2012}. The experimentally measured lifetimes of the states are indicated by the shading, with white shading indicating long lifetime. The $G_0W_0$ calculations are based on the optimized effective potential approach for exact exchange mentioned in Section \ref{sec:QSGW} that includes LDA correlation (OEPx(cLDA)). Figure adapted from \cite{Yan/etal:2011}. }}
\end{figure}

With $GW$, one can compute much more than only the band gap of a solid. A typical band structure computed with $GW$ is shown in Figure~\ref{fig:zno_bands} for ZnO. To visualize the results, $\mathbf{k}$ is allowed to vary on a linear path through the Brillouin zone. The quasiparticle energy as a function of $\mathbf{k}$ is also called the dispersion for the system. The $G_0W_0$ band structure for ZnO (red lines in Figure~\ref{fig:zno_bands}) is superimposed on the experimental photoemission results shown already in Figure~\ref{fig:PES_spec_BS}. Experiment and $G_0W_0$ agree very well both in terms of band positions as well as band curvatures.

\begin{figure} 
        \includegraphics[width=0.99\linewidth]{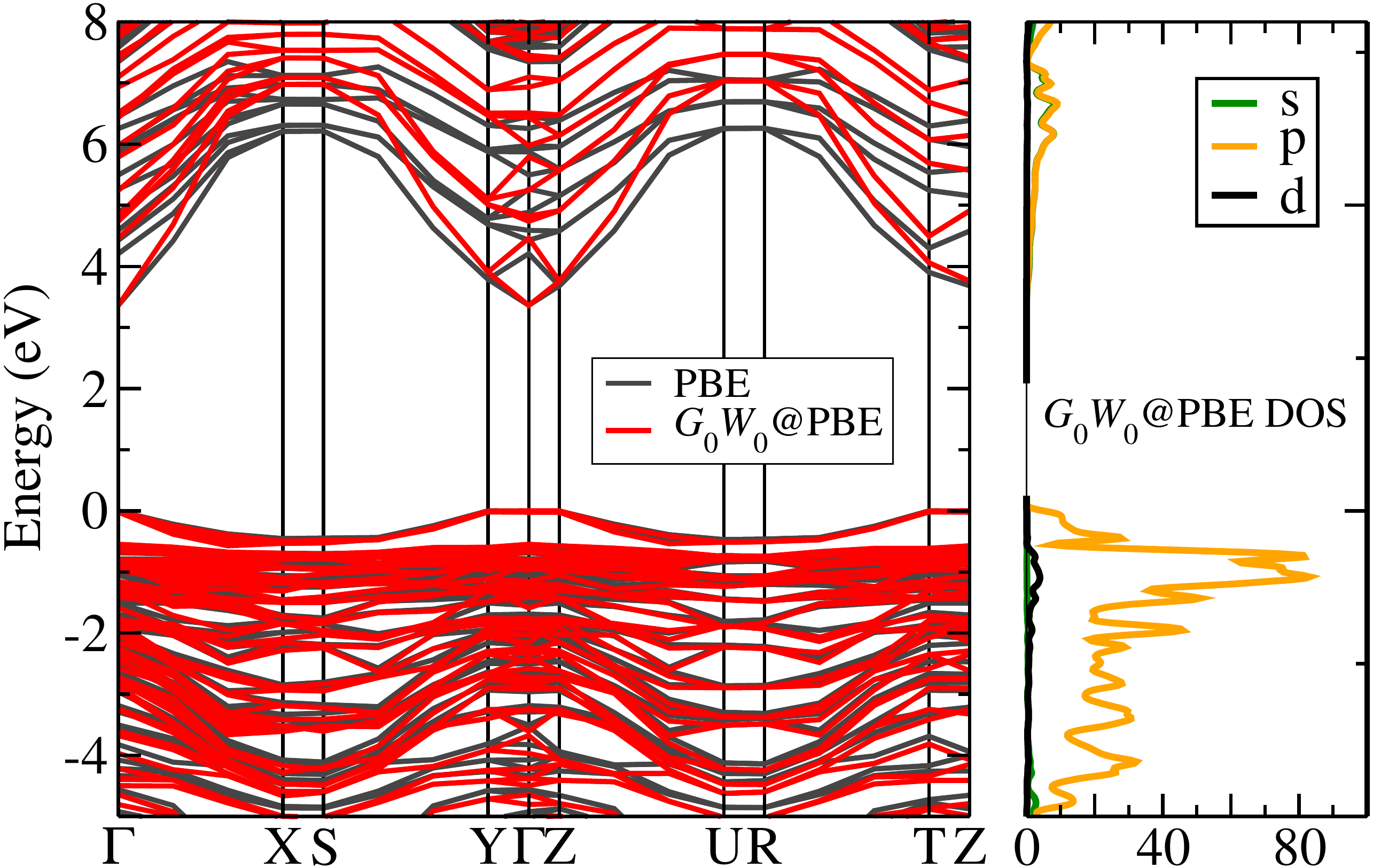}
	\caption{{\small \label{fig:BS_oxide}
              $G_0W_0$ band structure of K$_2$Sn$_3$O$_7$ \cite{McAuliffe/etal:2017}. The main panel illustrates the difference between the PBE (dark grey lines) and the $G_0W_0$@PBE (red lines) band structure. The unoccupied states of the PBE band structure have been shifted up in energy for better visibility so that the bottom of the conduction bands coincide in both band structures. The right panel shows the $G_0W_0$@PBE density of states (DOS) resolved into $s$, $p$ and $d$ angular momentum channels.}}
\end{figure}

Another example of a $G_0W_0$ band structure is shown in Figure~\ref{fig:BS_oxide} for K$_2$Sn$_3$O$_7$, a new prospective ion conductor or transparent conductor  \cite{McAuliffe/etal:2017}. The unoccupied states in the PBE band structure have been shifted up for the purposes of plotting so that the bottom of the conduction bands coincides in PBE and $G_0W_0@$PBE. This removes the PBE band gap problem from the comparison and makes it easier to spot differences in band curvatures. For the valence bands the PBE and $G_0W_0@$PBE band structures agree remarkably well for this material. Towards lower energies the deviations between the band structures become larger with $G_0W_0@$PBE generally giving lower band energies than PBE. This downward shift leads to a band width widening in $G_0W_0$ compared to PBE. For the conduction bands, the difference between PBE and $G_0W_0@$PBE is more pronounced. The band curvatures in $G_0W_0@$PBE are much steeper than in PBE, which subsequently leads to a significant underestimation of the PBE bands around the X, S, U and R points in the Brillouin zone. K$_2$Sn$_3$O$_7$ is another example of a material whose band gap and band structure were not known. The $G_0W_0@$PBE band gap amounts to 3.15~eV \cite{McAuliffe/etal:2017}, which now provides a reference value for this new material.

From the band structure, one can access the band gap, band widths, and estimate effective masses. If one models the dispersion at the band edges as parabolic, as is the case for a free particle, one can extract an effective mass from the band structure. The effective mass, labeled $m^*$, is
\begin{equation}
\label{eq:meff}
m^* = \hbar^2 \left[ \frac{d^2 E}{dk^2} \right]^{-1}
\end{equation}
so that the band edge dispersion is
\begin{equation}
E = \frac{ \hbar^2 k^2 }{ 2m^*} \label{eq:bandparabola}
\end{equation}
to mimic a free particle. In a real crystal, the effective mass is a tensor, not just a scalar. The effective mass model is closely related to the quasiparticle concept, and the renormalization factor $Z_s$ (Equation~\eqref{Eq:Z_s}) is one factor contributing to $m^*$. The quasiparticle effective mass is usually heavier than the free electron mass because of the drag induced by the surrounding electrons. Static mean-field theories like Kohn-Sham DFT also give an estimate of effective mass from their band structure. The $GW$ band structure is typically more ``curved'' or concave than the Kohn-Sham structure, as shown in Figure~\ref{fig:BS_oxide}, which means that $GW$ quasiparticles are ``lighter" than the KS particles. In silicon and methylammonium lead iodide, the $GW$ level of theory is necessary to predict effective masses in good agreement with experiment \cite{Filip/Verdi/Giustino:2015,Ponce/Margine/Giustino:2018}.

One can also compute the single-particle density of states (DOS), which in solids is analogous to taking horizontal slices through the band structure. This gives the total number of available states at the energy of that slice. In Green's function theory, the concept of the single-particle DOS is replaced by the spectral function. The spectral function is $\mathbf{k}$-dependent, so that the total effective DOS is obtained by adding up the spectral functions at all $\mathbf{k}$. However, computing the spectral function requires the solution of Dyson's equation, which is often not practical computationally. Instead, a $G_0W_0$ quasiparticle DOS is computed by summing up artifically broadened Gaussian peaks centered around each $G_0W_0$ energy.
The right panel of Figure~\ref{fig:BS_oxide} shows the $G_0W_0@$PBE DOS for K$_2$Sn$_3$O$_7$ \cite{McAuliffe/etal:2017}. In addition, this DOS is projected on the atomic angular momentum channels $s$, $p$ and $d$. Such information is usually extracted from DFT calculations and illustrates that both the valence band and the conduction band of K$_2$Sn$_3$O$_7$ is largely made up of $p$ states. 

Band parameters like effective masses are important characteristics of semiconductors and are key parameters for the semiconductor industry. $G_0W_0$ effective masses are more accurate than those computed with DFT.  Effective masses are either extracted directly from the $GW$ band structure by fitting Equation~\eqref{eq:meff} to a fine band structure path \cite{Schleife/etal:2009} or by fitting an effective $\mathbf{k} \cdot \mathbf{p}$ Hamiltonian to $GW$ quasiparticle energies \cite{Rinke/etal:2008}. In this way, important band parameters have been computed for silicon and silicon under strain \cite{Bouhassoune/Schindlmayr:2010,Ponce/Margine/Giustino:2018}, GaAs \cite{Cheiwchanchamnangij/Lambrecht:2011}, AlN, GaN, and InN \cite{Rinke/etal:2006,Rinke/etal:2008,Svane/etal:2010,Yan/etal:2011}, MgO, ZnO, and CdO \cite{Schleife/etal:2009,Yan/etal:2012} and more recently for perovskites and hybrid perovskites \cite{Filip/Verdi/Giustino:2015}. Such band parameters can then be used directly in device simulations to model, for example, charge carrier flows \cite{Kivisaari/etal:2017}. If one is interested in charge carrier mobilities and charge carrier densities, scattering due to phonons and impurities has to be taken into account \cite{Manos/etal:2010,Ponce/Margine/Giustino:2018}.  

$GW$ can further be used as one of the final steps in high-throughput screening studies for new materials. In a search for transparent $p$-type conductors, $G_0W_0$ calculations provided accurate band gaps and effective masses that screened out the final candidates \cite{Hautier/etal:2013}. 

\subsection{Lifetimes}
\label{sec:lifetimes}

Unlike mean-field theories, $GW$ also allows one to compute the lifetimes of states from first principles. The lifetime of the quasiparticle is the characteristic time over which the added particle decays into surrounding degrees of freedom. States which are ``closer'' to exact eigenstates of the system have longer lifetime. The  lifetime of a quasiparticle with corresponding energy $\epsilon_s$ is directly related to the non-Hermiticity of the self-energy and the magnitude of its imaginary part,
\begin{equation}
\tau_s^{-1} = 2 \, | \, \mathrm{Im} \, \Sigma(\epsilon_s) \, | .
\label{eq:lifetime}
\end{equation}
$\tau$ can be inferred from experimental spectra by its relation to the quasiparticle peak width, $\Gamma$, as $ \tau^{-1} = \Gamma/2 $ (not to be confused with the vertex function $\Gamma$). 

From simple arguments in Fermi liquid theory, lifetimes decrease as particle energy increases because the available phase space for scattering at a fixed energy grows with increasing energy. Studies of quasiparticle lifetimes in the $GW$ approximation for metals (Cu, Ag, Au) show good agreement with experiment at high energies \cite{Marini/etal:2002,Keyling/etal:2000,Bacelar/etal:2002,Yi/etal:2010}. At low energies, however, the agreement is not perfect. For example, $GW$ calculations cannot account for the sudden increase in experimental lifetimes of electrons in Cu at energies below 2 eV  \cite{Yi/etal:2010,Keyling/etal:2000}. These failures are attributed to the localized, short-range interactions of $d$-electrons in metals that are not described well by $GW$.

\subsection{More challenging solids}
\label{sec:oxides}

As computational power increased and the success of $GW$ became more widely known, studies were extended to more challenging materials like oxides, or $d$- and $f$-electron compounds. These materials are both a theoretical challenge for the $GW$ approximation and are numerically more difficult to compute. Broadly speaking, these materials suffer from a severe mean-field starting point problem and/or contain localized electrons which are not well described by $GW$.
Accordingly, studies of these materials required advances in the treatment of core electrons and the evaluation of the self-energy. The first studies of metals focused on the alkali metals Na and Al \cite{Northrup/etal:1987,Surh/etal:1988}. Metals served as a valuable test on the effects of self-consistency and vertex corrections \cite{Mahan/etal:1989,Shirley:1996}. Eventually, studies moved into oxides and materials with $d$-electrons \cite{Aryasetiawan/Karlsson:1996,Aryasetiawan/Gunnarsson:1995,Aryasetiawan:1992,Massidda/etal:1995,Massidda/etal:1997}. 

Already in the early nineties of the previous century Aryasetiawan tackled ferromagnetic nickel (Ni) with $G_0W_0$ \cite{Aryasetiawan:1992}. He found the quasiparticle band structure and the valence bandwidth to be in good agreement with angle-resolved photoemission data. However, the exchange splittings are not well reproduced by $G_0W_0$ and a satellite at 6~eV is missing. Later calculations for gadolinium (Gd) revealed similar observations \cite{Aryasetiawan/Karlsson:1996,Aryasetiawan:1997_2}. For Gd, satellites were seen in the $G_0W_0$ spectrum, but their spectral weight does not match experiment.

The previous millennium concluded with early explorations into transition metal oxides such as nickel oxide (NiO) and manganese oxide (MnO) \cite{Aryasetiawan/Karlsson:1996,Massidda/etal:1995,Massidda/etal:1997}. They, as well as iron and cobalt oxide (FeO and CoO, respectively), were then revisited with $GW$ in the 2000s \cite{Li/etal:2005,Kobayashi/etal:2008,Rodl/etal:2008,Roedl/etal:2009,Jiang/etal:2010}. These oxides present a challenge to $G_0W_0$ calculations because local and semi-local DFT starting points produce metallic states that then cannot be corrected into semiconductors by $G_0W_0$. Instead, DFT+$U$ and hybrid functionals were explored as alternative starting points \cite{Roedl/etal:2009,Jiang/etal:2010}. The resulting $G_0W_0$ DOSs are in good agreement with direct and inverse photoemission measurements for the low  temperature magnetically ordered phases. However, the $GW$ method cannot describe the DOS in the paramagnetic phase nor the transition to the paramagnetic phase.

The situation is similar in the less correlated copper oxide (Cu$_2$O) \cite{Bruneval/etal_2:2006}. $G_0W_0$@LDA again fails to give a proper account of the band structure, while QSGW provides good agreement with ARPES measurements. CuO poses more of a problem, as no starting point or self-consistency scheme produces a satisfying band gap or density of states \cite{Roedl/Sottile/Reining:2015,Roedl/etal:2017}.

The early 2000s saw other oxides gain rapid interest, as the semiconductor industry sought a replacement for silicon dioxide (SiO$_2$) in silicon-based microelectronic technology. To prevent gate leakage in ever-shrinking transistors, gate materials with a higher dielectric constant ($k$) than SiO$_2$ were required. Eventually hafnium dioxide won the race. During the development period, the electronic structure, in particular the band gap and the band offsets of so called high-$k$ materials were of enormous interest
\cite{Shaltaf/etal:2008,Gruening/Shaltaf/Rignanese:2010,Jiang/etal:2010_2,Benoit/etal:2018}. $G_0W_0$ calculations of the closely related compounds zirconium oxide (ZrO$_2$) and hafnium oxide (HfO$_2$) were performed \cite{Gruening/Shaltaf/Rignanese:2010,Jiang/etal:2010_2}.
Plane-wave and FLAPW $G_0W_0$ agree very well with each other for these materials. The all-electron calculations investigated the effect of the Hf $f$-electrons and found that they do not change the self-energy corrections in these materials \cite{Jiang/etal:2010_2}. The final band gap of monoclinic HfO$_2$, however, is still under debate. It was initially believed to lie around 5.8~eV and is now thought to be in excess of 6.3~eV \cite{Benoit/etal:2018}. What remains a challenge in strongly polarizable materials such as high-$k$ dielectrics, and could thus potentiall explain remaining discrepancies between $GW$ and experiment, is how to include ionic screening (i.e. screening due to nuclear motion) consistently in the dielectric function of a $GW$ calculation.

The list of interesting metal oxides and metallic, semiconducting, or insulating solids is long and the number of $GW$ calculations is steadily growing. Recent flagship applications even include defects, surface effects and solvents in their comparison to experiment \cite{Gerosa/etal:2018_2}. For a recent review on the performance of different $GW$ variants to metal oxides, we refer to \cite{Bruneval/Gatti:2014,Gerosa/etal:2018}.

While the $f$-electrons are relatively inert in HfO$_2$, they assume a much more prominent role in lanthanide and actinide metals and oxides. With the expection of early explorations into Gd, $GW$ calculations for $f$-electron compounds have only emerged fairly recently
\cite{Chantis/etal:2007,Jiang/etal:2009,Richter/etal:2011,Jiang/Rinke/Scheffler:2012,Sakuma/etal:2012,Kutepov/etal:2012,Jiang:2018}. These calculations are almost always performed with DFT+U starting points or some form of self-consistency, as local or semi-local DFT provides a poor description of the electronic structure. 

QS$GW$ calculations for the rare-earth metals Gd and Er and the rare-earth monopnictides GdN, EuN, YbN, GdAs, and ErAs place the occupied 4$f$ states in agreement with photoemission measurements, but then overestimate the position of the unoccupied $f$ states \cite{Chantis/etal:2007}. Also, upon closer inspection, multiplet splittings are not reproduced with $GW$ and require a beyond $GW$ treatment \cite{Richter/etal:2011}. For the lanthanide sesquioxide (Ln$_2$O$_3$) series, $G_0W_0$@LDA+U calculations reproduce the relative positions of the occupied and unoccupied lanthanide $f$ states across the series and confirm the experimental conjecture derived from phenomenological arguments \cite{Jiang/etal:2009,Jiang:2018}. 

Cerium (Ce) is another paradigmatic material. With only one $f$ electron per Ce atom, it should still be relatively easy to describe, but Ce turns out to be an intricate material full of surprises. The phase diagram exhibits an unusual iso-structural phase transition. Both the $\alpha$ and the $\gamma$ phase have an fcc crystal structure, but the $\alpha$ phase has a smaller equilibrium volume \cite{Amadon/etal:2006,Bieder/Amadon:2014,Devaux/etal:2015}. The different localization of the $f$-electrons in the two phases is believed to be the driving force for the phase transition \cite{Casadei/etal:2012,Devaux/etal:2015,Casadei/etal:2016}. Ce is traditionally thought to be a strongly correlated material that belongs to the realm of dynamical mean-field theory (DMFT), see Section~\ref{sec:beyond} for further details. However, the $\alpha$ and $\gamma$ phases are already captured by the random-phase approximation (see Section~\ref{sec:gs} for details). The $G_0W_0$ spectral function of the $\alpha$ phase is in good agreement with photo and inverse photoemission spectra \cite{Sakuma/etal:2012}. However, in the more correlated $\gamma$ phase, $G_0W_0$ produces a peak at the Fermi level that is absent in the experimental spectra \cite{Sakuma/etal:2012}. 

In conclusion of this section, we would like to reiterate that materials with $d$- or $f$-electrons remain one of the most challenging applications of $GW$. As a matter of principle, $GW$ cannot yield an insulator with an odd number of electrons per unit cell. Such Mott insulators \cite{rmp_40_mott} are a manifestation of strong electronic correlation. Modern approaches to describing such strongly-correlated, localized states often combine $GW$ with either a phenomenological or first-principles treatment of $d$- or $f$-electron correlation, a topic we discuss further in section \ref{sec:beyond}.

\subsection{Defects in solids}
\label{sec:defects}

\begin{figure}[t] 
						 \includegraphics[width=1.0\linewidth]{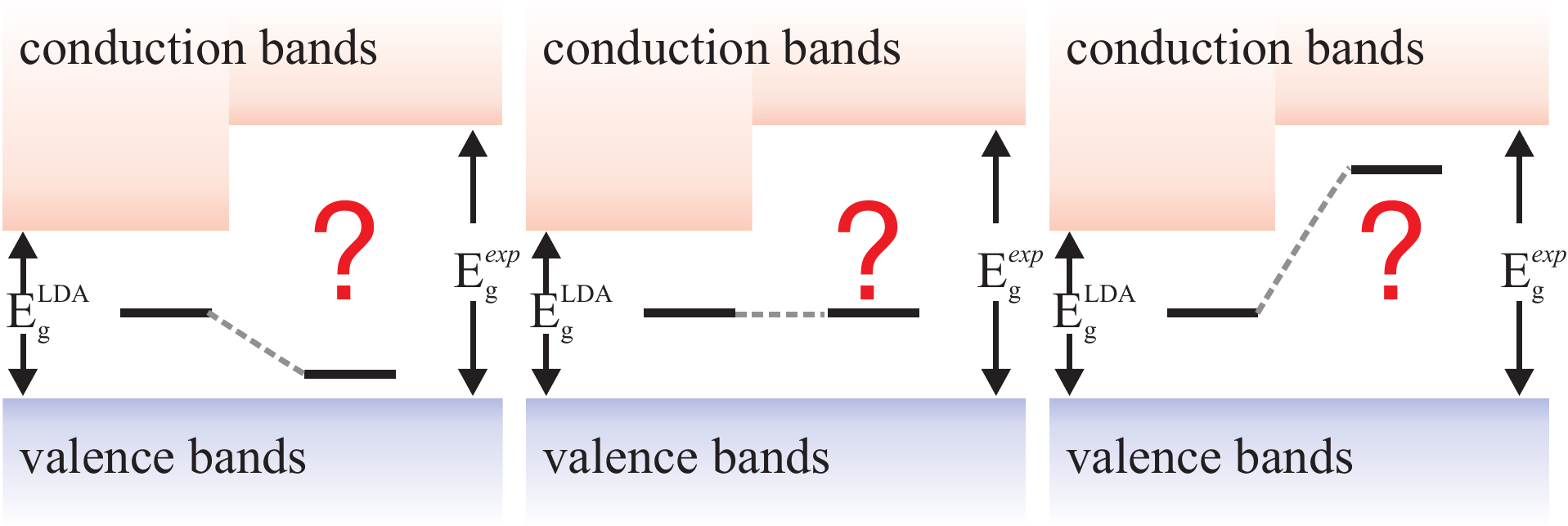}
         \caption{\label{fig:DFT_levels} For systems with mid-gap defect levels, computing the band gap alone is not enough to test the material for potential applications. The position of the defect level can also be computed with $GW$. In these cases, the absolute position of the VBM, CBM, and defect level are important.} 
\end{figure} 

So far, we have primarily discussed the performance of $GW$ for computing the band gap in solids, which does not depend on the absolute values of the band edges. However, the locations of the valence band maximum (VBM) and conduction band minimum (CBM) are essential for understanding defect level alignment in solids. The conceptual problem is illustrated in Figure \ref{fig:DFT_levels}. Assume an initial LDA calculation and then a $G_0W_0$ calculation of the band structure for a system with a defect level in the gap. We assume that the $G_0W_0$ band gap is in good agreement with experiment, but what about the defect level? The position of the defect level relative to the band edges and Fermi energy is critical for determining its occupancy when the system is put in contact with an electron reservoir. The band gap alone is no longer enough to assess the accuracy of the calculation. Similar to the defect problem, band alignment at semiconductor heterojunctions depends on the absolute position of the levels, a problem which is discussed more in Section~\ref{sec:surfaces}.

The accuracy of $GW$ for defect levels is still under investigation \cite{
Hedstroem/Schindlmayr/Scheffler:2002,%
Hedstroem/etal:2006,%
Weber/Janotti/Rinke/VandeWalle:2007,%
Ma/Rohlfing:2008,%
Rinke/etal:2009,%
Bruneval:2009,%
Ma/Rohlfing/Gali:2010,%
Bockstedte/etal:2010%
} and its comparison to hybrid functionals is summarized in the review of Chen and Pasquarello \cite{Chen/Pasquarello:2015}.
With a suitable choice of reference values to align the calculation with experiment, the accuracy of $G_0W_0$ is similar to that of hybrid functionals for predicting defect energy levels \cite{Chen/Pasquarello:2015}. Their major difference in performance can be attributed to their shift in the VBM, which has a direct effect on the defect level alignment and the calculated ionization potential. Hybrid functionals tend to symmetrically shift the VBM and CBM, while $G_0W_0$ mostly shifts the VBM down in energy which can worsen agreement with experiment for ionization potentials \cite{Chen/Pasquarello:2015}.

\subsection{Outlook on solids}
\label{sec:outlook}

As large-scale $GW$ implementations became more common and parallelism increased, $GW$ calculations became an indispensable tool for \textit{ab-initio} predictions in solids. Today, there are too many $GW$ calculations for solids to count. Even so, comparing $GW$ calculations to experiment must be done with some care because there are additional effects in the experiment that are not included in ordinary $GW$. For example, electron-phonon coupling can have a significant effect on the band gap in real materials \cite{Botti/Marques:2013,Kawai/etal:2014,Giustino/etal:2010,Antonius/etal:2014,Cannuccia/Marini:2011}. The effect of the electron-phonon interaction can also be described by a self-energy and calculated with perturbation theory \cite{Smondyrev:1986,Cederbaum/Domcke:1974}. Experimental spectroscopies are also surface sensitive techniques, as mentioned briefly in Section~\ref{sec:exp_spec}, which means that the measured band structure may not correspond perfectly to the bulk states. These considerations aside, the impressive success of $GW$ in solids encouraged studies of other systems, including surfaces and molecules.

\section{Surfaces}
	\label{sec:surfaces}

\begin{figure} 
        \includegraphics[width=0.99\columnwidth]{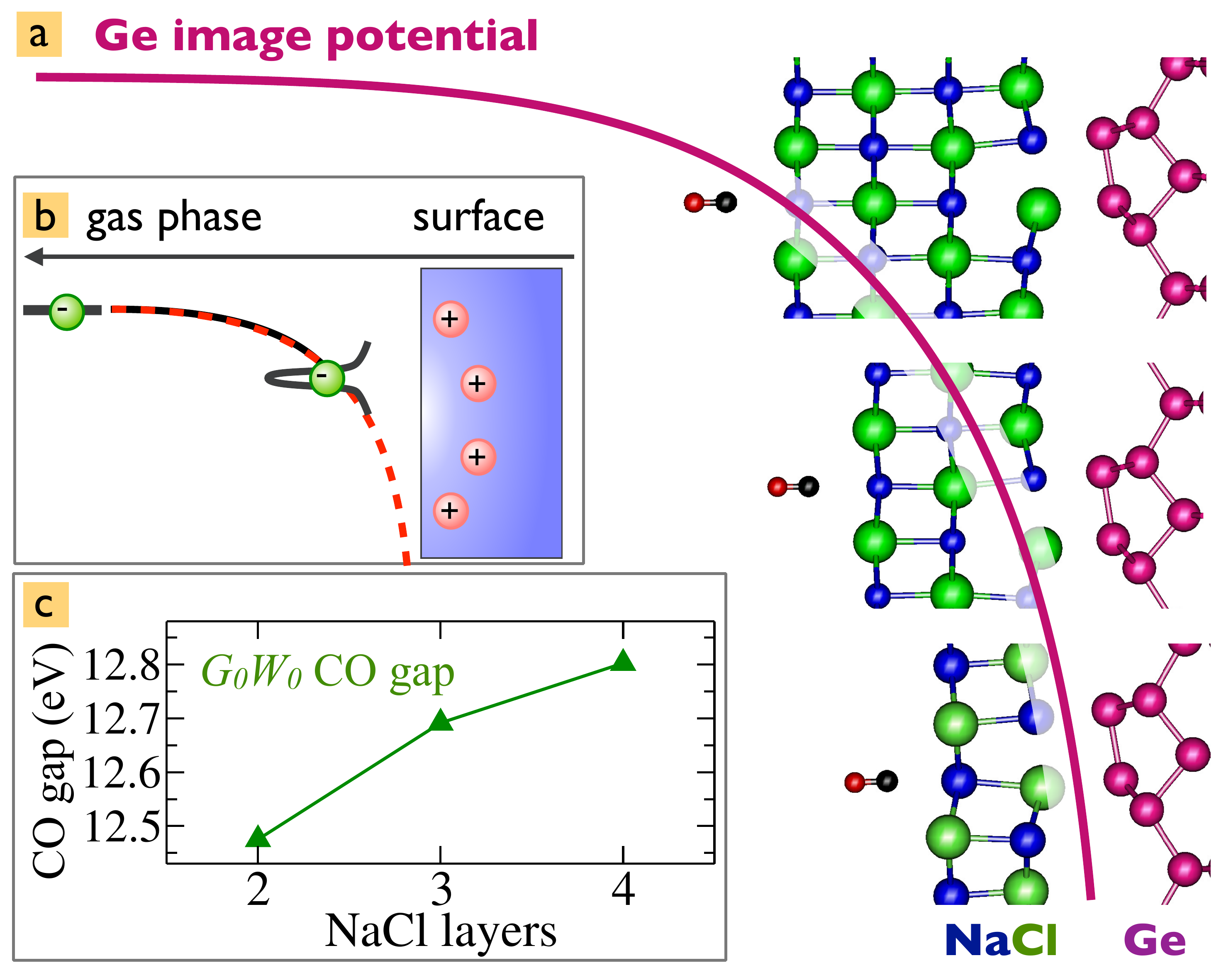}
	\caption{{\small \label{fig:impot} Illustration of the image effect: panel (b) shows the image charge and image potential induced by an additional electron (e.g. anionic charge on a molecule) outside a surface. Panel (a) provides a graphic illustration how the image potential of a germanium (Ge) surface could be probed with a carbon monoxide (CO) test molecule. By adding thicker and thicker sodium chloride (NaCl) layers between CO and Ge, the CO molecule moves along the Ge image potential. The resulting CO gap will then depend on the NaCl layer thickness, which is indeed the case as panel (c) illustrates. Figure adapted from \onlinecite{Freysoldt/Rinke/Scheffler:2009}.}}
\end{figure}

The application of $GW$ to surfaces and interfaces is not as common because these systems tend to have large unit cells with a number of atoms beyond the tractability of many $GW$ codes. However, what makes surfaces particularly interesting from the $GW$ perspective is a long-range polarization effect contained in the screened Coulomb interaction that is absent for bulk materials: the image effect. As illustrated schematically in Figure~\ref{fig:impot}, an additional charge (hole created in the photoemission process or added electron in inverse photoemission) outside a surface induces an image charge in the surface \cite{Deisz/Eguiluz/Hanke:1993}. This charge gives rise to an additional potential, the image potential, that renormalizes the energy of the electron or hole state. For metallic and dielectric surfaces it is easy to show from simple electrostatic considerations that the image potential should decay with the inverse distance from the surface. For other geometries, e.g., quantum dots or nanostructures, this decay behaviour is modified \cite{ClusterImStates:2004}.

That the $GW$ self-energy encompasses the image effect was first shown by extracting the image potential from the $GW$ self-energy for the Al(111) surface \cite{White/Godby/Rieger/Needs:1997}. Later, image resonances \cite{Fratesi/Brivio/Rinke/Godby} and image states for semiconductors, insulators \cite{Kutschera/etal:2007,Rohlfing/Wang/Krueger/Pollmann:2003} and nanoclusters \cite{ClusterImStates:2004} were calculated with $GW$. \onlinecite{Freysoldt/Rinke/Scheffler:2009} showed that the image potential can also be probed by monitoring the excitation energies of a test molecule (see Figure~\ref{fig:impot}). The test molecule (CO) can be moved along the image potential by introducing insulating spacer layers between molecule and surface. The energy of the CO states gets renormalized stronger the closer it is to the surface, i.e., the smaller the spacer layer is.  \onlinecite{Freysoldt/Rinke/Scheffler:2009} also showed that the energy of  semi-core states in different NaCl layers is affected by the image potential in the same way, a result that was later corroborated by \onlinecite{Strange/Thygesen:2012} in a model study.

The most significant effect of the image potential is that it renormalizes the energy of adsorbates such as organic molecules \cite{,Freysoldt/Rinke/Scheffler:2009,Thygesen/Rubio:2009,Garica-Lastra/etal:2009,Puschnig/etal:2012}. The energetic position of molecular states near or on the surface is different from the molecule in the gas phase. During the excitation process, an electron or hole is added at the molecule. The additional correlation energy due to the polarization of the surface further stabilizes the added charge. As result, occupied states move up in energy and unoccupied states down and the HOMO-LUMO gaps reduce consequently in size, see Figure~\ref{fig:impot}(b). The renormalization depends on the dielectric constant of the surface. The larger the dielectric constant, the larger the renormalization. Already for surfaces of insulators the HOMO-LUMO gap renormalization is of the order of 1~eV and can reach more than  3~eV for metallic surfaces \cite{Thygesen/Rubio:2009,Garica-Lastra/etal:2009}.

Apart from the HOMO-LUMO gap, the position of adsorbate states relative to the substrate's Fermi level or relative to the band edges is of significant interest in surface and interface science. This relative positioning of adsorbate to substrate states is commonly referred to as \emph{level alignment}. $GW$ calculations are currently considered to be the holy grail for an accurate determination of the level alignment. 
However, due to the aforementioned computational reasons (i.e. very large supercells) most $GW$ level alignment calculations reported in the literature are not converged. Careful $GW$ cluster calculations \cite{Patrick/Giustino:2012,Wippermann/etal:2014,Govoni/etal:2015} and very large scale $GW$ calculations report good agreement with experiment. For physisorbed molecules, whose electronic states do not couple strongly to the substrate, the $GW$ self-energy can be split into a surface and a molecular contribution. Such a simplified $GW$ polarization model has been  developed by \onlinecite{Neaton2006} for weakly interacting molecules at metallic surfaces. This model has been used to compute $GW$ estimates for the level alignment of amine-gold junctions and interfaces \cite{Quek/etal:2007,Tamblyn/etal:2011} as well as $\pi$-conjugated polymers at Au(111)\cite{DiGiovannantonio2018}.

\begin{figure} 
        \includegraphics[width=0.99\columnwidth]{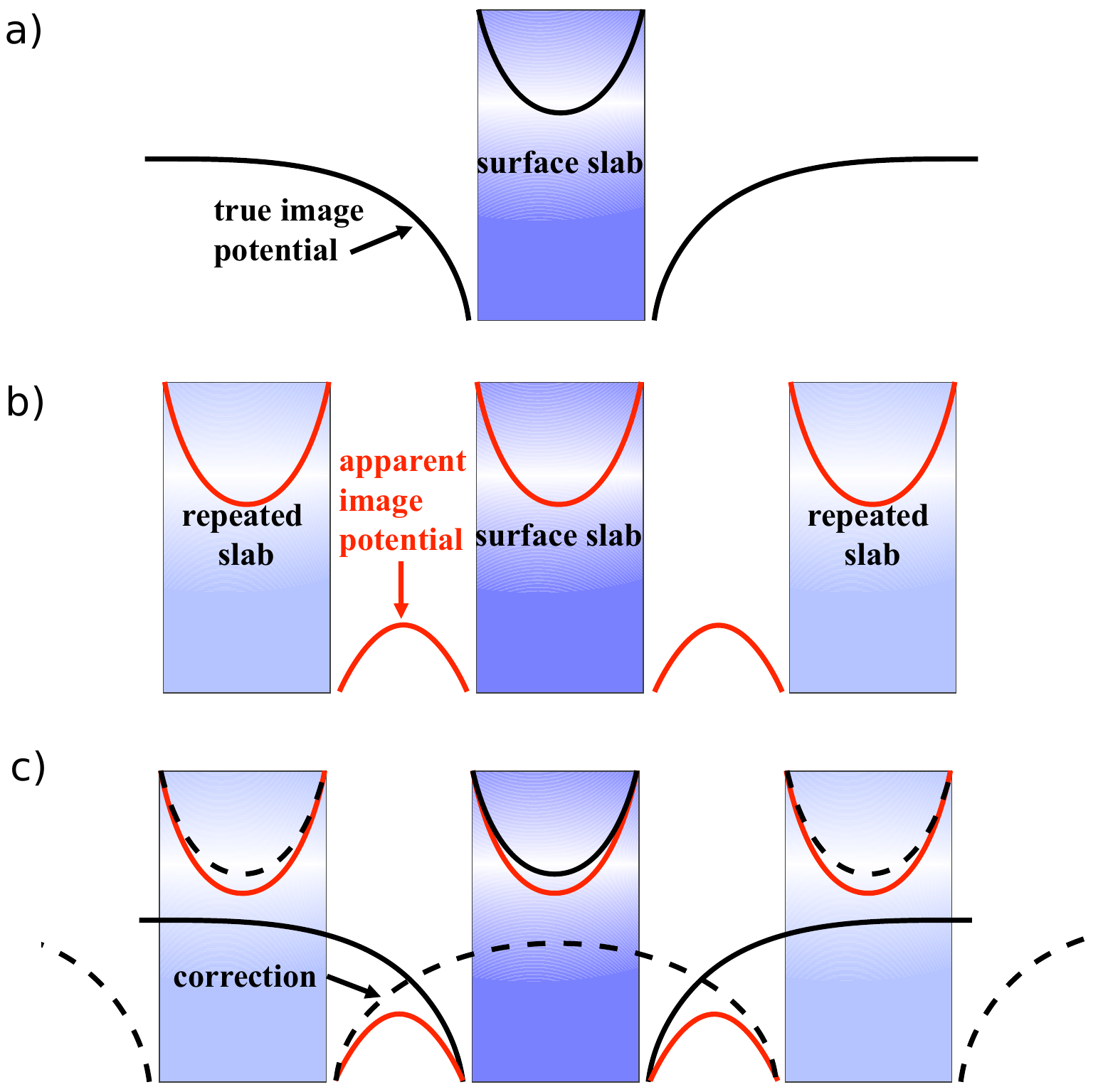}
	\caption{{\small \label{fig:im_rs} The image potential of a repeated slab system (b) differs from that of an isolated surface (a). The dashed lines in (c) mark the difference that can be computed with a suitable correction scheme \cite{slabgw}. As the charge moves across the interface, the ratio of dielectric constants for the ``charged" and ``uncharged" regions changes. As a result, the image potential changes sign. 
	 }}
\end{figure}

$GW$ calculations for surfaces and interfaces are not only challenging because of the large supercells. An additional complication is the vacuum spacing in the common repeated slab model.
In $GW$ calculations that apply periodic boundary conditions, the surface is modelled as a slab of material that is periodic in two dimensions and finite in the third. The rest of the supercell in the direction perpendicular to the surface is filled with vacuum. Since the periodic boundary conditions apply also in the dimension perpendicular to the surface, the final system is a heterostructure of repeated blocks of material and vacuum (see Figure~\ref{fig:im_rs}). $GW$ now couples these repeated slab images because the $GW$ interaction is long-ranged. The image potential decays only with the inverse distance between the slabs (see image effect discussion above) and not exponentially fast, as local or semi-local DFT functionals do. As a result, image potential tails generated by the repeated slab images reach into the surface region we would like to model with the slab model (see Figure~\ref{fig:im_rs}) and obscure the actual image potential.

In a $GW$ calculation, the image potential is always present, even if we are not explicitly interested in image states. Due to the long range of the interaction, the vacuum spacing cannot be converged out in any $GW$ implementation that has to place basis functions in the vacuum region (as for example plane waves) \cite{slabgw,Huser/etal:2013}. Two prevalent solutions to this problem have emerged: 1) to cut the interaction range and use an effective short-range interaction or 2) to apply post-processing corrections. The easiest way to limit the range is to impose a spherical cutoff on the Coulomb interaction every time it is used in the $GW$ equations \cite{Onida/etal:1995,Spataru/etal:2004,Rozzi/etal:2006,Sohrab:2006}. The largest disadvantage of this approach is that the spherical cutoff also limits the range of the $GW$ interaction inside the material and in the two directions parallel to the surface. The cutoff radius should therefore at least be as large as the slab is thick. This implies that the vacuum separation should at least be equal to the slab thickness, which increases the computation time again for thicker slabs. 

A computationally more efficient way is to apply post-processing corrections to a normal $GW$ calculation that does not modify the range of the Coulomb interaction \cite{slabgw}. Care has to be taken, however, that the $GW$ implementation correctly includes the dielectric tensor \cite{GW_space-time_method_surf:2007}. Otherwise, the $GW$ calculation will not converge with respect to $\vk$-points \cite{Huser/etal:2013,GW_space-time_method_surf:2007}.  Such a post-processing correction has been derived from an electrostatic model \cite{slabgw} and is depicted in Figure~\ref{fig:im_rs}. The true image potential is shown for two scenarios in Figure~\ref{fig:im_rs}(a): for a charge located outside or inside the slab.  
As the charge moves from outside the slab to inside, the image potential changes sign, as now the dielectric constant in the region where the charge resides (i.e. in the slab) is larger than where the charge is not (i.e. in the vacuum). Figure~\ref{fig:im_rs}(b) shows the image potential for a periodic array of slabs in the repeated slab approach. It is notably different from the image potential of a single slab in Figure~\ref{fig:im_rs}(a). The correction derived by~\onlinecite{Freysoldt/Rinke/Scheffler:2009} is shown as black dashed lines in Figure~\ref{fig:im_rs}(c) and resotes the correct behavior for a single slab. The corrections can be several tens of eV large and yield converged results already for small vacuum thicknesses \cite{slabgw}.

At surfaces, the DFT wave functions are sometimes poor approximations of certain surface states and image states. In such cases, it is desirable to calculate quasiparticle wave functions. This can be done by solving the full quasiparticle equation \eqref{Eq:qp_eq} in a suitable basis. If this solution is performed iteratively in energy, new quasiparticle states, such as image states, can be found that are absent from the DFT spectrum. Examples where the quasiparticle wave functions differ notably from the LDA or PBE wave functions are GaAs(110) \cite{Pulci/etal:1999} and the C(111) surface \cite{Marsili/etal:2005} as well as image states \cite{White/Godby/Rieger/Needs:1997,Rohlfing/Wang/Krueger/Pollmann:2003,Kutschera/etal:2007}.

Early $GW$ calculations for surfaces focused on surface states of simple semiconductors such as  silicon \cite{Rohlfing/Krueger/Pollmann:1995,Rohlfing/Krueger/Pollmann:1997,Rohlfing/Louie:1999,Hahn/Schmidt/Bechstedt:2001,Weinelt/etal:2004}, germanium \cite{Rohlfing/Krueger/Pollmann:1996}, silicon carbide \cite{Rohlfing:2001}, gallium phosphite \cite{Schmidt/etal:1999}, indium phosphite \cite{Schmidt/etal:2000,Hedstroem/etal:2006}  and insulators such as diamond \cite{Marsili/etal:2005}, lithium fluoride \cite{Wang/etal:2003} and sodium chloride \cite{Freysoldt/Rinke/Scheffler:2009}. Frequently, the $GW$ quasiparticle energies are taken as input for optical absorption or reflectance anisotropy spectroscopy (RAS) studies \cite{Pulci/etal:1998,Schmidt/etal:2000,Hahn/Schmidt/Bechstedt:2001}. The surface band structure and dispersion of surface states is in good agreement with available photoemission studies. Also, computed optical and RAS spectra agree well with experimental spectra for these systems. Later calculations for more complex surfaces or surface adsorbates have to be taken with a grain of salt, since they may not be fully converged with respect to all computational parameters, unless plasmon pole models, other model dielectric functions, or cluster models were used \cite{Giorgi/etal:2011,Patrick/Giustino:2012,Alves-Santos/etal:2014}.

\section{Two-dimensional materials}
	\label{sec:2d}
		
\begin{figure} 
    \includegraphics[width=0.99\columnwidth]{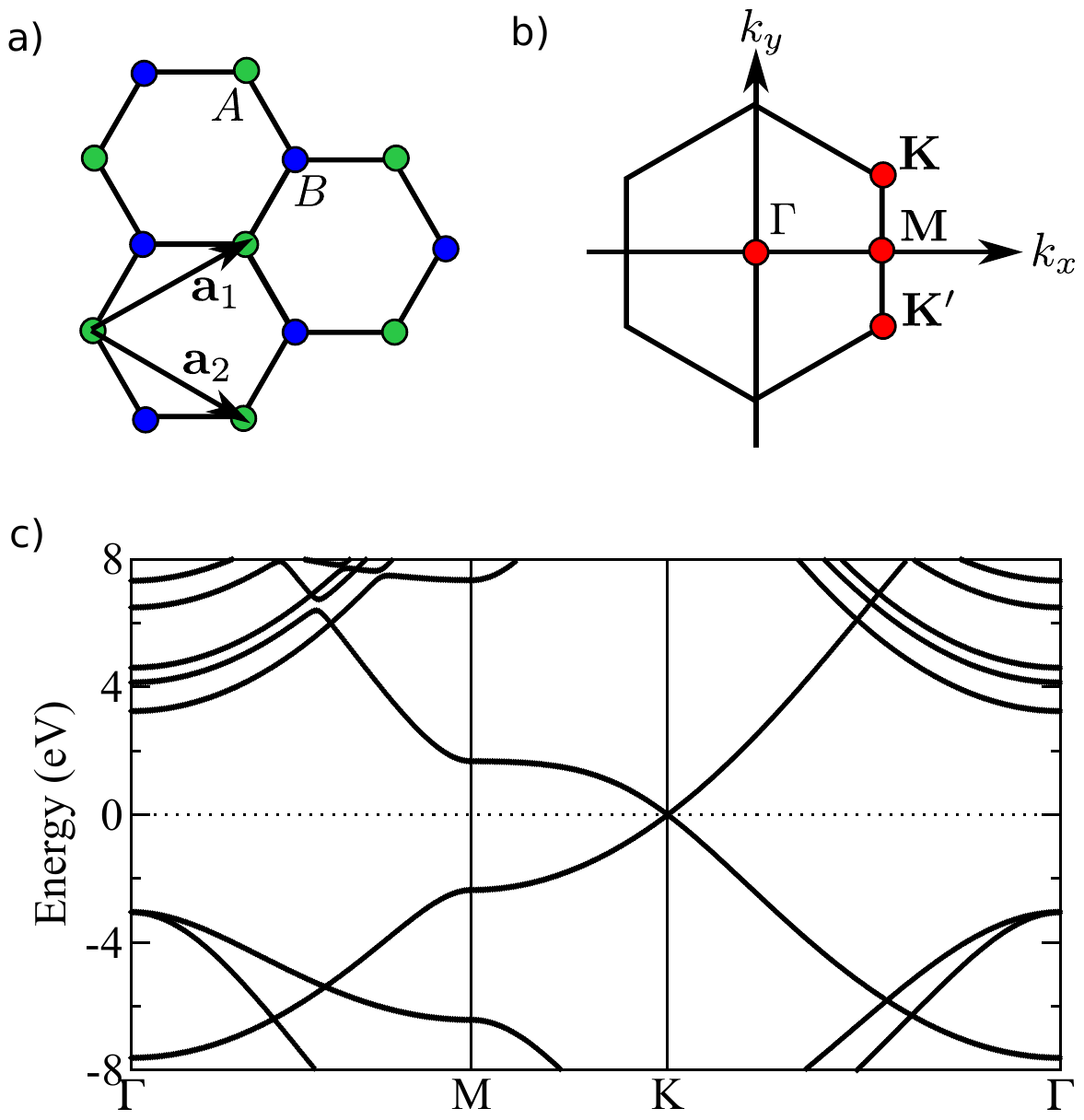}
	\caption{{\small \label{fig:graphene} (a) Graphene has two hexagonal sublattices ($A$ and $B$) in its honeycomb structure with translation vectors $\mathbf{a}_1$ and $\mathbf{a}_2$. (b) The Brillouin zone is hexagonal with two symmetry inequivalent corners labeled $\mathbf{K}$ and $\mathbf{K}'$. (c) Near the Dirac points at $\mathbf{K}$ and $\mathbf{K}'$, the dispersion is linear. The band structure is computed at the PBE level and taken from the Computational 2D Materials Database \cite{Haastrup_2018} with Fermi energy set to zero.
	 }}
\end{figure}

Research in two-dimensional materials developed rapidly after the isolation of graphene in 2004 \cite{novoselov_sci_306}. The crystal structure and Brillouin zone of graphene are shown in Figure~\ref{fig:graphene}. Two-dimensional materials have gained great fame for their interesting electronic structures, which include phenomena like Dirac fermions and topological insulators \cite{geim_nm_6,castro_neto_rmp_81,bhimanapati_acsn_9}. Models of these effects are largely in the single particle $-$ or single quasiparticle $-$ picture. $GW$ serves an important purpose to parameterize such models from a fully \textit{ab-initio} perspective. 

Two-dimensional materials often exist at the size and interaction strength that is ideally suited for $GW$. They are too large (the required Brillouin zone sampling is too dense) for more expensive wave function or beyond-$GW$ Green's function methods, but their correlation is usually weak enough that $GW$ gives a good description of their electronic structure. Similar to $GW$ calculations on surfaces (Section~\ref{sec:surfaces}) or molecules (Section~\ref{sec:molecules}), two-dimensional materials can show enhanced interaction effects from reduced dimensionality and decreased screening compared to bulk solids. Technical aspects of $GW$ calculations of two-dimensional materials include the truncation of the screened Coulomb interaction between layers (similar to surfaces) and slow convergence with respect to $\mathbf{k}$-points \cite{thygesen_2d_4,rasmussen_prb_94,Qiu2016}.

The band structure of graphene (and many other two-dimensional materials) is characterized by a zero band gap and linear dispersion near the Fermi energy, $E(\mathbf{q})_{\pm} \approx  \pm v_{\rm F} | \mathbf{q} |$ where the $+$ ($-$) sign refers to electrons (holes) and $\mathbf{q}$ is the wave vector relative to the $\mathbf{K}$ or $\mathbf{K}'$ points of the Brillouin zone, see Figure~\ref{fig:graphene}. $v_{\rm F}$ is called the Fermi velocity and is the slope of the dispersion at the band edges. This linear dispersion is strikingly different than the parabolic dispersion in Equation~\eqref{eq:bandparabola}, which is the case for most materials. Not long after its discovery, $GW$ was applied to graphene to calculate the band structure and $v_{\rm F}$ from first principles \cite{trevisanutto_prl_101,siegel_pnas_108,park_nl_9}. Compared with calculations based on the local density approximation, $GW$ preserves band closure at the Fermi energy and increases the Fermi velocity by $\sim 17 \%$ to give a value ($1.1 \times 10^6$ m/s) which is in good agreement with experiment. These studies also found kinks which appear in the low energy band structure from electron-phonon coupling and doping level dependent kinks of purely electronic origin.

Replacing carbon with a different group IV element creates a family of graphene-like materials. By preserving the honeycomb lattice of graphene, the materials still host Dirac fermions, but their chemistry and Fermi velocities depend on the specific element. For example, $GW$ calculations of the Fermi velocity of planar silicon, called silicene, give a value of $\sim 7.7 \times 10^5$ m/s \cite{huang_apl_102}. Because of silicon's tendency for $sp^3$ hybridization, silicene also has a buckled structure which preserves linear dispersion at the $GW$ level of theory \cite{wei_pccp_15}. As with graphene, the electronic structures of silicene and germanene (monolayer Ge) subject to hydrogenation, strain, and hybridization with other materials have been studied with $GW$ \cite{drissi_pe_68,yan_prb_91,wei_prb_88,wang_cpb_24,wang_pccp_2017,drissi_pe_74}. 

As one goes down the group IV elements, they become heavier; this has great significance for spin-orbit coupling (SOC) in two-dimensional materials. A SOC induced band gap in two-dimensional materials is critical to the topological character of their electronic structure. Stanene is a group IV monolayer (Sn) that has a sizable band gap due to SOC \cite{lu_sr_7,lu_jmc_22}.

A number of functionalizations or structural modifications to graphene have been proposed for modifying its electronic structure. Much of the research in functionalized graphene is directed towards achieving semiconducting graphene or, more generally, two-dimensional semiconductors. As interesting as Dirac fermions are, semiconducting layers are necessary to build many layered electronic devices like field-effect transistors. For example, passivating graphene with hydrogen transforms it from a sheet of $sp^2$ bonded carbon to $sp^3$. The passivated structure, called graphane, has a $GW$ band gap of $\sim 5$ eV \cite{karlicky_jctc_9,leenaerts_prb_82,lebegue_prb_79,hadipour_epjb_88} and could be useful as a two-dimensional semiconductor. Other passivated graphenes also open a band gap \cite{wei_prb_87,klintenberg_prb_81}. One can also apply strain, poke holes, or form other planar carbon allotropes by rearranging carbon bonds, many of which open an appreciable band gap ($\sim 1$ eV) in graphene at the $GW$ level \cite{liang_jmr_27,Dvorak2015,appelhans_prb_82,appelhans_njp_12,nisar_nano_2012}.
\begin{figure} 
    \includegraphics[width=0.99\columnwidth]{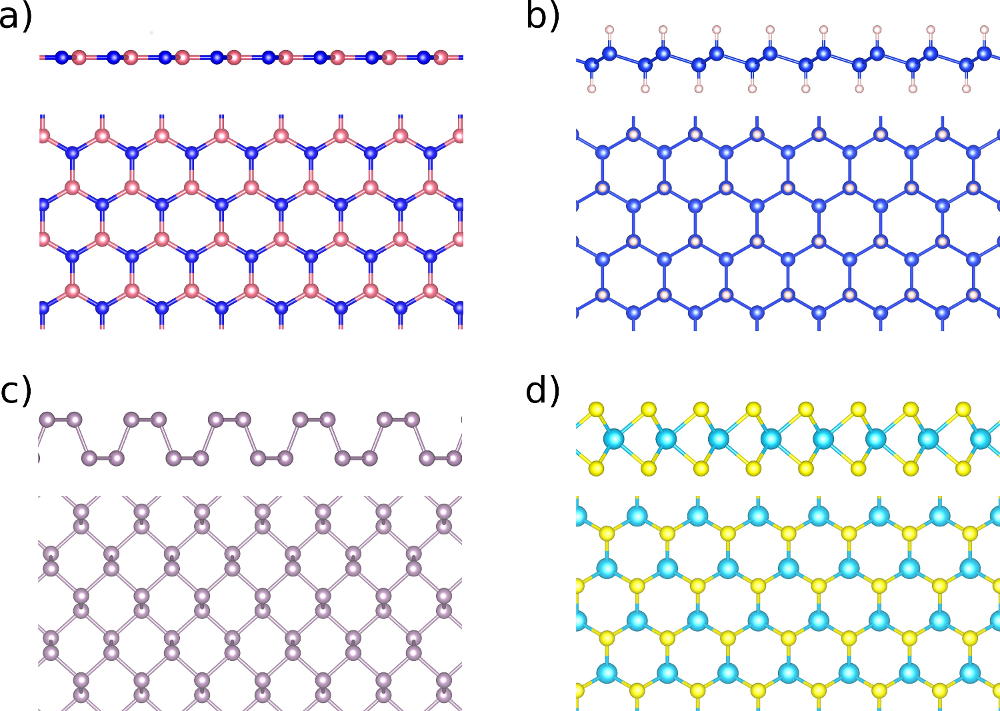}
	\caption{{\small \label{fig:2d} Top and side views of (a) hexagonal boron nitride, (b) hydrogenated silicene (silicane), (c) phosphorene, (d) and 2H-MoS$_2$. Structures taken from the Computational 2D Materials Database \cite{Haastrup_2018}.
	 }}
\end{figure}

To fill the need for two-dimensional semiconductors, one can move away from graphene and consider materials that are intrinsically semiconducting. Elements from the third and fifth groups of the periodic table (III-V compounds) often form a semiconducting monolayer, see Figure~\ref{fig:2d}. This is largely because of the $A$-$B$ sublattice imbalance in these materials, which opens a band gap at the tight-binding level of theory \cite{wallace_pr_71}. A monolayer of hexagonal boron nitride (hBN) is one possibility, with a $GW$ band gap of $\sim 7.5$ eV  \cite{prb_berseneva_87,wirtz_aip_786}. Because graphene and hBN have similar lattice constants, they can be layered or hybridized easily, which gives additional tunability of the electronic properties \cite{bernardi_prl_108}. GaAs is another example, with a $GW$ band gap of $\sim 3$ eV \cite{fakhrabad_pe_59}. Other III-V monolayers are also stable and have been studied with $GW$ \cite{ciraci_prb_80,wang_ssc_150,fakhrabad_sm_79,prete_apl_110}. Phosphorene is a somewhat unusual case, as a monatomic group V material. This is reflected in its unusual structure, which has armchair-like vertical buckling, shown in Figure~\ref{fig:2d}(c). Phosphorene is attractive because it has a smaller band gap than many other two-dimensional semiconductors, computed to be $\sim 2$ eV with $GW$ \cite{steinkasserer_prb_94,tran_prb_89,ferreira_prb_96,li_nn,rudenko_prb_92}, which is well-suited for applications.

\begin{figure} 
    \includegraphics[width=0.96\columnwidth]{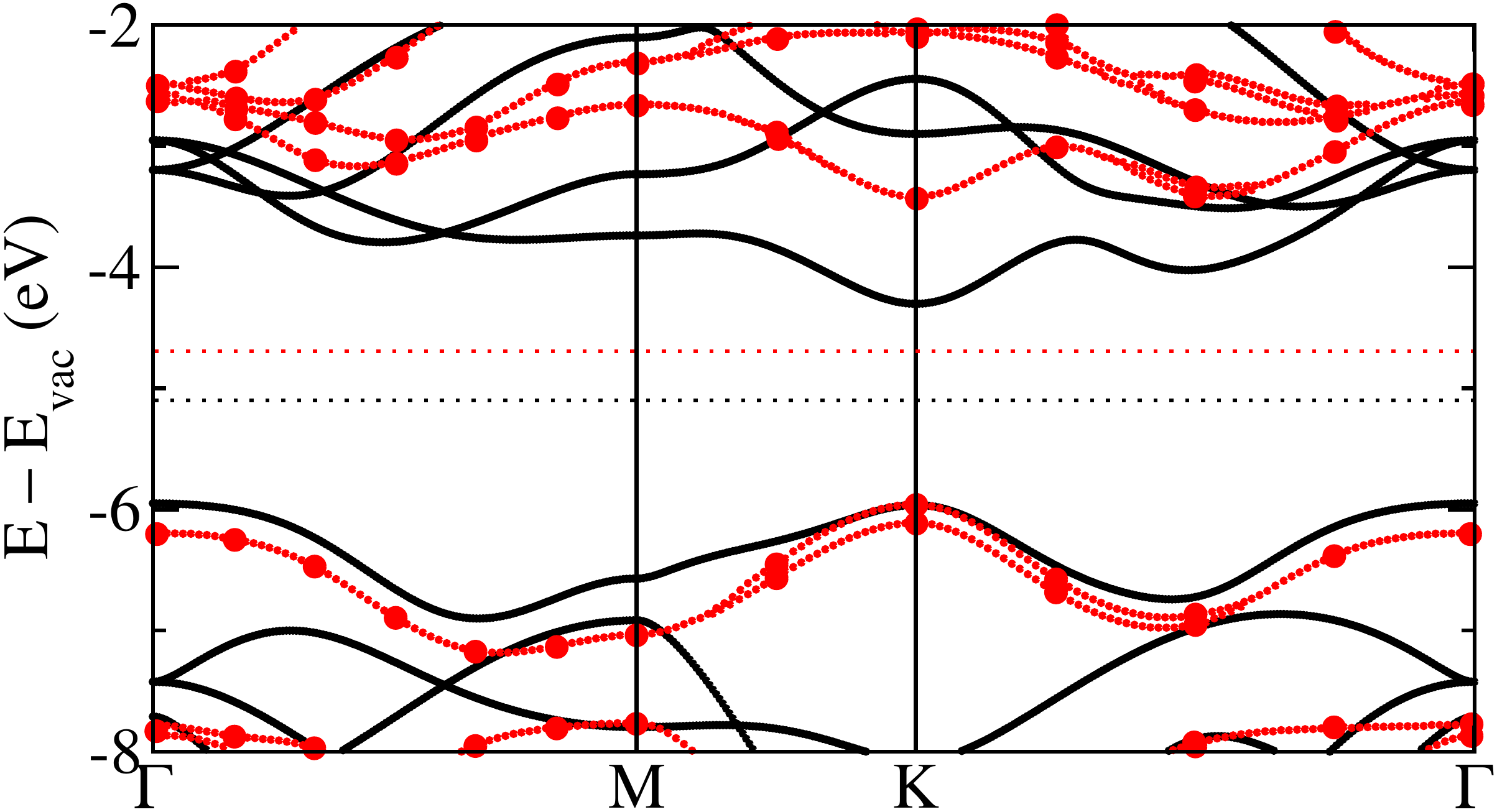}
	\caption{{\small \label{fig:mos2} Band structure of MoS$_2$ in the 2H phase at the PBE (black) and $G_0W_0$ (red) levels. The $G_0W_0$ bands include spin-orbit coupling but the PBE bands do not. The Fermi energies for each case are indicated by horizontal dotted lines. Data taken from the Computational 2D Materials Database \cite{Haastrup_2018}.
	 }}
\end{figure}
Finally, we get to the transition metal dichalcogenides (TMDs). TMDs have the chemical formula MX$_2$ where M is a transition metal and X is a chalcogen, commonly S, Se, or Te. In their stable two-dimensional phase, TMDs usually form a three-layered structure with the transition metal atoms in a central layer between the chalcogens (called the 2H phase). MoS$_2$, MoSe$_2$, WS$_2$, and WSe$_2$ are all semiconductors with $G_0W_0$@LDA band gaps from 2.0$-$2.5 eV when including SOC \cite{rasmussen_jpcc_119}. The band structure of MoS$_2$ is shown in Figure~\ref{fig:mos2}. TMDs feature unusual electronic structures derived from strong SOC and lack of inversion symmetry (see~\onlinecite{manzeli_nrm_2}). $GW$ calculations at either the perturbative or partially self-consistent levels improve the agreement with experiment for fundamental band gaps \cite{debbichi_prb_89,Shi/etal:2013,Komsa/etal:2012,Qiu2016,robert_prb_94,espejo_prb_87,lee_2dm_4,Huser/etal:2013,molina_prb_88,Ramasubramaniam/etal:2012,cheiwchanchamnangij_prb_85,Ugeda/etal:2014}. However, conclusions from different $GW$ studies on the magnitude and character (direct or indirect) of the band gap in MoS$_2$ are not entirely consistent. Depending on the level of self-consistency, truncation of Coulomb interaction, treatment of frequency dependence, and $\mathbf{k}$-point sampling, the $GW$ quasiparticle band gap of MoS$_2$ can vary by $\sim 0.44$ eV \cite{Qiu2016}. 

The MoS$_2$ case study highlights the importance of carefully converging $GW$ calculations and the difficulties of two-dimensional materials, in particular. In two-dimensional semiconductors, the dielectric function is a linear function of $\mathbf{q}$ which results in very slow $\mathbf{k}$-point convergence \cite{rasmussen_prb_94}. TMDs are also commonly stacked in layered materials called van der Waals heterostructures, which allow one to tune the electronic structure for device applications \cite{winther_2dm_4,zhang_2dm_4,arora_natcom_8}. $GW$ allows one to predict band alignment in these heterostructures from first principles \cite{ganesan_apl_108}.

\section{Molecules}
	\label{sec:molecules}
The application of $GW$ to molecules is a relatively new field of research that has developed rapidly over the last decade. The electronic screening is much weaker in molecules than in extended systems. The low charge density in molecules does not naturally fit a screening interpretation of correlation which is intrinsic to $GW$ and replacing the bare Coulomb potential with the dynamically screened Coulomb interaction $W$ might not be the obvious choice. Even so, a rigorous test of $GW$ for the He atom, with only two electrons, found excellent agreement with numerically exact results \cite{li_prl_118}. In addition, the first exploratory $G_0W_0$ studies on molecular systems revealed that the inclusion of screening at the $GW$ level substantially improves  electron removal and addition energies  \cite{Grossman/etal:2001,Dori2006,Ma/Rohlfing/Molteni:2009,Ma/Rohlfing/Molteni:2010,Niehaus2005,Rostgaard/Jacobsen/Thygesen:2010,Blase/Attaccalite/Olevano:2011,KeSanhuang:2011}.  
%
%
\subsection{First ionization potentials and electron affinities}
In molecules, the single-particle states $\{\phi_s^0\}$ correspond to molecular orbitals (MO) with discrete energies. The energy to remove an electron from an MO is referred to as ionization potential. The negative of the electron affinity (EA) corresponds to the energy needed to add an electron to the LUMO of the neutral system (-$\text{EA}_{\text{ LUMO}}$=$\epsilon_{\text{ LUMO}}$), see also Equations~\eqref{Eq:def:Ei} and \eqref{Eq:def:Ef}. $G_0W_0$ provides access to both quantities. 
Furthermore, we can calculate the fundamental gap from the first ionization potential, $\text{IP}_{\text{HOMO}}$, and the electron affinity
\begin{equation}
 \Delta_{\text{fgap}} =\text{IP}_{\text{HOMO}}-\text{EA}_{\text{LUMO}}.
\end{equation}
The fundamental gap should not be confused with the optical gap $\Delta_{\text{ogap}}$, which is the energy needed for the charge neutral excitation from the HOMO to the LUMO. The optical gap is lower in energy than $\Delta_{\text{fgap}}$ and can \textit{not} be obtained from $GW$. It defines the threshold for photons to be absorbed and for the formation of a bound electron-hole pair (exciton). Conversely, the fundamental gap is the energy threshold for the formation of a separate electron-hole pair, which is not bound together. It can be considered as the molecular equivalent to the band gap, see also Refs.~\cite{Bredas2014,Baerends2013}.\par
$GW$ has been mainly applied to compute the IP for the HOMO and the electron affinity for $\pi$-conjugated molecules with potential for organic photovoltaic applications \cite{KeSanhuang:2011,Blase/Attaccalite/Olevano:2011,Faber2011,Faber/etal:2012,Gallandi2015,Gallandi2016,Knight2016,Wilhelm2016,Marom:2017}. Examples for relevant $\pi$-conjugated organic molecules are linear acenes (linearly fused benzene rings), quinones, aromatic nitriles, anhydrides, porphyrins, and thiophene polymers. These classes of molecules are particularly suited as organic semiconductor because their EA is often positive\footnote{Note that different sign conventions are used for EA in literature. We define EA as the energy required to detach an electron from a negatively charged species. If EA is defined as the energy required to add an electron to a neutral atom, the sign swaps. EA refers always to the LUMO. The label for the state is therefore dropped in the following. } \cite{Richard2016}, i.e., they are electron acceptors and their fundamental gap is much smaller than in inorganic molecules. For example, smaller acenes have gaps between $6.0-7.0$~eV \cite{Richard2016}, whereas the fundamental gap of a small inorganic molecule like water is larger than 14.0~eV \cite{Setten2015}. \par
The fundamental gap, $\text{IP}_{\text{HOMO}}$ and EA are critical parameters for the charge transport in organic semiconductors. 
Over the last years it has been shown that $GW$ predicts these properties well. Using an appropriate starting point (see Section~\ref{subsec:g0w0start}), the reported mean absolute deviations (MADs) of $\text{IP}_{\text{HOMO}}$ and EA are less than 0.2~eV  from the CCSD(T) reference \cite{Gallandi2016,Knight2016}. The MAD of $\text{IP}_{\text{HOMO}}$ with respect to experiment can be even reduced to $< 0.1$~eV when including also vibrational effects in the
$GW$ spectra \cite{Gallandi2015}.\par
 The electronic properties of $\pi$-conjugated molecular structures can be tuned by, e.g., increasing the chain length. It has been shown that $GW$ correctly predicts the decrease of $\text{IP}_{\text{HOMO}}$ in trans-polyacetylene with increasing chain length \cite{Pinheiro2015,Bois2017}. Similar $GW$ studies were conducted for band gaps of linear acenes \cite{Wilhelm2016}. The photovoltaic properties can be further modulated by using two different organic semiconductors in the cell: a molecule with a low $\text{IP}_{\text{HOMO}}$ (electron donor) and molecule with electron-acceptor character, i.e. with a high EA~\cite{Kippelen2009}. The level alignment of  such donor-acceptor systems has been studied with $GW$ for tetrathiafulvalene (TTF) and tetracyanoethylene (TCNE) or tetracyanoquinodimethane (TCNQ) dimers, demonstrating the importance of well-chosen starting points or self-consistent schemes \cite{Caruso/etal:2014,Gallandi2015}.\par
 The accurate prediction of charged excitations is not only important for organic semiconductors, but also for DNA and RNA nucleobases in order to study their damage following exposure to ionizing radiation. IPs and EAs for these molecules have been reported at the $G_0W_0$ level in good agreement with experiment and quantum chemistry methods  \cite{Qian/Umari/Marzari:2011,Faber2011,Gallandi2015}. 
 
%
\subsection{Ionization spectra}
\begin{figure} 
        \includegraphics[width=0.99\linewidth]{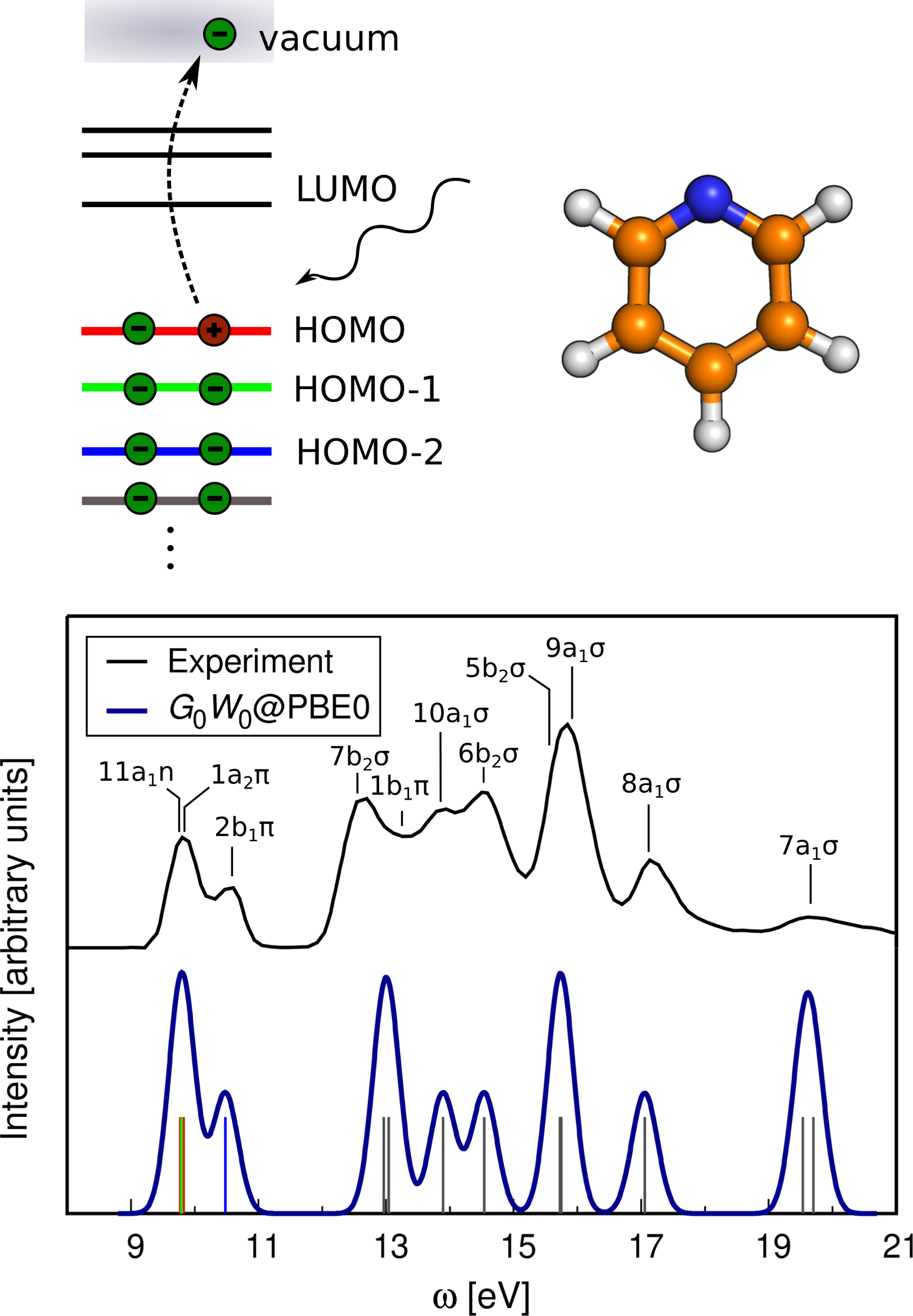}
	\caption{{\small \label{fig:IP_pyridine}
        Ionization spectrum of pyridine. $G_0W_0@$PBE0 QP energies compared to the experimental photoemission spectrum \cite{Liu2011}. The calculated spectrum has been artificially broadened; the position of the QP energies is indicated with vertical bars. All QP energies are extrapolated using the cc-pV$n$Z ($n$=3-6) basis sets, see Appendix~\ref{app:computational_details} for further computational details. The QPs of the first valence states are colored in red, green and blue.
    }}
\end{figure}
The $GW$ approximation has also been applied to calculate excitations of deeper valence states for small organic molecules \cite{Marom/etal:2012,Koerzdoerfer/Marom:2012,Egger2014,Ren/etal:2015,Caruso/etal:2013_tech} and also medium-sized $\pi$-conjugated molecules \cite{Dori2006}. An example is shown in Figure~\ref{fig:IP_pyridine}, where the ionization spectrum of pyridine is displayed for the first 12 valence states. Compared are the $G_0W_0$@PBE0 spectrum and the experimental PES. The positions of the peaks are in good agreement, in particular for the first three valence excitations. Benchmark studies for azabenzenes showed that a HF starting point yields distorted spectra, while hybrid DFT functionals and self-consistent schemes yield spectra that agree well with experiment \cite{Marom/etal:2012}. However, it has been found that the energy spacings and positions are not always reproduced satisfactorily. For example for benzene, the spacing of the HOMO-1 and HOMO-2 is vanishingly small for all starting points and also sc$GW$ \cite{Ren/etal:2015}. The exact spacing is larger than 0.5~eV. It has been demonstrated that a beyond $GW$ scheme, so-called ``vertex corrections,'' are necessary to separate these two peaks \cite{Ren/etal:2015}.\par
The deeper valence states are generally less valuable for characterization and chemical analysis. Core excitations energies, on the other hand, are a powerful tool to investigate the chemical structure of complex molecules and materials. They are element-specific, but are also sensitive to the atomic environment, such as covalent bonding, hybridization or the oxidation state \cite{Bagus2013,Bagus1999,Siegbahn1969all}. The application of $GW$ to core states is more difficult than to valence states, as we will explain in  more  detail in Section~\ref{sec:beyond}. Core excitations in $GW$ are an emerging research field and appropriate numerical algorithms have only been developed recently \cite{Golze2018}.

Lastly, we will briefly address peak broadening in $GW$ spectra.  The $G_0W_0$ spectrum in Figure~\ref{fig:IP_pyridine} has been artificially broadened to facilitate comparison with experiment. This broadening mimics vibrational, experimental resolution and finite lifetime effects. With regard to electronic lifetimes, also quasiparticle (QP) excitations in  molecules have finite lifetimes accompanied by a finite broadening. Such a finite broadening would be revealed in  the full spectral function $A(\omega)$, see Equation~\eqref{Eq:def_A}. The peaks close to the Fermi energy are usually sharp delta-like peaks, while higher energy excitations may decay through the formation of electron-hole pairs or collective excitations resulting in broader peaks, see \onlinecite{Caruso/etal:2013_tech} for a detailed discussion of lifetimes of  quasiparticles in molecules.


\begin{figure} 
        \includegraphics[width=0.99\linewidth]{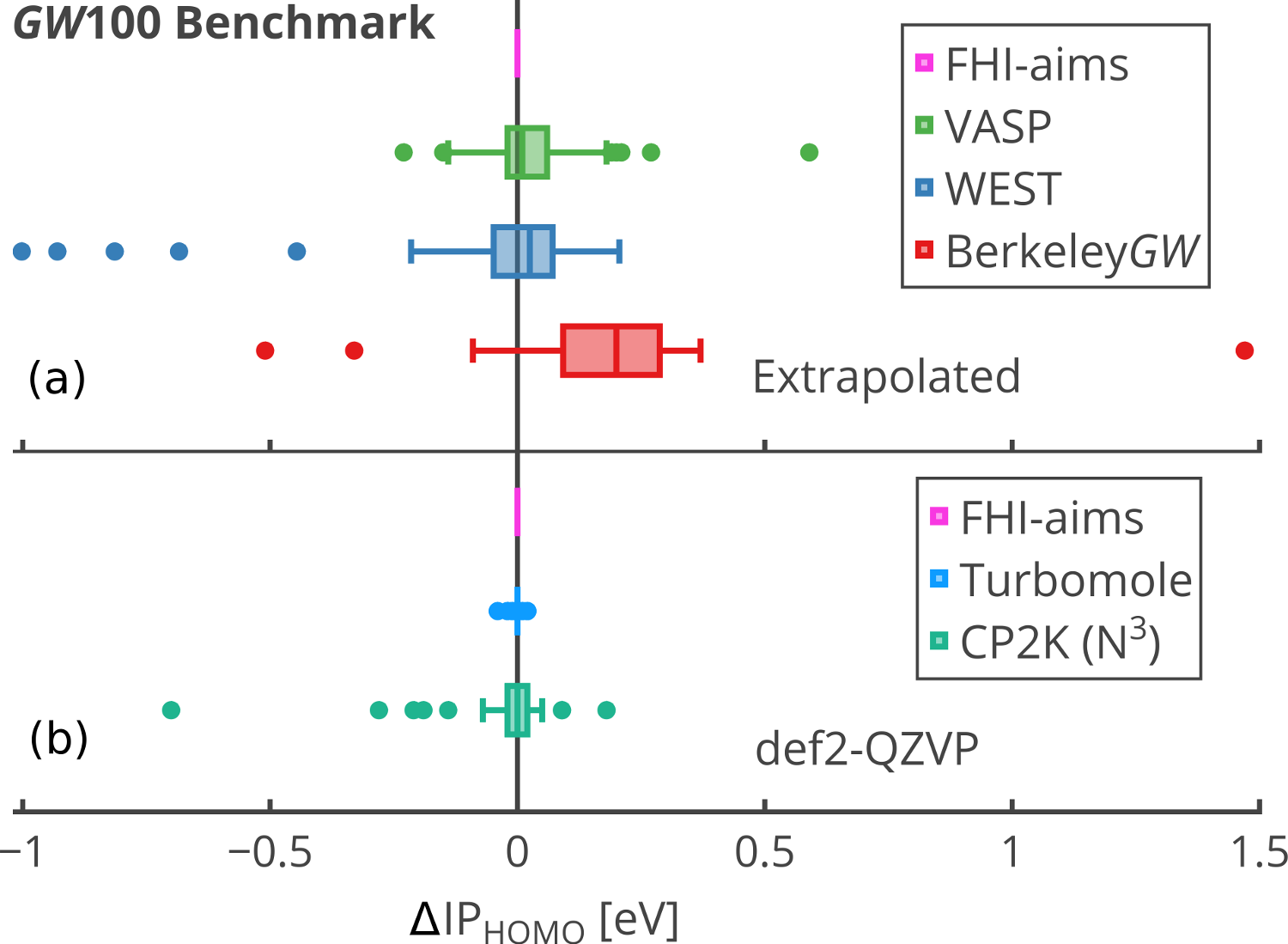}
	\caption{{\small \label{fig:gw100}
       $GW$100 benchmark comparing IP$_{\text{HOMO}}$ energies computed at the $G_0W_0$@PBE level. FHI-aims is set as reference: $\Delta\text{IP}_{\text{HOMO}}$ = $\text{IP}_{\text{HOMO}}$(FHI-aims)$-\text{IP}_{\text{HOMO}}$(X). (a) Comparison of extrapolated/converged results for VASP \cite{Maggio2017b}, WEST \cite{Govoni2018}, Berkeley$GW$ \cite{Setten2015}. Shown are the results from full-frequency treatments and iterative solutions of the QP equation. (b) Comparison of localized basis set codes using the Gaussian basis set def2-QZVP \cite{Weigend2005} for Turbomole (no-RI) \cite{Setten2015} and the $N^3$ implementation in CP2K \cite{Wilhelm2018}. Note that BN, O$_3$, MgO, BeO and CuCN are excluded for WEST, VASP and CP2K and that the Berkeley$GW$ and Turbomole data contain only a subset of 19 and 70 molecules, respectively. Box plot: Outliers represented by dots; boxes indicate the ``interquartile range'' measuring where the bulk of the data are.}}
\end{figure}
\subsection{The $ \bm{GW}$100 benchmark set}
\label{subsec:gw100}

An important aspect in electronic structure theory is benchmarking. Benchmark sets are very common in quantum chemistry, but have not found their way into $GW$ until recently. Molecules offer a distinct advantage compared to solids for benchmarking because accurate reference energies can be computed with high-level quantum chemical methods. For this purpose, sets of small molecules are beneficial since they are computationally tractable. Moreover, they provide diversity in the electronic structure due to different types of covalent bonding.\par
The first systematic benchmarks were performed using a small set of 34 molecules \cite{Rostgaard/Jacobsen/Thygesen:2010,Bruneval/Marques:2013}. Van Setten \textit{et al.} took this idea further and proposed the $GW$100 benchmark set \cite{Setten2015}, which is currently the largest and most popular $GW$ benchmark set. It contains 100 molecules that feature a variety of elements from the periodic table. The original $GW100$ paper reports HOMO and LUMO quasiparticle energies computed at the  $G_0W_0$@PBE level and the corresponding experimental references. Van Setten \textit{et al.} used the test set for a quantitative comparison of the different $GW$ methodologies implemented in the program packages Turbomole, FHI-aims and Berkeley$GW$. They compared the performance of different basis sets (plane wave vs. localized), handling of core and valence electrons (all-electron vs. pseudopotentials) and different frequency integration techniques. The codes with localized basis sets (FHI-aims and Turbomole) agree to a precision of 1~meV for most molecules. The deviation of the Berkeley$GW$ plane wave code to the basis-set-extrapolated FHI-aims and Turbomole results is in the range of 200~meV. These numbers refer to the IPs obtained from full-frequency integration techniques available in all three codes. Based on this, van Setten \textit{et al.} identified the basis set size as one important aspect for the accuracy of $GW$ calculations. \par
The test set served later as a benchmark for the PAW $G_0W_0$ implementation in VASP \cite{Maggio2017b}. This comparison established that the carefully converged PAW plane wave $G_0W_0$ calculations agree very well with the extrapolated results from the localized basis set codes. The MAD from the FHI-aims reference values is 60~meV. $GW$100 investigations with the WEST code gave similar results and highlighted the need for a re-evaluation of the pseudopotentials for some elements \cite{Govoni2018}. Moreover, the $GW$100 test set has been used to validate the accuracy of the low-scaling $GW$ algorithm based in CP2K \cite{Wilhelm2018}. A comparison between the different codes is reported in Figure~\ref{fig:gw100}. Extrapolated values are represented in Figure~\ref{fig:gw100}(a) comparing plane wave codes to FHI-aims, whereas the comparison in Figure~\ref{fig:gw100}(b) is restricted to codes with localized functions. A list of all codes that ran the $GW$100 benchmark can be found in Ref.~\cite{GW100}.\par
The $GW100$ test set was not only used to validate the reliability of numerical techniques in $G_0W_0$ implementations. It has been also used for a comprehensive assessment of different self-consistent $GW$ methodologies: sc$GW$, QS$GW$ and sc$GW_0$ \cite{Caruso2016}. The results were compared to CCDS(T) at the polarized triple-zeta level reporting the smallest discrepancies for QS$GW$. A comparison of basis set extrapolated CCSD(T) and $GW$ schemes was performed shortly afterwards for a smaller, more specialized benchmark set of 24 organic electron-acceptor molecules, where $G_0W_0$ based on long-range corrected hybrid functionals emerged as the best $GW$ method \cite{Knight2016,Gallandi2016}. Since then, also equation of motion (EOM) coupled cluster benchmark sets have been published that provide reference spectra (and not just HOMO or LUMO energies) for molecules \cite{Lange/etal:2018,Ranasinghe/etal:2019}.

\subsection{Molecular crystals}
\begin{figure} 
        \includegraphics[width=0.99\linewidth]{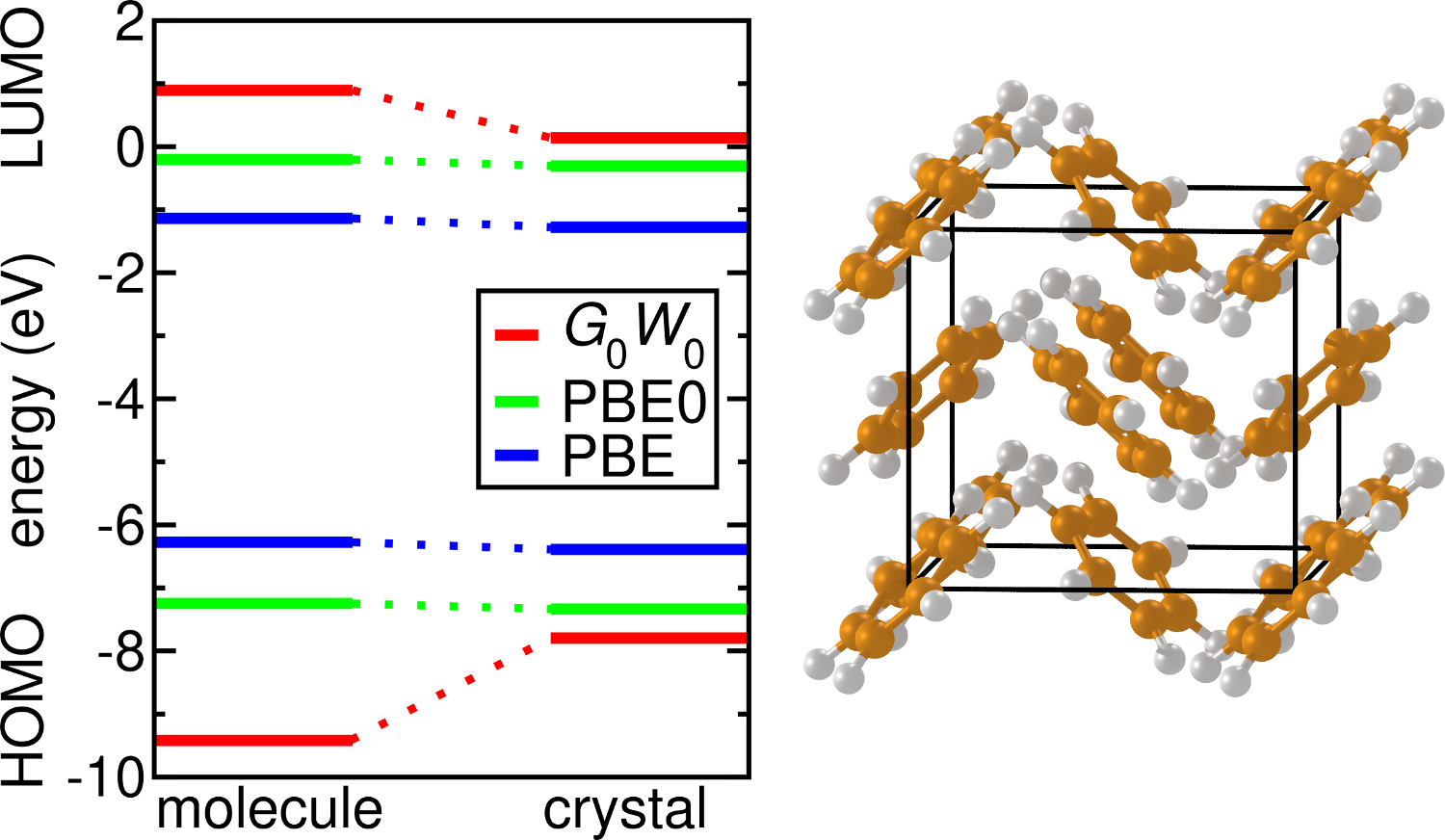}
	\caption{{\small \label{fig:benzene_crystal}
       Fundamental gaps of gas-phase benzene and band gap of the benzene crystal (space group Pbca). PBE was used as starting point for the $G_0W_0$ calculations.  Data retrieved from Ref.~\cite{Refaely-Abramson/etal:2013}.}}
\end{figure}
Modern applications of $GW$ comprise not only isolated molecules, but also molecules in the  condensed-phase, such as organic molecular crystals. These materials are composed of weakly bonded molecular units held together by, e.g., van-der-Waals interactions, dipole-dipole interactions or hydrogen bonds. Here, we summarize only some key application of $GW$ to molecular solids. A more comprehensive discussion can be found in the specialized review by \onlinecite{Kronik2016}.\par
Molecular solids exhibit a band gap renormalization similar to molecular adsorbates discussed in Section~\ref{sec:surfaces}. The band gap of molecular solids is significantly smaller than the fundamental gap of the isolated molecules \cite{Sato1981}. As for molecular adsorbates, the gap renormalization is a direct consequence of polarization effects. It is also present when there is no wave-function overlap between neighboring molecular units. If an electron is added to or removed from a certain molecule, the new charge carrier is screened not only by the molecule it was added to, but also by the surrounding molecules. This renormalization effect is shown in Figure~\ref{fig:benzene_crystal} for the benzene crystal. The HOMO level moves up in energy with respect to its position in the gas phase molecule, whereas the LUMO moves down resulting in a gap reduction. 

The gap renormalization typically lies in the range of 2 to 6~eV \cite{Kronik2016} and has been studied with $GW$ for benzene \cite{Refaely-Abramson/etal:2013},
corannulene-based materials \cite{Zoppi2011},  C$_{60}$ \cite{Refaely-Abramson/etal:2013}, pentacene \cite{Sharifzadeh2012,Refaely-Abramson/etal:2013}, perylenetetracarboxylic dianhydride (PTCDA) \cite{Sharifzadeh2012},  octaethylporphyrin (H$_2$OEP) \cite{Marsili2014}, 6,13-bis(triisopropylsilylethynyl)-pentacene (TIPS-pentacene) \cite{Sharifzadeh2015} and oligoacenes \cite{Rangel2016}. The gap reduction is not captured by standard DFT calculations \cite{Refaely-Abramson/etal:2013}, see also Figure~\ref{fig:benzene_crystal}. In fact, the DFT gap remains almost unchanged when transitioning from the gas to the crystalline phase because  the long-range polarization effects responsible for the gap renormalization are not included in conventional DFT functionals.\par
The molecular orbitals of molecular crystals resemble those of an isolated molecule. However, the overlap between neighboring molecules is not zero resulting in a $\mathbf{k}$ dependence (dispersion) of the energy levels. Starting with early work on C$_{60}$ \cite{Shirley1993}, $GW$ band structures have been reported for a wide range of organic crystals \cite{Tiago2003,Sharifzadeh2012,Refaely-Abramson/etal:2013,Fonari2014,Sharifzadeh2015,Refaely-Abramson2015,Rangel2018,Yanagisawa2017,Cocchi2018}. As for inorganic semiconductors, $GW$ opens the band gap and increases the band with with respect to DFT. For example, $GW$ band widths reported for pentacene \cite{Sharifzadeh2012,Tiago2003}, PTCDA \cite{Sharifzadeh2012}, rubrene \cite{Yanagisawa/etal:2013} or picene  \cite{Yanagisawa2014} are larger by more than 15~\%. The bands of molecular crystals are relatively flat compared to inorganic semiconductors (see Section \ref{sec:solids}). For example, $GW$-computed band widths for pentacene are only 0.4~eV for the valence and 0.7~eV for the conduction band \cite{Sharifzadeh2012}.\par
Molecular crystals are an ideal testbed for $GW$ embedding schemes since the band gap of molecular solids is mainly determined by polarization effects and significantly less by dispersion. In the spirit of quantum mechanics/molecular mechanics (QM/MM) embedding schemes the molecular crystal is partitioned into a small part that is calculated with $GW$ and a much larger MM part. In the embedding scheme proposed by Blase and co-workers, the small part to which $GW$ is applied consists of one or more molecules, while a continuum polarization model is used to include the response of the MM system \cite{Li2016,Duchemin2016}. They reported $GW$/MM gaps for pentacene and perfluoropentacene that are in close agreement with the bulk reference \cite{Li2018}. Such embedding schemes are often computationally more efficient than periodic boundary condition calculations, especially for local orbital basis set codes.

\section{Total energy and the electronic ground state}
	\label{sec:gs}
		In addition to the quasiparticle spectrum, the Green's function also provides information on the electronic ground state. Both the ground state density and the ground state total energy are accessible. However, very few studies have explored ground state properties with $GW$. Since this review mainly addresses spectroscopic properties, we will only briefly address $GW$ ground state calculations here.

\subsection{Electron density}

The ground-state density $ n({\bf r})$ follows directly from the Green's function \cite{Fetter/Walecka} 
\begin{equation}\label{eq:density}
    n({\bf r})=-i \sum_\sigma G^{\sigma}({\bf r},{\bf r},t=0^-) \quad.
\end{equation}
The total electron number contained in $G$ can be obtained through integration of the density. For a self-consistent $G$ that has been obtained from a converged solution of Dyson's equation \cite{Schindlmayr:1997}, this number should then equal the total number of electrons $N$ in the system. Also sc$GW_0$ satisfies this particle number conservation law, but all other approximate self-consistency schemes as well as $G_0W_0$ violate particle number conservation.
%
%
\begin{figure} 
        \includegraphics[width=0.99\columnwidth]{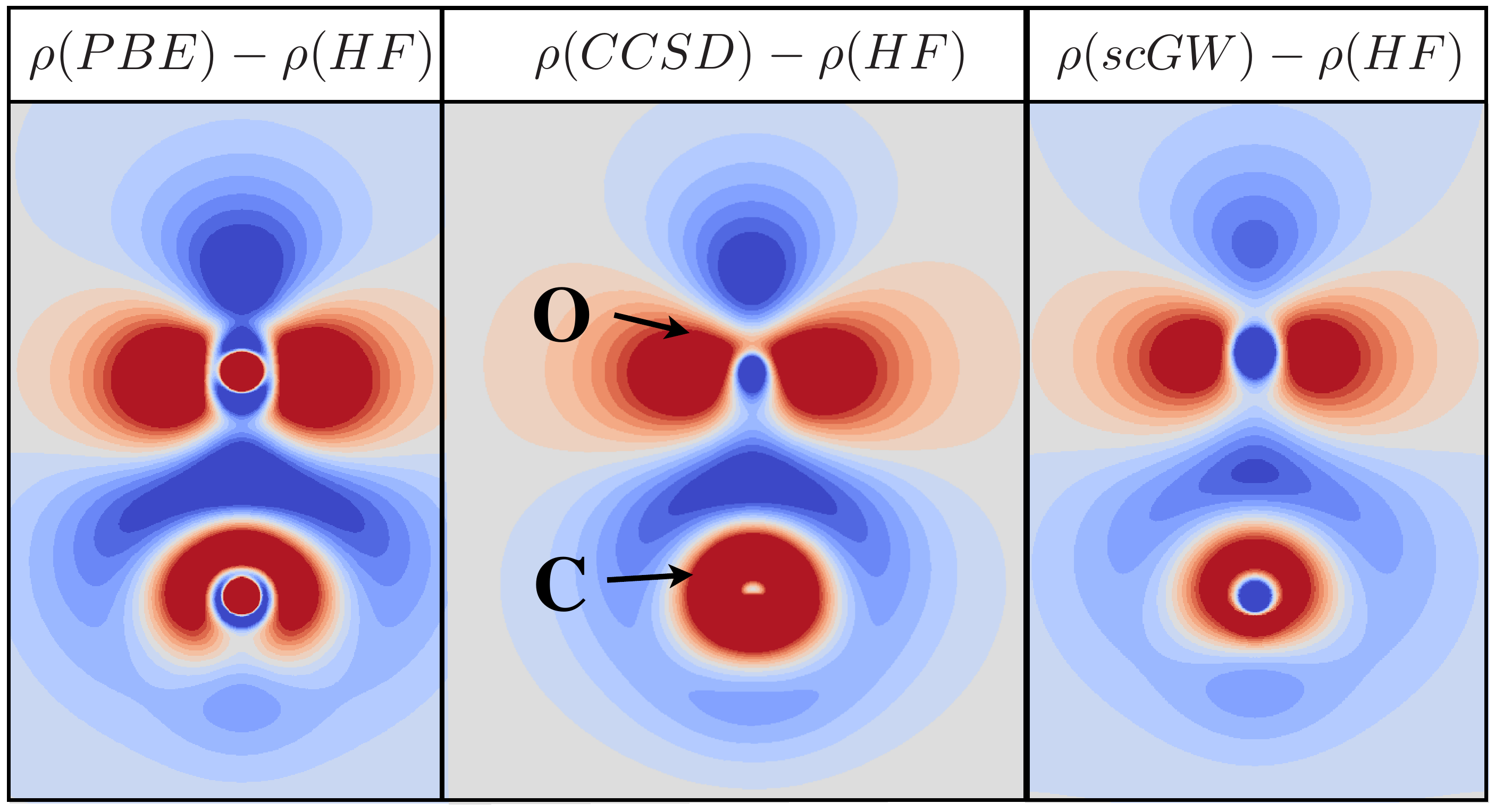}
	\caption{{\small \label{fig:CO_density} Density difference for the CO molecule between Hartree-Fock (HF) and PBE (left), coupled cluster singles-doubles (CCSD) and self-consistent $GW$ (right). Charge depletion in the three methods is encoded by blue and charge accumulation by red colors. The same computational settings as in Ref.~\cite{Caruso/etal:2013_tech} have been used.}}
\end{figure}

Figure~\ref{fig:CO_density} shows density differences compared to the Hartree-Fock method for PBE, coupled cluster singles-doubles (CCSD), and self-consistent $GW$ for the CO molecule. Overall the pattern is similar. All three methods remove charge from the bonding region and the top of the oxygen atom and focus it on the carbon atom and a $p$ orbital of the oxygen atom. The charge density difference pattern between CCSD, a high-level quantum chemistry method, and sc$GW$ is very similar. This indicates that the $GW$ density is of high quality. 

From the density, the dipole moment of CO can be calculated. In PBE the dipole moment amounts to 0.2 Debye, in HF to -0.13 Debye and from sc$GW$ we obtain 0.07 Debye \cite{Caruso/etal:2012}. The CCSD dipole moment is 0.06 Debye \cite{Caruso2012COdip}. All values were computed at the equilibrium bond-length of the respective method and the experimental dipole moment is 0.11 Debye \cite{NIST_database}. CCSD and sc$GW$ again agree closely and also match experiment reasonably well, whereas PBE overestimates the dipole moment and HF gives the wrong sign. The good agreement between sc$GW$ and CCSD and experiment is further testimony for the quality of the $GW$ density.

Since fully self-consistent $GW$ calculations are numerically quite involved and can currently only be performed for small systems, DFT densities are still used in the majority of $GW$ studies. However, in situations in which the underlying DFT Kohn-Sham spectrum has the wrong order of states, erroneous charge transfer can occur in the DFT calculation. This is, for example, frequently the case in molecular complexes, if the HOMO of one molecule erroneously ends up above the LUMO of another. 
%
%
The corresponding $G_0$ will not reflect the true ground state density of the complex and the subsequent $G_0W_0$ calculation will be wrong. $G_0W_0$ itself cannot rectify this situation because it has no access to the density. Only self-consistent schemes can correct the density and the Green's function. Examples of such molecular complexes are dimers of tetrathiafulvalene (TTF) with tetracyanoethylene (TCNE),  tetracyanoquinodimethane (TCNQ) and p-chloranil. In all cases, sc$GW$ stops the erroneous charge transfer that occurs in PBE and in hybrid functionals with a low amount of exact exchange  \cite{Caruso/etal:2014}. The resulting charge density reflects the molecular charge densities that are slightly perturbed where the molecules are closest to each other.

\subsection{Total energy}

The total electronic energy can be obtained from the single-particle Green's function $G$ via the Galitskii-Migdal (GM) formula:\cite{Galitskii/Migdal:1958,Fetter/Walecka}
\begin{equation}
E_{\rm GM}=-{i} \sum_\sigma \int d{\bf r} \,d t 
\lim_{\substack {{\bf r'}\rightarrow {\bf r} \\ {t' \rightarrow t^+} }} 
\left[i\frac{\partial }{\partial t} + \hat{h}^0
\right] \label{eq:mig1}
G^\sigma({\bf r} t,{\bf r'}t')\quad,
\end{equation}
where $\hat{h}_0$ contains the kinetic energy operator and the external potential. This equation can be recast into a more familiar looking form \cite{Strinati:1988,Caruso/etal:2013_tech}
\begin{align}
E_{\rm tot}[G] =T[G]+E_{\rm ext}[G]+E_{\rm H}[G]+E_{\rm xc}[G],
\label{eq:etot}
\end{align}
in which $T$ denotes the kinetic energy, $E_{\rm ext}$ the external potential energy, and $E_{\rm H}$ the Hartree 
energy. The exchange-correlation (xc) energy 
\begin{equation}
\label{eq:gm-xc}
 E_{\rm xc}[G]=\int_0^\infty\! \frac{d\omega}{2\pi}{\rm Tr} \{\Sigma(i\omega)G(i\omega)\},
\end{equation}
is given by the self-energy, $\Sigma$, and the Green's function. Equation~\eqref{eq:etot} is appealing because it contains the same terms as the DFT total energy. Notable differences are that the kinetic energy is the fully interacting kinetic energy and not that of an auxiliary non-interacting system. Correspondingly, the exchange-correlation energy is purely due to electronic exchange and correlation and does not need to also approximate the difference between the interacting and the non-interacting kinetic energy as in Kohn-Sham DFT.

The $GW$ total energy is closely related to the popular random-phase approximation (RPA) in DFT \cite{Langreth:1977,Hesselmann/Gorling:2010,Eshuis/Bates/Furche:2012,RPAreview}. The xc energy in $GW$ and RPA can be represented in terms of topologically identical Feynman diagrams \cite{Hellgren/vonBarth:2007,Caruso/etal:2013_H2} and thus have a total energy expression with the same functional dependence on the Green's function \cite{Klein:1961,Dahlen/Leeuwen/vonBarth:2006,Hellgren/vonBarth:2007}. However, the RPA energy is evaluated with a non-interacting Green's function (originating from a local Kohn-Sham potential) and the $GW$ energy with a fully interacting Green's function. In fact, the Dyson equation results as stationary equation from the optimization of the $GW$ total energy with respect to the Green's function in the Klein or Luttinger-Ward functionals.

\begin{figure} 
        \includegraphics[width=0.99\columnwidth]{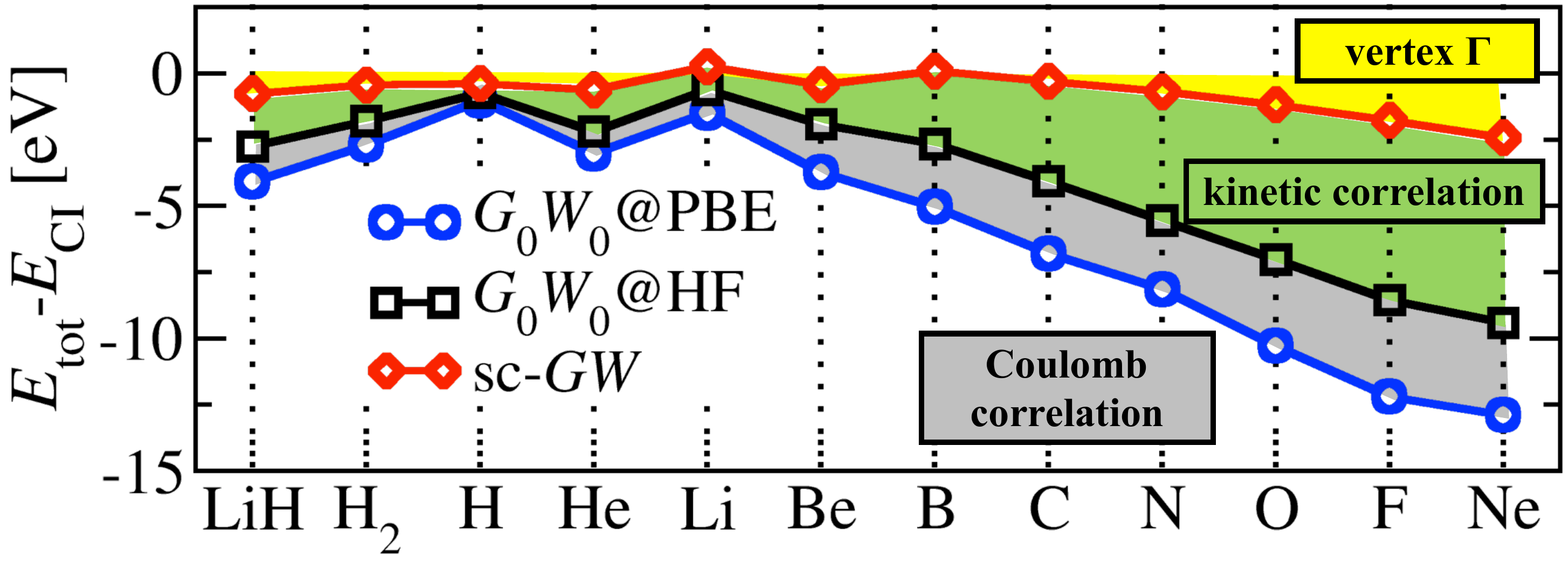}
	\caption{{\small \label{fig:Atoms_Etot} Total energy of atoms computed with three different $GW$ variants for atoms and small molecules plotted as a difference to the essentially exact Configuration Interaction (CI) results. Figure adapted from \cite{Caruso/etal:2012}. }}
\end{figure}

Early $GW$ calculation for the homogeneous electron gas found the total energy to be in good agreement with Quantum Monte Carlo calculations \cite{Holm/vonBarth:1996,Holm/vonBarth:1998,Holm:1999,Holm:2000,Garcia-Gonzalez/Godby:2001}. $GW$ also captures van der Waals interactions as exemplified by the total energy curve between two jellium slabs  \cite{Garcia-Gonzalez/Godby:2002} and by changes in the $GW$ density of the argon dimer \cite{Ferri/etal:2015}. More recently it was shown that the lattice constants and bulk moduli of simple solids agree much better between experiment and $GW$ than with LDAs, GGAs or HF \cite{Kutepov/etal:2009}. However, $GW$ total energy calculations for atoms (see Figure~\ref{fig:Atoms_Etot}) and small molecules show the opposite \cite{Stan/etal:2009,Caruso/etal:2012,Caruso/etal:2013_tech}. Presumably due to the low amount of screening self-consistent $GW$ calculations are outperformed by high-level quantum chemistry methods and even simple DFT functionals. 

Further analysis \cite{Hellgren/etal:2015} reveals that the difference between $G_0W_0$@HF and sc$GW$ can be ascribed to the difference in the kinetic energy (termed here kinetic correlation in analogy with DFT) because their Coulomb correlation energies are almost identical for the small systems shown in Figure~\ref{fig:Atoms_Etot}. Conversely, the difference between $G_0W_0$@PBE and sc$GW$ is almost entirely due to Coulomb correlation. Both Coulomb and kinetic correlation are large, as illustrated in Figure~\ref{fig:Atoms_Etot}. Once included, the remaining difference between sc$GW$ and full configuration interaction (the essentially exact solution) must be due to missing vertex corrections. This contribution is much smaller than the two correlation contributions.

\section{Current challenges and beyond $\bm{GW}$}
	\label{sec:beyond}
		\subsection{Challenges}
As successful as the $GW$ approximation is for describing quasiparticle excitations, there are still technical and theoretical challenges to overcome. Core-level spectroscopy is a valuable tool for chemical analysis and characterizing materials. The operating principle is the same as PES and IPES discussed in Section \ref{sec:TS}, though at higher incident energies using X-rays. The technique is then referred to as X-ray photoelectron spectroscopy (XPS). Core levels of the same type, for example, different carbon 1s states, are element-specific, but are also sensitive to the local chemical environment, i.e., bonding, hybridization or the oxidation state \cite{Egelhoff1987,Bagus1999}. However, these so-called chemical shifts are for second-row elements often smaller than 1~eV \cite{Siegbahn1969all}. The energetic differences are particularly minute for carbon with XPS peaks that are separated by less than 0.5~eV. Such spectra are hard to resolve and interpret. Theoretical spectroscopy can be a valuable tool to aid the interpretation of experimental results.\par
For core levels, however, the simple, single quasiparticle picture can break down. The incident photon in PES may produce spectral features away from the single-particle peak. If the additional peak is broad, these so-called satellites can be attributed to the collective excitation of the system after the electron is excited. If the additional peak is narrow or, equivalently, has a long lifetime, the electron has spectral weight divided between multiple particle-like eigenstates of the system \cite{Golze2018}. This effect can also appear when probing the multiplet structure of open-shell systems \cite{Lischner/etal:2012}. In these cases, the quasiparticle equation can have multiple solutions, making both the $GW$ calculation and interpretation of the result more difficult. The problem also appears for more conventional valence states of small molecules, and recent work has shown that these multiple solutions lead to unphysical discontinuities in quasiparticle energies and that ev$GW$ can exacerbate the problem \cite{loos_jctc_14,veril_jctc_14}. Just as for experimental spectroscopy, the sensitivity of core states to the local environment makes the $GW$ calculation more challenging than for conventional valence states. Due to its value for chemical analysis and dearly needed support for XPS experiments, $GW$ for core levels can yield useful insight and is an ongoing topic of research \cite{Zhou/etal:2015,Golze2018}.\par

Spin dependence in $GW$ calculations is important for understanding magnetic systems and is critical to the electronic structure of topological insulators. Already in the case of collinear spin, when the spin quantum number is either up or down, spin polarization has an effect on the excitation spectrum of MnO \cite{Rodl/etal:2008}. By including spin-orbit coupling (SOC) in the one electron Hamiltonian, single-particle states become noncollinear and can no longer be decomposed into up or down. Noncollinear calculations are important in relativistic systems with strong SOC or when describing magnetic effects \cite{Kuhn/etal:2015,Sakuma/etal:2011,Kutepov/etal:2012,Scherpelz/etal:2016,Ahmed/etal:2014}. For materials with heavy elements, energy shifts due to spin-orbit coupling must be included for good agreement with experiment on band gaps \cite{Scherpelz/etal:2016}. Topological insulators commonly contain heavy elements (Se, Te, Bi, Sb) and depend on spin-orbit coupling for band inversion \cite{Nechaev/etal:2015,Aguilera/etal:2013,Aguilera/etal:2013b,Aguilera/etal:2015,Nechaev/etal:2013}. For a detailed review of $GW$+SOC calculations, see~\onlinecite{gw_review_soc}. To describe spin-dependent \textit{interactions} between particles, one must generalize Hedin's equations beyond the Coulomb interaction, which has no spin dependence. This generalization was recently completed \cite{Aryasetiawan/Biermann:2009,Aryasetiawan/etal:2008} and allows one to treat magnetic dipole-dipole interactions, for example.

\subsection{Quantum chemistry}
Quantum chemistry offers an established, albeit expensive, route to compute particle addition/removal energies in molecules. Ionization energies and electron affinities can be computed as the difference of total energies between the neutral molecule and the ion. In fact, $GW$ calculations on small systems are often compared with coupled cluster results as a benchmark. A direct comparison between correlated wave-function and Green's function methods to determine the level of correlation described by each is somewhat challenging. In certain cases, it is possible: recent work compares diagrams included in $GW$ with those included in equation-of-motion coupled cluster theory \cite{Lange/etal:2018}. 

Generally, nonperturbative wave-function methods are considered beyond $GW$, even if they rely on an ansatz or other approximation. In Green's function embedding theories, quantum chemistry (either full or truncated configuration interaction) can be used as a high accuracy Green's function solver in a subspace \cite{zgid/etal:2012}. After computing the subspace wave function, one directly computes the amplitudes in Equation~\eqref{Eq:def_es} for the subspace Green's function. With $G$ and $G_0$ in hand, it is then trivial to compute the self-energy \cite{Pavlyukh/etal:2007}. In this subspace, the Green's function is computed from accurate many-body wave functions so that correlation is treated beyond $GW$. The subspace Green's function can be self-consistently iterated with the remaining degrees of freedom described at the $GW$ level of theory \cite{martin_reining_ceperley_2016}. Other routes to combine $GW$ with quantum chemistry are an emerging field. A newly developed method combines $GW$ with configuration interaction by embedding a wave function calculation inside of a Green's function calculation \cite{Dvorak/etal:2018,Dvorak2/etal:2018}. These developments offer valuable insight to merge these disciplines in the future. 

Green's functions are also directly studied in quantum chemistry, where they are more commonly called propagators. There certainly is some overlap between the two communities in their treatment of $GW$ or $GW$-like approximations. Because we primarily focus on $GW$ and Hedin's equations in physics, we refer the interested reader to the work of \onlinecite{Cederbaum/Domcke:1977} and \onlinecite{Ortiz:2012} for a perspective of propagators in chemistry.

\subsection{Non-equilibrium Green's functions}
\label{subsec:non-equilibrium}

The $GW$ approach has also been applied to systems in strong external fields. These include quantum transport calculations \cite{Thygesen/Rubio:2008,Spataru/Hybertsen/Louie:2009} and semiconductors in strong laser fields \cite{Spataru/Benedict/Louie:2004}. The problem of describing quantum transport is similar to that of level alignment at a molecule/metal interface discussed in Section~\ref{sec:surfaces}. First, the alignment of molecular states in the contact region relative to the Fermi level of the metal leads determines the overall conductance. Second, for applied biases, charge will flow from the lead into the molecule or molecules in the contact region. This charge flow will alter the electron density of the system and therefore the quantum mechanical interactions.

Self-consistent $GW$ calculations \cite{Thygesen/Rubio:2008} take charge transfer and the associated change in screening (e.g. image effect) and the many-body interactions into account correctly. sc$GW$ is an appropriate tool for finite, small bias quantum transport calculations, as benchmarked for instance for thiol- and amine-linked benzene/gold \cite{Strange/etal:2011} and alkane/gold junctions \cite{Strange/Thygesen:2011}. Strong correlation effects in quantum transport, such as the Coulomb blockade or the Kondo effect, can, however, not be captured with the $GW$ approach and require a beyond $GW$ treatment of correlation \cite{Spataru/Hybertsen/Louie:2009,Thoss/Evers:2018}. 

For strong external biases in quantum transport and systems in other strong external fields, the electron distribution is perturbed so strongly that it can no longer be described by an equilibrium Fermi function at a finite temperature. In such non-equilibrium cases, the Green's function theory described in this review article is not applicable anymore. Non-equilibrium scenarios can be incorporated into the Green's function formalism, by switching to non-equilibrium Green's functions defined on the Keldysh contour \cite{Dahlen/Stan/Leeuwen:2006}. These  non-equilibrium Green's functions obey the Kadanoff-Baym and not Hedin's equations. The Keldysh contour formalism goes beyond the scope of this review article and we refer the interested reader to two excellent recent books \cite{MBPT_book_Stefanucci/Leeuwen:2013,Karlsson2018}.

In one application of this non-equilibrium formalism highly excited GaAs was investigated. It had been hypothesised that GaAs could become metallic if enough electrons could be promoted from the valence to the conduction band with a strong laser. Non-equilibrium $GW$ calculations showed that the band gap was indeed decreasing with increasing laser power, but would not close completely, falsifying the hypothesis \cite{Spataru/Benedict/Louie:2004}.

\subsection{Vertex corrections}
\begin{figure} 
       \includegraphics[width=\columnwidth]{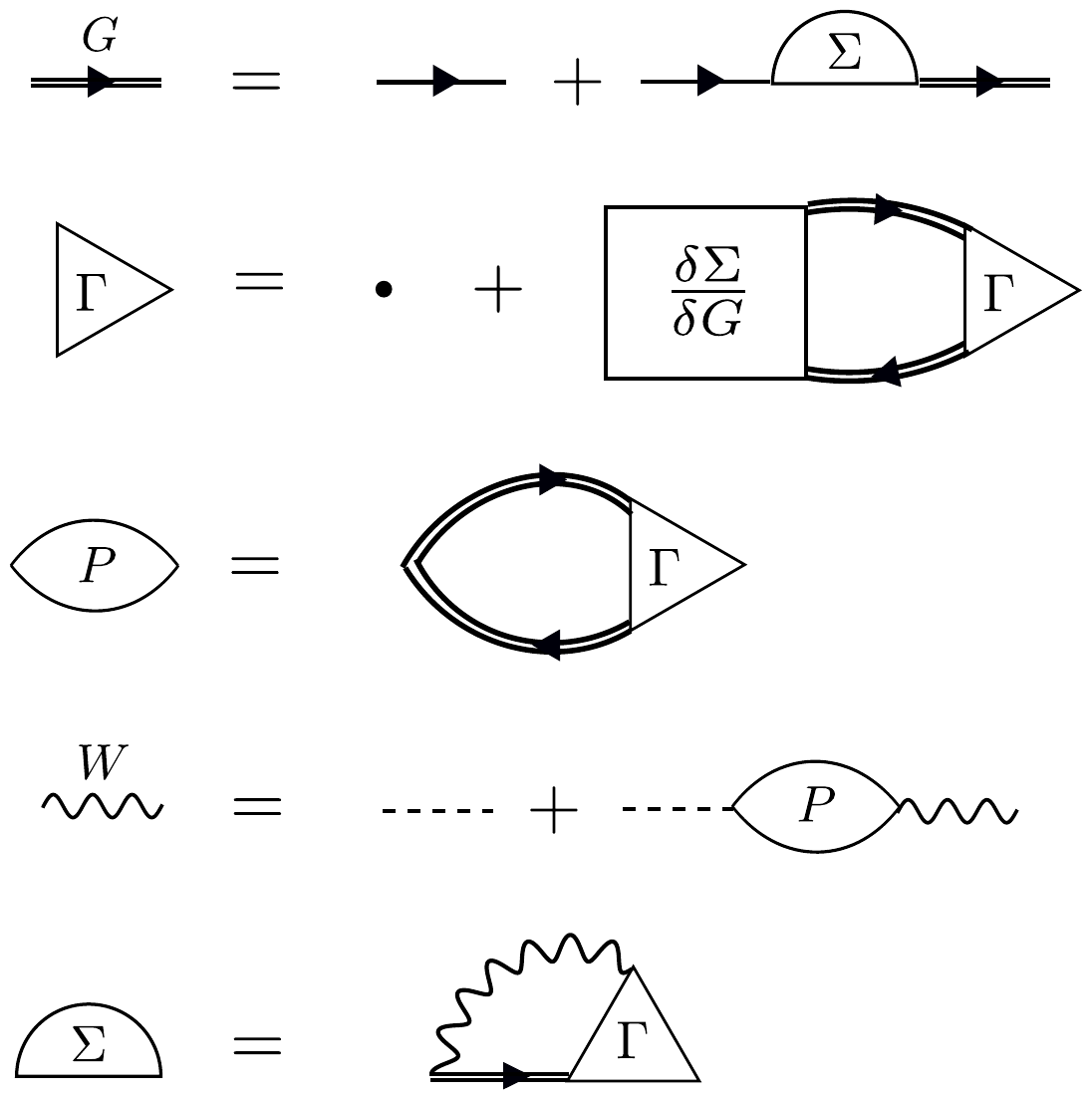}
	\caption{{\small \label{fig:hedin_diagram}
        	 Diagrammatic representation of Hedin's equations. All 5 quantities are coupled to all others. Here, we omit the Hartree potential from the $G$ diagram, though it must also be included when translating the diagrams to the equations in Appendix~\ref{sec:MB}.}}
\end{figure}

To go beyond the $GW$ approximation, one must include vertex corrections. The full set of Hedin's equations, including the vertex, are shown diagramatically in Figure~\ref{fig:hedin_diagram}, which can be directly compared to Figure~\ref{fig:gw_diagram}. The mathematical equations are in Appendix~\ref{sec:Hedins_eqs}.
By comparison to the $GW$ diagrams, we see that treating the vertex, $\Gamma$, beyond a single spacetime point significantly complicates the equations, as demonstrated for a single diagram in Figure~\ref{fig:vertex}. The vertex contributions beyond $\Gamma(1,2,3) = \delta(1,2) \delta(1,3)$ are commonly called vertex corrections. The exact vertex requires the functional derivative $\delta \Sigma / \delta G$. This functional derivative cannot be computed numerically and must be derived analytically. Any resulting vertex is extremely expensive to compute because it now depends on three spatial, spin, and time indices. There are a few reasonable options for reducing the expense: using an approximate $\Sigma$ from a different theory to simplify the derivative, using a diagrammatic but simplified $\Gamma$, or using an exact $\Gamma$ in only a small subspace. As in $GW$, $\Gamma$ can be selectively iterated or computed in a single-shot way to further lower the cost.

The earliest approaches to vertex corrections used an approximate vertex based on the local density approximation (LDA) to Kohn-Sham DFT \cite{DelSole/Reining/Godby:1994,Hybertsen/Louie:1986}. In these approximations, the LDA exchange-correlation functional is used in place of the self-energy to compute the functional derivative. The approximate vertex enters the polarizability and interaction as
\begin{eqnarray}
f_{xc} (1,2) &=& \frac{\delta v_{xc}(1) }{ \delta n(2)}  \nonumber  \\
\widetilde{W} &=& v [ 1 - \chi_0(v+f_{xc})]^{-1}
\end{eqnarray}
where $n$ is the electron density and $f_{xc}$ determines the vertex correction. The advantage of the LDA is that $f_{xc}$ can be calculated analytically.

These approaches are computationally much lighter than the true vertex and, for that reason, are still used \cite{Thygesen/etal:2017}. Approximate vertex corrections can also be extended beyond the LDA to recover a more realistic behavior \cite{Chen/etal:2015}. The inclusion of $f_{xc}$ has its roots in time-dependent density functional theory (TDDFT) and is somewhat inconsistent with the Green's function formalism. The final results of such calculations can improve band gaps compared to $G_0W_0$ \cite{Chen/etal:2015} or band centers compared to $G_0W_0$ \cite{Thygesen/etal:2017}. Shaltaf \textit{et al.} found that a GGA-based vertex correction had a negligible effect on band offsets in the Si/SiO$_2$ interface \cite{Shaltaf/etal:2008}.

Diagrammatic vertex corrections, instead of those based on a density functional, are a more consistent and formal extension to $GW$. One can build upon $GW$ in a fully diagrammatic framework by including vertex corrections in the perturbative, single-shot self-energy. These methods are analogous to the $G_0W_0$ approach in that the vertex correction is computed only one time, and the quasiparticle energies are usually computed in a diagonal approximation. For example, second order screened exchange (SOSEX) and related approximations include a subset of exchange-type diagrams which are a vertex correction to $GW$ \cite{Shirley/Martin:1993_2,Bobbert/vanHaeringen:1994}. In the approximation of \onlinecite{Ren/etal:2015}, the diagonal matrix elements of $GW+$SOSEX are
\begin{eqnarray}
& & \bra{\phi_s^0} \Sigma_{\mathrm{c}}^{GW}(i\omega) + \Sigma_{\mathrm{c}}^{\mathrm{SOSEX}}(i\omega) \ket{\phi_s^0}   \\
&=& -\frac{1}{2\pi} \int_{-\infty}^{\infty} d \omega ' \sum_{qrp} \frac{ (f_q - f_r ) \bra{sr}\ket{qp}\bra{qp}W(i\omega')\ket{sr} } { (i \omega + i \omega' - \epsilon_p^0 ) ( i\omega' + \epsilon_q^0 - \epsilon_r^0 ) } , \nonumber
\end{eqnarray}
where $f_q$ and $f_r$ are Fermi occupation factors and $\bra{sr}\ket{qp}$ is defined in Appendix~\ref{sec:notation}.
Notice extra matrix elements in the numerator and factors in the denominator compared to the equation for the $GW$ self-energy in Figure \ref{fig:g0w0procedure}. SOSEX-type approximations generally improve upon $GW$ band gaps in molecules \cite{Ren/etal:2015}. In the perturbative approach, these calculations are relatively lightweight but have the same starting point dependence of $G_0W_0$. 

A systematic bridge between diagrammatic vertex corrections and TDDFT was developed by Del Sole, Reining, and others \cite{Bruneval/etal:2005,Sottile/etal:2003,Reining/etal:2002,Adragna/etal:2003,DelSole/etal:2003,streitenberger_pssb_125,streitenberger_pla_106}. In this approach, the kernel to construct the irreducible polarisability is cast as only a two-point function. This is in contrast to the exact vertex, $\Gamma$, which depends on four spacetime coordinates to compute, making it much more expensive (four coordinates for the derivative $\delta \Sigma(1,2) / \delta G(4,5)$, see Appendix~\ref{sec:Hedins_eqs}). This two-point kernel can only be used inside of $W$. Outside of $W$, the three-point nature of the vertex is unavoidable. The simplified many-body approach retains the simplicity of a TDDFT kernel but has its foundation in many-body theory. By adopting the $GW$ approximation to $\Sigma$, an approximate, analytic $f_{xc}^{\mathrm{QP}}$ exists. Calculations of the dielectric function in Si and GaAs show good agreement between the $f_{xc}^{\mathrm{QP}}$ approach and a solution for the full vertex \cite{Adragna/etal:2003}.

More recent work has included diagrammatic vertex corrections to solve Hedin's equations at some level of self-consistency, though still at an approximate level \cite{Kutepov:2016,Kutepov:2017,Gruneis/etal:2014,Maggio/Kresse:2017}. The greatest conceptual and computational difficulty to these calculations is how to update $\Gamma$. Because $\Gamma$ enters in both $\chi_0$ and $\Sigma$, and because full self-consistency is so expensive, it is advantageous to update $\Gamma$ in only one portion of the calculation. For example, one could evaluate the interaction $W$ with a diagrammatic $\Gamma$ once at the beginning of the calculation. Keeping $W$ fixed, only $G$ and $\Sigma$ are updated through Dyson's equation in the self-consistency cycle. While this procedure is only partially self-consistent, it incorporates a diagrammatic $\Gamma$ while keeping the $GW$ level of complexity through the self-consistency cycle. When applied to semiconductors and insulators, and with some practical restrictions on $\Gamma$, solutions of Hedin's full equations give noticeably better band gaps than $GW$ \cite{Kutepov:2016,Kutepov:2017}. Full, self-consistent solutions of Hedin's exact equations remain out of reach in real systems, and even partially self-consistent schemes are a technical challenge.

\begin{figure} 
       \includegraphics[width=\columnwidth]{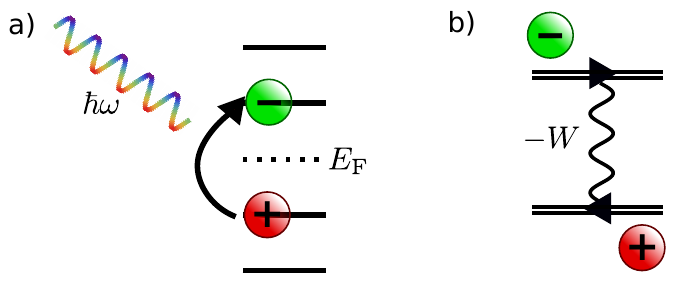}
	\caption{{\small \label{fig:bse}
        	 (a) Schematic representation of optical absorption. (b) Diagrammatic representation of $GW$/BSE. The electron and hole are represented by $G$ lines, computed in the $GW$ approximation, and their direct interaction is through the screened Coulomb interaction.}}
\end{figure}
\subsection{Optical properties}
Calculations of the many-body vertex have another application beyond particle addition/removal energies. Optical processes, such as photon absorption, can also be modeled in the Green's function formalism. In such a case, the relevant correlation function is the time-ordered two-particle correlation function, $L$. $L$ obeys a Dyson equation like $G$, except that the role of the self-energy is instead played by $\delta \Sigma / \delta G$. The Dyson series for the full vertex to determine $L$ is called the Bethe-Salpeter equation (BSE) in physics \cite{Salpeter/etal:1951,Held/etal:2011}. BSE calculations describe a different process than ordinary $GW$, so they are not beyond $GW$ in the same sense as including a vertex in the self-energy. Even so, they are closely related. The common implementation of the BSE for materials relies on the $GW$ approximation to the self-energy. In these $GW$/BSE calculations, the excited electron and hole instantaneously interact via $W$. 

The effective two-particle Hamiltonian for correlated optical excitations \cite{Rohlfing/etal:2000,Strinati:1982,Strinati:1984}, called excitons, is
\begin{eqnarray}
( \epsilon_a - \epsilon_i ) A_{ia}^P &+& \sum_{i'a'} \bra{ia} K \ket{i' a'} A_{i'a'} = \Omega^P A_{ia}  \nonumber  \\
\bra{ia} K \ket{i'a'}  &=&  -W_{iai'a'} + v_{i i' a a'}  \label{bse}
\end{eqnarray}
where $i$ and $a$ denote again occupied and empty states, respectively, $\epsilon_{i/a}$ is a quasiparticle energy, $A_{ia}^P$ is the $P^{\mathrm{th}}$ exciton wave function in the single-particle basis, and $\Omega^P$ is the $P^{\mathrm{th}}$ excitation energy. Equation~\eqref{bse} makes the common Tamm-Dancoff approximation (TDA), which ignores backward propagating electron-hole pairs that are present in the exact BSE. Matrix elements of $K$, $\delta \Sigma / \delta G$, include a direct screened interaction ($W$) and repulsive exchange ($v$). Schematic and diagrammatic representations of the optical process are shown in Figure~\ref{fig:bse}. BSE calculations can be considered a first iteration of $\Gamma$ in Hedin's equations to go beyond $GW$ for particle addition/removal energies, if the resulting polarizability is reinserted into $W$. 

The first BSE calculations included only the bare electron-hole exchange \cite{hanke_prb_12} in a semi-empirical basis and were then extended to include the direct, screened interaction \cite{hanke_prb_21,hanke_prl_43}. \textit{Ab-initio} $GW$/BSE calculations focused on semiconductors like Si, GaAs, and Li$_2$O where $GW$/BSE produces optical absorption spectra and exciton binding energies in close agreement with experiment \cite{albrecht_pssa_170,albrecht_prl_80,albrecht_prb_55,onida_prl_75,Benedict/etal:1998b,Benedict/etal:1998,Benedict/etal:1999,Rohlfing/etal:2000,Rohlfing/etal:1998}. Similar to the proliferation of $GW$ since its early successes, $GW$/BSE has been applied extensively to solids \cite{Schleife/etal:2011,Schleife/etal:2018,Rinke/etal:2012,Erhart/etal:2014}, molecules \cite{botti_jctc_10,bruneval_jcp_142,Jacquemin2015}, surfaces \cite{Palummo/etal:2004}, and two-dimensional materials \cite{Ugeda/etal:2014,Huser/etal:2013,Qiu/etal:2013,Komsa/etal:2012,Qiu2016,Ramasubramaniam/etal:2012,Shi/etal:2013,Dvorak2015}. As with Dyson's equation, equations with the Bethe-Salpeter form appear in different contexts in many-body theory. For example, a Bethe-Salpeter equation can also describe spin-flip excitations in magnetic materials \cite{friedrich/etal:2016}.

\subsection{$T$-matrix}
\begin{figure} 
       \includegraphics[width=0.7\columnwidth]{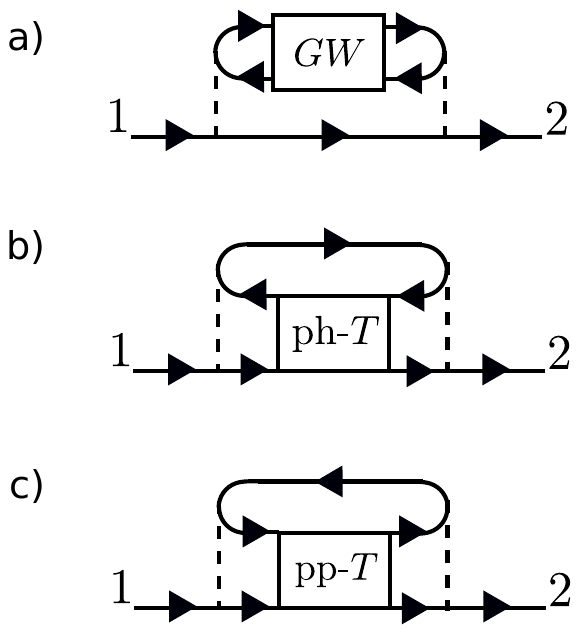}
	\caption{{\small \label{fig:t_matrix}
        	 Schematic representation of diagrams included with (a) $GW$, (b) particle-hole $T$-matrix, and (c) particle-particle $T$-matrix. In each case, channels going into the box are correlated further with additional interactions.}}
\end{figure}

The framework of Hedin's equations, and $GW$ in particular, places great emphasis on the screened Coulomb interaction.  Indeed, many of the approximate schemes presented here frame the exact vertex as a correction (hence the term ``vertex correction'') to a $GW$ calculation of the self-energy. In certain systems, it may be necessary to abandon this picture entirely. For example, in systems with low electron density and a similarly low number of electron-hole screening channels, as in very small atoms or molecules, screening of the Coulomb interaction may be insignificant. Roughly speaking, diagrams of the vertex type could be more important than screening diagrams included in $GW$. For these systems, we should adopt a different formalism which does not rely on screening and directly emphasizes other correlation channels. One such formalism is the $T$-matrix \cite{Zhang/Su/Yang:2017,Romaniello/etal:2012,Zhukov/Chulkov/Echenique:2004}, in which the self-energy is written as a product of $G$ with a four-point kernel, $T$,
\begin{equation}
\label{eq:t_matrix}
\Sigma(1,2) = -i \int d3 \; d4 \; G(4,3) T(1,3,2,4).
\end{equation}
The precise form of $T$ depends on the choice of the particle-particle (pp) or particle-hole (ph) $T$-matrix, which determines the channels that are correlated alongside a third propagating channel. $T$ obeys its own Dyson series and physically corresponds to repeated interactions, or scattering, between the selected channels (pp or ph). There are advantages of the $T$-matrix approach: it is exact up to second order in the bare interaction and includes many more exchange diagrams than $GW$, making it useful for magnetic systems. At first glance, the ph $T$-matrix may sound like $GW$. However, the two approximations correlate \textit{different} particle and hole channels in the self-energy diagram. A schematic representation of the correlated channels in $GW$ and $T$-matrix is shown in Figure~\ref{fig:t_matrix}. Notice the different topologies of $G_0$ lines correlated in $GW$ and ph-$T$. The $T$-matrix approach has been applied to understand the role of spin-flip excitations in metals \cite{friedrich/etal:2018,mlynczak_nc_10,Zhukov/etal:2006,Zhukov/Chulkov/Echenique:2004,Zhukov/etal:2005}, double ionizations and Auger spectroscopy \cite{Noguchi/etal_2:2008,Noguchi/etal_2010,Noguchi/etal:2007}, as well as satellites in metals \cite{Springer/etal:1998}.

\subsection{Cumulant expansion}
One long-standing problem with the $GW$ approximation is its description of plasmon satellites in, for example, Si, Na and Al \cite{Aryasetiawan/etal:1996,Guzzo/etal:2011,Guzzo/etal:2014,Zhou/etal:2015}. Plasmon satellites are peaks in the spectral function which are attributed not to a single quasiparticle, but to the coupling between a hole (in the particle removal case) and the collective excitation of the remaining electrons. This coupling leads to a quasiparticle peak in the spectral function and a series of progressively weaker plasmon replicas separated by the plasmon energy. 

A proven route to improve the plasmon description compared to $GW$ is the cumulant expansion to the Green's function, which has been tested on Na, Al, graphene, Si, and the electron gas   \cite{Aryasetiawan/etal:1996,Lischner/etal:2013,Zhou/etal:2018,Guzzo/etal:2011,Gattiz/Guzzo:2013,Lischner/etal:2014b,Caruso/etal:2015b,Caruso/etal:2015,Caruso/etal:2016}. Based on an exponential ansatz, somewhat analogous to the coupled cluster expansion for the wave function, the cumulant Green's function for a hole takes the form \cite{Lischner/etal:2013,Guzzo/etal:2011,Aryasetiawan/etal:1996}
\begin{equation}
\label{eq:cumulant}
G_s(t) = \Theta(-t) e^{ -i \epsilon_s^0 t + C_s(t) }
\end{equation}
where $\epsilon_s^0$ is the mean-field energy that enters $G_0$ for state $s$ and $C_s(t)$ is the cumulant. The exact form of the cumulant $C_s(t)$ depends on the chosen approximation to the self-energy. If one Taylor expands Equation~\eqref{eq:cumulant} in powers of $C_s$ and compares it to Dyson's equation with the $GW$ self-energy, an approximate closed form for the cumulant exists. This is called the $GW$+C method. The cumulant includes vertex corrections beyond the $GW$ self-energy at the same computational expense as ordinary $GW$. These vertex corrections generally improve the description of satellites over $GW$ when compared with experiment. 

The cumulant appears to be a tremendous success $-$ it miraculously provides vertex corrections for the same cost as $GW$. However, it does not improve the description of valence quasiparticle energies. The quasiparticle energy is still determined by ordinary $GW$. Furthermore, the cumulant ansatz in Equation~\eqref{eq:cumulant} relies on the separation of electron and hole branches of the Green's function and produces satellites only below the Fermi energy. In general, this separation is not correct, and it becomes a worse approximation closer to the Fermi energy \cite{martin_reining_ceperley_2016,guzzo_prb_89}. The formal connection between $GW$ and the cumulant is presented in \onlinecite{Gumhalter/etal:2016}.

\begin{figure} 
       \includegraphics[width=\columnwidth]{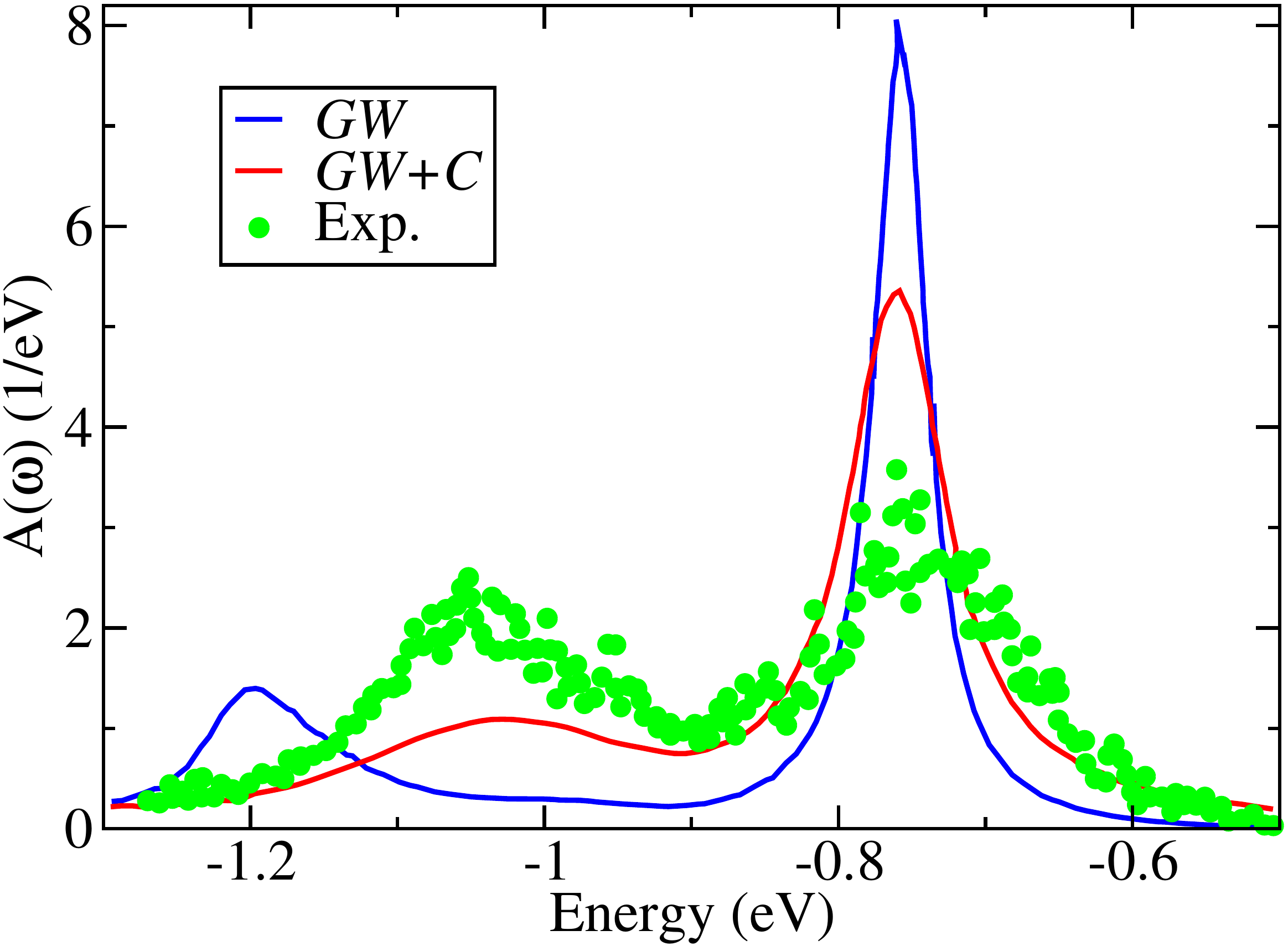}
	\caption{{\small \label{fig:cumulant}
        	 Spectral function $A(\omega)$ of graphene on SiC at the Dirac point for electron doping density of $n = -5.9 \times 10^{13}$ $\mathrm{ cm}^{-2}$. Data taken from~\onlinecite{Lischner/etal:2013}.}}
\end{figure}

As a case study of the cumulant, we highlight the study of doped graphene by Lischner and co-authors \cite{Lischner/etal:2013}. The spectral function of doped graphene on a SiC(0001) surface displays a quasiparticle peak and satellite, shown in Figure~\ref{fig:cumulant}. With ordinary $G_0W_0$, the splitting between the quasiparticle and satellite is 0.44 eV, which overestimates the experimental value of 0.3 eV. By including vertex corrections to the hole Green's function with the cumulant, the $GW$+C calculation reduces the splitting to 0.27 eV. $GW$+C also redistributes spectral weight away from the quasiparticle and to the satellite. Additionally, $GW$+C eliminates a spurious \textit{plasmaron} $-$ coupling between a hole and plasmon $-$ solution that appears in $GW$.

\subsection{Local vertex}
 The treatment of localized electrons in physics has become its own subfield \cite{Tomczak/etal:2017,Hirayama/etal:2017,Held/etal:2011}. In strongly-correlated physics, localized $d$- or $f$-electrons usually indicate a need to go beyond $GW$ \cite{Gattiz/Guzzo:2013,Nohara/etal:2009,Sakuma/Werner/Aryasetiawan:2013}, with transition metal oxides being typical test cases. Much of the previous discussion applies just as well to localized electrons as any others $-$ Hedin's equations are exact. However, the localized nature of these states lends themselves to model Hamiltonians, particularly the Hubbard model \cite{Hubbard:1963}, which describes localized interactions by a parameter $U$. $U$ is a measure of the repulsive interaction, or energetic cost, for electrons occupying the same spatial orbital. When combined with the LDA, the LDA+$U$ method can improve upon the poor description of localized states by mean-field and $GW$ theories \cite{Jiang:2018}. 
 
 In the Green's function formalism, including diagrams beyond $GW$ is made tractable by approximating the true Coulomb interaction by $U$. The combination of $GW$ with dynamical mean-field theory (DMFT) \cite{georges/etal:1992,kotliar/etal:2006}, the $GW$+DMFT method \cite{Biermann/etal:2003,Biermann:2014}, is a rigorous way of combining diagrams of higher order with the $GW$ approximation. $GW$+DMFT describes long-range correlation with $GW$ and local $d$- or $f$-electron correlation via the Hubbard $U$. $GW$+DMFT correlates a small set of states (the $d$- or $f$-electrons) using a non-perturbative vertex, called a ``local'' vertex since single site DMFT only includes diagrams which are local in orbital space. The first studies of SrVO$_3$ with $GW$+DMFT demonstrated its potential for predicting spectral properties of strongly-correlated solids \cite{Tomczak/etal:2012b,Tomczak/etal:2014}. The $GW$+DMFT method continues to be developed \cite{Biermann:2014,Choi/etal:2016}.

\section{Conclusion}
    \label{sec:conclusion}
        Photoelectron and inverse photoelectron spectroscopy will remain some of the most powerful probes of matter available to scientists. While experimental spectroscopy gives the ``answer'' in the form of the measured spectrum, it may not give the full understanding of the underlying physics. In this regard, theoretical spectroscopy plays a huge role as a complement to experimental techniques.

We have introduced the Green's function formalism and many-body theory from a perturbation theory perspective. The formalism is exact, in principle, and allows one to calculate both ground and excited state properties. From the Feynman diagram construction, we have given a heuristic motivation for Hedin's equations, which are themselves nonperturbative. Hedin's equations place emphasis on the screened Coulomb interaction. The intuitive nature of screening $-$ the simple idea that charges rearrange themselves, or respond, to an added charge $-$ is the major reason behind the appeal and success of the $GW$ approximation. The time or frequency dependence of the screened Coulomb interaction is largely what sets $GW$ apart from density functional or Hartree-Fock theories.

Impressive advances in code development and computing resources have pushed $GW$ calculations to a new scale. At the computationally lowest level of theory, $G_0W_0$ calculations remain the most widely used and can be routinely applied to systems with hundreds of atoms. Within the $G_0W_0$ approach, we have outlined the practical considerations presented to the user before performing any calculation: starting point, basis set, evaluation of the self-energy, and convergence are all for the user to decide. For a broad class of systems, $G_0W_0$ already gives electron addition and removal energies in good agreement with experiment. These successes give $GW$ an impressive ranking in computational value, or accuracy for computational cost, on any list of first-principles electronic structure methods. The versatility of $GW$ assures that new applications in physics, chemistry, and materials science will continue to emerge in the future.

\section{Acknowledgements}
This work is supported by the Academy of Finland through grant Nos. 284621, 305632, 316347 and 316168. The authors acknowledge the CSC-IT Center for Science, Finland, for generous computational resources and the Aalto University School of Science ``Science-IT'' project for computational resources. We would like to thank Fabio Caruso, Jan Wilhelm, Leeor Kronik and Lucia Reining for fruitful discussions and valuable feedback.

\appendix
\section{Integral notation}
	\label{sec:notation}
		We adapt the following integral notation

\begin{equation}
    \bracket{\phi_i}{\hat{O}}{\phi_j} = \iint d\bfr d\bfrp \phi_i^*(\bfr) \hat{O}(\bfr,\bfrp) \phi_j(\bfrp)
    \label{eq:twocenterint}
\end{equation}
\begin{equation}
    \begin{split}
         \bracket{\phi_i\phi_j}{\hat{O}}{\phi_k\phi_l} = \iint &d\bfr d\bfrp \phi_i^*(\bfr)\phi_j^*(\bfrp) \hat{O}(\bfr,\bfrp)\\&\times  \phi_k(\bfr)\phi_l(\bfrp)
    \end{split}
\end{equation}
where $\hat{O}(\bfr,\bfrp)$ is an operator that depends on the spatial variables $\bfr=(x,y,z)$ and $\bfrp=(x',y',z')$. Furthermore the following notation for Coulomb integrals are used
\begin{align}
    \braket{\phi_i\phi_j| \phi_k\phi_l} &= \braket{ij|kl} \\
    &=  \iint d\bfr d\bfrp \phi_i^*(\bfr)\phi_j^*(\bfrp) v(\bfr,\bfrp)   \phi_k(\bfr)\phi_l(\bfrp)\nonumber ,
\end{align}
where $v(\bfr,\bfrp)=1/|\bfr-\bfrp|$ is the Coulomb operator. The antisymmetrized Coulomb integrals are defined as
\begin{align}
   \braket{\phi_i\phi_j||\phi_k\phi_l} &=  \bra{ij} \ket{kl} \\
    &= \braket{\phi_i\phi_j|\phi_k\phi_l} -  \braket{\phi_i\phi_j|\phi_l\phi_k}\nonumber .
\end{align}
\section{The many-body problem}
	\label{sec:MB}
		In first principles electronic structure theory, the aim is to solve the Schr\"odinger equation. For simplicity we focus on the non-relativistic time-independent Schr\"odinger equation.  For a system of $N$ electrons and $M$ nuclei, the Schr\"odinger equation is given by
\begin{equation}
\label{eq:SE}
 \hat{H}\Psi=E\Psi
\end{equation}
with the many-body Hamiltonian
\begin{align}
 \hat{H}=&\underbrace{-\sum^{N}_{i=1}\frac{\nabla_i^2}{2}+\frac{1}{2}\sum^N_{i \ne j}\frac{1}{\left|\textbf{r}_i-\textbf{r}_j\right|}-
   \sum^N_{i=1}\sum^M_{a=1}\frac{Z_a}{\left|\textbf{r}_i-\textbf{R}_a\right|}}_{\hat{H}_{elec}} \nonumber   \\ 
   &-\sum^{M}_{a=1}\frac{\nabla_a^2}{2M_a}+\sum^M_{a=1}\sum^M_{b>a}\frac{Z_aZ_b}{\left|\textbf{R}_a-\textbf{R}_b\right|} \label{eq:Hmb} .
\end{align}
$r_i$ and $R_a$ are the positions of the electrons and the nuclei, respectively, and $ Z_a$  is the charge of the nuclei.

To make this system of coupled electrons and nuclei more tractable, the Born-Oppenheimer (BO) approximation of clamped nuclei is frequently introduced. In this case, we only need to consider the electronic Hamiltonian by itself,
\begin{equation}
\label{eq:SEe}
 \hat{H}_{elec}\Psi_{elec}=E_{elec}\Psi_{elec},
\end{equation}
in which the many-electron wave function $\Psi_{elec}$ depends parametrically on the position of the nuclei. The electronic Hamiltonian is then given by
\begin{align}
\label{eq:Helec1} 
 \hat{H}_{elec}&=\sum^{N}_{i=1}\left[-\frac{\nabla_i^2}{2}+v_{\rm ext}(\bfr_i) \right]+\frac{1}{2}\sum^N_{i \ne j}\frac{1}{\left|\textbf{r}_i-\textbf{r}_j \right|} \\
\label{eq:Helec2} 
         &=\sum^{N}_{i=1}\hat{h}^0(\bfr_i) +\frac{1}{2}\sum^N_{i \ne j}v(\bfr_i,\bfr_j) \, .
\end{align}
We use $v$ to denote the bare Coulomb interaction and the external potential is the same for every electron 
\begin{equation}
\label{eq:vext}
  v_{\rm ext}(\bfr)=-\sum^M_{a=1}\frac{Z_a}{\left|\bfr-\textbf{R}_a\right|} \, .
\end{equation}

The kinetic energy and external potential are grouped together in $\hat{h}^0(\mathbf{r}_i)$ in Equation~\eqref{eq:Helec2}. Approaches to solve the electronic Schr\"odinger equation (\ref{eq:SEe}) can be largely grouped according to the basic variable they work with: the many-body wave function, the density-matrix, the density, or Green's functions. Each choice has its advantages and disadvantages and no consensus has been reached on the optimal choice. Green's functions have a natural connection, however, to the particle addition/removal problem and theoretical spectroscopy.

	\subsection{Green's function formalism}
		\label{sec:G}
			To derive equations for the one-particle Green's function that are more amenable to approximations than definitions in Equations~\eqref{def:G} and \eqref{G:Leh},  we start from the equation of motion for the field operators, which relates the time derivative of $\hat{\psi}$ to the commutator of $\hat{\psi}$ and $\hat{H}_{elec}$ \cite{Gross/Runge:MPT,Fetter/Walecka}:
\begin{equation}
\label{eqm:fo}
  i\frac{\partial}{\partial t} \hat{\psi}(\vx,t)=\left[\hat{\psi}(\vx,t),\hat{H}_{elec}\right],
\end{equation}
where $\mathbf{x}$ contains also the spin variable $\sigma$, i.e., $\mathbf{x}=(\mathbf{r},\sigma)$.
Evaluating the commutator in the Heisenberg picture  and applying the anti-commutation relations   yields for the equation of motion
\begin{multline}
\label{eqm:fo2}
  i\frac{\partial}{\partial t} \hat{\psi}(\vx,t)= \\
      \left[\hat{h}^0 (\bfr)+\int \hat{\psi}^\dagger(\vxp,t)v(\bfr,\bfrp)  \hat{\psi}(\vxp,t) d\vxp \right]\hat{\psi}(\vx,t) \, .
\end{multline}
The equation of motion for the Green's function \refeq{def:G} then follows from \refeq{eqm:fo2}
\begin{multline}
  \label{eq:Geqmotfo}
  \left[ i \frac{\partial}{\partial t} -\hat{h}^0(\bfr)\right]
  G(\vx t,\vxp t')=    \\
 \delta(t-t') \delta(\vx-\vxp) -i \int \! d\vxpp \, v(\bfr,\bfrpp) \times \\
  \bracket{N}{\hat{T}\left[\hat{\psi}^\dagger(\vxpp,t)\hat{\psi}(\vxpp,t) \hat{\psi}(\vx,t)\hat{\psi}^\dagger(\vxp,t')\right]}{N}
\end{multline}
The term under the integral contains the two-particle Green's function, $G_2(\vx t,\vxpp t,\vxpp t^+,\vxp t')$, and includes all two-body correlations in the system. In order to calculate  the one-particle Green's function we would therefore require the equation of motion for the two-particle Green's function, which in turn introduces the three-particle Green's function. Applied iteratively, this procedure creates an infinite series of  higher order Green's functions and thus describes all the possible many-body  interactions in the system. In practice, however, the resulting recurrence  relation for the $n^{\mathrm{th}}$ order Green's function is impossible to solve for large $n$. Instead we introduce the non-local, time-dependent self-energy $\bar{\Sigma}(\vx t,\vxp t')$ 
\begin{align}
  -i \int \! d\vxpp \,  & v(\bfr,\bfrpp) G_2(\vx t,\vxpp t,\vxpp t^+,\vxp t') \equiv \nonumber \\
  & \int dt'' \! \int \! d\vxpp \, \bar{\Sigma}(\vx t,\vxpp t'') G(\vxpp t'',\vxp t') \,.
  \label{eq:mass_def}
\end{align}
Analogous to other electronic structure methods, we separate out the most dominant term,
the Hartree potential
\begin{align}
  \label{eq:VH_def}
  v_{\mathrm{H}}(\bfr)&=\int d \bfrp v(\bfr,\bfrp) \bracket{N}{\hat{\psi}^\dagger(\bfrp,t)\hat{\psi}(\bfrp,t)}{N} \\
                 &= \int d\bfrp v(\bfr,\bfrp)n(\bfrp) \, ,
\end{align}
where $n(\bfr)$ is the electron density. With this definition, the equation of motion for the Green's function \refeq{eq:Geqmotfo} adopts the much more convenient form of an integral equation involving the self-energy, $\Sigma=\bar{\Sigma}-v_{\mathrm{H}}$:
\begin{multline}
  \label{eq:Geqmot}
  \left[ i \frac{\partial}{\partial t} -\hat{h}^0(\bfr) -v_{\mathrm{H}}(\bfr) \right]
  G(\vx t,\vxp t')= \delta(t-t') \delta(\bfr-\bfrp) \\
  + \int dt'' \! \int \! d\vxpp \, \Sigma(\vx t,\vxpp t'') G(\vxpp t'',\vxp t') \, .
\end{multline}

If $G_0$ now denotes the Green's function that is a solution to $\hat{h} = \hat{h}^0 + v_{\mathrm{H}}(\mathbf{r})$ (the kinetic energy, the external potential, and the Hartree potential), then Equation~\eqref{eq:Geqmot} can be further rewritten as
\begin{align}
  \label{def:dyson2}
  G(1,2)&= G_0(1,2)  \\ 
   & +\int G_0(1,3)   \Sigma(3,4) G(4,2)d(3,4) \, .  \nonumber
\end{align}
Here we changed to the abbreviated notation $(1,2,\ldots)$ for the set of  position, time and spin variables $(\vx_1 t_1,\vx_2 t_2, \ldots)$. Accordingly $\int d(1)$ is a shorthand notation for the
integration in all three variables of the corresponding triple(s). In this context $1^+$ implies the addition of a positive infinitesimal to the time argument $1$. Equation~\eqref{def:dyson2} is again Dyson's equation \cite{Dyson1:1949,Dyson2:1949} (see also Equations~\eqref{eq:dyson} and \eqref{eq:g1}) and links the non-interacting Green's function, $G_0$, to the interacting one, $G$. Dyson's equation gives a physical interpretation to the self-energy instead of simply its definition by Equation~\eqref{eq:mass_def}. The self-energy quantifies the difference between a bare and a fully interacting electron, or \emph{quasielectron}. This brings us back to our phenomenological consideration in Section~\ref{sec:TS}. An additional electron or hole drags a dynamically adjusting polarization cloud through the system that alters its energy. Hence the name self-energy. If instead $G_0$ is the solution to the mean-field Hamiltonian $\hat{h}^{\mathrm{MF}}$ (e.g. of Kohn-Sham density-functional theory \cite{Kohn/Sham:1965}) then the self-energy, $\Sigma$, embodies the difference between a quasielectron and an electron in the static mean-field.

At this stage in the derivation, the self-energy as well as the Green's function are still exact and contain the electron-electron interaction to all orders. A full solution of Equations~\eqref{eq:Geqmot} and \eqref{def:dyson2} is not tractable and approximations are required.


\subsection{Hedin's equations}
\label{sec:Hedins_eqs}

In 1965 Hedin expanded the Green's function and the self-energy in terms of the screened instead of the bare Coulomb interaction \cite{Hedin:1965}. Introducing a small perturbing field $\varphi$ that will later be set to zero, the operator identity due to Schwinger \cite{Schwinger:1951_1}
\begin{equation}
\label{eq:Schwinger_ID}
       G_2(1,3,2,3^+)=G(1,2)G(3,3^+)-\frac{\delta G(1,2)}{\delta \varphi(3)}
\end{equation}
can be used to eliminate the two particle Green's function from Equation~\eqref{eq:mass_def} for the self-energy. Multiplying the resulting equation with $G^{-1}$ leads to the following expression for the self-energy
\begin{align}
\label{eq:self-energy_phi_1}
       \bar{\Sigma}(1,2)&=\delta(1,2)\int d(3)v(1,3)G(3,3^+) \nonumber \\
                                   &  \quad   -i\int d(3,4)v(1,3)G(1,4)\frac{\delta G^{-1}(4,2)}{\delta \varphi(3)} \\
                                   &= \delta(1,2) v_{\mathrm{H}}(1)   \nonumber  \\
                                   & \quad -i\int d(3,4)v(1,3)G(1,4)\frac{\delta G^{-1}(4,2)}{\delta \varphi(3)} \nonumber \, .
\end{align}
where $\bar{\Sigma}$ is defined to include the Hartree potential $v_{\mathrm{H}}$, unlike $\Sigma$. Using the following definitions \\

\noindent \emph{total potential:}        
\begin{equation}
 V(1)= \varphi(1)+v_{\mathrm{H}}(1)
\end{equation}

\noindent \emph{3-point vertex:}        
\begin{equation}
 \Gamma(1,2,3)= -\frac{\delta G^{-1}(1,2)}{\delta V(3)}
\end{equation}

\noindent \emph{dielectric function:}        
\begin{equation}
\label{eq:epsilon_inv}
 \varepsilon^{-1}(1,2)= \frac{\delta V(1)}{\delta \varphi(2)}
\end{equation}

\noindent \emph{screened Coulomb interaction:}        
\begin{equation}
W(1,2)= \int \! \! d(3) \; \varepsilon^{-1}(1,3)v(3,2)
\end{equation}

\noindent \emph{irreducible polarizability:}        
\begin{equation}
P(1,2)=-i\frac{\delta G(1,1^+)}{\delta V(2)}=\frac{\delta n(1)}{\delta V(2)}
\end{equation}

\noindent \emph{reducible polarizability:}        
\begin{equation}
\chi(1,2)=-i\frac{\delta G(1,1^+)}{\delta \varphi(2)}=\frac{\delta n(1)}{\delta \varphi(2)}
\end{equation}
we arrive at Hedin's  equations \cite{Hedin:1965}
\begin{align}
\label{eq:P}
  P(1,2)&=-i \! \! \int \! \! d(3,4) G(4,2)G(2,3)\Gamma(3,4,1) \\ 
\label{eq:W}
  W(1,2)&=v(1,2)+\! \! \int \!  \! d(3,4) v(1,3)P(3,4)W(4,2)  \\ 
  \Sigma(1,2)&=i\int \!  \! d(3,4) G(1,4)W(1^+,3)\Gamma(4,2,3) \\ 
\label{eq:gamma}
  \Gamma(1,2,3)&=\delta(1,2)\delta(1,3)+  \\
	 & \hspace{-0.4cm} \! \! \int \! \! d(4,5,6,7) \frac{\delta \Sigma(1,2)}{\delta G(4,5)} G(4,6)G(7,5)
	  \Gamma(6,7,3) . \nonumber
\end{align}
Dyson's equation (\ref{def:dyson2}) closes this set of integro-differential equations, which is shown pictorially in Figure~\ref{fig:Hedin_loops}(a). 

The benefit of Hedin's equations is that the self-energy is given in terms of an effective, or screened, rather than the bare Coulomb interaction
\begin{equation}
\label{eq:W_it_epsinv}
     W(1,2)=\int \! \! d(3) \:  \varepsilon^{-1}(1,3)v(3,2) \: .
\end{equation}
The screening 
\begin{equation}
\label{eq:epsinv_repeat_chi}
     \varepsilon^{-1}(1,2)= \delta(1,2)+\int \! \! d(3) \: v(1,3) \chi(3,2)
\end{equation}
follows from the reducible polarizability or density-density response function  
\begin{align}
   \chi(1,2)&= \chi_0(1,2)+ \int \! \! d(3,4) \chi_0(1,3)v(3,4)\chi(4,2)  \nonumber  \\
\label{eq:dens-dens_reponse_fc_N}
                &= \bracket{N}{\hat{T}\left[\delta \hat{n}(1)\delta \hat{n}(2)\right]}{N} \: .
\end{align}
Here $\delta \hat{n}(1)$ is a density fluctuation $\delta \hat{n}(1)= \hat{n}(1) - n(1)$ of the density around its average value, where $\hat{n}(1)=\sum_i^N \delta (r_1-r_i)$. For a polarizable medium, like a solid, screening is large and the screened Coulomb interaction will differ considerably from the bare one. It therefore makes sense to build a perturbation series on $W$ instead of $v$. Hedin's equations are physically transparent in this sense. Electron-hole pairs are created in the polarizability (one Green's function for the electron, a separate Green's function for the hole). They interact through the vertex function, which is determined by the change in potential upon excitation. The polarizability, in turn, determines the dielectric function, which screens the Coulomb interaction. The self-energy quantifies the energy contribution that the added electron or hole experiences through the interaction with its surrounding.

\section{Computational details}
 \label{app:computational_details}
  Figures \ref{fig:sigma_homo}, \ref{fig:zshot}, \ref{fig:freq_treat}, \ref{fig:basis_convergence}(b), \ref{fig:spd_H2O}, \ref{fig:sigma_evgw0_deltaE} and \ref{fig:IP_pyridine} present original content. All calculations are performed with the FHI-aims program package \cite{Blum2009,Havu2009}, which expands the all-electron KS equations in numeric-atom-centered orbitals (NAOs), see Section~\ref{sec:basis_sets}. The structures have been optimized at the DFT level using the Perdew-Burke-Ernzerhof (PBE) functional \cite{Perdew/Burke/Ernzerhof:1996} to model the XC potential and NAOs of \textit{tier 2} quality \cite{Blum2009} to represent core and valence electrons. Dispersion corrections are accounted for by employing the Tkatchenko-Scheffler van der Waals correction \cite{Tkatchenko2009}.\par
$G_0W_0$ calculations have been performed with the contour-deformation (CD) technique \cite{Golze2018} if not indicated otherwise. Calculations with the analytic continuation (AC) \cite{Xinguo/implem_full_author_list} have been conducted for the $G_0W_0$ self-energies in Figure~\ref{fig:freq_treat} and for the sc$GW$ result in Figure~\ref{fig:spd_H2O}. For both methods, CD and AC, a modified Gauss-Legendre grid with 200 grid points is used for the numerical integration of the integral over the imaginary frequency axis. In case of the AC approach, the same set of grid points $\{i\omega\}$ is employed to calculate $\Sigma^c_s(i\omega)$, which has been fitted to a Pad\'{e} approximant with at least 16 parameters \cite{Vidberg1977}, unless stated otherwise.\par
Quasiparticle energies have been computed by iteratively solving Equation~\eqref{Eq:qpe}. Furthermore, all quasiparticle energies have been extrapolated to the complete basis set limit using the Dunning basis set family cc-pV$n$Z $(n=3-6)$ \cite{Dunning1989,Wilson1996}. The cc-pV$n$Z basis sets are all-electron Gaussian basis sets, which can be considered as a special case of an NAO and are treated numerically in FHI-aims. The extrapolation has been performed by a linear regression against the inverse of the total number of basis functions.
  
\bibliography{GW_review}

\begin{thebibliography}{613}%
\makeatletter
\providecommand \@ifxundefined [1]{%
 \@ifx{#1\undefined}
}%
\providecommand \@ifnum [1]{%
 \ifnum #1\expandafter \@firstoftwo
 \else \expandafter \@secondoftwo
 \fi
}%
\providecommand \@ifx [1]{%
 \ifx #1\expandafter \@firstoftwo
 \else \expandafter \@secondoftwo
 \fi
}%
\providecommand \natexlab [1]{#1}%
\providecommand \enquote  [1]{``#1''}%
\providecommand \bibnamefont  [1]{#1}%
\providecommand \bibfnamefont [1]{#1}%
\providecommand \citenamefont [1]{#1}%
\providecommand \href@noop [0]{\@secondoftwo}%
\providecommand \href [0]{\begingroup \@sanitize@url \@href}%
\providecommand \@href[1]{\@@startlink{#1}\@@href}%
\providecommand \@@href[1]{\endgroup#1\@@endlink}%
\providecommand \@sanitize@url [0]{\catcode `\\12\catcode `\$12\catcode
  `\&12\catcode `\#12\catcode `\^12\catcode `\_12\catcode `\%12\relax}%
\providecommand \@@startlink[1]{}%
\providecommand \@@endlink[0]{}%
\providecommand \url  [0]{\begingroup\@sanitize@url \@url }%
\providecommand \@url [1]{\endgroup\@href {#1}{\urlprefix }}%
\providecommand \urlprefix  [0]{URL }%
\providecommand \Eprint [0]{\href }%
\providecommand \doibase [0]{http://dx.doi.org/}%
\providecommand \selectlanguage [0]{\@gobble}%
\providecommand \bibinfo  [0]{\@secondoftwo}%
\providecommand \bibfield  [0]{\@secondoftwo}%
\providecommand \translation [1]{[#1]}%
\providecommand \BibitemOpen [0]{}%
\providecommand \bibitemStop [0]{}%
\providecommand \bibitemNoStop [0]{.\EOS\space}%
\providecommand \EOS [0]{\spacefactor3000\relax}%
\providecommand \BibitemShut  [1]{\csname bibitem#1\endcsname}%
\let\auto@bib@innerbib\@empty
\bibitem [{\citenamefont {Adamo}\ and\ \citenamefont {Barone}(1999)}]{PBE0_3}%
  \BibitemOpen
  \bibfield  {author} {\bibinfo {author} {\bibnamefont {Adamo}, \bibfnamefont
  {C}}, \ and\ \bibinfo {author} {\bibfnamefont {V.}~\bibnamefont {Barone}}}
  (\bibinfo {year} {1999}),\ \bibfield  {title} {\enquote {\bibinfo {title}
  {{T}owards reliable density functional methods without adjustable parameters:
  {T}he {P}{B}{E}0 model},}\ }\href@noop {} {\bibfield  {journal} {\bibinfo
  {journal} {J.\ Chem.\ Phys}\ }\textbf {\bibinfo {volume} {110}},\ \bibinfo
  {pages} {6158}}\BibitemShut {NoStop}%
\bibitem [{\citenamefont {Adler}(1962)}]{Adler:1962}%
  \BibitemOpen
  \bibfield  {author} {\bibinfo {author} {\bibnamefont {Adler}, \bibfnamefont
  {S~L}}} (\bibinfo {year} {1962}),\ \bibfield  {title} {\enquote {\bibinfo
  {title} {{Q}uantum theory of the dielectric constant in real solids},}\
  }\href@noop {} {\bibfield  {journal} {\bibinfo  {journal} {Phys. Rev.}\
  }\textbf {\bibinfo {volume} {126}},\ \bibinfo {pages} {413}}\BibitemShut
  {NoStop}%
\bibitem [{\citenamefont {Adragna}\ \emph {et~al.}(2003)\citenamefont
  {Adragna}, \citenamefont {{Del Sole}},\ and\ \citenamefont
  {Marini}}]{Adragna/etal:2003}%
  \BibitemOpen
  \bibfield  {author} {\bibinfo {author} {\bibnamefont {Adragna}, \bibfnamefont
  {G}}, \bibinfo {author} {\bibfnamefont {R.}~\bibnamefont {{Del Sole}}}, \
  and\ \bibinfo {author} {\bibfnamefont {A.}~\bibnamefont {Marini}}} (\bibinfo
  {year} {2003}),\ \bibfield  {title} {\enquote {\bibinfo {title} {{A}b initio
  calculation of the exchange-correlation kernel in extended systems},}\
  }\href@noop {} {\bibfield  {journal} {\bibinfo  {journal} {Phys. Rev. B}\
  }\textbf {\bibinfo {volume} {68}},\ \bibinfo {pages} {165108}}\BibitemShut
  {NoStop}%
\bibitem [{\citenamefont {Aguilera}\ \emph
  {et~al.}(2013{\natexlab{a}})\citenamefont {Aguilera}, \citenamefont
  {Friedrich}, \citenamefont {Bihlmayer},\ and\ \citenamefont
  {Bl\"ugel}}]{Aguilera/etal:2013b}%
  \BibitemOpen
  \bibfield  {author} {\bibinfo {author} {\bibnamefont {Aguilera},
  \bibfnamefont {I}}, \bibinfo {author} {\bibfnamefont {C.}~\bibnamefont
  {Friedrich}}, \bibinfo {author} {\bibfnamefont {G.}~\bibnamefont
  {Bihlmayer}}, \ and\ \bibinfo {author} {\bibfnamefont {S.}~\bibnamefont
  {Bl\"ugel}}} (\bibinfo {year} {2013}{\natexlab{a}}),\ \bibfield  {title}
  {\enquote {\bibinfo {title} {${G}{W}$ study of topological insulators
  {B}i${}_{2}${S}e${}_{3}$, {B}i${}_{2}${T}e${}_{3}$, and
  {S}b${}_{2}${T}e${}_{3}$: {B}eyond the perturbative one-shot approach},}\
  }\href@noop {} {\bibfield  {journal} {\bibinfo  {journal} {Phys. Rev. B}\
  }\textbf {\bibinfo {volume} {88}},\ \bibinfo {pages} {045206}}\BibitemShut
  {NoStop}%
\bibitem [{\citenamefont {Aguilera}\ \emph
  {et~al.}(2013{\natexlab{b}})\citenamefont {Aguilera}, \citenamefont
  {Friedrich},\ and\ \citenamefont {Bl\"ugel}}]{Aguilera/etal:2013}%
  \BibitemOpen
  \bibfield  {author} {\bibinfo {author} {\bibnamefont {Aguilera},
  \bibfnamefont {I}}, \bibinfo {author} {\bibfnamefont {C.}~\bibnamefont
  {Friedrich}}, \ and\ \bibinfo {author} {\bibfnamefont {S.}~\bibnamefont
  {Bl\"ugel}}} (\bibinfo {year} {2013}{\natexlab{b}}),\ \bibfield  {title}
  {\enquote {\bibinfo {title} {{S}pin-orbit coupling in quasiparticle studies
  of topological insulators},}\ }\href@noop {} {\bibfield  {journal} {\bibinfo
  {journal} {Phys. Rev. B}\ }\textbf {\bibinfo {volume} {88}},\ \bibinfo
  {pages} {165136}}\BibitemShut {NoStop}%
\bibitem [{\citenamefont {Aguilera}\ \emph
  {et~al.}(2015{\natexlab{a}})\citenamefont {Aguilera}, \citenamefont
  {Friedrich},\ and\ \citenamefont {Bl\"ugel}}]{Aguilera/etal:2015}%
  \BibitemOpen
  \bibfield  {author} {\bibinfo {author} {\bibnamefont {Aguilera},
  \bibfnamefont {I}}, \bibinfo {author} {\bibfnamefont {C.}~\bibnamefont
  {Friedrich}}, \ and\ \bibinfo {author} {\bibfnamefont {S.}~\bibnamefont
  {Bl\"ugel}}} (\bibinfo {year} {2015}{\natexlab{a}}),\ \bibfield  {title}
  {\enquote {\bibinfo {title} {{E}lectronic phase transitions of bismuth under
  strain from relativistic self-consistent ${G}{W}$ calculations},}\
  }\href@noop {} {\bibfield  {journal} {\bibinfo  {journal} {Phys. Rev. B}\
  }\textbf {\bibinfo {volume} {91}},\ \bibinfo {pages} {125129}}\BibitemShut
  {NoStop}%
\bibitem [{\citenamefont {Aguilera}\ \emph
  {et~al.}(2015{\natexlab{b}})\citenamefont {Aguilera}, \citenamefont
  {Nechaev}, \citenamefont {Friedrich}, \citenamefont {Bl\"ugel},\ and\
  \citenamefont {Chulkov}}]{gw_review_soc}%
  \BibitemOpen
  \bibfield  {author} {\bibinfo {author} {\bibnamefont {Aguilera},
  \bibfnamefont {I}}, \bibinfo {author} {\bibfnamefont {I.~A.}\ \bibnamefont
  {Nechaev}}, \bibinfo {author} {\bibfnamefont {C.}~\bibnamefont {Friedrich}},
  \bibinfo {author} {\bibfnamefont {S.}~\bibnamefont {Bl\"ugel}}, \ and\
  \bibinfo {author} {\bibfnamefont {E.~V.}\ \bibnamefont {Chulkov}}} (\bibinfo
  {year} {2015}{\natexlab{b}}),\ \enquote {\bibinfo {title} {{M}any-{B}ody
  {E}ffects in the {E}lectronic {S}tructure of {T}opological {I}nsulators},}\
  in\ \href {\doibase 10.1002/9783527681594.ch7} {\emph {\bibinfo {booktitle}
  {{T}opological {I}nsulators}}},\ Chap.~\bibinfo {chapter} {7}\ (\bibinfo
  {publisher} {John Wiley \& Sons, Ltd})\ pp.\ \bibinfo {pages}
  {161--189}\BibitemShut {NoStop}%
\bibitem [{\citenamefont {Aguilera}\ \emph {et~al.}(2011)\citenamefont
  {Aguilera}, \citenamefont {Vidal}, \citenamefont {Wahn\'on}, \citenamefont
  {Reining},\ and\ \citenamefont {Botti}}]{Aguilera/etal:2011}%
  \BibitemOpen
  \bibfield  {author} {\bibinfo {author} {\bibnamefont {Aguilera},
  \bibfnamefont {I}}, \bibinfo {author} {\bibfnamefont {J.}~\bibnamefont
  {Vidal}}, \bibinfo {author} {\bibfnamefont {P.}~\bibnamefont {Wahn\'on}},
  \bibinfo {author} {\bibfnamefont {L.}~\bibnamefont {Reining}}, \ and\
  \bibinfo {author} {\bibfnamefont {S.}~\bibnamefont {Botti}}} (\bibinfo {year}
  {2011}),\ \bibfield  {title} {\enquote {\bibinfo {title} {{F}irst-principles
  study of the band structure and optical absorption of {C}u{G}a{S}${}_{2}$},}\
  }\href@noop {} {\bibfield  {journal} {\bibinfo  {journal} {Phys. Rev. B}\
  }\textbf {\bibinfo {volume} {84}},\ \bibinfo {pages} {085145}}\BibitemShut
  {NoStop}%
\bibitem [{\citenamefont {Ahmed}\ \emph {et~al.}(2014)\citenamefont {Ahmed},
  \citenamefont {Albers}, \citenamefont {Balatsky}, \citenamefont {Friedrich},\
  and\ \citenamefont {Zhu}}]{Ahmed/etal:2014}%
  \BibitemOpen
  \bibfield  {author} {\bibinfo {author} {\bibnamefont {Ahmed}, \bibfnamefont
  {T}}, \bibinfo {author} {\bibfnamefont {R.~C.}\ \bibnamefont {Albers}},
  \bibinfo {author} {\bibfnamefont {A.~V.}\ \bibnamefont {Balatsky}}, \bibinfo
  {author} {\bibfnamefont {C.}~\bibnamefont {Friedrich}}, \ and\ \bibinfo
  {author} {\bibfnamefont {J.-X.}\ \bibnamefont {Zhu}}} (\bibinfo {year}
  {2014}),\ \bibfield  {title} {\enquote {\bibinfo {title} {${G}{W}$
  quasiparticle calculations with spin-orbit coupling for the light
  actinides},}\ }\href@noop {} {\bibfield  {journal} {\bibinfo  {journal}
  {Phys. Rev. B}\ }\textbf {\bibinfo {volume} {89}},\ \bibinfo {pages}
  {035104}}\BibitemShut {NoStop}%
\bibitem [{\citenamefont {Albrecht}\ \emph {et~al.}(1997)\citenamefont
  {Albrecht}, \citenamefont {Onida},\ and\ \citenamefont
  {Reining}}]{albrecht_prb_55}%
  \BibitemOpen
  \bibfield  {author} {\bibinfo {author} {\bibnamefont {Albrecht},
  \bibfnamefont {S}}, \bibinfo {author} {\bibfnamefont {G.}~\bibnamefont
  {Onida}}, \ and\ \bibinfo {author} {\bibfnamefont {L.}~\bibnamefont
  {Reining}}} (\bibinfo {year} {1997}),\ \bibfield  {title} {\enquote {\bibinfo
  {title} {{A}b initio calculation of the quasiparticle spectrum and excitonic
  effects in {L}i$_{2}${O}},}\ }\href {\doibase 10.1103/PhysRevB.55.10278}
  {\bibfield  {journal} {\bibinfo  {journal} {Phys. Rev. B}\ }\textbf {\bibinfo
  {volume} {55}},\ \bibinfo {pages} {10278--10281}}\BibitemShut {NoStop}%
\bibitem [{\citenamefont {Albrecht}\ \emph
  {et~al.}(1998{\natexlab{a}})\citenamefont {Albrecht}, \citenamefont
  {Reining}, \citenamefont {Sole},\ and\ \citenamefont
  {Onida}}]{albrecht_prl_80}%
  \BibitemOpen
  \bibfield  {author} {\bibinfo {author} {\bibnamefont {Albrecht},
  \bibfnamefont {S}}, \bibinfo {author} {\bibfnamefont {L.}~\bibnamefont
  {Reining}}, \bibinfo {author} {\bibfnamefont {R.~Del}\ \bibnamefont {Sole}},
  \ and\ \bibinfo {author} {\bibfnamefont {G.}~\bibnamefont {Onida}}} (\bibinfo
  {year} {1998}{\natexlab{a}}),\ \bibfield  {title} {\enquote {\bibinfo {title}
  {{A}b {I}nitio {C}alculation of {E}xcitonic {E}ffects in the {O}ptical
  {S}pectra of {S}emiconductors},}\ }\href {\doibase
  10.1103/PhysRevLett.80.4510} {\bibfield  {journal} {\bibinfo  {journal}
  {Phys. Rev. Lett.}\ }\textbf {\bibinfo {volume} {80}},\ \bibinfo {pages}
  {4510--4513}}\BibitemShut {NoStop}%
\bibitem [{\citenamefont {Albrecht}\ \emph
  {et~al.}(1998{\natexlab{b}})\citenamefont {Albrecht}, \citenamefont
  {Reining}, \citenamefont {Sole},\ and\ \citenamefont
  {Onida}}]{albrecht_pssa_170}%
  \BibitemOpen
  \bibfield  {author} {\bibinfo {author} {\bibnamefont {Albrecht},
  \bibfnamefont {S}}, \bibinfo {author} {\bibfnamefont {L.}~\bibnamefont
  {Reining}}, \bibinfo {author} {\bibfnamefont {R.~Del}\ \bibnamefont {Sole}},
  \ and\ \bibinfo {author} {\bibfnamefont {G.}~\bibnamefont {Onida}}} (\bibinfo
  {year} {1998}{\natexlab{b}}),\ \bibfield  {title} {\enquote {\bibinfo {title}
  {{E}xcitonic {E}ffects in the {O}ptical {P}roperties},}\ }\href {\doibase
  10.1002/(SICI)1521-396X(199812)170:2<189::AID-PSSA189>3.0.CO;2-3} {\bibfield
  {journal} {\bibinfo  {journal} {Phys. Status Solidi A}\ }\textbf {\bibinfo
  {volume} {170}}~(\bibinfo {number} {2}),\ \bibinfo {pages}
  {189--197}}\BibitemShut {NoStop}%
\bibitem [{\citenamefont {Almbladh}\ and\ \citenamefont {von
  Barth}(1985)}]{Almbladh/Barth:1985}%
  \BibitemOpen
  \bibfield  {author} {\bibinfo {author} {\bibnamefont {Almbladh},
  \bibfnamefont {C~O}}, \ and\ \bibinfo {author} {\bibfnamefont
  {U.}~\bibnamefont {von Barth}}} (\bibinfo {year} {1985}),\ \bibfield  {title}
  {\enquote {\bibinfo {title} {{E}xact results for the charge and spin
  densities, exchange-correlation potentials, and density-functional
  eigenvalues},}\ }\href@noop {} {\bibfield  {journal} {\bibinfo  {journal}
  {Phys.\ Rev.\ B}\ }\textbf {\bibinfo {volume} {31}},\ \bibinfo {pages}
  {3231}}\BibitemShut {NoStop}%
\bibitem [{\citenamefont {Almbladh}\ and\ \citenamefont
  {Hedin}(1983)}]{Almbladh/Hedin:1983}%
  \BibitemOpen
  \bibfield  {author} {\bibinfo {author} {\bibnamefont {Almbladh},
  \bibfnamefont {C~O}}, \ and\ \bibinfo {author} {\bibfnamefont
  {L.}~\bibnamefont {Hedin}}} (\bibinfo {year} {1983}),\ \bibfield  {title}
  {\enquote {\bibinfo {title} {{B}eyond the {O}ne-{E}lectron {M}odel.
  {M}any-{B}ody {E}ffects in {A}toms, {M}olecules, and {S}olids},}\ }in\
  \href@noop {} {\emph {\bibinfo {booktitle} {{H}andbook on {S}ynchroton
  {R}adiation}}},\ Vol.~\bibinfo {volume} {1},\ \bibinfo {editor} {edited by\
  \bibinfo {editor} {\bibfnamefont {E.~E.}\ \bibnamefont {Koch}}}\ (\bibinfo
  {publisher} {North-Holland, Amsterdam})\ pp.\ \bibinfo {pages}
  {607--904}\BibitemShut {NoStop}%
\bibitem [{\citenamefont {Alves-Santos}\ \emph {et~al.}(2014)\citenamefont
  {Alves-Santos}, \citenamefont {Jorge}, \citenamefont {Caldas},\ and\
  \citenamefont {Varsano}}]{Alves-Santos/etal:2014}%
  \BibitemOpen
  \bibfield  {author} {\bibinfo {author} {\bibnamefont {Alves-Santos},
  \bibfnamefont {M}}, \bibinfo {author} {\bibfnamefont {L.~M.~M.}\ \bibnamefont
  {Jorge}}, \bibinfo {author} {\bibfnamefont {M.~J.}\ \bibnamefont {Caldas}}, \
  and\ \bibinfo {author} {\bibfnamefont {D.}~\bibnamefont {Varsano}}} (\bibinfo
  {year} {2014}),\ \bibfield  {title} {\enquote {\bibinfo {title} {{E}lectronic
  {S}tructure of {I}nterfaces between {T}hiophene and {T}i{O}$_2$
  {N}anostructures},}\ }\href {\doibase 10.1021/jp407275e} {\bibfield
  {journal} {\bibinfo  {journal} {J. Phys. Chem. C}\ }\textbf {\bibinfo
  {volume} {118}}~(\bibinfo {number} {25}),\ \bibinfo {pages}
  {13539--13544}}\BibitemShut {NoStop}%
\bibitem [{\citenamefont {Amadon}\ \emph {et~al.}(2006)\citenamefont {Amadon},
  \citenamefont {Biermann}, \citenamefont {Georges},\ and\ \citenamefont
  {Aryasetiawan}}]{Amadon/etal:2006}%
  \BibitemOpen
  \bibfield  {author} {\bibinfo {author} {\bibnamefont {Amadon}, \bibfnamefont
  {B}}, \bibinfo {author} {\bibfnamefont {S.}~\bibnamefont {Biermann}},
  \bibinfo {author} {\bibfnamefont {A.}~\bibnamefont {Georges}}, \ and\
  \bibinfo {author} {\bibfnamefont {F.}~\bibnamefont {Aryasetiawan}}} (\bibinfo
  {year} {2006}),\ \bibfield  {title} {\enquote {\bibinfo {title} {{T}he
  $\ensuremath{\alpha}\mathrm{\text{\ensuremath{-}}}\ensuremath{\gamma}$
  {T}ransition of {C}erium {I}s {E}ntropy {D}riven},}\ }\href@noop {}
  {\bibfield  {journal} {\bibinfo  {journal} {Phys. Rev. Lett.}\ }\textbf
  {\bibinfo {volume} {96}},\ \bibinfo {pages} {066402}}\BibitemShut {NoStop}%
\bibitem [{\citenamefont {Andersen}(1975)}]{Andersen1975}%
  \BibitemOpen
  \bibfield  {author} {\bibinfo {author} {\bibnamefont {Andersen},
  \bibfnamefont {O~K}}} (\bibinfo {year} {1975}),\ \bibfield  {title} {\enquote
  {\bibinfo {title} {{L}inear methods in band theory},}\ }\href {\doibase
  10.1103/PhysRevB.12.3060} {\bibfield  {journal} {\bibinfo  {journal} {Phys.
  Rev. B}\ }\textbf {\bibinfo {volume} {12}},\ \bibinfo {pages}
  {3060--3083}}\BibitemShut {NoStop}%
\bibitem [{\citenamefont {Antonius}\ \emph {et~al.}(2014)\citenamefont
  {Antonius}, \citenamefont {Ponc\'e}, \citenamefont {Boulanger}, \citenamefont
  {C\^ot\'e},\ and\ \citenamefont {X.}}]{Antonius/etal:2014}%
  \BibitemOpen
  \bibfield  {author} {\bibinfo {author} {\bibnamefont {Antonius},
  \bibfnamefont {G}}, \bibinfo {author} {\bibfnamefont {S.}~\bibnamefont
  {Ponc\'e}}, \bibinfo {author} {\bibfnamefont {P.}~\bibnamefont {Boulanger}},
  \bibinfo {author} {\bibfnamefont {M.}~\bibnamefont {C\^ot\'e}}, \ and\
  \bibinfo {author} {\bibnamefont {X.}}} (\bibinfo {year} {2014}),\ \bibfield
  {title} {\enquote {\bibinfo {title} {{M}any-{B}ody {E}ffects on the
  {Z}ero-{P}oint {R}enormalization of the {B}and {S}tructure},}\ }\href@noop {}
  {\bibfield  {journal} {\bibinfo  {journal} {Phys. Rev. Lett.}\ }\textbf
  {\bibinfo {volume} {112}},\ \bibinfo {pages} {215501}}\BibitemShut {NoStop}%
\bibitem [{\citenamefont {Appelhans}\ \emph
  {et~al.}(2010{\natexlab{a}})\citenamefont {Appelhans}, \citenamefont {Carr},\
  and\ \citenamefont {Lusk}}]{appelhans_njp_12}%
  \BibitemOpen
  \bibfield  {author} {\bibinfo {author} {\bibnamefont {Appelhans},
  \bibfnamefont {D~J}}, \bibinfo {author} {\bibfnamefont {L.~D.}\ \bibnamefont
  {Carr}}, \ and\ \bibinfo {author} {\bibfnamefont {M.~T.}\ \bibnamefont
  {Lusk}}} (\bibinfo {year} {2010}{\natexlab{a}}),\ \bibfield  {title}
  {\enquote {\bibinfo {title} {{E}mbedded ribbons of graphene allotropes: an
  extended defect perspective},}\ }\href {\doibase
  10.1088/1367-2630/12/12/125006} {\bibfield  {journal} {\bibinfo  {journal}
  {New J. Phys.}\ }\textbf {\bibinfo {volume} {12}}~(\bibinfo {number} {12}),\
  \bibinfo {pages} {125006}}\BibitemShut {NoStop}%
\bibitem [{\citenamefont {Appelhans}\ \emph
  {et~al.}(2010{\natexlab{b}})\citenamefont {Appelhans}, \citenamefont {Lin},\
  and\ \citenamefont {Lusk}}]{appelhans_prb_82}%
  \BibitemOpen
  \bibfield  {author} {\bibinfo {author} {\bibnamefont {Appelhans},
  \bibfnamefont {D~J}}, \bibinfo {author} {\bibfnamefont {Z.}~\bibnamefont
  {Lin}}, \ and\ \bibinfo {author} {\bibfnamefont {M.~T.}\ \bibnamefont
  {Lusk}}} (\bibinfo {year} {2010}{\natexlab{b}}),\ \bibfield  {title}
  {\enquote {\bibinfo {title} {{T}wo-dimensional carbon semiconductor:
  {D}ensity functional theory calculations},}\ }\href {\doibase
  10.1103/PhysRevB.82.073410} {\bibfield  {journal} {\bibinfo  {journal} {Phys.
  Rev. B}\ }\textbf {\bibinfo {volume} {82}},\ \bibinfo {pages}
  {073410}}\BibitemShut {NoStop}%
\bibitem [{\citenamefont {Arora}\ \emph {et~al.}(2017)\citenamefont {Arora},
  \citenamefont {Dr\"uppel}, \citenamefont {Schmidt}, \citenamefont {Deilmann},
  \citenamefont {Schneider}, \citenamefont {Molas}, \citenamefont {Marauhn},
  \citenamefont {de~Vasconcellos}, \citenamefont {Potemski}, \citenamefont
  {Rohlfing},\ and\ \citenamefont {Bratschitsch}}]{arora_natcom_8}%
  \BibitemOpen
  \bibfield  {author} {\bibinfo {author} {\bibnamefont {Arora}, \bibfnamefont
  {A}}, \bibinfo {author} {\bibfnamefont {M.}~\bibnamefont {Dr\"uppel}},
  \bibinfo {author} {\bibfnamefont {R.}~\bibnamefont {Schmidt}}, \bibinfo
  {author} {\bibfnamefont {T.}~\bibnamefont {Deilmann}}, \bibinfo {author}
  {\bibfnamefont {R.}~\bibnamefont {Schneider}}, \bibinfo {author}
  {\bibfnamefont {M.~R.}\ \bibnamefont {Molas}}, \bibinfo {author}
  {\bibfnamefont {P.}~\bibnamefont {Marauhn}}, \bibinfo {author} {\bibfnamefont
  {S.~Michaelis}\ \bibnamefont {de~Vasconcellos}}, \bibinfo {author}
  {\bibfnamefont {M.}~\bibnamefont {Potemski}}, \bibinfo {author}
  {\bibfnamefont {M.}~\bibnamefont {Rohlfing}}, \ and\ \bibinfo {author}
  {\bibfnamefont {R.}~\bibnamefont {Bratschitsch}}} (\bibinfo {year} {2017}),\
  \bibfield  {title} {\enquote {\bibinfo {title} {{I}nterlayer excitons in a
  bulk van der {W}aals semiconductor},}\ }\href@noop {} {\bibfield  {journal}
  {\bibinfo  {journal} {Nat. Comm.}\ }\textbf {\bibinfo {volume} {8}},\
  \bibinfo {pages} {639}}\BibitemShut {NoStop}%
\bibitem [{\citenamefont {Aryasetiawan}(1992)}]{Aryasetiawan:1992}%
  \BibitemOpen
  \bibfield  {author} {\bibinfo {author} {\bibnamefont {Aryasetiawan},
  \bibfnamefont {F}}} (\bibinfo {year} {1992}),\ \bibfield  {title} {\enquote
  {\bibinfo {title} {{S}elf-energy of ferromagnetic nickel in the ${G}{W}$
  approximation},}\ }\href {\doibase 10.1103/PhysRevB.46.13051} {\bibfield
  {journal} {\bibinfo  {journal} {Phys. Rev. B}\ }\textbf {\bibinfo {volume}
  {46}}~(\bibinfo {number} {20}),\ \bibinfo {pages} {13051--13064}}\BibitemShut
  {NoStop}%
\bibitem [{\citenamefont {Aryasetiawan}(1997)}]{Aryasetiawan:1997_2}%
  \BibitemOpen
  \bibfield  {author} {\bibinfo {author} {\bibnamefont {Aryasetiawan},
  \bibfnamefont {F}}} (\bibinfo {year} {1997}),\ \bibfield  {title} {\enquote
  {\bibinfo {title} {${G}{W}$ method for 3d and 4f systems},}\ }\href@noop {}
  {\bibfield  {journal} {\bibinfo  {journal} {Physica B}\ }\textbf {\bibinfo
  {volume} {237-238}},\ \bibinfo {pages} {321 -- 323}},\ \bibinfo {note}
  {{P}roceedings of the Yamada Conference XLV, the International Conference on
  the Physics of Transition Metals}\BibitemShut {NoStop}%
\bibitem [{\citenamefont {Aryasetiawan}\ and\ \citenamefont
  {Biermann}(2008)}]{Aryasetiawan/etal:2008}%
  \BibitemOpen
  \bibfield  {author} {\bibinfo {author} {\bibnamefont {Aryasetiawan},
  \bibfnamefont {F}}, \ and\ \bibinfo {author} {\bibfnamefont {S.}~\bibnamefont
  {Biermann}}} (\bibinfo {year} {2008}),\ \bibfield  {title} {\enquote
  {\bibinfo {title} {{G}eneralized {H}edin's {E}quations for {Q}uantum
  {M}any-{B}ody {S}ystems with {S}pin-{D}ependent {I}nteractions},}\ }\href
  {\doibase 10.1103/PhysRevLett.100.116402} {\bibfield  {journal} {\bibinfo
  {journal} {Phys. Rev. Lett.}\ }\textbf {\bibinfo {volume} {100}},\ \bibinfo
  {pages} {116402}}\BibitemShut {NoStop}%
\bibitem [{\citenamefont {Aryasetiawan}\ and\ \citenamefont
  {Biermann}(2009)}]{Aryasetiawan/Biermann:2009}%
  \BibitemOpen
  \bibfield  {author} {\bibinfo {author} {\bibnamefont {Aryasetiawan},
  \bibfnamefont {F}}, \ and\ \bibinfo {author} {\bibfnamefont {S.}~\bibnamefont
  {Biermann}}} (\bibinfo {year} {2009}),\ \bibfield  {title} {\enquote
  {\bibinfo {title} {{G}eneralized {H}edin equations and ${G}{W}$ approximation
  for quantum many-body systems with spin-dependent interactions},}\
  }\href@noop {} {\bibfield  {journal} {\bibinfo  {journal} {J. Phys.: Condens.
  Matter}\ }\textbf {\bibinfo {volume} {21}}~(\bibinfo {number} {6}),\ \bibinfo
  {pages} {064232}}\BibitemShut {NoStop}%
\bibitem [{\citenamefont {Aryasetiawan}\ and\ \citenamefont
  {Gunnarsson}(1995)}]{Aryasetiawan/Gunnarsson:1995}%
  \BibitemOpen
  \bibfield  {author} {\bibinfo {author} {\bibnamefont {Aryasetiawan},
  \bibfnamefont {F}}, \ and\ \bibinfo {author} {\bibfnamefont {O.}~\bibnamefont
  {Gunnarsson}}} (\bibinfo {year} {1995}),\ \bibfield  {title} {\enquote
  {\bibinfo {title} {{E}lectronic {S}tructure of {N}i{O} in the ${G}{W}$
  {A}pproximation},}\ }\href@noop {} {\bibfield  {journal} {\bibinfo  {journal}
  {Phys. Rev. Lett.}\ }\textbf {\bibinfo {volume} {74}},\ \bibinfo {pages}
  {3221}}\BibitemShut {NoStop}%
\bibitem [{\citenamefont {Aryasetiawan}\ and\ \citenamefont
  {Gunnarsson}(1998)}]{Aryasetiawan/Gunnarsson:1998}%
  \BibitemOpen
  \bibfield  {author} {\bibinfo {author} {\bibnamefont {Aryasetiawan},
  \bibfnamefont {F}}, \ and\ \bibinfo {author} {\bibfnamefont {O.}~\bibnamefont
  {Gunnarsson}}} (\bibinfo {year} {1998}),\ \bibfield  {title} {\enquote
  {\bibinfo {title} {{T}he ${G}{W}$ method},}\ }\href@noop {} {\bibfield
  {journal} {\bibinfo  {journal} {Rep. Prog. Phys.}\ }\textbf {\bibinfo
  {volume} {61}},\ \bibinfo {pages} {237}}\BibitemShut {NoStop}%
\bibitem [{\citenamefont {Aryasetiawan}\ \emph {et~al.}(1996)\citenamefont
  {Aryasetiawan}, \citenamefont {Hedin},\ and\ \citenamefont
  {Karlsson}}]{Aryasetiawan/etal:1996}%
  \BibitemOpen
  \bibfield  {author} {\bibinfo {author} {\bibnamefont {Aryasetiawan},
  \bibfnamefont {F}}, \bibinfo {author} {\bibfnamefont {L.}~\bibnamefont
  {Hedin}}, \ and\ \bibinfo {author} {\bibfnamefont {K.}~\bibnamefont
  {Karlsson}}} (\bibinfo {year} {1996}),\ \bibfield  {title} {\enquote
  {\bibinfo {title} {{M}ultiple {P}lasmon {S}atellites in {N}a and {A}l
  {S}pectral {F}unctions from {A}b {I}nitio {C}umulant {E}xpansion},}\
  }\href@noop {} {\bibfield  {journal} {\bibinfo  {journal} {Phys. Rev. Lett.}\
  }\textbf {\bibinfo {volume} {77}},\ \bibinfo {pages}
  {2268--2271}}\BibitemShut {NoStop}%
\bibitem [{\citenamefont {Aryasetiawan}\ and\ \citenamefont
  {Karlsson}(1996)}]{Aryasetiawan/Karlsson:1996}%
  \BibitemOpen
  \bibfield  {author} {\bibinfo {author} {\bibnamefont {Aryasetiawan},
  \bibfnamefont {F}}, \ and\ \bibinfo {author} {\bibfnamefont {K.}~\bibnamefont
  {Karlsson}}} (\bibinfo {year} {1996}),\ \bibfield  {title} {\enquote
  {\bibinfo {title} {${G}{W}$ spectral functions of {G}d and {N}i{O}},}\
  }\href@noop {} {\bibfield  {journal} {\bibinfo  {journal} {Phys. Rev. B}\
  }\textbf {\bibinfo {volume} {54}},\ \bibinfo {pages} {5353}}\BibitemShut
  {NoStop}%
\bibitem [{\citenamefont {Atalla}\ \emph {et~al.}(2013)\citenamefont {Atalla},
  \citenamefont {Yoon}, \citenamefont {Caruso}, \citenamefont {Rinke},\ and\
  \citenamefont {Scheffler}}]{Atalla2013}%
  \BibitemOpen
  \bibfield  {author} {\bibinfo {author} {\bibnamefont {Atalla}, \bibfnamefont
  {V}}, \bibinfo {author} {\bibfnamefont {M.}~\bibnamefont {Yoon}}, \bibinfo
  {author} {\bibfnamefont {F.}~\bibnamefont {Caruso}}, \bibinfo {author}
  {\bibfnamefont {P.}~\bibnamefont {Rinke}}, \ and\ \bibinfo {author}
  {\bibfnamefont {M.}~\bibnamefont {Scheffler}}} (\bibinfo {year} {2013}),\
  \bibfield  {title} {\enquote {\bibinfo {title} {{H}ybrid density functional
  theory meets quasiparticle calculations: {A} consistent electronic structure
  approach},}\ }\href {\doibase 10.1103/PhysRevB.88.165122} {\bibfield
  {journal} {\bibinfo  {journal} {Phys. Rev. B}\ }\textbf {\bibinfo {volume}
  {88}},\ \bibinfo {pages} {165122}}\BibitemShut {NoStop}%
\bibitem [{\citenamefont {Aulbur}\ \emph {et~al.}(2000)\citenamefont {Aulbur},
  \citenamefont {J\"onsson},\ and\ \citenamefont
  {Wilkins}}]{Aulbur/Jonsson/Wilkins:2000}%
  \BibitemOpen
  \bibfield  {author} {\bibinfo {author} {\bibnamefont {Aulbur}, \bibfnamefont
  {W~G}}, \bibinfo {author} {\bibfnamefont {L.}~\bibnamefont {J\"onsson}}, \
  and\ \bibinfo {author} {\bibfnamefont {J.~W.}\ \bibnamefont {Wilkins}}}
  (\bibinfo {year} {2000}),\ \bibfield  {title} {\enquote {\bibinfo {title}
  {{Q}uasiparticle {C}alculations in {S}olids},}\ }in\ \href@noop {} {\emph
  {\bibinfo {booktitle} {{S}olid {S}tate {P}hysics}}},\ Vol.~\bibinfo {volume}
  {54},\ \bibinfo {editor} {edited by\ \bibinfo {editor} {\bibfnamefont
  {Henry}\ \bibnamefont {Ehrenreich}}\ and\ \bibinfo {editor} {\bibfnamefont
  {Frans}\ \bibnamefont {Spaepen}}}\ (\bibinfo  {publisher} {Academic Press})\
  pp.\ \bibinfo {pages} {1 -- 218}\BibitemShut {NoStop}%
\bibitem [{\citenamefont {Bacelar}\ \emph {et~al.}(2002)\citenamefont
  {Bacelar}, \citenamefont {Sch\"one}, \citenamefont {Keyling},\ and\
  \citenamefont {Ekardt}}]{Bacelar/etal:2002}%
  \BibitemOpen
  \bibfield  {author} {\bibinfo {author} {\bibnamefont {Bacelar}, \bibfnamefont
  {M~Rui}}, \bibinfo {author} {\bibfnamefont {W.-D.}\ \bibnamefont {Sch\"one}},
  \bibinfo {author} {\bibfnamefont {R.}~\bibnamefont {Keyling}}, \ and\
  \bibinfo {author} {\bibfnamefont {W.}~\bibnamefont {Ekardt}}} (\bibinfo
  {year} {2002}),\ \bibfield  {title} {\enquote {\bibinfo {title} {{L}ifetime
  of excited electrons in transition metals},}\ }\href@noop {} {\bibfield
  {journal} {\bibinfo  {journal} {Phys. Rev. B}\ }\textbf {\bibinfo {volume}
  {66}},\ \bibinfo {pages} {153101}}\BibitemShut {NoStop}%
\bibitem [{\citenamefont {Bachelet}\ \emph {et~al.}(1982)\citenamefont
  {Bachelet}, \citenamefont {Hamann},\ and\ \citenamefont
  {Schl\"uter}}]{Bachelet1982}%
  \BibitemOpen
  \bibfield  {author} {\bibinfo {author} {\bibnamefont {Bachelet},
  \bibfnamefont {G~B}}, \bibinfo {author} {\bibfnamefont {D.~R.}\ \bibnamefont
  {Hamann}}, \ and\ \bibinfo {author} {\bibfnamefont {M.}~\bibnamefont
  {Schl\"uter}}} (\bibinfo {year} {1982}),\ \bibfield  {title} {\enquote
  {\bibinfo {title} {{P}seudopotentials that work: {F}rom {H} to {P}u},}\
  }\href {\doibase 10.1103/PhysRevB.26.4199} {\bibfield  {journal} {\bibinfo
  {journal} {Phys. Rev. B}\ }\textbf {\bibinfo {volume} {26}},\ \bibinfo
  {pages} {4199--4228}}\BibitemShut {NoStop}%
\bibitem [{\citenamefont {Baerends}\ \emph {et~al.}(2013)\citenamefont
  {Baerends}, \citenamefont {Gritsenko},\ and\ \citenamefont {van
  Meer}}]{Baerends2013}%
  \BibitemOpen
  \bibfield  {author} {\bibinfo {author} {\bibnamefont {Baerends},
  \bibfnamefont {E~J}}, \bibinfo {author} {\bibfnamefont {O.~V.}\ \bibnamefont
  {Gritsenko}}, \ and\ \bibinfo {author} {\bibfnamefont {R.}~\bibnamefont {van
  Meer}}} (\bibinfo {year} {2013}),\ \bibfield  {title} {\enquote {\bibinfo
  {title} {{T}he {K}ohn–{S}ham gap{,} the fundamental gap and the optical
  gap: the physical meaning of occupied and virtual {K}ohn–{S}ham orbital
  energies},}\ }\href {\doibase 10.1039/C3CP52547C} {\bibfield  {journal}
  {\bibinfo  {journal} {Phys. Chem. Chem. Phys.}\ }\textbf {\bibinfo {volume}
  {15}},\ \bibinfo {pages} {16408--16425}}\BibitemShut {NoStop}%
\bibitem [{\citenamefont {Bagus}\ \emph {et~al.}(1999)\citenamefont {Bagus},
  \citenamefont {Illas}, \citenamefont {Pacchioni},\ and\ \citenamefont
  {Parmigiani}}]{Bagus1999}%
  \BibitemOpen
  \bibfield  {author} {\bibinfo {author} {\bibnamefont {Bagus}, \bibfnamefont
  {P~S}}, \bibinfo {author} {\bibfnamefont {F.}~\bibnamefont {Illas}}, \bibinfo
  {author} {\bibfnamefont {G.}~\bibnamefont {Pacchioni}}, \ and\ \bibinfo
  {author} {\bibfnamefont {F.}~\bibnamefont {Parmigiani}}} (\bibinfo {year}
  {1999}),\ \bibfield  {title} {\enquote {\bibinfo {title} {{M}echanisms
  responsible for chemical shifts of core-level binding energies and their
  relationship to chemical bonding},}\ }\href {\doibase
  https://doi.org/10.1016/S0368-2048(99)00048-1} {\bibfield  {journal}
  {\bibinfo  {journal} {J. Electron Spectrosc. Relat. Phenom.}\ }\textbf
  {\bibinfo {volume} {100}}~(\bibinfo {number} {1}),\ \bibinfo {pages}
  {215--236}}\BibitemShut {NoStop}%
\bibitem [{\citenamefont {Bagus}\ \emph {et~al.}(2013)\citenamefont {Bagus},
  \citenamefont {Ilton},\ and\ \citenamefont {Nelin}}]{Bagus2013}%
  \BibitemOpen
  \bibfield  {author} {\bibinfo {author} {\bibnamefont {Bagus}, \bibfnamefont
  {P~S}}, \bibinfo {author} {\bibfnamefont {E.~S.}\ \bibnamefont {Ilton}}, \
  and\ \bibinfo {author} {\bibfnamefont {C.~J.}\ \bibnamefont {Nelin}}}
  (\bibinfo {year} {2013}),\ \bibfield  {title} {\enquote {\bibinfo {title}
  {{T}he interpretation of {X}{P}{S} spectra: {I}nsights into materials
  properties},}\ }\href {\doibase
  https://doi.org/10.1016/j.surfrep.2013.03.001} {\bibfield  {journal}
  {\bibinfo  {journal} {Surf. Sci. Rep.}\ }\textbf {\bibinfo {volume}
  {68}}~(\bibinfo {number} {2}),\ \bibinfo {pages} {273--304}}\BibitemShut
  {NoStop}%
\bibitem [{\citenamefont {Baroni}\ \emph {et~al.}(1987)\citenamefont {Baroni},
  \citenamefont {Giannozzi},\ and\ \citenamefont {Testa}}]{Baroni1987}%
  \BibitemOpen
  \bibfield  {author} {\bibinfo {author} {\bibnamefont {Baroni}, \bibfnamefont
  {S}}, \bibinfo {author} {\bibfnamefont {P.}~\bibnamefont {Giannozzi}}, \ and\
  \bibinfo {author} {\bibfnamefont {A.}~\bibnamefont {Testa}}} (\bibinfo {year}
  {1987}),\ \bibfield  {title} {\enquote {\bibinfo {title} {{G}reen's-function
  approach to linear response in solids},}\ }\href {\doibase
  10.1103/PhysRevLett.58.1861} {\bibfield  {journal} {\bibinfo  {journal}
  {Phys. Rev. Lett.}\ }\textbf {\bibinfo {volume} {58}},\ \bibinfo {pages}
  {1861--1864}}\BibitemShut {NoStop}%
\bibitem [{\citenamefont {Baroni}\ \emph {et~al.}(2001)\citenamefont {Baroni},
  \citenamefont {Gironcoli}, \citenamefont {Corso},\ and\ \citenamefont
  {Giannozzi}}]{Baroni/RevModPhys:2001}%
  \BibitemOpen
  \bibfield  {author} {\bibinfo {author} {\bibnamefont {Baroni}, \bibfnamefont
  {S}}, \bibinfo {author} {\bibfnamefont {S.~D.}\ \bibnamefont {Gironcoli}},
  \bibinfo {author} {\bibfnamefont {A.~D.}\ \bibnamefont {Corso}}, \ and\
  \bibinfo {author} {\bibfnamefont {P.}~\bibnamefont {Giannozzi}}} (\bibinfo
  {year} {2001}),\ \bibfield  {title} {\enquote {\bibinfo {title} {{P}honons
  and related crystal properties from density-functional perturbation
  theory},}\ }\href@noop {} {\bibfield  {journal} {\bibinfo  {journal} {Rev.\
  Mod.\ Phys.}\ }\textbf {\bibinfo {volume} {73}},\ \bibinfo {pages}
  {515}}\BibitemShut {NoStop}%
\bibitem [{\citenamefont {von Barth}\ and\ \citenamefont
  {Holm}(1996)}]{Holm/vonBarth:1996}%
  \BibitemOpen
  \bibfield  {author} {\bibinfo {author} {\bibnamefont {von Barth},
  \bibfnamefont {U}}, \ and\ \bibinfo {author} {\bibfnamefont {B.}~\bibnamefont
  {Holm}}} (\bibinfo {year} {1996}),\ \bibfield  {title} {\enquote {\bibinfo
  {title} {{S}elf-consistent ${G}{W}$ results for the electron gas: {F}ixed
  screened potential ${W}$ within the random-phase approximation},}\ }\href
  {\doibase 10.1103/PhysRevB.54.8411} {\bibfield  {journal} {\bibinfo
  {journal} {Phys. Rev. B}\ }\textbf {\bibinfo {volume} {54}}~(\bibinfo
  {number} {12}),\ \bibinfo {pages} {8411--8419}}\BibitemShut {NoStop}%
\bibitem [{\citenamefont {Bechstedt}(2014)}]{bechstedt_2014}%
  \BibitemOpen
  \bibfield  {author} {\bibinfo {author} {\bibnamefont {Bechstedt},
  \bibfnamefont {F}}} (\bibinfo {year} {2014}),\ \href
  {https://books.google.fi/books?id=\_PyyoAEACAAJ} {\emph {\bibinfo {title}
  {{M}any-{B}ody {A}pproach to {E}lectronic {E}xcitations: {C}oncepts and
  {A}pplications}}},\ Springer Series in Solid-State Sciences\ (\bibinfo
  {publisher} {Springer Berlin Heidelberg})\BibitemShut {NoStop}%
\bibitem [{\citenamefont {Bechstedt}\ and\ \citenamefont
  {Furthm\"uller}(2002)}]{Bechstedt/Furthmueller:2002}%
  \BibitemOpen
  \bibfield  {author} {\bibinfo {author} {\bibnamefont {Bechstedt},
  \bibfnamefont {F}}, \ and\ \bibinfo {author} {\bibfnamefont {J.}~\bibnamefont
  {Furthm\"uller}}} (\bibinfo {year} {2002}),\ \bibfield  {title} {\enquote
  {\bibinfo {title} {{D}o we know the fundamental energy gap of {I}n{N}?}}\
  }\href@noop {} {\bibfield  {journal} {\bibinfo  {journal} {J.\ Cryst.\
  Growth}\ }\textbf {\bibinfo {volume} {246}},\ \bibinfo {pages}
  {315}}\BibitemShut {NoStop}%
\bibitem [{\citenamefont {Becke}(1988)}]{Becke:1988a}%
  \BibitemOpen
  \bibfield  {author} {\bibinfo {author} {\bibnamefont {Becke}, \bibfnamefont
  {A}}} (\bibinfo {year} {1988}),\ \bibfield  {title} {\enquote {\bibinfo
  {title} {{D}ensity-functional exchange-energy approximation with correct
  asymptotic behavior},}\ }\href@noop {} {\bibfield  {journal} {\bibinfo
  {journal} {Phys. Rev. A}\ }\textbf {\bibinfo {volume} {38}},\ \bibinfo
  {pages} {3098}}\BibitemShut {NoStop}%
\bibitem [{\citenamefont {Benedict}\ and\ \citenamefont
  {Shirley}(1999)}]{Benedict/etal:1999}%
  \BibitemOpen
  \bibfield  {author} {\bibinfo {author} {\bibnamefont {Benedict},
  \bibfnamefont {L~X}}, \ and\ \bibinfo {author} {\bibfnamefont {E.~L.}\
  \bibnamefont {Shirley}}} (\bibinfo {year} {1999}),\ \bibfield  {title}
  {\enquote {\bibinfo {title} {{A}b initio calculation of
  ${\ensuremath{\epsilon}}_{2}(\ensuremath{\omega})$ including the
  electron-hole interaction: {A}pplication to {G}a{N} and {C}a{F}$_2$},}\
  }\href@noop {} {\bibfield  {journal} {\bibinfo  {journal} {Phys. Rev. B}\
  }\textbf {\bibinfo {volume} {59}},\ \bibinfo {pages}
  {5441--5451}}\BibitemShut {NoStop}%
\bibitem [{\citenamefont {Benedict}\ \emph
  {et~al.}(1998{\natexlab{a}})\citenamefont {Benedict}, \citenamefont
  {Shirley},\ and\ \citenamefont {Bohn}}]{Benedict/etal:1998b}%
  \BibitemOpen
  \bibfield  {author} {\bibinfo {author} {\bibnamefont {Benedict},
  \bibfnamefont {L~X}}, \bibinfo {author} {\bibfnamefont {E.~L.}\ \bibnamefont
  {Shirley}}, \ and\ \bibinfo {author} {\bibfnamefont {R.~B.}\ \bibnamefont
  {Bohn}}} (\bibinfo {year} {1998}{\natexlab{a}}),\ \bibfield  {title}
  {\enquote {\bibinfo {title} {{O}ptical {A}bsorption of {I}nsulators and the
  {E}lectron-{H}ole {I}nteraction: {A}n {A}b {I}nitio {C}alculation},}\
  }\href@noop {} {\bibfield  {journal} {\bibinfo  {journal} {Phys. Rev. Lett.}\
  }\textbf {\bibinfo {volume} {80}},\ \bibinfo {pages}
  {4514--4517}}\BibitemShut {NoStop}%
\bibitem [{\citenamefont {Benedict}\ \emph
  {et~al.}(1998{\natexlab{b}})\citenamefont {Benedict}, \citenamefont
  {Shirley},\ and\ \citenamefont {Bohn}}]{Benedict/etal:1998}%
  \BibitemOpen
  \bibfield  {author} {\bibinfo {author} {\bibnamefont {Benedict},
  \bibfnamefont {L~X}}, \bibinfo {author} {\bibfnamefont {E.~L.}\ \bibnamefont
  {Shirley}}, \ and\ \bibinfo {author} {\bibfnamefont {R.~B.}\ \bibnamefont
  {Bohn}}} (\bibinfo {year} {1998}{\natexlab{b}}),\ \bibfield  {title}
  {\enquote {\bibinfo {title} {{T}heory of optical absorption in diamond, {S}i,
  {G}e, and {G}a{A}s},}\ }\href@noop {} {\bibfield  {journal} {\bibinfo
  {journal} {Phys. Rev. B}\ }\textbf {\bibinfo {volume} {57}},\ \bibinfo
  {pages} {R9385--R9387}}\BibitemShut {NoStop}%
\bibitem [{\citenamefont {Berger}\ \emph {et~al.}(2010)\citenamefont {Berger},
  \citenamefont {Reining},\ and\ \citenamefont {Sottile}}]{Berger2010}%
  \BibitemOpen
  \bibfield  {author} {\bibinfo {author} {\bibnamefont {Berger}, \bibfnamefont
  {J~A}}, \bibinfo {author} {\bibfnamefont {L.}~\bibnamefont {Reining}}, \ and\
  \bibinfo {author} {\bibfnamefont {F.}~\bibnamefont {Sottile}}} (\bibinfo
  {year} {2010}),\ \bibfield  {title} {\enquote {\bibinfo {title} {{A}b initio
  calculations of electronic excitations: {C}ollapsing spectral sums},}\ }\href
  {\doibase 10.1103/PhysRevB.82.041103} {\bibfield  {journal} {\bibinfo
  {journal} {Phys. Rev. B}\ }\textbf {\bibinfo {volume} {82}},\ \bibinfo
  {pages} {041103}}\BibitemShut {NoStop}%
\bibitem [{\citenamefont {Berger}\ \emph {et~al.}(2014)\citenamefont {Berger},
  \citenamefont {Romaniello}, \citenamefont {Tandetzky}, \citenamefont
  {Mendoza}, \citenamefont {Brouder},\ and\ \citenamefont
  {Reining}}]{Berger_2014}%
  \BibitemOpen
  \bibfield  {author} {\bibinfo {author} {\bibnamefont {Berger}, \bibfnamefont
  {J~A}}, \bibinfo {author} {\bibfnamefont {P.}~\bibnamefont {Romaniello}},
  \bibinfo {author} {\bibfnamefont {F.}~\bibnamefont {Tandetzky}}, \bibinfo
  {author} {\bibfnamefont {B.~S.}\ \bibnamefont {Mendoza}}, \bibinfo {author}
  {\bibfnamefont {C.}~\bibnamefont {Brouder}}, \ and\ \bibinfo {author}
  {\bibfnamefont {L.}~\bibnamefont {Reining}}} (\bibinfo {year} {2014}),\
  \bibfield  {title} {\enquote {\bibinfo {title} {{S}olution to the many-body
  problem in one point},}\ }\href {\doibase 10.1088/1367-2630/16/11/113025}
  {\bibfield  {journal} {\bibinfo  {journal} {New J. Phys.}\ }\textbf {\bibinfo
  {volume} {16}}~(\bibinfo {number} {11}),\ \bibinfo {pages}
  {113025}}\BibitemShut {NoStop}%
\bibitem [{\citenamefont {Bernardi}\ \emph {et~al.}(2012)\citenamefont
  {Bernardi}, \citenamefont {Palummo},\ and\ \citenamefont
  {Grossman}}]{bernardi_prl_108}%
  \BibitemOpen
  \bibfield  {author} {\bibinfo {author} {\bibnamefont {Bernardi},
  \bibfnamefont {M}}, \bibinfo {author} {\bibfnamefont {M.}~\bibnamefont
  {Palummo}}, \ and\ \bibinfo {author} {\bibfnamefont {J.~C.}\ \bibnamefont
  {Grossman}}} (\bibinfo {year} {2012}),\ \bibfield  {title} {\enquote
  {\bibinfo {title} {{O}ptoelectronic {P}roperties in {M}onolayers of
  {H}ybridized {G}raphene and {H}exagonal {B}oron {N}itride},}\ }\href
  {\doibase 10.1103/PhysRevLett.108.226805} {\bibfield  {journal} {\bibinfo
  {journal} {Phys. Rev. Lett.}\ }\textbf {\bibinfo {volume} {108}},\ \bibinfo
  {pages} {226805}}\BibitemShut {NoStop}%
\bibitem [{\citenamefont {Berseneva}\ \emph {et~al.}(2013)\citenamefont
  {Berseneva}, \citenamefont {Gulans}, \citenamefont {Krasheninnikov},\ and\
  \citenamefont {Nieminen}}]{prb_berseneva_87}%
  \BibitemOpen
  \bibfield  {author} {\bibinfo {author} {\bibnamefont {Berseneva},
  \bibfnamefont {N}}, \bibinfo {author} {\bibfnamefont {A.}~\bibnamefont
  {Gulans}}, \bibinfo {author} {\bibfnamefont {A.~V.}\ \bibnamefont
  {Krasheninnikov}}, \ and\ \bibinfo {author} {\bibfnamefont {R.~M.}\
  \bibnamefont {Nieminen}}} (\bibinfo {year} {2013}),\ \bibfield  {title}
  {\enquote {\bibinfo {title} {{E}lectronic structure of boron nitride sheets
  doped with carbon from first-principles calculations},}\ }\href {\doibase
  10.1103/PhysRevB.87.035404} {\bibfield  {journal} {\bibinfo  {journal} {Phys.
  Rev. B}\ }\textbf {\bibinfo {volume} {87}},\ \bibinfo {pages}
  {035404}}\BibitemShut {NoStop}%
\bibitem [{\citenamefont {Betzinger}\ \emph {et~al.}(2013)\citenamefont
  {Betzinger}, \citenamefont {Friedrich},\ and\ \citenamefont
  {Bl\"ugel}}]{Betzinger2013}%
  \BibitemOpen
  \bibfield  {author} {\bibinfo {author} {\bibnamefont {Betzinger},
  \bibfnamefont {M}}, \bibinfo {author} {\bibfnamefont {C.}~\bibnamefont
  {Friedrich}}, \ and\ \bibinfo {author} {\bibfnamefont {S.}~\bibnamefont
  {Bl\"ugel}}} (\bibinfo {year} {2013}),\ \bibfield  {title} {\enquote
  {\bibinfo {title} {{P}recise response functions in all-electron methods:
  {G}eneralization to nonspherical perturbations and application to {N}i{O}},}\
  }\href {\doibase 10.1103/PhysRevB.88.075130} {\bibfield  {journal} {\bibinfo
  {journal} {Phys. Rev. B}\ }\textbf {\bibinfo {volume} {88}},\ \bibinfo
  {pages} {075130}}\BibitemShut {NoStop}%
\bibitem [{\citenamefont {Betzinger}\ \emph {et~al.}(2012)\citenamefont
  {Betzinger}, \citenamefont {Friedrich}, \citenamefont {G\"orling},\ and\
  \citenamefont {Bl\"ugel}}]{Betzinger2012}%
  \BibitemOpen
  \bibfield  {author} {\bibinfo {author} {\bibnamefont {Betzinger},
  \bibfnamefont {M}}, \bibinfo {author} {\bibfnamefont {C.}~\bibnamefont
  {Friedrich}}, \bibinfo {author} {\bibfnamefont {A.}~\bibnamefont
  {G\"orling}}, \ and\ \bibinfo {author} {\bibfnamefont {S.}~\bibnamefont
  {Bl\"ugel}}} (\bibinfo {year} {2012}),\ \bibfield  {title} {\enquote
  {\bibinfo {title} {{P}recise response functions in all-electron methods:
  {A}pplication to the optimized-effective-potential approach},}\ }\href
  {\doibase 10.1103/PhysRevB.85.245124} {\bibfield  {journal} {\bibinfo
  {journal} {Phys. Rev. B}\ }\textbf {\bibinfo {volume} {85}},\ \bibinfo
  {pages} {245124}}\BibitemShut {NoStop}%
\bibitem [{\citenamefont {Betzinger}\ \emph {et~al.}(2015)\citenamefont
  {Betzinger}, \citenamefont {Friedrich}, \citenamefont {G\"orling},\ and\
  \citenamefont {Bl\"ugel}}]{Betzinger2015}%
  \BibitemOpen
  \bibfield  {author} {\bibinfo {author} {\bibnamefont {Betzinger},
  \bibfnamefont {M}}, \bibinfo {author} {\bibfnamefont {C.}~\bibnamefont
  {Friedrich}}, \bibinfo {author} {\bibfnamefont {A.}~\bibnamefont
  {G\"orling}}, \ and\ \bibinfo {author} {\bibfnamefont {S.}~\bibnamefont
  {Bl\"ugel}}} (\bibinfo {year} {2015}),\ \bibfield  {title} {\enquote
  {\bibinfo {title} {{P}recise all-electron dynamical response functions:
  {A}pplication to {C}{O}{H}{S}{E}{X} and the {R}{P}{A} correlation energy},}\
  }\href {\doibase 10.1103/PhysRevB.92.245101} {\bibfield  {journal} {\bibinfo
  {journal} {Phys. Rev. B}\ }\textbf {\bibinfo {volume} {92}},\ \bibinfo
  {pages} {245101}}\BibitemShut {NoStop}%
\bibitem [{\citenamefont {Bhimanapati}\ \emph {et~al.}(2015)\citenamefont
  {Bhimanapati}, \citenamefont {Lin}, \citenamefont {Meunier}, \citenamefont
  {Jung}, \citenamefont {Cha}, \citenamefont {Das}, \citenamefont {Xiao},
  \citenamefont {Son}, \citenamefont {Strano}, \citenamefont {Cooper},
  \citenamefont {Liang}, \citenamefont {Louie}, \citenamefont {Ringe},
  \citenamefont {Zhou}, \citenamefont {Kim}, \citenamefont {Naik},
  \citenamefont {Sumpter}, \citenamefont {Terrones}, \citenamefont {Xia},
  \citenamefont {Wang}, \citenamefont {Zhu}, \citenamefont {Akinwande},
  \citenamefont {Alem}, \citenamefont {Schuller}, \citenamefont {Schaak},
  \citenamefont {Terrones},\ and\ \citenamefont
  {Robinson}}]{bhimanapati_acsn_9}%
  \BibitemOpen
  \bibfield  {author} {\bibinfo {author} {\bibnamefont {Bhimanapati},
  \bibfnamefont {G~R}}, \bibinfo {author} {\bibfnamefont {Z.}~\bibnamefont
  {Lin}}, \bibinfo {author} {\bibfnamefont {V.}~\bibnamefont {Meunier}},
  \bibinfo {author} {\bibfnamefont {Y.}~\bibnamefont {Jung}}, \bibinfo {author}
  {\bibfnamefont {J.}~\bibnamefont {Cha}}, \bibinfo {author} {\bibfnamefont
  {S.}~\bibnamefont {Das}}, \bibinfo {author} {\bibfnamefont {D.}~\bibnamefont
  {Xiao}}, \bibinfo {author} {\bibfnamefont {Y.}~\bibnamefont {Son}}, \bibinfo
  {author} {\bibfnamefont {M.~S.}\ \bibnamefont {Strano}}, \bibinfo {author}
  {\bibfnamefont {V.~R.}\ \bibnamefont {Cooper}}, \bibinfo {author}
  {\bibfnamefont {L.}~\bibnamefont {Liang}}, \bibinfo {author} {\bibfnamefont
  {S.~G.}\ \bibnamefont {Louie}}, \bibinfo {author} {\bibfnamefont
  {E.}~\bibnamefont {Ringe}}, \bibinfo {author} {\bibfnamefont
  {W.}~\bibnamefont {Zhou}}, \bibinfo {author} {\bibfnamefont {S.~S.}\
  \bibnamefont {Kim}}, \bibinfo {author} {\bibfnamefont {R.~R.}\ \bibnamefont
  {Naik}}, \bibinfo {author} {\bibfnamefont {B.~G.}\ \bibnamefont {Sumpter}},
  \bibinfo {author} {\bibfnamefont {H.}~\bibnamefont {Terrones}}, \bibinfo
  {author} {\bibfnamefont {F.}~\bibnamefont {Xia}}, \bibinfo {author}
  {\bibfnamefont {Y.}~\bibnamefont {Wang}}, \bibinfo {author} {\bibfnamefont
  {J.}~\bibnamefont {Zhu}}, \bibinfo {author} {\bibfnamefont {D.}~\bibnamefont
  {Akinwande}}, \bibinfo {author} {\bibfnamefont {N.}~\bibnamefont {Alem}},
  \bibinfo {author} {\bibfnamefont {J.~A.}\ \bibnamefont {Schuller}}, \bibinfo
  {author} {\bibfnamefont {R.~E.}\ \bibnamefont {Schaak}}, \bibinfo {author}
  {\bibfnamefont {M.}~\bibnamefont {Terrones}}, \ and\ \bibinfo {author}
  {\bibfnamefont {J.~A.}\ \bibnamefont {Robinson}}} (\bibinfo {year} {2015}),\
  \bibfield  {title} {\enquote {\bibinfo {title} {{R}ecent {A}dvances in
  {T}wo-{D}imensional {M}aterials beyond {G}raphene},}\ }\href@noop {}
  {\bibfield  {journal} {\bibinfo  {journal} {ACS Nano}\ }\textbf {\bibinfo
  {volume} {9}}~(\bibinfo {number} {12}),\ \bibinfo {pages}
  {11509--11539}}\BibitemShut {NoStop}%
\bibitem [{\citenamefont {Bieder}\ and\ \citenamefont
  {Amadon}(2014)}]{Bieder/Amadon:2014}%
  \BibitemOpen
  \bibfield  {author} {\bibinfo {author} {\bibnamefont {Bieder}, \bibfnamefont
  {J}}, \ and\ \bibinfo {author} {\bibfnamefont {B.}~\bibnamefont {Amadon}}}
  (\bibinfo {year} {2014}),\ \bibfield  {title} {\enquote {\bibinfo {title}
  {{T}hermodynamics of the $\ensuremath{\alpha}$-$\ensuremath{\gamma}$
  transition in cerium from first principles},}\ }\href@noop {} {\bibfield
  {journal} {\bibinfo  {journal} {Phys. Rev. B}\ }\textbf {\bibinfo {volume}
  {89}},\ \bibinfo {pages} {195132}}\BibitemShut {NoStop}%
\bibitem [{\citenamefont {Biermann}(2014)}]{Biermann:2014}%
  \BibitemOpen
  \bibfield  {author} {\bibinfo {author} {\bibnamefont {Biermann},
  \bibfnamefont {S}}} (\bibinfo {year} {2014}),\ \bibfield  {title} {\enquote
  {\bibinfo {title} {{D}ynamical screening effects in correlated electron
  materials—a progress report on combined many-body perturbation and
  dynamical mean field theory: ‘${G}{W}$ + {D}{M}{F}{T}’},}\ }\href
  {http://stacks.iop.org/0953-8984/26/i=17/a=173202} {\bibfield  {journal}
  {\bibinfo  {journal} {J. Phys.: Condens. Matter}\ }\textbf {\bibinfo {volume}
  {26}}~(\bibinfo {number} {17}),\ \bibinfo {pages} {173202}}\BibitemShut
  {NoStop}%
\bibitem [{\citenamefont {Biermann}\ \emph {et~al.}(2003)\citenamefont
  {Biermann}, \citenamefont {Aryasetiawan},\ and\ \citenamefont
  {Georges}}]{Biermann/etal:2003}%
  \BibitemOpen
  \bibfield  {author} {\bibinfo {author} {\bibnamefont {Biermann},
  \bibfnamefont {S}}, \bibinfo {author} {\bibfnamefont {F.}~\bibnamefont
  {Aryasetiawan}}, \ and\ \bibinfo {author} {\bibfnamefont {A.}~\bibnamefont
  {Georges}}} (\bibinfo {year} {2003}),\ \bibfield  {title} {\enquote {\bibinfo
  {title} {{F}irst-{P}rinciples {A}pproach to the {E}lectronic {S}tructure of
  {S}trongly {C}orrelated {S}ystems: {C}ombining the ${G}{W}$ {A}pproximation
  and {D}ynamical {M}ean-{F}ield {T}heory},}\ }\href@noop {} {\bibfield
  {journal} {\bibinfo  {journal} {Phys. Rev. Lett.}\ }\textbf {\bibinfo
  {volume} {90}},\ \bibinfo {pages} {086402}}\BibitemShut {NoStop}%
\bibitem [{\citenamefont {Blase}\ \emph {et~al.}(2011)\citenamefont {Blase},
  \citenamefont {Attaccalite},\ and\ \citenamefont
  {Olevano}}]{Blase/Attaccalite/Olevano:2011}%
  \BibitemOpen
  \bibfield  {author} {\bibinfo {author} {\bibnamefont {Blase}, \bibfnamefont
  {X}}, \bibinfo {author} {\bibfnamefont {C.}~\bibnamefont {Attaccalite}}, \
  and\ \bibinfo {author} {\bibfnamefont {V.}~\bibnamefont {Olevano}}} (\bibinfo
  {year} {2011}),\ \bibfield  {title} {\enquote {\bibinfo {title}
  {{F}irst-principles ${G}{W}$ calculations for fullerenes, porphyrins,
  phtalocyanine, and other molecules of interest for organic photovoltaic
  applications},}\ }\href@noop {} {\bibfield  {journal} {\bibinfo  {journal}
  {Phys. Rev. B}\ }\textbf {\bibinfo {volume} {83}},\ \bibinfo {pages}
  {115103}}\BibitemShut {NoStop}%
\bibitem [{\citenamefont {Blase}\ \emph {et~al.}(1995)\citenamefont {Blase},
  \citenamefont {Rubio}, \citenamefont {Louie},\ and\ \citenamefont
  {Cohen}}]{Blase/etal:1995}%
  \BibitemOpen
  \bibfield  {author} {\bibinfo {author} {\bibnamefont {Blase}, \bibfnamefont
  {X}}, \bibinfo {author} {\bibfnamefont {A.}~\bibnamefont {Rubio}}, \bibinfo
  {author} {\bibfnamefont {S.~G.}\ \bibnamefont {Louie}}, \ and\ \bibinfo
  {author} {\bibfnamefont {M.~L.}\ \bibnamefont {Cohen}}} (\bibinfo {year}
  {1995}),\ \bibfield  {title} {\enquote {\bibinfo {title} {{Q}uasiparticle
  band structure of bulk hexagonal boron nitride and related systems},}\
  }\href@noop {} {\bibfield  {journal} {\bibinfo  {journal} {Phys. Rev. B}\
  }\textbf {\bibinfo {volume} {51}},\ \bibinfo {pages}
  {6868--6875}}\BibitemShut {NoStop}%
\bibitem [{\citenamefont {Bloch}(1929)}]{Bloch1929}%
  \BibitemOpen
  \bibfield  {author} {\bibinfo {author} {\bibnamefont {Bloch}, \bibfnamefont
  {F}}} (\bibinfo {year} {1929}),\ \bibfield  {title} {\enquote {\bibinfo
  {title} {{\"U}ber die {Q}uantenmechanik der {E}lektronen in
  {K}ristallgittern},}\ }\href {\doibase 10.1007/BF01339455} {\bibfield
  {journal} {\bibinfo  {journal} {Z. Phys.}\ }\textbf {\bibinfo {volume}
  {52}}~(\bibinfo {number} {7}),\ \bibinfo {pages} {555--600}}\BibitemShut
  {NoStop}%
\bibitem [{\citenamefont {Bl\"ochl}(1994)}]{Bloechl1994}%
  \BibitemOpen
  \bibfield  {author} {\bibinfo {author} {\bibnamefont {Bl\"ochl},
  \bibfnamefont {P~E}}} (\bibinfo {year} {1994}),\ \bibfield  {title} {\enquote
  {\bibinfo {title} {{P}rojector augmented-wave method},}\ }\href {\doibase
  10.1103/PhysRevB.50.17953} {\bibfield  {journal} {\bibinfo  {journal} {Phys.
  Rev. B}\ }\textbf {\bibinfo {volume} {50}},\ \bibinfo {pages}
  {17953--17979}}\BibitemShut {NoStop}%
\bibitem [{\citenamefont {Blum}\ \emph {et~al.}(2009)\citenamefont {Blum},
  \citenamefont {Gehrke}, \citenamefont {Hanke}, \citenamefont {Havu},
  \citenamefont {Havu}, \citenamefont {Ren}, \citenamefont {Reuter},\ and\
  \citenamefont {Scheffler}}]{Blum2009}%
  \BibitemOpen
  \bibfield  {author} {\bibinfo {author} {\bibnamefont {Blum}, \bibfnamefont
  {V}}, \bibinfo {author} {\bibfnamefont {R.}~\bibnamefont {Gehrke}}, \bibinfo
  {author} {\bibfnamefont {F.}~\bibnamefont {Hanke}}, \bibinfo {author}
  {\bibfnamefont {P.}~\bibnamefont {Havu}}, \bibinfo {author} {\bibfnamefont
  {V.}~\bibnamefont {Havu}}, \bibinfo {author} {\bibfnamefont {X.}~\bibnamefont
  {Ren}}, \bibinfo {author} {\bibfnamefont {K.}~\bibnamefont {Reuter}}, \ and\
  \bibinfo {author} {\bibfnamefont {M.}~\bibnamefont {Scheffler}}} (\bibinfo
  {year} {2009}),\ \bibfield  {title} {\enquote {\bibinfo {title} {{A}b initio
  molecular simulations with numeric atom-centered orbitals},}\ }\href@noop {}
  {\bibfield  {journal} {\bibinfo  {journal} {Comput. Phys. Commun.}\ }\textbf
  {\bibinfo {volume} {180}}~(\bibinfo {number} {11}),\ \bibinfo {pages}
  {2175--2196}}\BibitemShut {NoStop}%
\bibitem [{\citenamefont {Bobbert}\ and\ \citenamefont {van
  Haeringen}(1994)}]{Bobbert/vanHaeringen:1994}%
  \BibitemOpen
  \bibfield  {author} {\bibinfo {author} {\bibnamefont {Bobbert}, \bibfnamefont
  {P~A}}, \ and\ \bibinfo {author} {\bibfnamefont {W.}~\bibnamefont {van
  Haeringen}}} (\bibinfo {year} {1994}),\ \bibfield  {title} {\enquote
  {\bibinfo {title} {{L}owest-order vertex-correction contribution to the
  direct gap of silicon},}\ }\href@noop {} {\bibfield  {journal} {\bibinfo
  {journal} {Phys. Rev. B}\ }\textbf {\bibinfo {volume} {49}},\ \bibinfo
  {pages} {10326--10331}}\BibitemShut {NoStop}%
\bibitem [{\citenamefont {Bockstedte}\ \emph {et~al.}(2010)\citenamefont
  {Bockstedte}, \citenamefont {Marini}, \citenamefont {Pankratov},\ and\
  \citenamefont {Rubio}}]{Bockstedte/etal:2010}%
  \BibitemOpen
  \bibfield  {author} {\bibinfo {author} {\bibnamefont {Bockstedte},
  \bibfnamefont {M}}, \bibinfo {author} {\bibfnamefont {A.}~\bibnamefont
  {Marini}}, \bibinfo {author} {\bibfnamefont {O.}~\bibnamefont {Pankratov}}, \
  and\ \bibinfo {author} {\bibfnamefont {A.}~\bibnamefont {Rubio}}} (\bibinfo
  {year} {2010}),\ \bibfield  {title} {\enquote {\bibinfo {title}
  {{M}any-{B}ody {E}ffects in the {E}xcitation {S}pectrum of a {D}efect in
  {S}i{C}},}\ }\href@noop {} {\bibfield  {journal} {\bibinfo  {journal} {Phys.\
  Rev.\ Lett.}\ }\textbf {\bibinfo {volume} {105}},\ \bibinfo {pages}
  {026401}}\BibitemShut {NoStop}%
\bibitem [{\citenamefont {Bois}\ and\ \citenamefont
  {K\"orzd\"orfer}(2017)}]{Bois2017}%
  \BibitemOpen
  \bibfield  {author} {\bibinfo {author} {\bibnamefont {Bois}, \bibfnamefont
  {J}}, \ and\ \bibinfo {author} {\bibfnamefont {T.}~\bibnamefont
  {K\"orzd\"orfer}}} (\bibinfo {year} {2017}),\ \bibfield  {title} {\enquote
  {\bibinfo {title} {{S}ize-{D}ependence of {N}onempirically {T}uned {D}{F}{T}
  {S}tarting {P}oints for ${G}_0{W}_0$ {A}pplied to $\pi$-{C}onjugated
  {M}olecular {C}hains},}\ }\href@noop {} {\bibfield  {journal} {\bibinfo
  {journal} {J. Chem. Theory Comput.}\ }\textbf {\bibinfo {volume}
  {13}}~(\bibinfo {number} {10}),\ \bibinfo {pages} {4962--4971}}\BibitemShut
  {NoStop}%
\bibitem [{\citenamefont {Botti}\ and\ \citenamefont
  {Marques}(2013)}]{Botti/Marques:2013}%
  \BibitemOpen
  \bibfield  {author} {\bibinfo {author} {\bibnamefont {Botti}, \bibfnamefont
  {S}}, \ and\ \bibinfo {author} {\bibfnamefont {M.~A.~L.}\ \bibnamefont
  {Marques}}} (\bibinfo {year} {2013}),\ \bibfield  {title} {\enquote {\bibinfo
  {title} {{S}trong {R}enormalization of the {E}lectronic {B}and {G}ap due to
  {L}attice {P}olarization in the ${G}{W}$ {F}ormalism},}\ }\href@noop {}
  {\bibfield  {journal} {\bibinfo  {journal} {Phys. Rev. Lett.}\ }\textbf
  {\bibinfo {volume} {110}},\ \bibinfo {pages} {226404}}\BibitemShut {NoStop}%
\bibitem [{\citenamefont {Bouhassoune}\ and\ \citenamefont
  {Schindlmayr}(2010)}]{Bouhassoune/Schindlmayr:2010}%
  \BibitemOpen
  \bibfield  {author} {\bibinfo {author} {\bibnamefont {Bouhassoune},
  \bibfnamefont {M}}, \ and\ \bibinfo {author} {\bibfnamefont {A.}~\bibnamefont
  {Schindlmayr}}} (\bibinfo {year} {2010}),\ \bibfield  {title} {\enquote
  {\bibinfo {title} {{E}lectronic structure and effective masses in strained
  silicon},}\ }\href@noop {} {\bibfield  {journal} {\bibinfo  {journal} {Phys.
  Status Solidi C}\ }\textbf {\bibinfo {volume} {7}}~(\bibinfo {number} {2}),\
  \bibinfo {pages} {460--463}}\BibitemShut {NoStop}%
\bibitem [{\citenamefont {Bredas}(2014)}]{Bredas2014}%
  \BibitemOpen
  \bibfield  {author} {\bibinfo {author} {\bibnamefont {Bredas}, \bibfnamefont
  {J-L}}} (\bibinfo {year} {2014}),\ \bibfield  {title} {\enquote {\bibinfo
  {title} {{M}ind the gap!}}\ }\href {\doibase 10.1039/C3MH00098B} {\bibfield
  {journal} {\bibinfo  {journal} {Mater. Horiz.}\ }\textbf {\bibinfo {volume}
  {1}},\ \bibinfo {pages} {17--19}}\BibitemShut {NoStop}%
\bibitem [{\citenamefont {Brivio}\ \emph {et~al.}(2014)\citenamefont {Brivio},
  \citenamefont {Butler}, \citenamefont {Walsh},\ and\ \citenamefont {van
  Schilfgaarde}}]{Brivio/etal:2014}%
  \BibitemOpen
  \bibfield  {author} {\bibinfo {author} {\bibnamefont {Brivio}, \bibfnamefont
  {F}}, \bibinfo {author} {\bibfnamefont {K.~T.}\ \bibnamefont {Butler}},
  \bibinfo {author} {\bibfnamefont {A.}~\bibnamefont {Walsh}}, \ and\ \bibinfo
  {author} {\bibfnamefont {M.}~\bibnamefont {van Schilfgaarde}}} (\bibinfo
  {year} {2014}),\ \bibfield  {title} {\enquote {\bibinfo {title}
  {{R}elativistic quasiparticle self-consistent electronic structure of hybrid
  halide perovskite photovoltaic absorbers},}\ }\href@noop {} {\bibfield
  {journal} {\bibinfo  {journal} {Phys. Rev. B}\ }\textbf {\bibinfo {volume}
  {89}},\ \bibinfo {pages} {155204}}\BibitemShut {NoStop}%
\bibitem [{\citenamefont {Bruneval}(2009)}]{Bruneval:2009}%
  \BibitemOpen
  \bibfield  {author} {\bibinfo {author} {\bibnamefont {Bruneval},
  \bibfnamefont {F}}} (\bibinfo {year} {2009}),\ \bibfield  {title} {\enquote
  {\bibinfo {title} {${G}{W}$ {A}pproximation of the {M}any-{B}ody {P}roblem
  and {C}hanges in the {P}article {N}umber},}\ }\href@noop {} {\bibfield
  {journal} {\bibinfo  {journal} {Phys.\ Rev.\ Lett.}\ }\textbf {\bibinfo
  {volume} {103}},\ \bibinfo {pages} {176403}}\BibitemShut {NoStop}%
\bibitem [{\citenamefont {Bruneval}(2012)}]{Bruneval:2012}%
  \BibitemOpen
  \bibfield  {author} {\bibinfo {author} {\bibnamefont {Bruneval},
  \bibfnamefont {F}}} (\bibinfo {year} {2012}),\ \bibfield  {title} {\enquote
  {\bibinfo {title} {{I}onization energy of atoms obtained from ${G}{W}$
  self-energy or from random phase approximation total energies},}\ }\href@noop
  {} {\bibfield  {journal} {\bibinfo  {journal} {J. Chem. Phys.}\ }\textbf
  {\bibinfo {volume} {136}},\ \bibinfo {pages} {194107}}\BibitemShut {NoStop}%
\bibitem [{\citenamefont {Bruneval}(2016)}]{Bruneval2016a}%
  \BibitemOpen
  \bibfield  {author} {\bibinfo {author} {\bibnamefont {Bruneval},
  \bibfnamefont {F}}} (\bibinfo {year} {2016}),\ \bibfield  {title} {\enquote
  {\bibinfo {title} {{O}ptimized virtual orbital subspace for faster ${G}{W}$
  calculations in localized basis},}\ }\href@noop {} {\bibfield  {journal}
  {\bibinfo  {journal} {J. Chem. Phys.}\ }\textbf {\bibinfo {volume}
  {145}}~(\bibinfo {number} {23}),\ \bibinfo {pages} {234110}}\BibitemShut
  {NoStop}%
\bibitem [{\citenamefont {Bruneval}\ and\ \citenamefont
  {Gatti}(2014)}]{Bruneval/Gatti:2014}%
  \BibitemOpen
  \bibfield  {author} {\bibinfo {author} {\bibnamefont {Bruneval},
  \bibfnamefont {F}}, \ and\ \bibinfo {author} {\bibfnamefont {M.}~\bibnamefont
  {Gatti}}} (\bibinfo {year} {2014}),\ \bibfield  {title} {\enquote {\bibinfo
  {title} {{Q}uasiparticle {S}elf-{C}onsistent ${G}{W}$ {M}ethod for the
  {S}pectral {P}roperties of {C}omplex {M}aterials},}\ }in\ \href@noop {}
  {\emph {\bibinfo {booktitle} {{F}irst {P}rinciples {A}pproaches to
  {S}pectroscopic {P}roperties of {C}omplex {M}aterials}}},\ \bibinfo {editor}
  {edited by\ \bibinfo {editor} {\bibfnamefont {C.~Di}\ \bibnamefont
  {Valentin}}, \bibinfo {editor} {\bibfnamefont {S.}~\bibnamefont {Botti}}, \
  and\ \bibinfo {editor} {\bibfnamefont {M.}~\bibnamefont {Cococcioni}}}\
  (\bibinfo  {publisher} {Springer Berlin Heidelberg},\ \bibinfo {address}
  {Berlin, Heidelberg})\ pp.\ \bibinfo {pages} {99--135}\BibitemShut {NoStop}%
\bibitem [{\citenamefont {Bruneval}\ and\ \citenamefont
  {Gonze}(2008)}]{Bruneval2008}%
  \BibitemOpen
  \bibfield  {author} {\bibinfo {author} {\bibnamefont {Bruneval},
  \bibfnamefont {F}}, \ and\ \bibinfo {author} {\bibfnamefont {X.}~\bibnamefont
  {Gonze}}} (\bibinfo {year} {2008}),\ \bibfield  {title} {\enquote {\bibinfo
  {title} {{A}ccurate ${G}{W}$ self-energies in a plane-wave basis using only a
  few empty states: {T}owards large systems},}\ }\href {\doibase
  10.1103/PhysRevB.78.085125} {\bibfield  {journal} {\bibinfo  {journal} {Phys.
  Rev. B}\ }\textbf {\bibinfo {volume} {78}},\ \bibinfo {pages}
  {085125}}\BibitemShut {NoStop}%
\bibitem [{\citenamefont {Bruneval}\ \emph {et~al.}(2015)\citenamefont
  {Bruneval}, \citenamefont {Hamed},\ and\ \citenamefont
  {Neaton}}]{bruneval_jcp_142}%
  \BibitemOpen
  \bibfield  {author} {\bibinfo {author} {\bibnamefont {Bruneval},
  \bibfnamefont {F}}, \bibinfo {author} {\bibfnamefont {S.~M.}\ \bibnamefont
  {Hamed}}, \ and\ \bibinfo {author} {\bibfnamefont {J.~B.}\ \bibnamefont
  {Neaton}}} (\bibinfo {year} {2015}),\ \bibfield  {title} {\enquote {\bibinfo
  {title} {{A} systematic benchmark of the ab initio {B}ethe-{S}alpeter
  equation approach for low-lying optical excitations of small organic
  molecules},}\ }\href@noop {} {\bibfield  {journal} {\bibinfo  {journal} {J.
  Chem. Phys.}\ }\textbf {\bibinfo {volume} {142}}~(\bibinfo {number} {24}),\
  \bibinfo {pages} {244101}}\BibitemShut {NoStop}%
\bibitem [{\citenamefont {Bruneval}\ and\ \citenamefont
  {Marques}(2013)}]{Bruneval/Marques:2013}%
  \BibitemOpen
  \bibfield  {author} {\bibinfo {author} {\bibnamefont {Bruneval},
  \bibfnamefont {F}}, \ and\ \bibinfo {author} {\bibfnamefont {M.~A.~L.}\
  \bibnamefont {Marques}}} (\bibinfo {year} {2013}),\ \bibfield  {title}
  {\enquote {\bibinfo {title} {{B}enchmarking the {S}tarting {P}oints of the
  ${G}{W}$ {A}pproximation for {M}olecules},}\ }\href {\doibase
  10.1021/ct300835h} {\bibfield  {journal} {\bibinfo  {journal} {J. Chem.
  Theory Comput.}\ }\textbf {\bibinfo {volume} {9}}~(\bibinfo {number} {1}),\
  \bibinfo {pages} {324--329}}\BibitemShut {NoStop}%
\bibitem [{\citenamefont {Bruneval}\ \emph {et~al.}(2016)\citenamefont
  {Bruneval}, \citenamefont {Rangel}, \citenamefont {Hamed}, \citenamefont
  {Shao}, \citenamefont {Yang},\ and\ \citenamefont {Neaton}}]{Bruneval2016}%
  \BibitemOpen
  \bibfield  {author} {\bibinfo {author} {\bibnamefont {Bruneval},
  \bibfnamefont {F}}, \bibinfo {author} {\bibfnamefont {T.}~\bibnamefont
  {Rangel}}, \bibinfo {author} {\bibfnamefont {S.~M.}\ \bibnamefont {Hamed}},
  \bibinfo {author} {\bibfnamefont {M.}~\bibnamefont {Shao}}, \bibinfo {author}
  {\bibfnamefont {C.}~\bibnamefont {Yang}}, \ and\ \bibinfo {author}
  {\bibfnamefont {J.~B.}\ \bibnamefont {Neaton}}} (\bibinfo {year} {2016}),\
  \bibfield  {title} {\enquote {\bibinfo {title} {{M}{O}{L}{G}{W}1: {M}any-body
  perturbation theory software for atoms, molecules, and clusters},}\ }\href
  {\doibase https://doi.org/10.1016/j.cpc.2016.06.019} {\bibfield  {journal}
  {\bibinfo  {journal} {Comput. Phys. Commun.}\ }\textbf {\bibinfo {volume}
  {208}},\ \bibinfo {pages} {149--161}}\BibitemShut {NoStop}%
\bibitem [{\citenamefont {Bruneval}\ \emph {et~al.}(2005)\citenamefont
  {Bruneval}, \citenamefont {Sottile}, \citenamefont {Olevano}, \citenamefont
  {{Del Sole}},\ and\ \citenamefont {Reining}}]{Bruneval/etal:2005}%
  \BibitemOpen
  \bibfield  {author} {\bibinfo {author} {\bibnamefont {Bruneval},
  \bibfnamefont {F}}, \bibinfo {author} {\bibfnamefont {F.}~\bibnamefont
  {Sottile}}, \bibinfo {author} {\bibfnamefont {V.}~\bibnamefont {Olevano}},
  \bibinfo {author} {\bibfnamefont {R.}~\bibnamefont {{Del Sole}}}, \ and\
  \bibinfo {author} {\bibfnamefont {L.}~\bibnamefont {Reining}}} (\bibinfo
  {year} {2005}),\ \bibfield  {title} {\enquote {\bibinfo {title} {{M}any-body
  perturbation theory using the density-functional concept: beyond the ${G}{W}$
  approximation},}\ }\href@noop {} {\bibfield  {journal} {\bibinfo  {journal}
  {Phys.\ Rev.\ Lett.}\ }\textbf {\bibinfo {volume} {94}},\ \bibinfo {pages}
  {186402}}\BibitemShut {NoStop}%
\bibitem [{\citenamefont {Bruneval}\ \emph {et~al.}(2006)\citenamefont
  {Bruneval}, \citenamefont {Vast}, \citenamefont {Reining}, \citenamefont
  {Izquierdo}, \citenamefont {Sirotti},\ and\ \citenamefont
  {Barrett}}]{Bruneval/etal_2:2006}%
  \BibitemOpen
  \bibfield  {author} {\bibinfo {author} {\bibnamefont {Bruneval},
  \bibfnamefont {F}}, \bibinfo {author} {\bibfnamefont {N.}~\bibnamefont
  {Vast}}, \bibinfo {author} {\bibfnamefont {L.}~\bibnamefont {Reining}},
  \bibinfo {author} {\bibfnamefont {M.}~\bibnamefont {Izquierdo}}, \bibinfo
  {author} {\bibfnamefont {F.}~\bibnamefont {Sirotti}}, \ and\ \bibinfo
  {author} {\bibfnamefont {N.}~\bibnamefont {Barrett}}} (\bibinfo {year}
  {2006}),\ \bibfield  {title} {\enquote {\bibinfo {title} {{E}xchange and
  {C}orrelation {E}ffects in {E}lectronic {E}xcitations of {C}u$_2$0},}\
  }\href@noop {} {\bibfield  {journal} {\bibinfo  {journal} {Phys.\ Rev.\
  Lett.}\ }\textbf {\bibinfo {volume} {97}},\ \bibinfo {pages}
  {267601}}\BibitemShut {NoStop}%
\bibitem [{\citenamefont {Butcher}\ and\ \citenamefont
  {Tansley}(2005)}]{Butcher/Tansley:2005}%
  \BibitemOpen
  \bibfield  {author} {\bibinfo {author} {\bibnamefont {Butcher}, \bibfnamefont
  {K~S~A}}, \ and\ \bibinfo {author} {\bibfnamefont {T.~L.}\ \bibnamefont
  {Tansley}}} (\bibinfo {year} {2005}),\ \bibfield  {title} {\enquote {\bibinfo
  {title} {{I}n{N}, latest development and a review of the band-gap
  controversy},}\ }\href@noop {} {\bibfield  {journal} {\bibinfo  {journal}
  {Superlattices and Microstructures}\ }\textbf {\bibinfo {volume} {38}},\
  \bibinfo {pages} {1}}\BibitemShut {NoStop}%
\bibitem [{\citenamefont {Cannuccia}\ and\ \citenamefont
  {Marini}(2011)}]{Cannuccia/Marini:2011}%
  \BibitemOpen
  \bibfield  {author} {\bibinfo {author} {\bibnamefont {Cannuccia},
  \bibfnamefont {E}}, \ and\ \bibinfo {author} {\bibfnamefont {A.}~\bibnamefont
  {Marini}}} (\bibinfo {year} {2011}),\ \bibfield  {title} {\enquote {\bibinfo
  {title} {{E}ffect of the {Q}uantum {Z}ero-{P}oint {A}tomic {M}otion on the
  {O}ptical and {E}lectronic {P}roperties of {D}iamond and
  {T}rans-{P}olyacetylene},}\ }\href@noop {} {\bibfield  {journal} {\bibinfo
  {journal} {Phys. Rev. Lett.}\ }\textbf {\bibinfo {volume} {107}},\ \bibinfo
  {pages} {255501}}\BibitemShut {NoStop}%
\bibitem [{\citenamefont {Caruso}\ \emph {et~al.}(2014)\citenamefont {Caruso},
  \citenamefont {Atalla}, \citenamefont {Ren}, \citenamefont {Rubio},
  \citenamefont {Scheffler},\ and\ \citenamefont {Rinke}}]{Caruso/etal:2014}%
  \BibitemOpen
  \bibfield  {author} {\bibinfo {author} {\bibnamefont {Caruso}, \bibfnamefont
  {F}}, \bibinfo {author} {\bibfnamefont {V.}~\bibnamefont {Atalla}}, \bibinfo
  {author} {\bibfnamefont {X.}~\bibnamefont {Ren}}, \bibinfo {author}
  {\bibfnamefont {A.}~\bibnamefont {Rubio}}, \bibinfo {author} {\bibfnamefont
  {M.}~\bibnamefont {Scheffler}}, \ and\ \bibinfo {author} {\bibfnamefont
  {P.}~\bibnamefont {Rinke}}} (\bibinfo {year} {2014}),\ \bibfield  {title}
  {\enquote {\bibinfo {title} {{F}irst-principles description of charge
  transfer in donor-acceptor compounds from self-consistent many-body
  perturbation theory},}\ }\href@noop {} {\bibfield  {journal} {\bibinfo
  {journal} {Phys. Rev. B}\ }\textbf {\bibinfo {volume} {90}},\ \bibinfo
  {pages} {085141}}\BibitemShut {NoStop}%
\bibitem [{\citenamefont {Caruso}\ \emph {et~al.}(2016)\citenamefont {Caruso},
  \citenamefont {Dauth}, \citenamefont {van Setten},\ and\ \citenamefont
  {Rinke}}]{Caruso2016}%
  \BibitemOpen
  \bibfield  {author} {\bibinfo {author} {\bibnamefont {Caruso}, \bibfnamefont
  {F}}, \bibinfo {author} {\bibfnamefont {M.}~\bibnamefont {Dauth}}, \bibinfo
  {author} {\bibfnamefont {M.~J.}\ \bibnamefont {van Setten}}, \ and\ \bibinfo
  {author} {\bibfnamefont {P.}~\bibnamefont {Rinke}}} (\bibinfo {year}
  {2016}),\ \bibfield  {title} {\enquote {\bibinfo {title} {{B}enchmark of
  ${G}{W}$ {A}pproaches for the ${G}{W}$100 {T}est {S}et},}\ }\href@noop {}
  {\bibfield  {journal} {\bibinfo  {journal} {J. Chem. Theory Comput.}\
  }\textbf {\bibinfo {volume} {12}}~(\bibinfo {number} {10}),\ \bibinfo {pages}
  {5076--5087}}\BibitemShut {NoStop}%
\bibitem [{\citenamefont {Caruso}\ and\ \citenamefont
  {Giustino}(2015)}]{Caruso/etal:2015}%
  \BibitemOpen
  \bibfield  {author} {\bibinfo {author} {\bibnamefont {Caruso}, \bibfnamefont
  {F}}, \ and\ \bibinfo {author} {\bibfnamefont {F.}~\bibnamefont {Giustino}}}
  (\bibinfo {year} {2015}),\ \bibfield  {title} {\enquote {\bibinfo {title}
  {{S}pectral fingerprints of electron-plasmon coupling},}\ }\href@noop {}
  {\bibfield  {journal} {\bibinfo  {journal} {Phys. Rev. B}\ }\textbf {\bibinfo
  {volume} {92}},\ \bibinfo {pages} {045123}}\BibitemShut {NoStop}%
\bibitem [{\citenamefont {Caruso}\ and\ \citenamefont
  {Giustino}(2016)}]{Caruso/etal:2016}%
  \BibitemOpen
  \bibfield  {author} {\bibinfo {author} {\bibnamefont {Caruso}, \bibfnamefont
  {F}}, \ and\ \bibinfo {author} {\bibfnamefont {F.}~\bibnamefont {Giustino}}}
  (\bibinfo {year} {2016}),\ \bibfield  {title} {\enquote {\bibinfo {title}
  {{T}he ${G}{W}$ plus cumulant method and plasmonic polarons: application to
  the homogeneous electron gas},}\ }\href@noop {} {\bibfield  {journal}
  {\bibinfo  {journal} {Eur. Phys. J. B}\ }\textbf {\bibinfo {volume}
  {89}}~(\bibinfo {number} {11}),\ \bibinfo {pages} {238}}\BibitemShut
  {NoStop}%
\bibitem [{\citenamefont {Caruso}\ \emph {et~al.}(2015)\citenamefont {Caruso},
  \citenamefont {Lambert},\ and\ \citenamefont {Giustino}}]{Caruso/etal:2015b}%
  \BibitemOpen
  \bibfield  {author} {\bibinfo {author} {\bibnamefont {Caruso}, \bibfnamefont
  {F}}, \bibinfo {author} {\bibfnamefont {H.}~\bibnamefont {Lambert}}, \ and\
  \bibinfo {author} {\bibfnamefont {F.}~\bibnamefont {Giustino}}} (\bibinfo
  {year} {2015}),\ \bibfield  {title} {\enquote {\bibinfo {title} {{B}and
  {S}tructures of {P}lasmonic {P}olarons},}\ }\href@noop {} {\bibfield
  {journal} {\bibinfo  {journal} {Phys. Rev. Lett.}\ }\textbf {\bibinfo
  {volume} {114}},\ \bibinfo {pages} {146404}}\BibitemShut {NoStop}%
\bibitem [{\citenamefont {Caruso}\ \emph
  {et~al.}(2013{\natexlab{a}})\citenamefont {Caruso}, \citenamefont {Rinke},
  \citenamefont {Ren}, \citenamefont {Rubio},\ and\ \citenamefont
  {Scheffler}}]{Caruso/etal:2013_tech}%
  \BibitemOpen
  \bibfield  {author} {\bibinfo {author} {\bibnamefont {Caruso}, \bibfnamefont
  {F}}, \bibinfo {author} {\bibfnamefont {P.}~\bibnamefont {Rinke}}, \bibinfo
  {author} {\bibfnamefont {X.}~\bibnamefont {Ren}}, \bibinfo {author}
  {\bibfnamefont {A.}~\bibnamefont {Rubio}}, \ and\ \bibinfo {author}
  {\bibfnamefont {M.}~\bibnamefont {Scheffler}}} (\bibinfo {year}
  {2013}{\natexlab{a}}),\ \bibfield  {title} {\enquote {\bibinfo {title}
  {{S}elf-consistent ${G}{W}$: {A}ll-electron implementation with localized
  basis functions},}\ }\href@noop {} {\bibfield  {journal} {\bibinfo  {journal}
  {Phys. Rev. B}\ }\textbf {\bibinfo {volume} {88}},\ \bibinfo {pages}
  {075105}}\BibitemShut {NoStop}%
\bibitem [{\citenamefont {Caruso}\ \emph
  {et~al.}(2012{\natexlab{a}})\citenamefont {Caruso}, \citenamefont {Rinke},
  \citenamefont {Ren}, \citenamefont {Scheffler},\ and\ \citenamefont
  {Rubio}}]{Caruso/etal:2012}%
  \BibitemOpen
  \bibfield  {author} {\bibinfo {author} {\bibnamefont {Caruso}, \bibfnamefont
  {F}}, \bibinfo {author} {\bibfnamefont {P.}~\bibnamefont {Rinke}}, \bibinfo
  {author} {\bibfnamefont {X.}~\bibnamefont {Ren}}, \bibinfo {author}
  {\bibfnamefont {M.}~\bibnamefont {Scheffler}}, \ and\ \bibinfo {author}
  {\bibfnamefont {A.}~\bibnamefont {Rubio}}} (\bibinfo {year}
  {2012}{\natexlab{a}}),\ \bibfield  {title} {\enquote {\bibinfo {title}
  {{U}nified description of ground and excited states of finite systems: {T}he
  self-consistent ${G}{W}$ approach},}\ }\href@noop {} {\bibfield  {journal}
  {\bibinfo  {journal} {Phys. Rev. B}\ }\textbf {\bibinfo {volume} {86}},\
  \bibinfo {pages} {081102(R)}}\BibitemShut {NoStop}%
\bibitem [{\citenamefont {Caruso}\ \emph
  {et~al.}(2012{\natexlab{b}})\citenamefont {Caruso}, \citenamefont {Rinke},
  \citenamefont {Ren}, \citenamefont {Scheffler},\ and\ \citenamefont
  {Rubio}}]{Caruso2012COdip}%
  \BibitemOpen
  \bibfield  {author} {\bibinfo {author} {\bibnamefont {Caruso}, \bibfnamefont
  {F}}, \bibinfo {author} {\bibfnamefont {P.}~\bibnamefont {Rinke}}, \bibinfo
  {author} {\bibfnamefont {X.}~\bibnamefont {Ren}}, \bibinfo {author}
  {\bibfnamefont {M.}~\bibnamefont {Scheffler}}, \ and\ \bibinfo {author}
  {\bibfnamefont {A.}~\bibnamefont {Rubio}}} (\bibinfo {year}
  {2012}{\natexlab{b}}),\ \href@noop {} {\enquote {\bibinfo {title}
  {Unpublished result from calculations for {R}ef. {P}{R}{B} 86, 081102},}\
  }\bibinfo {note} {CCSD dipole moment calculated at the aug-cc-pVTZ level
  using the settings described in SI of {P}{R}{B} 86, 081102}\BibitemShut
  {NoStop}%
\bibitem [{\citenamefont {Caruso}\ \emph
  {et~al.}(2013{\natexlab{b}})\citenamefont {Caruso}, \citenamefont {Rohr},
  \citenamefont {Hellgren}, \citenamefont {Ren}, \citenamefont {Rinke},
  \citenamefont {Rubio},\ and\ \citenamefont
  {Scheffler}}]{Caruso/etal:2013_H2}%
  \BibitemOpen
  \bibfield  {author} {\bibinfo {author} {\bibnamefont {Caruso}, \bibfnamefont
  {F}}, \bibinfo {author} {\bibfnamefont {D.~R.}\ \bibnamefont {Rohr}},
  \bibinfo {author} {\bibfnamefont {M.}~\bibnamefont {Hellgren}}, \bibinfo
  {author} {\bibfnamefont {X.}~\bibnamefont {Ren}}, \bibinfo {author}
  {\bibfnamefont {P.}~\bibnamefont {Rinke}}, \bibinfo {author} {\bibfnamefont
  {A.}~\bibnamefont {Rubio}}, \ and\ \bibinfo {author} {\bibfnamefont
  {M.}~\bibnamefont {Scheffler}}} (\bibinfo {year} {2013}{\natexlab{b}}),\
  \bibfield  {title} {\enquote {\bibinfo {title} {{B}ond {B}reaking and {B}ond
  {F}ormation: {H}ow {E}lectron {C}orrelation is {C}aptured in {M}any-{B}ody
  {P}erturbation {T}heory and {D}ensity-{F}unctional {T}heory},}\ }\href@noop
  {} {\bibfield  {journal} {\bibinfo  {journal} {Phys. Rev. Lett.}\ }\textbf
  {\bibinfo {volume} {110}},\ \bibinfo {pages} {146403}}\BibitemShut {NoStop}%
\bibitem [{\citenamefont {Casadei}\ \emph {et~al.}(2012)\citenamefont
  {Casadei}, \citenamefont {Ren}, \citenamefont {Rinke}, \citenamefont
  {Rubio},\ and\ \citenamefont {Scheffler}}]{Casadei/etal:2012}%
  \BibitemOpen
  \bibfield  {author} {\bibinfo {author} {\bibnamefont {Casadei}, \bibfnamefont
  {M}}, \bibinfo {author} {\bibfnamefont {X.}~\bibnamefont {Ren}}, \bibinfo
  {author} {\bibfnamefont {P.}~\bibnamefont {Rinke}}, \bibinfo {author}
  {\bibfnamefont {A.}~\bibnamefont {Rubio}}, \ and\ \bibinfo {author}
  {\bibfnamefont {M.}~\bibnamefont {Scheffler}}} (\bibinfo {year} {2012}),\
  \bibfield  {title} {\enquote {\bibinfo {title} {{D}ensity-{F}unctional
  {T}heory for $f$-{E}lectron {S}ystems: {T}he $\alpha$-$\gamma$ {P}hase
  {T}ransition in {C}erium},}\ }\href@noop {} {\bibfield  {journal} {\bibinfo
  {journal} {Phys. Rev. Lett.}\ }\textbf {\bibinfo {volume} {109}},\ \bibinfo
  {pages} {146402}}\BibitemShut {NoStop}%
\bibitem [{\citenamefont {Casadei}\ \emph {et~al.}(2016)\citenamefont
  {Casadei}, \citenamefont {Ren}, \citenamefont {Rinke}, \citenamefont
  {Rubio},\ and\ \citenamefont {Scheffler}}]{Casadei/etal:2016}%
  \BibitemOpen
  \bibfield  {author} {\bibinfo {author} {\bibnamefont {Casadei}, \bibfnamefont
  {M}}, \bibinfo {author} {\bibfnamefont {X.}~\bibnamefont {Ren}}, \bibinfo
  {author} {\bibfnamefont {P.}~\bibnamefont {Rinke}}, \bibinfo {author}
  {\bibfnamefont {A.}~\bibnamefont {Rubio}}, \ and\ \bibinfo {author}
  {\bibfnamefont {M.}~\bibnamefont {Scheffler}}} (\bibinfo {year} {2016}),\
  \bibfield  {title} {\enquote {\bibinfo {title} {{D}ensity functional theory
  study of the $\ensuremath{\alpha}\text{-}\ensuremath{\gamma}$ phase
  transition in cerium: {R}ole of electron correlation and $f$-orbital
  localization},}\ }\href@noop {} {\bibfield  {journal} {\bibinfo  {journal}
  {Phys. Rev. B}\ }\textbf {\bibinfo {volume} {93}},\ \bibinfo {pages}
  {075153}}\BibitemShut {NoStop}%
\bibitem [{\citenamefont {Casida}(1995{\natexlab{a}})}]{Casida:1995}%
  \BibitemOpen
  \bibfield  {author} {\bibinfo {author} {\bibnamefont {Casida}, \bibfnamefont
  {M~E}}} (\bibinfo {year} {1995}{\natexlab{a}}),\ \bibfield  {title} {\enquote
  {\bibinfo {title} {{G}eneralization of the optimized-effective-potential
  model to include electron correlation: {A} variational derivation of the
  {S}ham-{S}chl\"uter equation for the exact exchange-correlation potential},}\
  }\href@noop {} {\bibfield  {journal} {\bibinfo  {journal} {Phys.\ Rev.\ A}\
  }\textbf {\bibinfo {volume} {51}},\ \bibinfo {pages} {2005}}\BibitemShut
  {NoStop}%
\bibitem [{\citenamefont {Casida}(1995{\natexlab{b}})}]{Casida1995}%
  \BibitemOpen
  \bibfield  {author} {\bibinfo {author} {\bibnamefont {Casida}, \bibfnamefont
  {M~E}}} (\bibinfo {year} {1995}{\natexlab{b}}),\ \enquote {\bibinfo {title}
  {{R}ecent advances in {D}ensity {F}unctional {T}heory},}\ in\ \href {\doibase
  10.1142/9789812830586_0005} {\emph {\bibinfo {booktitle} {{R}ecent {A}dvances
  in {D}ensity {F}unctional {M}ethods}}},\ \bibinfo {editor} {edited by\
  \bibinfo {editor} {\bibfnamefont {Delano~P.}\ \bibnamefont {Chong}}},\ Chap.\
  \bibinfo {chapter} {Time-Dependent Density Functional Response Theory for
  Molecules}\ (\bibinfo  {publisher} {World Scientific: Singapore})\ pp.\
  \bibinfo {pages} {155--192}\BibitemShut {NoStop}%
\bibitem [{\citenamefont {Cazzaniga}(2012)}]{Cazzaniga:2012}%
  \BibitemOpen
  \bibfield  {author} {\bibinfo {author} {\bibnamefont {Cazzaniga},
  \bibfnamefont {M}}} (\bibinfo {year} {2012}),\ \bibfield  {title} {\enquote
  {\bibinfo {title} {${G}{W}$ and beyond approaches to quasiparticle properties
  in metals},}\ }\href@noop {} {\bibfield  {journal} {\bibinfo  {journal}
  {Phys. Rev. B}\ }\textbf {\bibinfo {volume} {86}},\ \bibinfo {pages}
  {035120}}\BibitemShut {NoStop}%
\bibitem [{\citenamefont {Cederbaum}\ and\ \citenamefont
  {Domcke}(1974)}]{Cederbaum/Domcke:1974}%
  \BibitemOpen
  \bibfield  {author} {\bibinfo {author} {\bibnamefont {Cederbaum},
  \bibfnamefont {L~S}}, \ and\ \bibinfo {author} {\bibfnamefont
  {W.}~\bibnamefont {Domcke}}} (\bibinfo {year} {1974}),\ \bibfield  {title}
  {\enquote {\bibinfo {title} {{O}n the vibrational structure in photoelectron
  spectra by the method of {G}reen's functions},}\ }\href {\doibase
  10.1063/1.1681457} {\bibfield  {journal} {\bibinfo  {journal} {J. Chem.
  Phys.}\ }\textbf {\bibinfo {volume} {60}}~(\bibinfo {number} {7}),\ \bibinfo
  {pages} {2878--2889}}\BibitemShut {NoStop}%
\bibitem [{\citenamefont {Cederbaum}\ and\ \citenamefont
  {Domcke}(2007)}]{Cederbaum/Domcke:1977}%
  \BibitemOpen
  \bibfield  {author} {\bibinfo {author} {\bibnamefont {Cederbaum},
  \bibfnamefont {L~S}}, \ and\ \bibinfo {author} {\bibfnamefont
  {W.}~\bibnamefont {Domcke}}} (\bibinfo {year} {2007}),\ \bibfield  {title}
  {\enquote {\bibinfo {title} {{T}heoretical {A}spects of {I}onization
  {P}otentials and {P}hotoelectron {S}pectroscopy: {A} {G}reen's {F}unction
  {A}pproach},}\ }in\ \href@noop {} {\emph {\bibinfo {booktitle} {{A}dvances in
  {C}hemical {P}hysics}}},\ \bibinfo {editor} {edited by\ \bibinfo {editor}
  {\bibfnamefont {I.}~\bibnamefont {Prigogine}}\ and\ \bibinfo {editor}
  {\bibfnamefont {S.~A.}\ \bibnamefont {Rice}}}\ (\bibinfo  {publisher} {John
  Wiley \& Sons, Ltd})\ pp.\ \bibinfo {pages} {205--344}\BibitemShut {NoStop}%
\bibitem [{\citenamefont {Chantis}\ \emph {et~al.}(2007)\citenamefont
  {Chantis}, \citenamefont {{van Schilfgaarde}},\ and\ \citenamefont
  {Kotani}}]{Chantis/etal:2007}%
  \BibitemOpen
  \bibfield  {author} {\bibinfo {author} {\bibnamefont {Chantis}, \bibfnamefont
  {A~N}}, \bibinfo {author} {\bibfnamefont {M.}~\bibnamefont {{van
  Schilfgaarde}}}, \ and\ \bibinfo {author} {\bibfnamefont {T.}~\bibnamefont
  {Kotani}}} (\bibinfo {year} {2007}),\ \bibfield  {title} {\enquote {\bibinfo
  {title} {{Q}uasiparticle self-consistent ${G}{W}$ method applied to localized
  4f electron systems},}\ }\href@noop {} {\bibfield  {journal} {\bibinfo
  {journal} {Phys. Rev. B}\ }\textbf {\bibinfo {volume} {76}},\ \bibinfo
  {pages} {165126}}\BibitemShut {NoStop}%
\bibitem [{\citenamefont {Cheiwchanchamnangij}\ and\ \citenamefont
  {Lambrecht}(2011)}]{Cheiwchanchamnangij/Lambrecht:2011}%
  \BibitemOpen
  \bibfield  {author} {\bibinfo {author} {\bibnamefont {Cheiwchanchamnangij},
  \bibfnamefont {T}}, \ and\ \bibinfo {author} {\bibfnamefont {W.~R.~L.}\
  \bibnamefont {Lambrecht}}} (\bibinfo {year} {2011}),\ \bibfield  {title}
  {\enquote {\bibinfo {title} {{B}and structure parameters of wurtzite and
  zinc-blende {G}a{A}s under strain in the ${G}{W}$ approximation},}\
  }\href@noop {} {\bibfield  {journal} {\bibinfo  {journal} {Phys. Rev. B}\
  }\textbf {\bibinfo {volume} {84}},\ \bibinfo {pages} {035203}}\BibitemShut
  {NoStop}%
\bibitem [{\citenamefont {Cheiwchanchamnangij}\ and\ \citenamefont
  {Lambrecht}(2012)}]{cheiwchanchamnangij_prb_85}%
  \BibitemOpen
  \bibfield  {author} {\bibinfo {author} {\bibnamefont {Cheiwchanchamnangij},
  \bibfnamefont {T}}, \ and\ \bibinfo {author} {\bibfnamefont {W.~R.~L.}\
  \bibnamefont {Lambrecht}}} (\bibinfo {year} {2012}),\ \bibfield  {title}
  {\enquote {\bibinfo {title} {{Q}uasiparticle band structure calculation of
  monolayer, bilayer, and bulk {M}o{S}${}_{2}$},}\ }\href@noop {} {\bibfield
  {journal} {\bibinfo  {journal} {Phys. Rev. B}\ }\textbf {\bibinfo {volume}
  {85}},\ \bibinfo {pages} {205302}}\BibitemShut {NoStop}%
\bibitem [{\citenamefont {Chen}\ and\ \citenamefont
  {Pasquarello}(2015{\natexlab{a}})}]{Chen/etal:2015}%
  \BibitemOpen
  \bibfield  {author} {\bibinfo {author} {\bibnamefont {Chen}, \bibfnamefont
  {W}}, \ and\ \bibinfo {author} {\bibfnamefont {A.}~\bibnamefont
  {Pasquarello}}} (\bibinfo {year} {2015}{\natexlab{a}}),\ \bibfield  {title}
  {\enquote {\bibinfo {title} {{A}ccurate band gaps of extended systems via
  efficient vertex corrections in ${G}{W}$},}\ }\href {\doibase
  10.1103/PhysRevB.92.041115} {\bibfield  {journal} {\bibinfo  {journal} {Phys.
  Rev. B}\ }\textbf {\bibinfo {volume} {92}},\ \bibinfo {pages}
  {041115}}\BibitemShut {NoStop}%
\bibitem [{\citenamefont {Chen}\ and\ \citenamefont
  {Pasquarello}(2015{\natexlab{b}})}]{Chen/Pasquarello:2015}%
  \BibitemOpen
  \bibfield  {author} {\bibinfo {author} {\bibnamefont {Chen}, \bibfnamefont
  {W}}, \ and\ \bibinfo {author} {\bibfnamefont {A.}~\bibnamefont
  {Pasquarello}}} (\bibinfo {year} {2015}{\natexlab{b}}),\ \bibfield  {title}
  {\enquote {\bibinfo {title} {{F}irst-principles determination of defect
  energy levels through hybrid density functionals and ${G}{W}$},}\ }\href
  {http://stacks.iop.org/0953-8984/27/i=13/a=133202} {\bibfield  {journal}
  {\bibinfo  {journal} {J. Phys.: Condens. Matter}\ }\textbf {\bibinfo {volume}
  {27}}~(\bibinfo {number} {13}),\ \bibinfo {pages} {133202}}\BibitemShut
  {NoStop}%
\bibitem [{\citenamefont {Choi}\ \emph {et~al.}(2016)\citenamefont {Choi},
  \citenamefont {Kutepov}, \citenamefont {Haule}, \citenamefont {van
  Schilfgaarde},\ and\ \citenamefont {Kotliar}}]{Choi/etal:2016}%
  \BibitemOpen
  \bibfield  {author} {\bibinfo {author} {\bibnamefont {Choi}, \bibfnamefont
  {S}}, \bibinfo {author} {\bibfnamefont {A.}~\bibnamefont {Kutepov}}, \bibinfo
  {author} {\bibfnamefont {K.}~\bibnamefont {Haule}}, \bibinfo {author}
  {\bibfnamefont {M.}~\bibnamefont {van Schilfgaarde}}, \ and\ \bibinfo
  {author} {\bibfnamefont {G.}~\bibnamefont {Kotliar}}} (\bibinfo {year}
  {2016}),\ \bibfield  {title} {\enquote {\bibinfo {title} {{F}irst-principles
  treatment of {M}ott insulators: linearized {Q}{S}${G}{W}$+{D}{M}{F}{T}
  approach},}\ }\href@noop {} {\bibfield  {journal} {\bibinfo  {journal} {NPJ
  Quantum Mat.}\ }\textbf {\bibinfo {volume} {1}},\ \bibinfo {pages}
  {16001}}\BibitemShut {NoStop}%
\bibitem [{\citenamefont {Cocchi}\ \emph {et~al.}(2018)\citenamefont {Cocchi},
  \citenamefont {Breuer}, \citenamefont {Witte},\ and\ \citenamefont
  {Draxl}}]{Cocchi2018}%
  \BibitemOpen
  \bibfield  {author} {\bibinfo {author} {\bibnamefont {Cocchi}, \bibfnamefont
  {C}}, \bibinfo {author} {\bibfnamefont {T.}~\bibnamefont {Breuer}}, \bibinfo
  {author} {\bibfnamefont {G.}~\bibnamefont {Witte}}, \ and\ \bibinfo {author}
  {\bibfnamefont {C.}~\bibnamefont {Draxl}}} (\bibinfo {year} {2018}),\
  \bibfield  {title} {\enquote {\bibinfo {title} {{P}olarized absorbance and
  {D}avydov splitting in bulk and thin-film pentacene polymorphs},}\ }\href
  {\doibase 10.1039/C8CP06384B} {\bibfield  {journal} {\bibinfo  {journal}
  {Phys. Chem. Chem. Phys.}\ }\textbf {\bibinfo {volume} {20}},\ \bibinfo
  {pages} {29724--29736}}\BibitemShut {NoStop}%
\bibitem [{\citenamefont {\c{S}ahin}\ \emph {et~al.}(2009)\citenamefont
  {\c{S}ahin}, \citenamefont {Cahangirov}, \citenamefont {Topsakal},
  \citenamefont {Bekaroglu}, \citenamefont {Akturk}, \citenamefont {Senger},\
  and\ \citenamefont {Ciraci}}]{ciraci_prb_80}%
  \BibitemOpen
  \bibfield  {author} {\bibinfo {author} {\bibnamefont {\c{S}ahin},
  \bibfnamefont {H}}, \bibinfo {author} {\bibfnamefont {S.}~\bibnamefont
  {Cahangirov}}, \bibinfo {author} {\bibfnamefont {M.}~\bibnamefont
  {Topsakal}}, \bibinfo {author} {\bibfnamefont {E.}~\bibnamefont {Bekaroglu}},
  \bibinfo {author} {\bibfnamefont {E.}~\bibnamefont {Akturk}}, \bibinfo
  {author} {\bibfnamefont {R.~T.}\ \bibnamefont {Senger}}, \ and\ \bibinfo
  {author} {\bibfnamefont {S.}~\bibnamefont {Ciraci}}} (\bibinfo {year}
  {2009}),\ \bibfield  {title} {\enquote {\bibinfo {title} {{M}onolayer
  honeycomb structures of group-{I}{V} elements and {I}{I}{I}-{V} binary
  compounds: {F}irst-principles calculations},}\ }\href {\doibase
  10.1103/PhysRevB.80.155453} {\bibfield  {journal} {\bibinfo  {journal} {Phys.
  Rev. B}\ }\textbf {\bibinfo {volume} {80}},\ \bibinfo {pages}
  {155453}}\BibitemShut {NoStop}%
\bibitem [{\citenamefont {Dahlen}\ \emph
  {et~al.}(2006{\natexlab{a}})\citenamefont {Dahlen}, \citenamefont {van
  Leeuwen},\ and\ \citenamefont {von Barth}}]{Dahlen/Leeuwen/vonBarth:2006}%
  \BibitemOpen
  \bibfield  {author} {\bibinfo {author} {\bibnamefont {Dahlen}, \bibfnamefont
  {N~E}}, \bibinfo {author} {\bibfnamefont {R.}~\bibnamefont {van Leeuwen}}, \
  and\ \bibinfo {author} {\bibfnamefont {U.}~\bibnamefont {von Barth}}}
  (\bibinfo {year} {2006}{\natexlab{a}}),\ \bibfield  {title} {\enquote
  {\bibinfo {title} {{V}ariational energy functionals of the {G}reen function
  and of the density tested on molecules},}\ }\href {\doibase
  10.1103/PhysRevA.73.012511} {\bibfield  {journal} {\bibinfo  {journal} {Phys.
  Rev. A}\ }\textbf {\bibinfo {volume} {73}}~(\bibinfo {number} {1}),\ \bibinfo
  {pages} {012511}}\BibitemShut {NoStop}%
\bibitem [{\citenamefont {Dahlen}\ \emph
  {et~al.}(2006{\natexlab{b}})\citenamefont {Dahlen}, \citenamefont {Stan},\
  and\ \citenamefont {Leeuwen}}]{Dahlen/Stan/Leeuwen:2006}%
  \BibitemOpen
  \bibfield  {author} {\bibinfo {author} {\bibnamefont {Dahlen}, \bibfnamefont
  {N~E}}, \bibinfo {author} {\bibfnamefont {A.}~\bibnamefont {Stan}}, \ and\
  \bibinfo {author} {\bibfnamefont {R.}~\bibnamefont {Leeuwen}}} (\bibinfo
  {year} {2006}{\natexlab{b}}),\ \bibfield  {title} {\enquote {\bibinfo {title}
  {{N}onequilibrium {G}reen function theory for excitation and transport in
  atoms and molecules},}\ }\href@noop {} {\bibfield  {journal} {\bibinfo
  {journal} {J. Phys. Conf. Ser.}\ }\textbf {\bibinfo {volume} {35}},\ \bibinfo
  {pages} {324--339}}\BibitemShut {NoStop}%
\bibitem [{\citenamefont {Dauth}\ \emph {et~al.}(2016)\citenamefont {Dauth},
  \citenamefont {Caruso}, \citenamefont {K\"ummel},\ and\ \citenamefont
  {Rinke}}]{Dauth2016}%
  \BibitemOpen
  \bibfield  {author} {\bibinfo {author} {\bibnamefont {Dauth}, \bibfnamefont
  {M}}, \bibinfo {author} {\bibfnamefont {F.}~\bibnamefont {Caruso}}, \bibinfo
  {author} {\bibfnamefont {S.}~\bibnamefont {K\"ummel}}, \ and\ \bibinfo
  {author} {\bibfnamefont {P.}~\bibnamefont {Rinke}}} (\bibinfo {year}
  {2016}),\ \bibfield  {title} {\enquote {\bibinfo {title} {{P}iecewise
  linearity in the ${G}{W}$ approximation for accurate quasiparticle energy
  predictions},}\ }\href {\doibase 10.1103/PhysRevB.93.121115} {\bibfield
  {journal} {\bibinfo  {journal} {Phys. Rev. B}\ }\textbf {\bibinfo {volume}
  {93}},\ \bibinfo {pages} {121115}}\BibitemShut {NoStop}%
\bibitem [{\citenamefont {Debbichi}\ \emph {et~al.}(2014)\citenamefont
  {Debbichi}, \citenamefont {Eriksson},\ and\ \citenamefont
  {Leb\`egue}}]{debbichi_prb_89}%
  \BibitemOpen
  \bibfield  {author} {\bibinfo {author} {\bibnamefont {Debbichi},
  \bibfnamefont {L}}, \bibinfo {author} {\bibfnamefont {O.}~\bibnamefont
  {Eriksson}}, \ and\ \bibinfo {author} {\bibfnamefont {S.}~\bibnamefont
  {Leb\`egue}}} (\bibinfo {year} {2014}),\ \bibfield  {title} {\enquote
  {\bibinfo {title} {{E}lectronic structure of two-dimensional transition metal
  dichalcogenide bilayers from ab initio theory},}\ }\href {\doibase
  10.1103/PhysRevB.89.205311} {\bibfield  {journal} {\bibinfo  {journal} {Phys.
  Rev. B}\ }\textbf {\bibinfo {volume} {89}},\ \bibinfo {pages}
  {205311}}\BibitemShut {NoStop}%
\bibitem [{\citenamefont {Deisz}\ \emph {et~al.}(1993)\citenamefont {Deisz},
  \citenamefont {Eguiluz},\ and\ \citenamefont
  {Hanke}}]{Deisz/Eguiluz/Hanke:1993}%
  \BibitemOpen
  \bibfield  {author} {\bibinfo {author} {\bibnamefont {Deisz}, \bibfnamefont
  {J~J}}, \bibinfo {author} {\bibfnamefont {A.~G.}\ \bibnamefont {Eguiluz}}, \
  and\ \bibinfo {author} {\bibfnamefont {W.}~\bibnamefont {Hanke}}} (\bibinfo
  {year} {1993}),\ \bibfield  {title} {\enquote {\bibinfo {title}
  {{Q}uasiparticle theory versus density-functional theory at a metal
  surface},}\ }\href@noop {} {\bibfield  {journal} {\bibinfo  {journal} {Phys.
  Rev. Lett.}\ }\textbf {\bibinfo {volume} {71}},\ \bibinfo {pages}
  {2793--2796}}\BibitemShut {NoStop}%
\bibitem [{\citenamefont {{Del Ben}}\ \emph {et~al.}(2019)\citenamefont {{Del
  Ben}}, \citenamefont {da~Jornada}, \citenamefont {Canning}, \citenamefont
  {Wichmann}, \citenamefont {Raman}, \citenamefont {Sasanka}, \citenamefont
  {Yang}, \citenamefont {Louie},\ and\ \citenamefont {Deslippe}}]{DelBen2019}%
  \BibitemOpen
  \bibfield  {author} {\bibinfo {author} {\bibnamefont {{Del Ben}},
  \bibfnamefont {M}}, \bibinfo {author} {\bibfnamefont {F.~H.}\ \bibnamefont
  {da~Jornada}}, \bibinfo {author} {\bibfnamefont {A.}~\bibnamefont {Canning}},
  \bibinfo {author} {\bibfnamefont {N.}~\bibnamefont {Wichmann}}, \bibinfo
  {author} {\bibfnamefont {K.}~\bibnamefont {Raman}}, \bibinfo {author}
  {\bibfnamefont {R.}~\bibnamefont {Sasanka}}, \bibinfo {author} {\bibfnamefont
  {C.}~\bibnamefont {Yang}}, \bibinfo {author} {\bibfnamefont {S.~G.}\
  \bibnamefont {Louie}}, \ and\ \bibinfo {author} {\bibfnamefont
  {J.}~\bibnamefont {Deslippe}}} (\bibinfo {year} {2019}),\ \bibfield  {title}
  {\enquote {\bibinfo {title} {{L}arge-scale ${G}{W}$ calculations on
  pre-exascale {H}{P}{C} systems},}\ }\href {\doibase
  https://doi.org/10.1016/j.cpc.2018.09.003} {\bibfield  {journal} {\bibinfo
  {journal} {Comput. Phys. Commun.}\ }\textbf {\bibinfo {volume} {235}},\
  \bibinfo {pages} {187 -- 195}}\BibitemShut {NoStop}%
\bibitem [{\citenamefont {{Del Sole}}\ \emph {et~al.}(2003)\citenamefont {{Del
  Sole}}, \citenamefont {Adragna}, \citenamefont {Olevano},\ and\ \citenamefont
  {Reining}}]{DelSole/etal:2003}%
  \BibitemOpen
  \bibfield  {author} {\bibinfo {author} {\bibnamefont {{Del Sole}},
  \bibfnamefont {R}}, \bibinfo {author} {\bibfnamefont {G.}~\bibnamefont
  {Adragna}}, \bibinfo {author} {\bibfnamefont {V.}~\bibnamefont {Olevano}}, \
  and\ \bibinfo {author} {\bibfnamefont {L.}~\bibnamefont {Reining}}} (\bibinfo
  {year} {2003}),\ \bibfield  {title} {\enquote {\bibinfo {title} {{L}ong-range
  behavior and frequency dependence of exchange-correlation kernels in
  solids},}\ }\href@noop {} {\bibfield  {journal} {\bibinfo  {journal} {Phys.
  Rev. B}\ }\textbf {\bibinfo {volume} {67}},\ \bibinfo {pages}
  {045207}}\BibitemShut {NoStop}%
\bibitem [{\citenamefont {{Del Sole}}\ \emph {et~al.}(1994)\citenamefont {{Del
  Sole}}, \citenamefont {Reining},\ and\ \citenamefont
  {Godby}}]{DelSole/Reining/Godby:1994}%
  \BibitemOpen
  \bibfield  {author} {\bibinfo {author} {\bibnamefont {{Del Sole}},
  \bibfnamefont {R}}, \bibinfo {author} {\bibfnamefont {L.}~\bibnamefont
  {Reining}}, \ and\ \bibinfo {author} {\bibfnamefont {R.~W.}\ \bibnamefont
  {Godby}}} (\bibinfo {year} {1994}),\ \bibfield  {title} {\enquote {\bibinfo
  {title} {${G}{W}{\Gamma}$ approximation for electron self-energies in
  semiconductors and insulators},}\ }\href@noop {} {\bibfield  {journal}
  {\bibinfo  {journal} {Phys.\ Rev.\ B}\ }\textbf {\bibinfo {volume} {49}},\
  \bibinfo {pages} {8024}}\BibitemShut {NoStop}%
\bibitem [{\citenamefont {Delaney}\ \emph {et~al.}(2004)\citenamefont
  {Delaney}, \citenamefont {Garc\'{\i}a-Gonz\'alez}, \citenamefont {Rubio},
  \citenamefont {Rinke},\ and\ \citenamefont
  {Godby}}]{Delaney/Garcia-Gonzalez/Rubio/Rinke/Godby:2004}%
  \BibitemOpen
  \bibfield  {author} {\bibinfo {author} {\bibnamefont {Delaney}, \bibfnamefont
  {K}}, \bibinfo {author} {\bibfnamefont {P.}~\bibnamefont
  {Garc\'{\i}a-Gonz\'alez}}, \bibinfo {author} {\bibfnamefont {A.}~\bibnamefont
  {Rubio}}, \bibinfo {author} {\bibfnamefont {P.}~\bibnamefont {Rinke}}, \ and\
  \bibinfo {author} {\bibfnamefont {R.~W.}\ \bibnamefont {Godby}}} (\bibinfo
  {year} {2004}),\ \bibfield  {title} {\enquote {\bibinfo {title} {{C}omment on
  "{B}and-{G}ap {P}roblem in {S}emiconductors {R}evisited: {E}ffects of {C}ore
  {S}tates and {M}any-{B}ody {S}elf-{C}onsistency"},}\ }\href@noop {}
  {\bibfield  {journal} {\bibinfo  {journal} {Phys. Rev. Lett.}\ }\textbf
  {\bibinfo {volume} {93}},\ \bibinfo {pages} {249701}}\BibitemShut {NoStop}%
\bibitem [{\citenamefont {Deslippe}\ \emph {et~al.}(2012)\citenamefont
  {Deslippe}, \citenamefont {Samsonidze}, \citenamefont {Strubbe},
  \citenamefont {Jain}, \citenamefont {Cohen},\ and\ \citenamefont
  {Louie}}]{Deslippe/etal:2012}%
  \BibitemOpen
  \bibfield  {author} {\bibinfo {author} {\bibnamefont {Deslippe},
  \bibfnamefont {J}}, \bibinfo {author} {\bibfnamefont {G.}~\bibnamefont
  {Samsonidze}}, \bibinfo {author} {\bibfnamefont {D.~A.}\ \bibnamefont
  {Strubbe}}, \bibinfo {author} {\bibfnamefont {M.}~\bibnamefont {Jain}},
  \bibinfo {author} {\bibfnamefont {M.~L.}\ \bibnamefont {Cohen}}, \ and\
  \bibinfo {author} {\bibfnamefont {S.~G.}\ \bibnamefont {Louie}}} (\bibinfo
  {year} {2012}),\ \bibfield  {title} {\enquote {\bibinfo {title}
  {{B}erkeley{G}{W}: {A} massively parallel computer package for the
  calculation of the quasiparticle and optical properties of materials and
  nanostructures},}\ }\href@noop {} {\bibfield  {journal} {\bibinfo  {journal}
  {Comput. Phys. Commun.}\ }\textbf {\bibinfo {volume} {183}}~(\bibinfo
  {number} {6}),\ \bibinfo {pages} {1269 -- 1289}}\BibitemShut {NoStop}%
\bibitem [{\citenamefont {Devaux}\ \emph {et~al.}(2015)\citenamefont {Devaux},
  \citenamefont {Casula}, \citenamefont {Decremps},\ and\ \citenamefont
  {Sorella}}]{Devaux/etal:2015}%
  \BibitemOpen
  \bibfield  {author} {\bibinfo {author} {\bibnamefont {Devaux}, \bibfnamefont
  {N}}, \bibinfo {author} {\bibfnamefont {M.}~\bibnamefont {Casula}}, \bibinfo
  {author} {\bibfnamefont {F.}~\bibnamefont {Decremps}}, \ and\ \bibinfo
  {author} {\bibfnamefont {S.}~\bibnamefont {Sorella}}} (\bibinfo {year}
  {2015}),\ \bibfield  {title} {\enquote {\bibinfo {title} {{E}lectronic origin
  of the volume collapse in cerium},}\ }\href@noop {} {\bibfield  {journal}
  {\bibinfo  {journal} {Phys. Rev. B}\ }\textbf {\bibinfo {volume} {91}},\
  \bibinfo {pages} {081101}}\BibitemShut {NoStop}%
\bibitem [{\citenamefont {Dori}\ \emph {et~al.}(2006)\citenamefont {Dori},
  \citenamefont {Menon}, \citenamefont {Kilian}, \citenamefont {Sokolowski},
  \citenamefont {Kronik},\ and\ \citenamefont {Umbach}}]{Dori2006}%
  \BibitemOpen
  \bibfield  {author} {\bibinfo {author} {\bibnamefont {Dori}, \bibfnamefont
  {N}}, \bibinfo {author} {\bibfnamefont {M.}~\bibnamefont {Menon}}, \bibinfo
  {author} {\bibfnamefont {L.}~\bibnamefont {Kilian}}, \bibinfo {author}
  {\bibfnamefont {M.}~\bibnamefont {Sokolowski}}, \bibinfo {author}
  {\bibfnamefont {L.}~\bibnamefont {Kronik}}, \ and\ \bibinfo {author}
  {\bibfnamefont {E.}~\bibnamefont {Umbach}}} (\bibinfo {year} {2006}),\
  \bibfield  {title} {\enquote {\bibinfo {title} {{V}alence electronic
  structure of gas-phase 3,4,9,10-perylene tetracarboxylic acid dianhydride:
  {E}xperiment and theory},}\ }\href {\doibase 10.1103/PhysRevB.73.195208}
  {\bibfield  {journal} {\bibinfo  {journal} {Phys. Rev. B}\ }\textbf {\bibinfo
  {volume} {73}},\ \bibinfo {pages} {195208}}\BibitemShut {NoStop}%
\bibitem [{\citenamefont {Dose}(1985)}]{Dose:1985}%
  \BibitemOpen
  \bibfield  {author} {\bibinfo {author} {\bibnamefont {Dose}, \bibfnamefont
  {V}}} (\bibinfo {year} {1985}),\ \bibfield  {title} {\enquote {\bibinfo
  {title} {{M}omentum-{R}esolved {I}nverse {P}hotoemission},}\ }\href@noop {}
  {\bibfield  {journal} {\bibinfo  {journal} {Surf. Sci. Rep.}\ }\textbf
  {\bibinfo {volume} {5}},\ \bibinfo {pages} {337}}\BibitemShut {NoStop}%
\bibitem [{\citenamefont {Drissi}\ and\ \citenamefont
  {Ramadan}(2015{\natexlab{a}})}]{drissi_pe_74}%
  \BibitemOpen
  \bibfield  {author} {\bibinfo {author} {\bibnamefont {Drissi}, \bibfnamefont
  {L}}, \ and\ \bibinfo {author} {\bibfnamefont {F.}~\bibnamefont {Ramadan}}}
  (\bibinfo {year} {2015}{\natexlab{a}}),\ \bibfield  {title} {\enquote
  {\bibinfo {title} {{E}xcitonic effects in {G}e{C} hybrid: {M}any-body
  {G}reen's function calculations},}\ }\href {\doibase
  https://doi.org/10.1016/j.physe.2015.07.030} {\bibfield  {journal} {\bibinfo
  {journal} {Physica E}\ }\textbf {\bibinfo {volume} {74}},\ \bibinfo {pages}
  {377 -- 381}}\BibitemShut {NoStop}%
\bibitem [{\citenamefont {Drissi}\ and\ \citenamefont
  {Ramadan}(2015{\natexlab{b}})}]{drissi_pe_68}%
  \BibitemOpen
  \bibfield  {author} {\bibinfo {author} {\bibnamefont {Drissi}, \bibfnamefont
  {L}}, \ and\ \bibinfo {author} {\bibfnamefont {F.}~\bibnamefont {Ramadan}}}
  (\bibinfo {year} {2015}{\natexlab{b}}),\ \bibfield  {title} {\enquote
  {\bibinfo {title} {{M}any body effects study of electronic and optical
  properties of silicene–graphene hybrid},}\ }\href {\doibase
  https://doi.org/10.1016/j.physe.2014.12.009} {\bibfield  {journal} {\bibinfo
  {journal} {Physica E}\ }\textbf {\bibinfo {volume} {68}},\ \bibinfo {pages}
  {38 -- 41}}\BibitemShut {NoStop}%
\bibitem [{\citenamefont {Dr\"uppel}\ \emph {et~al.}(2018)\citenamefont
  {Dr\"uppel}, \citenamefont {Deilmann}, \citenamefont {Noky}, \citenamefont
  {Marauhn}, \citenamefont {Kr\"uger},\ and\ \citenamefont
  {Rohlfing}}]{Drueppel2018}%
  \BibitemOpen
  \bibfield  {author} {\bibinfo {author} {\bibnamefont {Dr\"uppel},
  \bibfnamefont {M}}, \bibinfo {author} {\bibfnamefont {T.}~\bibnamefont
  {Deilmann}}, \bibinfo {author} {\bibfnamefont {J.}~\bibnamefont {Noky}},
  \bibinfo {author} {\bibfnamefont {P.}~\bibnamefont {Marauhn}}, \bibinfo
  {author} {\bibfnamefont {P.}~\bibnamefont {Kr\"uger}}, \ and\ \bibinfo
  {author} {\bibfnamefont {M.}~\bibnamefont {Rohlfing}}} (\bibinfo {year}
  {2018}),\ \bibfield  {title} {\enquote {\bibinfo {title} {{E}lectronic
  excitations in transition metal dichalcogenide monolayers from an
  $\mathrm{L{D}A}+\mathit{GdW}$ approach},}\ }\href {\doibase
  10.1103/PhysRevB.98.155433} {\bibfield  {journal} {\bibinfo  {journal} {Phys.
  Rev. B}\ }\textbf {\bibinfo {volume} {98}},\ \bibinfo {pages}
  {155433}}\BibitemShut {NoStop}%
\bibitem [{\citenamefont {Duchemin}\ \emph {et~al.}(2016)\citenamefont
  {Duchemin}, \citenamefont {Jacquemin},\ and\ \citenamefont
  {Blase}}]{Duchemin2016}%
  \BibitemOpen
  \bibfield  {author} {\bibinfo {author} {\bibnamefont {Duchemin},
  \bibfnamefont {I}}, \bibinfo {author} {\bibfnamefont {D.}~\bibnamefont
  {Jacquemin}}, \ and\ \bibinfo {author} {\bibfnamefont {X.}~\bibnamefont
  {Blase}}} (\bibinfo {year} {2016}),\ \bibfield  {title} {\enquote {\bibinfo
  {title} {{C}ombining the ${G}{W}$ formalism with the polarizable continuum
  model: {A} state-specific non-equilibrium approach},}\ }\href {\doibase
  10.1063/1.4946778} {\bibfield  {journal} {\bibinfo  {journal} {J. Chem.
  Phys.}\ }\textbf {\bibinfo {volume} {144}}~(\bibinfo {number} {16}),\
  \bibinfo {pages} {164106}}\BibitemShut {NoStop}%
\bibitem [{\citenamefont {Dunning}(1989)}]{Dunning1989}%
  \BibitemOpen
  \bibfield  {author} {\bibinfo {author} {\bibnamefont {Dunning}, \bibfnamefont
  {T~H}}} (\bibinfo {year} {1989}),\ \bibfield  {title} {\enquote {\bibinfo
  {title} {{G}aussian basis sets for use in correlated molecular calculations.
  {I}. {T}he atoms boron through neon and hydrogen},}\ }\href@noop {}
  {\bibfield  {journal} {\bibinfo  {journal} {J. Chem. Phys.}\ }\textbf
  {\bibinfo {volume} {90}}~(\bibinfo {number} {2}),\ \bibinfo {pages}
  {1007--1023}}\BibitemShut {NoStop}%
\bibitem [{\citenamefont {Dvorak}\ \emph {et~al.}(2014)\citenamefont {Dvorak},
  \citenamefont {Chen},\ and\ \citenamefont {Wu}}]{Dvorak/etal:2014}%
  \BibitemOpen
  \bibfield  {author} {\bibinfo {author} {\bibnamefont {Dvorak}, \bibfnamefont
  {M}}, \bibinfo {author} {\bibfnamefont {X.-J.}\ \bibnamefont {Chen}}, \ and\
  \bibinfo {author} {\bibfnamefont {Z.}~\bibnamefont {Wu}}} (\bibinfo {year}
  {2014}),\ \bibfield  {title} {\enquote {\bibinfo {title} {{Q}uasiparticle
  energies and excitonic effects in dense solid hydrogen near metallization},}\
  }\href@noop {} {\bibfield  {journal} {\bibinfo  {journal} {Phys. Rev. B}\
  }\textbf {\bibinfo {volume} {90}},\ \bibinfo {pages} {035103}}\BibitemShut
  {NoStop}%
\bibitem [{\citenamefont {Dvorak}\ \emph {et~al.}(2018)\citenamefont {Dvorak},
  \citenamefont {Golze},\ and\ \citenamefont {Rinke}}]{Dvorak2/etal:2018}%
  \BibitemOpen
  \bibfield  {author} {\bibinfo {author} {\bibnamefont {Dvorak}, \bibfnamefont
  {M}}, \bibinfo {author} {\bibfnamefont {D.}~\bibnamefont {Golze}}, \ and\
  \bibinfo {author} {\bibfnamefont {P.}~\bibnamefont {Rinke}}} (\bibinfo {year}
  {2018}),\ \bibfield  {title} {\enquote {\bibinfo {title} {{A} quantum
  embedding theory in the screened {C}oulomb interaction: {C}ombining
  configuration interaction with ${G}{W}$/{B}{S}{E}},}\ }\href@noop {}
  {\bibinfo  {journal} {arXiv:1810.12005}\ }\BibitemShut {NoStop}%
\bibitem [{\citenamefont {Dvorak}\ and\ \citenamefont
  {Rinke}(2019)}]{Dvorak/etal:2018}%
  \BibitemOpen
\bibfield  {journal} {  }\bibfield  {author} {\bibinfo {author} {\bibnamefont
  {Dvorak}, \bibfnamefont {M}}, \ and\ \bibinfo {author} {\bibfnamefont
  {P.}~\bibnamefont {Rinke}}} (\bibinfo {year} {2019}),\ \bibfield  {title}
  {\enquote {\bibinfo {title} {{D}ynamical configuration interaction: {Q}uantum
  embedding that combines wave functions and {G}reen's functions},}\ }\href
  {\doibase 10.1103/PhysRevB.99.115134} {\bibfield  {journal} {\bibinfo
  {journal} {Phys. Rev. B}\ }\textbf {\bibinfo {volume} {99}},\ \bibinfo
  {pages} {115134}}\BibitemShut {NoStop}%
\bibitem [{\citenamefont {Dvorak}\ and\ \citenamefont {Wu}(2015)}]{Dvorak2015}%
  \BibitemOpen
  \bibfield  {author} {\bibinfo {author} {\bibnamefont {Dvorak}, \bibfnamefont
  {M}}, \ and\ \bibinfo {author} {\bibfnamefont {Z.}~\bibnamefont {Wu}}}
  (\bibinfo {year} {2015}),\ \bibfield  {title} {\enquote {\bibinfo {title}
  {{T}unable many-body interactions in semiconducting graphene: {G}iant
  excitonic effect and strong optical absorption},}\ }\href {\doibase
  10.1103/PhysRevB.92.035422} {\bibfield  {journal} {\bibinfo  {journal} {Phys.
  Rev. B}\ }\textbf {\bibinfo {volume} {92}},\ \bibinfo {pages}
  {035422}}\BibitemShut {NoStop}%
\bibitem [{\citenamefont {Dyson}(1949{\natexlab{a}})}]{Dyson1:1949}%
  \BibitemOpen
  \bibfield  {author} {\bibinfo {author} {\bibnamefont {Dyson}, \bibfnamefont
  {F~J}}} (\bibinfo {year} {1949}{\natexlab{a}}),\ \bibfield  {title} {\enquote
  {\bibinfo {title} {{T}he {R}adiation {T}heories of {T}omonaga, {S}chwinger,
  and {F}eynman},}\ }\href@noop {} {\bibfield  {journal} {\bibinfo  {journal}
  {Phys. Rev.}\ }\textbf {\bibinfo {volume} {75}},\ \bibinfo {pages}
  {486}}\BibitemShut {NoStop}%
\bibitem [{\citenamefont {Dyson}(1949{\natexlab{b}})}]{Dyson2:1949}%
  \BibitemOpen
  \bibfield  {author} {\bibinfo {author} {\bibnamefont {Dyson}, \bibfnamefont
  {F~J}}} (\bibinfo {year} {1949}{\natexlab{b}}),\ \bibfield  {title} {\enquote
  {\bibinfo {title} {{T}he {S} {M}atrix in {Q}uantum {E}lectrodynamics},}\
  }\href@noop {} {\bibfield  {journal} {\bibinfo  {journal} {Phys. Rev.}\
  }\textbf {\bibinfo {volume} {75}},\ \bibinfo {pages} {1736}}\BibitemShut
  {NoStop}%
\bibitem [{\citenamefont {Egelhoff}(1987)}]{Egelhoff1987}%
  \BibitemOpen
  \bibfield  {author} {\bibinfo {author} {\bibnamefont {Egelhoff},
  \bibfnamefont {W~F}}} (\bibinfo {year} {1987}),\ \bibfield  {title} {\enquote
  {\bibinfo {title} {{C}ore-level binding-energy shifts at surfaces and in
  solids},}\ }\href@noop {} {\bibfield  {journal} {\bibinfo  {journal} {Surf.
  Sci. Rep.}\ }\textbf {\bibinfo {volume} {6}}~(\bibinfo {number} {6}),\
  \bibinfo {pages} {253--415}}\BibitemShut {NoStop}%
\bibitem [{\citenamefont {Egger}\ \emph {et~al.}(2014)\citenamefont {Egger},
  \citenamefont {Weissman}, \citenamefont {Refaely-Abramson}, \citenamefont
  {Sharifzadeh}, \citenamefont {Dauth}, \citenamefont {Baer}, \citenamefont
  {K\"ummel}, \citenamefont {Neaton}, \citenamefont {Zojer},\ and\
  \citenamefont {Kronik}}]{Egger2014}%
  \BibitemOpen
  \bibfield  {author} {\bibinfo {author} {\bibnamefont {Egger}, \bibfnamefont
  {D~A}}, \bibinfo {author} {\bibfnamefont {S.}~\bibnamefont {Weissman}},
  \bibinfo {author} {\bibfnamefont {S.}~\bibnamefont {Refaely-Abramson}},
  \bibinfo {author} {\bibfnamefont {S.}~\bibnamefont {Sharifzadeh}}, \bibinfo
  {author} {\bibfnamefont {M.}~\bibnamefont {Dauth}}, \bibinfo {author}
  {\bibfnamefont {R.}~\bibnamefont {Baer}}, \bibinfo {author} {\bibfnamefont
  {S.}~\bibnamefont {K\"ummel}}, \bibinfo {author} {\bibfnamefont {J.~B.}\
  \bibnamefont {Neaton}}, \bibinfo {author} {\bibfnamefont {E.}~\bibnamefont
  {Zojer}}, \ and\ \bibinfo {author} {\bibfnamefont {L.}~\bibnamefont
  {Kronik}}} (\bibinfo {year} {2014}),\ \bibfield  {title} {\enquote {\bibinfo
  {title} {{O}uter-valence {E}lectron {S}pectra of {P}rototypical {A}romatic
  {H}eterocycles from an {O}ptimally {T}uned {R}ange-{S}eparated {H}ybrid
  {F}unctional},}\ }\href@noop {} {\bibfield  {journal} {\bibinfo  {journal}
  {J. Chem. Theory Comput.}\ }\textbf {\bibinfo {volume} {10}}~(\bibinfo
  {number} {5}),\ \bibinfo {pages} {1934--1952}}\BibitemShut {NoStop}%
\bibitem [{\citenamefont {Engel}\ and\ \citenamefont
  {Farid}(1993)}]{Engel1993}%
  \BibitemOpen
  \bibfield  {author} {\bibinfo {author} {\bibnamefont {Engel}, \bibfnamefont
  {G~E}}, \ and\ \bibinfo {author} {\bibfnamefont {B.}~\bibnamefont {Farid}}}
  (\bibinfo {year} {1993}),\ \bibfield  {title} {\enquote {\bibinfo {title}
  {{G}eneralized plasmon-pole model and plasmon band structures of crystals},}\
  }\href {\doibase 10.1103/PhysRevB.47.15931} {\bibfield  {journal} {\bibinfo
  {journal} {Phys. Rev. B}\ }\textbf {\bibinfo {volume} {47}},\ \bibinfo
  {pages} {15931--15934}}\BibitemShut {NoStop}%
\bibitem [{\citenamefont {Enkovaara}\ \emph {et~al.}(2010)\citenamefont
  {Enkovaara}, \citenamefont {Rostgaard}, \citenamefont {Mortensen},
  \citenamefont {Chen}, \citenamefont {Du{\l}ak}, \citenamefont {Ferrighi},
  \citenamefont {Gavnholt}, \citenamefont {Glinsvad}, \citenamefont {Haikola},
  \citenamefont {Hansen}, \citenamefont {Kristoffersen}, \citenamefont
  {Kuisma}, \citenamefont {Larsen}, \citenamefont {Lehtovaara}, \citenamefont
  {Ljungberg}, \citenamefont {Lopez-Acevedo}, \citenamefont {Moses},
  \citenamefont {Ojanen}, \citenamefont {Olsen}, \citenamefont {Petzold},
  \citenamefont {Romero}, \citenamefont {Stausholm-M{\o}ller}, \citenamefont
  {Strange}, \citenamefont {Tritsaris}, \citenamefont {Vanin}, \citenamefont
  {Walter}, \citenamefont {Hammer}, \citenamefont {H\"akkinen}, \citenamefont
  {Madsen}, \citenamefont {Nieminen}, \citenamefont {N{\o}rskov}, \citenamefont
  {Puska}, \citenamefont {Rantala}, \citenamefont {Schi{\o}tz}, \citenamefont
  {Thygesen},\ and\ \citenamefont {Jacobsen}}]{Enkovaara2010}%
  \BibitemOpen
  \bibfield  {author} {\bibinfo {author} {\bibnamefont {Enkovaara},
  \bibfnamefont {J}}, \bibinfo {author} {\bibfnamefont {C.}~\bibnamefont
  {Rostgaard}}, \bibinfo {author} {\bibfnamefont {J.~J.}\ \bibnamefont
  {Mortensen}}, \bibinfo {author} {\bibfnamefont {J.}~\bibnamefont {Chen}},
  \bibinfo {author} {\bibfnamefont {M.}~\bibnamefont {Du{\l}ak}}, \bibinfo
  {author} {\bibfnamefont {L.}~\bibnamefont {Ferrighi}}, \bibinfo {author}
  {\bibfnamefont {J.}~\bibnamefont {Gavnholt}}, \bibinfo {author}
  {\bibfnamefont {C.}~\bibnamefont {Glinsvad}}, \bibinfo {author}
  {\bibfnamefont {V.}~\bibnamefont {Haikola}}, \bibinfo {author} {\bibfnamefont
  {H.~A.}\ \bibnamefont {Hansen}}, \bibinfo {author} {\bibfnamefont {H.~H.}\
  \bibnamefont {Kristoffersen}}, \bibinfo {author} {\bibfnamefont
  {M.}~\bibnamefont {Kuisma}}, \bibinfo {author} {\bibfnamefont {A.~H.}\
  \bibnamefont {Larsen}}, \bibinfo {author} {\bibfnamefont {L.}~\bibnamefont
  {Lehtovaara}}, \bibinfo {author} {\bibfnamefont {M.}~\bibnamefont
  {Ljungberg}}, \bibinfo {author} {\bibfnamefont {O.}~\bibnamefont
  {Lopez-Acevedo}}, \bibinfo {author} {\bibfnamefont {P.~G.}\ \bibnamefont
  {Moses}}, \bibinfo {author} {\bibfnamefont {J.}~\bibnamefont {Ojanen}},
  \bibinfo {author} {\bibfnamefont {T.}~\bibnamefont {Olsen}}, \bibinfo
  {author} {\bibfnamefont {V.}~\bibnamefont {Petzold}}, \bibinfo {author}
  {\bibfnamefont {N.~A.}\ \bibnamefont {Romero}}, \bibinfo {author}
  {\bibfnamefont {J.}~\bibnamefont {Stausholm-M{\o}ller}}, \bibinfo {author}
  {\bibfnamefont {M.}~\bibnamefont {Strange}}, \bibinfo {author} {\bibfnamefont
  {G.~A.}\ \bibnamefont {Tritsaris}}, \bibinfo {author} {\bibfnamefont
  {M.}~\bibnamefont {Vanin}}, \bibinfo {author} {\bibfnamefont
  {M.}~\bibnamefont {Walter}}, \bibinfo {author} {\bibfnamefont
  {B.}~\bibnamefont {Hammer}}, \bibinfo {author} {\bibfnamefont
  {H.}~\bibnamefont {H\"akkinen}}, \bibinfo {author} {\bibfnamefont {G.~K.~H.}\
  \bibnamefont {Madsen}}, \bibinfo {author} {\bibfnamefont {R.~M.}\
  \bibnamefont {Nieminen}}, \bibinfo {author} {\bibfnamefont {J.~K.}\
  \bibnamefont {N{\o}rskov}}, \bibinfo {author} {\bibfnamefont
  {M.}~\bibnamefont {Puska}}, \bibinfo {author} {\bibfnamefont {T.~T.}\
  \bibnamefont {Rantala}}, \bibinfo {author} {\bibfnamefont {J.}~\bibnamefont
  {Schi{\o}tz}}, \bibinfo {author} {\bibfnamefont {K.~S.}\ \bibnamefont
  {Thygesen}}, \ and\ \bibinfo {author} {\bibfnamefont {K.~W.}\ \bibnamefont
  {Jacobsen}}} (\bibinfo {year} {2010}),\ \bibfield  {title} {\enquote
  {\bibinfo {title} {{E}lectronic structure calculations with {G}{P}{A}{W}: a
  real-space implementation of the projector augmented-wave method},}\ }\href
  {http://stacks.iop.org/0953-8984/22/i=25/a=253202} {\bibfield  {journal}
  {\bibinfo  {journal} {J. Phys.: Condens. Matter}\ }\textbf {\bibinfo {volume}
  {22}}~(\bibinfo {number} {25}),\ \bibinfo {pages} {253202}}\BibitemShut
  {NoStop}%
\bibitem [{\citenamefont {Erhart}\ \emph {et~al.}(2014)\citenamefont {Erhart},
  \citenamefont {Schleife}, \citenamefont {Sadigh},\ and\ \citenamefont
  {\AA{}berg}}]{Erhart/etal:2014}%
  \BibitemOpen
  \bibfield  {author} {\bibinfo {author} {\bibnamefont {Erhart}, \bibfnamefont
  {P}}, \bibinfo {author} {\bibfnamefont {A.}~\bibnamefont {Schleife}},
  \bibinfo {author} {\bibfnamefont {B.}~\bibnamefont {Sadigh}}, \ and\ \bibinfo
  {author} {\bibfnamefont {D.}~\bibnamefont {\AA{}berg}}} (\bibinfo {year}
  {2014}),\ \bibfield  {title} {\enquote {\bibinfo {title} {{Q}uasiparticle
  spectra, absorption spectra, and excitonic properties of {N}a{I} and
  {S}r{I}$_2$ from many-body perturbation theory},}\ }\href {\doibase
  10.1103/PhysRevB.89.075132} {\bibfield  {journal} {\bibinfo  {journal} {Phys.
  Rev. B}\ }\textbf {\bibinfo {volume} {89}},\ \bibinfo {pages}
  {075132}}\BibitemShut {NoStop}%
\bibitem [{\citenamefont {Ernzerhof}\ and\ \citenamefont
  {Scuseria}(1999)}]{PBE0_2}%
  \BibitemOpen
  \bibfield  {author} {\bibinfo {author} {\bibnamefont {Ernzerhof},
  \bibfnamefont {M}}, \ and\ \bibinfo {author} {\bibfnamefont {G.~E.}\
  \bibnamefont {Scuseria}}} (\bibinfo {year} {1999}),\ \bibfield  {title}
  {\enquote {\bibinfo {title} {{A}ssessment of the {P}erdew-{B}urke-{E}rnzerhof
  exchange-correlation functional},}\ }\href@noop {} {\bibfield  {journal}
  {\bibinfo  {journal} {J.\ Chem.\ Phys}\ }\textbf {\bibinfo {volume} {110}},\
  \bibinfo {pages} {5029}}\BibitemShut {NoStop}%
\bibitem [{\citenamefont {Eshuis}\ \emph {et~al.}(2012)\citenamefont {Eshuis},
  \citenamefont {Bates},\ and\ \citenamefont
  {Furche}}]{Eshuis/Bates/Furche:2012}%
  \BibitemOpen
  \bibfield  {author} {\bibinfo {author} {\bibnamefont {Eshuis}, \bibfnamefont
  {H}}, \bibinfo {author} {\bibfnamefont {J.~E.}\ \bibnamefont {Bates}}, \ and\
  \bibinfo {author} {\bibfnamefont {F.}~\bibnamefont {Furche}}} (\bibinfo
  {year} {2012}),\ \bibfield  {title} {\enquote {\bibinfo {title} {{E}lectron
  {C}orrelation {M}ethods {B}ased on the {R}andom {P}hase {A}pproximation},}\
  }\href@noop {} {\bibfield  {journal} {\bibinfo  {journal} {Theor. Chem.
  Acc.}\ }\textbf {\bibinfo {volume} {131}},\ \bibinfo {pages}
  {1084}}\BibitemShut {NoStop}%
\bibitem [{\citenamefont {Espejo}\ \emph {et~al.}(2013)\citenamefont {Espejo},
  \citenamefont {Rangel}, \citenamefont {Romero}, \citenamefont {X.},\ and\
  \citenamefont {Rignanese}}]{espejo_prb_87}%
  \BibitemOpen
  \bibfield  {author} {\bibinfo {author} {\bibnamefont {Espejo}, \bibfnamefont
  {C}}, \bibinfo {author} {\bibfnamefont {T.}~\bibnamefont {Rangel}}, \bibinfo
  {author} {\bibfnamefont {A.~H.}\ \bibnamefont {Romero}}, \bibinfo {author}
  {\bibnamefont {X.}}, \ and\ \bibinfo {author} {\bibfnamefont {G.-M.}\
  \bibnamefont {Rignanese}}} (\bibinfo {year} {2013}),\ \bibfield  {title}
  {\enquote {\bibinfo {title} {{B}and structure tunability in {M}o{S}${}_{2}$
  under interlayer compression: {A} {D}{F}{T} and ${G}{W}$ study},}\ }\href
  {\doibase 10.1103/PhysRevB.87.245114} {\bibfield  {journal} {\bibinfo
  {journal} {Phys. Rev. B}\ }\textbf {\bibinfo {volume} {87}},\ \bibinfo
  {pages} {245114}}\BibitemShut {NoStop}%
\bibitem [{\citenamefont {Faber}\ \emph {et~al.}(2011)\citenamefont {Faber},
  \citenamefont {Attaccalite}, \citenamefont {Olevano}, \citenamefont {Runge},\
  and\ \citenamefont {Blase}}]{Faber2011}%
  \BibitemOpen
  \bibfield  {author} {\bibinfo {author} {\bibnamefont {Faber}, \bibfnamefont
  {C}}, \bibinfo {author} {\bibfnamefont {C.}~\bibnamefont {Attaccalite}},
  \bibinfo {author} {\bibfnamefont {V.}~\bibnamefont {Olevano}}, \bibinfo
  {author} {\bibfnamefont {E.}~\bibnamefont {Runge}}, \ and\ \bibinfo {author}
  {\bibfnamefont {X.}~\bibnamefont {Blase}}} (\bibinfo {year} {2011}),\
  \bibfield  {title} {\enquote {\bibinfo {title} {{F}irst-principles
  $\mathit{GW}$ calculations for {D}{N}{A} and {R}{N}{A} nucleobases},}\ }\href
  {\doibase 10.1103/PhysRevB.83.115123} {\bibfield  {journal} {\bibinfo
  {journal} {Phys. Rev. B}\ }\textbf {\bibinfo {volume} {83}},\ \bibinfo
  {pages} {115123}}\BibitemShut {NoStop}%
\bibitem [{\citenamefont {Faber}\ \emph {et~al.}(2014)\citenamefont {Faber},
  \citenamefont {Boulanger}, \citenamefont {Attaccalite}, \citenamefont
  {Duchemin},\ and\ \citenamefont {Blase}}]{Faber/etal:2014}%
  \BibitemOpen
  \bibfield  {author} {\bibinfo {author} {\bibnamefont {Faber}, \bibfnamefont
  {C}}, \bibinfo {author} {\bibfnamefont {P.}~\bibnamefont {Boulanger}},
  \bibinfo {author} {\bibfnamefont {C.}~\bibnamefont {Attaccalite}}, \bibinfo
  {author} {\bibfnamefont {I.}~\bibnamefont {Duchemin}}, \ and\ \bibinfo
  {author} {\bibfnamefont {X.}~\bibnamefont {Blase}}} (\bibinfo {year}
  {2014}),\ \bibfield  {title} {\enquote {\bibinfo {title} {{E}xcited states
  properties of organic molecules: from density functional theory to the
  ${G}{W}$ and {B}ethe-{S}alpeter {G}reen's function formalisms},}\ }\href@noop
  {} {\bibfield  {journal} {\bibinfo  {journal} {Philos. Trans. Royal Soc. A}\
  }\textbf {\bibinfo {volume} {372}}~(\bibinfo {number} {2011}),\ \bibinfo
  {pages} {20130271}}\BibitemShut {NoStop}%
\bibitem [{\citenamefont {Faber}\ \emph {et~al.}(2012)\citenamefont {Faber},
  \citenamefont {Duchemin}, \citenamefont {Deutsch},\ and\ \citenamefont
  {Blase}}]{Faber/etal:2012}%
  \BibitemOpen
  \bibfield  {author} {\bibinfo {author} {\bibnamefont {Faber}, \bibfnamefont
  {C}}, \bibinfo {author} {\bibfnamefont {I.}~\bibnamefont {Duchemin}},
  \bibinfo {author} {\bibfnamefont {T.}~\bibnamefont {Deutsch}}, \ and\
  \bibinfo {author} {\bibfnamefont {X.}~\bibnamefont {Blase}}} (\bibinfo {year}
  {2012}),\ \bibfield  {title} {\enquote {\bibinfo {title} {{M}any-body
  {G}reen's function study of coumarins for dye-sensitized solar cells},}\
  }\href@noop {} {\bibfield  {journal} {\bibinfo  {journal} {Phys. Rev. B}\
  }\textbf {\bibinfo {volume} {86}},\ \bibinfo {pages} {155315}}\BibitemShut
  {NoStop}%
\bibitem [{\citenamefont {Fakhrabad}\ \emph {et~al.}(2015)\citenamefont
  {Fakhrabad}, \citenamefont {Shahtahmasebi},\ and\ \citenamefont
  {Ashhadi}}]{fakhrabad_sm_79}%
  \BibitemOpen
  \bibfield  {author} {\bibinfo {author} {\bibnamefont {Fakhrabad},
  \bibfnamefont {D~V}}, \bibinfo {author} {\bibfnamefont {N.}~\bibnamefont
  {Shahtahmasebi}}, \ and\ \bibinfo {author} {\bibfnamefont {M.}~\bibnamefont
  {Ashhadi}}} (\bibinfo {year} {2015}),\ \bibfield  {title} {\enquote {\bibinfo
  {title} {{O}ptical excitations and quasiparticle energies in the {A}l{N}
  monolayer honeycomb structure},}\ }\href {\doibase
  https://doi.org/10.1016/j.spmi.2014.12.012} {\bibfield  {journal} {\bibinfo
  {journal} {Superlattices and Microstruct.}\ }\textbf {\bibinfo {volume}
  {79}},\ \bibinfo {pages} {38 -- 44}}\BibitemShut {NoStop}%
\bibitem [{\citenamefont {Fakhrabad}\ \emph {et~al.}(2014)\citenamefont
  {Fakhrabad}, \citenamefont {Shahtamasebi},\ and\ \citenamefont
  {Ashhadi}}]{fakhrabad_pe_59}%
  \BibitemOpen
  \bibfield  {author} {\bibinfo {author} {\bibnamefont {Fakhrabad},
  \bibfnamefont {D~V}}, \bibinfo {author} {\bibfnamefont {N.}~\bibnamefont
  {Shahtamasebi}}, \ and\ \bibinfo {author} {\bibfnamefont {M.}~\bibnamefont
  {Ashhadi}}} (\bibinfo {year} {2014}),\ \bibfield  {title} {\enquote {\bibinfo
  {title} {{Q}uasiparticle energies and optical excitations in the {G}a{A}s
  monolayer},}\ }\href {\doibase https://doi.org/10.1016/j.physe.2013.12.019}
  {\bibfield  {journal} {\bibinfo  {journal} {Physica E}\ }\textbf {\bibinfo
  {volume} {59}},\ \bibinfo {pages} {107 -- 109}}\BibitemShut {NoStop}%
\bibitem [{\citenamefont {Faleev}\ \emph {et~al.}(2004)\citenamefont {Faleev},
  \citenamefont {van Schilfgaarde},\ and\ \citenamefont
  {Kotani}}]{Faleev/Schilfgaarde/Kotani:2004}%
  \BibitemOpen
  \bibfield  {author} {\bibinfo {author} {\bibnamefont {Faleev}, \bibfnamefont
  {S~V}}, \bibinfo {author} {\bibfnamefont {M.}~\bibnamefont {van
  Schilfgaarde}}, \ and\ \bibinfo {author} {\bibfnamefont {T.}~\bibnamefont
  {Kotani}}} (\bibinfo {year} {2004}),\ \bibfield  {title} {\enquote {\bibinfo
  {title} {{A}ll-{E}lectron {S}elf-{C}onsistent ${G}{W}$ {A}pproximation:
  {A}pplication to {S}i, {M}n{O}, and {N}i{O}},}\ }\href@noop {} {\bibfield
  {journal} {\bibinfo  {journal} {Phys. Rev. Lett.}\ }\textbf {\bibinfo
  {volume} {93}},\ \bibinfo {pages} {126406}}\BibitemShut {NoStop}%
\bibitem [{\citenamefont {Ferreira}\ and\ \citenamefont
  {Ribeiro}(2017)}]{ferreira_prb_96}%
  \BibitemOpen
  \bibfield  {author} {\bibinfo {author} {\bibnamefont {Ferreira},
  \bibfnamefont {F}}, \ and\ \bibinfo {author} {\bibfnamefont {R.~M.}\
  \bibnamefont {Ribeiro}}} (\bibinfo {year} {2017}),\ \bibfield  {title}
  {\enquote {\bibinfo {title} {{I}mprovements in the ${G}{W}$ and
  {B}ethe-{S}alpeter-equation calculations on phosphorene},}\ }\href {\doibase
  10.1103/PhysRevB.96.115431} {\bibfield  {journal} {\bibinfo  {journal} {Phys.
  Rev. B}\ }\textbf {\bibinfo {volume} {96}},\ \bibinfo {pages}
  {115431}}\BibitemShut {NoStop}%
\bibitem [{\citenamefont {Ferri}\ \emph {et~al.}(2015)\citenamefont {Ferri},
  \citenamefont {DiStasio}, \citenamefont {Ambrosetti}, \citenamefont {Car},\
  and\ \citenamefont {Tkatchenko}}]{Ferri/etal:2015}%
  \BibitemOpen
  \bibfield  {author} {\bibinfo {author} {\bibnamefont {Ferri}, \bibfnamefont
  {N}}, \bibinfo {author} {\bibfnamefont {R.~A.}\ \bibnamefont {DiStasio}},
  \bibinfo {author} {\bibfnamefont {A.}~\bibnamefont {Ambrosetti}}, \bibinfo
  {author} {\bibfnamefont {R.}~\bibnamefont {Car}}, \ and\ \bibinfo {author}
  {\bibfnamefont {A.}~\bibnamefont {Tkatchenko}}} (\bibinfo {year} {2015}),\
  \bibfield  {title} {\enquote {\bibinfo {title} {{E}lectronic {P}roperties of
  {M}olecules and {S}urfaces with a {S}elf-{C}onsistent {I}nteratomic van der
  {W}aals {D}ensity {F}unctional},}\ }\href@noop {} {\bibfield  {journal}
  {\bibinfo  {journal} {Phys. Rev. Lett.}\ }\textbf {\bibinfo {volume} {114}},\
  \bibinfo {pages} {176802}}\BibitemShut {NoStop}%
\bibitem [{\citenamefont {Fetter}\ and\ \citenamefont
  {Walecka}(1971)}]{Fetter/Walecka}%
  \BibitemOpen
  \bibfield  {author} {\bibinfo {author} {\bibnamefont {Fetter}, \bibfnamefont
  {A~L}}, \ and\ \bibinfo {author} {\bibfnamefont {J.~D.}\ \bibnamefont
  {Walecka}}} (\bibinfo {year} {1971}),\ \href@noop {} {\emph {\bibinfo {title}
  {{Q}uantum {T}heory of {M}any-{P}article {S}ystems}}}\ (\bibinfo  {publisher}
  {Courier Dover Publications})\BibitemShut {NoStop}%
\bibitem [{\citenamefont {Filip}\ \emph {et~al.}(2015)\citenamefont {Filip},
  \citenamefont {Verdi},\ and\ \citenamefont
  {Giustino}}]{Filip/Verdi/Giustino:2015}%
  \BibitemOpen
  \bibfield  {author} {\bibinfo {author} {\bibnamefont {Filip}, \bibfnamefont
  {M~R}}, \bibinfo {author} {\bibfnamefont {C.}~\bibnamefont {Verdi}}, \ and\
  \bibinfo {author} {\bibfnamefont {F.}~\bibnamefont {Giustino}}} (\bibinfo
  {year} {2015}),\ \bibfield  {title} {\enquote {\bibinfo {title} {${G}{W}$
  {B}and {S}tructures and {C}arrier {E}ffective {M}asses of
  {C}{H}$_3${N}{H}$_3${P}b{I}$_3$ and {H}ypothetical {P}erovskites of the
  {T}ype {A}{P}b{I}$_3$: {A} = {N}{H}$_4$, {P}{H}$_4$, {A}s{H}$_4$, and
  {S}b{H}$_4$},}\ }\href@noop {} {\bibfield  {journal} {\bibinfo  {journal} {J.
  Phys. Chem. C}\ }\textbf {\bibinfo {volume} {119}}~(\bibinfo {number} {45}),\
  \bibinfo {pages} {25209--25219}}\BibitemShut {NoStop}%
\bibitem [{\citenamefont {Fleszar}\ and\ \citenamefont
  {Hanke}(2005)}]{Fleszar/Hanke:2005}%
  \BibitemOpen
  \bibfield  {author} {\bibinfo {author} {\bibnamefont {Fleszar}, \bibfnamefont
  {A}}, \ and\ \bibinfo {author} {\bibfnamefont {W.}~\bibnamefont {Hanke}}}
  (\bibinfo {year} {2005}),\ \bibfield  {title} {\enquote {\bibinfo {title}
  {{E}lectronic structure of {I}{I}{B}-{V}{I} semiconductors in the ${G}{W}$
  approximation},}\ }\href@noop {} {\bibfield  {journal} {\bibinfo  {journal}
  {Phys. Rev. B}\ }\textbf {\bibinfo {volume} {71}},\ \bibinfo {pages}
  {045207}}\BibitemShut {NoStop}%
\bibitem [{\citenamefont {Foerster}\ \emph {et~al.}(2011)\citenamefont
  {Foerster}, \citenamefont {Koval},\ and\ \citenamefont
  {S{\'a}nchez-Portal}}]{Foerster/etal:2011}%
  \BibitemOpen
  \bibfield  {author} {\bibinfo {author} {\bibnamefont {Foerster},
  \bibfnamefont {D}}, \bibinfo {author} {\bibfnamefont {P.}~\bibnamefont
  {Koval}}, \ and\ \bibinfo {author} {\bibfnamefont {D.}~\bibnamefont
  {S{\'a}nchez-Portal}}} (\bibinfo {year} {2011}),\ \bibfield  {title}
  {\enquote {\bibinfo {title} {{A}n {O}({N}$^3$) implementation of {H}edin's
  \textit{GW} approximation for molecules},}\ }\href@noop {} {\bibfield
  {journal} {\bibinfo  {journal} {J. Chem. Phys.}\ }\textbf {\bibinfo {volume}
  {135}},\ \bibinfo {pages} {074105}}\BibitemShut {NoStop}%
\bibitem [{\citenamefont {Fonari}\ \emph {et~al.}(2014)\citenamefont {Fonari},
  \citenamefont {Sutton}, \citenamefont {Br\'edas},\ and\ \citenamefont
  {Coropceanu}}]{Fonari2014}%
  \BibitemOpen
  \bibfield  {author} {\bibinfo {author} {\bibnamefont {Fonari}, \bibfnamefont
  {A}}, \bibinfo {author} {\bibfnamefont {C.}~\bibnamefont {Sutton}}, \bibinfo
  {author} {\bibfnamefont {J.-L.}\ \bibnamefont {Br\'edas}}, \ and\ \bibinfo
  {author} {\bibfnamefont {V.}~\bibnamefont {Coropceanu}}} (\bibinfo {year}
  {2014}),\ \bibfield  {title} {\enquote {\bibinfo {title} {{I}mpact of exact
  exchange in the description of the electronic structure of organic
  charge-transfer molecular crystals},}\ }\href {\doibase
  10.1103/PhysRevB.90.165205} {\bibfield  {journal} {\bibinfo  {journal} {Phys.
  Rev. B}\ }\textbf {\bibinfo {volume} {90}},\ \bibinfo {pages}
  {165205}}\BibitemShut {NoStop}%
\bibitem [{\citenamefont {Fratesi}\ \emph {et~al.}(2003)\citenamefont
  {Fratesi}, \citenamefont {Brivio}, \citenamefont {Rinke},\ and\ \citenamefont
  {Godby}}]{Fratesi/Brivio/Rinke/Godby}%
  \BibitemOpen
  \bibfield  {author} {\bibinfo {author} {\bibnamefont {Fratesi}, \bibfnamefont
  {G}}, \bibinfo {author} {\bibfnamefont {G.~P.}\ \bibnamefont {Brivio}},
  \bibinfo {author} {\bibfnamefont {P.}~\bibnamefont {Rinke}}, \ and\ \bibinfo
  {author} {\bibfnamefont {R.}~\bibnamefont {Godby}}} (\bibinfo {year}
  {2003}),\ \bibfield  {title} {\enquote {\bibinfo {title} {{I}mage resonance
  in the many-body density of states at a metal surface},}\ }\href@noop {}
  {\bibfield  {journal} {\bibinfo  {journal} {Phys.\ Rev.\ B}\ }\textbf
  {\bibinfo {volume} {68}},\ \bibinfo {pages} {195404}}\BibitemShut {NoStop}%
\bibitem [{\citenamefont {Freysoldt}\ \emph {et~al.}(2007)\citenamefont
  {Freysoldt}, \citenamefont {Eggert}, \citenamefont {Rinke}, \citenamefont
  {Schindlmayr}, \citenamefont {Godby},\ and\ \citenamefont
  {Scheffler}}]{GW_space-time_method_surf:2007}%
  \BibitemOpen
  \bibfield  {author} {\bibinfo {author} {\bibnamefont {Freysoldt},
  \bibfnamefont {C}}, \bibinfo {author} {\bibfnamefont {P.}~\bibnamefont
  {Eggert}}, \bibinfo {author} {\bibfnamefont {P.}~\bibnamefont {Rinke}},
  \bibinfo {author} {\bibfnamefont {A.}~\bibnamefont {Schindlmayr}}, \bibinfo
  {author} {\bibfnamefont {R.~W.}\ \bibnamefont {Godby}}, \ and\ \bibinfo
  {author} {\bibfnamefont {M.}~\bibnamefont {Scheffler}}} (\bibinfo {year}
  {2007}),\ \bibfield  {title} {\enquote {\bibinfo {title} {{D}ielectric
  anisotropy in the \textit{GW} space-time method},}\ }\href@noop {} {\bibfield
   {journal} {\bibinfo  {journal} {Comput. Phys. Commun.}\ }\textbf {\bibinfo
  {volume} {176}},\ \bibinfo {pages} {1}}\BibitemShut {NoStop}%
\bibitem [{\citenamefont {Freysoldt}\ \emph {et~al.}(2008)\citenamefont
  {Freysoldt}, \citenamefont {Eggert}, \citenamefont {Rinke}, \citenamefont
  {Schindlmayr},\ and\ \citenamefont {Scheffler}}]{slabgw}%
  \BibitemOpen
  \bibfield  {author} {\bibinfo {author} {\bibnamefont {Freysoldt},
  \bibfnamefont {C}}, \bibinfo {author} {\bibfnamefont {P.}~\bibnamefont
  {Eggert}}, \bibinfo {author} {\bibfnamefont {P.}~\bibnamefont {Rinke}},
  \bibinfo {author} {\bibfnamefont {A.}~\bibnamefont {Schindlmayr}}, \ and\
  \bibinfo {author} {\bibfnamefont {M.}~\bibnamefont {Scheffler}}} (\bibinfo
  {year} {2008}),\ \bibfield  {title} {\enquote {\bibinfo {title} {{S}creening
  in 2{D}: \textit{GW} calculations for surfaces and thin films using the
  repeated-slab approach},}\ }\href@noop {} {\bibfield  {journal} {\bibinfo
  {journal} {Phys. Rev. B}\ }\textbf {\bibinfo {volume} {77}},\ \bibinfo
  {pages} {235428}}\BibitemShut {NoStop}%
\bibitem [{\citenamefont {Freysoldt}\ \emph {et~al.}(2009)\citenamefont
  {Freysoldt}, \citenamefont {Rinke},\ and\ \citenamefont
  {Scheffler}}]{Freysoldt/Rinke/Scheffler:2009}%
  \BibitemOpen
  \bibfield  {author} {\bibinfo {author} {\bibnamefont {Freysoldt},
  \bibfnamefont {C}}, \bibinfo {author} {\bibfnamefont {P.}~\bibnamefont
  {Rinke}}, \ and\ \bibinfo {author} {\bibfnamefont {M.}~\bibnamefont
  {Scheffler}}} (\bibinfo {year} {2009}),\ \bibfield  {title} {\enquote
  {\bibinfo {title} {{C}ontrolling {P}olarization at {I}nsulating {S}urfaces:
  {Q}uasiparticle {C}alculations for {M}olecules {A}dsorbed on {I}nsulator
  {F}ilms},}\ }\href@noop {} {\bibfield  {journal} {\bibinfo  {journal} {Phys.\
  Rev.\ Lett.}\ }\textbf {\bibinfo {volume} {103}},\ \bibinfo {pages}
  {056803}}\BibitemShut {NoStop}%
\bibitem [{\citenamefont {Friedrich}\ \emph {et~al.}(2012)\citenamefont
  {Friedrich}, \citenamefont {Betzinger}, \citenamefont {Schlipf},
  \citenamefont {Bl\"ugel},\ and\ \citenamefont
  {Schindlmayr}}]{Friedrich/etal:2012}%
  \BibitemOpen
  \bibfield  {author} {\bibinfo {author} {\bibnamefont {Friedrich},
  \bibfnamefont {C}}, \bibinfo {author} {\bibfnamefont {M.}~\bibnamefont
  {Betzinger}}, \bibinfo {author} {\bibfnamefont {M.}~\bibnamefont {Schlipf}},
  \bibinfo {author} {\bibfnamefont {S.}~\bibnamefont {Bl\"ugel}}, \ and\
  \bibinfo {author} {\bibfnamefont {A.}~\bibnamefont {Schindlmayr}}} (\bibinfo
  {year} {2012}),\ \bibfield  {title} {\enquote {\bibinfo {title} {{H}ybrid
  functionals and ${G}{W}$ approximation in the {F}{L}{A}{P}{W} method},}\
  }\href@noop {} {\bibfield  {journal} {\bibinfo  {journal} {J. Phys.: Condens.
  Matter}\ }\textbf {\bibinfo {volume} {24}}~(\bibinfo {number} {29}),\
  \bibinfo {pages} {293201}}\BibitemShut {NoStop}%
\bibitem [{\citenamefont {Friedrich}\ \emph {et~al.}(2010)\citenamefont
  {Friedrich}, \citenamefont {Bl\"ugel},\ and\ \citenamefont
  {Schindlmayr}}]{Friedrich2010}%
  \BibitemOpen
  \bibfield  {author} {\bibinfo {author} {\bibnamefont {Friedrich},
  \bibfnamefont {C}}, \bibinfo {author} {\bibfnamefont {S.}~\bibnamefont
  {Bl\"ugel}}, \ and\ \bibinfo {author} {\bibfnamefont {A.}~\bibnamefont
  {Schindlmayr}}} (\bibinfo {year} {2010}),\ \bibfield  {title} {\enquote
  {\bibinfo {title} {{E}fficient implementation of the ${G}{W}$ approximation
  within the all-electron {F}{L}{A}{P}{W} method},}\ }\href {\doibase
  10.1103/PhysRevB.81.125102} {\bibfield  {journal} {\bibinfo  {journal} {Phys.
  Rev. B}\ }\textbf {\bibinfo {volume} {81}},\ \bibinfo {pages}
  {125102}}\BibitemShut {NoStop}%
\bibitem [{\citenamefont {Friedrich}\ \emph {et~al.}(2011)\citenamefont
  {Friedrich}, \citenamefont {M\"uller},\ and\ \citenamefont
  {Bl\"ugel}}]{Friedrich2011}%
  \BibitemOpen
  \bibfield  {author} {\bibinfo {author} {\bibnamefont {Friedrich},
  \bibfnamefont {C}}, \bibinfo {author} {\bibfnamefont {M.~C.}\ \bibnamefont
  {M\"uller}}, \ and\ \bibinfo {author} {\bibfnamefont {S.}~\bibnamefont
  {Bl\"ugel}}} (\bibinfo {year} {2011}),\ \bibfield  {title} {\enquote
  {\bibinfo {title} {{B}and convergence and linearization error correction of
  all-electron $\mathit{GW}$ calculations: {T}he extreme case of zinc oxide},}\
  }\href {\doibase 10.1103/PhysRevB.83.081101} {\bibfield  {journal} {\bibinfo
  {journal} {Phys. Rev. B}\ }\textbf {\bibinfo {volume} {83}},\ \bibinfo
  {pages} {081101}}\BibitemShut {NoStop}%
\bibitem [{\citenamefont {Friedrich}\ \emph {et~al.}(2006)\citenamefont
  {Friedrich}, \citenamefont {Schindlmayr}, \citenamefont {Bl\"ugel},\ and\
  \citenamefont {Kotani}}]{Friedrich/etal:2006}%
  \BibitemOpen
  \bibfield  {author} {\bibinfo {author} {\bibnamefont {Friedrich},
  \bibfnamefont {C}}, \bibinfo {author} {\bibfnamefont {A.}~\bibnamefont
  {Schindlmayr}}, \bibinfo {author} {\bibfnamefont {S.}~\bibnamefont
  {Bl\"ugel}}, \ and\ \bibinfo {author} {\bibfnamefont {T.}~\bibnamefont
  {Kotani}}} (\bibinfo {year} {2006}),\ \bibfield  {title} {\enquote {\bibinfo
  {title} {{E}limination of the linearization error in ${G}{W}$ calculations
  based on the linearized augmented-plane-wave method},}\ }\href {\doibase
  10.1103/PhysRevB.74.045104} {\bibfield  {journal} {\bibinfo  {journal} {Phys.
  Rev. B}\ }\textbf {\bibinfo {volume} {74}},\ \bibinfo {pages}
  {045104}}\BibitemShut {NoStop}%
\bibitem [{\citenamefont {Fuchs}\ \emph {et~al.}(2007)\citenamefont {Fuchs},
  \citenamefont {Furthm\"uller}, \citenamefont {Bechstedt}, \citenamefont
  {Shishkin},\ and\ \citenamefont {Kresse}}]{Fuchs/etal:2007}%
  \BibitemOpen
  \bibfield  {author} {\bibinfo {author} {\bibnamefont {Fuchs}, \bibfnamefont
  {F}}, \bibinfo {author} {\bibfnamefont {J.}~\bibnamefont {Furthm\"uller}},
  \bibinfo {author} {\bibfnamefont {F.}~\bibnamefont {Bechstedt}}, \bibinfo
  {author} {\bibfnamefont {M.}~\bibnamefont {Shishkin}}, \ and\ \bibinfo
  {author} {\bibfnamefont {G.}~\bibnamefont {Kresse}}} (\bibinfo {year}
  {2007}),\ \bibfield  {title} {\enquote {\bibinfo {title} {{Q}uasiparticle
  band structure based on a generalized {K}ohn-{S}ham scheme},}\ }\href@noop {}
  {\bibfield  {journal} {\bibinfo  {journal} {Phys.\ Rev.\ B}\ }\textbf
  {\bibinfo {volume} {76}},\ \bibinfo {pages} {115109}}\BibitemShut {NoStop}%
\bibitem [{\citenamefont {Fuchs}\ and\ \citenamefont
  {Scheffler}(1999)}]{fhi98PP}%
  \BibitemOpen
  \bibfield  {author} {\bibinfo {author} {\bibnamefont {Fuchs}, \bibfnamefont
  {M}}, \ and\ \bibinfo {author} {\bibfnamefont {M.}~\bibnamefont {Scheffler}}}
  (\bibinfo {year} {1999}),\ \bibfield  {title} {\enquote {\bibinfo {title}
  {{A}b initio pseudopotentials for electronic structure calculations of
  poly-atomic systems using density-functional theory},}\ }\href@noop {}
  {\bibfield  {journal} {\bibinfo  {journal} {Comput.\ Phys.\ Commun.}\
  }\textbf {\bibinfo {volume} {119}},\ \bibinfo {pages} {67}}\BibitemShut
  {NoStop}%
\bibitem [{\citenamefont {Fuggle}\ and\ \citenamefont
  {Inglesfield}(1992)}]{IPES:1992}%
  \BibitemOpen
  \bibinfo {editor} {\bibnamefont {Fuggle}, \bibfnamefont {J~C}}, \ and\
  \bibinfo {editor} {\bibfnamefont {J.~E.}\ \bibnamefont {Inglesfield}},\ Eds.
  (\bibinfo {year} {1992}),\ \href@noop {} {\emph {\bibinfo {title}
  {{U}noccupied {E}lectronic {S}tates}}}\ (\bibinfo  {publisher}
  {Springer-Verlag})\BibitemShut {NoStop}%
\bibitem [{\citenamefont {Furthm\"uller}\ \emph {et~al.}(2005)\citenamefont
  {Furthm\"uller}, \citenamefont {Hahn}, \citenamefont {Fuchs},\ and\
  \citenamefont {Bechstedt}}]{Furthmueller/etal:2005}%
  \BibitemOpen
  \bibfield  {author} {\bibinfo {author} {\bibnamefont {Furthm\"uller},
  \bibfnamefont {J}}, \bibinfo {author} {\bibfnamefont {P.~H.}\ \bibnamefont
  {Hahn}}, \bibinfo {author} {\bibfnamefont {F.}~\bibnamefont {Fuchs}}, \ and\
  \bibinfo {author} {\bibfnamefont {F.}~\bibnamefont {Bechstedt}}} (\bibinfo
  {year} {2005}),\ \bibfield  {title} {\enquote {\bibinfo {title} {{B}and
  structures and optical spectra of {I}n{N} polymorphs: {I}nfluence of
  quasiparticle and excitonic effects},}\ }\href@noop {} {\bibfield  {journal}
  {\bibinfo  {journal} {Phys.\ Rev.\ B}\ }\textbf {\bibinfo {volume} {72}},\
  \bibinfo {pages} {205106}}\BibitemShut {NoStop}%
\bibitem [{\citenamefont {Galitskii}\ and\ \citenamefont
  {Migdal}(1958)}]{Galitskii/Migdal:1958}%
  \BibitemOpen
  \bibfield  {author} {\bibinfo {author} {\bibnamefont {Galitskii},
  \bibfnamefont {V}}, \ and\ \bibinfo {author} {\bibfnamefont {A.}~\bibnamefont
  {Migdal}}} (\bibinfo {year} {1958}),\ \bibfield  {title} {\enquote {\bibinfo
  {title} {{A}pplication of quantum field theory methods to the many body
  problem},}\ }\href@noop {} {\bibfield  {journal} {\bibinfo  {journal} {Sov.
  Phys. JETP}\ }\textbf {\bibinfo {volume} {7}}~(\bibinfo {number} {7}),\
  \bibinfo {pages} {96}}\BibitemShut {NoStop}%
\bibitem [{\citenamefont {Gallandi}\ and\ \citenamefont
  {K\"orzd\"orfer}(2015)}]{Gallandi2015}%
  \BibitemOpen
  \bibfield  {author} {\bibinfo {author} {\bibnamefont {Gallandi},
  \bibfnamefont {L}}, \ and\ \bibinfo {author} {\bibfnamefont {T.}~\bibnamefont
  {K\"orzd\"orfer}}} (\bibinfo {year} {2015}),\ \bibfield  {title} {\enquote
  {\bibinfo {title} {{L}ong-{R}ange {C}orrected {D}{F}{T} {M}eets ${G}{W}$:
  {V}ibrationally {R}esolved {P}hotoelectron {S}pectra from {F}irst
  {P}rinciples},}\ }\href@noop {} {\bibfield  {journal} {\bibinfo  {journal}
  {J. Chem. Theory Comput.}\ }\textbf {\bibinfo {volume} {11}}~(\bibinfo
  {number} {11}),\ \bibinfo {pages} {5391--5400}}\BibitemShut {NoStop}%
\bibitem [{\citenamefont {Gallandi}\ \emph {et~al.}(2016)\citenamefont
  {Gallandi}, \citenamefont {Marom}, \citenamefont {Rinke},\ and\ \citenamefont
  {K\"orzd\"orfer}}]{Gallandi2016}%
  \BibitemOpen
  \bibfield  {author} {\bibinfo {author} {\bibnamefont {Gallandi},
  \bibfnamefont {L}}, \bibinfo {author} {\bibfnamefont {N.}~\bibnamefont
  {Marom}}, \bibinfo {author} {\bibfnamefont {P.}~\bibnamefont {Rinke}}, \ and\
  \bibinfo {author} {\bibfnamefont {T.}~\bibnamefont {K\"orzd\"orfer}}}
  (\bibinfo {year} {2016}),\ \bibfield  {title} {\enquote {\bibinfo {title}
  {{A}ccurate {I}onization {P}otentials and {E}lectron {A}ffinities of
  {A}cceptor {M}olecules {I}{I}: {N}on-{E}mpirically {T}uned {L}ong-{R}ange
  {C}orrected {H}ybrid {F}unctionals},}\ }\href@noop {} {\bibfield  {journal}
  {\bibinfo  {journal} {J. Chem. Theory Comput.}\ }\textbf {\bibinfo {volume}
  {12}}~(\bibinfo {number} {2}),\ \bibinfo {pages} {605--614}}\BibitemShut
  {NoStop}%
\bibitem [{\citenamefont {Ganesan}\ \emph {et~al.}(2016)\citenamefont
  {Ganesan}, \citenamefont {Linghu}, \citenamefont {Zhang}, \citenamefont
  {Feng},\ and\ \citenamefont {Shen}}]{ganesan_apl_108}%
  \BibitemOpen
  \bibfield  {author} {\bibinfo {author} {\bibnamefont {Ganesan}, \bibfnamefont
  {V~D~S}}, \bibinfo {author} {\bibfnamefont {J.}~\bibnamefont {Linghu}},
  \bibinfo {author} {\bibfnamefont {C.}~\bibnamefont {Zhang}}, \bibinfo
  {author} {\bibfnamefont {Y.~P.}\ \bibnamefont {Feng}}, \ and\ \bibinfo
  {author} {\bibfnamefont {L.}~\bibnamefont {Shen}}} (\bibinfo {year} {2016}),\
  \bibfield  {title} {\enquote {\bibinfo {title} {{H}eterostructures of
  phosphorene and transition metal dichalcogenides for excitonic solar cells:
  {A} first-principles study},}\ }\href {\doibase 10.1063/1.4944642} {\bibfield
   {journal} {\bibinfo  {journal} {Appl. Phys. Lett.}\ }\textbf {\bibinfo
  {volume} {108}}~(\bibinfo {number} {12}),\ \bibinfo {pages}
  {122105}}\BibitemShut {NoStop}%
\bibitem [{\citenamefont {Gao}\ \emph {et~al.}(2016)\citenamefont {Gao},
  \citenamefont {Xia}, \citenamefont {Gao},\ and\ \citenamefont
  {Zhang}}]{Gao2016}%
  \BibitemOpen
  \bibfield  {author} {\bibinfo {author} {\bibnamefont {Gao}, \bibfnamefont
  {W}}, \bibinfo {author} {\bibfnamefont {W.}~\bibnamefont {Xia}}, \bibinfo
  {author} {\bibfnamefont {X.}~\bibnamefont {Gao}}, \ and\ \bibinfo {author}
  {\bibfnamefont {P.}~\bibnamefont {Zhang}}} (\bibinfo {year} {2016}),\
  \bibfield  {title} {\enquote {\bibinfo {title} {{S}peeding up ${G}{W}$
  {C}alculations to {M}eet the {C}hallenge of {L}arge {S}cale {Q}uasiparticle
  {P}redictions},}\ }\href {https://doi.org/10.1038/srep36849} {\bibfield
  {journal} {\bibinfo  {journal} {Sci. Rep.}\ }\textbf {\bibinfo {volume}
  {6}},\ \bibinfo {pages} {36849}}\BibitemShut {NoStop}%
\bibitem [{\citenamefont {Garc{\'i}a}\ \emph {et~al.}(2018)\citenamefont
  {Garc{\'i}a}, \citenamefont {Verstraete}, \citenamefont {Pouillon},\ and\
  \citenamefont {Junquera}}]{Garcia/etal:2018}%
  \BibitemOpen
  \bibfield  {author} {\bibinfo {author} {\bibnamefont {Garc{\'i}a},
  \bibfnamefont {A}}, \bibinfo {author} {\bibfnamefont {M.~J.}\ \bibnamefont
  {Verstraete}}, \bibinfo {author} {\bibfnamefont {Y.}~\bibnamefont
  {Pouillon}}, \ and\ \bibinfo {author} {\bibfnamefont {J.}~\bibnamefont
  {Junquera}}} (\bibinfo {year} {2018}),\ \bibfield  {title} {\enquote
  {\bibinfo {title} {{T}he {P}{S}{M}{L} format and library for norm-conserving
  pseudopotential data curation and interoperability},}\ }\href@noop {}
  {\bibfield  {journal} {\bibinfo  {journal} {Comput. Phys. Commun.}\ }\textbf
  {\bibinfo {volume} {227}},\ \bibinfo {pages} {51 -- 71}}\BibitemShut
  {NoStop}%
\bibitem [{\citenamefont {Garc{\'\i}a-Gonz\'alez}\ and\ \citenamefont
  {Godby}(2002)}]{Garcia-Gonzalez/Godby:2002}%
  \BibitemOpen
  \bibfield  {author} {\bibinfo {author} {\bibnamefont
  {Garc{\'\i}a-Gonz\'alez}, \bibfnamefont {P}}, \ and\ \bibinfo {author}
  {\bibfnamefont {R.~W.}\ \bibnamefont {Godby}}} (\bibinfo {year} {2002}),\
  \bibfield  {title} {\enquote {\bibinfo {title} {{M}any-{B}ody $\mathit{GW}$
  {C}alculations of {G}round-{S}tate {P}roperties: {Q}uasi-2{D} {E}lectron
  {S}ystems and van der {W}aals {F}orces},}\ }\href@noop {} {\bibfield
  {journal} {\bibinfo  {journal} {Phys. Rev. Lett.}\ }\textbf {\bibinfo
  {volume} {88}},\ \bibinfo {pages} {056406}}\BibitemShut {NoStop}%
\bibitem [{\citenamefont {Garc\'{\i}a-Gonz\'alez}\ and\ \citenamefont
  {Godby}(2001)}]{Garcia-Gonzalez/Godby:2001}%
  \BibitemOpen
  \bibfield  {author} {\bibinfo {author} {\bibnamefont
  {Garc\'{\i}a-Gonz\'alez}, \bibfnamefont {P~P}}, \ and\ \bibinfo {author}
  {\bibfnamefont {R.~W.}\ \bibnamefont {Godby}}} (\bibinfo {year} {2001}),\
  \bibfield  {title} {\enquote {\bibinfo {title} {{S}elf-consistent calculation
  of total energies of the electron gas using many-body perturbation theory},}\
  }\href@noop {} {\bibfield  {journal} {\bibinfo  {journal} {Phys. Rev. B}\
  }\textbf {\bibinfo {volume} {63}},\ \bibinfo {pages} {075112}}\BibitemShut
  {NoStop}%
\bibitem [{\citenamefont {Garc{\'i}a-Lastra}\ \emph {et~al.}(2009)\citenamefont
  {Garc{\'i}a-Lastra}, \citenamefont {Rostgaard}, \citenamefont {Rubio},\ and\
  \citenamefont {Thygesen}}]{Garica-Lastra/etal:2009}%
  \BibitemOpen
  \bibfield  {author} {\bibinfo {author} {\bibnamefont {Garc{\'i}a-Lastra},
  \bibfnamefont {J~M}}, \bibinfo {author} {\bibfnamefont {C.}~\bibnamefont
  {Rostgaard}}, \bibinfo {author} {\bibfnamefont {A.}~\bibnamefont {Rubio}}, \
  and\ \bibinfo {author} {\bibfnamefont {K.~S.}\ \bibnamefont {Thygesen}}}
  (\bibinfo {year} {2009}),\ \bibfield  {title} {\enquote {\bibinfo {title}
  {{P}olarization-induced renormalization of molecular levels at metallic and
  semiconducting surfaces},}\ }\href@noop {} {\bibfield  {journal} {\bibinfo
  {journal} {Phys. Rev. B}\ }\textbf {\bibinfo {volume} {80}},\ \bibinfo
  {pages} {245427}}\BibitemShut {NoStop}%
\bibitem [{\citenamefont {Gatti}\ \emph {et~al.}(2007)\citenamefont {Gatti},
  \citenamefont {Bruneval}, \citenamefont {Olevano},\ and\ \citenamefont
  {Reining}}]{Gatti/etal:2007}%
  \BibitemOpen
  \bibfield  {author} {\bibinfo {author} {\bibnamefont {Gatti}, \bibfnamefont
  {M}}, \bibinfo {author} {\bibfnamefont {F.}~\bibnamefont {Bruneval}},
  \bibinfo {author} {\bibfnamefont {V.}~\bibnamefont {Olevano}}, \ and\
  \bibinfo {author} {\bibfnamefont {L.}~\bibnamefont {Reining}}} (\bibinfo
  {year} {2007}),\ \bibfield  {title} {\enquote {\bibinfo {title}
  {{U}nderstanding correlations in vanadium dioxide from first principles},}\
  }\href@noop {} {\bibfield  {journal} {\bibinfo  {journal} {Phys.\ Rev.\
  Lett.}\ }\textbf {\bibinfo {volume} {99}},\ \bibinfo {pages}
  {266402}}\BibitemShut {NoStop}%
\bibitem [{\citenamefont {Gatti}\ and\ \citenamefont
  {Guzzo}(2013)}]{Gattiz/Guzzo:2013}%
  \BibitemOpen
  \bibfield  {author} {\bibinfo {author} {\bibnamefont {Gatti}, \bibfnamefont
  {M}}, \ and\ \bibinfo {author} {\bibfnamefont {M.}~\bibnamefont {Guzzo}}}
  (\bibinfo {year} {2013}),\ \bibfield  {title} {\enquote {\bibinfo {title}
  {{D}ynamical screening in correlated metals: {S}pectral properties of
  {S}r{V}{O}${}_{3}$ in the ${G}{W}$ approximation and beyond},}\ }\href
  {\doibase 10.1103/PhysRevB.87.155147} {\bibfield  {journal} {\bibinfo
  {journal} {Phys. Rev. B}\ }\textbf {\bibinfo {volume} {87}},\ \bibinfo
  {pages} {155147}}\BibitemShut {NoStop}%
\bibitem [{\citenamefont {Gatti}\ \emph {et~al.}(2015)\citenamefont {Gatti},
  \citenamefont {Panaccione},\ and\ \citenamefont {Reining}}]{Gatti2015}%
  \BibitemOpen
  \bibfield  {author} {\bibinfo {author} {\bibnamefont {Gatti}, \bibfnamefont
  {M}}, \bibinfo {author} {\bibfnamefont {G.}~\bibnamefont {Panaccione}}, \
  and\ \bibinfo {author} {\bibfnamefont {L.}~\bibnamefont {Reining}}} (\bibinfo
  {year} {2015}),\ \bibfield  {title} {\enquote {\bibinfo {title} {{E}ffects of
  {L}ow-{E}nergy {E}xcitations on {S}pectral {P}roperties at {H}igher {B}inding
  {E}nergy: {T}he {M}etal-{I}nsulator {T}ransition of {V}{O}$_{2}$},}\ }\href
  {\doibase 10.1103/PhysRevLett.114.116402} {\bibfield  {journal} {\bibinfo
  {journal} {Phys. Rev. Lett.}\ }\textbf {\bibinfo {volume} {114}},\ \bibinfo
  {pages} {116402}}\BibitemShut {NoStop}%
\bibitem [{\citenamefont {Geim}\ and\ \citenamefont
  {Novoselov}(2007)}]{geim_nm_6}%
  \BibitemOpen
  \bibfield  {author} {\bibinfo {author} {\bibnamefont {Geim}, \bibfnamefont
  {A~K}}, \ and\ \bibinfo {author} {\bibfnamefont {K.~S.}\ \bibnamefont
  {Novoselov}}} (\bibinfo {year} {2007}),\ \bibfield  {title} {\enquote
  {\bibinfo {title} {{T}he rise of graphene},}\ }\href@noop {} {\bibfield
  {journal} {\bibinfo  {journal} {Nat. Mat.}\ }\textbf {\bibinfo {volume}
  {6}},\ \bibinfo {pages} {183--191}}\BibitemShut {NoStop}%
\bibitem [{\citenamefont {Georges}\ and\ \citenamefont
  {Kotliar}(1992)}]{georges/etal:1992}%
  \BibitemOpen
  \bibfield  {author} {\bibinfo {author} {\bibnamefont {Georges}, \bibfnamefont
  {A}}, \ and\ \bibinfo {author} {\bibfnamefont {G.}~\bibnamefont {Kotliar}}}
  (\bibinfo {year} {1992}),\ \bibfield  {title} {\enquote {\bibinfo {title}
  {{H}ubbard model in infinite dimensions},}\ }\href@noop {} {\bibfield
  {journal} {\bibinfo  {journal} {Phys. Rev. B}\ }\textbf {\bibinfo {volume}
  {45}},\ \bibinfo {pages} {6479--6483}}\BibitemShut {NoStop}%
\bibitem [{\citenamefont {Gerosa}\ \emph
  {et~al.}(2018{\natexlab{a}})\citenamefont {Gerosa}, \citenamefont {Bottani},
  \citenamefont {Valentin}, \citenamefont {Onida},\ and\ \citenamefont
  {Pacchioni}}]{Gerosa/etal:2018}%
  \BibitemOpen
  \bibfield  {author} {\bibinfo {author} {\bibnamefont {Gerosa}, \bibfnamefont
  {M}}, \bibinfo {author} {\bibfnamefont {C.~E.}\ \bibnamefont {Bottani}},
  \bibinfo {author} {\bibfnamefont {C.~D.}\ \bibnamefont {Valentin}}, \bibinfo
  {author} {\bibfnamefont {G.}~\bibnamefont {Onida}}, \ and\ \bibinfo {author}
  {\bibfnamefont {G.}~\bibnamefont {Pacchioni}}} (\bibinfo {year}
  {2018}{\natexlab{a}}),\ \bibfield  {title} {\enquote {\bibinfo {title}
  {{A}ccuracy of dielectric-dependent hybrid functionals in the prediction of
  optoelectronic properties of metal oxide semiconductors: a comprehensive
  comparison with many-body ${G}{W}$ and experiments},}\ }\href@noop {}
  {\bibfield  {journal} {\bibinfo  {journal} {J. Phys.: Condens. Matter}\
  }\textbf {\bibinfo {volume} {30}}~(\bibinfo {number} {4}),\ \bibinfo {pages}
  {044003}}\BibitemShut {NoStop}%
\bibitem [{\citenamefont {Gerosa}\ \emph
  {et~al.}(2018{\natexlab{b}})\citenamefont {Gerosa}, \citenamefont {Gygi},
  \citenamefont {Govoni},\ and\ \citenamefont {Galli}}]{Gerosa/etal:2018_2}%
  \BibitemOpen
  \bibfield  {author} {\bibinfo {author} {\bibnamefont {Gerosa}, \bibfnamefont
  {M}}, \bibinfo {author} {\bibfnamefont {F.}~\bibnamefont {Gygi}}, \bibinfo
  {author} {\bibfnamefont {M.}~\bibnamefont {Govoni}}, \ and\ \bibinfo {author}
  {\bibfnamefont {G.}~\bibnamefont {Galli}}} (\bibinfo {year}
  {2018}{\natexlab{b}}),\ \bibfield  {title} {\enquote {\bibinfo {title} {{T}he
  role of defects and excess surface charges at finite temperature for
  optimizing oxide photoabsorbers},}\ }\href@noop {} {\bibfield  {journal}
  {\bibinfo  {journal} {Nat. Mater.}\ }\textbf {\bibinfo {volume}
  {17}}~(\bibinfo {number} {12}),\ \bibinfo {pages} {1122--1127}}\BibitemShut
  {NoStop}%
\bibitem [{\citenamefont {Giantomassi}\ \emph {et~al.}(2011)\citenamefont
  {Giantomassi}, \citenamefont {Stankovski}, \citenamefont {Shaltaf},
  \citenamefont {Gr\"uning}, \citenamefont {Bruneval}, \citenamefont {Rinke},\
  and\ \citenamefont {Rignanese}}]{Giantomassi/etal:2011}%
  \BibitemOpen
  \bibfield  {author} {\bibinfo {author} {\bibnamefont {Giantomassi},
  \bibfnamefont {M}}, \bibinfo {author} {\bibfnamefont {M.}~\bibnamefont
  {Stankovski}}, \bibinfo {author} {\bibfnamefont {R.}~\bibnamefont {Shaltaf}},
  \bibinfo {author} {\bibfnamefont {M.}~\bibnamefont {Gr\"uning}}, \bibinfo
  {author} {\bibfnamefont {F.}~\bibnamefont {Bruneval}}, \bibinfo {author}
  {\bibfnamefont {P.}~\bibnamefont {Rinke}}, \ and\ \bibinfo {author}
  {\bibfnamefont {G.-M.}\ \bibnamefont {Rignanese}}} (\bibinfo {year} {2011}),\
  \enquote {\bibinfo {title} {{E}lectronic {P}roperties of {I}nterfaces and
  {D}efects from {M}any-{B}ody {P}erturbation {T}heory: {R}ecent {D}evelopments
  and {A}pplications},}\ in\ \href@noop {} {\emph {\bibinfo {booktitle}
  {{A}dvanced {C}alculations for {D}efects in {M}aterials}}},\ Chap.~\bibinfo
  {chapter} {3}\ (\bibinfo  {publisher} {John Wiley \& Sons, Ltd})\ pp.\
  \bibinfo {pages} {33--60}\BibitemShut {NoStop}%
\bibitem [{\citenamefont {Giorgi}\ \emph {et~al.}(2011)\citenamefont {Giorgi},
  \citenamefont {Palummo}, \citenamefont {Chiodo},\ and\ \citenamefont
  {Yamashita}}]{Giorgi/etal:2011}%
  \BibitemOpen
  \bibfield  {author} {\bibinfo {author} {\bibnamefont {Giorgi}, \bibfnamefont
  {G}}, \bibinfo {author} {\bibfnamefont {M.}~\bibnamefont {Palummo}}, \bibinfo
  {author} {\bibfnamefont {L.}~\bibnamefont {Chiodo}}, \ and\ \bibinfo {author}
  {\bibfnamefont {K.}~\bibnamefont {Yamashita}}} (\bibinfo {year} {2011}),\
  \bibfield  {title} {\enquote {\bibinfo {title} {{E}xcitons at the (001)
  surface of anatase: {S}patial behavior and optical signatures},}\ }\href@noop
  {} {\bibfield  {journal} {\bibinfo  {journal} {Phys. Rev. B}\ }\textbf
  {\bibinfo {volume} {84}},\ \bibinfo {pages} {073404}}\BibitemShut {NoStop}%
\bibitem [{\citenamefont {Giovannantonio}\ \emph {et~al.}(2018)\citenamefont
  {Giovannantonio}, \citenamefont {Urgel}, \citenamefont {Beser}, \citenamefont
  {Yakutovich}, \citenamefont {Wilhelm}, \citenamefont {Pignedoli},
  \citenamefont {Ruffieux}, \citenamefont {Narita}, \citenamefont {M\"ullen},\
  and\ \citenamefont {Fasel}}]{DiGiovannantonio2018}%
  \BibitemOpen
  \bibfield  {author} {\bibinfo {author} {\bibnamefont {Giovannantonio},
  \bibfnamefont {M~Di}}, \bibinfo {author} {\bibfnamefont {J.~I.}\ \bibnamefont
  {Urgel}}, \bibinfo {author} {\bibfnamefont {U.}~\bibnamefont {Beser}},
  \bibinfo {author} {\bibfnamefont {A.~V.}\ \bibnamefont {Yakutovich}},
  \bibinfo {author} {\bibfnamefont {J.}~\bibnamefont {Wilhelm}}, \bibinfo
  {author} {\bibfnamefont {C.~A.}\ \bibnamefont {Pignedoli}}, \bibinfo {author}
  {\bibfnamefont {P.}~\bibnamefont {Ruffieux}}, \bibinfo {author}
  {\bibfnamefont {A.}~\bibnamefont {Narita}}, \bibinfo {author} {\bibfnamefont
  {K.}~\bibnamefont {M\"ullen}}, \ and\ \bibinfo {author} {\bibfnamefont
  {R.}~\bibnamefont {Fasel}}} (\bibinfo {year} {2018}),\ \bibfield  {title}
  {\enquote {\bibinfo {title} {{O}n-{S}urface {S}ynthesis of {I}ndenofluorene
  {P}olymers by {O}xidative {F}ive-{M}embered {R}ing {F}ormation},}\
  }\href@noop {} {\bibfield  {journal} {\bibinfo  {journal} {J. Am. Chem.
  Soc.}\ }\textbf {\bibinfo {volume} {140}}~(\bibinfo {number} {10}),\ \bibinfo
  {pages} {3532--3536}}\BibitemShut {NoStop}%
\bibitem [{\citenamefont {Giustino}\ \emph
  {et~al.}(2010{\natexlab{a}})\citenamefont {Giustino}, \citenamefont {Cohen},\
  and\ \citenamefont {Louie}}]{Giustino2010}%
  \BibitemOpen
  \bibfield  {author} {\bibinfo {author} {\bibnamefont {Giustino},
  \bibfnamefont {F}}, \bibinfo {author} {\bibfnamefont {M.~L.}\ \bibnamefont
  {Cohen}}, \ and\ \bibinfo {author} {\bibfnamefont {S.~G.}\ \bibnamefont
  {Louie}}} (\bibinfo {year} {2010}{\natexlab{a}}),\ \bibfield  {title}
  {\enquote {\bibinfo {title} {${G}{W}$ method with the self-consistent
  {S}ternheimer equation},}\ }\href {\doibase 10.1103/PhysRevB.81.115105}
  {\bibfield  {journal} {\bibinfo  {journal} {Phys. Rev. B}\ }\textbf {\bibinfo
  {volume} {81}},\ \bibinfo {pages} {115105}}\BibitemShut {NoStop}%
\bibitem [{\citenamefont {Giustino}\ \emph
  {et~al.}(2010{\natexlab{b}})\citenamefont {Giustino}, \citenamefont {Louie},\
  and\ \citenamefont {Cohen}}]{Giustino/etal:2010}%
  \BibitemOpen
  \bibfield  {author} {\bibinfo {author} {\bibnamefont {Giustino},
  \bibfnamefont {F}}, \bibinfo {author} {\bibfnamefont {S.~G.}\ \bibnamefont
  {Louie}}, \ and\ \bibinfo {author} {\bibfnamefont {M.~L.}\ \bibnamefont
  {Cohen}}} (\bibinfo {year} {2010}{\natexlab{b}}),\ \bibfield  {title}
  {\enquote {\bibinfo {title} {{E}lectron-{P}honon {R}enormalization of the
  {D}irect {B}and {G}ap of {D}iamond},}\ }\href@noop {} {\bibfield  {journal}
  {\bibinfo  {journal} {Phys. Rev. Lett.}\ }\textbf {\bibinfo {volume} {105}},\
  \bibinfo {pages} {265501}}\BibitemShut {NoStop}%
\bibitem [{\citenamefont {Godby}\ and\ \citenamefont
  {Needs}(1989)}]{Godby1989}%
  \BibitemOpen
  \bibfield  {author} {\bibinfo {author} {\bibnamefont {Godby}, \bibfnamefont
  {R~W}}, \ and\ \bibinfo {author} {\bibfnamefont {R.~J.}\ \bibnamefont
  {Needs}}} (\bibinfo {year} {1989}),\ \bibfield  {title} {\enquote {\bibinfo
  {title} {{M}etal-insulator transition in {K}ohn-{S}ham theory and
  quasiparticle theory},}\ }\href {\doibase 10.1103/PhysRevLett.62.1169}
  {\bibfield  {journal} {\bibinfo  {journal} {Phys. Rev. Lett.}\ }\textbf
  {\bibinfo {volume} {62}},\ \bibinfo {pages} {1169--1172}}\BibitemShut
  {NoStop}%
\bibitem [{\citenamefont {Godby}\ \emph {et~al.}(1986)\citenamefont {Godby},
  \citenamefont {Schl\"uter},\ and\ \citenamefont {Sham}}]{Godby/etal:1986}%
  \BibitemOpen
  \bibfield  {author} {\bibinfo {author} {\bibnamefont {Godby}, \bibfnamefont
  {R~W}}, \bibinfo {author} {\bibfnamefont {M.}~\bibnamefont {Schl\"uter}}, \
  and\ \bibinfo {author} {\bibfnamefont {L.~J.}\ \bibnamefont {Sham}}}
  (\bibinfo {year} {1986}),\ \bibfield  {title} {\enquote {\bibinfo {title}
  {{A}ccurate {E}xchange-{C}orrelation {P}otential for {S}ilicon and {I}ts
  {D}iscontinuity on {A}ddition of an {E}lectron},}\ }\href@noop {} {\bibfield
  {journal} {\bibinfo  {journal} {Phys. Rev. Lett.}\ }\textbf {\bibinfo
  {volume} {56}},\ \bibinfo {pages} {2415--2418}}\BibitemShut {NoStop}%
\bibitem [{\citenamefont {Godby}\ \emph
  {et~al.}(1987{\natexlab{a}})\citenamefont {Godby}, \citenamefont
  {Schl\"uter},\ and\ \citenamefont {Sham}}]{Godby/etal:1987b}%
  \BibitemOpen
  \bibfield  {author} {\bibinfo {author} {\bibnamefont {Godby}, \bibfnamefont
  {R~W}}, \bibinfo {author} {\bibfnamefont {M.}~\bibnamefont {Schl\"uter}}, \
  and\ \bibinfo {author} {\bibfnamefont {L.~J.}\ \bibnamefont {Sham}}}
  (\bibinfo {year} {1987}{\natexlab{a}}),\ \bibfield  {title} {\enquote
  {\bibinfo {title} {{Q}uasiparticle energies in {G}a{A}s and {A}l{A}s},}\
  }\href@noop {} {\bibfield  {journal} {\bibinfo  {journal} {Phys. Rev. B}\
  }\textbf {\bibinfo {volume} {35}},\ \bibinfo {pages}
  {4170--4171}}\BibitemShut {NoStop}%
\bibitem [{\citenamefont {Godby}\ \emph
  {et~al.}(1987{\natexlab{b}})\citenamefont {Godby}, \citenamefont
  {Schl\"uter},\ and\ \citenamefont {Sham}}]{Godby/etal:1987}%
  \BibitemOpen
  \bibfield  {author} {\bibinfo {author} {\bibnamefont {Godby}, \bibfnamefont
  {R~W}}, \bibinfo {author} {\bibfnamefont {M.}~\bibnamefont {Schl\"uter}}, \
  and\ \bibinfo {author} {\bibfnamefont {L.~J.}\ \bibnamefont {Sham}}}
  (\bibinfo {year} {1987}{\natexlab{b}}),\ \bibfield  {title} {\enquote
  {\bibinfo {title} {{T}rends in self-energy operators and their corresponding
  exchange-correlation potentials},}\ }\href@noop {} {\bibfield  {journal}
  {\bibinfo  {journal} {Phys. Rev. B}\ }\textbf {\bibinfo {volume} {36}},\
  \bibinfo {pages} {6497--6500}}\BibitemShut {NoStop}%
\bibitem [{\citenamefont {Godby}\ \emph {et~al.}(1988)\citenamefont {Godby},
  \citenamefont {Schl\"uter},\ and\ \citenamefont
  {Sham}}]{Godby/Schlueter/Sham:1988}%
  \BibitemOpen
  \bibfield  {author} {\bibinfo {author} {\bibnamefont {Godby}, \bibfnamefont
  {R~W}}, \bibinfo {author} {\bibfnamefont {M.}~\bibnamefont {Schl\"uter}}, \
  and\ \bibinfo {author} {\bibfnamefont {L.~J.}\ \bibnamefont {Sham}}}
  (\bibinfo {year} {1988}),\ \bibfield  {title} {\enquote {\bibinfo {title}
  {{S}elf-energy operators and exchange-correlation potentials in
  semiconductors},}\ }\href@noop {} {\bibfield  {journal} {\bibinfo  {journal}
  {Phys.\ Rev.\ B}\ }\textbf {\bibinfo {volume} {37}},\ \bibinfo {pages}
  {10159}}\BibitemShut {NoStop}%
\bibitem [{\citenamefont {Goedecker}\ \emph {et~al.}(1996)\citenamefont
  {Goedecker}, \citenamefont {Teter},\ and\ \citenamefont
  {Hutter}}]{Goedecker1996}%
  \BibitemOpen
  \bibfield  {author} {\bibinfo {author} {\bibnamefont {Goedecker},
  \bibfnamefont {S}}, \bibinfo {author} {\bibfnamefont {M.}~\bibnamefont
  {Teter}}, \ and\ \bibinfo {author} {\bibfnamefont {J.}~\bibnamefont
  {Hutter}}} (\bibinfo {year} {1996}),\ \bibfield  {title} {\enquote {\bibinfo
  {title} {{S}eparable dual-space {G}aussian pseudopotentials},}\ }\href
  {\doibase 10.1103/PhysRevB.54.1703} {\bibfield  {journal} {\bibinfo
  {journal} {Phys. Rev. B}\ }\textbf {\bibinfo {volume} {54}},\ \bibinfo
  {pages} {1703--1710}}\BibitemShut {NoStop}%
\bibitem [{\citenamefont {Golze}\ \emph {et~al.}(2018)\citenamefont {Golze},
  \citenamefont {Wilhelm}, \citenamefont {van Setten},\ and\ \citenamefont
  {Rinke}}]{Golze2018}%
  \BibitemOpen
  \bibfield  {author} {\bibinfo {author} {\bibnamefont {Golze}, \bibfnamefont
  {D}}, \bibinfo {author} {\bibfnamefont {J.}~\bibnamefont {Wilhelm}}, \bibinfo
  {author} {\bibfnamefont {M.~J.}\ \bibnamefont {van Setten}}, \ and\ \bibinfo
  {author} {\bibfnamefont {P.}~\bibnamefont {Rinke}}} (\bibinfo {year}
  {2018}),\ \bibfield  {title} {\enquote {\bibinfo {title} {{C}ore-{L}evel
  {B}inding {E}nergies from ${G}{W}$: {A}n {E}fficient {F}ull-{F}requency
  {A}pproach within a {L}ocalized {B}asis},}\ }\href@noop {} {\bibfield
  {journal} {\bibinfo  {journal} {J. Chem. Theory Comput.}\ }\textbf {\bibinfo
  {volume} {14}}~(\bibinfo {number} {9}),\ \bibinfo {pages}
  {4856--4869}}\BibitemShut {NoStop}%
\bibitem [{\citenamefont {Gonze}(1995)}]{Gonze1995}%
  \BibitemOpen
  \bibfield  {author} {\bibinfo {author} {\bibnamefont {Gonze}, \bibfnamefont
  {X}}} (\bibinfo {year} {1995}),\ \bibfield  {title} {\enquote {\bibinfo
  {title} {{A}diabatic density-functional perturbation theory},}\ }\href
  {\doibase 10.1103/PhysRevA.52.1096} {\bibfield  {journal} {\bibinfo
  {journal} {Phys. Rev. A}\ }\textbf {\bibinfo {volume} {52}},\ \bibinfo
  {pages} {1096--1114}}\BibitemShut {NoStop}%
\bibitem [{\citenamefont {Gonze}(1997)}]{Gonze1997}%
  \BibitemOpen
  \bibfield  {author} {\bibinfo {author} {\bibnamefont {Gonze}, \bibfnamefont
  {X}}} (\bibinfo {year} {1997}),\ \bibfield  {title} {\enquote {\bibinfo
  {title} {{F}irst-principles responses of solids to atomic displacements and
  homogeneous electric fields: {I}mplementation of a conjugate-gradient
  algorithm},}\ }\href {\doibase 10.1103/PhysRevB.55.10337} {\bibfield
  {journal} {\bibinfo  {journal} {Phys. Rev. B}\ }\textbf {\bibinfo {volume}
  {55}},\ \bibinfo {pages} {10337--10354}}\BibitemShut {NoStop}%
\bibitem [{\citenamefont {Gonze}\ \emph {et~al.}(2009)\citenamefont {Gonze},
  \citenamefont {Amadon}, \citenamefont {Anglade}, \citenamefont {Beuken},
  \citenamefont {Bottin}, \citenamefont {Boulanger}, \citenamefont {Bruneval},
  \citenamefont {Caliste}, \citenamefont {Caracas}, \citenamefont
  {C\^{o}t\'{e}}, \citenamefont {Deutsch}, \citenamefont {Genovese},
  \citenamefont {Ghosez}, \citenamefont {Giantomassi}, \citenamefont
  {Goedecker}, \citenamefont {Hamann}, \citenamefont {Hermet}, \citenamefont
  {Jollet}, \citenamefont {Jomard}, \citenamefont {Leroux}, \citenamefont
  {Mancini}, \citenamefont {Mazevet}, \citenamefont {Oliveira}, \citenamefont
  {Onida}, \citenamefont {Pouillon}, \citenamefont {Rangel}, \citenamefont
  {Rignanese}, \citenamefont {Sangalli}, \citenamefont {Shaltaf}, \citenamefont
  {Torrent}, \citenamefont {Verstraete}, \citenamefont {Zerah},\ and\
  \citenamefont {Zwanziger}}]{Gonze2009}%
  \BibitemOpen
  \bibfield  {author} {\bibinfo {author} {\bibnamefont {Gonze}, \bibfnamefont
  {X}}, \bibinfo {author} {\bibfnamefont {B.}~\bibnamefont {Amadon}}, \bibinfo
  {author} {\bibfnamefont {P.-M.}\ \bibnamefont {Anglade}}, \bibinfo {author}
  {\bibfnamefont {J.-M.}\ \bibnamefont {Beuken}}, \bibinfo {author}
  {\bibfnamefont {F.}~\bibnamefont {Bottin}}, \bibinfo {author} {\bibfnamefont
  {P.}~\bibnamefont {Boulanger}}, \bibinfo {author} {\bibfnamefont
  {F.}~\bibnamefont {Bruneval}}, \bibinfo {author} {\bibfnamefont
  {D.}~\bibnamefont {Caliste}}, \bibinfo {author} {\bibfnamefont
  {R.}~\bibnamefont {Caracas}}, \bibinfo {author} {\bibfnamefont
  {M.}~\bibnamefont {C\^{o}t\'{e}}}, \bibinfo {author} {\bibfnamefont
  {T.}~\bibnamefont {Deutsch}}, \bibinfo {author} {\bibfnamefont
  {L.}~\bibnamefont {Genovese}}, \bibinfo {author} {\bibfnamefont
  {P.}~\bibnamefont {Ghosez}}, \bibinfo {author} {\bibfnamefont
  {M.}~\bibnamefont {Giantomassi}}, \bibinfo {author} {\bibfnamefont
  {S.}~\bibnamefont {Goedecker}}, \bibinfo {author} {\bibfnamefont {D.~R.}\
  \bibnamefont {Hamann}}, \bibinfo {author} {\bibfnamefont {P.}~\bibnamefont
  {Hermet}}, \bibinfo {author} {\bibfnamefont {F.}~\bibnamefont {Jollet}},
  \bibinfo {author} {\bibfnamefont {G.}~\bibnamefont {Jomard}}, \bibinfo
  {author} {\bibfnamefont {S.}~\bibnamefont {Leroux}}, \bibinfo {author}
  {\bibfnamefont {M.}~\bibnamefont {Mancini}}, \bibinfo {author} {\bibfnamefont
  {S.}~\bibnamefont {Mazevet}}, \bibinfo {author} {\bibfnamefont {M.~J.~T.}\
  \bibnamefont {Oliveira}}, \bibinfo {author} {\bibfnamefont {G.}~\bibnamefont
  {Onida}}, \bibinfo {author} {\bibfnamefont {Y.}~\bibnamefont {Pouillon}},
  \bibinfo {author} {\bibfnamefont {T.}~\bibnamefont {Rangel}}, \bibinfo
  {author} {\bibfnamefont {G.-M.}\ \bibnamefont {Rignanese}}, \bibinfo {author}
  {\bibfnamefont {D.}~\bibnamefont {Sangalli}}, \bibinfo {author}
  {\bibfnamefont {R.}~\bibnamefont {Shaltaf}}, \bibinfo {author} {\bibfnamefont
  {M.}~\bibnamefont {Torrent}}, \bibinfo {author} {\bibfnamefont {M.~J.}\
  \bibnamefont {Verstraete}}, \bibinfo {author} {\bibfnamefont
  {G.}~\bibnamefont {Zerah}}, \ and\ \bibinfo {author} {\bibfnamefont {J.~W.}\
  \bibnamefont {Zwanziger}}} (\bibinfo {year} {2009}),\ \bibfield  {title}
  {\enquote {\bibinfo {title} {{A}{B}{I}{N}{I}{T}: {F}irst-principles approach
  to material and nanosystem properties},}\ }\href {\doibase
  https://doi.org/10.1016/j.cpc.2009.07.007} {\bibfield  {journal} {\bibinfo
  {journal} {Comput. Phys. Commun.}\ }\textbf {\bibinfo {volume}
  {180}}~(\bibinfo {number} {12}),\ \bibinfo {pages} {2582 --
  2615}}\BibitemShut {NoStop}%
\bibitem [{\citenamefont {Gonze}\ \emph {et~al.}(1990)\citenamefont {Gonze},
  \citenamefont {K{\"a}ckell},\ and\ \citenamefont
  {Scheffler}}]{Gonze/Kaeckell/Scheffler:1990}%
  \BibitemOpen
  \bibfield  {author} {\bibinfo {author} {\bibnamefont {Gonze}, \bibfnamefont
  {X}}, \bibinfo {author} {\bibfnamefont {P.}~\bibnamefont {K{\"a}ckell}}, \
  and\ \bibinfo {author} {\bibfnamefont {M.}~\bibnamefont {Scheffler}}}
  (\bibinfo {year} {1990}),\ \bibfield  {title} {\enquote {\bibinfo {title}
  {{G}host states for separable, norm-conserving, \textit{ab initio}
  pseudopotentials},}\ }\href@noop {} {\bibfield  {journal} {\bibinfo
  {journal} {Phys. Rev. B}\ }\textbf {\bibinfo {volume} {41}},\ \bibinfo
  {pages} {12264--12267}}\BibitemShut {NoStop}%
\bibitem [{\citenamefont {Govoni}\ and\ \citenamefont
  {Galli}(2015)}]{Govoni/etal:2015}%
  \BibitemOpen
  \bibfield  {author} {\bibinfo {author} {\bibnamefont {Govoni}, \bibfnamefont
  {M}}, \ and\ \bibinfo {author} {\bibfnamefont {G.}~\bibnamefont {Galli}}}
  (\bibinfo {year} {2015}),\ \bibfield  {title} {\enquote {\bibinfo {title}
  {{L}arge {S}cale ${G}{W}$ {C}alculations},}\ }\href@noop {} {\bibfield
  {journal} {\bibinfo  {journal} {J. Chem. Theory Comput.}\ }\textbf {\bibinfo
  {volume} {11}}~(\bibinfo {number} {6}),\ \bibinfo {pages}
  {2680--2696}}\BibitemShut {NoStop}%
\bibitem [{\citenamefont {Govoni}\ and\ \citenamefont
  {Galli}(2018)}]{Govoni2018}%
  \BibitemOpen
  \bibfield  {author} {\bibinfo {author} {\bibnamefont {Govoni}, \bibfnamefont
  {M}}, \ and\ \bibinfo {author} {\bibfnamefont {G.}~\bibnamefont {Galli}}}
  (\bibinfo {year} {2018}),\ \bibfield  {title} {\enquote {\bibinfo {title}
  {${G}{W}$100: {C}omparison of {M}ethods and {A}ccuracy of {R}esults
  {O}btained with the {W}{E}{S}{T} {C}ode},}\ }\href@noop {} {\bibfield
  {journal} {\bibinfo  {journal} {J. Chem. Theory Comput.}\ }\textbf {\bibinfo
  {volume} {14}}~(\bibinfo {number} {4}),\ \bibinfo {pages}
  {1895--1909}}\BibitemShut {NoStop}%
\bibitem [{\citenamefont {Gross}\ \emph {et~al.}(1991)\citenamefont {Gross},
  \citenamefont {Runge},\ and\ \citenamefont {Heinonen}}]{Gross/Runge:MPT}%
  \BibitemOpen
  \bibfield  {author} {\bibinfo {author} {\bibnamefont {Gross}, \bibfnamefont
  {E~K~U}}, \bibinfo {author} {\bibfnamefont {E.}~\bibnamefont {Runge}}, \ and\
  \bibinfo {author} {\bibfnamefont {O.}~\bibnamefont {Heinonen}}} (\bibinfo
  {year} {1991}),\ \href@noop {} {\emph {\bibinfo {title} {{M}any-{P}article
  {T}heory}}}\ (\bibinfo  {publisher} {Adam Hilger})\BibitemShut {NoStop}%
\bibitem [{\citenamefont {Grossman}\ \emph {et~al.}(2001)\citenamefont
  {Grossman}, \citenamefont {Rohlfing}, \citenamefont {Mitas}, \citenamefont
  {Louie},\ and\ \citenamefont {Cohen}}]{Grossman/etal:2001}%
  \BibitemOpen
  \bibfield  {author} {\bibinfo {author} {\bibnamefont {Grossman},
  \bibfnamefont {J~C}}, \bibinfo {author} {\bibfnamefont {M.}~\bibnamefont
  {Rohlfing}}, \bibinfo {author} {\bibfnamefont {L.}~\bibnamefont {Mitas}},
  \bibinfo {author} {\bibfnamefont {S.~G.}\ \bibnamefont {Louie}}, \ and\
  \bibinfo {author} {\bibfnamefont {M.~L.}\ \bibnamefont {Cohen}}} (\bibinfo
  {year} {2001}),\ \bibfield  {title} {\enquote {\bibinfo {title} {{H}igh
  {A}ccuracy {M}any-{B}ody {C}alculational {A}pproaches for {E}xcitations in
  {M}olecules},}\ }\href@noop {} {\bibfield  {journal} {\bibinfo  {journal}
  {Phys. Rev. Lett.}\ }\textbf {\bibinfo {volume} {86}},\ \bibinfo {pages}
  {472--475}}\BibitemShut {NoStop}%
\bibitem [{\citenamefont {Grumet}\ \emph {et~al.}(2018)\citenamefont {Grumet},
  \citenamefont {Liu}, \citenamefont {Kaltak}, \citenamefont {Klime{\^s}},\
  and\ \citenamefont {Kresse}}]{Kresse/etal:2018}%
  \BibitemOpen
  \bibfield  {author} {\bibinfo {author} {\bibnamefont {Grumet}, \bibfnamefont
  {M}}, \bibinfo {author} {\bibfnamefont {P.}~\bibnamefont {Liu}}, \bibinfo
  {author} {\bibfnamefont {M.}~\bibnamefont {Kaltak}}, \bibinfo {author}
  {\bibfnamefont {J.}~\bibnamefont {Klime{\^s}}}, \ and\ \bibinfo {author}
  {\bibfnamefont {G.}~\bibnamefont {Kresse}}} (\bibinfo {year} {2018}),\
  \bibfield  {title} {\enquote {\bibinfo {title} {{B}eyond the quasiparticle
  approximation: {F}ully self-consistent ${G}{W}$ calculations},}\ }\href@noop
  {} {\bibfield  {journal} {\bibinfo  {journal} {Phys. Rev. B}\ }\textbf
  {\bibinfo {volume} {98}},\ \bibinfo {pages} {155143}}\BibitemShut {NoStop}%
\bibitem [{\citenamefont {Gr\"uneis}\ \emph {et~al.}(2014)\citenamefont
  {Gr\"uneis}, \citenamefont {Kresse}, \citenamefont {Hinuma},\ and\
  \citenamefont {Oba}}]{Gruneis/etal:2014}%
  \BibitemOpen
  \bibfield  {author} {\bibinfo {author} {\bibnamefont {Gr\"uneis},
  \bibfnamefont {A}}, \bibinfo {author} {\bibfnamefont {G.}~\bibnamefont
  {Kresse}}, \bibinfo {author} {\bibfnamefont {Y.}~\bibnamefont {Hinuma}}, \
  and\ \bibinfo {author} {\bibfnamefont {F.}~\bibnamefont {Oba}}} (\bibinfo
  {year} {2014}),\ \bibfield  {title} {\enquote {\bibinfo {title} {{I}onization
  {P}otentials of {S}olids: {T}he {I}mportance of {V}ertex {C}orrections},}\
  }\href@noop {} {\bibfield  {journal} {\bibinfo  {journal} {Phys.\ Rev.\
  Lett.}\ }\textbf {\bibinfo {volume} {112}},\ \bibinfo {pages}
  {096401}}\BibitemShut {NoStop}%
\bibitem [{\citenamefont {Gr\"uning}\ \emph {et~al.}(2010)\citenamefont
  {Gr\"uning}, \citenamefont {Shaltaf},\ and\ \citenamefont
  {Rignanese}}]{Gruening/Shaltaf/Rignanese:2010}%
  \BibitemOpen
  \bibfield  {author} {\bibinfo {author} {\bibnamefont {Gr\"uning},
  \bibfnamefont {M}}, \bibinfo {author} {\bibfnamefont {R.}~\bibnamefont
  {Shaltaf}}, \ and\ \bibinfo {author} {\bibfnamefont {G.-M.}\ \bibnamefont
  {Rignanese}}} (\bibinfo {year} {2010}),\ \bibfield  {title} {\enquote
  {\bibinfo {title} {{Q}uasiparticle calculations of the electronic properties
  of {Z}r{O}$_{2}$ and {H}f{O}$_{2}$ polymorphs and their interface with
  {S}i},}\ }\href {\doibase 10.1103/PhysRevB.81.035330} {\bibfield  {journal}
  {\bibinfo  {journal} {Phys. Rev. B}\ }\textbf {\bibinfo {volume} {81}},\
  \bibinfo {pages} {035330}}\BibitemShut {NoStop}%
\bibitem [{\citenamefont {Gulans}\ \emph {et~al.}(2014)\citenamefont {Gulans},
  \citenamefont {Kontur}, \citenamefont {Meisenbichler}, \citenamefont {Nabok},
  \citenamefont {Pavone}, \citenamefont {Rigamonti}, \citenamefont
  {Sagmeister}, \citenamefont {Werner},\ and\ \citenamefont
  {Draxl}}]{Exciting:2014}%
  \BibitemOpen
  \bibfield  {author} {\bibinfo {author} {\bibnamefont {Gulans}, \bibfnamefont
  {A}}, \bibinfo {author} {\bibfnamefont {S.}~\bibnamefont {Kontur}}, \bibinfo
  {author} {\bibfnamefont {C.}~\bibnamefont {Meisenbichler}}, \bibinfo {author}
  {\bibfnamefont {D.}~\bibnamefont {Nabok}}, \bibinfo {author} {\bibfnamefont
  {P.}~\bibnamefont {Pavone}}, \bibinfo {author} {\bibfnamefont
  {S.}~\bibnamefont {Rigamonti}}, \bibinfo {author} {\bibfnamefont
  {S.}~\bibnamefont {Sagmeister}}, \bibinfo {author} {\bibfnamefont
  {U.}~\bibnamefont {Werner}}, \ and\ \bibinfo {author} {\bibfnamefont
  {C.}~\bibnamefont {Draxl}}} (\bibinfo {year} {2014}),\ \bibfield  {title}
  {\enquote {\bibinfo {title} {exciting: a full-potential all-electron package
  implementing density-functional theory and many-body perturbation theory},}\
  }\href {http://stacks.iop.org/0953-8984/26/i=36/a=363202} {\bibfield
  {journal} {\bibinfo  {journal} {J. Phys.: Condens. Matter}\ }\textbf
  {\bibinfo {volume} {26}}~(\bibinfo {number} {36}),\ \bibinfo {pages}
  {363202}}\BibitemShut {NoStop}%
\bibitem [{\citenamefont {Gumhalter}\ \emph {et~al.}(2016)\citenamefont
  {Gumhalter}, \citenamefont {Kova\v{c}}, \citenamefont {Caruso}, \citenamefont
  {Lambert},\ and\ \citenamefont {Giustino}}]{Gumhalter/etal:2016}%
  \BibitemOpen
  \bibfield  {author} {\bibinfo {author} {\bibnamefont {Gumhalter},
  \bibfnamefont {B}}, \bibinfo {author} {\bibfnamefont {V.}~\bibnamefont
  {Kova\v{c}}}, \bibinfo {author} {\bibfnamefont {F.}~\bibnamefont {Caruso}},
  \bibinfo {author} {\bibfnamefont {H.}~\bibnamefont {Lambert}}, \ and\
  \bibinfo {author} {\bibfnamefont {F.}~\bibnamefont {Giustino}}} (\bibinfo
  {year} {2016}),\ \bibfield  {title} {\enquote {\bibinfo {title} {{O}n the
  combined use of ${G}{W}$ approximation and cumulant expansion in the
  calculations of quasiparticle spectra: {T}he paradigm of {S}i valence
  bands},}\ }\href@noop {} {\bibfield  {journal} {\bibinfo  {journal} {Phys.
  Rev. B}\ }\textbf {\bibinfo {volume} {94}},\ \bibinfo {pages}
  {035103}}\BibitemShut {NoStop}%
\bibitem [{\citenamefont {Gunnarsson}\ \emph {et~al.}(2017)\citenamefont
  {Gunnarsson}, \citenamefont {Rohringer}, \citenamefont {Sch\"afer},
  \citenamefont {Sangiovanni},\ and\ \citenamefont
  {Toschi}}]{prl_gunnarsson_119}%
  \BibitemOpen
  \bibfield  {author} {\bibinfo {author} {\bibnamefont {Gunnarsson},
  \bibfnamefont {O}}, \bibinfo {author} {\bibfnamefont {G.}~\bibnamefont
  {Rohringer}}, \bibinfo {author} {\bibfnamefont {T.}~\bibnamefont
  {Sch\"afer}}, \bibinfo {author} {\bibfnamefont {G.}~\bibnamefont
  {Sangiovanni}}, \ and\ \bibinfo {author} {\bibfnamefont {A.}~\bibnamefont
  {Toschi}}} (\bibinfo {year} {2017}),\ \bibfield  {title} {\enquote {\bibinfo
  {title} {{B}reakdown of {T}raditional {M}any-{B}ody {T}heories for
  {C}orrelated {E}lectrons},}\ }\href@noop {} {\bibfield  {journal} {\bibinfo
  {journal} {Phys. Rev. Lett.}\ }\textbf {\bibinfo {volume} {119}},\ \bibinfo
  {pages} {056402}}\BibitemShut {NoStop}%
\bibitem [{\citenamefont {Guzzo}\ \emph
  {et~al.}(2014{\natexlab{a}})\citenamefont {Guzzo}, \citenamefont {Kas},
  \citenamefont {Sponza}, \citenamefont {Giorgetti}, \citenamefont {Sottile},
  \citenamefont {Pierucci}, \citenamefont {Silly}, \citenamefont {Sirotti},
  \citenamefont {Rehr},\ and\ \citenamefont {Reining}}]{Guzzo/etal:2014}%
  \BibitemOpen
  \bibfield  {author} {\bibinfo {author} {\bibnamefont {Guzzo}, \bibfnamefont
  {M}}, \bibinfo {author} {\bibfnamefont {J.~J.}\ \bibnamefont {Kas}}, \bibinfo
  {author} {\bibfnamefont {L.}~\bibnamefont {Sponza}}, \bibinfo {author}
  {\bibfnamefont {C.}~\bibnamefont {Giorgetti}}, \bibinfo {author}
  {\bibfnamefont {F.}~\bibnamefont {Sottile}}, \bibinfo {author} {\bibfnamefont
  {D.}~\bibnamefont {Pierucci}}, \bibinfo {author} {\bibfnamefont {M.~G.}\
  \bibnamefont {Silly}}, \bibinfo {author} {\bibfnamefont {F.}~\bibnamefont
  {Sirotti}}, \bibinfo {author} {\bibfnamefont {J.~J.}\ \bibnamefont {Rehr}}, \
  and\ \bibinfo {author} {\bibfnamefont {L.}~\bibnamefont {Reining}}} (\bibinfo
  {year} {2014}{\natexlab{a}}),\ \bibfield  {title} {\enquote {\bibinfo {title}
  {{M}ultiple satellites in materials with complex plasmon spectra: {F}rom
  graphite to graphene},}\ }\href@noop {} {\bibfield  {journal} {\bibinfo
  {journal} {Phys. Rev. B}\ }\textbf {\bibinfo {volume} {89}},\ \bibinfo
  {pages} {085425}}\BibitemShut {NoStop}%
\bibitem [{\citenamefont {Guzzo}\ \emph
  {et~al.}(2014{\natexlab{b}})\citenamefont {Guzzo}, \citenamefont {Kas},
  \citenamefont {Sponza}, \citenamefont {Giorgetti}, \citenamefont {Sottile},
  \citenamefont {Pierucci}, \citenamefont {Silly}, \citenamefont {Sirotti},
  \citenamefont {Rehr},\ and\ \citenamefont {Reining}}]{guzzo_prb_89}%
  \BibitemOpen
  \bibfield  {author} {\bibinfo {author} {\bibnamefont {Guzzo}, \bibfnamefont
  {M}}, \bibinfo {author} {\bibfnamefont {J.~J.}\ \bibnamefont {Kas}}, \bibinfo
  {author} {\bibfnamefont {L.}~\bibnamefont {Sponza}}, \bibinfo {author}
  {\bibfnamefont {C.}~\bibnamefont {Giorgetti}}, \bibinfo {author}
  {\bibfnamefont {F.}~\bibnamefont {Sottile}}, \bibinfo {author} {\bibfnamefont
  {D.}~\bibnamefont {Pierucci}}, \bibinfo {author} {\bibfnamefont {M.~G.}\
  \bibnamefont {Silly}}, \bibinfo {author} {\bibfnamefont {F.}~\bibnamefont
  {Sirotti}}, \bibinfo {author} {\bibfnamefont {J.~J.}\ \bibnamefont {Rehr}}, \
  and\ \bibinfo {author} {\bibfnamefont {L.}~\bibnamefont {Reining}}} (\bibinfo
  {year} {2014}{\natexlab{b}}),\ \bibfield  {title} {\enquote {\bibinfo {title}
  {{M}ultiple satellites in materials with complex plasmon spectra: {F}rom
  graphite to graphene},}\ }\href {\doibase 10.1103/PhysRevB.89.085425}
  {\bibfield  {journal} {\bibinfo  {journal} {Phys. Rev. B}\ }\textbf {\bibinfo
  {volume} {89}},\ \bibinfo {pages} {085425}}\BibitemShut {NoStop}%
\bibitem [{\citenamefont {Guzzo}\ \emph {et~al.}(2011)\citenamefont {Guzzo},
  \citenamefont {Lani}, \citenamefont {Sottile}, \citenamefont {Romaniello},
  \citenamefont {Gatti}, \citenamefont {Kas}, \citenamefont {Rehr},
  \citenamefont {Silly}, \citenamefont {Sirotti},\ and\ \citenamefont
  {Reining}}]{Guzzo/etal:2011}%
  \BibitemOpen
  \bibfield  {author} {\bibinfo {author} {\bibnamefont {Guzzo}, \bibfnamefont
  {M}}, \bibinfo {author} {\bibfnamefont {G.}~\bibnamefont {Lani}}, \bibinfo
  {author} {\bibfnamefont {F.}~\bibnamefont {Sottile}}, \bibinfo {author}
  {\bibfnamefont {P.}~\bibnamefont {Romaniello}}, \bibinfo {author}
  {\bibfnamefont {M.}~\bibnamefont {Gatti}}, \bibinfo {author} {\bibfnamefont
  {J.~J.}\ \bibnamefont {Kas}}, \bibinfo {author} {\bibfnamefont {J.~J.}\
  \bibnamefont {Rehr}}, \bibinfo {author} {\bibfnamefont {M.~G.}\ \bibnamefont
  {Silly}}, \bibinfo {author} {\bibfnamefont {F.}~\bibnamefont {Sirotti}}, \
  and\ \bibinfo {author} {\bibfnamefont {L.}~\bibnamefont {Reining}}} (\bibinfo
  {year} {2011}),\ \bibfield  {title} {\enquote {\bibinfo {title} {{V}alence
  {E}lectron {P}hotoemission {S}pectrum of {S}emiconductors:
  \textit{Ab~{I}nitio} {D}escription of {M}ultiple {S}atellites},}\ }\href@noop
  {} {\bibfield  {journal} {\bibinfo  {journal} {Phys. Rev. Lett.}\ }\textbf
  {\bibinfo {volume} {107}},\ \bibinfo {pages} {166401}}\BibitemShut {NoStop}%
\bibitem [{\citenamefont {GW100}(2018)}]{GW100}%
  \BibitemOpen
  \bibfield  {author} {\bibinfo {author} {\bibnamefont {GW100},}} (\bibinfo
  {year} {2018}),\ \href@noop {} {\bibinfo  {journal}
  {https://gw100.wordpress.com}\ }\BibitemShut {NoStop}%
\bibitem [{\citenamefont {Haastrup}\ \emph {et~al.}(2018)\citenamefont
  {Haastrup}, \citenamefont {Strange}, \citenamefont {Pandey}, \citenamefont
  {Deilmann}, \citenamefont {Schmidt}, \citenamefont {Hinsche}, \citenamefont
  {Gjerding}, \citenamefont {Torelli}, \citenamefont {Larsen}, \citenamefont
  {Riis-Jensen}, \citenamefont {Gath}, \citenamefont {Jacobsen}, \citenamefont
  {Mortensen}, \citenamefont {Olsen},\ and\ \citenamefont
  {Thygesen}}]{Haastrup_2018}%
  \BibitemOpen
\bibfield  {journal} {  }\bibfield  {author} {\bibinfo {author} {\bibnamefont
  {Haastrup}, \bibfnamefont {S}}, \bibinfo {author} {\bibfnamefont
  {M.}~\bibnamefont {Strange}}, \bibinfo {author} {\bibfnamefont
  {M.}~\bibnamefont {Pandey}}, \bibinfo {author} {\bibfnamefont
  {T.}~\bibnamefont {Deilmann}}, \bibinfo {author} {\bibfnamefont {P.~S.}\
  \bibnamefont {Schmidt}}, \bibinfo {author} {\bibfnamefont {N.~F.}\
  \bibnamefont {Hinsche}}, \bibinfo {author} {\bibfnamefont {M.~N.}\
  \bibnamefont {Gjerding}}, \bibinfo {author} {\bibfnamefont {D.}~\bibnamefont
  {Torelli}}, \bibinfo {author} {\bibfnamefont {P.~M.}\ \bibnamefont {Larsen}},
  \bibinfo {author} {\bibfnamefont {A.~C.}\ \bibnamefont {Riis-Jensen}},
  \bibinfo {author} {\bibfnamefont {J.}~\bibnamefont {Gath}}, \bibinfo {author}
  {\bibfnamefont {K.~W.}\ \bibnamefont {Jacobsen}}, \bibinfo {author}
  {\bibfnamefont {J.~J.}\ \bibnamefont {Mortensen}}, \bibinfo {author}
  {\bibfnamefont {T.}~\bibnamefont {Olsen}}, \ and\ \bibinfo {author}
  {\bibfnamefont {K.~S.}\ \bibnamefont {Thygesen}}} (\bibinfo {year} {2018}),\
  \bibfield  {title} {\enquote {\bibinfo {title} {{T}he {C}omputational 2{D}
  {M}aterials {D}atabase: high-throughput modeling and discovery of atomically
  thin crystals},}\ }\href {\doibase 10.1088/2053-1583/aacfc1} {\bibfield
  {journal} {\bibinfo  {journal} {2D Mater.}\ }\textbf {\bibinfo {volume}
  {5}}~(\bibinfo {number} {4}),\ \bibinfo {pages} {042002}}\BibitemShut
  {NoStop}%
\bibitem [{\citenamefont {Hadipour}\ and\ \citenamefont
  {Jafari}(2015)}]{hadipour_epjb_88}%
  \BibitemOpen
  \bibfield  {author} {\bibinfo {author} {\bibnamefont {Hadipour},
  \bibfnamefont {H}}, \ and\ \bibinfo {author} {\bibfnamefont {S.~A.}\
  \bibnamefont {Jafari}}} (\bibinfo {year} {2015}),\ \bibfield  {title}
  {\enquote {\bibinfo {title} {{T}he importance of electron correlation in
  graphene and hydrogenated graphene},}\ }\href {\doibase
  10.1140/epjb/e2015-60454-1} {\bibfield  {journal} {\bibinfo  {journal} {Eur.
  Phys. J. B}\ }\textbf {\bibinfo {volume} {88}}~(\bibinfo {number} {10}),\
  \bibinfo {pages} {270}}\BibitemShut {NoStop}%
\bibitem [{\citenamefont {Hahn}\ \emph {et~al.}(2001)\citenamefont {Hahn},
  \citenamefont {Schmidt},\ and\ \citenamefont
  {Bechstedt}}]{Hahn/Schmidt/Bechstedt:2001}%
  \BibitemOpen
  \bibfield  {author} {\bibinfo {author} {\bibnamefont {Hahn}, \bibfnamefont
  {P~H}}, \bibinfo {author} {\bibfnamefont {W.~G.}\ \bibnamefont {Schmidt}}, \
  and\ \bibinfo {author} {\bibfnamefont {F.}~\bibnamefont {Bechstedt}}}
  (\bibinfo {year} {2001}),\ \bibfield  {title} {\enquote {\bibinfo {title}
  {{B}ulk {E}xcitonic {E}ffects in {S}urface {O}ptical {S}pectra},}\
  }\href@noop {} {\bibfield  {journal} {\bibinfo  {journal} {Phys. Rev. Lett.}\
  }\textbf {\bibinfo {volume} {88}},\ \bibinfo {pages} {016402}}\BibitemShut
  {NoStop}%
\bibitem [{\citenamefont {Hamann}\ \emph {et~al.}(1979)\citenamefont {Hamann},
  \citenamefont {Schl\"uter},\ and\ \citenamefont {Chiang}}]{Hamann1979}%
  \BibitemOpen
  \bibfield  {author} {\bibinfo {author} {\bibnamefont {Hamann}, \bibfnamefont
  {D~R}}, \bibinfo {author} {\bibfnamefont {M.}~\bibnamefont {Schl\"uter}}, \
  and\ \bibinfo {author} {\bibfnamefont {C.}~\bibnamefont {Chiang}}} (\bibinfo
  {year} {1979}),\ \bibfield  {title} {\enquote {\bibinfo {title}
  {{N}orm-{C}onserving {P}seudopotentials},}\ }\href {\doibase
  10.1103/PhysRevLett.43.1494} {\bibfield  {journal} {\bibinfo  {journal}
  {Phys. Rev. Lett.}\ }\textbf {\bibinfo {volume} {43}},\ \bibinfo {pages}
  {1494--1497}}\BibitemShut {NoStop}%
\bibitem [{\citenamefont {Hanke}\ and\ \citenamefont
  {Sham}(1975)}]{hanke_prb_12}%
  \BibitemOpen
  \bibfield  {author} {\bibinfo {author} {\bibnamefont {Hanke}, \bibfnamefont
  {W}}, \ and\ \bibinfo {author} {\bibfnamefont {L.~J.}\ \bibnamefont {Sham}}}
  (\bibinfo {year} {1975}),\ \bibfield  {title} {\enquote {\bibinfo {title}
  {{L}ocal-field and excitonic effects in the optical spectrum of a covalent
  crystal},}\ }\href {\doibase 10.1103/PhysRevB.12.4501} {\bibfield  {journal}
  {\bibinfo  {journal} {Phys. Rev. B}\ }\textbf {\bibinfo {volume} {12}},\
  \bibinfo {pages} {4501--4511}}\BibitemShut {NoStop}%
\bibitem [{\citenamefont {Hanke}\ and\ \citenamefont
  {Sham}(1979)}]{hanke_prl_43}%
  \BibitemOpen
  \bibfield  {author} {\bibinfo {author} {\bibnamefont {Hanke}, \bibfnamefont
  {W}}, \ and\ \bibinfo {author} {\bibfnamefont {L.~J.}\ \bibnamefont {Sham}}}
  (\bibinfo {year} {1979}),\ \bibfield  {title} {\enquote {\bibinfo {title}
  {{M}any-{P}article {E}ffects in the {O}ptical {E}xcitations of a
  {S}emiconductor},}\ }\href {\doibase 10.1103/PhysRevLett.43.387} {\bibfield
  {journal} {\bibinfo  {journal} {Phys. Rev. Lett.}\ }\textbf {\bibinfo
  {volume} {43}},\ \bibinfo {pages} {387--390}}\BibitemShut {NoStop}%
\bibitem [{\citenamefont {Hanke}\ and\ \citenamefont
  {Sham}(1980)}]{hanke_prb_21}%
  \BibitemOpen
  \bibfield  {author} {\bibinfo {author} {\bibnamefont {Hanke}, \bibfnamefont
  {W}}, \ and\ \bibinfo {author} {\bibfnamefont {L.~J.}\ \bibnamefont {Sham}}}
  (\bibinfo {year} {1980}),\ \bibfield  {title} {\enquote {\bibinfo {title}
  {{M}any-particle effects in the optical spectrum of a semiconductor},}\
  }\href {\doibase 10.1103/PhysRevB.21.4656} {\bibfield  {journal} {\bibinfo
  {journal} {Phys. Rev. B}\ }\textbf {\bibinfo {volume} {21}},\ \bibinfo
  {pages} {4656--4673}}\BibitemShut {NoStop}%
\bibitem [{\citenamefont {Hautier}\ \emph {et~al.}(2013)\citenamefont
  {Hautier}, \citenamefont {Miglio}, \citenamefont {Ceder}, \citenamefont
  {Rignanese},\ and\ \citenamefont {X.}}]{Hautier/etal:2013}%
  \BibitemOpen
  \bibfield  {author} {\bibinfo {author} {\bibnamefont {Hautier}, \bibfnamefont
  {G}}, \bibinfo {author} {\bibfnamefont {A.}~\bibnamefont {Miglio}}, \bibinfo
  {author} {\bibfnamefont {G.}~\bibnamefont {Ceder}}, \bibinfo {author}
  {\bibfnamefont {G.-M.}\ \bibnamefont {Rignanese}}, \ and\ \bibinfo {author}
  {\bibnamefont {X.}}} (\bibinfo {year} {2013}),\ \bibfield  {title} {\enquote
  {\bibinfo {title} {{I}dentification and design principles of low hole
  effective mass p-type transparent conducting oxides},}\ }\href@noop {}
  {\bibfield  {journal} {\bibinfo  {journal} {Nat. Commun.}\ }\textbf {\bibinfo
  {volume} {4}},\ \bibinfo {pages} {2292}}\BibitemShut {NoStop}%
\bibitem [{\citenamefont {Havu}\ \emph {et~al.}(2009)\citenamefont {Havu},
  \citenamefont {Blum}, \citenamefont {Havu},\ and\ \citenamefont
  {Scheffler}}]{Havu2009}%
  \BibitemOpen
  \bibfield  {author} {\bibinfo {author} {\bibnamefont {Havu}, \bibfnamefont
  {V}}, \bibinfo {author} {\bibfnamefont {V.}~\bibnamefont {Blum}}, \bibinfo
  {author} {\bibfnamefont {P.}~\bibnamefont {Havu}}, \ and\ \bibinfo {author}
  {\bibfnamefont {M.}~\bibnamefont {Scheffler}}} (\bibinfo {year} {2009}),\
  \bibfield  {title} {\enquote {\bibinfo {title} {{E}fficient integration for
  all-electron electronic structure calculation using numeric basis
  functions},}\ }\href@noop {} {\bibfield  {journal} {\bibinfo  {journal} {J.
  Comput. Phys.}\ }\textbf {\bibinfo {volume} {228}}~(\bibinfo {number} {22}),\
  \bibinfo {pages} {8367--8379}}\BibitemShut {NoStop}%
\bibitem [{\citenamefont {Hedin}(1965)}]{Hedin:1965}%
  \BibitemOpen
  \bibfield  {author} {\bibinfo {author} {\bibnamefont {Hedin}, \bibfnamefont
  {L}}} (\bibinfo {year} {1965}),\ \bibfield  {title} {\enquote {\bibinfo
  {title} {{N}ew {M}ethod for {C}alculating the {O}ne-{P}article {G}reen's
  {F}unction with {A}pplication to the {E}lectron-{G}as {P}roblem},}\
  }\href@noop {} {\bibfield  {journal} {\bibinfo  {journal} {Phys.\ Rev.}\
  }\textbf {\bibinfo {volume} {139}},\ \bibinfo {pages} {A796}}\BibitemShut
  {NoStop}%
\bibitem [{\citenamefont {Hedin}(1999)}]{Hedin:1999}%
  \BibitemOpen
  \bibfield  {author} {\bibinfo {author} {\bibnamefont {Hedin}, \bibfnamefont
  {L}}} (\bibinfo {year} {1999}),\ \bibfield  {title} {\enquote {\bibinfo
  {title} {{O}n correlation effects in electron spectroscopies and the ${G}{W}$
  approximation},}\ }\href@noop {} {\bibfield  {journal} {\bibinfo  {journal}
  {J. Phys.: Condens. Matter}\ }\textbf {\bibinfo {volume} {11}},\ \bibinfo
  {pages} {R489}}\BibitemShut {NoStop}%
\bibitem [{\citenamefont {Hedin}\ and\ \citenamefont {Lee}(2002)}]{Hedin2002}%
  \BibitemOpen
  \bibfield  {author} {\bibinfo {author} {\bibnamefont {Hedin}, \bibfnamefont
  {L}}, \ and\ \bibinfo {author} {\bibfnamefont {J.}~\bibnamefont {Lee}}}
  (\bibinfo {year} {2002}),\ \bibfield  {title} {\enquote {\bibinfo {title}
  {{S}udden approximation in photoemission and beyond},}\ }\href@noop {}
  {\bibfield  {journal} {\bibinfo  {journal} {J. Electron Spectrosc. Relat.
  Phenom.}\ }\textbf {\bibinfo {volume} {124}}~(\bibinfo {number} {2}),\
  \bibinfo {pages} {289--315}},\ \bibinfo {note} {frontiers in photoemission
  spectroscopy of solids and surfaces}\BibitemShut {NoStop}%
\bibitem [{\citenamefont {Hedin}\ and\ \citenamefont
  {Lundqvist}(1970)}]{Hedin/Lundqvist:GW}%
  \BibitemOpen
  \bibfield  {author} {\bibinfo {author} {\bibnamefont {Hedin}, \bibfnamefont
  {L}}, \ and\ \bibinfo {author} {\bibfnamefont {S.}~\bibnamefont {Lundqvist}}}
  (\bibinfo {year} {1970}),\ \bibfield  {title} {\enquote {\bibinfo {title}
  {{E}ffects of {E}lectron-{E}lectron and {E}lectron-{P}honon {I}nteractions on
  the {O}ne-{E}lectron {S}tates of solids},}\ }in\ \href {\doibase
  https://doi.org/10.1016/S0081-1947(08)60615-3} {\emph {\bibinfo {booktitle}
  {{S}olid State {P}hysics}}},\ Vol.~\bibinfo {volume} {23},\ \bibinfo {editor}
  {edited by\ \bibinfo {editor} {\bibfnamefont {H.}~\bibnamefont {Ehrenreich}},
  \bibinfo {editor} {\bibfnamefont {F.}~\bibnamefont {Seitz}}, \ and\ \bibinfo
  {editor} {\bibfnamefont {D.}~\bibnamefont {Turnbull}}}\ (\bibinfo
  {publisher} {Academic Press, New York})\ pp.\ \bibinfo {pages}
  {1--181}\BibitemShut {NoStop}%
\bibitem [{\citenamefont {Hedstr\"om}\ \emph {et~al.}(2002)\citenamefont
  {Hedstr\"om}, \citenamefont {Schindlmayr},\ and\ \citenamefont
  {Scheffler}}]{Hedstroem/Schindlmayr/Scheffler:2002}%
  \BibitemOpen
  \bibfield  {author} {\bibinfo {author} {\bibnamefont {Hedstr\"om},
  \bibfnamefont {M}}, \bibinfo {author} {\bibfnamefont {A.}~\bibnamefont
  {Schindlmayr}}, \ and\ \bibinfo {author} {\bibfnamefont {M.}~\bibnamefont
  {Scheffler}}} (\bibinfo {year} {2002}),\ \bibfield  {title} {\enquote
  {\bibinfo {title} {{Q}uasiparticle calculations for point defects on
  semiconductor surfaces},}\ }\href@noop {} {\bibfield  {journal} {\bibinfo
  {journal} {Phys. Status Solidi}\ }\textbf {\bibinfo {volume} {234}},\
  \bibinfo {pages} {346}}\BibitemShut {NoStop}%
\bibitem [{\citenamefont {Hedstr\"om}\ \emph {et~al.}(2006)\citenamefont
  {Hedstr\"om}, \citenamefont {Schindlmayr}, \citenamefont {Schwarz},\ and\
  \citenamefont {Scheffler}}]{Hedstroem/etal:2006}%
  \BibitemOpen
  \bibfield  {author} {\bibinfo {author} {\bibnamefont {Hedstr\"om},
  \bibfnamefont {M}}, \bibinfo {author} {\bibfnamefont {A.}~\bibnamefont
  {Schindlmayr}}, \bibinfo {author} {\bibfnamefont {G.}~\bibnamefont
  {Schwarz}}, \ and\ \bibinfo {author} {\bibfnamefont {M.}~\bibnamefont
  {Scheffler}}} (\bibinfo {year} {2006}),\ \bibfield  {title} {\enquote
  {\bibinfo {title} {{Q}uasiparticle {C}orrections to the {E}lectronic
  {P}roperties of {A}nion {V}acancies at {G}a{A}s(110) and {I}n{P}(110)},}\
  }\href@noop {} {\bibfield  {journal} {\bibinfo  {journal} {Phys.\ Rev.\
  Lett.}\ }\textbf {\bibinfo {volume} {97}},\ \bibinfo {pages}
  {226401}}\BibitemShut {NoStop}%
\bibitem [{\citenamefont {Held}\ \emph {et~al.}(2011)\citenamefont {Held},
  \citenamefont {Taranto}, \citenamefont {Rohringer},\ and\ \citenamefont
  {Toschi}}]{Held/etal:2011}%
  \BibitemOpen
  \bibfield  {author} {\bibinfo {author} {\bibnamefont {Held}, \bibfnamefont
  {K}}, \bibinfo {author} {\bibfnamefont {C.}~\bibnamefont {Taranto}}, \bibinfo
  {author} {\bibfnamefont {G.}~\bibnamefont {Rohringer}}, \ and\ \bibinfo
  {author} {\bibfnamefont {A.}~\bibnamefont {Toschi}}} (\bibinfo {year}
  {2011}),\ \bibfield  {title} {\enquote {\bibinfo {title} {{H}edin equations,
  ${G}{W}$, ${G}{W}$+{D}{M}{F}{T}, and all that},}\ }\href@noop {} {\bibinfo
  {journal} {arXiv:1109.3972}\ }\BibitemShut {NoStop}%
\bibitem [{\citenamefont {Helgaker}\ \emph {et~al.}(1997)\citenamefont
  {Helgaker}, \citenamefont {Klopper}, \citenamefont {Koch},\ and\
  \citenamefont {Noga}}]{Helgaker1997}%
  \BibitemOpen
\bibfield  {journal} {  }\bibfield  {author} {\bibinfo {author} {\bibnamefont
  {Helgaker}, \bibfnamefont {T}}, \bibinfo {author} {\bibfnamefont
  {W.}~\bibnamefont {Klopper}}, \bibinfo {author} {\bibfnamefont
  {H.}~\bibnamefont {Koch}}, \ and\ \bibinfo {author} {\bibfnamefont
  {J.}~\bibnamefont {Noga}}} (\bibinfo {year} {1997}),\ \bibfield  {title}
  {\enquote {\bibinfo {title} {{B}asis-set convergence of correlated
  calculations on water},}\ }\href {\doibase 10.1063/1.473863} {\bibfield
  {journal} {\bibinfo  {journal} {J. Chem. Phys.}\ }\textbf {\bibinfo {volume}
  {106}}~(\bibinfo {number} {23}),\ \bibinfo {pages} {9639--9646}}\BibitemShut
  {NoStop}%
\bibitem [{\citenamefont {Hellgren}\ and\ \citenamefont {von
  Barth}(2007)}]{Hellgren/vonBarth:2007}%
  \BibitemOpen
  \bibfield  {author} {\bibinfo {author} {\bibnamefont {Hellgren},
  \bibfnamefont {M}}, \ and\ \bibinfo {author} {\bibfnamefont {U.}~\bibnamefont
  {von Barth}}} (\bibinfo {year} {2007}),\ \bibfield  {title} {\enquote
  {\bibinfo {title} {{C}orrelation potential in density functional theory at
  the {G}{W}{A} level: {S}pherical atoms},}\ }\href@noop {} {\bibfield
  {journal} {\bibinfo  {journal} {Phys. Rev. B}\ }\textbf {\bibinfo {volume}
  {76}},\ \bibinfo {pages} {075107}}\BibitemShut {NoStop}%
\bibitem [{\citenamefont {Hellgren}\ \emph {et~al.}(2015)\citenamefont
  {Hellgren}, \citenamefont {Caruso}, \citenamefont {Rohr}, \citenamefont
  {Ren}, \citenamefont {Rubio}, \citenamefont {Scheffler},\ and\ \citenamefont
  {Rinke}}]{Hellgren/etal:2015}%
  \BibitemOpen
  \bibfield  {author} {\bibinfo {author} {\bibnamefont {Hellgren},
  \bibfnamefont {M}}, \bibinfo {author} {\bibfnamefont {F.}~\bibnamefont
  {Caruso}}, \bibinfo {author} {\bibfnamefont {D.~R.}\ \bibnamefont {Rohr}},
  \bibinfo {author} {\bibfnamefont {X.}~\bibnamefont {Ren}}, \bibinfo {author}
  {\bibfnamefont {A.}~\bibnamefont {Rubio}}, \bibinfo {author} {\bibfnamefont
  {M.}~\bibnamefont {Scheffler}}, \ and\ \bibinfo {author} {\bibfnamefont
  {P.}~\bibnamefont {Rinke}}} (\bibinfo {year} {2015}),\ \bibfield  {title}
  {\enquote {\bibinfo {title} {{S}tatic correlation and electron localization
  in molecular dimers from the self-consistent {R}{P}{A} and ${G}{W}$
  approximation},}\ }\href@noop {} {\bibfield  {journal} {\bibinfo  {journal}
  {Phys. Rev. B}\ }\textbf {\bibinfo {volume} {91}},\ \bibinfo {pages}
  {165110}}\BibitemShut {NoStop}%
\bibitem [{\citenamefont {Hesselmann}\ and\ \citenamefont
  {G\"orling}(2010)}]{Hesselmann/Gorling:2010}%
  \BibitemOpen
  \bibfield  {author} {\bibinfo {author} {\bibnamefont {Hesselmann},
  \bibfnamefont {A}}, \ and\ \bibinfo {author} {\bibfnamefont {A.}~\bibnamefont
  {G\"orling}}} (\bibinfo {year} {2010}),\ \bibfield  {title} {\enquote
  {\bibinfo {title} {{R}andom phase approximation correlation energies with
  exact {K}ohn–{S}ham exchange},}\ }\href@noop {} {\bibfield  {journal}
  {\bibinfo  {journal} {Mol. Phys.}\ }\textbf {\bibinfo {volume} {108}},\
  \bibinfo {pages} {359}}\BibitemShut {NoStop}%
\bibitem [{\citenamefont {Himpsel}(1983)}]{Himpsel:1983}%
  \BibitemOpen
  \bibfield  {author} {\bibinfo {author} {\bibnamefont {Himpsel}, \bibfnamefont
  {F~J}}} (\bibinfo {year} {1983}),\ \bibfield  {title} {\enquote {\bibinfo
  {title} {{A}ngle-{R}esolved {M}easurements of the {P}hotoemission of
  {E}lectrons in the {S}tudy of {S}olids},}\ }\href@noop {} {\bibfield
  {journal} {\bibinfo  {journal} {Adv.\ Physics}\ }\textbf {\bibinfo {volume}
  {32}},\ \bibinfo {pages} {1}}\BibitemShut {NoStop}%
\bibitem [{\citenamefont {Hirayama}\ \emph {et~al.}(2017)\citenamefont
  {Hirayama}, \citenamefont {Miyake}, \citenamefont {Imada},\ and\
  \citenamefont {Biermann}}]{Hirayama/etal:2017}%
  \BibitemOpen
  \bibfield  {author} {\bibinfo {author} {\bibnamefont {Hirayama},
  \bibfnamefont {M}}, \bibinfo {author} {\bibfnamefont {T.}~\bibnamefont
  {Miyake}}, \bibinfo {author} {\bibfnamefont {M.}~\bibnamefont {Imada}}, \
  and\ \bibinfo {author} {\bibfnamefont {S.}~\bibnamefont {Biermann}}}
  (\bibinfo {year} {2017}),\ \bibfield  {title} {\enquote {\bibinfo {title}
  {{L}ow-energy effective {H}amiltonians for correlated electron systems beyond
  density functional theory},}\ }\href@noop {} {\bibfield  {journal} {\bibinfo
  {journal} {Phys. Rev. B}\ }\textbf {\bibinfo {volume} {96}},\ \bibinfo
  {pages} {075102}}\BibitemShut {NoStop}%
\bibitem [{\citenamefont {Hogan}\ \emph {et~al.}(2013)\citenamefont {Hogan},
  \citenamefont {Palummo}, \citenamefont {Gierschner},\ and\ \citenamefont
  {Rubio}}]{Hogan/etal:2013}%
  \BibitemOpen
  \bibfield  {author} {\bibinfo {author} {\bibnamefont {Hogan}, \bibfnamefont
  {C}}, \bibinfo {author} {\bibfnamefont {M.}~\bibnamefont {Palummo}}, \bibinfo
  {author} {\bibfnamefont {J.}~\bibnamefont {Gierschner}}, \ and\ \bibinfo
  {author} {\bibfnamefont {A.}~\bibnamefont {Rubio}}} (\bibinfo {year}
  {2013}),\ \bibfield  {title} {\enquote {\bibinfo {title} {{C}orrelation
  effects in the optical spectra of porphyrin oligomer chains: {E}xciton
  confinement and length dependence},}\ }\href@noop {} {\bibfield  {journal}
  {\bibinfo  {journal} {J. Chem. Phys.}\ }\textbf {\bibinfo {volume}
  {138}}~(\bibinfo {number} {2}),\ \bibinfo {eid} {024312}}\BibitemShut
  {NoStop}%
\bibitem [{\citenamefont {Hohenberg}\ and\ \citenamefont
  {Kohn}(1964)}]{Hohenberg/Kohn:1964}%
  \BibitemOpen
  \bibfield  {author} {\bibinfo {author} {\bibnamefont {Hohenberg},
  \bibfnamefont {P}}, \ and\ \bibinfo {author} {\bibfnamefont {W.}~\bibnamefont
  {Kohn}}} (\bibinfo {year} {1964}),\ \bibfield  {title} {\enquote {\bibinfo
  {title} {{I}nhomogeneous {E}lectron {G}as},}\ }\href@noop {} {\bibfield
  {journal} {\bibinfo  {journal} {Phys.\ Rev.}\ }\textbf {\bibinfo {volume}
  {136}},\ \bibinfo {pages} {B864}}\BibitemShut {NoStop}%
\bibitem [{\citenamefont {Holm}(1999)}]{Holm:1999}%
  \BibitemOpen
  \bibfield  {author} {\bibinfo {author} {\bibnamefont {Holm}, \bibfnamefont
  {B}}} (\bibinfo {year} {1999}),\ \bibfield  {title} {\enquote {\bibinfo
  {title} {{T}otal {E}nergies from ${G}{W}$ {C}alculations},}\ }\href {\doibase
  10.1103/PhysRevLett.83.788} {\bibfield  {journal} {\bibinfo  {journal} {Phys.
  Rev. Lett.}\ }\textbf {\bibinfo {volume} {83}}~(\bibinfo {number} {4}),\
  \bibinfo {pages} {788--791}}\BibitemShut {NoStop}%
\bibitem [{\citenamefont {Holm}\ and\ \citenamefont
  {Aryasetiawan}(2000)}]{Holm:2000}%
  \BibitemOpen
  \bibfield  {author} {\bibinfo {author} {\bibnamefont {Holm}, \bibfnamefont
  {B}}, \ and\ \bibinfo {author} {\bibfnamefont {F.}~\bibnamefont
  {Aryasetiawan}}} (\bibinfo {year} {2000}),\ \bibfield  {title} {\enquote
  {\bibinfo {title} {{T}otal energy from the {G}alitskii-{M}igdal formula using
  realistic spectral functions},}\ }\href {\doibase 10.1103/PhysRevB.62.4858}
  {\bibfield  {journal} {\bibinfo  {journal} {Phys. Rev. B}\ }\textbf {\bibinfo
  {volume} {62}}~(\bibinfo {number} {8}),\ \bibinfo {pages}
  {4858--4865}}\BibitemShut {NoStop}%
\bibitem [{\citenamefont {Holm}\ and\ \citenamefont {von
  Barth}(1998)}]{Holm/vonBarth:1998}%
  \BibitemOpen
  \bibfield  {author} {\bibinfo {author} {\bibnamefont {Holm}, \bibfnamefont
  {B}}, \ and\ \bibinfo {author} {\bibfnamefont {U.}~\bibnamefont {von Barth}}}
  (\bibinfo {year} {1998}),\ \bibfield  {title} {\enquote {\bibinfo {title}
  {{F}ully self-consistent ${G}{W}$ self-energy of the electron gas},}\
  }\href@noop {} {\bibfield  {journal} {\bibinfo  {journal} {Phys. Rev. B}\
  }\textbf {\bibinfo {volume} {57}},\ \bibinfo {pages} {2108}}\BibitemShut
  {NoStop}%
\bibitem [{\citenamefont {Hora}\ and\ \citenamefont
  {Scheffler}(1984)}]{Hora/Scheffler:1984}%
  \BibitemOpen
  \bibfield  {author} {\bibinfo {author} {\bibnamefont {Hora}, \bibfnamefont
  {R}}, \ and\ \bibinfo {author} {\bibfnamefont {M.}~\bibnamefont {Scheffler}}}
  (\bibinfo {year} {1984}),\ \bibfield  {title} {\enquote {\bibinfo {title}
  {{A}ngle-resolved photoemission and the electronic structure of {P}d(111)},}\
  }\href@noop {} {\bibfield  {journal} {\bibinfo  {journal} {Phys.\ Rev.\ B}\
  }\textbf {\bibinfo {volume} {29}},\ \bibinfo {pages} {692}}\BibitemShut
  {NoStop}%
\bibitem [{\citenamefont {Huang}\ \emph {et~al.}(2013)\citenamefont {Huang},
  \citenamefont {Kang},\ and\ \citenamefont {Yang}}]{huang_apl_102}%
  \BibitemOpen
  \bibfield  {author} {\bibinfo {author} {\bibnamefont {Huang}, \bibfnamefont
  {S}}, \bibinfo {author} {\bibfnamefont {W.}~\bibnamefont {Kang}}, \ and\
  \bibinfo {author} {\bibfnamefont {L.}~\bibnamefont {Yang}}} (\bibinfo {year}
  {2013}),\ \bibfield  {title} {\enquote {\bibinfo {title} {{E}lectronic
  structure and quasiparticle bandgap of silicene structures},}\ }\href
  {\doibase 10.1063/1.4801309} {\bibfield  {journal} {\bibinfo  {journal}
  {Appl. Phys. Lett.}\ }\textbf {\bibinfo {volume} {102}}~(\bibinfo {number}
  {13}),\ \bibinfo {pages} {133106}}\BibitemShut {NoStop}%
\bibitem [{\citenamefont {Hubbard}(1963)}]{Hubbard:1963}%
  \BibitemOpen
  \bibfield  {author} {\bibinfo {author} {\bibnamefont {Hubbard}, \bibfnamefont
  {J}}} (\bibinfo {year} {1963}),\ \bibfield  {title} {\enquote {\bibinfo
  {title} {{E}lectron correlations in narrow energy bands},}\ }\href {\doibase
  10.1098/rspa.1963.0204} {\bibfield  {journal} {\bibinfo  {journal} {Proc.
  Roy. Soc. London A: Math., Phys. and Eng. Sci.}\ }\textbf {\bibinfo {volume}
  {276}}~(\bibinfo {number} {1365}),\ \bibinfo {pages} {238--257}}\BibitemShut
  {NoStop}%
\bibitem [{\citenamefont {H\"ubener}\ \emph
  {et~al.}(2012{\natexlab{a}})\citenamefont {H\"ubener}, \citenamefont
  {P\'erez-Osorio}, \citenamefont {Ordej\'on},\ and\ \citenamefont
  {Giustino}}]{Huebener2012a}%
  \BibitemOpen
  \bibfield  {author} {\bibinfo {author} {\bibnamefont {H\"ubener},
  \bibfnamefont {H}}, \bibinfo {author} {\bibfnamefont {M.~A.}\ \bibnamefont
  {P\'erez-Osorio}}, \bibinfo {author} {\bibfnamefont {P.}~\bibnamefont
  {Ordej\'on}}, \ and\ \bibinfo {author} {\bibfnamefont {F.}~\bibnamefont
  {Giustino}}} (\bibinfo {year} {2012}{\natexlab{a}}),\ \bibfield  {title}
  {\enquote {\bibinfo {title} {{D}ielectric screening in extended systems using
  the self-consistent {S}ternheimer equation and localized basis sets},}\
  }\href {\doibase 10.1103/PhysRevB.85.245125} {\bibfield  {journal} {\bibinfo
  {journal} {Phys. Rev. B}\ }\textbf {\bibinfo {volume} {85}},\ \bibinfo
  {pages} {245125}}\BibitemShut {NoStop}%
\bibitem [{\citenamefont {H\"ubener}\ \emph
  {et~al.}(2012{\natexlab{b}})\citenamefont {H\"ubener}, \citenamefont
  {P{\'e}rez-Osorio}, \citenamefont {Ordej{\'o}n},\ and\ \citenamefont
  {Giustino}}]{Huebener2012b}%
  \BibitemOpen
  \bibfield  {author} {\bibinfo {author} {\bibnamefont {H\"ubener},
  \bibfnamefont {H}}, \bibinfo {author} {\bibfnamefont {M.~A.}\ \bibnamefont
  {P{\'e}rez-Osorio}}, \bibinfo {author} {\bibfnamefont {P.}~\bibnamefont
  {Ordej{\'o}n}}, \ and\ \bibinfo {author} {\bibfnamefont {F.}~\bibnamefont
  {Giustino}}} (\bibinfo {year} {2012}{\natexlab{b}}),\ \bibfield  {title}
  {\enquote {\bibinfo {title} {{P}erformance of local orbital basis sets in the
  self-consistent {S}ternheimer method for dielectric matrices of extended
  systems},}\ }\href@noop {} {\bibfield  {journal} {\bibinfo  {journal} {Eur.
  Phys. J. B}\ }\textbf {\bibinfo {volume} {85}}~(\bibinfo {number} {9}),\
  \bibinfo {pages} {321}}\BibitemShut {NoStop}%
\bibitem [{\citenamefont {H\"user}\ \emph
  {et~al.}(2013{\natexlab{a}})\citenamefont {H\"user}, \citenamefont {Olsen},\
  and\ \citenamefont {Thygesen}}]{Huser/etal:2013}%
  \BibitemOpen
  \bibfield  {author} {\bibinfo {author} {\bibnamefont {H\"user}, \bibfnamefont
  {F}}, \bibinfo {author} {\bibfnamefont {T.}~\bibnamefont {Olsen}}, \ and\
  \bibinfo {author} {\bibfnamefont {K.~S.}\ \bibnamefont {Thygesen}}} (\bibinfo
  {year} {2013}{\natexlab{a}}),\ \bibfield  {title} {\enquote {\bibinfo {title}
  {{H}ow dielectric screening in two-dimensional crystals affects the
  convergence of excited-state calculations: {M}onolayer {M}o{S}${}_{2}$},}\
  }\href@noop {} {\bibfield  {journal} {\bibinfo  {journal} {Phys. Rev. B}\
  }\textbf {\bibinfo {volume} {88}},\ \bibinfo {pages} {245309}}\BibitemShut
  {NoStop}%
\bibitem [{\citenamefont {H\"user}\ \emph
  {et~al.}(2013{\natexlab{b}})\citenamefont {H\"user}, \citenamefont {Olsen},\
  and\ \citenamefont {Thygesen}}]{Hueser/Olsen/Thygesen:2013}%
  \BibitemOpen
  \bibfield  {author} {\bibinfo {author} {\bibnamefont {H\"user}, \bibfnamefont
  {F}}, \bibinfo {author} {\bibfnamefont {T.}~\bibnamefont {Olsen}}, \ and\
  \bibinfo {author} {\bibfnamefont {K.~S.}\ \bibnamefont {Thygesen}}} (\bibinfo
  {year} {2013}{\natexlab{b}}),\ \bibfield  {title} {\enquote {\bibinfo {title}
  {{Q}uasiparticle ${G}{W}$ calculations for solids, molecules, and
  two-dimensional materials},}\ }\href@noop {} {\bibfield  {journal} {\bibinfo
  {journal} {Phys. Rev. B}\ }\textbf {\bibinfo {volume} {87}},\ \bibinfo
  {pages} {235132}}\BibitemShut {NoStop}%
\bibitem [{\citenamefont {Hybertsen}\ and\ \citenamefont
  {Louie}(1985)}]{Hybertsen/Louie:1985}%
  \BibitemOpen
  \bibfield  {author} {\bibinfo {author} {\bibnamefont {Hybertsen},
  \bibfnamefont {M~S}}, \ and\ \bibinfo {author} {\bibfnamefont {S.~G.}\
  \bibnamefont {Louie}}} (\bibinfo {year} {1985}),\ \bibfield  {title}
  {\enquote {\bibinfo {title} {{F}irst-{P}rinciples {T}heory of
  {Q}uasiparticles: {C}alculation of {B}and {G}aps in {S}emiconductors and
  {I}nsulators},}\ }\href@noop {} {\bibfield  {journal} {\bibinfo  {journal}
  {Phys. Rev. Lett.}\ }\textbf {\bibinfo {volume} {55}},\ \bibinfo {pages}
  {1418--1421}}\BibitemShut {NoStop}%
\bibitem [{\citenamefont {Hybertsen}\ and\ \citenamefont
  {Louie}(1986)}]{Hybertsen/Louie:1986}%
  \BibitemOpen
  \bibfield  {author} {\bibinfo {author} {\bibnamefont {Hybertsen},
  \bibfnamefont {M~S}}, \ and\ \bibinfo {author} {\bibfnamefont {S.~G.}\
  \bibnamefont {Louie}}} (\bibinfo {year} {1986}),\ \bibfield  {title}
  {\enquote {\bibinfo {title} {{E}lectron correlation in semiconductors and
  insulators: {B}and gaps and quasiparticle energies},}\ }\href@noop {}
  {\bibfield  {journal} {\bibinfo  {journal} {Phys.\ Rev.\ B}\ }\textbf
  {\bibinfo {volume} {34}},\ \bibinfo {pages} {5390}}\BibitemShut {NoStop}%
\bibitem [{\citenamefont {Ismail-Beigi}(2006)}]{Sohrab:2006}%
  \BibitemOpen
  \bibfield  {author} {\bibinfo {author} {\bibnamefont {Ismail-Beigi},
  \bibfnamefont {S}}} (\bibinfo {year} {2006}),\ \bibfield  {title} {\enquote
  {\bibinfo {title} {{T}runcation of periodic image interactions for confined
  systems},}\ }\href@noop {} {\bibfield  {journal} {\bibinfo  {journal} {Phys.
  Rev. B}\ }\textbf {\bibinfo {volume} {73}},\ \bibinfo {pages}
  {233103}}\BibitemShut {NoStop}%
\bibitem [{\citenamefont {Isseroff}\ and\ \citenamefont
  {Carter}(2012)}]{Isseroff/Carter:2012}%
  \BibitemOpen
  \bibfield  {author} {\bibinfo {author} {\bibnamefont {Isseroff},
  \bibfnamefont {L~Y}}, \ and\ \bibinfo {author} {\bibfnamefont {E.~A.}\
  \bibnamefont {Carter}}} (\bibinfo {year} {2012}),\ \bibfield  {title}
  {\enquote {\bibinfo {title} {{I}mportance of reference {H}amiltonians
  containing exact exchange for accurate one-shot ${G}{W}$ calculations of
  {C}u${}_{2}${O}},}\ }\href@noop {} {\bibfield  {journal} {\bibinfo  {journal}
  {Phys. Rev. B}\ }\textbf {\bibinfo {volume} {85}},\ \bibinfo {pages}
  {235142}}\BibitemShut {NoStop}%
\bibitem [{\citenamefont {Jacquemin}\ \emph {et~al.}(2015)\citenamefont
  {Jacquemin}, \citenamefont {Duchemin},\ and\ \citenamefont
  {Blase}}]{Jacquemin2015}%
  \BibitemOpen
  \bibfield  {author} {\bibinfo {author} {\bibnamefont {Jacquemin},
  \bibfnamefont {D}}, \bibinfo {author} {\bibfnamefont {I.}~\bibnamefont
  {Duchemin}}, \ and\ \bibinfo {author} {\bibfnamefont {X.}~\bibnamefont
  {Blase}}} (\bibinfo {year} {2015}),\ \bibfield  {title} {\enquote {\bibinfo
  {title} {{B}enchmarking the {B}ethe-{S}alpeter {F}ormalism on a {S}tandard
  {O}rganic {M}olecular {S}et},}\ }\href@noop {} {\bibfield  {journal}
  {\bibinfo  {journal} {J. Chem. Theory Comput.}\ }\textbf {\bibinfo {volume}
  {11}}~(\bibinfo {number} {7}),\ \bibinfo {pages} {3290--3304}}\BibitemShut
  {NoStop}%
\bibitem [{\citenamefont {Jain}\ \emph {et~al.}(2011)\citenamefont {Jain},
  \citenamefont {Chelikowsky},\ and\ \citenamefont
  {Louie}}]{Jain/Chelikowsky/Louie:2011}%
  \BibitemOpen
  \bibfield  {author} {\bibinfo {author} {\bibnamefont {Jain}, \bibfnamefont
  {M}}, \bibinfo {author} {\bibfnamefont {J.~R.}\ \bibnamefont {Chelikowsky}},
  \ and\ \bibinfo {author} {\bibfnamefont {S.~G.}\ \bibnamefont {Louie}}}
  (\bibinfo {year} {2011}),\ \bibfield  {title} {\enquote {\bibinfo {title}
  {{Q}uasiparticle {E}xcitations and {C}harge {T}ransition {L}evels of {O}xygen
  {V}acancies in {H}afnia},}\ }\href@noop {} {\bibfield  {journal} {\bibinfo
  {journal} {Phys. Rev. Lett.}\ }\textbf {\bibinfo {volume} {107}},\ \bibinfo
  {pages} {216803}}\BibitemShut {NoStop}%
\bibitem [{\citenamefont {Jain}\ \emph {et~al.}(2014)\citenamefont {Jain},
  \citenamefont {Deslippe}, \citenamefont {Samsonidze}, \citenamefont {Cohen},
  \citenamefont {Chelikowsky},\ and\ \citenamefont {Louie}}]{Jain/etal:2014}%
  \BibitemOpen
  \bibfield  {author} {\bibinfo {author} {\bibnamefont {Jain}, \bibfnamefont
  {M}}, \bibinfo {author} {\bibfnamefont {J.}~\bibnamefont {Deslippe}},
  \bibinfo {author} {\bibfnamefont {G.}~\bibnamefont {Samsonidze}}, \bibinfo
  {author} {\bibfnamefont {M.~L.}\ \bibnamefont {Cohen}}, \bibinfo {author}
  {\bibfnamefont {J.~R.}\ \bibnamefont {Chelikowsky}}, \ and\ \bibinfo {author}
  {\bibfnamefont {S.~G.}\ \bibnamefont {Louie}}} (\bibinfo {year} {2014}),\
  \bibfield  {title} {\enquote {\bibinfo {title} {{I}mproved quasiparticle wave
  functions and mean field for ${G}_{0}{W}_{0}$ calculations: {I}nitialization
  with the {C}{O}{H}{S}{E}{X} operator},}\ }\href@noop {} {\bibfield  {journal}
  {\bibinfo  {journal} {Phys. Rev. B}\ }\textbf {\bibinfo {volume} {90}},\
  \bibinfo {pages} {115148}}\BibitemShut {NoStop}%
\bibitem [{\citenamefont {Jiang}(2018)}]{Jiang:2018}%
  \BibitemOpen
  \bibfield  {author} {\bibinfo {author} {\bibnamefont {Jiang}, \bibfnamefont
  {H}}} (\bibinfo {year} {2018}),\ \bibfield  {title} {\enquote {\bibinfo
  {title} {{R}evisiting the ${G}{W}$ approach to $d$- and $f$-electron
  oxides},}\ }\href {\doibase 10.1103/PhysRevB.97.245132} {\bibfield  {journal}
  {\bibinfo  {journal} {Phys. Rev. B}\ }\textbf {\bibinfo {volume} {97}},\
  \bibinfo {pages} {245132}}\BibitemShut {NoStop}%
\bibitem [{\citenamefont {Jiang}\ and\ \citenamefont
  {Blaha}(2016)}]{Jiang/Blaha:2016}%
  \BibitemOpen
  \bibfield  {author} {\bibinfo {author} {\bibnamefont {Jiang}, \bibfnamefont
  {H}}, \ and\ \bibinfo {author} {\bibfnamefont {P.}~\bibnamefont {Blaha}}}
  (\bibinfo {year} {2016}),\ \bibfield  {title} {\enquote {\bibinfo {title}
  {${G}{W}$ with linearized augmented plane waves extended by high-energy local
  orbitals},}\ }\href@noop {} {\bibfield  {journal} {\bibinfo  {journal} {Phys.
  Rev. B}\ }\textbf {\bibinfo {volume} {93}},\ \bibinfo {pages}
  {115203}}\BibitemShut {NoStop}%
\bibitem [{\citenamefont {Jiang}\ \emph {et~al.}(2009)\citenamefont {Jiang},
  \citenamefont {G\'omez-Abal}, \citenamefont {Rinke},\ and\ \citenamefont
  {Scheffler}}]{Jiang/etal:2009}%
  \BibitemOpen
  \bibfield  {author} {\bibinfo {author} {\bibnamefont {Jiang}, \bibfnamefont
  {H}}, \bibinfo {author} {\bibfnamefont {R.}~\bibnamefont {G\'omez-Abal}},
  \bibinfo {author} {\bibfnamefont {P.}~\bibnamefont {Rinke}}, \ and\ \bibinfo
  {author} {\bibfnamefont {M.}~\bibnamefont {Scheffler}}} (\bibinfo {year}
  {2009}),\ \bibfield  {title} {\enquote {\bibinfo {title} {{L}ocalized and
  itinerant states in lanthanide oxides united by ${G}{W}$@{L}{D}{A}+{U}},}\
  }\href@noop {} {\bibfield  {journal} {\bibinfo  {journal} {Phys. Rev. Lett.}\
  }\textbf {\bibinfo {volume} {102}},\ \bibinfo {pages} {126403}}\BibitemShut
  {NoStop}%
\bibitem [{\citenamefont {Jiang}\ \emph {et~al.}(2013)\citenamefont {Jiang},
  \citenamefont {G\'omez-Abal}, \citenamefont {Li}, \citenamefont
  {Meisenbichler}, \citenamefont {Ambrosch-Draxl},\ and\ \citenamefont
  {Scheffler}}]{Jiang/etal:2013}%
  \BibitemOpen
  \bibfield  {author} {\bibinfo {author} {\bibnamefont {Jiang}, \bibfnamefont
  {H}}, \bibinfo {author} {\bibfnamefont {R.~I.}\ \bibnamefont {G\'omez-Abal}},
  \bibinfo {author} {\bibfnamefont {X.-Z.}\ \bibnamefont {Li}}, \bibinfo
  {author} {\bibfnamefont {C.}~\bibnamefont {Meisenbichler}}, \bibinfo {author}
  {\bibfnamefont {C.}~\bibnamefont {Ambrosch-Draxl}}, \ and\ \bibinfo {author}
  {\bibfnamefont {M.}~\bibnamefont {Scheffler}}} (\bibinfo {year} {2013}),\
  \bibfield  {title} {\enquote {\bibinfo {title} {{F}{H}{I}-gap: {A} code based
  on the all-electron augmented plane wave method},}\ }\href@noop {} {\bibfield
   {journal} {\bibinfo  {journal} {Comput. Phys. Commun.}\ }\textbf {\bibinfo
  {volume} {184}}~(\bibinfo {number} {2}),\ \bibinfo {pages} {348 --
  366}}\BibitemShut {NoStop}%
\bibitem [{\citenamefont {Jiang}\ \emph
  {et~al.}(2010{\natexlab{a}})\citenamefont {Jiang}, \citenamefont
  {Gomez-Abal}, \citenamefont {Rinke},\ and\ \citenamefont
  {Scheffler}}]{Jiang/etal:2010_2}%
  \BibitemOpen
  \bibfield  {author} {\bibinfo {author} {\bibnamefont {Jiang}, \bibfnamefont
  {H}}, \bibinfo {author} {\bibfnamefont {R.~I.}\ \bibnamefont {Gomez-Abal}},
  \bibinfo {author} {\bibfnamefont {P.}~\bibnamefont {Rinke}}, \ and\ \bibinfo
  {author} {\bibfnamefont {M.}~\bibnamefont {Scheffler}}} (\bibinfo {year}
  {2010}{\natexlab{a}}),\ \bibfield  {title} {\enquote {\bibinfo {title}
  {{E}lectronic band structure of zirconia and hafnia polymorphs from the
  ${G}{W}$ perspective},}\ }\href@noop {} {\bibfield  {journal} {\bibinfo
  {journal} {Phys. Rev. B}\ }\textbf {\bibinfo {volume} {81}},\ \bibinfo
  {pages} {085119}}\BibitemShut {NoStop}%
\bibitem [{\citenamefont {Jiang}\ \emph
  {et~al.}(2010{\natexlab{b}})\citenamefont {Jiang}, \citenamefont
  {G\'omez-Abal}, \citenamefont {Rinke},\ and\ \citenamefont
  {Scheffler}}]{Jiang/etal:2010}%
  \BibitemOpen
  \bibfield  {author} {\bibinfo {author} {\bibnamefont {Jiang}, \bibfnamefont
  {H}}, \bibinfo {author} {\bibfnamefont {R.~I.}\ \bibnamefont {G\'omez-Abal}},
  \bibinfo {author} {\bibfnamefont {P.}~\bibnamefont {Rinke}}, \ and\ \bibinfo
  {author} {\bibfnamefont {M.}~\bibnamefont {Scheffler}}} (\bibinfo {year}
  {2010}{\natexlab{b}}),\ \bibfield  {title} {\enquote {\bibinfo {title}
  {{F}irst-principles modeling of localized $d$ states with the
  ${G}{W}@\text{L{D}A}+{U}$ approach},}\ }\href@noop {} {\bibfield  {journal}
  {\bibinfo  {journal} {Phys. Rev. B}\ }\textbf {\bibinfo {volume} {82}},\
  \bibinfo {pages} {045108}}\BibitemShut {NoStop}%
\bibitem [{\citenamefont {Jiang}\ \emph {et~al.}(2012)\citenamefont {Jiang},
  \citenamefont {Rinke},\ and\ \citenamefont
  {Scheffler}}]{Jiang/Rinke/Scheffler:2012}%
  \BibitemOpen
  \bibfield  {author} {\bibinfo {author} {\bibnamefont {Jiang}, \bibfnamefont
  {H}}, \bibinfo {author} {\bibfnamefont {P.}~\bibnamefont {Rinke}}, \ and\
  \bibinfo {author} {\bibfnamefont {M.}~\bibnamefont {Scheffler}}} (\bibinfo
  {year} {2012}),\ \bibfield  {title} {\enquote {\bibinfo {title} {{E}lectronic
  properties of lanthanide oxides from the ${G}{W}$ perspective},}\ }\href@noop
  {} {\bibfield  {journal} {\bibinfo  {journal} {Phys. Rev. B}\ }\textbf
  {\bibinfo {volume} {86}},\ \bibinfo {pages} {125115}}\BibitemShut {NoStop}%
\bibitem [{\citenamefont {Jin}\ \emph {et~al.}(2016)\citenamefont {Jin},
  \citenamefont {Chen}, \citenamefont {Cui}, \citenamefont {Mao}, \citenamefont
  {Zhang}, \citenamefont {Zhuang}, \citenamefont {Bao}, \citenamefont {Zhou},
  \citenamefont {Liu}, \citenamefont {Zhou},\ and\ \citenamefont
  {He}}]{Jin/etal:2016}%
  \BibitemOpen
  \bibfield  {author} {\bibinfo {author} {\bibnamefont {Jin}, \bibfnamefont
  {X}}, \bibinfo {author} {\bibfnamefont {X.-J.}\ \bibnamefont {Chen}},
  \bibinfo {author} {\bibfnamefont {T.}~\bibnamefont {Cui}}, \bibinfo {author}
  {\bibfnamefont {H.-K.}\ \bibnamefont {Mao}}, \bibinfo {author} {\bibfnamefont
  {H.}~\bibnamefont {Zhang}}, \bibinfo {author} {\bibfnamefont
  {Q.}~\bibnamefont {Zhuang}}, \bibinfo {author} {\bibfnamefont
  {K.}~\bibnamefont {Bao}}, \bibinfo {author} {\bibfnamefont {D.}~\bibnamefont
  {Zhou}}, \bibinfo {author} {\bibfnamefont {B.}~\bibnamefont {Liu}}, \bibinfo
  {author} {\bibfnamefont {Q.}~\bibnamefont {Zhou}}, \ and\ \bibinfo {author}
  {\bibfnamefont {Z.}~\bibnamefont {He}}} (\bibinfo {year} {2016}),\ \bibfield
  {title} {\enquote {\bibinfo {title} {{C}rossover from metal to insulator in
  dense lithium-rich compound {C}{L}i$_4$},}\ }\href {\doibase
  10.1073/pnas.1525412113} {\bibfield  {journal} {\bibinfo  {journal} {Proc.
  Natl. Acad. Sci. USA}\ }\textbf {\bibinfo {volume} {113}}~(\bibinfo {number}
  {9}),\ \bibinfo {pages} {2366--2369}}\BibitemShut {NoStop}%
\bibitem [{\citenamefont {Jin}\ \emph {et~al.}(2019)\citenamefont {Jin},
  \citenamefont {Su},\ and\ \citenamefont {Yang}}]{Jin2019}%
  \BibitemOpen
  \bibfield  {author} {\bibinfo {author} {\bibnamefont {Jin}, \bibfnamefont
  {Y}}, \bibinfo {author} {\bibfnamefont {N.~Q.}\ \bibnamefont {Su}}, \ and\
  \bibinfo {author} {\bibfnamefont {W.}~\bibnamefont {Yang}}} (\bibinfo {year}
  {2019}),\ \bibfield  {title} {\enquote {\bibinfo {title} {{R}enormalized
  {S}ingles {G}reen’s {F}unction for {Q}uasi-{P}article {C}alculations beyond
  the ${G}_0{W}_0$ {A}pproximation},}\ }\href {\doibase
  10.1021/acs.jpclett.8b03337} {\bibfield  {journal} {\bibinfo  {journal} {J.
  Phys. Chem. Lett.}\ }\textbf {\bibinfo {volume} {10}}~(\bibinfo {number}
  {3}),\ \bibinfo {pages} {447--452}}\BibitemShut {NoStop}%
\bibitem [{\citenamefont {Johnson}(1974)}]{Johnson1974}%
  \BibitemOpen
  \bibfield  {author} {\bibinfo {author} {\bibnamefont {Johnson}, \bibfnamefont
  {D~L}}} (\bibinfo {year} {1974}),\ \bibfield  {title} {\enquote {\bibinfo
  {title} {{L}ocal field effects and the dielectric response matrix of
  insulators: {A} model},}\ }\href {\doibase 10.1103/PhysRevB.9.4475}
  {\bibfield  {journal} {\bibinfo  {journal} {Phys. Rev. B}\ }\textbf {\bibinfo
  {volume} {9}},\ \bibinfo {pages} {4475--4484}}\BibitemShut {NoStop}%
\bibitem [{\citenamefont {Kang}\ \emph {et~al.}(2019)\citenamefont {Kang},
  \citenamefont {Kononov}, \citenamefont {Lee}, \citenamefont {Leveillee},
  \citenamefont {Shapera}, \citenamefont {Zhang},\ and\ \citenamefont
  {Schleife}}]{Kang/etal:2019}%
  \BibitemOpen
  \bibfield  {author} {\bibinfo {author} {\bibnamefont {Kang}, \bibfnamefont
  {K}}, \bibinfo {author} {\bibfnamefont {A.}~\bibnamefont {Kononov}}, \bibinfo
  {author} {\bibfnamefont {C.-W.}\ \bibnamefont {Lee}}, \bibinfo {author}
  {\bibfnamefont {J.~A.}\ \bibnamefont {Leveillee}}, \bibinfo {author}
  {\bibfnamefont {E.~P.}\ \bibnamefont {Shapera}}, \bibinfo {author}
  {\bibfnamefont {X.}~\bibnamefont {Zhang}}, \ and\ \bibinfo {author}
  {\bibfnamefont {A.}~\bibnamefont {Schleife}}} (\bibinfo {year} {2019}),\
  \bibfield  {title} {\enquote {\bibinfo {title} {{P}ushing the frontiers of
  modeling excited electronic states and dynamics to accelerate materials
  engineering and design},}\ }\href@noop {} {\bibfield  {journal} {\bibinfo
  {journal} {Comput. Mater. Sci.}\ }\textbf {\bibinfo {volume} {160}},\
  \bibinfo {pages} {207--216}}\BibitemShut {NoStop}%
\bibitem [{\citenamefont {Karlick\'{y}}\ and\ \citenamefont
  {Otyepka}(2013)}]{karlicky_jctc_9}%
  \BibitemOpen
  \bibfield  {author} {\bibinfo {author} {\bibnamefont {Karlick\'{y}},
  \bibfnamefont {F}}, \ and\ \bibinfo {author} {\bibfnamefont {M.}~\bibnamefont
  {Otyepka}}} (\bibinfo {year} {2013}),\ \bibfield  {title} {\enquote {\bibinfo
  {title} {{B}and {G}aps and {O}ptical {S}pectra of {C}hlorographene,
  {F}luorographene and {G}raphane from ${G}_0{W}_0$, ${G}{W}_0$ and ${G}{W}$
  {C}alculations on {T}op of {P}{B}{E} and {H}{S}{E}06 {O}rbitals},}\ }\href
  {\doibase 10.1021/ct400476r} {\bibfield  {journal} {\bibinfo  {journal} {J.
  Chem. Theory Comput.}\ }\textbf {\bibinfo {volume} {9}}~(\bibinfo {number}
  {9}),\ \bibinfo {pages} {4155--4164}}\BibitemShut {NoStop}%
\bibitem [{\citenamefont {Karlsson}\ and\ \citenamefont
  {Leeuwen}(2018)}]{Karlsson2018}%
  \BibitemOpen
  \bibfield  {author} {\bibinfo {author} {\bibnamefont {Karlsson},
  \bibfnamefont {D}}, \ and\ \bibinfo {author} {\bibfnamefont {R.~V.}\
  \bibnamefont {Leeuwen}}} (\bibinfo {year} {2018}),\ \enquote {\bibinfo
  {title} {{N}on-equilibrium {G}reen's {F}unctions for {C}oupled
  {F}ermion-{B}oson {S}ystems},}\ in\ \href {\doibase
  10.1007/978-3-319-42913-7_8-1} {\emph {\bibinfo {booktitle} {{H}andbook of
  {M}aterials {M}odeling : {M}ethods: {T}heory and {M}odeling}}},\ \bibinfo
  {editor} {edited by\ \bibinfo {editor} {\bibfnamefont {Wanda}\ \bibnamefont
  {Andreoni}}\ and\ \bibinfo {editor} {\bibfnamefont {Sidney}\ \bibnamefont
  {Yip}}}\ (\bibinfo  {publisher} {Springer International Publishing},\
  \bibinfo {address} {Cham})\ pp.\ \bibinfo {pages} {1--29}\BibitemShut
  {NoStop}%
\bibitem [{\citenamefont {Kas}\ \emph {et~al.}(2007)\citenamefont {Kas},
  \citenamefont {Sorini}, \citenamefont {Prange}, \citenamefont {Cambell},
  \citenamefont {Soininen},\ and\ \citenamefont {Rehr}}]{Kas2007}%
  \BibitemOpen
  \bibfield  {author} {\bibinfo {author} {\bibnamefont {Kas}, \bibfnamefont
  {J~J}}, \bibinfo {author} {\bibfnamefont {A.~P.}\ \bibnamefont {Sorini}},
  \bibinfo {author} {\bibfnamefont {M.~P.}\ \bibnamefont {Prange}}, \bibinfo
  {author} {\bibfnamefont {L.~W.}\ \bibnamefont {Cambell}}, \bibinfo {author}
  {\bibfnamefont {J.~A.}\ \bibnamefont {Soininen}}, \ and\ \bibinfo {author}
  {\bibfnamefont {J.~J.}\ \bibnamefont {Rehr}}} (\bibinfo {year} {2007}),\
  \bibfield  {title} {\enquote {\bibinfo {title} {{M}any-pole model of
  inelastic losses in x-ray absorption spectra},}\ }\href {\doibase
  10.1103/PhysRevB.76.195116} {\bibfield  {journal} {\bibinfo  {journal} {Phys.
  Rev. B}\ }\textbf {\bibinfo {volume} {76}},\ \bibinfo {pages}
  {195116}}\BibitemShut {NoStop}%
\bibitem [{\citenamefont {Kawai}\ \emph {et~al.}(2014)\citenamefont {Kawai},
  \citenamefont {Yamashita}, \citenamefont {Cannuccia},\ and\ \citenamefont
  {Marini}}]{Kawai/etal:2014}%
  \BibitemOpen
  \bibfield  {author} {\bibinfo {author} {\bibnamefont {Kawai}, \bibfnamefont
  {H}}, \bibinfo {author} {\bibfnamefont {K.}~\bibnamefont {Yamashita}},
  \bibinfo {author} {\bibfnamefont {E.}~\bibnamefont {Cannuccia}}, \ and\
  \bibinfo {author} {\bibfnamefont {A.}~\bibnamefont {Marini}}} (\bibinfo
  {year} {2014}),\ \bibfield  {title} {\enquote {\bibinfo {title}
  {{E}lectron-electron and electron-phonon correlation effects on the
  finite-temperature electronic and optical properties of zinc-blende
  {G}a{N}},}\ }\href@noop {} {\bibfield  {journal} {\bibinfo  {journal} {Phys.
  Rev. B}\ }\textbf {\bibinfo {volume} {89}},\ \bibinfo {pages}
  {085202}}\BibitemShut {NoStop}%
\bibitem [{\citenamefont {Ke}(2011)}]{KeSanhuang:2011}%
  \BibitemOpen
  \bibfield  {author} {\bibinfo {author} {\bibnamefont {Ke}, \bibfnamefont
  {S-H}}} (\bibinfo {year} {2011}),\ \bibfield  {title} {\enquote {\bibinfo
  {title} {{A}ll-electron ${G}{W}$ methods implemented in molecular orbital
  space: {I}onization energy and electron affinity of conjugated molecules},}\
  }\href@noop {} {\bibfield  {journal} {\bibinfo  {journal} {Phys. Rev. B}\
  }\textbf {\bibinfo {volume} {84}},\ \bibinfo {pages} {205415}}\BibitemShut
  {NoStop}%
\bibitem [{\citenamefont {Kendall}\ \emph {et~al.}(1992)\citenamefont
  {Kendall}, \citenamefont {Dunning},\ and\ \citenamefont
  {Harrison}}]{Kendall1992}%
  \BibitemOpen
  \bibfield  {author} {\bibinfo {author} {\bibnamefont {Kendall}, \bibfnamefont
  {R~A}}, \bibinfo {author} {\bibfnamefont {T.~H.}\ \bibnamefont {Dunning}}, \
  and\ \bibinfo {author} {\bibfnamefont {R.~J.}\ \bibnamefont {Harrison}}}
  (\bibinfo {year} {1992}),\ \bibfield  {title} {\enquote {\bibinfo {title}
  {{E}lectron affinities of the first‐row atoms revisited. {S}ystematic basis
  sets and wave functions},}\ }\href@noop {} {\bibfield  {journal} {\bibinfo
  {journal} {J. Chem. Phys.}\ }\textbf {\bibinfo {volume} {96}}~(\bibinfo
  {number} {9}),\ \bibinfo {pages} {6796--6806}}\BibitemShut {NoStop}%
\bibitem [{\citenamefont {Kevan}(1992)}]{ARPES:1992}%
  \BibitemOpen
  \bibinfo {editor} {\bibnamefont {Kevan}, \bibfnamefont {S~D}},\ Ed. (\bibinfo
  {year} {1992}),\ \href@noop {} {\emph {\bibinfo {title} {{A}ngle-{R}esolved
  {P}hotoemission: {Th}eory and {C}urrent {A}pplications}}}\ (\bibinfo
  {publisher} {Elsevier, Amsterdam})\BibitemShut {NoStop}%
\bibitem [{\citenamefont {Keyling}\ \emph {et~al.}(2000)\citenamefont
  {Keyling}, \citenamefont {Sch\"one},\ and\ \citenamefont
  {Ekardt}}]{Keyling/etal:2000}%
  \BibitemOpen
  \bibfield  {author} {\bibinfo {author} {\bibnamefont {Keyling}, \bibfnamefont
  {R}}, \bibinfo {author} {\bibfnamefont {W.-D.}\ \bibnamefont {Sch\"one}}, \
  and\ \bibinfo {author} {\bibfnamefont {W.}~\bibnamefont {Ekardt}}} (\bibinfo
  {year} {2000}),\ \bibfield  {title} {\enquote {\bibinfo {title} {{C}omparison
  of the lifetime of excited electrons in noble metals},}\ }\href@noop {}
  {\bibfield  {journal} {\bibinfo  {journal} {Phys. Rev. B}\ }\textbf {\bibinfo
  {volume} {61}},\ \bibinfo {pages} {1670--1673}}\BibitemShut {NoStop}%
\bibitem [{\citenamefont {Khairallah}\ and\ \citenamefont
  {Militzer}(2008)}]{Khairallah/etal:2008}%
  \BibitemOpen
  \bibfield  {author} {\bibinfo {author} {\bibnamefont {Khairallah},
  \bibfnamefont {S~A}}, \ and\ \bibinfo {author} {\bibfnamefont
  {B.}~\bibnamefont {Militzer}}} (\bibinfo {year} {2008}),\ \bibfield  {title}
  {\enquote {\bibinfo {title} {{F}irst-{P}rinciples {S}tudies of the
  {M}etallization and the {E}quation of {S}tate of {S}olid {H}elium},}\
  }\href@noop {} {\bibfield  {journal} {\bibinfo  {journal} {Phys. Rev. Lett.}\
  }\textbf {\bibinfo {volume} {101}},\ \bibinfo {pages} {106407}}\BibitemShut
  {NoStop}%
\bibitem [{\citenamefont {Kioupakis}\ \emph {et~al.}(2010)\citenamefont
  {Kioupakis}, \citenamefont {Rinke}, \citenamefont {Schleife}, \citenamefont
  {Bechstedt},\ and\ \citenamefont {{Van de Walle}}}]{Manos/etal:2010}%
  \BibitemOpen
  \bibfield  {author} {\bibinfo {author} {\bibnamefont {Kioupakis},
  \bibfnamefont {E}}, \bibinfo {author} {\bibfnamefont {P.}~\bibnamefont
  {Rinke}}, \bibinfo {author} {\bibfnamefont {A.}~\bibnamefont {Schleife}},
  \bibinfo {author} {\bibfnamefont {F.}~\bibnamefont {Bechstedt}}, \ and\
  \bibinfo {author} {\bibfnamefont {C.~G.}\ \bibnamefont {{Van de Walle}}}}
  (\bibinfo {year} {2010}),\ \bibfield  {title} {\enquote {\bibinfo {title}
  {{F}ree-carrier absorption in nitrides from first principles},}\ }\href@noop
  {} {\bibfield  {journal} {\bibinfo  {journal} {Phys.\ Rev.\ B}\ }\textbf
  {\bibinfo {volume} {81}},\ \bibinfo {pages} {241201(R)}}\BibitemShut
  {NoStop}%
\bibitem [{\citenamefont {Kippelen}\ and\ \citenamefont
  {Br\'{e}das}(2009)}]{Kippelen2009}%
  \BibitemOpen
  \bibfield  {author} {\bibinfo {author} {\bibnamefont {Kippelen},
  \bibfnamefont {B}}, \ and\ \bibinfo {author} {\bibfnamefont {J.-L.}\
  \bibnamefont {Br\'{e}das}}} (\bibinfo {year} {2009}),\ \bibfield  {title}
  {\enquote {\bibinfo {title} {{O}rganic photovoltaics},}\ }\href {\doibase
  10.1039/B812502N} {\bibfield  {journal} {\bibinfo  {journal} {Energy Environ.
  Sci.}\ }\textbf {\bibinfo {volume} {2}},\ \bibinfo {pages}
  {251--261}}\BibitemShut {NoStop}%
\bibitem [{\citenamefont {Kivisaari}\ \emph {et~al.}(2017)\citenamefont
  {Kivisaari}, \citenamefont {Sadi}, \citenamefont {Li}, \citenamefont
  {Rinke},\ and\ \citenamefont {Oksanen}}]{Kivisaari/etal:2017}%
  \BibitemOpen
  \bibfield  {author} {\bibinfo {author} {\bibnamefont {Kivisaari},
  \bibfnamefont {P}}, \bibinfo {author} {\bibfnamefont {T.}~\bibnamefont
  {Sadi}}, \bibinfo {author} {\bibfnamefont {J.}~\bibnamefont {Li}}, \bibinfo
  {author} {\bibfnamefont {P.}~\bibnamefont {Rinke}}, \ and\ \bibinfo {author}
  {\bibfnamefont {J.}~\bibnamefont {Oksanen}}} (\bibinfo {year} {2017}),\
  \bibfield  {title} {\enquote {\bibinfo {title} {{O}n the {M}onte {C}arlo
  {D}escription of {H}ot {C}arrier {E}ffects and {D}evice {C}haracteristics of
  {I}{I}{I}-{N} {L}{E}{D}s},}\ }\href@noop {} {\bibfield  {journal} {\bibinfo
  {journal} {Adv. Electron. Mater.}\ }\textbf {\bibinfo {volume} {3}}~(\bibinfo
  {number} {6}),\ \bibinfo {pages} {1600494}}\BibitemShut {NoStop}%
\bibitem [{\citenamefont {Klein}(1961)}]{Klein:1961}%
  \BibitemOpen
  \bibfield  {author} {\bibinfo {author} {\bibnamefont {Klein}, \bibfnamefont
  {A}}} (\bibinfo {year} {1961}),\ \bibfield  {title} {\enquote {\bibinfo
  {title} {{P}erturbation {T}heory for an {I}nfinite {M}edium of {F}ermions.
  {I}{I}},}\ }\href@noop {} {\bibfield  {journal} {\bibinfo  {journal} {Phys.
  Rev.}\ }\textbf {\bibinfo {volume} {121}}~(\bibinfo {number} {4}),\ \bibinfo
  {pages} {950--956}}\BibitemShut {NoStop}%
\bibitem [{\citenamefont {Klime{\^s}}\ \emph {et~al.}(2014)\citenamefont
  {Klime{\^s}}, \citenamefont {Kaltak},\ and\ \citenamefont
  {Kresse}}]{Klimes/Kaltak/Kresse:2014}%
  \BibitemOpen
  \bibfield  {author} {\bibinfo {author} {\bibnamefont {Klime{\^s}},
  \bibfnamefont {J}}, \bibinfo {author} {\bibfnamefont {M.}~\bibnamefont
  {Kaltak}}, \ and\ \bibinfo {author} {\bibfnamefont {G.}~\bibnamefont
  {Kresse}}} (\bibinfo {year} {2014}),\ \bibfield  {title} {\enquote {\bibinfo
  {title} {{P}redictive ${G}{W}$ calculations using plane waves and
  pseudopotentials},}\ }\href@noop {} {\bibfield  {journal} {\bibinfo
  {journal} {Phys. Rev. B}\ }\textbf {\bibinfo {volume} {90}},\ \bibinfo
  {pages} {075125}}\BibitemShut {NoStop}%
\bibitem [{\citenamefont {Klintenberg}\ \emph {et~al.}(2010)\citenamefont
  {Klintenberg}, \citenamefont {Leb\`egue}, \citenamefont {Katsnelson},\ and\
  \citenamefont {Eriksson}}]{klintenberg_prb_81}%
  \BibitemOpen
  \bibfield  {author} {\bibinfo {author} {\bibnamefont {Klintenberg},
  \bibfnamefont {M}}, \bibinfo {author} {\bibfnamefont {S.}~\bibnamefont
  {Leb\`egue}}, \bibinfo {author} {\bibfnamefont {M.~I.}\ \bibnamefont
  {Katsnelson}}, \ and\ \bibinfo {author} {\bibfnamefont {O.}~\bibnamefont
  {Eriksson}}} (\bibinfo {year} {2010}),\ \bibfield  {title} {\enquote
  {\bibinfo {title} {{T}heoretical analysis of the chemical bonding and
  electronic structure of graphene interacting with {G}roup {I}{A} and {G}roup
  {V}{I}{I}{A} elements},}\ }\href {\doibase 10.1103/PhysRevB.81.085433}
  {\bibfield  {journal} {\bibinfo  {journal} {Phys. Rev. B}\ }\textbf {\bibinfo
  {volume} {81}},\ \bibinfo {pages} {085433}}\BibitemShut {NoStop}%
\bibitem [{\citenamefont {Klopper}\ \emph {et~al.}(1999)\citenamefont
  {Klopper}, \citenamefont {Bak}, \citenamefont {Jørgensen}, \citenamefont
  {Olsen},\ and\ \citenamefont {Helgaker}}]{Klopper1999}%
  \BibitemOpen
  \bibfield  {author} {\bibinfo {author} {\bibnamefont {Klopper}, \bibfnamefont
  {W}}, \bibinfo {author} {\bibfnamefont {K.~L.}\ \bibnamefont {Bak}}, \bibinfo
  {author} {\bibfnamefont {P.}~\bibnamefont {Jørgensen}}, \bibinfo {author}
  {\bibfnamefont {J.}~\bibnamefont {Olsen}}, \ and\ \bibinfo {author}
  {\bibfnamefont {T.}~\bibnamefont {Helgaker}}} (\bibinfo {year} {1999}),\
  \bibfield  {title} {\enquote {\bibinfo {title} {{H}ighly accurate
  calculations of molecular electronic structure},}\ }\href
  {http://stacks.iop.org/0953-4075/32/i=13/a=201} {\bibfield  {journal}
  {\bibinfo  {journal} {J. Phys. B: At., Mol. Opt. Phys.}\ }\textbf {\bibinfo
  {volume} {32}}~(\bibinfo {number} {13}),\ \bibinfo {pages}
  {R103}}\BibitemShut {NoStop}%
\bibitem [{\citenamefont {Knight}\ \emph {et~al.}(2016)\citenamefont {Knight},
  \citenamefont {Wang}, \citenamefont {Gallandi}, \citenamefont
  {Dolgounitcheva}, \citenamefont {Ren}, \citenamefont {Ortiz}, \citenamefont
  {Rinke}, \citenamefont {K\"orzd\"orfer},\ and\ \citenamefont
  {Marom}}]{Knight2016}%
  \BibitemOpen
  \bibfield  {author} {\bibinfo {author} {\bibnamefont {Knight}, \bibfnamefont
  {J~W}}, \bibinfo {author} {\bibfnamefont {X.}~\bibnamefont {Wang}}, \bibinfo
  {author} {\bibfnamefont {L.}~\bibnamefont {Gallandi}}, \bibinfo {author}
  {\bibfnamefont {O.}~\bibnamefont {Dolgounitcheva}}, \bibinfo {author}
  {\bibfnamefont {X.}~\bibnamefont {Ren}}, \bibinfo {author} {\bibfnamefont
  {J.~V.}\ \bibnamefont {Ortiz}}, \bibinfo {author} {\bibfnamefont
  {P.}~\bibnamefont {Rinke}}, \bibinfo {author} {\bibfnamefont
  {T.}~\bibnamefont {K\"orzd\"orfer}}, \ and\ \bibinfo {author} {\bibfnamefont
  {N.}~\bibnamefont {Marom}}} (\bibinfo {year} {2016}),\ \bibfield  {title}
  {\enquote {\bibinfo {title} {{A}ccurate {I}onization {P}otentials and
  {E}lectron {A}ffinities of {A}cceptor {M}olecules {I}{I}{I}: {A} {B}enchmark
  of ${G}{W}$ {M}ethods},}\ }\href@noop {} {\bibfield  {journal} {\bibinfo
  {journal} {J. Chem. Theory Comput.}\ }\textbf {\bibinfo {volume}
  {12}}~(\bibinfo {number} {2}),\ \bibinfo {pages} {615--626}}\BibitemShut
  {NoStop}%
\bibitem [{\citenamefont {Kobayashi}(2013)}]{Kobayashi:privcom}%
  \BibitemOpen
  \bibfield  {author} {\bibinfo {author} {\bibnamefont {Kobayashi},
  \bibfnamefont {M}}} (\bibinfo {year} {2013}),\ \href@noop {} {}\bibinfo
  {note} {Private communication}\BibitemShut {NoStop}%
\bibitem [{\citenamefont {Kobayashi}\ \emph {et~al.}(2009)\citenamefont
  {Kobayashi}, \citenamefont {Song}, \citenamefont {Kataoka}, \citenamefont
  {Sakamoto}, \citenamefont {Fujimori}, \citenamefont {Ohkochi}, \citenamefont
  {Takeda}, \citenamefont {Okane}, \citenamefont {Saitoh}, \citenamefont
  {Yamagami}, \citenamefont {Yamahara}, \citenamefont {Saeki}, \citenamefont
  {Kawai},\ and\ \citenamefont {Tabata}}]{Kobayashi:2009}%
  \BibitemOpen
  \bibfield  {author} {\bibinfo {author} {\bibnamefont {Kobayashi},
  \bibfnamefont {M}}, \bibinfo {author} {\bibfnamefont {G.~S.}\ \bibnamefont
  {Song}}, \bibinfo {author} {\bibfnamefont {T.}~\bibnamefont {Kataoka}},
  \bibinfo {author} {\bibfnamefont {Y.}~\bibnamefont {Sakamoto}}, \bibinfo
  {author} {\bibfnamefont {A.}~\bibnamefont {Fujimori}}, \bibinfo {author}
  {\bibfnamefont {T.}~\bibnamefont {Ohkochi}}, \bibinfo {author} {\bibfnamefont
  {Y.}~\bibnamefont {Takeda}}, \bibinfo {author} {\bibfnamefont
  {T.}~\bibnamefont {Okane}}, \bibinfo {author} {\bibfnamefont
  {Y.}~\bibnamefont {Saitoh}}, \bibinfo {author} {\bibfnamefont
  {H.}~\bibnamefont {Yamagami}}, \bibinfo {author} {\bibfnamefont
  {H.}~\bibnamefont {Yamahara}}, \bibinfo {author} {\bibfnamefont
  {H.}~\bibnamefont {Saeki}}, \bibinfo {author} {\bibfnamefont
  {T.}~\bibnamefont {Kawai}}, \ and\ \bibinfo {author} {\bibfnamefont
  {H.}~\bibnamefont {Tabata}}} (\bibinfo {year} {2009}),\ \bibfield  {title}
  {\enquote {\bibinfo {title} {{E}xperimental observation of bulk band
  dispersions in the oxide semiconductor {Z}n{O} using soft x-ray
  angle-resolved photoemission spectroscopy},}\ }\href {\doibase
  10.1063/1.3116223} {\bibfield  {journal} {\bibinfo  {journal} {J. Appl.
  Phys.}\ }\textbf {\bibinfo {volume} {105}}~(\bibinfo {number} {12}),\
  \bibinfo {pages} {122403}}\BibitemShut {NoStop}%
\bibitem [{\citenamefont {Kobayashi}\ \emph {et~al.}(2008)\citenamefont
  {Kobayashi}, \citenamefont {Nohara}, \citenamefont {Yamamoto},\ and\
  \citenamefont {Fujiwara}}]{Kobayashi/etal:2008}%
  \BibitemOpen
  \bibfield  {author} {\bibinfo {author} {\bibnamefont {Kobayashi},
  \bibfnamefont {S}}, \bibinfo {author} {\bibfnamefont {Y.}~\bibnamefont
  {Nohara}}, \bibinfo {author} {\bibfnamefont {S.}~\bibnamefont {Yamamoto}}, \
  and\ \bibinfo {author} {\bibfnamefont {T.}~\bibnamefont {Fujiwara}}}
  (\bibinfo {year} {2008}),\ \bibfield  {title} {\enquote {\bibinfo {title}
  {${G}{W}$ approximation with {L}{S}{D}{A}+{U} method and applications to
  {N}i{O}, {M}n{O}, and {V}$_2${O}$_3$},}\ }\href@noop {} {\bibfield  {journal}
  {\bibinfo  {journal} {Phys.\ Rev.\ B}\ }\textbf {\bibinfo {volume} {78}},\
  \bibinfo {pages} {155112}}\BibitemShut {NoStop}%
\bibitem [{\citenamefont {Kohn}\ and\ \citenamefont
  {Sham}(1965)}]{Kohn/Sham:1965}%
  \BibitemOpen
  \bibfield  {author} {\bibinfo {author} {\bibnamefont {Kohn}, \bibfnamefont
  {W}}, \ and\ \bibinfo {author} {\bibfnamefont {K.~J.}\ \bibnamefont {Sham}}}
  (\bibinfo {year} {1965}),\ \bibfield  {title} {\enquote {\bibinfo {title}
  {{S}elf-{C}onsistent {E}quations {I}ncluding {E}xchange and {C}orrelation
  {E}ffects},}\ }\href@noop {} {\bibfield  {journal} {\bibinfo  {journal}
  {Phys.\ Rev.}\ }\textbf {\bibinfo {volume} {140}},\ \bibinfo {pages}
  {A1133}}\BibitemShut {NoStop}%
\bibitem [{\citenamefont {Komsa}\ and\ \citenamefont
  {Krasheninnikov}(2012)}]{Komsa/etal:2012}%
  \BibitemOpen
  \bibfield  {author} {\bibinfo {author} {\bibnamefont {Komsa}, \bibfnamefont
  {H-P}}, \ and\ \bibinfo {author} {\bibfnamefont {A.~V.}\ \bibnamefont
  {Krasheninnikov}}} (\bibinfo {year} {2012}),\ \bibfield  {title} {\enquote
  {\bibinfo {title} {{E}ffects of confinement and environment on the electronic
  structure and exciton binding energy of {M}o{S}${}_{2}$ from first
  principles},}\ }\href@noop {} {\bibfield  {journal} {\bibinfo  {journal}
  {Phys. Rev. B}\ }\textbf {\bibinfo {volume} {86}},\ \bibinfo {pages}
  {241201}}\BibitemShut {NoStop}%
\bibitem [{\citenamefont {K\"orbel}\ \emph {et~al.}(2014)\citenamefont
  {K\"orbel}, \citenamefont {Boulanger}, \citenamefont {Duchemin},
  \citenamefont {Blase}, \citenamefont {Marques},\ and\ \citenamefont
  {Botti}}]{botti_jctc_10}%
  \BibitemOpen
  \bibfield  {author} {\bibinfo {author} {\bibnamefont {K\"orbel},
  \bibfnamefont {S}}, \bibinfo {author} {\bibfnamefont {P.}~\bibnamefont
  {Boulanger}}, \bibinfo {author} {\bibfnamefont {I.}~\bibnamefont {Duchemin}},
  \bibinfo {author} {\bibfnamefont {X.}~\bibnamefont {Blase}}, \bibinfo
  {author} {\bibfnamefont {M.~A.~L.}\ \bibnamefont {Marques}}, \ and\ \bibinfo
  {author} {\bibfnamefont {S.}~\bibnamefont {Botti}}} (\bibinfo {year}
  {2014}),\ \bibfield  {title} {\enquote {\bibinfo {title} {{B}enchmark
  {M}any-{B}ody ${G}{W}$ and {B}ethe-{S}alpeter {C}alculations for {S}mall
  {T}ransition {M}etal {M}olecules},}\ }\href@noop {} {\bibfield  {journal}
  {\bibinfo  {journal} {J. Chem. Theory Comput.}\ }\textbf {\bibinfo {volume}
  {10}}~(\bibinfo {number} {9}),\ \bibinfo {pages} {3934--3943}}\BibitemShut
  {NoStop}%
\bibitem [{\citenamefont {K\"orzd\"orfer}\ \emph {et~al.}(2009)\citenamefont
  {K\"orzd\"orfer}, \citenamefont {K\"ummel}, \citenamefont {Marom},\ and\
  \citenamefont {Kronik}}]{Koerzdoerfer2009}%
  \BibitemOpen
  \bibfield  {author} {\bibinfo {author} {\bibnamefont {K\"orzd\"orfer},
  \bibfnamefont {T}}, \bibinfo {author} {\bibfnamefont {S.}~\bibnamefont
  {K\"ummel}}, \bibinfo {author} {\bibfnamefont {N.}~\bibnamefont {Marom}}, \
  and\ \bibinfo {author} {\bibfnamefont {L.}~\bibnamefont {Kronik}}} (\bibinfo
  {year} {2009}),\ \bibfield  {title} {\enquote {\bibinfo {title} {{W}hen to
  trust photoelectron spectra from {K}ohn-{S}ham eigenvalues: {T}he case of
  organic semiconductors},}\ }\href {\doibase 10.1103/PhysRevB.79.201205}
  {\bibfield  {journal} {\bibinfo  {journal} {Phys. Rev. B}\ }\textbf {\bibinfo
  {volume} {79}},\ \bibinfo {pages} {201205}}\BibitemShut {NoStop}%
\bibitem [{\citenamefont {K\"orzd\"orfer}\ and\ \citenamefont
  {Marom}(2012)}]{Koerzdoerfer/Marom:2012}%
  \BibitemOpen
  \bibfield  {author} {\bibinfo {author} {\bibnamefont {K\"orzd\"orfer},
  \bibfnamefont {T}}, \ and\ \bibinfo {author} {\bibfnamefont {N.}~\bibnamefont
  {Marom}}} (\bibinfo {year} {2012}),\ \bibfield  {title} {\enquote {\bibinfo
  {title} {{S}trategy for finding a reliable starting point for
  ${G}_{0}{W}_{0}$ demonstrated for molecules},}\ }\href@noop {} {\bibfield
  {journal} {\bibinfo  {journal} {Phys. Rev. B}\ }\textbf {\bibinfo {volume}
  {86}},\ \bibinfo {pages} {041110}}\BibitemShut {NoStop}%
\bibitem [{\citenamefont {K\"orzd\"orfer}\ \emph {et~al.}(2012)\citenamefont
  {K\"orzd\"orfer}, \citenamefont {Parrish}, \citenamefont {Marom},
  \citenamefont {Sears}, \citenamefont {Sherrill},\ and\ \citenamefont
  {Br\'edas}}]{Koerzdoerfer/Marom:2012_2}%
  \BibitemOpen
  \bibfield  {author} {\bibinfo {author} {\bibnamefont {K\"orzd\"orfer},
  \bibfnamefont {T}}, \bibinfo {author} {\bibfnamefont {R.~M.}\ \bibnamefont
  {Parrish}}, \bibinfo {author} {\bibfnamefont {N.}~\bibnamefont {Marom}},
  \bibinfo {author} {\bibfnamefont {J.~S.}\ \bibnamefont {Sears}}, \bibinfo
  {author} {\bibfnamefont {C.~D.}\ \bibnamefont {Sherrill}}, \ and\ \bibinfo
  {author} {\bibfnamefont {J.-L.}\ \bibnamefont {Br\'edas}}} (\bibinfo {year}
  {2012}),\ \bibfield  {title} {\enquote {\bibinfo {title} {{A}ssessment of the
  performance of tuned range-separated hybrid density functionals in predicting
  accurate quasiparticle spectra},}\ }\href@noop {} {\bibfield  {journal}
  {\bibinfo  {journal} {Phys. Rev. B}\ }\textbf {\bibinfo {volume} {86}},\
  \bibinfo {pages} {205110}}\BibitemShut {NoStop}%
\bibitem [{\citenamefont {Kotani}\ \emph
  {et~al.}(2007{\natexlab{a}})\citenamefont {Kotani}, \citenamefont {van
  Schilfgaarde}, \citenamefont {Faleev},\ and\ \citenamefont
  {Chantis}}]{Kotani/etal:2007}%
  \BibitemOpen
  \bibfield  {author} {\bibinfo {author} {\bibnamefont {Kotani}, \bibfnamefont
  {T}}, \bibinfo {author} {\bibfnamefont {M.}~\bibnamefont {van Schilfgaarde}},
  \bibinfo {author} {\bibfnamefont {S.~V.}\ \bibnamefont {Faleev}}, \ and\
  \bibinfo {author} {\bibfnamefont {A.}~\bibnamefont {Chantis}}} (\bibinfo
  {year} {2007}{\natexlab{a}}),\ \bibfield  {title} {\enquote {\bibinfo {title}
  {{Q}uasiparticle self-consistent ${G}{W}$ method: a short summary},}\
  }\href@noop {} {\bibfield  {journal} {\bibinfo  {journal} {J.\ Phys.:
  Condens.\ Matter}\ }\textbf {\bibinfo {volume} {19}},\ \bibinfo {pages}
  {365236}}\BibitemShut {NoStop}%
\bibitem [{\citenamefont {Kotani}\ \emph
  {et~al.}(2007{\natexlab{b}})\citenamefont {Kotani}, \citenamefont {{van
  Schilfgaarde}},\ and\ \citenamefont {Faleev}}]{Kotani/etal_2:2007}%
  \BibitemOpen
  \bibfield  {author} {\bibinfo {author} {\bibnamefont {Kotani}, \bibfnamefont
  {T}}, \bibinfo {author} {\bibfnamefont {M.}~\bibnamefont {{van
  Schilfgaarde}}}, \ and\ \bibinfo {author} {\bibfnamefont {S.~V.}\
  \bibnamefont {Faleev}}} (\bibinfo {year} {2007}{\natexlab{b}}),\ \bibfield
  {title} {\enquote {\bibinfo {title} {{Q}uasiparticle self-consistent ${G}{W}$
  method: {A} basis for the independent-particle approximation},}\ }\href@noop
  {} {\bibfield  {journal} {\bibinfo  {journal} {Phys. Rev. B}\ }\textbf
  {\bibinfo {volume} {76}},\ \bibinfo {pages} {165106}}\BibitemShut {NoStop}%
\bibitem [{\citenamefont {Kotliar}\ \emph {et~al.}(2006)\citenamefont
  {Kotliar}, \citenamefont {Savrasov}, \citenamefont {Haule}, \citenamefont
  {Oudovenko}, \citenamefont {Parcollet},\ and\ \citenamefont
  {Marianetti}}]{kotliar/etal:2006}%
  \BibitemOpen
  \bibfield  {author} {\bibinfo {author} {\bibnamefont {Kotliar}, \bibfnamefont
  {G}}, \bibinfo {author} {\bibfnamefont {S.~Y.}\ \bibnamefont {Savrasov}},
  \bibinfo {author} {\bibfnamefont {K.}~\bibnamefont {Haule}}, \bibinfo
  {author} {\bibfnamefont {V.~S.}\ \bibnamefont {Oudovenko}}, \bibinfo {author}
  {\bibfnamefont {O.}~\bibnamefont {Parcollet}}, \ and\ \bibinfo {author}
  {\bibfnamefont {C.~A.}\ \bibnamefont {Marianetti}}} (\bibinfo {year}
  {2006}),\ \bibfield  {title} {\enquote {\bibinfo {title} {{E}lectronic
  structure calculations with dynamical mean-field theory},}\ }\href@noop {}
  {\bibfield  {journal} {\bibinfo  {journal} {Rev. Mod. Phys.}\ }\textbf
  {\bibinfo {volume} {78}},\ \bibinfo {pages} {865--951}}\BibitemShut {NoStop}%
\bibitem [{\citenamefont {Kozik}\ \emph {et~al.}(2015)\citenamefont {Kozik},
  \citenamefont {Ferrero},\ and\ \citenamefont {Georges}}]{kozik_prl_114}%
  \BibitemOpen
  \bibfield  {author} {\bibinfo {author} {\bibnamefont {Kozik}, \bibfnamefont
  {E}}, \bibinfo {author} {\bibfnamefont {M.}~\bibnamefont {Ferrero}}, \ and\
  \bibinfo {author} {\bibfnamefont {A.}~\bibnamefont {Georges}}} (\bibinfo
  {year} {2015}),\ \bibfield  {title} {\enquote {\bibinfo {title}
  {{N}onexistence of the {L}uttinger-{W}ard {F}unctional and {M}isleading
  {C}onvergence of {S}keleton {D}iagrammatic {S}eries for {H}ubbard-{L}ike
  {M}odels},}\ }\href@noop {} {\bibfield  {journal} {\bibinfo  {journal} {Phys.
  Rev. Lett.}\ }\textbf {\bibinfo {volume} {114}},\ \bibinfo {pages}
  {156402}}\BibitemShut {NoStop}%
\bibitem [{\citenamefont {Kraisler}\ and\ \citenamefont
  {Kronik}(2013)}]{Kraisler2013}%
  \BibitemOpen
  \bibfield  {author} {\bibinfo {author} {\bibnamefont {Kraisler},
  \bibfnamefont {E}}, \ and\ \bibinfo {author} {\bibfnamefont {L.}~\bibnamefont
  {Kronik}}} (\bibinfo {year} {2013}),\ \bibfield  {title} {\enquote {\bibinfo
  {title} {{P}iecewise {L}inearity of {A}pproximate {D}ensity {F}unctionals
  {R}evisited: {I}mplications for {F}rontier {O}rbital {E}nergies},}\ }\href
  {\doibase 10.1103/PhysRevLett.110.126403} {\bibfield  {journal} {\bibinfo
  {journal} {Phys. Rev. Lett.}\ }\textbf {\bibinfo {volume} {110}},\ \bibinfo
  {pages} {126403}}\BibitemShut {NoStop}%
\bibitem [{\citenamefont {Kresse}\ and\ \citenamefont
  {Joubert}(1999)}]{Kresse1999}%
  \BibitemOpen
  \bibfield  {author} {\bibinfo {author} {\bibnamefont {Kresse}, \bibfnamefont
  {G}}, \ and\ \bibinfo {author} {\bibfnamefont {D.}~\bibnamefont {Joubert}}}
  (\bibinfo {year} {1999}),\ \bibfield  {title} {\enquote {\bibinfo {title}
  {{F}rom ultrasoft pseudopotentials to the projector augmented-wave method},}\
  }\href {\doibase 10.1103/PhysRevB.59.1758} {\bibfield  {journal} {\bibinfo
  {journal} {Phys. Rev. B}\ }\textbf {\bibinfo {volume} {59}},\ \bibinfo
  {pages} {1758--1775}}\BibitemShut {NoStop}%
\bibitem [{\citenamefont {Kronik}\ and\ \citenamefont
  {Neaton}(2016)}]{Kronik2016}%
  \BibitemOpen
  \bibfield  {author} {\bibinfo {author} {\bibnamefont {Kronik}, \bibfnamefont
  {L}}, \ and\ \bibinfo {author} {\bibfnamefont {J.~B.}\ \bibnamefont
  {Neaton}}} (\bibinfo {year} {2016}),\ \bibfield  {title} {\enquote {\bibinfo
  {title} {{E}xcited-{S}tate {P}roperties of {M}olecular {S}olids from {F}irst
  {P}rinciples},}\ }\href {\doibase 10.1146/annurev-physchem-040214-121351}
  {\bibfield  {journal} {\bibinfo  {journal} {Annu. Rev. Phys. Chem.}\ }\textbf
  {\bibinfo {volume} {67}}~(\bibinfo {number} {1}),\ \bibinfo {pages}
  {587--616}}\BibitemShut {NoStop}%
\bibitem [{\citenamefont {Ku}\ and\ \citenamefont
  {Eguiluz}(2002)}]{Ku/Eguiluz:2002}%
  \BibitemOpen
  \bibfield  {author} {\bibinfo {author} {\bibnamefont {Ku}, \bibfnamefont
  {W}}, \ and\ \bibinfo {author} {\bibfnamefont {A.~G.}\ \bibnamefont
  {Eguiluz}}} (\bibinfo {year} {2002}),\ \bibfield  {title} {\enquote {\bibinfo
  {title} {{B}and-{G}ap {P}roblem in {S}emiconductors {R}evisited: {E}ffects of
  {C}ore {S}tates and {M}any-{B}ody {S}elf-{C}onsistency},}\ }\href@noop {}
  {\bibfield  {journal} {\bibinfo  {journal} {Phys. Rev. Lett.}\ }\textbf
  {\bibinfo {volume} {89}},\ \bibinfo {pages} {126401}}\BibitemShut {NoStop}%
\bibitem [{\citenamefont {K\"uhn}\ and\ \citenamefont
  {Weigend}(2015)}]{Kuhn/etal:2015}%
  \BibitemOpen
  \bibfield  {author} {\bibinfo {author} {\bibnamefont {K\"uhn}, \bibfnamefont
  {M}}, \ and\ \bibinfo {author} {\bibfnamefont {F.}~\bibnamefont {Weigend}}}
  (\bibinfo {year} {2015}),\ \bibfield  {title} {\enquote {\bibinfo {title}
  {{O}ne-{E}lectron {E}nergies from the {T}wo-{C}omponent ${G}{W}$ {M}ethod},}\
  }\href@noop {} {\bibfield  {journal} {\bibinfo  {journal} {J. Chem. Theory
  Comput.}\ }\textbf {\bibinfo {volume} {11}}~(\bibinfo {number} {3}),\
  \bibinfo {pages} {969--979}}\BibitemShut {NoStop}%
\bibitem [{\citenamefont {K\"ummel}\ and\ \citenamefont
  {Kronik}(2008)}]{Kuemmel/Kronik:2008}%
  \BibitemOpen
  \bibfield  {author} {\bibinfo {author} {\bibnamefont {K\"ummel},
  \bibfnamefont {S}}, \ and\ \bibinfo {author} {\bibfnamefont {L.}~\bibnamefont
  {Kronik}}} (\bibinfo {year} {2008}),\ \bibfield  {title} {\enquote {\bibinfo
  {title} {{O}rbital-dependent functionals: theory and applications},}\
  }\href@noop {} {\bibfield  {journal} {\bibinfo  {journal} {Rev.\ Mod.\
  Phys.}\ }\textbf {\bibinfo {volume} {80}},\ \bibinfo {pages} {3}}\BibitemShut
  {NoStop}%
\bibitem [{\citenamefont {Kutepov}\ \emph {et~al.}(2012)\citenamefont
  {Kutepov}, \citenamefont {Haule}, \citenamefont {Savrasov},\ and\
  \citenamefont {Kotliar}}]{Kutepov/etal:2012}%
  \BibitemOpen
  \bibfield  {author} {\bibinfo {author} {\bibnamefont {Kutepov}, \bibfnamefont
  {A}}, \bibinfo {author} {\bibfnamefont {K.}~\bibnamefont {Haule}}, \bibinfo
  {author} {\bibfnamefont {S.~Y.}\ \bibnamefont {Savrasov}}, \ and\ \bibinfo
  {author} {\bibfnamefont {G.}~\bibnamefont {Kotliar}}} (\bibinfo {year}
  {2012}),\ \bibfield  {title} {\enquote {\bibinfo {title} {{E}lectronic
  structure of {P}u and {A}m metals by self-consistent relativistic ${G}{W}$
  method},}\ }\href@noop {} {\bibfield  {journal} {\bibinfo  {journal} {Phys.
  Rev. B}\ }\textbf {\bibinfo {volume} {85}},\ \bibinfo {pages}
  {155129}}\BibitemShut {NoStop}%
\bibitem [{\citenamefont {Kutepov}\ \emph {et~al.}(2009)\citenamefont
  {Kutepov}, \citenamefont {Savrasov},\ and\ \citenamefont
  {Kotliar}}]{Kutepov/etal:2009}%
  \BibitemOpen
  \bibfield  {author} {\bibinfo {author} {\bibnamefont {Kutepov}, \bibfnamefont
  {A}}, \bibinfo {author} {\bibfnamefont {S.~Y.}\ \bibnamefont {Savrasov}}, \
  and\ \bibinfo {author} {\bibfnamefont {G.}~\bibnamefont {Kotliar}}} (\bibinfo
  {year} {2009}),\ \bibfield  {title} {\enquote {\bibinfo {title}
  {{G}round-state properties of simple elements from ${G}{W}$ calculations},}\
  }\href@noop {} {\bibfield  {journal} {\bibinfo  {journal} {Phys. Rev. B}\
  }\textbf {\bibinfo {volume} {80}}~(\bibinfo {number} {4}),\ \bibinfo {pages}
  {041103}}\BibitemShut {NoStop}%
\bibitem [{\citenamefont {Kutepov}(2016)}]{Kutepov:2016}%
  \BibitemOpen
  \bibfield  {author} {\bibinfo {author} {\bibnamefont {Kutepov}, \bibfnamefont
  {A~L}}} (\bibinfo {year} {2016}),\ \bibfield  {title} {\enquote {\bibinfo
  {title} {{E}lectronic structure of {N}a, {K}, {S}i, and {L}i{F} from
  self-consistent solution of {H}edin's equations including vertex
  corrections},}\ }\href@noop {} {\bibfield  {journal} {\bibinfo  {journal}
  {Phys.\ Rev.\ B}\ }\textbf {\bibinfo {volume} {94}},\ \bibinfo {pages}
  {155101}}\BibitemShut {NoStop}%
\bibitem [{\citenamefont {Kutepov}(2017)}]{Kutepov:2017}%
  \BibitemOpen
  \bibfield  {author} {\bibinfo {author} {\bibnamefont {Kutepov}, \bibfnamefont
  {A~L}}} (\bibinfo {year} {2017}),\ \bibfield  {title} {\enquote {\bibinfo
  {title} {{S}elf-consistent solution of {H}edin's equations: {S}emiconductors
  and insulators},}\ }\href@noop {} {\bibfield  {journal} {\bibinfo  {journal}
  {Phys.\ Rev.\ B}\ }\textbf {\bibinfo {volume} {95}},\ \bibinfo {pages}
  {195120}}\BibitemShut {NoStop}%
\bibitem [{\citenamefont {Kutschera}\ \emph {et~al.}(2007)\citenamefont
  {Kutschera}, \citenamefont {Weinelt}, \citenamefont {Rohlfing},\ and\
  \citenamefont {Fauster}}]{Kutschera/etal:2007}%
  \BibitemOpen
  \bibfield  {author} {\bibinfo {author} {\bibnamefont {Kutschera},
  \bibfnamefont {M}}, \bibinfo {author} {\bibfnamefont {M.}~\bibnamefont
  {Weinelt}}, \bibinfo {author} {\bibfnamefont {M.}~\bibnamefont {Rohlfing}}, \
  and\ \bibinfo {author} {\bibfnamefont {T.}~\bibnamefont {Fauster}}} (\bibinfo
  {year} {2007}),\ \bibfield  {title} {{\selectlanguage {English}\enquote
  {\bibinfo {title} {{I}mage-potential-induced surface state at {S}i(100)},}\
  }}\href {\doibase 10.1007/s00339-007-4074-x} {\bibfield  {journal} {\bibinfo
  {journal} {Appl. Phys. A}\ }\textbf {\bibinfo {volume} {88}}~(\bibinfo
  {number} {3}),\ \bibinfo {pages} {519--526}}\BibitemShut {NoStop}%
\bibitem [{\citenamefont {Kutzelnigg}\ and\ \citenamefont
  {Morgan}(1992)}]{Kutzelnigg1992}%
  \BibitemOpen
  \bibfield  {author} {\bibinfo {author} {\bibnamefont {Kutzelnigg},
  \bibfnamefont {W}}, \ and\ \bibinfo {author} {\bibfnamefont {J.~D.}\
  \bibnamefont {Morgan}}} (\bibinfo {year} {1992}),\ \bibfield  {title}
  {\enquote {\bibinfo {title} {{R}ates of convergence of the partial‐wave
  expansions of atomic correlation energies},}\ }\href@noop {} {\bibfield
  {journal} {\bibinfo  {journal} {J. Chem. Phys.}\ }\textbf {\bibinfo {volume}
  {96}}~(\bibinfo {number} {6}),\ \bibinfo {pages} {4484--4508}}\BibitemShut
  {NoStop}%
\bibitem [{\citenamefont {Laasner}(2014)}]{Laasner2014}%
  \BibitemOpen
  \bibfield  {author} {\bibinfo {author} {\bibnamefont {Laasner}, \bibfnamefont
  {R}}} (\bibinfo {year} {2014}),\ \bibfield  {title} {\enquote {\bibinfo
  {title} {${G}_0{W}_0$ band structure of {C}d{W}{O}$_4$},}\ }\href
  {http://stacks.iop.org/0953-8984/26/i=12/a=125503} {\bibfield  {journal}
  {\bibinfo  {journal} {J. Phys.: Condens. Matter}\ }\textbf {\bibinfo {volume}
  {26}}~(\bibinfo {number} {12}),\ \bibinfo {pages} {125503}}\BibitemShut
  {NoStop}%
\bibitem [{\citenamefont {Lambert}\ and\ \citenamefont
  {Giustino}(2013)}]{Lambert/Giustino:2013}%
  \BibitemOpen
  \bibfield  {author} {\bibinfo {author} {\bibnamefont {Lambert}, \bibfnamefont
  {H}}, \ and\ \bibinfo {author} {\bibfnamefont {F.}~\bibnamefont {Giustino}}}
  (\bibinfo {year} {2013}),\ \bibfield  {title} {\enquote {\bibinfo {title}
  {\textit{Ab initio} {S}ternheimer-${G}{W}$ method for quasiparticle
  calculations using plane waves},}\ }\href@noop {} {\bibfield  {journal}
  {\bibinfo  {journal} {Phys. Rev. B}\ }\textbf {\bibinfo {volume} {88}},\
  \bibinfo {pages} {075117}}\BibitemShut {NoStop}%
\bibitem [{\citenamefont {Landau}\ \emph {et~al.}(1980)\citenamefont {Landau},
  \citenamefont {Lifshitz},\ and\ \citenamefont {Pitaevskij}}]{Landau:1980}%
  \BibitemOpen
  \bibfield  {author} {\bibinfo {author} {\bibnamefont {Landau}, \bibfnamefont
  {L~D}}, \bibinfo {author} {\bibfnamefont {E.~M.}\ \bibnamefont {Lifshitz}}, \
  and\ \bibinfo {author} {\bibfnamefont {L.~P.}\ \bibnamefont {Pitaevskij}}}
  (\bibinfo {year} {1980}),\ \href
  {https://books.google.fi/books?id=dEVtKQEACAAJ} {\emph {\bibinfo {title}
  {{S}tatistical {P}hysics: {P}art 2 : {T}heory of {C}ondensed {S}tate}}},\
  Landau and Lifshitz Course of theoretical physics\ (\bibinfo  {publisher}
  {Oxford})\BibitemShut {NoStop}%
\bibitem [{\citenamefont {Lange}\ and\ \citenamefont
  {Berkelbach}(2018)}]{Lange/etal:2018}%
  \BibitemOpen
  \bibfield  {author} {\bibinfo {author} {\bibnamefont {Lange}, \bibfnamefont
  {M~F}}, \ and\ \bibinfo {author} {\bibfnamefont {T.~C.}\ \bibnamefont
  {Berkelbach}}} (\bibinfo {year} {2018}),\ \bibfield  {title} {\enquote
  {\bibinfo {title} {{O}n the {R}elation between {E}quation-of-{M}otion
  {C}oupled-{C}luster {T}heory and the ${G}{W}$ {A}pproximation},}\ }\href@noop
  {} {\bibfield  {journal} {\bibinfo  {journal} {J.\ Chem.\ Theory\ Comput.}\
  }\textbf {\bibinfo {volume} {14}}~(\bibinfo {number} {8}),\ \bibinfo {pages}
  {4224--4236}}\BibitemShut {NoStop}%
\bibitem [{\citenamefont {Langreth}\ and\ \citenamefont
  {Perdew}(1977)}]{Langreth:1977}%
  \BibitemOpen
  \bibfield  {author} {\bibinfo {author} {\bibnamefont {Langreth},
  \bibfnamefont {D~C}}, \ and\ \bibinfo {author} {\bibfnamefont {J.~P.}\
  \bibnamefont {Perdew}}} (\bibinfo {year} {1977}),\ \bibfield  {title}
  {\enquote {\bibinfo {title} {{E}xchange-correlation energy of a metallic
  surface: {W}ave-vector analysis},}\ }\href {\doibase
  10.1103/PhysRevB.15.2884} {\bibfield  {journal} {\bibinfo  {journal} {Phys.
  Rev. B}\ }\textbf {\bibinfo {volume} {15}},\ \bibinfo {pages}
  {2884--2901}}\BibitemShut {NoStop}%
\bibitem [{\citenamefont {Lani}\ \emph {et~al.}(2012)\citenamefont {Lani},
  \citenamefont {Romaniello},\ and\ \citenamefont {Reining}}]{Lani_2012}%
  \BibitemOpen
  \bibfield  {author} {\bibinfo {author} {\bibnamefont {Lani}, \bibfnamefont
  {G}}, \bibinfo {author} {\bibfnamefont {P.}~\bibnamefont {Romaniello}}, \
  and\ \bibinfo {author} {\bibfnamefont {L.}~\bibnamefont {Reining}}} (\bibinfo
  {year} {2012}),\ \bibfield  {title} {\enquote {\bibinfo {title}
  {{A}pproximations for many-body {G}reen's functions: insights from the
  fundamental equations},}\ }\href {\doibase 10.1088/1367-2630/14/1/013056}
  {\bibfield  {journal} {\bibinfo  {journal} {New J. Phys.}\ }\textbf {\bibinfo
  {volume} {14}}~(\bibinfo {number} {1}),\ \bibinfo {pages}
  {013056}}\BibitemShut {NoStop}%
\bibitem [{\citenamefont {Larson}\ \emph {et~al.}(2013)\citenamefont {Larson},
  \citenamefont {Dvorak},\ and\ \citenamefont {Wu}}]{Larson2013}%
  \BibitemOpen
  \bibfield  {author} {\bibinfo {author} {\bibnamefont {Larson}, \bibfnamefont
  {P}}, \bibinfo {author} {\bibfnamefont {M.}~\bibnamefont {Dvorak}}, \ and\
  \bibinfo {author} {\bibfnamefont {Z.}~\bibnamefont {Wu}}} (\bibinfo {year}
  {2013}),\ \bibfield  {title} {\enquote {\bibinfo {title} {{R}ole of the
  plasmon-pole model in the ${G}{W}$ approximation},}\ }\href {\doibase
  10.1103/PhysRevB.88.125205} {\bibfield  {journal} {\bibinfo  {journal} {Phys.
  Rev. B}\ }\textbf {\bibinfo {volume} {88}},\ \bibinfo {pages}
  {125205}}\BibitemShut {NoStop}%
\bibitem [{\citenamefont {Leb{\`e}gue}\ \emph {et~al.}(2012)\citenamefont
  {Leb{\`e}gue}, \citenamefont {Araujo}, \citenamefont {Kim}, \citenamefont
  {Ramzan}, \citenamefont {Mao},\ and\ \citenamefont
  {Ahuja}}]{Lebegue/etal:2012}%
  \BibitemOpen
  \bibfield  {author} {\bibinfo {author} {\bibnamefont {Leb{\`e}gue},
  \bibfnamefont {S}}, \bibinfo {author} {\bibfnamefont {C.~M.}\ \bibnamefont
  {Araujo}}, \bibinfo {author} {\bibfnamefont {D.~Y.}\ \bibnamefont {Kim}},
  \bibinfo {author} {\bibfnamefont {M.}~\bibnamefont {Ramzan}}, \bibinfo
  {author} {\bibfnamefont {H.-K.}\ \bibnamefont {Mao}}, \ and\ \bibinfo
  {author} {\bibfnamefont {R.}~\bibnamefont {Ahuja}}} (\bibinfo {year}
  {2012}),\ \bibfield  {title} {\enquote {\bibinfo {title} {{S}emimetallic
  dense hydrogen above 260~{G}{P}a},}\ }\href {\doibase
  10.1073/pnas.1207065109} {\bibfield  {journal} {\bibinfo  {journal} {Proc.
  Natl. Acad. Sci. USA}\ }\textbf {\bibinfo {volume} {109}}~(\bibinfo {number}
  {25}),\ \bibinfo {pages} {9766--9769}}\BibitemShut {NoStop}%
\bibitem [{\citenamefont {Leb\`egue}\ \emph {et~al.}(2003)\citenamefont
  {Leb\`egue}, \citenamefont {Arnaud}, \citenamefont {Alouani},\ and\
  \citenamefont {Bloechl}}]{Lebegue2003}%
  \BibitemOpen
  \bibfield  {author} {\bibinfo {author} {\bibnamefont {Leb\`egue},
  \bibfnamefont {S}}, \bibinfo {author} {\bibfnamefont {B.}~\bibnamefont
  {Arnaud}}, \bibinfo {author} {\bibfnamefont {M.}~\bibnamefont {Alouani}}, \
  and\ \bibinfo {author} {\bibfnamefont {P.~E.}\ \bibnamefont {Bloechl}}}
  (\bibinfo {year} {2003}),\ \bibfield  {title} {\enquote {\bibinfo {title}
  {{I}mplementation of an all-electron ${G}{W}$ approximation based on the
  projector augmented wave method without plasmon pole approximation:
  {A}pplication to {S}i, {S}i{C}, {A}l{A}s, {I}n{A}s, {N}a{H}, and {K}{H}},}\
  }\href {\doibase 10.1103/PhysRevB.67.155208} {\bibfield  {journal} {\bibinfo
  {journal} {Phys. Rev. B}\ }\textbf {\bibinfo {volume} {67}},\ \bibinfo
  {pages} {155208}}\BibitemShut {NoStop}%
\bibitem [{\citenamefont {Leb\`egue}\ \emph {et~al.}(2009)\citenamefont
  {Leb\`egue}, \citenamefont {Klintenberg}, \citenamefont {Eriksson},\ and\
  \citenamefont {Katsnelson}}]{lebegue_prb_79}%
  \BibitemOpen
  \bibfield  {author} {\bibinfo {author} {\bibnamefont {Leb\`egue},
  \bibfnamefont {S}}, \bibinfo {author} {\bibfnamefont {M.}~\bibnamefont
  {Klintenberg}}, \bibinfo {author} {\bibfnamefont {O.}~\bibnamefont
  {Eriksson}}, \ and\ \bibinfo {author} {\bibfnamefont {M.~I.}\ \bibnamefont
  {Katsnelson}}} (\bibinfo {year} {2009}),\ \bibfield  {title} {\enquote
  {\bibinfo {title} {{A}ccurate electronic band gap of pure and functionalized
  graphane from ${G}{W}$ calculations},}\ }\href {\doibase
  10.1103/PhysRevB.79.245117} {\bibfield  {journal} {\bibinfo  {journal} {Phys.
  Rev. B}\ }\textbf {\bibinfo {volume} {79}},\ \bibinfo {pages}
  {245117}}\BibitemShut {NoStop}%
\bibitem [{\citenamefont {Lee}\ \emph {et~al.}(1988)\citenamefont {Lee},
  \citenamefont {Yang},\ and\ \citenamefont {Parr}}]{Lee/Yang/Parr:1992}%
  \BibitemOpen
  \bibfield  {author} {\bibinfo {author} {\bibnamefont {Lee}, \bibfnamefont
  {C}}, \bibinfo {author} {\bibfnamefont {W.}~\bibnamefont {Yang}}, \ and\
  \bibinfo {author} {\bibfnamefont {R.~G.}\ \bibnamefont {Parr}}} (\bibinfo
  {year} {1988}),\ \bibfield  {title} {\enquote {\bibinfo {title}
  {{D}evelopment of the {C}olle-{S}alvetti correlation-energy formula into a
  functional of the electron density},}\ }\href {\doibase
  10.1103/PhysRevB.37.785} {\bibfield  {journal} {\bibinfo  {journal} {Phys.
  Rev. B}\ }\textbf {\bibinfo {volume} {37}}~(\bibinfo {number} {2}),\ \bibinfo
  {pages} {785--789}}\BibitemShut {NoStop}%
\bibitem [{\citenamefont {Lee}\ \emph {et~al.}(2017)\citenamefont {Lee},
  \citenamefont {Huang}, \citenamefont {Sumpter},\ and\ \citenamefont
  {Yoon}}]{lee_2dm_4}%
  \BibitemOpen
  \bibfield  {author} {\bibinfo {author} {\bibnamefont {Lee}, \bibfnamefont
  {J}}, \bibinfo {author} {\bibfnamefont {J.}~\bibnamefont {Huang}}, \bibinfo
  {author} {\bibfnamefont {B.~G.}\ \bibnamefont {Sumpter}}, \ and\ \bibinfo
  {author} {\bibfnamefont {M.}~\bibnamefont {Yoon}}} (\bibinfo {year} {2017}),\
  \bibfield  {title} {\enquote {\bibinfo {title} {{S}train-engineered
  optoelectronic properties of 2{D} transition metal dichalcogenide lateral
  heterostructures},}\ }\href {\doibase 10.1088/2053-1583/aa5542} {\bibfield
  {journal} {\bibinfo  {journal} {2D Mater.}\ }\textbf {\bibinfo {volume}
  {4}}~(\bibinfo {number} {2}),\ \bibinfo {pages} {021016}}\BibitemShut
  {NoStop}%
\bibitem [{\citenamefont {Lee}\ \emph {et~al.}(1999)\citenamefont {Lee},
  \citenamefont {Gunnarsson},\ and\ \citenamefont
  {Hedin}}]{Lee/Gunnarsson/Hedin:1999}%
  \BibitemOpen
  \bibfield  {author} {\bibinfo {author} {\bibnamefont {Lee}, \bibfnamefont
  {J~D}}, \bibinfo {author} {\bibfnamefont {O.}~\bibnamefont {Gunnarsson}}, \
  and\ \bibinfo {author} {\bibfnamefont {L.}~\bibnamefont {Hedin}}} (\bibinfo
  {year} {1999}),\ \bibfield  {title} {\enquote {\bibinfo {title} {{T}ransition
  from the adiabatic to the sudden limit in core-level photoemission: a model
  study of a localized system},}\ }\href@noop {} {\bibfield  {journal}
  {\bibinfo  {journal} {Phys.\ Rev.\ B}\ }\textbf {\bibinfo {volume} {60}},\
  \bibinfo {pages} {8034}}\BibitemShut {NoStop}%
\bibitem [{\citenamefont {Lee}\ and\ \citenamefont
  {Needs}(2003)}]{Lee/Needs:2003}%
  \BibitemOpen
  \bibfield  {author} {\bibinfo {author} {\bibnamefont {Lee}, \bibfnamefont
  {Y}}, \ and\ \bibinfo {author} {\bibfnamefont {R.~J.}\ \bibnamefont {Needs}}}
  (\bibinfo {year} {2003}),\ \bibfield  {title} {\enquote {\bibinfo {title}
  {{C}ore-polarization potentials for {S}i and {T}i},}\ }\href@noop {}
  {\bibfield  {journal} {\bibinfo  {journal} {Phys.\ Rev.\ B}\ }\textbf
  {\bibinfo {volume} {67}},\ \bibinfo {pages} {035121}}\BibitemShut {NoStop}%
\bibitem [{\citenamefont {Leenaerts}\ \emph {et~al.}(2010)\citenamefont
  {Leenaerts}, \citenamefont {Peelaers}, \citenamefont {Hern\'andez-Nieves},
  \citenamefont {Partoens},\ and\ \citenamefont {Peeters}}]{leenaerts_prb_82}%
  \BibitemOpen
  \bibfield  {author} {\bibinfo {author} {\bibnamefont {Leenaerts},
  \bibfnamefont {O}}, \bibinfo {author} {\bibfnamefont {H.}~\bibnamefont
  {Peelaers}}, \bibinfo {author} {\bibfnamefont {A.~D.}\ \bibnamefont
  {Hern\'andez-Nieves}}, \bibinfo {author} {\bibfnamefont {B.}~\bibnamefont
  {Partoens}}, \ and\ \bibinfo {author} {\bibfnamefont {F.~M.}\ \bibnamefont
  {Peeters}}} (\bibinfo {year} {2010}),\ \bibfield  {title} {\enquote {\bibinfo
  {title} {{F}irst-principles investigation of graphene fluoride and
  graphane},}\ }\href {\doibase 10.1103/PhysRevB.82.195436} {\bibfield
  {journal} {\bibinfo  {journal} {Phys. Rev. B}\ }\textbf {\bibinfo {volume}
  {82}},\ \bibinfo {pages} {195436}}\BibitemShut {NoStop}%
\bibitem [{\citenamefont {Lejaeghere}\ \emph {et~al.}(2016)\citenamefont
  {Lejaeghere}, \citenamefont {Bihlmayer}, \citenamefont {Bj{\"o}rkman},
  \citenamefont {Blaha}, \citenamefont {Bl{\"u}gel}, \citenamefont {Blum},
  \citenamefont {Caliste}, \citenamefont {Castelli}, \citenamefont {Clark},
  \citenamefont {Corso}, \citenamefont {de~Gironcoli}, \citenamefont {Deutsch},
  \citenamefont {Dewhurst}, \citenamefont {Marco}, \citenamefont {Draxl},
  \citenamefont {Du{\l}ak}, \citenamefont {Eriksson}, \citenamefont
  {Flores-Livas}, \citenamefont {Garrity}, \citenamefont {Genovese},
  \citenamefont {Giannozzi}, \citenamefont {Giantomassi}, \citenamefont
  {Goedecker}, \citenamefont {X.}, \citenamefont {Gr{\r a}n{\"a}s},
  \citenamefont {Gross}, \citenamefont {Gulans}, \citenamefont {Gygi},
  \citenamefont {Hamann}, \citenamefont {Hasnip}, \citenamefont {Holzwarth},
  \citenamefont {Iu{\c s}an}, \citenamefont {Jochym}, \citenamefont {Jollet},
  \citenamefont {Jones}, \citenamefont {Kresse}, \citenamefont {Koepernik},
  \citenamefont {K{\"u}{\c c}{\"u}kbenli}, \citenamefont {Kvashnin},
  \citenamefont {Locht}, \citenamefont {Lubeck}, \citenamefont {Marsman},
  \citenamefont {Marzari}, \citenamefont {Nitzsche}, \citenamefont
  {Nordstr{\"o}m}, \citenamefont {Ozaki}, \citenamefont {Paulatto},
  \citenamefont {Pickard}, \citenamefont {Poelmans}, \citenamefont {Probert},
  \citenamefont {Refson}, \citenamefont {Richter}, \citenamefont {Rignanese},
  \citenamefont {Saha}, \citenamefont {Scheffler}, \citenamefont {Schlipf},
  \citenamefont {Schwarz}, \citenamefont {Sharma}, \citenamefont {Tavazza},
  \citenamefont {Thunstr{\"o}m}, \citenamefont {Tkatchenko}, \citenamefont
  {Torrent}, \citenamefont {Vanderbilt}, \citenamefont {van Setten},
  \citenamefont {Speybroeck}, \citenamefont {Wills}, \citenamefont {Yates},
  \citenamefont {Zhang},\ and\ \citenamefont
  {Cottenier}}]{Lejaeghere/etal:2016}%
  \BibitemOpen
  \bibfield  {author} {\bibinfo {author} {\bibnamefont {Lejaeghere},
  \bibfnamefont {K}}, \bibinfo {author} {\bibfnamefont {G.}~\bibnamefont
  {Bihlmayer}}, \bibinfo {author} {\bibfnamefont {T.}~\bibnamefont
  {Bj{\"o}rkman}}, \bibinfo {author} {\bibfnamefont {P.}~\bibnamefont {Blaha}},
  \bibinfo {author} {\bibfnamefont {S.}~\bibnamefont {Bl{\"u}gel}}, \bibinfo
  {author} {\bibfnamefont {V.}~\bibnamefont {Blum}}, \bibinfo {author}
  {\bibfnamefont {D.}~\bibnamefont {Caliste}}, \bibinfo {author} {\bibfnamefont
  {I.~E.}\ \bibnamefont {Castelli}}, \bibinfo {author} {\bibfnamefont {S.~J.}\
  \bibnamefont {Clark}}, \bibinfo {author} {\bibfnamefont {A.~Dal}\
  \bibnamefont {Corso}}, \bibinfo {author} {\bibfnamefont {S.}~\bibnamefont
  {de~Gironcoli}}, \bibinfo {author} {\bibfnamefont {T.}~\bibnamefont
  {Deutsch}}, \bibinfo {author} {\bibfnamefont {J.~K.}\ \bibnamefont
  {Dewhurst}}, \bibinfo {author} {\bibfnamefont {I.~Di}\ \bibnamefont {Marco}},
  \bibinfo {author} {\bibfnamefont {C.}~\bibnamefont {Draxl}}, \bibinfo
  {author} {\bibfnamefont {M.}~\bibnamefont {Du{\l}ak}}, \bibinfo {author}
  {\bibfnamefont {O.}~\bibnamefont {Eriksson}}, \bibinfo {author}
  {\bibfnamefont {J.~A.}\ \bibnamefont {Flores-Livas}}, \bibinfo {author}
  {\bibfnamefont {K.~F.}\ \bibnamefont {Garrity}}, \bibinfo {author}
  {\bibfnamefont {L.}~\bibnamefont {Genovese}}, \bibinfo {author}
  {\bibfnamefont {P.}~\bibnamefont {Giannozzi}}, \bibinfo {author}
  {\bibfnamefont {M.}~\bibnamefont {Giantomassi}}, \bibinfo {author}
  {\bibfnamefont {S.}~\bibnamefont {Goedecker}}, \bibinfo {author}
  {\bibnamefont {X.}}, \bibinfo {author} {\bibfnamefont {O.}~\bibnamefont
  {Gr{\r a}n{\"a}s}}, \bibinfo {author} {\bibfnamefont {E.~K.~U.}\ \bibnamefont
  {Gross}}, \bibinfo {author} {\bibfnamefont {A.}~\bibnamefont {Gulans}},
  \bibinfo {author} {\bibfnamefont {F.~C.}\ \bibnamefont {Gygi}}, \bibinfo
  {author} {\bibfnamefont {D.~R.}\ \bibnamefont {Hamann}}, \bibinfo {author}
  {\bibfnamefont {P.~J.}\ \bibnamefont {Hasnip}}, \bibinfo {author}
  {\bibfnamefont {N.~A.~W.}\ \bibnamefont {Holzwarth}}, \bibinfo {author}
  {\bibfnamefont {D.}~\bibnamefont {Iu{\c s}an}}, \bibinfo {author}
  {\bibfnamefont {D.~B.}\ \bibnamefont {Jochym}}, \bibinfo {author}
  {\bibfnamefont {F.~C.}\ \bibnamefont {Jollet}}, \bibinfo {author}
  {\bibfnamefont {D.}~\bibnamefont {Jones}}, \bibinfo {author} {\bibfnamefont
  {G.}~\bibnamefont {Kresse}}, \bibinfo {author} {\bibfnamefont
  {K.}~\bibnamefont {Koepernik}}, \bibinfo {author} {\bibfnamefont
  {E.}~\bibnamefont {K{\"u}{\c c}{\"u}kbenli}}, \bibinfo {author}
  {\bibfnamefont {Y.~O.}\ \bibnamefont {Kvashnin}}, \bibinfo {author}
  {\bibfnamefont {I.~L.~M.}\ \bibnamefont {Locht}}, \bibinfo {author}
  {\bibfnamefont {S.}~\bibnamefont {Lubeck}}, \bibinfo {author} {\bibfnamefont
  {M.}~\bibnamefont {Marsman}}, \bibinfo {author} {\bibfnamefont
  {N.}~\bibnamefont {Marzari}}, \bibinfo {author} {\bibfnamefont
  {U.}~\bibnamefont {Nitzsche}}, \bibinfo {author} {\bibfnamefont
  {L.}~\bibnamefont {Nordstr{\"o}m}}, \bibinfo {author} {\bibfnamefont
  {T.}~\bibnamefont {Ozaki}}, \bibinfo {author} {\bibfnamefont
  {L.}~\bibnamefont {Paulatto}}, \bibinfo {author} {\bibfnamefont {C.~J.}\
  \bibnamefont {Pickard}}, \bibinfo {author} {\bibfnamefont {W.}~\bibnamefont
  {Poelmans}}, \bibinfo {author} {\bibfnamefont {M.~I.~J.}\ \bibnamefont
  {Probert}}, \bibinfo {author} {\bibfnamefont {K.}~\bibnamefont {Refson}},
  \bibinfo {author} {\bibfnamefont {M.}~\bibnamefont {Richter}}, \bibinfo
  {author} {\bibfnamefont {G.-M.}\ \bibnamefont {Rignanese}}, \bibinfo {author}
  {\bibfnamefont {S.}~\bibnamefont {Saha}}, \bibinfo {author} {\bibfnamefont
  {M.}~\bibnamefont {Scheffler}}, \bibinfo {author} {\bibfnamefont
  {M.}~\bibnamefont {Schlipf}}, \bibinfo {author} {\bibfnamefont
  {K.}~\bibnamefont {Schwarz}}, \bibinfo {author} {\bibfnamefont
  {S.}~\bibnamefont {Sharma}}, \bibinfo {author} {\bibfnamefont
  {F.}~\bibnamefont {Tavazza}}, \bibinfo {author} {\bibfnamefont
  {P.}~\bibnamefont {Thunstr{\"o}m}}, \bibinfo {author} {\bibfnamefont
  {A.}~\bibnamefont {Tkatchenko}}, \bibinfo {author} {\bibfnamefont
  {M.}~\bibnamefont {Torrent}}, \bibinfo {author} {\bibfnamefont
  {D.}~\bibnamefont {Vanderbilt}}, \bibinfo {author} {\bibfnamefont {M.~J.}\
  \bibnamefont {van Setten}}, \bibinfo {author} {\bibfnamefont {V.~Van}\
  \bibnamefont {Speybroeck}}, \bibinfo {author} {\bibfnamefont {J.~M.}\
  \bibnamefont {Wills}}, \bibinfo {author} {\bibfnamefont {J.~R.}\ \bibnamefont
  {Yates}}, \bibinfo {author} {\bibfnamefont {G.-X.}\ \bibnamefont {Zhang}}, \
  and\ \bibinfo {author} {\bibfnamefont {S.}~\bibnamefont {Cottenier}}}
  (\bibinfo {year} {2016}),\ \bibfield  {title} {\enquote {\bibinfo {title}
  {{R}eproducibility in density functional theory calculations of solids},}\
  }\href@noop {} {\bibfield  {journal} {\bibinfo  {journal} {Science}\ }\textbf
  {\bibinfo {volume} {351}}~(\bibinfo {number} {6280}),\ \bibinfo {pages}
  {1415}}\BibitemShut {NoStop}%
\bibitem [{\citenamefont {Lejaeghere}\ \emph {et~al.}(2014)\citenamefont
  {Lejaeghere}, \citenamefont {Speybroeck}, \citenamefont {Oost},\ and\
  \citenamefont {Cottenier}}]{Lejaeghere/etal:2014}%
  \BibitemOpen
  \bibfield  {author} {\bibinfo {author} {\bibnamefont {Lejaeghere},
  \bibfnamefont {K}}, \bibinfo {author} {\bibfnamefont {V.~Van}\ \bibnamefont
  {Speybroeck}}, \bibinfo {author} {\bibfnamefont {G.~Van}\ \bibnamefont
  {Oost}}, \ and\ \bibinfo {author} {\bibfnamefont {S.}~\bibnamefont
  {Cottenier}}} (\bibinfo {year} {2014}),\ \bibfield  {title} {\enquote
  {\bibinfo {title} {{E}rror {E}stimates for {S}olid-{S}tate
  {D}ensity-{F}unctional {T}heory {P}redictions: {A}n {O}verview by {M}eans of
  the {G}round-{S}tate {E}lemental {C}rystals},}\ }\href@noop {} {\bibfield
  {journal} {\bibinfo  {journal} {Crit. Rev. Solid State Mater. Sci.}\ }\textbf
  {\bibinfo {volume} {39}}~(\bibinfo {number} {1}),\ \bibinfo {pages}
  {1--24}}\BibitemShut {NoStop}%
\bibitem [{\citenamefont {Levy}\ \emph {et~al.}(1984)\citenamefont {Levy},
  \citenamefont {Perdew},\ and\ \citenamefont
  {Sahni}}]{Levy/Perdew/Sahni:1984}%
  \BibitemOpen
  \bibfield  {author} {\bibinfo {author} {\bibnamefont {Levy}, \bibfnamefont
  {M}}, \bibinfo {author} {\bibfnamefont {J.~P.}\ \bibnamefont {Perdew}}, \
  and\ \bibinfo {author} {\bibfnamefont {V.}~\bibnamefont {Sahni}}} (\bibinfo
  {year} {1984}),\ \bibfield  {title} {\enquote {\bibinfo {title} {{E}xact
  differential equation for the density and ionization energy of a
  many-particle system},}\ }\href@noop {} {\bibfield  {journal} {\bibinfo
  {journal} {Phys.\ Rev.\ A}\ }\textbf {\bibinfo {volume} {30}},\ \bibinfo
  {pages} {2745}}\BibitemShut {NoStop}%
\bibitem [{\citenamefont {Li}\ \emph {et~al.}(2016{\natexlab{a}})\citenamefont
  {Li}, \citenamefont {D'Avino}, \citenamefont {Duchemin}, \citenamefont
  {Beljonne},\ and\ \citenamefont {Blase}}]{Li2016}%
  \BibitemOpen
  \bibfield  {author} {\bibinfo {author} {\bibnamefont {Li}, \bibfnamefont
  {J}}, \bibinfo {author} {\bibfnamefont {G.}~\bibnamefont {D'Avino}}, \bibinfo
  {author} {\bibfnamefont {I.}~\bibnamefont {Duchemin}}, \bibinfo {author}
  {\bibfnamefont {D.}~\bibnamefont {Beljonne}}, \ and\ \bibinfo {author}
  {\bibfnamefont {X.}~\bibnamefont {Blase}}} (\bibinfo {year}
  {2016}{\natexlab{a}}),\ \bibfield  {title} {\enquote {\bibinfo {title}
  {{C}ombining the {M}any-{B}ody ${G}{W}$ {F}ormalism with {C}lassical
  {P}olarizable {M}odels: {I}nsights on the {E}lectronic {S}tructure of
  {M}olecular {S}olids},}\ }\href {\doibase 10.1021/acs.jpclett.6b01302}
  {\bibfield  {journal} {\bibinfo  {journal} {J. Phys. Chem. Lett.}\ }\textbf
  {\bibinfo {volume} {7}}~(\bibinfo {number} {14}),\ \bibinfo {pages}
  {2814--2820}}\BibitemShut {NoStop}%
\bibitem [{\citenamefont {Li}\ \emph {et~al.}(2018)\citenamefont {Li},
  \citenamefont {D'Avino}, \citenamefont {Duchemin}, \citenamefont {Beljonne},\
  and\ \citenamefont {Blase}}]{Li2018}%
  \BibitemOpen
  \bibfield  {author} {\bibinfo {author} {\bibnamefont {Li}, \bibfnamefont
  {J}}, \bibinfo {author} {\bibfnamefont {G.}~\bibnamefont {D'Avino}}, \bibinfo
  {author} {\bibfnamefont {I.}~\bibnamefont {Duchemin}}, \bibinfo {author}
  {\bibfnamefont {D.}~\bibnamefont {Beljonne}}, \ and\ \bibinfo {author}
  {\bibfnamefont {X.}~\bibnamefont {Blase}}} (\bibinfo {year} {2018}),\
  \bibfield  {title} {\enquote {\bibinfo {title} {{A}ccurate description of
  charged excitations in molecular solids from embedded many-body perturbation
  theory},}\ }\href {\doibase 10.1103/PhysRevB.97.035108} {\bibfield  {journal}
  {\bibinfo  {journal} {Phys. Rev. B}\ }\textbf {\bibinfo {volume} {97}},\
  \bibinfo {pages} {035108}}\BibitemShut {NoStop}%
\bibitem [{\citenamefont {Li}\ \emph {et~al.}(2017)\citenamefont {Li},
  \citenamefont {Holzmann}, \citenamefont {Duchemin}, \citenamefont {Blase},\
  and\ \citenamefont {Olevano}}]{li_prl_118}%
  \BibitemOpen
  \bibfield  {author} {\bibinfo {author} {\bibnamefont {Li}, \bibfnamefont
  {J}}, \bibinfo {author} {\bibfnamefont {M.}~\bibnamefont {Holzmann}},
  \bibinfo {author} {\bibfnamefont {I.}~\bibnamefont {Duchemin}}, \bibinfo
  {author} {\bibfnamefont {X.}~\bibnamefont {Blase}}, \ and\ \bibinfo {author}
  {\bibfnamefont {V.}~\bibnamefont {Olevano}}} (\bibinfo {year} {2017}),\
  \bibfield  {title} {\enquote {\bibinfo {title} {{H}elium {A}tom {E}xcitations
  by the ${G}{W}$ and {B}ethe-{S}alpeter {M}any-{B}ody {F}ormalism},}\ }\href
  {\doibase 10.1103/PhysRevLett.118.163001} {\bibfield  {journal} {\bibinfo
  {journal} {Phys. Rev. Lett.}\ }\textbf {\bibinfo {volume} {118}},\ \bibinfo
  {pages} {163001}}\BibitemShut {NoStop}%
\bibitem [{\citenamefont {Li}\ \emph {et~al.}(2005)\citenamefont {Li},
  \citenamefont {Rignanese},\ and\ \citenamefont {Louie}}]{Li/etal:2005}%
  \BibitemOpen
  \bibfield  {author} {\bibinfo {author} {\bibnamefont {Li}, \bibfnamefont
  {J-L}}, \bibinfo {author} {\bibfnamefont {G.-M.}\ \bibnamefont {Rignanese}},
  \ and\ \bibinfo {author} {\bibfnamefont {S.~G.}\ \bibnamefont {Louie}}}
  (\bibinfo {year} {2005}),\ \bibfield  {title} {\enquote {\bibinfo {title}
  {{Q}uasiparticle energy bands of {N}i{O} in the ${G}{W}$ approximation},}\
  }\href@noop {} {\bibfield  {journal} {\bibinfo  {journal} {Phys.\ Rev.\ B}\
  }\textbf {\bibinfo {volume} {71}},\ \bibinfo {pages} {193102}}\BibitemShut
  {NoStop}%
\bibitem [{\citenamefont {Li}\ \emph {et~al.}(2016{\natexlab{b}})\citenamefont
  {Li}, \citenamefont {Kim}, \citenamefont {Jin}, \citenamefont {Ye},
  \citenamefont {Qiu}, \citenamefont {da~Jornada}, \citenamefont {Shi},
  \citenamefont {Chen}, \citenamefont {Zhang}, \citenamefont {Yang},
  \citenamefont {Watanabe}, \citenamefont {Taniguchi}, \citenamefont {Ren},
  \citenamefont {Louie}, \citenamefont {Chen}, \citenamefont {Zhang},\ and\
  \citenamefont {Wang}}]{li_nn}%
  \BibitemOpen
  \bibfield  {author} {\bibinfo {author} {\bibnamefont {Li}, \bibfnamefont
  {L}}, \bibinfo {author} {\bibfnamefont {J.}~\bibnamefont {Kim}}, \bibinfo
  {author} {\bibfnamefont {C.}~\bibnamefont {Jin}}, \bibinfo {author}
  {\bibfnamefont {G.~J.}\ \bibnamefont {Ye}}, \bibinfo {author} {\bibfnamefont
  {D.~Y.}\ \bibnamefont {Qiu}}, \bibinfo {author} {\bibfnamefont {F.~H.}\
  \bibnamefont {da~Jornada}}, \bibinfo {author} {\bibfnamefont
  {Z.}~\bibnamefont {Shi}}, \bibinfo {author} {\bibfnamefont {L.}~\bibnamefont
  {Chen}}, \bibinfo {author} {\bibfnamefont {Z.}~\bibnamefont {Zhang}},
  \bibinfo {author} {\bibfnamefont {F.}~\bibnamefont {Yang}}, \bibinfo {author}
  {\bibfnamefont {K.}~\bibnamefont {Watanabe}}, \bibinfo {author}
  {\bibfnamefont {T.}~\bibnamefont {Taniguchi}}, \bibinfo {author}
  {\bibfnamefont {W.}~\bibnamefont {Ren}}, \bibinfo {author} {\bibfnamefont
  {S.~G.}\ \bibnamefont {Louie}}, \bibinfo {author} {\bibfnamefont {X.~H.}\
  \bibnamefont {Chen}}, \bibinfo {author} {\bibfnamefont {Y.}~\bibnamefont
  {Zhang}}, \ and\ \bibinfo {author} {\bibfnamefont {F.}~\bibnamefont {Wang}}}
  (\bibinfo {year} {2016}{\natexlab{b}}),\ \bibfield  {title} {\enquote
  {\bibinfo {title} {{D}irect observation of the layer-dependent electronic
  structure in phosphorene},}\ }\href@noop {} {\bibfield  {journal} {\bibinfo
  {journal} {Nat. Nano.}\ }\textbf {\bibinfo {volume} {12}},\ \bibinfo {pages}
  {21--25}}\BibitemShut {NoStop}%
\bibitem [{\citenamefont {Liang}\ \emph {et~al.}(2012)\citenamefont {Liang},
  \citenamefont {Huang},\ and\ \citenamefont {Yang}}]{liang_jmr_27}%
  \BibitemOpen
  \bibfield  {author} {\bibinfo {author} {\bibnamefont {Liang}, \bibfnamefont
  {Y}}, \bibinfo {author} {\bibfnamefont {S.}~\bibnamefont {Huang}}, \ and\
  \bibinfo {author} {\bibfnamefont {L.}~\bibnamefont {Yang}}} (\bibinfo {year}
  {2012}),\ \bibfield  {title} {\enquote {\bibinfo {title} {{M}any-electron
  effects on optical absorption spectra of strained graphene},}\ }\href
  {\doibase 10.1557/jmr.2011.412} {\bibfield  {journal} {\bibinfo  {journal}
  {J. Mater. Res.}\ }\textbf {\bibinfo {volume} {27}}~(\bibinfo {number} {2}),\
  \bibinfo {pages} {403–409}}\BibitemShut {NoStop}%
\bibitem [{\citenamefont {Liao}\ and\ \citenamefont
  {Carter}(2011)}]{Liao/Carter:2011}%
  \BibitemOpen
  \bibfield  {author} {\bibinfo {author} {\bibnamefont {Liao}, \bibfnamefont
  {P}}, \ and\ \bibinfo {author} {\bibfnamefont {E.~A.}\ \bibnamefont
  {Carter}}} (\bibinfo {year} {2011}),\ \bibfield  {title} {\enquote {\bibinfo
  {title} {{T}esting variations of the ${G}{W}$ approximation on strongly
  correlated transition metal oxides: hematite ($\alpha$-{F}e$_2${O}$_3$) as a
  benchmark},}\ }\href@noop {} {\bibfield  {journal} {\bibinfo  {journal}
  {Phys. Chem. Chem. Phys.}\ }\textbf {\bibinfo {volume} {13}},\ \bibinfo
  {pages} {15189--15199}}\BibitemShut {NoStop}%
\bibitem [{\citenamefont {Lias}\ and\ \citenamefont {Liebman}(2003)}]{NIST}%
  \BibitemOpen
  \bibfield  {author} {\bibinfo {author} {\bibnamefont {Lias}, \bibfnamefont
  {S~G}}, \ and\ \bibinfo {author} {\bibfnamefont {J.~F.}\ \bibnamefont
  {Liebman}}} (\bibinfo {year} {2003}),\ \enquote {\bibinfo {title}
  {{N}{I}{S}{T} {C}hemistry {W}eb{B}ook, {N}{I}{S}{T} {S}tandard {R}eference
  {D}atabase {N}umber 69},}\ Chap.\ \bibinfo {chapter} {Ion Energetics Data}\
  (\bibinfo  {publisher} {Institute of Standards and Technology, Gaithersburg
  MD})\ \bibinfo {note} {http://webbook.nist.gov}\BibitemShut {NoStop}%
\bibitem [{\citenamefont {von~der Linden}\ and\ \citenamefont
  {Horsch}(1988)}]{Linden1988}%
  \BibitemOpen
  \bibfield  {author} {\bibinfo {author} {\bibnamefont {von~der Linden},
  \bibfnamefont {W}}, \ and\ \bibinfo {author} {\bibfnamefont {P.}~\bibnamefont
  {Horsch}}} (\bibinfo {year} {1988}),\ \bibfield  {title} {\enquote {\bibinfo
  {title} {{P}recise quasiparticle energies and {H}artree-{F}ock bands of
  semiconductors and insulators},}\ }\href {\doibase 10.1103/PhysRevB.37.8351}
  {\bibfield  {journal} {\bibinfo  {journal} {Phys. Rev. B}\ }\textbf {\bibinfo
  {volume} {37}},\ \bibinfo {pages} {8351--8362}}\BibitemShut {NoStop}%
\bibitem [{\citenamefont {Lischner}\ \emph {et~al.}(2012)\citenamefont
  {Lischner}, \citenamefont {Deslippe}, \citenamefont {Jain},\ and\
  \citenamefont {Louie}}]{Lischner/etal:2012}%
  \BibitemOpen
  \bibfield  {author} {\bibinfo {author} {\bibnamefont {Lischner},
  \bibfnamefont {J}}, \bibinfo {author} {\bibfnamefont {J.}~\bibnamefont
  {Deslippe}}, \bibinfo {author} {\bibfnamefont {M.}~\bibnamefont {Jain}}, \
  and\ \bibinfo {author} {\bibfnamefont {S.~G.}\ \bibnamefont {Louie}}}
  (\bibinfo {year} {2012}),\ \bibfield  {title} {\enquote {\bibinfo {title}
  {{F}irst-{P}rinciples {C}alculations of {Q}uasiparticle {E}xcitations of
  {O}pen-{S}hell {C}ondensed {M}atter {S}ystems},}\ }\href@noop {} {\bibfield
  {journal} {\bibinfo  {journal} {Phys. Rev. Lett.}\ }\textbf {\bibinfo
  {volume} {109}},\ \bibinfo {pages} {036406}}\BibitemShut {NoStop}%
\bibitem [{\citenamefont {Lischner}\ \emph {et~al.}(2013)\citenamefont
  {Lischner}, \citenamefont {Vigil-Fowler},\ and\ \citenamefont
  {Louie}}]{Lischner/etal:2013}%
  \BibitemOpen
  \bibfield  {author} {\bibinfo {author} {\bibnamefont {Lischner},
  \bibfnamefont {J}}, \bibinfo {author} {\bibfnamefont {D.}~\bibnamefont
  {Vigil-Fowler}}, \ and\ \bibinfo {author} {\bibfnamefont {S.~G.}\
  \bibnamefont {Louie}}} (\bibinfo {year} {2013}),\ \bibfield  {title}
  {\enquote {\bibinfo {title} {{P}hysical {O}rigin of {S}atellites in
  {P}hotoemission of {D}oped {G}raphene: {A}n {A}b {I}nitio ${G}{W}$ {P}lus
  {C}umulant {S}tudy},}\ }\href@noop {} {\bibfield  {journal} {\bibinfo
  {journal} {Phys. Rev. Lett.}\ }\textbf {\bibinfo {volume} {110}},\ \bibinfo
  {pages} {146801}}\BibitemShut {NoStop}%
\bibitem [{\citenamefont {Lischner}\ \emph {et~al.}(2014)\citenamefont
  {Lischner}, \citenamefont {Vigil-Fowler},\ and\ \citenamefont
  {Louie}}]{Lischner/etal:2014b}%
  \BibitemOpen
  \bibfield  {author} {\bibinfo {author} {\bibnamefont {Lischner},
  \bibfnamefont {J}}, \bibinfo {author} {\bibfnamefont {D.}~\bibnamefont
  {Vigil-Fowler}}, \ and\ \bibinfo {author} {\bibfnamefont {S.~G.}\
  \bibnamefont {Louie}}} (\bibinfo {year} {2014}),\ \bibfield  {title}
  {\enquote {\bibinfo {title} {{S}atellite structures in the spectral functions
  of the two-dimensional electron gas in semiconductor quantum wells: {A}
  ${G}{W}$ plus cumulant study},}\ }\href@noop {} {\bibfield  {journal}
  {\bibinfo  {journal} {Phys. Rev. B}\ }\textbf {\bibinfo {volume} {89}},\
  \bibinfo {pages} {125430}}\BibitemShut {NoStop}%
\bibitem [{\citenamefont {Liu}\ \emph {et~al.}(2016)\citenamefont {Liu},
  \citenamefont {Kaltak}, \citenamefont {Klime\v{s}},\ and\ \citenamefont
  {Kresse}}]{Liu2016}%
  \BibitemOpen
  \bibfield  {author} {\bibinfo {author} {\bibnamefont {Liu}, \bibfnamefont
  {P}}, \bibinfo {author} {\bibfnamefont {M.}~\bibnamefont {Kaltak}}, \bibinfo
  {author} {\bibfnamefont {J.}~\bibnamefont {Klime\v{s}}}, \ and\ \bibinfo
  {author} {\bibfnamefont {G.}~\bibnamefont {Kresse}}} (\bibinfo {year}
  {2016}),\ \bibfield  {title} {\enquote {\bibinfo {title} {{C}ubic scaling
  ${G}{W}$: {T}owards fast quasiparticle calculations},}\ }\href {\doibase
  10.1103/PhysRevB.94.165109} {\bibfield  {journal} {\bibinfo  {journal} {Phys.
  Rev. B}\ }\textbf {\bibinfo {volume} {94}},\ \bibinfo {pages}
  {165109}}\BibitemShut {NoStop}%
\bibitem [{\citenamefont {Liu}\ \emph {et~al.}(2011)\citenamefont {Liu},
  \citenamefont {Alnama}, \citenamefont {Matsumoto}, \citenamefont {Nishizawa},
  \citenamefont {Kohguchi}, \citenamefont {Lee},\ and\ \citenamefont
  {Suzuki}}]{Liu2011}%
  \BibitemOpen
  \bibfield  {author} {\bibinfo {author} {\bibnamefont {Liu}, \bibfnamefont
  {S-Y}}, \bibinfo {author} {\bibfnamefont {K.}~\bibnamefont {Alnama}},
  \bibinfo {author} {\bibfnamefont {J.}~\bibnamefont {Matsumoto}}, \bibinfo
  {author} {\bibfnamefont {K.}~\bibnamefont {Nishizawa}}, \bibinfo {author}
  {\bibfnamefont {H.}~\bibnamefont {Kohguchi}}, \bibinfo {author}
  {\bibfnamefont {Y.-P.}\ \bibnamefont {Lee}}, \ and\ \bibinfo {author}
  {\bibfnamefont {T.}~\bibnamefont {Suzuki}}} (\bibinfo {year} {2011}),\
  \bibfield  {title} {\enquote {\bibinfo {title} {{H}e {I} {U}ltraviolet
  {P}hotoelectron {S}pectroscopy of {B}enzene and {P}yridine in {S}upersonic
  {M}olecular {B}eams {U}sing {P}hotoelectron {I}maging},}\ }\href@noop {}
  {\bibfield  {journal} {\bibinfo  {journal} {J. Phys. Chem. A}\ }\textbf
  {\bibinfo {volume} {115}}~(\bibinfo {number} {14}),\ \bibinfo {pages}
  {2953--2965}}\BibitemShut {NoStop}%
\bibitem [{\citenamefont {Loos}\ \emph {et~al.}(2018)\citenamefont {Loos},
  \citenamefont {Romaniello},\ and\ \citenamefont {Berger}}]{loos_jctc_14}%
  \BibitemOpen
  \bibfield  {author} {\bibinfo {author} {\bibnamefont {Loos}, \bibfnamefont
  {P-F}}, \bibinfo {author} {\bibfnamefont {P.}~\bibnamefont {Romaniello}}, \
  and\ \bibinfo {author} {\bibfnamefont {J.~A.}\ \bibnamefont {Berger}}}
  (\bibinfo {year} {2018}),\ \bibfield  {title} {\enquote {\bibinfo {title}
  {{G}reen {F}unctions and {S}elf-{C}onsistency: {I}nsights {F}rom the
  {S}pherium {M}odel},}\ }\href@noop {} {\bibfield  {journal} {\bibinfo
  {journal} {J. Chem. Theory Comput.}\ }\textbf {\bibinfo {volume}
  {14}}~(\bibinfo {number} {6}),\ \bibinfo {pages} {3071--3082}}\BibitemShut
  {NoStop}%
\bibitem [{\citenamefont {L\"oser}\ \emph {et~al.}(2012)\citenamefont
  {L\"oser}, \citenamefont {Wenderoth}, \citenamefont {Spaeth}, \citenamefont
  {Garleff}, \citenamefont {Ulbrich}, \citenamefont {P\"otter},\ and\
  \citenamefont {Rohlfing}}]{Loeser/etal:2012}%
  \BibitemOpen
  \bibfield  {author} {\bibinfo {author} {\bibnamefont {L\"oser}, \bibfnamefont
  {K}}, \bibinfo {author} {\bibfnamefont {M.}~\bibnamefont {Wenderoth}},
  \bibinfo {author} {\bibfnamefont {T.~K.~A.}\ \bibnamefont {Spaeth}}, \bibinfo
  {author} {\bibfnamefont {J.~K.}\ \bibnamefont {Garleff}}, \bibinfo {author}
  {\bibfnamefont {R.~G.}\ \bibnamefont {Ulbrich}}, \bibinfo {author}
  {\bibfnamefont {M.}~\bibnamefont {P\"otter}}, \ and\ \bibinfo {author}
  {\bibfnamefont {M.}~\bibnamefont {Rohlfing}}} (\bibinfo {year} {2012}),\
  \bibfield  {title} {\enquote {\bibinfo {title} {{S}pectroscopy of positively
  and negatively buckled domains on
  {S}i(111)-$2\ifmmode\times\else\texttimes\fi{}1$},}\ }\href@noop {}
  {\bibfield  {journal} {\bibinfo  {journal} {Phys. Rev. B}\ }\textbf {\bibinfo
  {volume} {86}},\ \bibinfo {pages} {085303}}\BibitemShut {NoStop}%
\bibitem [{\citenamefont {Lu}\ \emph {et~al.}(2017)\citenamefont {Lu},
  \citenamefont {Wu}, \citenamefont {Yang}, \citenamefont {Liang},
  \citenamefont {Quhe}, \citenamefont {Guan},\ and\ \citenamefont
  {Wang}}]{lu_sr_7}%
  \BibitemOpen
  \bibfield  {author} {\bibinfo {author} {\bibnamefont {Lu}, \bibfnamefont
  {P}}, \bibinfo {author} {\bibfnamefont {L.}~\bibnamefont {Wu}}, \bibinfo
  {author} {\bibfnamefont {C.}~\bibnamefont {Yang}}, \bibinfo {author}
  {\bibfnamefont {D.}~\bibnamefont {Liang}}, \bibinfo {author} {\bibfnamefont
  {R.}~\bibnamefont {Quhe}}, \bibinfo {author} {\bibfnamefont {P.}~\bibnamefont
  {Guan}}, \ and\ \bibinfo {author} {\bibfnamefont {S.}~\bibnamefont {Wang}}}
  (\bibinfo {year} {2017}),\ \bibfield  {title} {\enquote {\bibinfo {title}
  {{Q}uasiparticle and optical properties of strained stanene and stanane},}\
  }\href@noop {} {\bibfield  {journal} {\bibinfo  {journal} {Sci. Rep.}\
  }\textbf {\bibinfo {volume} {7}},\ \bibinfo {pages} {3912}}\BibitemShut
  {NoStop}%
\bibitem [{\citenamefont {L\"u}\ \emph {et~al.}(2012)\citenamefont {L\"u},
  \citenamefont {Liao}, \citenamefont {Wang},\ and\ \citenamefont
  {Zheng}}]{lu_jmc_22}%
  \BibitemOpen
  \bibfield  {author} {\bibinfo {author} {\bibnamefont {L\"u}, \bibfnamefont
  {T-Y}}, \bibinfo {author} {\bibfnamefont {X.-X.}\ \bibnamefont {Liao}},
  \bibinfo {author} {\bibfnamefont {H.-Q.}\ \bibnamefont {Wang}}, \ and\
  \bibinfo {author} {\bibfnamefont {J.-C.}\ \bibnamefont {Zheng}}} (\bibinfo
  {year} {2012}),\ \bibfield  {title} {\enquote {\bibinfo {title} {{T}uning the
  indirect–direct band gap transition of {S}i{C}, {G}e{C} and {S}n{C}
  monolayer in a graphene-like honeycomb structure by strain engineering: a
  quasiparticle ${G}{W}$ study},}\ }\href {\doibase 10.1039/C2JM30915G}
  {\bibfield  {journal} {\bibinfo  {journal} {J. Mater. Chem.}\ }\textbf
  {\bibinfo {volume} {22}},\ \bibinfo {pages} {10062--10068}}\BibitemShut
  {NoStop}%
\bibitem [{\citenamefont {L\"uder}\ \emph {et~al.}(2016)\citenamefont
  {L\"uder}, \citenamefont {Puglia}, \citenamefont {Ottosson}, \citenamefont
  {Eriksson}, \citenamefont {Sanyal},\ and\ \citenamefont
  {Brena}}]{Lueder2016}%
  \BibitemOpen
  \bibfield  {author} {\bibinfo {author} {\bibnamefont {L\"uder}, \bibfnamefont
  {J}}, \bibinfo {author} {\bibfnamefont {C.}~\bibnamefont {Puglia}}, \bibinfo
  {author} {\bibfnamefont {H.}~\bibnamefont {Ottosson}}, \bibinfo {author}
  {\bibfnamefont {O.}~\bibnamefont {Eriksson}}, \bibinfo {author}
  {\bibfnamefont {B.}~\bibnamefont {Sanyal}}, \ and\ \bibinfo {author}
  {\bibfnamefont {B.}~\bibnamefont {Brena}}} (\bibinfo {year} {2016}),\
  \bibfield  {title} {\enquote {\bibinfo {title} {{M}any-body effects and
  excitonic features in 2{D} biphenylene carbon},}\ }\href@noop {} {\bibfield
  {journal} {\bibinfo  {journal} {J. Chem. Phys.}\ }\textbf {\bibinfo {volume}
  {144}}~(\bibinfo {number} {2}),\ \bibinfo {pages} {024702}}\BibitemShut
  {NoStop}%
\bibitem [{\citenamefont {Luo}\ \emph {et~al.}(2002)\citenamefont {Luo},
  \citenamefont {Ismail-Beigi}, \citenamefont {Cohen},\ and\ \citenamefont
  {Louie}}]{Luo/Louie:2002}%
  \BibitemOpen
  \bibfield  {author} {\bibinfo {author} {\bibnamefont {Luo}, \bibfnamefont
  {W}}, \bibinfo {author} {\bibfnamefont {S.}~\bibnamefont {Ismail-Beigi}},
  \bibinfo {author} {\bibfnamefont {M.~L.}\ \bibnamefont {Cohen}}, \ and\
  \bibinfo {author} {\bibfnamefont {S.~G.}\ \bibnamefont {Louie}}} (\bibinfo
  {year} {2002}),\ \bibfield  {title} {\enquote {\bibinfo {title}
  {{Q}uasiparticle band structures of {Z}n{S} and {Z}n{S}e},}\ }\href@noop {}
  {\bibfield  {journal} {\bibinfo  {journal} {Phys. Rev. B}\ }\textbf {\bibinfo
  {volume} {66}},\ \bibinfo {pages} {195215}}\BibitemShut {NoStop}%
\bibitem [{\citenamefont {Ma}\ and\ \citenamefont
  {Rohlfing}(2008)}]{Ma/Rohlfing:2008}%
  \BibitemOpen
  \bibfield  {author} {\bibinfo {author} {\bibnamefont {Ma}, \bibfnamefont
  {Y}}, \ and\ \bibinfo {author} {\bibfnamefont {M.}~\bibnamefont {Rohlfing}}}
  (\bibinfo {year} {2008}),\ \bibfield  {title} {\enquote {\bibinfo {title}
  {{O}ptical excitation of deep defect levels in insulators within many-body
  perturbation theory: {T}he {F} center in calcium fluoride},}\ }\href@noop {}
  {\bibfield  {journal} {\bibinfo  {journal} {Phys.\ Rev.\ B}\ }\textbf
  {\bibinfo {volume} {77}},\ \bibinfo {pages} {115118}}\BibitemShut {NoStop}%
\bibitem [{\citenamefont {Ma}\ \emph {et~al.}(2010{\natexlab{a}})\citenamefont
  {Ma}, \citenamefont {Rohlfing},\ and\ \citenamefont
  {Gali}}]{Ma/Rohlfing/Gali:2010}%
  \BibitemOpen
  \bibfield  {author} {\bibinfo {author} {\bibnamefont {Ma}, \bibfnamefont
  {Y}}, \bibinfo {author} {\bibfnamefont {M.}~\bibnamefont {Rohlfing}}, \ and\
  \bibinfo {author} {\bibfnamefont {A.}~\bibnamefont {Gali}}} (\bibinfo {year}
  {2010}{\natexlab{a}}),\ \bibfield  {title} {\enquote {\bibinfo {title}
  {{E}xcited states of the negatively charged nitrogen-vacancy color center in
  diamond},}\ }\href@noop {} {\bibfield  {journal} {\bibinfo  {journal} {Phys.\
  Rev.\ B}\ }\textbf {\bibinfo {volume} {81}},\ \bibinfo {pages}
  {041204(R)}}\BibitemShut {NoStop}%
\bibitem [{\citenamefont {Ma}\ \emph {et~al.}(2009)\citenamefont {Ma},
  \citenamefont {Rohlfing},\ and\ \citenamefont
  {Molteni}}]{Ma/Rohlfing/Molteni:2009}%
  \BibitemOpen
  \bibfield  {author} {\bibinfo {author} {\bibnamefont {Ma}, \bibfnamefont
  {Y}}, \bibinfo {author} {\bibfnamefont {M.}~\bibnamefont {Rohlfing}}, \ and\
  \bibinfo {author} {\bibfnamefont {C.}~\bibnamefont {Molteni}}} (\bibinfo
  {year} {2009}),\ \bibfield  {title} {\enquote {\bibinfo {title} {{E}xcited
  states of biological chromophores studied using many-body perturbation
  theory: {E}ffects of resonant-antiresonant coupling and dynamical
  screening},}\ }\href@noop {} {\bibfield  {journal} {\bibinfo  {journal}
  {Phys. Rev. B}\ }\textbf {\bibinfo {volume} {80}},\ \bibinfo {pages}
  {241405}}\BibitemShut {NoStop}%
\bibitem [{\citenamefont {Ma}\ \emph {et~al.}(2010{\natexlab{b}})\citenamefont
  {Ma}, \citenamefont {Rohlfing},\ and\ \citenamefont
  {Molteni}}]{Ma/Rohlfing/Molteni:2010}%
  \BibitemOpen
  \bibfield  {author} {\bibinfo {author} {\bibnamefont {Ma}, \bibfnamefont
  {Y}}, \bibinfo {author} {\bibfnamefont {M.}~\bibnamefont {Rohlfing}}, \ and\
  \bibinfo {author} {\bibfnamefont {C.}~\bibnamefont {Molteni}}} (\bibinfo
  {year} {2010}{\natexlab{b}}),\ \bibfield  {title} {\enquote {\bibinfo {title}
  {{M}odeling the {E}xcited {S}tates of {B}iological {C}hromophores within
  {M}any-{B}ody {G}reen's {F}unction {T}heory},}\ }\href {\doibase
  10.1021/ct900528h} {\bibfield  {journal} {\bibinfo  {journal} {J. Chem.
  Theory Comput.}\ }\textbf {\bibinfo {volume} {6}}~(\bibinfo {number} {1}),\
  \bibinfo {pages} {257--265}}\BibitemShut {NoStop}%
\bibitem [{\citenamefont {Maggio}\ and\ \citenamefont
  {Kresse}(2017)}]{Maggio/Kresse:2017}%
  \BibitemOpen
  \bibfield  {author} {\bibinfo {author} {\bibnamefont {Maggio}, \bibfnamefont
  {E}}, \ and\ \bibinfo {author} {\bibfnamefont {G.}~\bibnamefont {Kresse}}}
  (\bibinfo {year} {2017}),\ \bibfield  {title} {\enquote {\bibinfo {title}
  {${G}{W}$ {V}ertex {C}orrected {C}alculations for {M}olecular {S}ystems},}\
  }\href@noop {} {\bibfield  {journal} {\bibinfo  {journal} {J.\ Chem.\ Theory\
  Comput.}\ }\textbf {\bibinfo {volume} {13}}~(\bibinfo {number} {10}),\
  \bibinfo {pages} {4765--4778}}\BibitemShut {NoStop}%
\bibitem [{\citenamefont {Maggio}\ \emph {et~al.}(2017)\citenamefont {Maggio},
  \citenamefont {Liu}, \citenamefont {van Setten},\ and\ \citenamefont
  {Kresse}}]{Maggio2017b}%
  \BibitemOpen
  \bibfield  {author} {\bibinfo {author} {\bibnamefont {Maggio}, \bibfnamefont
  {E}}, \bibinfo {author} {\bibfnamefont {P.}~\bibnamefont {Liu}}, \bibinfo
  {author} {\bibfnamefont {M.~J.}\ \bibnamefont {van Setten}}, \ and\ \bibinfo
  {author} {\bibfnamefont {G.}~\bibnamefont {Kresse}}} (\bibinfo {year}
  {2017}),\ \bibfield  {title} {\enquote {\bibinfo {title} {${G}{W}$100: {A}
  {P}lane {W}ave {P}erspective for {S}mall {M}olecules},}\ }\href@noop {}
  {\bibfield  {journal} {\bibinfo  {journal} {J. Chem. Theory Comput.}\
  }\textbf {\bibinfo {volume} {13}}~(\bibinfo {number} {2}),\ \bibinfo {pages}
  {635--648}}\BibitemShut {NoStop}%
\bibitem [{\citenamefont {Mahan}\ and\ \citenamefont
  {Sernelius}(1989)}]{Mahan/etal:1989}%
  \BibitemOpen
  \bibfield  {author} {\bibinfo {author} {\bibnamefont {Mahan}, \bibfnamefont
  {G~D}}, \ and\ \bibinfo {author} {\bibfnamefont {B.~E.}\ \bibnamefont
  {Sernelius}}} (\bibinfo {year} {1989}),\ \bibfield  {title} {\enquote
  {\bibinfo {title} {{E}lectron-electron interactions and the bandwidth of
  metals},}\ }\href@noop {} {\bibfield  {journal} {\bibinfo  {journal} {Phys.
  Rev. Lett.}\ }\textbf {\bibinfo {volume} {62}},\ \bibinfo {pages}
  {2718--2720}}\BibitemShut {NoStop}%
\bibitem [{\citenamefont {Manzeli}\ \emph {et~al.}(2017)\citenamefont
  {Manzeli}, \citenamefont {Ovchinnikov}, \citenamefont {Pasquier},
  \citenamefont {Yazyev},\ and\ \citenamefont {Kis}}]{manzeli_nrm_2}%
  \BibitemOpen
  \bibfield  {author} {\bibinfo {author} {\bibnamefont {Manzeli}, \bibfnamefont
  {S}}, \bibinfo {author} {\bibfnamefont {D.}~\bibnamefont {Ovchinnikov}},
  \bibinfo {author} {\bibfnamefont {D.}~\bibnamefont {Pasquier}}, \bibinfo
  {author} {\bibfnamefont {O.~V.}\ \bibnamefont {Yazyev}}, \ and\ \bibinfo
  {author} {\bibfnamefont {A.}~\bibnamefont {Kis}}} (\bibinfo {year} {2017}),\
  \bibfield  {title} {\enquote {\bibinfo {title} {2{D} transition metal
  dichalcogenides},}\ }\href@noop {} {\bibfield  {journal} {\bibinfo  {journal}
  {Nat. Rev. Mat.}\ }\textbf {\bibinfo {volume} {2}},\ \bibinfo {pages}
  {17033}}\BibitemShut {NoStop}%
\bibitem [{\citenamefont {Marini}\ \emph {et~al.}(2009)\citenamefont {Marini},
  \citenamefont {Hogan}, \citenamefont {Gr{\"u}ning},\ and\ \citenamefont
  {Varsano}}]{Marini/etal:2009}%
  \BibitemOpen
  \bibfield  {author} {\bibinfo {author} {\bibnamefont {Marini}, \bibfnamefont
  {A}}, \bibinfo {author} {\bibfnamefont {C.}~\bibnamefont {Hogan}}, \bibinfo
  {author} {\bibfnamefont {M.}~\bibnamefont {Gr{\"u}ning}}, \ and\ \bibinfo
  {author} {\bibfnamefont {D.}~\bibnamefont {Varsano}}} (\bibinfo {year}
  {2009}),\ \bibfield  {title} {\enquote {\bibinfo {title} {yambo: {A}n ab
  initio tool for excited state calculations},}\ }\href@noop {} {\bibfield
  {journal} {\bibinfo  {journal} {Comput. Phys. Commun.}\ }\textbf {\bibinfo
  {volume} {180}}~(\bibinfo {number} {8}),\ \bibinfo {pages} {1392 --
  1403}}\BibitemShut {NoStop}%
\bibitem [{\citenamefont {Marini}\ \emph {et~al.}(2002)\citenamefont {Marini},
  \citenamefont {Sole}, \citenamefont {Rubio},\ and\ \citenamefont
  {Onida}}]{Marini/etal:2002}%
  \BibitemOpen
  \bibfield  {author} {\bibinfo {author} {\bibnamefont {Marini}, \bibfnamefont
  {A}}, \bibinfo {author} {\bibfnamefont {R.~Del}\ \bibnamefont {Sole}},
  \bibinfo {author} {\bibfnamefont {A.}~\bibnamefont {Rubio}}, \ and\ \bibinfo
  {author} {\bibfnamefont {G.}~\bibnamefont {Onida}}} (\bibinfo {year}
  {2002}),\ \bibfield  {title} {\enquote {\bibinfo {title} {{Q}uasiparticle
  band-structure effects on the $d$ hole lifetimes of copper within the
  ${G}{W}$ approximation},}\ }\href@noop {} {\bibfield  {journal} {\bibinfo
  {journal} {Phys. Rev. B}\ }\textbf {\bibinfo {volume} {66}},\ \bibinfo
  {pages} {161104}}\BibitemShut {NoStop}%
\bibitem [{\citenamefont {Marom}(2017)}]{Marom:2017}%
  \BibitemOpen
  \bibfield  {author} {\bibinfo {author} {\bibnamefont {Marom}, \bibfnamefont
  {N}}} (\bibinfo {year} {2017}),\ \bibfield  {title} {\enquote {\bibinfo
  {title} {{A}ccurate description of the electronic structure of organic
  semiconductors by ${G}{W}$ methods},}\ }\href@noop {} {\bibfield  {journal}
  {\bibinfo  {journal} {J. Phys.: Condens. Matter}\ }\textbf {\bibinfo {volume}
  {29}}~(\bibinfo {number} {10}),\ \bibinfo {pages} {103003}}\BibitemShut
  {NoStop}%
\bibitem [{\citenamefont {Marom}\ \emph {et~al.}(2012)\citenamefont {Marom},
  \citenamefont {Caruso}, \citenamefont {Ren}, \citenamefont {Hofmann},
  \citenamefont {K\"orzd\"orfer}, \citenamefont {Chelikowsky}, \citenamefont
  {Rubio}, \citenamefont {Scheffler},\ and\ \citenamefont
  {Rinke}}]{Marom/etal:2012}%
  \BibitemOpen
  \bibfield  {author} {\bibinfo {author} {\bibnamefont {Marom}, \bibfnamefont
  {N}}, \bibinfo {author} {\bibfnamefont {F.}~\bibnamefont {Caruso}}, \bibinfo
  {author} {\bibfnamefont {X.}~\bibnamefont {Ren}}, \bibinfo {author}
  {\bibfnamefont {O.~T.}\ \bibnamefont {Hofmann}}, \bibinfo {author}
  {\bibfnamefont {T.}~\bibnamefont {K\"orzd\"orfer}}, \bibinfo {author}
  {\bibfnamefont {J.~R.}\ \bibnamefont {Chelikowsky}}, \bibinfo {author}
  {\bibfnamefont {A.}~\bibnamefont {Rubio}}, \bibinfo {author} {\bibfnamefont
  {M.}~\bibnamefont {Scheffler}}, \ and\ \bibinfo {author} {\bibfnamefont
  {P.}~\bibnamefont {Rinke}}} (\bibinfo {year} {2012}),\ \bibfield  {title}
  {\enquote {\bibinfo {title} {{B}enchmark of ${G}{W}$ methods for
  azabenzenes},}\ }\href@noop {} {\bibfield  {journal} {\bibinfo  {journal}
  {Phys. Rev. B}\ }\textbf {\bibinfo {volume} {86}},\ \bibinfo {pages}
  {245127}}\BibitemShut {NoStop}%
\bibitem [{\citenamefont {Marom}\ \emph {et~al.}(2011)\citenamefont {Marom},
  \citenamefont {Ren}, \citenamefont {Moussa}, \citenamefont {Chelikowsky},\
  and\ \citenamefont {Kronik}}]{Marom2011}%
  \BibitemOpen
  \bibfield  {author} {\bibinfo {author} {\bibnamefont {Marom}, \bibfnamefont
  {N}}, \bibinfo {author} {\bibfnamefont {X.}~\bibnamefont {Ren}}, \bibinfo
  {author} {\bibfnamefont {J.~E.}\ \bibnamefont {Moussa}}, \bibinfo {author}
  {\bibfnamefont {J.~R.}\ \bibnamefont {Chelikowsky}}, \ and\ \bibinfo {author}
  {\bibfnamefont {L.}~\bibnamefont {Kronik}}} (\bibinfo {year} {2011}),\
  \bibfield  {title} {\enquote {\bibinfo {title} {{E}lectronic structure of
  copper phthalocyanine from ${G}_{0}{W}_{0}$ calculations},}\ }\href {\doibase
  10.1103/PhysRevB.84.195143} {\bibfield  {journal} {\bibinfo  {journal} {Phys.
  Rev. B}\ }\textbf {\bibinfo {volume} {84}},\ \bibinfo {pages}
  {195143}}\BibitemShut {NoStop}%
\bibitem [{\citenamefont {Marsili}\ \emph {et~al.}(2005)\citenamefont
  {Marsili}, \citenamefont {Pulci}, \citenamefont {Bechstedt},\ and\
  \citenamefont {{Del Sole}}}]{Marsili/etal:2005}%
  \BibitemOpen
  \bibfield  {author} {\bibinfo {author} {\bibnamefont {Marsili}, \bibfnamefont
  {M}}, \bibinfo {author} {\bibfnamefont {O.}~\bibnamefont {Pulci}}, \bibinfo
  {author} {\bibfnamefont {F.}~\bibnamefont {Bechstedt}}, \ and\ \bibinfo
  {author} {\bibfnamefont {R.}~\bibnamefont {{Del Sole}}}} (\bibinfo {year}
  {2005}),\ \bibfield  {title} {\enquote {\bibinfo {title} {{E}lectronic
  structure of the {C}(111) surface: {S}olution by self-consistent many-body
  calculations},}\ }\href@noop {} {\bibfield  {journal} {\bibinfo  {journal}
  {Phys.\ Rev.\ B}\ }\textbf {\bibinfo {volume} {72}},\ \bibinfo {pages}
  {115415}}\BibitemShut {NoStop}%
\bibitem [{\citenamefont {Marsili}\ \emph {et~al.}(2014)\citenamefont
  {Marsili}, \citenamefont {Umari}, \citenamefont {Santo}, \citenamefont
  {Caputo}, \citenamefont {Panighel}, \citenamefont {Goldoni}, \citenamefont
  {Kumar},\ and\ \citenamefont {Pedio}}]{Marsili2014}%
  \BibitemOpen
  \bibfield  {author} {\bibinfo {author} {\bibnamefont {Marsili}, \bibfnamefont
  {M}}, \bibinfo {author} {\bibfnamefont {P.}~\bibnamefont {Umari}}, \bibinfo
  {author} {\bibfnamefont {G.~Di}\ \bibnamefont {Santo}}, \bibinfo {author}
  {\bibfnamefont {M.}~\bibnamefont {Caputo}}, \bibinfo {author} {\bibfnamefont
  {M.}~\bibnamefont {Panighel}}, \bibinfo {author} {\bibfnamefont
  {A.}~\bibnamefont {Goldoni}}, \bibinfo {author} {\bibfnamefont
  {M.}~\bibnamefont {Kumar}}, \ and\ \bibinfo {author} {\bibfnamefont
  {M.}~\bibnamefont {Pedio}}} (\bibinfo {year} {2014}),\ \bibfield  {title}
  {\enquote {\bibinfo {title} {{S}olid state effects on the electronic
  structure of {H}$_2${O}{E}{P}},}\ }\href {\doibase 10.1039/C4CP03450C}
  {\bibfield  {journal} {\bibinfo  {journal} {Phys. Chem. Chem. Phys.}\
  }\textbf {\bibinfo {volume} {16}},\ \bibinfo {pages}
  {27104--27111}}\BibitemShut {NoStop}%
\bibitem [{\citenamefont {Martin}\ \emph {et~al.}(2016)\citenamefont {Martin},
  \citenamefont {Reining},\ and\ \citenamefont
  {Ceperley}}]{martin_reining_ceperley_2016}%
  \BibitemOpen
  \bibfield  {author} {\bibinfo {author} {\bibnamefont {Martin}, \bibfnamefont
  {R~M}}, \bibinfo {author} {\bibfnamefont {L.}~\bibnamefont {Reining}}, \ and\
  \bibinfo {author} {\bibfnamefont {D.~M.}\ \bibnamefont {Ceperley}}} (\bibinfo
  {year} {2016}),\ \href {\doibase 10.1017/CBO9781139050807} {\emph {\bibinfo
  {title} {{I}nteracting {E}lectrons: {T}heory and {C}omputational
  {A}pproaches}}}\ (\bibinfo  {publisher} {Cambridge University
  Press})\BibitemShut {NoStop}%
\bibitem [{\citenamefont {Martin}(2004)}]{MartinBook}%
  \BibitemOpen
  \bibfield  {author} {\bibinfo {author} {\bibnamefont {Martin}, \bibfnamefont
  {R~R}}} (\bibinfo {year} {2004}),\ \href@noop {} {\emph {\bibinfo {title}
  {{E}lectronic {S}tructure: {B}asic {T}heory and {P}ractical {M}ethods}}}\
  (\bibinfo  {publisher} {Cambridge University Press})\BibitemShut {NoStop}%
\bibitem [{\citenamefont {Martin-Samos}\ and\ \citenamefont
  {Bussi}(2009)}]{MartinSamos/Bussi:2009}%
  \BibitemOpen
  \bibfield  {author} {\bibinfo {author} {\bibnamefont {Martin-Samos},
  \bibfnamefont {L}}, \ and\ \bibinfo {author} {\bibfnamefont {G.}~\bibnamefont
  {Bussi}}} (\bibinfo {year} {2009}),\ \bibfield  {title} {\enquote {\bibinfo
  {title} {{S}a{X}: {A}n open source package for electronic-structure and
  optical-properties calculations in the ${G}{W}$ approximation},}\ }\href
  {\doibase 10.1016/j.cpc.2009.02.005} {\bibfield  {journal} {\bibinfo
  {journal} {Comput. Phys. Commun.}\ }\textbf {\bibinfo {volume}
  {180}}~(\bibinfo {number} {8}),\ \bibinfo {pages} {1416 -- 1425}}\BibitemShut
  {NoStop}%
\bibitem [{\citenamefont {Marx}\ and\ \citenamefont
  {Hutter}(2009)}]{HutterBook}%
  \BibitemOpen
  \bibfield  {author} {\bibinfo {author} {\bibnamefont {Marx}, \bibfnamefont
  {D}}, \ and\ \bibinfo {author} {\bibfnamefont {J.}~\bibnamefont {Hutter}}}
  (\bibinfo {year} {2009}),\ \href@noop {} {\emph {\bibinfo {title} {{A}b
  {I}nitio {M}olecular {D}ynamics}}}\ (\bibinfo  {publisher} {Cambridge
  University Press},\ \bibinfo {address} {New York})\BibitemShut {NoStop}%
\bibitem [{\citenamefont {Massidda}\ \emph {et~al.}(1995)\citenamefont
  {Massidda}, \citenamefont {Continenza}, \citenamefont {Posternak},\ and\
  \citenamefont {Baldereschi}}]{Massidda/etal:1995}%
  \BibitemOpen
  \bibfield  {author} {\bibinfo {author} {\bibnamefont {Massidda},
  \bibfnamefont {S}}, \bibinfo {author} {\bibfnamefont {A.}~\bibnamefont
  {Continenza}}, \bibinfo {author} {\bibfnamefont {M.}~\bibnamefont
  {Posternak}}, \ and\ \bibinfo {author} {\bibfnamefont {A.}~\bibnamefont
  {Baldereschi}}} (\bibinfo {year} {1995}),\ \bibfield  {title} {\enquote
  {\bibinfo {title} {{B}and-structure picture for {M}n{O} reexplored: a model
  ${G}{W}$ calculation},}\ }\href@noop {} {\bibfield  {journal} {\bibinfo
  {journal} {Phys. Rev. Lett.}\ }\textbf {\bibinfo {volume} {74}},\ \bibinfo
  {pages} {2323}}\BibitemShut {NoStop}%
\bibitem [{\citenamefont {Massidda}\ \emph {et~al.}(1997)\citenamefont
  {Massidda}, \citenamefont {Continenza}, \citenamefont {Posternak},\ and\
  \citenamefont {Baldereschi}}]{Massidda/etal:1997}%
  \BibitemOpen
  \bibfield  {author} {\bibinfo {author} {\bibnamefont {Massidda},
  \bibfnamefont {S}}, \bibinfo {author} {\bibfnamefont {A.}~\bibnamefont
  {Continenza}}, \bibinfo {author} {\bibfnamefont {M.}~\bibnamefont
  {Posternak}}, \ and\ \bibinfo {author} {\bibfnamefont {A.}~\bibnamefont
  {Baldereschi}}} (\bibinfo {year} {1997}),\ \bibfield  {title} {\enquote
  {\bibinfo {title} {{Q}uasiparticle energy bands of transition-metal oxides
  within a model ${G}{W}$ scheme},}\ }\href@noop {} {\bibfield  {journal}
  {\bibinfo  {journal} {Phys. Rev. B}\ }\textbf {\bibinfo {volume} {55}},\
  \bibinfo {pages} {13494}}\BibitemShut {NoStop}%
\bibitem [{\citenamefont {Mattuck}(1992)}]{Mattuck:1992}%
  \BibitemOpen
  \bibfield  {author} {\bibinfo {author} {\bibnamefont {Mattuck}, \bibfnamefont
  {R~D}}} (\bibinfo {year} {1992}),\ \href
  {https://books.google.fi/books?id=pe-v8zfxE68C} {\emph {\bibinfo {title} {{A}
  {G}uide to {F}eynman {D}iagrams in the {M}any-body {P}roblem}}},\ Dover Books
  on Physics Series\ (\bibinfo  {publisher} {Dover Publications})\BibitemShut
  {NoStop}%
\bibitem [{\citenamefont {McAuliffe}\ \emph {et~al.}(2017)\citenamefont
  {McAuliffe}, \citenamefont {Miller}, \citenamefont {Zhang}, \citenamefont
  {Hulbert}, \citenamefont {Huq}, \citenamefont {dela Cruz}, \citenamefont
  {Schleife},\ and\ \citenamefont {Shoemaker}}]{McAuliffe/etal:2017}%
  \BibitemOpen
  \bibfield  {author} {\bibinfo {author} {\bibnamefont {McAuliffe},
  \bibfnamefont {R~D}}, \bibinfo {author} {\bibfnamefont {C.~A.}\ \bibnamefont
  {Miller}}, \bibinfo {author} {\bibfnamefont {X.}~\bibnamefont {Zhang}},
  \bibinfo {author} {\bibfnamefont {B.~S.}\ \bibnamefont {Hulbert}}, \bibinfo
  {author} {\bibfnamefont {A.}~\bibnamefont {Huq}}, \bibinfo {author}
  {\bibfnamefont {C.}~\bibnamefont {dela Cruz}}, \bibinfo {author}
  {\bibfnamefont {A.}~\bibnamefont {Schleife}}, \ and\ \bibinfo {author}
  {\bibfnamefont {D.~P.}\ \bibnamefont {Shoemaker}}} (\bibinfo {year} {2017}),\
  \bibfield  {title} {\enquote {\bibinfo {title} {{S}tructural, {E}lectronic,
  and {O}ptical {P}roperties of {K}$_2${S}n$_3${O}$_7$ with an {O}ffset
  {H}ollandite {S}tructure},}\ }\href@noop {} {\bibfield  {journal} {\bibinfo
  {journal} {Inorg. Chem.}\ }\textbf {\bibinfo {volume} {56}}~(\bibinfo
  {number} {5}),\ \bibinfo {pages} {2914--2918}}\BibitemShut {NoStop}%
\bibitem [{\citenamefont {McMinis}\ \emph {et~al.}(2015)\citenamefont
  {McMinis}, \citenamefont {Clay}, \citenamefont {Lee},\ and\ \citenamefont
  {Morales}}]{Mcminis/etal:2015}%
  \BibitemOpen
  \bibfield  {author} {\bibinfo {author} {\bibnamefont {McMinis}, \bibfnamefont
  {J}}, \bibinfo {author} {\bibfnamefont {R.~C.}\ \bibnamefont {Clay}},
  \bibinfo {author} {\bibfnamefont {D.}~\bibnamefont {Lee}}, \ and\ \bibinfo
  {author} {\bibfnamefont {M.~A.}\ \bibnamefont {Morales}}} (\bibinfo {year}
  {2015}),\ \bibfield  {title} {\enquote {\bibinfo {title} {{M}olecular to
  {A}tomic {P}hase {T}ransition in {H}ydrogen under {H}igh {P}ressure},}\
  }\href@noop {} {\bibfield  {journal} {\bibinfo  {journal} {Phys. Rev. Lett.}\
  }\textbf {\bibinfo {volume} {114}},\ \bibinfo {pages} {105305}}\BibitemShut
  {NoStop}%
\bibitem [{\citenamefont {Methfessel}\ \emph {et~al.}(2000)\citenamefont
  {Methfessel}, \citenamefont {van Schilfgaarde},\ and\ \citenamefont
  {Casali}}]{Methfessel2000}%
  \BibitemOpen
  \bibfield  {author} {\bibinfo {author} {\bibnamefont {Methfessel},
  \bibfnamefont {M}}, \bibinfo {author} {\bibfnamefont {M.}~\bibnamefont {van
  Schilfgaarde}}, \ and\ \bibinfo {author} {\bibfnamefont {R.~A.}\ \bibnamefont
  {Casali}}} (\bibinfo {year} {2000}),\ \bibfield  {title} {\enquote {\bibinfo
  {title} {{A} {F}ull-{P}otential {L}{M}{T}{O} {M}ethod {B}ased on {S}mooth
  {H}ankel {F}unctions},}\ }in\ \href@noop {} {\emph {\bibinfo {booktitle}
  {{E}lectronic {S}tructure and {P}hysical {P}roperies of {S}olids: {T}he
  {U}ses of the {L}{M}{T}{O} {M}ethod}}},\ \bibinfo {editor} {edited by\
  \bibinfo {editor} {\bibfnamefont {Hugues}\ \bibnamefont {Dreyss{\'e}}}}\
  (\bibinfo  {publisher} {Lecture Notes in Physics, Springer Berlin
  Heidelberg},\ \bibinfo {address} {Berlin, Heidelberg})\ pp.\ \bibinfo {pages}
  {114--147}\BibitemShut {NoStop}%
\bibitem [{\citenamefont {Miglio}\ \emph {et~al.}(2012)\citenamefont {Miglio},
  \citenamefont {Waroquiers}, \citenamefont {Antonius}, \citenamefont
  {Giantomassi}, \citenamefont {Stankovski}, \citenamefont {C{\^o}t{\'e}},
  \citenamefont {X.},\ and\ \citenamefont {Rignanese}}]{Miglio/etal:2012}%
  \BibitemOpen
  \bibfield  {author} {\bibinfo {author} {\bibnamefont {Miglio}, \bibfnamefont
  {A}}, \bibinfo {author} {\bibfnamefont {D.}~\bibnamefont {Waroquiers}},
  \bibinfo {author} {\bibfnamefont {G.}~\bibnamefont {Antonius}}, \bibinfo
  {author} {\bibfnamefont {M.}~\bibnamefont {Giantomassi}}, \bibinfo {author}
  {\bibfnamefont {M.}~\bibnamefont {Stankovski}}, \bibinfo {author}
  {\bibfnamefont {M.}~\bibnamefont {C{\^o}t{\'e}}}, \bibinfo {author}
  {\bibnamefont {X.}}, \ and\ \bibinfo {author} {\bibfnamefont {G.-M.}\
  \bibnamefont {Rignanese}}} (\bibinfo {year} {2012}),\ \bibfield  {title}
  {\enquote {\bibinfo {title} {{E}ffects of plasmon pole models on the
  ${G}_0{W}_0$ electronic structure of various oxides},}\ }\href@noop {}
  {\bibfield  {journal} {\bibinfo  {journal} {Eur. Phys. J. B}\ }\textbf
  {\bibinfo {volume} {85}}~(\bibinfo {number} {9}),\ \bibinfo {pages}
  {1--8}}\BibitemShut {NoStop}%
\bibitem [{\citenamefont {Min\'{a}r}\ \emph {et~al.}(2013)\citenamefont
  {Min\'{a}r}, \citenamefont {Braun},\ and\ \citenamefont {Ebert}}]{Minar2013}%
  \BibitemOpen
  \bibfield  {author} {\bibinfo {author} {\bibnamefont {Min\'{a}r},
  \bibfnamefont {J}}, \bibinfo {author} {\bibfnamefont {J.}~\bibnamefont
  {Braun}}, \ and\ \bibinfo {author} {\bibfnamefont {H.}~\bibnamefont {Ebert}}}
  (\bibinfo {year} {2013}),\ \bibfield  {title} {\enquote {\bibinfo {title}
  {{R}ecent developments in the theory of {H}{A}{R}{P}{E}{S}},}\ }\href@noop {}
  {\bibfield  {journal} {\bibinfo  {journal} {J. Electron Spectrosc. Relat.
  Phenom.}\ }\textbf {\bibinfo {volume} {190}},\ \bibinfo {pages} {159 --
  164}}\BibitemShut {NoStop}%
\bibitem [{\citenamefont {Min\'{a}r}\ \emph {et~al.}(2011)\citenamefont
  {Min\'{a}r}, \citenamefont {Braun}, \citenamefont {Mankovsky},\ and\
  \citenamefont {Ebert}}]{Minar2011}%
  \BibitemOpen
  \bibfield  {author} {\bibinfo {author} {\bibnamefont {Min\'{a}r},
  \bibfnamefont {J}}, \bibinfo {author} {\bibfnamefont {J.}~\bibnamefont
  {Braun}}, \bibinfo {author} {\bibfnamefont {S.}~\bibnamefont {Mankovsky}}, \
  and\ \bibinfo {author} {\bibfnamefont {H.}~\bibnamefont {Ebert}}} (\bibinfo
  {year} {2011}),\ \bibfield  {title} {\enquote {\bibinfo {title}
  {{C}alculation of angle-resolved photo emission spectra within the one-step
  model of photo emission - {R}ecent developments},}\ }\href@noop {} {\bibfield
   {journal} {\bibinfo  {journal} {J. Electron Spectrosc. Relat. Phenom.}\
  }\textbf {\bibinfo {volume} {184}}~(\bibinfo {number} {3}),\ \bibinfo {pages}
  {91 -- 99}},\ \bibinfo {note} {advances in Vacuum Ultraviolet and X-ray
  Physics}\BibitemShut {NoStop}%
\bibitem [{\citenamefont {M\l{}y\'{n}czak}\ \emph {et~al.}(2019)\citenamefont
  {M\l{}y\'{n}czak}, \citenamefont {M\"uller}, \citenamefont {Gospodari\v{c}},
  \citenamefont {Heider}, \citenamefont {Aguilera}, \citenamefont {Bihlmayer},
  \citenamefont {Gehlmann}, \citenamefont {Jugovac}, \citenamefont
  {Zamborlini}, \citenamefont {Tusche}, \citenamefont {Suga}, \citenamefont
  {Feyer}, \citenamefont {Plucinski}, \citenamefont {Friedrich}, \citenamefont
  {Bl\"ugel},\ and\ \citenamefont {Schneider}}]{mlynczak_nc_10}%
  \BibitemOpen
  \bibfield  {author} {\bibinfo {author} {\bibnamefont {M\l{}y\'{n}czak},
  \bibfnamefont {E}}, \bibinfo {author} {\bibfnamefont {M.~C. T.~D.}\
  \bibnamefont {M\"uller}}, \bibinfo {author} {\bibfnamefont {P.}~\bibnamefont
  {Gospodari\v{c}}}, \bibinfo {author} {\bibfnamefont {T.}~\bibnamefont
  {Heider}}, \bibinfo {author} {\bibfnamefont {I.}~\bibnamefont {Aguilera}},
  \bibinfo {author} {\bibfnamefont {G.}~\bibnamefont {Bihlmayer}}, \bibinfo
  {author} {\bibfnamefont {M.}~\bibnamefont {Gehlmann}}, \bibinfo {author}
  {\bibfnamefont {M.}~\bibnamefont {Jugovac}}, \bibinfo {author} {\bibfnamefont
  {G.}~\bibnamefont {Zamborlini}}, \bibinfo {author} {\bibfnamefont
  {C.}~\bibnamefont {Tusche}}, \bibinfo {author} {\bibfnamefont
  {S.}~\bibnamefont {Suga}}, \bibinfo {author} {\bibfnamefont {V.}~\bibnamefont
  {Feyer}}, \bibinfo {author} {\bibfnamefont {L.}~\bibnamefont {Plucinski}},
  \bibinfo {author} {\bibfnamefont {C.}~\bibnamefont {Friedrich}}, \bibinfo
  {author} {\bibfnamefont {S.}~\bibnamefont {Bl\"ugel}}, \ and\ \bibinfo
  {author} {\bibfnamefont {C.~M.}\ \bibnamefont {Schneider}}} (\bibinfo {year}
  {2019}),\ \bibfield  {title} {\enquote {\bibinfo {title} {{K}ink far below
  the {F}ermi level reveals new electron-magnon scattering channel in {F}e},}\
  }\href@noop {} {\bibfield  {journal} {\bibinfo  {journal} {Nat. Comm}\
  }\textbf {\bibinfo {volume} {10}},\ \bibinfo {pages} {505}}\BibitemShut
  {NoStop}%
\bibitem [{\citenamefont {Molina-S\'anchez}\ \emph {et~al.}(2013)\citenamefont
  {Molina-S\'anchez}, \citenamefont {Sangalli}, \citenamefont {Hummer},
  \citenamefont {Marini},\ and\ \citenamefont {Wirtz}}]{molina_prb_88}%
  \BibitemOpen
  \bibfield  {author} {\bibinfo {author} {\bibnamefont {Molina-S\'anchez},
  \bibfnamefont {A}}, \bibinfo {author} {\bibfnamefont {D.}~\bibnamefont
  {Sangalli}}, \bibinfo {author} {\bibfnamefont {K.}~\bibnamefont {Hummer}},
  \bibinfo {author} {\bibfnamefont {A.}~\bibnamefont {Marini}}, \ and\ \bibinfo
  {author} {\bibfnamefont {L.}~\bibnamefont {Wirtz}}} (\bibinfo {year}
  {2013}),\ \bibfield  {title} {\enquote {\bibinfo {title} {{E}ffect of
  spin-orbit interaction on the optical spectra of single-layer, double-layer,
  and bulk {M}o{S}${}_{2}$},}\ }\href@noop {} {\bibfield  {journal} {\bibinfo
  {journal} {Phys. Rev. B}\ }\textbf {\bibinfo {volume} {88}},\ \bibinfo
  {pages} {045412}}\BibitemShut {NoStop}%
\bibitem [{\citenamefont {Monkhorst}(2005)}]{Monkhorst:2005}%
  \BibitemOpen
  \bibfield  {author} {\bibinfo {author} {\bibnamefont {Monkhorst},
  \bibfnamefont {H~J}}} (\bibinfo {year} {2005}),\ \bibfield  {title} {\enquote
  {\bibinfo {title} {${G}{W}$ method for extended, periodic systems with a
  mixed {S}later-orbital/plane-wave basis and {F}ourier transform
  techniques},}\ }\href@noop {} {\bibfield  {journal} {\bibinfo  {journal}
  {Adv. Quant. Chem.}\ }\textbf {\bibinfo {volume} {48}},\ \bibinfo {pages}
  {35}}\BibitemShut {NoStop}%
\bibitem [{\citenamefont {Mori-S\'{a}nchez}\ \emph {et~al.}(2006)\citenamefont
  {Mori-S\'{a}nchez}, \citenamefont {Cohen},\ and\ \citenamefont
  {Yang}}]{Mori-Sanchez2006}%
  \BibitemOpen
  \bibfield  {author} {\bibinfo {author} {\bibnamefont {Mori-S\'{a}nchez},
  \bibfnamefont {P}}, \bibinfo {author} {\bibfnamefont {A.~J.}\ \bibnamefont
  {Cohen}}, \ and\ \bibinfo {author} {\bibfnamefont {W.}~\bibnamefont {Yang}}}
  (\bibinfo {year} {2006}),\ \bibfield  {title} {\enquote {\bibinfo {title}
  {{M}any-electron self-interaction error in approximate density
  functionals},}\ }\href@noop {} {\bibfield  {journal} {\bibinfo  {journal} {J.
  Chem. Phys.}\ }\textbf {\bibinfo {volume} {125}}~(\bibinfo {number} {20}),\
  \bibinfo {pages} {201102}}\BibitemShut {NoStop}%
\bibitem [{\citenamefont {Mortensen}\ \emph {et~al.}(2005)\citenamefont
  {Mortensen}, \citenamefont {Hansen},\ and\ \citenamefont
  {Jacobsen}}]{Mortensen2005}%
  \BibitemOpen
  \bibfield  {author} {\bibinfo {author} {\bibnamefont {Mortensen},
  \bibfnamefont {J~J}}, \bibinfo {author} {\bibfnamefont {L.~B.}\ \bibnamefont
  {Hansen}}, \ and\ \bibinfo {author} {\bibfnamefont {K.~W.}\ \bibnamefont
  {Jacobsen}}} (\bibinfo {year} {2005}),\ \bibfield  {title} {\enquote
  {\bibinfo {title} {{R}eal-space grid implementation of the projector
  augmented wave method},}\ }\href {\doibase 10.1103/PhysRevB.71.035109}
  {\bibfield  {journal} {\bibinfo  {journal} {Phys. Rev. B}\ }\textbf {\bibinfo
  {volume} {71}},\ \bibinfo {pages} {035109}}\BibitemShut {NoStop}%
\bibitem [{\citenamefont {Mott}(1968)}]{rmp_40_mott}%
  \BibitemOpen
  \bibfield  {author} {\bibinfo {author} {\bibnamefont {Mott}, \bibfnamefont
  {N~F}}} (\bibinfo {year} {1968}),\ \bibfield  {title} {\enquote {\bibinfo
  {title} {{M}etal-{I}nsulator {T}ransition},}\ }\href {\doibase
  10.1103/RevModPhys.40.677} {\bibfield  {journal} {\bibinfo  {journal} {Rev.
  Mod. Phys.}\ }\textbf {\bibinfo {volume} {40}},\ \bibinfo {pages}
  {677--683}}\BibitemShut {NoStop}%
\bibitem [{\citenamefont {M\"uller}\ \emph {et~al.}(2018)\citenamefont
  {M\"uller}, \citenamefont {Bl\"ugel},\ and\ \citenamefont
  {Friedrich}}]{friedrich/etal:2018}%
  \BibitemOpen
  \bibfield  {author} {\bibinfo {author} {\bibnamefont {M\"uller},
  \bibfnamefont {M~C T~D}}, \bibinfo {author} {\bibfnamefont {S.}~\bibnamefont
  {Bl\"ugel}}, \ and\ \bibinfo {author} {\bibfnamefont {C.}~\bibnamefont
  {Friedrich}}} (\bibinfo {year} {2018}),\ \bibfield  {title} {\enquote
  {\bibinfo {title} {{E}lectron-magnon scattering in elementary ferromagnets
  from first principles: lifetime broadening and kinks},}\ }\href@noop {}
  {\bibinfo  {journal} {arXiv:1809.02395}\ }\BibitemShut {NoStop}%
\bibitem [{\citenamefont {M\"uller}\ \emph {et~al.}(2016)\citenamefont
  {M\"uller}, \citenamefont {Friedrich},\ and\ \citenamefont
  {Bl\"ugel}}]{friedrich/etal:2016}%
  \BibitemOpen
\bibfield  {journal} {  }\bibfield  {author} {\bibinfo {author} {\bibnamefont
  {M\"uller}, \bibfnamefont {M~C T~D}}, \bibinfo {author} {\bibfnamefont
  {C.}~\bibnamefont {Friedrich}}, \ and\ \bibinfo {author} {\bibfnamefont
  {S.}~\bibnamefont {Bl\"ugel}}} (\bibinfo {year} {2016}),\ \bibfield  {title}
  {\enquote {\bibinfo {title} {{A}coustic magnons in the long-wavelength limit:
  {I}nvestigating the {G}oldstone violation in many-body perturbation
  theory},}\ }\href {\doibase 10.1103/PhysRevB.94.064433} {\bibfield  {journal}
  {\bibinfo  {journal} {Phys. Rev. B}\ }\textbf {\bibinfo {volume} {94}},\
  \bibinfo {pages} {064433}}\BibitemShut {NoStop}%
\bibitem [{\citenamefont {Nabok}\ \emph {et~al.}(2016)\citenamefont {Nabok},
  \citenamefont {Gulans},\ and\ \citenamefont {Draxl}}]{Nabok2016}%
  \BibitemOpen
  \bibfield  {author} {\bibinfo {author} {\bibnamefont {Nabok}, \bibfnamefont
  {D}}, \bibinfo {author} {\bibfnamefont {A.}~\bibnamefont {Gulans}}, \ and\
  \bibinfo {author} {\bibfnamefont {C.}~\bibnamefont {Draxl}}} (\bibinfo {year}
  {2016}),\ \bibfield  {title} {\enquote {\bibinfo {title} {{A}ccurate
  all-electron ${G}_{0}{W}_{0}$ quasiparticle energies employing the
  full-potential augmented plane-wave method},}\ }\href {\doibase
  10.1103/PhysRevB.94.035118} {\bibfield  {journal} {\bibinfo  {journal} {Phys.
  Rev. B}\ }\textbf {\bibinfo {volume} {94}},\ \bibinfo {pages}
  {035118}}\BibitemShut {NoStop}%
\bibitem [{\citenamefont {Neaton}\ \emph {et~al.}(2006)\citenamefont {Neaton},
  \citenamefont {Hybertsen},\ and\ \citenamefont {Louie}}]{Neaton2006}%
  \BibitemOpen
  \bibfield  {author} {\bibinfo {author} {\bibnamefont {Neaton}, \bibfnamefont
  {J~B}}, \bibinfo {author} {\bibfnamefont {M.~S.}\ \bibnamefont {Hybertsen}},
  \ and\ \bibinfo {author} {\bibfnamefont {S.~G.}\ \bibnamefont {Louie}}}
  (\bibinfo {year} {2006}),\ \bibfield  {title} {\enquote {\bibinfo {title}
  {{R}enormalization of {M}olecular {E}lectronic {L}evels at {M}etal-{M}olecule
  {I}nterfaces},}\ }\href@noop {} {\bibfield  {journal} {\bibinfo  {journal}
  {Phys. Rev. Lett.}\ }\textbf {\bibinfo {volume} {97}},\ \bibinfo {pages}
  {216405}}\BibitemShut {NoStop}%
\bibitem [{\citenamefont {Nechaev}\ \emph {et~al.}(2015)\citenamefont
  {Nechaev}, \citenamefont {Aguilera}, \citenamefont {Renzi}, \citenamefont
  {di~Bona}, \citenamefont {Rizzini}, \citenamefont {Mio}, \citenamefont
  {Nicotra}, \citenamefont {Politano}, \citenamefont {Scalese}, \citenamefont
  {Aliev}, \citenamefont {Babanly}, \citenamefont {Friedrich}, \citenamefont
  {Bl\"ugel},\ and\ \citenamefont {Chulkov}}]{Nechaev/etal:2015}%
  \BibitemOpen
  \bibfield  {author} {\bibinfo {author} {\bibnamefont {Nechaev}, \bibfnamefont
  {I~A}}, \bibinfo {author} {\bibfnamefont {I.}~\bibnamefont {Aguilera}},
  \bibinfo {author} {\bibfnamefont {V.~De}\ \bibnamefont {Renzi}}, \bibinfo
  {author} {\bibfnamefont {A.}~\bibnamefont {di~Bona}}, \bibinfo {author}
  {\bibfnamefont {A.~Lodi}\ \bibnamefont {Rizzini}}, \bibinfo {author}
  {\bibfnamefont {A.~M.}\ \bibnamefont {Mio}}, \bibinfo {author} {\bibfnamefont
  {G.}~\bibnamefont {Nicotra}}, \bibinfo {author} {\bibfnamefont
  {A.}~\bibnamefont {Politano}}, \bibinfo {author} {\bibfnamefont
  {S.}~\bibnamefont {Scalese}}, \bibinfo {author} {\bibfnamefont {Z.~S.}\
  \bibnamefont {Aliev}}, \bibinfo {author} {\bibfnamefont {M.~B.}\ \bibnamefont
  {Babanly}}, \bibinfo {author} {\bibfnamefont {C.}~\bibnamefont {Friedrich}},
  \bibinfo {author} {\bibfnamefont {S.}~\bibnamefont {Bl\"ugel}}, \ and\
  \bibinfo {author} {\bibfnamefont {E.~V.}\ \bibnamefont {Chulkov}}} (\bibinfo
  {year} {2015}),\ \bibfield  {title} {\enquote {\bibinfo {title}
  {{Q}uasiparticle spectrum and plasmonic excitations in the topological
  insulator {Sb}$_{2}${Te}$_{3}$},}\ }\href@noop {} {\bibfield  {journal}
  {\bibinfo  {journal} {Phys. Rev. B}\ }\textbf {\bibinfo {volume} {91}},\
  \bibinfo {pages} {245123}}\BibitemShut {NoStop}%
\bibitem [{\citenamefont {Nechaev}\ and\ \citenamefont
  {Chulkov}(2013)}]{Nechaev/etal:2013}%
  \BibitemOpen
  \bibfield  {author} {\bibinfo {author} {\bibnamefont {Nechaev}, \bibfnamefont
  {I~A}}, \ and\ \bibinfo {author} {\bibfnamefont {E.~V.}\ \bibnamefont
  {Chulkov}}} (\bibinfo {year} {2013}),\ \bibfield  {title} {\enquote {\bibinfo
  {title} {{Q}uasiparticle band gap in the topological insulator
  {B}i$_{2}${T}e$_{3}$},}\ }\href@noop {} {\bibfield  {journal} {\bibinfo
  {journal} {Phys. Rev. B}\ }\textbf {\bibinfo {volume} {88}},\ \bibinfo
  {pages} {165135}}\BibitemShut {NoStop}%
\bibitem [{\citenamefont {Neto}\ \emph {et~al.}(2009)\citenamefont {Neto},
  \citenamefont {Guinea}, \citenamefont {Peres}, \citenamefont {Novoselov},\
  and\ \citenamefont {Geim}}]{castro_neto_rmp_81}%
  \BibitemOpen
  \bibfield  {author} {\bibinfo {author} {\bibnamefont {Neto}, \bibfnamefont
  {A~H~Castro}}, \bibinfo {author} {\bibfnamefont {F.}~\bibnamefont {Guinea}},
  \bibinfo {author} {\bibfnamefont {N.~M.~R.}\ \bibnamefont {Peres}}, \bibinfo
  {author} {\bibfnamefont {K.~S.}\ \bibnamefont {Novoselov}}, \ and\ \bibinfo
  {author} {\bibfnamefont {A.~K.}\ \bibnamefont {Geim}}} (\bibinfo {year}
  {2009}),\ \bibfield  {title} {\enquote {\bibinfo {title} {{T}he electronic
  properties of graphene},}\ }\href {\doibase 10.1103/RevModPhys.81.109}
  {\bibfield  {journal} {\bibinfo  {journal} {Rev. Mod. Phys.}\ }\textbf
  {\bibinfo {volume} {81}},\ \bibinfo {pages} {109--162}}\BibitemShut {NoStop}%
\bibitem [{\citenamefont {Neuhauser}\ \emph {et~al.}(2014)\citenamefont
  {Neuhauser}, \citenamefont {Gao}, \citenamefont {Arntsen}, \citenamefont
  {Karshenas}, \citenamefont {Rabani},\ and\ \citenamefont
  {Baer}}]{Neuhauser/etal:2014}%
  \BibitemOpen
  \bibfield  {author} {\bibinfo {author} {\bibnamefont {Neuhauser},
  \bibfnamefont {D}}, \bibinfo {author} {\bibfnamefont {Y.}~\bibnamefont
  {Gao}}, \bibinfo {author} {\bibfnamefont {C.}~\bibnamefont {Arntsen}},
  \bibinfo {author} {\bibfnamefont {C.}~\bibnamefont {Karshenas}}, \bibinfo
  {author} {\bibfnamefont {E.}~\bibnamefont {Rabani}}, \ and\ \bibinfo {author}
  {\bibfnamefont {R.}~\bibnamefont {Baer}}} (\bibinfo {year} {2014}),\
  \bibfield  {title} {\enquote {\bibinfo {title} {{B}reaking the {T}heoretical
  {S}caling {L}imit for {P}redicting {Q}uasiparticle {E}nergies: {T}he
  {S}tochastic ${G}{W}$ {A}pproach},}\ }\href@noop {} {\bibfield  {journal}
  {\bibinfo  {journal} {Phys. Rev. Lett.}\ }\textbf {\bibinfo {volume} {113}},\
  \bibinfo {pages} {076402}}\BibitemShut {NoStop}%
\bibitem [{\citenamefont {Nguyen}\ \emph {et~al.}(2012)\citenamefont {Nguyen},
  \citenamefont {Pham}, \citenamefont {Rocca},\ and\ \citenamefont
  {Galli}}]{Nguyen2012}%
  \BibitemOpen
  \bibfield  {author} {\bibinfo {author} {\bibnamefont {Nguyen}, \bibfnamefont
  {H-V}}, \bibinfo {author} {\bibfnamefont {T.~A.}\ \bibnamefont {Pham}},
  \bibinfo {author} {\bibfnamefont {D.}~\bibnamefont {Rocca}}, \ and\ \bibinfo
  {author} {\bibfnamefont {G.}~\bibnamefont {Galli}}} (\bibinfo {year}
  {2012}),\ \bibfield  {title} {\enquote {\bibinfo {title} {{I}mproving
  accuracy and efficiency of calculations of photoemission spectra within the
  many-body perturbation theory},}\ }\href {\doibase
  10.1103/PhysRevB.85.081101} {\bibfield  {journal} {\bibinfo  {journal} {Phys.
  Rev. B}\ }\textbf {\bibinfo {volume} {85}},\ \bibinfo {pages}
  {081101}}\BibitemShut {NoStop}%
\bibitem [{\citenamefont {Niehaus}\ \emph {et~al.}(2005)\citenamefont
  {Niehaus}, \citenamefont {Rohlfing}, \citenamefont {Sala}, \citenamefont
  {Carlo},\ and\ \citenamefont {Frauenheim}}]{Niehaus2005}%
  \BibitemOpen
  \bibfield  {author} {\bibinfo {author} {\bibnamefont {Niehaus}, \bibfnamefont
  {T~A}}, \bibinfo {author} {\bibfnamefont {M.}~\bibnamefont {Rohlfing}},
  \bibinfo {author} {\bibfnamefont {F.~Della}\ \bibnamefont {Sala}}, \bibinfo
  {author} {\bibfnamefont {A.~Di}\ \bibnamefont {Carlo}}, \ and\ \bibinfo
  {author} {\bibfnamefont {Th.}\ \bibnamefont {Frauenheim}}} (\bibinfo {year}
  {2005}),\ \bibfield  {title} {\enquote {\bibinfo {title} {{Q}uasiparticle
  energies for large molecules: {A} tight-binding-based {G}reen's-function
  approach},}\ }\href {\doibase 10.1103/PhysRevA.71.022508} {\bibfield
  {journal} {\bibinfo  {journal} {Phys. Rev. A}\ }\textbf {\bibinfo {volume}
  {71}},\ \bibinfo {pages} {022508}}\BibitemShut {NoStop}%
\bibitem [{\citenamefont {Nisar}\ \emph {et~al.}(2012)\citenamefont {Nisar},
  \citenamefont {Jiang}, \citenamefont {Pathak}, \citenamefont {Zhao},
  \citenamefont {Kang},\ and\ \citenamefont {Ahuja}}]{nisar_nano_2012}%
  \BibitemOpen
  \bibfield  {author} {\bibinfo {author} {\bibnamefont {Nisar}, \bibfnamefont
  {J}}, \bibinfo {author} {\bibfnamefont {X.}~\bibnamefont {Jiang}}, \bibinfo
  {author} {\bibfnamefont {B.}~\bibnamefont {Pathak}}, \bibinfo {author}
  {\bibfnamefont {J.}~\bibnamefont {Zhao}}, \bibinfo {author} {\bibfnamefont
  {T.~W.}\ \bibnamefont {Kang}}, \ and\ \bibinfo {author} {\bibfnamefont
  {R.}~\bibnamefont {Ahuja}}} (\bibinfo {year} {2012}),\ \bibfield  {title}
  {\enquote {\bibinfo {title} {{S}emiconducting allotrope of graphene},}\
  }\href {\doibase 10.1088/0957-4484/23/38/385704} {\bibfield  {journal}
  {\bibinfo  {journal} {Nanotechnology}\ }\textbf {\bibinfo {volume}
  {23}}~(\bibinfo {number} {38}),\ \bibinfo {pages} {385704}}\BibitemShut
  {NoStop}%
\bibitem [{\citenamefont {NIST}(2019)}]{NIST_database}%
  \BibitemOpen
  \bibfield  {author} {\bibinfo {author} {\bibnamefont {NIST},}} (\bibinfo
  {year} {2019}),\ \href@noop {} {\bibinfo  {journal}
  {http://cccbdb.nist.gov/}\ }\BibitemShut {NoStop}%
\bibitem [{\citenamefont {Noguchi}\ \emph {et~al.}(2007)\citenamefont
  {Noguchi}, \citenamefont {Ishii},\ and\ \citenamefont
  {Ohno}}]{Noguchi/etal:2007}%
  \BibitemOpen
\bibfield  {journal} {  }\bibfield  {author} {\bibinfo {author} {\bibnamefont
  {Noguchi}, \bibfnamefont {Y}}, \bibinfo {author} {\bibfnamefont
  {S.}~\bibnamefont {Ishii}}, \ and\ \bibinfo {author} {\bibfnamefont
  {K.}~\bibnamefont {Ohno}}} (\bibinfo {year} {2007}),\ \bibfield  {title}
  {\enquote {\bibinfo {title} {{F}irst principles calculations of double
  ionization energy spectra and two-electron distribution function using
  {T}-matrix theory},}\ }\href {\doibase
  https://doi.org/10.1016/j.elspec.2006.11.025} {\bibfield  {journal} {\bibinfo
   {journal} {J. Electron Spectrosc. Relat. Phenom.}\ }\textbf {\bibinfo
  {volume} {156-158}},\ \bibinfo {pages} {155 -- 157}}\BibitemShut {NoStop}%
\bibitem [{\citenamefont {Noguchi}\ \emph {et~al.}(2008)\citenamefont
  {Noguchi}, \citenamefont {Ishii}, \citenamefont {Ohno}, \citenamefont
  {Solovyev},\ and\ \citenamefont {Sasaki}}]{Noguchi/etal_2:2008}%
  \BibitemOpen
  \bibfield  {author} {\bibinfo {author} {\bibnamefont {Noguchi}, \bibfnamefont
  {Y}}, \bibinfo {author} {\bibfnamefont {S.}~\bibnamefont {Ishii}}, \bibinfo
  {author} {\bibfnamefont {K.}~\bibnamefont {Ohno}}, \bibinfo {author}
  {\bibfnamefont {I.}~\bibnamefont {Solovyev}}, \ and\ \bibinfo {author}
  {\bibfnamefont {T.}~\bibnamefont {Sasaki}}} (\bibinfo {year} {2008}),\
  \bibfield  {title} {\enquote {\bibinfo {title} {{F}irst principles
  ${T}$-matrix calculations for {A}uger spectra of hydrocarbon systems},}\
  }\href {\doibase 10.1103/PhysRevB.77.035132} {\bibfield  {journal} {\bibinfo
  {journal} {Phys. Rev. B}\ }\textbf {\bibinfo {volume} {77}},\ \bibinfo
  {pages} {035132}}\BibitemShut {NoStop}%
\bibitem [{\citenamefont {Noguchi}\ \emph {et~al.}(2010)\citenamefont
  {Noguchi}, \citenamefont {Ohno}, \citenamefont {Solovyev},\ and\
  \citenamefont {Sasaki}}]{Noguchi/etal_2010}%
  \BibitemOpen
  \bibfield  {author} {\bibinfo {author} {\bibnamefont {Noguchi}, \bibfnamefont
  {Y}}, \bibinfo {author} {\bibfnamefont {K.}~\bibnamefont {Ohno}}, \bibinfo
  {author} {\bibfnamefont {I.}~\bibnamefont {Solovyev}}, \ and\ \bibinfo
  {author} {\bibfnamefont {T.}~\bibnamefont {Sasaki}}} (\bibinfo {year}
  {2010}),\ \bibfield  {title} {\enquote {\bibinfo {title} {{C}luster size
  dependence of double ionization energy spectra of spin-polarized aluminum and
  sodium clusters: {A}ll-electron spin-polarized ${G}{W}+{T}$-matrix method},}\
  }\href@noop {} {\bibfield  {journal} {\bibinfo  {journal} {Phys. Rev. B}\
  }\textbf {\bibinfo {volume} {81}},\ \bibinfo {pages} {165411}}\BibitemShut
  {NoStop}%
\bibitem [{\citenamefont {Nohara}\ \emph {et~al.}(2009)\citenamefont {Nohara},
  \citenamefont {Yamamoto},\ and\ \citenamefont {Fujiwara}}]{Nohara/etal:2009}%
  \BibitemOpen
  \bibfield  {author} {\bibinfo {author} {\bibnamefont {Nohara}, \bibfnamefont
  {Y}}, \bibinfo {author} {\bibfnamefont {S.}~\bibnamefont {Yamamoto}}, \ and\
  \bibinfo {author} {\bibfnamefont {T.}~\bibnamefont {Fujiwara}}} (\bibinfo
  {year} {2009}),\ \bibfield  {title} {\enquote {\bibinfo {title} {{E}lectronic
  structure of perovskite-type transition metal oxides {L}a${M}${O}$_3$
  ({M}={T}i$\sim${C}u) by {U}+${G}{W}$ approximation},}\ }\href@noop {}
  {\bibfield  {journal} {\bibinfo  {journal} {Phys.\ Rev.\ B}\ }\textbf
  {\bibinfo {volume} {79}},\ \bibinfo {pages} {195110}}\BibitemShut {NoStop}%
\bibitem [{\citenamefont {Northrup}\ \emph {et~al.}(1987)\citenamefont
  {Northrup}, \citenamefont {Hybertsen},\ and\ \citenamefont
  {Louie}}]{Northrup/etal:1987}%
  \BibitemOpen
  \bibfield  {author} {\bibinfo {author} {\bibnamefont {Northrup},
  \bibfnamefont {J~E}}, \bibinfo {author} {\bibfnamefont {M.~S.}\ \bibnamefont
  {Hybertsen}}, \ and\ \bibinfo {author} {\bibfnamefont {S.~G.}\ \bibnamefont
  {Louie}}} (\bibinfo {year} {1987}),\ \bibfield  {title} {\enquote {\bibinfo
  {title} {{T}heory of quasiparticle energies in alkali metals},}\ }\href@noop
  {} {\bibfield  {journal} {\bibinfo  {journal} {Phys. Rev. Lett.}\ }\textbf
  {\bibinfo {volume} {59}},\ \bibinfo {pages} {819--822}}\BibitemShut {NoStop}%
\bibitem [{\citenamefont {Novoselov}\ \emph {et~al.}(2004)\citenamefont
  {Novoselov}, \citenamefont {Geim}, \citenamefont {Morozov}, \citenamefont
  {Jiang}, \citenamefont {Zhang}, \citenamefont {Dubonos}, \citenamefont
  {Grigorieva},\ and\ \citenamefont {Firsov}}]{novoselov_sci_306}%
  \BibitemOpen
  \bibfield  {author} {\bibinfo {author} {\bibnamefont {Novoselov},
  \bibfnamefont {K~S}}, \bibinfo {author} {\bibfnamefont {A.~K.}\ \bibnamefont
  {Geim}}, \bibinfo {author} {\bibfnamefont {S.~V.}\ \bibnamefont {Morozov}},
  \bibinfo {author} {\bibfnamefont {D.}~\bibnamefont {Jiang}}, \bibinfo
  {author} {\bibfnamefont {Y.}~\bibnamefont {Zhang}}, \bibinfo {author}
  {\bibfnamefont {S.~V.}\ \bibnamefont {Dubonos}}, \bibinfo {author}
  {\bibfnamefont {I.~V.}\ \bibnamefont {Grigorieva}}, \ and\ \bibinfo {author}
  {\bibfnamefont {A.~A.}\ \bibnamefont {Firsov}}} (\bibinfo {year} {2004}),\
  \bibfield  {title} {\enquote {\bibinfo {title} {{E}lectric {F}ield {E}ffect
  in {A}tomically {T}hin {C}arbon {F}ilms},}\ }\href {\doibase
  10.1126/science.1102896} {\bibfield  {journal} {\bibinfo  {journal}
  {Science}\ }\textbf {\bibinfo {volume} {306}}~(\bibinfo {number} {5696}),\
  \bibinfo {pages} {666--669}}\BibitemShut {NoStop}%
\bibitem [{\citenamefont {Onida}\ \emph
  {et~al.}(1995{\natexlab{a}})\citenamefont {Onida}, \citenamefont {Reining},
  \citenamefont {Godby}, \citenamefont {Sole},\ and\ \citenamefont
  {Andreoni}}]{Onida/etal:1995}%
  \BibitemOpen
  \bibfield  {author} {\bibinfo {author} {\bibnamefont {Onida}, \bibfnamefont
  {G}}, \bibinfo {author} {\bibfnamefont {L.}~\bibnamefont {Reining}}, \bibinfo
  {author} {\bibfnamefont {R.~W.}\ \bibnamefont {Godby}}, \bibinfo {author}
  {\bibfnamefont {R.~Del}\ \bibnamefont {Sole}}, \ and\ \bibinfo {author}
  {\bibfnamefont {W.}~\bibnamefont {Andreoni}}} (\bibinfo {year}
  {1995}{\natexlab{a}}),\ \bibfield  {title} {\enquote {\bibinfo {title} {{A}b
  {I}nitio {C}alculations of the {Q}uasiparticle and {A}bsorption {S}pectra of
  {C}lusters: {T}he {S}odium {T}etramer},}\ }\href@noop {} {\bibfield
  {journal} {\bibinfo  {journal} {Phys. Rev. Lett.}\ }\textbf {\bibinfo
  {volume} {75}},\ \bibinfo {pages} {818}}\BibitemShut {NoStop}%
\bibitem [{\citenamefont {Onida}\ \emph
  {et~al.}(1995{\natexlab{b}})\citenamefont {Onida}, \citenamefont {Reining},
  \citenamefont {Godby}, \citenamefont {Sole},\ and\ \citenamefont
  {Andreoni}}]{onida_prl_75}%
  \BibitemOpen
  \bibfield  {author} {\bibinfo {author} {\bibnamefont {Onida}, \bibfnamefont
  {G}}, \bibinfo {author} {\bibfnamefont {L.}~\bibnamefont {Reining}}, \bibinfo
  {author} {\bibfnamefont {R.~W.}\ \bibnamefont {Godby}}, \bibinfo {author}
  {\bibfnamefont {R.~Del}\ \bibnamefont {Sole}}, \ and\ \bibinfo {author}
  {\bibfnamefont {W.}~\bibnamefont {Andreoni}}} (\bibinfo {year}
  {1995}{\natexlab{b}}),\ \bibfield  {title} {\enquote {\bibinfo {title} {{A}b
  {I}nitio {C}alculations of the {Q}uasiparticle and {A}bsorption {S}pectra of
  {C}lusters: {T}he {S}odium {T}etramer},}\ }\href {\doibase
  10.1103/PhysRevLett.75.818} {\bibfield  {journal} {\bibinfo  {journal} {Phys.
  Rev. Lett.}\ }\textbf {\bibinfo {volume} {75}},\ \bibinfo {pages}
  {818--821}}\BibitemShut {NoStop}%
\bibitem [{\citenamefont {Onida}\ \emph {et~al.}(2002)\citenamefont {Onida},
  \citenamefont {Reining},\ and\ \citenamefont
  {Rubio}}]{Onida/Reining/Rubio:2002}%
  \BibitemOpen
  \bibfield  {author} {\bibinfo {author} {\bibnamefont {Onida}, \bibfnamefont
  {G}}, \bibinfo {author} {\bibfnamefont {L.}~\bibnamefont {Reining}}, \ and\
  \bibinfo {author} {\bibfnamefont {A.}~\bibnamefont {Rubio}}} (\bibinfo {year}
  {2002}),\ \bibfield  {title} {\enquote {\bibinfo {title} {{E}lectronic
  excitations: density-functional versus many-body {G}reen's function
  approaches},}\ }\href@noop {} {\bibfield  {journal} {\bibinfo  {journal}
  {Rev. Mod. Phys.}\ }\textbf {\bibinfo {volume} {74}},\ \bibinfo {pages}
  {601}}\BibitemShut {NoStop}%
\bibitem [{\citenamefont {Ono}\ \emph {et~al.}(2015)\citenamefont {Ono},
  \citenamefont {Noguchi}, \citenamefont {Sahara}, \citenamefont {Kawazoe},\
  and\ \citenamefont {Ohno}}]{Ono/etal:2015}%
  \BibitemOpen
  \bibfield  {author} {\bibinfo {author} {\bibnamefont {Ono}, \bibfnamefont
  {S}}, \bibinfo {author} {\bibfnamefont {Y.}~\bibnamefont {Noguchi}}, \bibinfo
  {author} {\bibfnamefont {R.}~\bibnamefont {Sahara}}, \bibinfo {author}
  {\bibfnamefont {Y.}~\bibnamefont {Kawazoe}}, \ and\ \bibinfo {author}
  {\bibfnamefont {K.}~\bibnamefont {Ohno}}} (\bibinfo {year} {2015}),\
  \bibfield  {title} {\enquote {\bibinfo {title} {{T}{O}{M}{B}{O}:
  {A}ll-electron mixed-basis approach to condensed matter physics},}\ }\href
  {\doibase https://doi.org/10.1016/j.cpc.2014.11.012} {\bibfield  {journal}
  {\bibinfo  {journal} {Comput. Phys. Commun.}\ }\textbf {\bibinfo {volume}
  {189}},\ \bibinfo {pages} {20 -- 30}}\BibitemShut {NoStop}%
\bibitem [{\citenamefont {Ortiz}(2012)}]{Ortiz:2012}%
  \BibitemOpen
  \bibfield  {author} {\bibinfo {author} {\bibnamefont {Ortiz}, \bibfnamefont
  {J~V}}} (\bibinfo {year} {2012}),\ \bibfield  {title} {\enquote {\bibinfo
  {title} {{E}lectron propagator theory: an approach to prediction and
  interpretation in quantum chemistry},}\ }\href@noop {} {\bibfield  {journal}
  {\bibinfo  {journal} {Wiley Interdiscip. Rev. Comput. Mol. Sci.}\ }\textbf
  {\bibinfo {volume} {3}}~(\bibinfo {number} {2}),\ \bibinfo {pages}
  {123--142}}\BibitemShut {NoStop}%
\bibitem [{\citenamefont {Page}\ \emph {et~al.}(1988)\citenamefont {Page},
  \citenamefont {Larkin}, \citenamefont {Shen},\ and\ \citenamefont
  {Lee}}]{Page1988}%
  \BibitemOpen
  \bibfield  {author} {\bibinfo {author} {\bibnamefont {Page}, \bibfnamefont
  {R~H}}, \bibinfo {author} {\bibfnamefont {R.~J.}\ \bibnamefont {Larkin}},
  \bibinfo {author} {\bibfnamefont {Y.~R.}\ \bibnamefont {Shen}}, \ and\
  \bibinfo {author} {\bibfnamefont {Y.~T.}\ \bibnamefont {Lee}}} (\bibinfo
  {year} {1988}),\ \bibfield  {title} {\enquote {\bibinfo {title}
  {{H}igh‐resolution photoionization spectrum of water molecules in a
  supersonic beam},}\ }\href@noop {} {\bibfield  {journal} {\bibinfo  {journal}
  {J. Chem. Phys.}\ }\textbf {\bibinfo {volume} {88}}~(\bibinfo {number} {4}),\
  \bibinfo {pages} {2249--2263}}\BibitemShut {NoStop}%
\bibitem [{\citenamefont {Palummo}\ \emph {et~al.}(2004)\citenamefont
  {Palummo}, \citenamefont {Pulci}, \citenamefont {Sole}, \citenamefont
  {Marini}, \citenamefont {Hahn}, \citenamefont {Schmidt},\ and\ \citenamefont
  {Bechstedt}}]{Palummo/etal:2004}%
  \BibitemOpen
  \bibfield  {author} {\bibinfo {author} {\bibnamefont {Palummo}, \bibfnamefont
  {M}}, \bibinfo {author} {\bibfnamefont {O.}~\bibnamefont {Pulci}}, \bibinfo
  {author} {\bibfnamefont {R.~Del}\ \bibnamefont {Sole}}, \bibinfo {author}
  {\bibfnamefont {A.}~\bibnamefont {Marini}}, \bibinfo {author} {\bibfnamefont
  {P.}~\bibnamefont {Hahn}}, \bibinfo {author} {\bibfnamefont {W.~G.}\
  \bibnamefont {Schmidt}}, \ and\ \bibinfo {author} {\bibfnamefont
  {F.}~\bibnamefont {Bechstedt}}} (\bibinfo {year} {2004}),\ \bibfield  {title}
  {\enquote {\bibinfo {title} {{T}he {B}ethe–{S}alpeter equation: a
  first-principles approach for calculating surface optical spectra},}\ }\href
  {http://stacks.iop.org/0953-8984/16/i=39/a=006} {\bibfield  {journal}
  {\bibinfo  {journal} {J. Phys.: Condens. Matter}\ }\textbf {\bibinfo {volume}
  {16}}~(\bibinfo {number} {39}),\ \bibinfo {pages} {S4313}}\BibitemShut
  {NoStop}%
\bibitem [{\citenamefont {Park}\ \emph {et~al.}(2009)\citenamefont {Park},
  \citenamefont {Giustino}, \citenamefont {Spataru}, \citenamefont {Cohen},\
  and\ \citenamefont {Louie}}]{park_nl_9}%
  \BibitemOpen
  \bibfield  {author} {\bibinfo {author} {\bibnamefont {Park}, \bibfnamefont
  {C-H}}, \bibinfo {author} {\bibfnamefont {F.}~\bibnamefont {Giustino}},
  \bibinfo {author} {\bibfnamefont {C.~D.}\ \bibnamefont {Spataru}}, \bibinfo
  {author} {\bibfnamefont {M.~L.}\ \bibnamefont {Cohen}}, \ and\ \bibinfo
  {author} {\bibfnamefont {S.~G.}\ \bibnamefont {Louie}}} (\bibinfo {year}
  {2009}),\ \bibfield  {title} {\enquote {\bibinfo {title} {{A}ngle-{R}esolved
  {P}hotoemission {S}pectra of {G}raphene from {F}irst-{P}rinciples
  {C}alculations},}\ }\href@noop {} {\bibfield  {journal} {\bibinfo  {journal}
  {Nano Letters}\ }\textbf {\bibinfo {volume} {9}}~(\bibinfo {number} {12}),\
  \bibinfo {pages} {4234--4239}}\BibitemShut {NoStop}%
\bibitem [{\citenamefont {Patrick}\ and\ \citenamefont
  {Giustino}(2012)}]{Patrick/Giustino:2012}%
  \BibitemOpen
  \bibfield  {author} {\bibinfo {author} {\bibnamefont {Patrick}, \bibfnamefont
  {C~E}}, \ and\ \bibinfo {author} {\bibfnamefont {F.}~\bibnamefont
  {Giustino}}} (\bibinfo {year} {2012}),\ \bibfield  {title} {\enquote
  {\bibinfo {title} {{Q}uantitative {A}nalysis of {V}alence {P}hotoemission
  {S}pectra and {Q}uasiparticle {E}xcitations at {C}hromophore-{S}emiconductor
  {I}nterfaces},}\ }\href {\doibase 10.1103/PhysRevLett.109.116801} {\bibfield
  {journal} {\bibinfo  {journal} {Phys. Rev. Lett.}\ }\textbf {\bibinfo
  {volume} {109}},\ \bibinfo {pages} {116801}}\BibitemShut {NoStop}%
\bibitem [{\citenamefont {Pavlyukh}\ and\ \citenamefont
  {H\"ubner}(2007)}]{Pavlyukh/etal:2007}%
  \BibitemOpen
  \bibfield  {author} {\bibinfo {author} {\bibnamefont {Pavlyukh},
  \bibfnamefont {Y}}, \ and\ \bibinfo {author} {\bibfnamefont {W.}~\bibnamefont
  {H\"ubner}}} (\bibinfo {year} {2007}),\ \bibfield  {title} {\enquote
  {\bibinfo {title} {{C}onfiguration interaction approach for the computation
  of the electronic self-energy},}\ }\href@noop {} {\bibfield  {journal}
  {\bibinfo  {journal} {Phys. Rev. B}\ }\textbf {\bibinfo {volume} {75}},\
  \bibinfo {pages} {205129}}\BibitemShut {NoStop}%
\bibitem [{\citenamefont {Perdew}\ \emph
  {et~al.}(1996{\natexlab{a}})\citenamefont {Perdew}, \citenamefont {Burke},\
  and\ \citenamefont {Ernzerhof}}]{Perdew/Burke/Ernzerhof:1996}%
  \BibitemOpen
  \bibfield  {author} {\bibinfo {author} {\bibnamefont {Perdew}, \bibfnamefont
  {J~P}}, \bibinfo {author} {\bibfnamefont {K.}~\bibnamefont {Burke}}, \ and\
  \bibinfo {author} {\bibfnamefont {M.}~\bibnamefont {Ernzerhof}}} (\bibinfo
  {year} {1996}{\natexlab{a}}),\ \bibfield  {title} {\enquote {\bibinfo {title}
  {{G}eneralized {G}radient {A}pproximation {M}ade {S}imple},}\ }\href@noop {}
  {\bibfield  {journal} {\bibinfo  {journal} {Phys. Rev. Lett}\ }\textbf
  {\bibinfo {volume} {77}},\ \bibinfo {pages} {3865}}\BibitemShut {NoStop}%
\bibitem [{\citenamefont {Perdew}\ \emph
  {et~al.}(1996{\natexlab{b}})\citenamefont {Perdew}, \citenamefont
  {Ernzerhof},\ and\ \citenamefont {Burke}}]{PBE0_1}%
  \BibitemOpen
  \bibfield  {author} {\bibinfo {author} {\bibnamefont {Perdew}, \bibfnamefont
  {J~P}}, \bibinfo {author} {\bibfnamefont {M.}~\bibnamefont {Ernzerhof}}, \
  and\ \bibinfo {author} {\bibfnamefont {K.}~\bibnamefont {Burke}}} (\bibinfo
  {year} {1996}{\natexlab{b}}),\ \bibfield  {title} {\enquote {\bibinfo {title}
  {{R}ationale for mixing exact exchange with density functional
  approximations},}\ }\href@noop {} {\bibfield  {journal} {\bibinfo  {journal}
  {J.\ Chem.\ Phys}\ }\textbf {\bibinfo {volume} {105}},\ \bibinfo {pages}
  {9982}}\BibitemShut {NoStop}%
\bibitem [{\citenamefont {Perdew}\ \emph {et~al.}(1982)\citenamefont {Perdew},
  \citenamefont {Levy},\ and\ \citenamefont {Balduz}}]{Perdew/etal:1982}%
  \BibitemOpen
  \bibfield  {author} {\bibinfo {author} {\bibnamefont {Perdew}, \bibfnamefont
  {J~P}}, \bibinfo {author} {\bibfnamefont {M.}~\bibnamefont {Levy}}, \ and\
  \bibinfo {author} {\bibfnamefont {J.~L.}\ \bibnamefont {Balduz}}} (\bibinfo
  {year} {1982}),\ \bibfield  {title} {\enquote {\bibinfo {title}
  {{D}ensity-{F}unctional {T}heory for {F}ractional {P}article {N}umber:
  {D}erivative {D}iscontinuities of the {E}nergy},}\ }\href@noop {} {\bibfield
  {journal} {\bibinfo  {journal} {Phys.\ Rev.\ Lett.}\ }\textbf {\bibinfo
  {volume} {49}}~(\bibinfo {number} {23}),\ \bibinfo {pages}
  {1691}}\BibitemShut {NoStop}%
\bibitem [{\citenamefont {Perdew}\ and\ \citenamefont
  {Zunger}(1981)}]{Perdew/Zunger:1981}%
  \BibitemOpen
  \bibfield  {author} {\bibinfo {author} {\bibnamefont {Perdew}, \bibfnamefont
  {J~P}}, \ and\ \bibinfo {author} {\bibfnamefont {A.}~\bibnamefont {Zunger}}}
  (\bibinfo {year} {1981}),\ \bibfield  {title} {\enquote {\bibinfo {title}
  {{S}elf-{I}nteraction {C}orrection to {D}ensity-{F}unctional {A}pproximation
  for {M}any-{E}lectron {S}ystems},}\ }\href@noop {} {\bibfield  {journal}
  {\bibinfo  {journal} {Phys.\ Rev.\ B}\ }\textbf {\bibinfo {volume} {23}},\
  \bibinfo {pages} {5048}}\BibitemShut {NoStop}%
\bibitem [{\citenamefont {Pham}\ \emph {et~al.}(2013)\citenamefont {Pham},
  \citenamefont {Nguyen}, \citenamefont {Rocca},\ and\ \citenamefont
  {Galli}}]{Pham/etal:2013}%
  \BibitemOpen
  \bibfield  {author} {\bibinfo {author} {\bibnamefont {Pham}, \bibfnamefont
  {T~A}}, \bibinfo {author} {\bibfnamefont {H.-V.}\ \bibnamefont {Nguyen}},
  \bibinfo {author} {\bibfnamefont {D.}~\bibnamefont {Rocca}}, \ and\ \bibinfo
  {author} {\bibfnamefont {G.}~\bibnamefont {Galli}}} (\bibinfo {year}
  {2013}),\ \bibfield  {title} {\enquote {\bibinfo {title} {${G}{W}$
  calculations using the spectral decomposition of the dielectric matrix:
  {V}erification, validation, and comparison of methods},}\ }\href@noop {}
  {\bibfield  {journal} {\bibinfo  {journal} {Phys. Rev. B}\ }\textbf {\bibinfo
  {volume} {87}},\ \bibinfo {pages} {155148}}\BibitemShut {NoStop}%
\bibitem [{\citenamefont {Ping}\ \emph {et~al.}(2013)\citenamefont {Ping},
  \citenamefont {Rocca},\ and\ \citenamefont {Galli}}]{Ping/etal:2013}%
  \BibitemOpen
  \bibfield  {author} {\bibinfo {author} {\bibnamefont {Ping}, \bibfnamefont
  {Y}}, \bibinfo {author} {\bibfnamefont {D.}~\bibnamefont {Rocca}}, \ and\
  \bibinfo {author} {\bibfnamefont {G.}~\bibnamefont {Galli}}} (\bibinfo {year}
  {2013}),\ \bibfield  {title} {\enquote {\bibinfo {title} {{E}lectronic
  excitations in light absorbers for photoelectrochemical energy conversion:
  first principles calculations based on many body perturbation theory},}\
  }\href {\doibase 10.1039/C3CS00007A} {\bibfield  {journal} {\bibinfo
  {journal} {Chem. Soc. Rev.}\ }\textbf {\bibinfo {volume} {42}},\ \bibinfo
  {pages} {2437--2469}}\BibitemShut {NoStop}%
\bibitem [{\citenamefont {Pinheiro}\ \emph {et~al.}(2015)\citenamefont
  {Pinheiro}, \citenamefont {Caldas}, \citenamefont {Rinke}, \citenamefont
  {Blum},\ and\ \citenamefont {Scheffler}}]{Pinheiro2015}%
  \BibitemOpen
  \bibfield  {author} {\bibinfo {author} {\bibnamefont {Pinheiro},
  \bibfnamefont {M}}, \bibinfo {author} {\bibfnamefont {M.~J.}\ \bibnamefont
  {Caldas}}, \bibinfo {author} {\bibfnamefont {P.}~\bibnamefont {Rinke}},
  \bibinfo {author} {\bibfnamefont {V.}~\bibnamefont {Blum}}, \ and\ \bibinfo
  {author} {\bibfnamefont {M.}~\bibnamefont {Scheffler}}} (\bibinfo {year}
  {2015}),\ \bibfield  {title} {\enquote {\bibinfo {title} {{L}ength dependence
  of ionization potentials of transacetylenes: {I}nternally consistent
  {D}{F}{T}/${G}{W}$ approach},}\ }\href {\doibase 10.1103/PhysRevB.92.195134}
  {\bibfield  {journal} {\bibinfo  {journal} {Phys. Rev. B}\ }\textbf {\bibinfo
  {volume} {92}},\ \bibinfo {pages} {195134}}\BibitemShut {NoStop}%
\bibitem [{\citenamefont {Plummer}\ and\ \citenamefont
  {Eberhardt}(1982)}]{Plummer/Eberhardt:1982}%
  \BibitemOpen
  \bibfield  {author} {\bibinfo {author} {\bibnamefont {Plummer}, \bibfnamefont
  {E~W}}, \ and\ \bibinfo {author} {\bibfnamefont {W.}~\bibnamefont
  {Eberhardt}}} (\bibinfo {year} {1982}),\ \bibfield  {title} {\enquote
  {\bibinfo {title} {{A}ngle-{R}esolved {P}hotoemission as a {T}ool for the
  {S}tudy of {S}urfaces},}\ }\href@noop {} {\bibfield  {journal} {\bibinfo
  {journal} {Adv.\ Chem.\ Phys.}\ }\textbf {\bibinfo {volume} {49}},\ \bibinfo
  {pages} {533}}\BibitemShut {NoStop}%
\bibitem [{\citenamefont {Pollehn}\ \emph {et~al.}(1998)\citenamefont
  {Pollehn}, \citenamefont {Schindlmayr},\ and\ \citenamefont
  {Godby}}]{Pollehn1998}%
  \BibitemOpen
  \bibfield  {author} {\bibinfo {author} {\bibnamefont {Pollehn}, \bibfnamefont
  {T~J}}, \bibinfo {author} {\bibfnamefont {A.}~\bibnamefont {Schindlmayr}}, \
  and\ \bibinfo {author} {\bibfnamefont {R.~W.}\ \bibnamefont {Godby}}}
  (\bibinfo {year} {1998}),\ \bibfield  {title} {\enquote {\bibinfo {title}
  {{A}ssessment of the ${G}{W}$ approximation using {H}ubbard chains},}\ }\href
  {\doibase 10.1088/0953-8984/10/6/011} {\bibfield  {journal} {\bibinfo
  {journal} {J. Phys.: Condens. Matter}\ }\textbf {\bibinfo {volume}
  {10}}~(\bibinfo {number} {6}),\ \bibinfo {pages} {1273--1283}}\BibitemShut
  {NoStop}%
\bibitem [{\citenamefont {Ponc\'e}\ \emph {et~al.}(2018)\citenamefont
  {Ponc\'e}, \citenamefont {Margine},\ and\ \citenamefont
  {Giustino}}]{Ponce/Margine/Giustino:2018}%
  \BibitemOpen
  \bibfield  {author} {\bibinfo {author} {\bibnamefont {Ponc\'e}, \bibfnamefont
  {S}}, \bibinfo {author} {\bibfnamefont {E.~R.}\ \bibnamefont {Margine}}, \
  and\ \bibinfo {author} {\bibfnamefont {F.}~\bibnamefont {Giustino}}}
  (\bibinfo {year} {2018}),\ \bibfield  {title} {\enquote {\bibinfo {title}
  {{T}owards predictive many-body calculations of phonon-limited carrier
  mobilities in semiconductors},}\ }\href@noop {} {\bibfield  {journal}
  {\bibinfo  {journal} {Phys. Rev. B}\ }\textbf {\bibinfo {volume} {97}},\
  \bibinfo {pages} {121201}}\BibitemShut {NoStop}%
\bibitem [{\citenamefont {Prete}\ \emph {et~al.}(2017)\citenamefont {Prete},
  \citenamefont {Conte}, \citenamefont {Gori}, \citenamefont {Bechstedt},\ and\
  \citenamefont {Pulci}}]{prete_apl_110}%
  \BibitemOpen
  \bibfield  {author} {\bibinfo {author} {\bibnamefont {Prete}, \bibfnamefont
  {M~S}}, \bibinfo {author} {\bibfnamefont {A.~Mosca}\ \bibnamefont {Conte}},
  \bibinfo {author} {\bibfnamefont {P.}~\bibnamefont {Gori}}, \bibinfo {author}
  {\bibfnamefont {F.}~\bibnamefont {Bechstedt}}, \ and\ \bibinfo {author}
  {\bibfnamefont {O.}~\bibnamefont {Pulci}}} (\bibinfo {year} {2017}),\
  \bibfield  {title} {\enquote {\bibinfo {title} {{T}unable electronic
  properties of two-dimensional nitrides for light harvesting
  heterostructures},}\ }\href {\doibase 10.1063/1.4973753} {\bibfield
  {journal} {\bibinfo  {journal} {Appl. Phys. Lett.}\ }\textbf {\bibinfo
  {volume} {110}}~(\bibinfo {number} {1}),\ \bibinfo {pages}
  {012103}}\BibitemShut {NoStop}%
\bibitem [{\citenamefont {Pulci}\ \emph {et~al.}(1999)\citenamefont {Pulci},
  \citenamefont {Bechstedt}, \citenamefont {Onida}, \citenamefont {Sole},\ and\
  \citenamefont {Reining}}]{Pulci/etal:1999}%
  \BibitemOpen
  \bibfield  {author} {\bibinfo {author} {\bibnamefont {Pulci}, \bibfnamefont
  {O}}, \bibinfo {author} {\bibfnamefont {F.}~\bibnamefont {Bechstedt}},
  \bibinfo {author} {\bibfnamefont {G.}~\bibnamefont {Onida}}, \bibinfo
  {author} {\bibfnamefont {R.~Del}\ \bibnamefont {Sole}}, \ and\ \bibinfo
  {author} {\bibfnamefont {L.}~\bibnamefont {Reining}}} (\bibinfo {year}
  {1999}),\ \bibfield  {title} {\enquote {\bibinfo {title} {{S}tate mixing for
  quasiparticles at surfaces: {N}onperturbative ${G}{W}$ approximation},}\
  }\href@noop {} {\bibfield  {journal} {\bibinfo  {journal} {Phys. Rev. B}\
  }\textbf {\bibinfo {volume} {60}},\ \bibinfo {pages}
  {16758--16761}}\BibitemShut {NoStop}%
\bibitem [{\citenamefont {Pulci}\ \emph {et~al.}(1998)\citenamefont {Pulci},
  \citenamefont {Onida}, \citenamefont {Sole},\ and\ \citenamefont
  {Reining}}]{Pulci/etal:1998}%
  \BibitemOpen
  \bibfield  {author} {\bibinfo {author} {\bibnamefont {Pulci}, \bibfnamefont
  {O}}, \bibinfo {author} {\bibfnamefont {G.}~\bibnamefont {Onida}}, \bibinfo
  {author} {\bibfnamefont {R.~Del}\ \bibnamefont {Sole}}, \ and\ \bibinfo
  {author} {\bibfnamefont {L.}~\bibnamefont {Reining}}} (\bibinfo {year}
  {1998}),\ \bibfield  {title} {\enquote {\bibinfo {title} {{A}b {I}nitio
  {C}alculation of {S}elf-{E}nergy {E}ffects on {O}ptical {P}roperties of
  {G}a{A}s(110)},}\ }\href@noop {} {\bibfield  {journal} {\bibinfo  {journal}
  {Phys. Rev. Lett.}\ }\textbf {\bibinfo {volume} {81}},\ \bibinfo {pages}
  {5374--5377}}\BibitemShut {NoStop}%
\bibitem [{\citenamefont {Puschnig}\ \emph {et~al.}(2012)\citenamefont
  {Puschnig}, \citenamefont {Amiri},\ and\ \citenamefont
  {Draxl}}]{Puschnig/etal:2012}%
  \BibitemOpen
  \bibfield  {author} {\bibinfo {author} {\bibnamefont {Puschnig},
  \bibfnamefont {P}}, \bibinfo {author} {\bibfnamefont {P.}~\bibnamefont
  {Amiri}}, \ and\ \bibinfo {author} {\bibfnamefont {C.}~\bibnamefont {Draxl}}}
  (\bibinfo {year} {2012}),\ \bibfield  {title} {\enquote {\bibinfo {title}
  {{B}and renormalization of a polymer physisorbed on graphene investigated by
  many-body perturbation theory},}\ }\href@noop {} {\bibfield  {journal}
  {\bibinfo  {journal} {Phys. Rev. B}\ }\textbf {\bibinfo {volume} {86}},\
  \bibinfo {pages} {085107}}\BibitemShut {NoStop}%
\bibitem [{\citenamefont {Qian}\ \emph {et~al.}(2011)\citenamefont {Qian},
  \citenamefont {Umari},\ and\ \citenamefont
  {Marzari}}]{Qian/Umari/Marzari:2011}%
  \BibitemOpen
  \bibfield  {author} {\bibinfo {author} {\bibnamefont {Qian}, \bibfnamefont
  {X}}, \bibinfo {author} {\bibfnamefont {P.}~\bibnamefont {Umari}}, \ and\
  \bibinfo {author} {\bibfnamefont {N.}~\bibnamefont {Marzari}}} (\bibinfo
  {year} {2011}),\ \bibfield  {title} {\enquote {\bibinfo {title}
  {{P}hotoelectron properties of {D}{N}{A} and {R}{N}{A} bases from many-body
  perturbation theory},}\ }\href {\doibase 10.1103/PhysRevB.84.075103}
  {\bibfield  {journal} {\bibinfo  {journal} {Phys. Rev. B}\ }\textbf {\bibinfo
  {volume} {84}},\ \bibinfo {pages} {075103}}\BibitemShut {NoStop}%
\bibitem [{\citenamefont {Qiu}\ \emph {et~al.}(2013)\citenamefont {Qiu},
  \citenamefont {da~Jornada},\ and\ \citenamefont {Louie}}]{Qiu/etal:2013}%
  \BibitemOpen
  \bibfield  {author} {\bibinfo {author} {\bibnamefont {Qiu}, \bibfnamefont
  {D~Y}}, \bibinfo {author} {\bibfnamefont {F.~H.}\ \bibnamefont {da~Jornada}},
  \ and\ \bibinfo {author} {\bibfnamefont {S.~G.}\ \bibnamefont {Louie}}}
  (\bibinfo {year} {2013}),\ \bibfield  {title} {\enquote {\bibinfo {title}
  {{O}ptical {S}pectrum of ${\mathrm{mos}}_{2}$: {M}any-{B}ody {E}ffects and
  {D}iversity of {E}xciton {S}tates},}\ }\href {\doibase
  10.1103/PhysRevLett.111.216805} {\bibfield  {journal} {\bibinfo  {journal}
  {Phys. Rev. Lett.}\ }\textbf {\bibinfo {volume} {111}},\ \bibinfo {pages}
  {216805}}\BibitemShut {NoStop}%
\bibitem [{\citenamefont {Qiu}\ \emph {et~al.}(2016)\citenamefont {Qiu},
  \citenamefont {da~Jornada},\ and\ \citenamefont {Louie}}]{Qiu2016}%
  \BibitemOpen
  \bibfield  {author} {\bibinfo {author} {\bibnamefont {Qiu}, \bibfnamefont
  {D~Y}}, \bibinfo {author} {\bibfnamefont {F.~H.}\ \bibnamefont {da~Jornada}},
  \ and\ \bibinfo {author} {\bibfnamefont {S.~G.}\ \bibnamefont {Louie}}}
  (\bibinfo {year} {2016}),\ \bibfield  {title} {\enquote {\bibinfo {title}
  {{S}creening and many-body effects in two-dimensional crystals: {M}onolayer
  {M}o{S}$_{2}$},}\ }\href {\doibase 10.1103/PhysRevB.93.235435} {\bibfield
  {journal} {\bibinfo  {journal} {Phys. Rev. B}\ }\textbf {\bibinfo {volume}
  {93}},\ \bibinfo {pages} {235435}}\BibitemShut {NoStop}%
\bibitem [{\citenamefont {Qteish}\ \emph {et~al.}(2006)\citenamefont {Qteish},
  \citenamefont {Rinke}, \citenamefont {Neugebauer},\ and\ \citenamefont
  {Scheffler}}]{Qteish/etal:2006}%
  \BibitemOpen
  \bibfield  {author} {\bibinfo {author} {\bibnamefont {Qteish}, \bibfnamefont
  {A}}, \bibinfo {author} {\bibfnamefont {P.}~\bibnamefont {Rinke}}, \bibinfo
  {author} {\bibfnamefont {J.}~\bibnamefont {Neugebauer}}, \ and\ \bibinfo
  {author} {\bibfnamefont {M.}~\bibnamefont {Scheffler}}} (\bibinfo {year}
  {2006}),\ \bibfield  {title} {\enquote {\bibinfo {title} {{E}xact-exchange
  based quasiparticle energy calculations for the band gap, effective masses
  and deformation potentials of {S}c{N}},}\ }\href@noop {} {\bibfield
  {journal} {\bibinfo  {journal} {Phys.\ Rev.\ B}\ }\textbf {\bibinfo {volume}
  {74}},\ \bibinfo {pages} {245208}}\BibitemShut {NoStop}%
\bibitem [{\citenamefont {Quek}\ \emph {et~al.}(2007)\citenamefont {Quek},
  \citenamefont {Venkataraman}, \citenamefont {Choi}, \citenamefont {Louie},
  \citenamefont {Hybertsen},\ and\ \citenamefont {Neaton}}]{Quek/etal:2007}%
  \BibitemOpen
  \bibfield  {author} {\bibinfo {author} {\bibnamefont {Quek}, \bibfnamefont
  {S~Y}}, \bibinfo {author} {\bibfnamefont {L.}~\bibnamefont {Venkataraman}},
  \bibinfo {author} {\bibfnamefont {H.~J.}\ \bibnamefont {Choi}}, \bibinfo
  {author} {\bibfnamefont {S.~G.}\ \bibnamefont {Louie}}, \bibinfo {author}
  {\bibfnamefont {M.~S.}\ \bibnamefont {Hybertsen}}, \ and\ \bibinfo {author}
  {\bibfnamefont {J.~B.}\ \bibnamefont {Neaton}}} (\bibinfo {year} {2007}),\
  \bibfield  {title} {\enquote {\bibinfo {title} {{A}mine-{G}old {L}inked
  {S}ingle-{M}olecule {C}ircuits: {E}xperiment and {T}heory},}\ }\href@noop {}
  {\bibfield  {journal} {\bibinfo  {journal} {Nano Lett.}\ }\textbf {\bibinfo
  {volume} {7}}~(\bibinfo {number} {11}),\ \bibinfo {pages}
  {3477--3482}}\BibitemShut {NoStop}%
\bibitem [{\citenamefont {Questaal}(2018)}]{questaal}%
  \BibitemOpen
  \bibfield  {author} {\bibinfo {author} {\bibnamefont {Questaal},}} (\bibinfo
  {year} {2018}),\ \href@noop {} {\bibinfo  {journal}
  {https://www.questaal.org/}\ }\BibitemShut {NoStop}%
\bibitem [{\citenamefont {Ramasubramaniam}(2012)}]{Ramasubramaniam/etal:2012}%
  \BibitemOpen
\bibfield  {journal} {  }\bibfield  {author} {\bibinfo {author} {\bibnamefont
  {Ramasubramaniam}, \bibfnamefont {A}}} (\bibinfo {year} {2012}),\ \bibfield
  {title} {\enquote {\bibinfo {title} {{L}arge excitonic effects in monolayers
  of molybdenum and tungsten dichalcogenides},}\ }\href@noop {} {\bibfield
  {journal} {\bibinfo  {journal} {Phys. Rev. B}\ }\textbf {\bibinfo {volume}
  {86}},\ \bibinfo {pages} {115409}}\BibitemShut {NoStop}%
\bibitem [{\citenamefont {Ramzan}\ \emph {et~al.}(2010)\citenamefont {Ramzan},
  \citenamefont {Leb\`egue},\ and\ \citenamefont {Ahuja}}]{Ramzan/etal:2010}%
  \BibitemOpen
  \bibfield  {author} {\bibinfo {author} {\bibnamefont {Ramzan}, \bibfnamefont
  {M}}, \bibinfo {author} {\bibfnamefont {S.}~\bibnamefont {Leb\`egue}}, \ and\
  \bibinfo {author} {\bibfnamefont {R.}~\bibnamefont {Ahuja}}} (\bibinfo {year}
  {2010}),\ \bibfield  {title} {\enquote {\bibinfo {title} {{E}lectronic
  structure and metalization of a silane-hydrogen system under high pressure
  investigated using density functional and ${G}{W}$ calculations},}\
  }\href@noop {} {\bibfield  {journal} {\bibinfo  {journal} {Phys. Rev. B}\
  }\textbf {\bibinfo {volume} {81}},\ \bibinfo {pages} {233103}}\BibitemShut
  {NoStop}%
\bibitem [{\citenamefont {Ranasinghe}\ \emph {et~al.}(2019)\citenamefont
  {Ranasinghe}, \citenamefont {Margraf}, \citenamefont {Perera},\ and\
  \citenamefont {Bartlett}}]{Ranasinghe/etal:2019}%
  \BibitemOpen
  \bibfield  {author} {\bibinfo {author} {\bibnamefont {Ranasinghe},
  \bibfnamefont {D~S}}, \bibinfo {author} {\bibfnamefont {J.~T.}\ \bibnamefont
  {Margraf}}, \bibinfo {author} {\bibfnamefont {A.}~\bibnamefont {Perera}}, \
  and\ \bibinfo {author} {\bibfnamefont {R.~J.}\ \bibnamefont {Bartlett}}}
  (\bibinfo {year} {2019}),\ \bibfield  {title} {\enquote {\bibinfo {title}
  {{V}ertical valence ionization potential benchmarks from equation-of-motion
  coupled cluster theory and {Q}{T}{P} functionals},}\ }\href {\doibase
  10.1063/1.5084728} {\bibfield  {journal} {\bibinfo  {journal} {J. Chem.
  Phys.}\ }\textbf {\bibinfo {volume} {150}}~(\bibinfo {number} {7}),\ \bibinfo
  {pages} {074108}}\BibitemShut {NoStop}%
\bibitem [{\citenamefont {Rangel}\ \emph {et~al.}(2016)\citenamefont {Rangel},
  \citenamefont {Berland}, \citenamefont {Sharifzadeh}, \citenamefont
  {Brown-Altvater}, \citenamefont {Lee}, \citenamefont {Hyldgaard},
  \citenamefont {Kronik},\ and\ \citenamefont {Neaton}}]{Rangel2016}%
  \BibitemOpen
  \bibfield  {author} {\bibinfo {author} {\bibnamefont {Rangel}, \bibfnamefont
  {T}}, \bibinfo {author} {\bibfnamefont {K.}~\bibnamefont {Berland}}, \bibinfo
  {author} {\bibfnamefont {S.}~\bibnamefont {Sharifzadeh}}, \bibinfo {author}
  {\bibfnamefont {F.}~\bibnamefont {Brown-Altvater}}, \bibinfo {author}
  {\bibfnamefont {K.}~\bibnamefont {Lee}}, \bibinfo {author} {\bibfnamefont
  {P.}~\bibnamefont {Hyldgaard}}, \bibinfo {author} {\bibfnamefont
  {L.}~\bibnamefont {Kronik}}, \ and\ \bibinfo {author} {\bibfnamefont {J.~B.}\
  \bibnamefont {Neaton}}} (\bibinfo {year} {2016}),\ \bibfield  {title}
  {\enquote {\bibinfo {title} {{S}tructural and excited-state properties of
  oligoacene crystals from first principles},}\ }\href {\doibase
  10.1103/PhysRevB.93.115206} {\bibfield  {journal} {\bibinfo  {journal} {Phys.
  Rev. B}\ }\textbf {\bibinfo {volume} {93}},\ \bibinfo {pages}
  {115206}}\BibitemShut {NoStop}%
\bibitem [{\citenamefont {Rangel}\ \emph {et~al.}(2012)\citenamefont {Rangel},
  \citenamefont {Kecik}, \citenamefont {Trevisanutto}, \citenamefont
  {Rignanese}, \citenamefont {Swygenhoven},\ and\ \citenamefont
  {Olevano}}]{Rangel/etal:2012}%
  \BibitemOpen
  \bibfield  {author} {\bibinfo {author} {\bibnamefont {Rangel}, \bibfnamefont
  {T}}, \bibinfo {author} {\bibfnamefont {D.}~\bibnamefont {Kecik}}, \bibinfo
  {author} {\bibfnamefont {P.~E.}\ \bibnamefont {Trevisanutto}}, \bibinfo
  {author} {\bibfnamefont {G.-M.}\ \bibnamefont {Rignanese}}, \bibinfo {author}
  {\bibfnamefont {H.~Van}\ \bibnamefont {Swygenhoven}}, \ and\ \bibinfo
  {author} {\bibfnamefont {V.}~\bibnamefont {Olevano}}} (\bibinfo {year}
  {2012}),\ \bibfield  {title} {\enquote {\bibinfo {title} {{B}and structure of
  gold from many-body perturbation theory},}\ }\href@noop {} {\bibfield
  {journal} {\bibinfo  {journal} {Phys. Rev. B}\ }\textbf {\bibinfo {volume}
  {86}},\ \bibinfo {pages} {125125}}\BibitemShut {NoStop}%
\bibitem [{\citenamefont {Rangel}\ \emph {et~al.}(2018)\citenamefont {Rangel},
  \citenamefont {Rinn}, \citenamefont {Sharifzadeh}, \citenamefont
  {da~Jornada}, \citenamefont {Pick}, \citenamefont {Louie}, \citenamefont
  {Witte}, \citenamefont {Kronik}, \citenamefont {Neaton},\ and\ \citenamefont
  {Chatterjee}}]{Rangel2018}%
  \BibitemOpen
  \bibfield  {author} {\bibinfo {author} {\bibnamefont {Rangel}, \bibfnamefont
  {T}}, \bibinfo {author} {\bibfnamefont {A.}~\bibnamefont {Rinn}}, \bibinfo
  {author} {\bibfnamefont {S.}~\bibnamefont {Sharifzadeh}}, \bibinfo {author}
  {\bibfnamefont {F.~H.}\ \bibnamefont {da~Jornada}}, \bibinfo {author}
  {\bibfnamefont {A.}~\bibnamefont {Pick}}, \bibinfo {author} {\bibfnamefont
  {S.~G.}\ \bibnamefont {Louie}}, \bibinfo {author} {\bibfnamefont
  {G.}~\bibnamefont {Witte}}, \bibinfo {author} {\bibfnamefont
  {L.}~\bibnamefont {Kronik}}, \bibinfo {author} {\bibfnamefont {J.~B.}\
  \bibnamefont {Neaton}}, \ and\ \bibinfo {author} {\bibfnamefont
  {S.}~\bibnamefont {Chatterjee}}} (\bibinfo {year} {2018}),\ \bibfield
  {title} {\enquote {\bibinfo {title} {{L}ow-lying excited states in
  crystalline perylene},}\ }\href {\doibase 10.1073/pnas.1711126115} {\bibfield
   {journal} {\bibinfo  {journal} {Proc. Natl. Acad. Sci. USA}\ }\textbf
  {\bibinfo {volume} {115}}~(\bibinfo {number} {2}),\ \bibinfo {pages}
  {284--289}}\BibitemShut {NoStop}%
\bibitem [{\citenamefont {Rasmussen}\ \emph {et~al.}(2016)\citenamefont
  {Rasmussen}, \citenamefont {Schmidt}, \citenamefont {Winther},\ and\
  \citenamefont {Thygesen}}]{rasmussen_prb_94}%
  \BibitemOpen
  \bibfield  {author} {\bibinfo {author} {\bibnamefont {Rasmussen},
  \bibfnamefont {F~A}}, \bibinfo {author} {\bibfnamefont {P.~S.}\ \bibnamefont
  {Schmidt}}, \bibinfo {author} {\bibfnamefont {K.~T.}\ \bibnamefont
  {Winther}}, \ and\ \bibinfo {author} {\bibfnamefont {K.~S.}\ \bibnamefont
  {Thygesen}}} (\bibinfo {year} {2016}),\ \bibfield  {title} {\enquote
  {\bibinfo {title} {{E}fficient many-body calculations for two-dimensional
  materials using exact limits for the screened potential: {B}and gaps of
  {M}o{S}$_{2}, h$-{B}{N}, and phosphorene},}\ }\href {\doibase
  10.1103/PhysRevB.94.155406} {\bibfield  {journal} {\bibinfo  {journal} {Phys.
  Rev. B}\ }\textbf {\bibinfo {volume} {94}},\ \bibinfo {pages}
  {155406}}\BibitemShut {NoStop}%
\bibitem [{\citenamefont {Rasmussen}\ and\ \citenamefont
  {Thygesen}(2015)}]{rasmussen_jpcc_119}%
  \BibitemOpen
  \bibfield  {author} {\bibinfo {author} {\bibnamefont {Rasmussen},
  \bibfnamefont {F~A}}, \ and\ \bibinfo {author} {\bibfnamefont {K.~S.}\
  \bibnamefont {Thygesen}}} (\bibinfo {year} {2015}),\ \bibfield  {title}
  {\enquote {\bibinfo {title} {{C}omputational 2{D} {M}aterials {D}atabase:
  {E}lectronic {S}tructure of {T}ransition-{M}etal {D}ichalcogenides and
  {O}xides},}\ }\href {\doibase 10.1021/acs.jpcc.5b02950} {\bibfield  {journal}
  {\bibinfo  {journal} {J. Phys. Chem. C}\ }\textbf {\bibinfo {volume}
  {119}}~(\bibinfo {number} {23}),\ \bibinfo {pages}
  {13169--13183}}\BibitemShut {NoStop}%
\bibitem [{\citenamefont {Refaely-Abramson}\ \emph {et~al.}(2015)\citenamefont
  {Refaely-Abramson}, \citenamefont {Jain}, \citenamefont {Sharifzadeh},
  \citenamefont {Neaton},\ and\ \citenamefont {Kronik}}]{Refaely-Abramson2015}%
  \BibitemOpen
  \bibfield  {author} {\bibinfo {author} {\bibnamefont {Refaely-Abramson},
  \bibfnamefont {S}}, \bibinfo {author} {\bibfnamefont {M.}~\bibnamefont
  {Jain}}, \bibinfo {author} {\bibfnamefont {S.}~\bibnamefont {Sharifzadeh}},
  \bibinfo {author} {\bibfnamefont {J.~B.}\ \bibnamefont {Neaton}}, \ and\
  \bibinfo {author} {\bibfnamefont {L.}~\bibnamefont {Kronik}}} (\bibinfo
  {year} {2015}),\ \bibfield  {title} {\enquote {\bibinfo {title}
  {{S}olid-state optical absorption from optimally tuned time-dependent
  range-separated hybrid density functional theory},}\ }\href {\doibase
  10.1103/PhysRevB.92.081204} {\bibfield  {journal} {\bibinfo  {journal} {Phys.
  Rev. B}\ }\textbf {\bibinfo {volume} {92}},\ \bibinfo {pages}
  {081204}}\BibitemShut {NoStop}%
\bibitem [{\citenamefont {Refaely-Abramson}\ \emph {et~al.}(2012)\citenamefont
  {Refaely-Abramson}, \citenamefont {Sharifzadeh}, \citenamefont {Govind},
  \citenamefont {Autschbach}, \citenamefont {Neaton}, \citenamefont {Baer},\
  and\ \citenamefont {Kronik}}]{Refaely-Abramson2012}%
  \BibitemOpen
  \bibfield  {author} {\bibinfo {author} {\bibnamefont {Refaely-Abramson},
  \bibfnamefont {S}}, \bibinfo {author} {\bibfnamefont {S.}~\bibnamefont
  {Sharifzadeh}}, \bibinfo {author} {\bibfnamefont {N.}~\bibnamefont {Govind}},
  \bibinfo {author} {\bibfnamefont {J.}~\bibnamefont {Autschbach}}, \bibinfo
  {author} {\bibfnamefont {J.~B.}\ \bibnamefont {Neaton}}, \bibinfo {author}
  {\bibfnamefont {R.}~\bibnamefont {Baer}}, \ and\ \bibinfo {author}
  {\bibfnamefont {L.}~\bibnamefont {Kronik}}} (\bibinfo {year} {2012}),\
  \bibfield  {title} {\enquote {\bibinfo {title} {{Q}uasiparticle {S}pectra
  from a {N}onempirical {O}ptimally {T}uned {R}ange-{S}eparated {H}ybrid
  {D}ensity {F}unctional},}\ }\href {\doibase 10.1103/PhysRevLett.109.226405}
  {\bibfield  {journal} {\bibinfo  {journal} {Phys. Rev. Lett.}\ }\textbf
  {\bibinfo {volume} {109}},\ \bibinfo {pages} {226405}}\BibitemShut {NoStop}%
\bibitem [{\citenamefont {Refaely-Abramson}\ \emph {et~al.}(2013)\citenamefont
  {Refaely-Abramson}, \citenamefont {Sharifzadeh}, \citenamefont {Jain},
  \citenamefont {Baer}, \citenamefont {Neaton},\ and\ \citenamefont
  {Kronik}}]{Refaely-Abramson/etal:2013}%
  \BibitemOpen
  \bibfield  {author} {\bibinfo {author} {\bibnamefont {Refaely-Abramson},
  \bibfnamefont {S}}, \bibinfo {author} {\bibfnamefont {S.}~\bibnamefont
  {Sharifzadeh}}, \bibinfo {author} {\bibfnamefont {M.}~\bibnamefont {Jain}},
  \bibinfo {author} {\bibfnamefont {R.}~\bibnamefont {Baer}}, \bibinfo {author}
  {\bibfnamefont {J.~B.}\ \bibnamefont {Neaton}}, \ and\ \bibinfo {author}
  {\bibfnamefont {L.}~\bibnamefont {Kronik}}} (\bibinfo {year} {2013}),\
  \bibfield  {title} {\enquote {\bibinfo {title} {{G}ap renormalization of
  molecular crystals from density-functional theory},}\ }\href@noop {}
  {\bibfield  {journal} {\bibinfo  {journal} {Phys. Rev. B}\ }\textbf {\bibinfo
  {volume} {88}},\ \bibinfo {pages} {081204}}\BibitemShut {NoStop}%
\bibitem [{\citenamefont {Reining}(2017)}]{Reining2017}%
  \BibitemOpen
  \bibfield  {author} {\bibinfo {author} {\bibnamefont {Reining}, \bibfnamefont
  {L}}} (\bibinfo {year} {2017}),\ \bibfield  {title} {\enquote {\bibinfo
  {title} {{T}he ${G}{W}$ approximation: content, successes and limitations},}\
  }\href@noop {} {\bibfield  {journal} {\bibinfo  {journal} {WIREs Comput. Mol.
  Sci.}\ }\textbf {\bibinfo {volume} {8}}~(\bibinfo {number} {3}),\ \bibinfo
  {pages} {e1344}}\BibitemShut {NoStop}%
\bibitem [{\citenamefont {Reining}\ \emph {et~al.}(2002)\citenamefont
  {Reining}, \citenamefont {Olevano}, \citenamefont {Rubio},\ and\
  \citenamefont {Onida}}]{Reining/etal:2002}%
  \BibitemOpen
  \bibfield  {author} {\bibinfo {author} {\bibnamefont {Reining}, \bibfnamefont
  {L}}, \bibinfo {author} {\bibfnamefont {V.}~\bibnamefont {Olevano}}, \bibinfo
  {author} {\bibfnamefont {A.}~\bibnamefont {Rubio}}, \ and\ \bibinfo {author}
  {\bibfnamefont {G.}~\bibnamefont {Onida}}} (\bibinfo {year} {2002}),\
  \bibfield  {title} {\enquote {\bibinfo {title} {{E}xcitonic {E}ffects in
  {S}olids {D}escribed by {T}ime-{D}ependent {D}ensity-{F}unctional
  {T}heory},}\ }\href@noop {} {\bibfield  {journal} {\bibinfo  {journal} {Phys.
  Rev. Lett.}\ }\textbf {\bibinfo {volume} {88}},\ \bibinfo {pages}
  {066404}}\BibitemShut {NoStop}%
\bibitem [{\citenamefont {Reining}\ \emph {et~al.}(1997)\citenamefont
  {Reining}, \citenamefont {Onida},\ and\ \citenamefont {Godby}}]{Reining1997}%
  \BibitemOpen
  \bibfield  {author} {\bibinfo {author} {\bibnamefont {Reining}, \bibfnamefont
  {L}}, \bibinfo {author} {\bibfnamefont {G.}~\bibnamefont {Onida}}, \ and\
  \bibinfo {author} {\bibfnamefont {R.~W.}\ \bibnamefont {Godby}}} (\bibinfo
  {year} {1997}),\ \bibfield  {title} {\enquote {\bibinfo {title}
  {{E}limination of unoccupied-state summations in ab initio self-energy
  calculations for large supercells},}\ }\href {\doibase
  10.1103/PhysRevB.56.R4301} {\bibfield  {journal} {\bibinfo  {journal} {Phys.
  Rev. B}\ }\textbf {\bibinfo {volume} {56}},\ \bibinfo {pages}
  {R4301--R4304}}\BibitemShut {NoStop}%
\bibitem [{\citenamefont {Ren}\ \emph {et~al.}(2015)\citenamefont {Ren},
  \citenamefont {Marom}, \citenamefont {Caruso}, \citenamefont {Scheffler},\
  and\ \citenamefont {Rinke}}]{Ren/etal:2015}%
  \BibitemOpen
  \bibfield  {author} {\bibinfo {author} {\bibnamefont {Ren}, \bibfnamefont
  {X}}, \bibinfo {author} {\bibfnamefont {N.}~\bibnamefont {Marom}}, \bibinfo
  {author} {\bibfnamefont {F.}~\bibnamefont {Caruso}}, \bibinfo {author}
  {\bibfnamefont {M.}~\bibnamefont {Scheffler}}, \ and\ \bibinfo {author}
  {\bibfnamefont {P.}~\bibnamefont {Rinke}}} (\bibinfo {year} {2015}),\
  \bibfield  {title} {\enquote {\bibinfo {title} {{B}eyond the ${G}{W}$
  approximation: {A} second-order screened exchange correction},}\ }\href
  {\doibase 10.1103/PhysRevB.92.081104} {\bibfield  {journal} {\bibinfo
  {journal} {Phys.\ Rev.\ B}\ }\textbf {\bibinfo {volume} {92}},\ \bibinfo
  {pages} {081104}}\BibitemShut {NoStop}%
\bibitem [{\citenamefont {Ren}\ \emph {et~al.}(2012{\natexlab{a}})\citenamefont
  {Ren}, \citenamefont {Rinke}, \citenamefont {Blum}, \citenamefont
  {Wieferink}, \citenamefont {Tkatchenko}, \citenamefont {Sanfilippo},
  \citenamefont {Reuter},\ and\ \citenamefont
  {Scheffler}}]{Xinguo/implem_full_author_list}%
  \BibitemOpen
  \bibfield  {author} {\bibinfo {author} {\bibnamefont {Ren}, \bibfnamefont
  {X}}, \bibinfo {author} {\bibfnamefont {P.}~\bibnamefont {Rinke}}, \bibinfo
  {author} {\bibfnamefont {V.}~\bibnamefont {Blum}}, \bibinfo {author}
  {\bibfnamefont {J.}~\bibnamefont {Wieferink}}, \bibinfo {author}
  {\bibfnamefont {A.}~\bibnamefont {Tkatchenko}}, \bibinfo {author}
  {\bibfnamefont {A.}~\bibnamefont {Sanfilippo}}, \bibinfo {author}
  {\bibfnamefont {K.}~\bibnamefont {Reuter}}, \ and\ \bibinfo {author}
  {\bibfnamefont {M.}~\bibnamefont {Scheffler}}} (\bibinfo {year}
  {2012}{\natexlab{a}}),\ \bibfield  {title} {\enquote {\bibinfo {title}
  {{R}esolution-of-identity approach to {H}artree-{F}ock, hybrid density
  functionals, {R}{P}{A}, {M}{P}2 and ${G}{W}$ with numeric atom-centered
  orbital basis functions},}\ }\href@noop {} {\bibfield  {journal} {\bibinfo
  {journal} {New J. Phys.}\ }\textbf {\bibinfo {volume} {14}},\ \bibinfo
  {pages} {053020}}\BibitemShut {NoStop}%
\bibitem [{\citenamefont {Ren}\ \emph {et~al.}(2012{\natexlab{b}})\citenamefont
  {Ren}, \citenamefont {Rinke}, \citenamefont {Joas},\ and\ \citenamefont
  {Scheffler}}]{RPAreview}%
  \BibitemOpen
  \bibfield  {author} {\bibinfo {author} {\bibnamefont {Ren}, \bibfnamefont
  {X}}, \bibinfo {author} {\bibfnamefont {P.}~\bibnamefont {Rinke}}, \bibinfo
  {author} {\bibfnamefont {C.}~\bibnamefont {Joas}}, \ and\ \bibinfo {author}
  {\bibfnamefont {M.}~\bibnamefont {Scheffler}}} (\bibinfo {year}
  {2012}{\natexlab{b}}),\ \bibfield  {title} {\enquote {\bibinfo {title}
  {{R}andom-phase approximation and its applications in computational chemistry
  and materials science},}\ }\href@noop {} {\bibfield  {journal} {\bibinfo
  {journal} {J. Mater. Sci.}\ }\textbf {\bibinfo {volume} {47}},\ \bibinfo
  {pages} {21}}\BibitemShut {NoStop}%
\bibitem [{\citenamefont {Reyes-Lillo}\ \emph {et~al.}(2016)\citenamefont
  {Reyes-Lillo}, \citenamefont {Rangel}, \citenamefont {Bruneval},\ and\
  \citenamefont {Neaton}}]{Reyes-Lillo2016}%
  \BibitemOpen
  \bibfield  {author} {\bibinfo {author} {\bibnamefont {Reyes-Lillo},
  \bibfnamefont {S~E}}, \bibinfo {author} {\bibfnamefont {T.}~\bibnamefont
  {Rangel}}, \bibinfo {author} {\bibfnamefont {F.}~\bibnamefont {Bruneval}}, \
  and\ \bibinfo {author} {\bibfnamefont {J.~B.}\ \bibnamefont {Neaton}}}
  (\bibinfo {year} {2016}),\ \bibfield  {title} {\enquote {\bibinfo {title}
  {{E}ffects of quantum confinement on excited state properties of
  ${\mathrm{sr{t}io}}_{3}$ from ab initio many-body perturbation theory},}\
  }\href {\doibase 10.1103/PhysRevB.94.041107} {\bibfield  {journal} {\bibinfo
  {journal} {Phys. Rev. B}\ }\textbf {\bibinfo {volume} {94}},\ \bibinfo
  {pages} {041107}}\BibitemShut {NoStop}%
\bibitem [{\citenamefont {Richard}\ \emph {et~al.}(2016)\citenamefont
  {Richard}, \citenamefont {Marshall}, \citenamefont {Dolgounitcheva},
  \citenamefont {Ortiz}, \citenamefont {Br\'{e}das}, \citenamefont {Marom},\
  and\ \citenamefont {Sherrill}}]{Richard2016}%
  \BibitemOpen
  \bibfield  {author} {\bibinfo {author} {\bibnamefont {Richard}, \bibfnamefont
  {R~M}}, \bibinfo {author} {\bibfnamefont {M.~S.}\ \bibnamefont {Marshall}},
  \bibinfo {author} {\bibfnamefont {O.}~\bibnamefont {Dolgounitcheva}},
  \bibinfo {author} {\bibfnamefont {J.~V.}\ \bibnamefont {Ortiz}}, \bibinfo
  {author} {\bibfnamefont {J.-L.}\ \bibnamefont {Br\'{e}das}}, \bibinfo
  {author} {\bibfnamefont {N.}~\bibnamefont {Marom}}, \ and\ \bibinfo {author}
  {\bibfnamefont {C.~D.}\ \bibnamefont {Sherrill}}} (\bibinfo {year} {2016}),\
  \bibfield  {title} {\enquote {\bibinfo {title} {{A}ccurate {I}onization
  {P}otentials and {E}lectron {A}ffinities of {A}cceptor {M}olecules {I}.
  {R}eference {D}ata at the {C}{C}{S}{D}({T}) {C}omplete {B}asis {S}et
  {L}imit},}\ }\href@noop {} {\bibfield  {journal} {\bibinfo  {journal} {J.
  Chem. Theory Comput.}\ }\textbf {\bibinfo {volume} {12}}~(\bibinfo {number}
  {2}),\ \bibinfo {pages} {595--604}}\BibitemShut {NoStop}%
\bibitem [{\citenamefont {Richter}\ \emph {et~al.}(2011)\citenamefont
  {Richter}, \citenamefont {Ruck}, \citenamefont {Simpson}, \citenamefont
  {Natali}, \citenamefont {Plank}, \citenamefont {Azeem}, \citenamefont
  {Trodahl}, \citenamefont {Preston}, \citenamefont {Chen}, \citenamefont
  {McNulty}, \citenamefont {Smith}, \citenamefont {Tadich}, \citenamefont
  {Cowie}, \citenamefont {Svane}, \citenamefont {{van Schilfgaarde}},\ and\
  \citenamefont {Lambrecht}}]{Richter/etal:2011}%
  \BibitemOpen
  \bibfield  {author} {\bibinfo {author} {\bibnamefont {Richter}, \bibfnamefont
  {J~H}}, \bibinfo {author} {\bibfnamefont {B.~J.}\ \bibnamefont {Ruck}},
  \bibinfo {author} {\bibfnamefont {M.}~\bibnamefont {Simpson}}, \bibinfo
  {author} {\bibfnamefont {F.}~\bibnamefont {Natali}}, \bibinfo {author}
  {\bibfnamefont {N.~O.~V.}\ \bibnamefont {Plank}}, \bibinfo {author}
  {\bibfnamefont {M.}~\bibnamefont {Azeem}}, \bibinfo {author} {\bibfnamefont
  {H.~J.}\ \bibnamefont {Trodahl}}, \bibinfo {author} {\bibfnamefont
  {A.~R.~H.}\ \bibnamefont {Preston}}, \bibinfo {author} {\bibfnamefont
  {B.}~\bibnamefont {Chen}}, \bibinfo {author} {\bibfnamefont {J.}~\bibnamefont
  {McNulty}}, \bibinfo {author} {\bibfnamefont {K.~E.}\ \bibnamefont {Smith}},
  \bibinfo {author} {\bibfnamefont {A.}~\bibnamefont {Tadich}}, \bibinfo
  {author} {\bibfnamefont {B.}~\bibnamefont {Cowie}}, \bibinfo {author}
  {\bibfnamefont {A.}~\bibnamefont {Svane}}, \bibinfo {author} {\bibfnamefont
  {M.}~\bibnamefont {{van Schilfgaarde}}}, \ and\ \bibinfo {author}
  {\bibfnamefont {W.~R.~L.}\ \bibnamefont {Lambrecht}}} (\bibinfo {year}
  {2011}),\ \bibfield  {title} {\enquote {\bibinfo {title} {{E}lectronic
  structure of {E}u{N}: {G}rowth, spectroscopy, and theory},}\ }\href@noop {}
  {\bibfield  {journal} {\bibinfo  {journal} {Phys. Rev. B}\ }\textbf {\bibinfo
  {volume} {84}},\ \bibinfo {pages} {235120}}\BibitemShut {NoStop}%
\bibitem [{\citenamefont {Rieger}\ \emph {et~al.}(1999)\citenamefont {Rieger},
  \citenamefont {Steinbeck}, \citenamefont {White}, \citenamefont {Rojas},\
  and\ \citenamefont {Godby}}]{GW_space-time_method:1998}%
  \BibitemOpen
  \bibfield  {author} {\bibinfo {author} {\bibnamefont {Rieger}, \bibfnamefont
  {M~M}}, \bibinfo {author} {\bibfnamefont {L.}~\bibnamefont {Steinbeck}},
  \bibinfo {author} {\bibfnamefont {I.~D.}\ \bibnamefont {White}}, \bibinfo
  {author} {\bibfnamefont {H.~N.}\ \bibnamefont {Rojas}}, \ and\ \bibinfo
  {author} {\bibfnamefont {R.~W.}\ \bibnamefont {Godby}}} (\bibinfo {year}
  {1999}),\ \bibfield  {title} {\enquote {\bibinfo {title} {{T}he \textit{GW}
  space-time method for the self-energy of large systems},}\ }\href@noop {}
  {\bibfield  {journal} {\bibinfo  {journal} {Comput. Phys. Commun.}\ }\textbf
  {\bibinfo {volume} {117}},\ \bibinfo {pages} {211}}\BibitemShut {NoStop}%
\bibitem [{\citenamefont {Rinke}\ \emph {et~al.}(2004)\citenamefont {Rinke},
  \citenamefont {Delaney}, \citenamefont {Garc\'{\i}a-Gonz\'alez},\ and\
  \citenamefont {Godby}}]{ClusterImStates:2004}%
  \BibitemOpen
  \bibfield  {author} {\bibinfo {author} {\bibnamefont {Rinke}, \bibfnamefont
  {P}}, \bibinfo {author} {\bibfnamefont {K.}~\bibnamefont {Delaney}}, \bibinfo
  {author} {\bibfnamefont {P.}~\bibnamefont {Garc\'{\i}a-Gonz\'alez}}, \ and\
  \bibinfo {author} {\bibfnamefont {R.~W.}\ \bibnamefont {Godby}}} (\bibinfo
  {year} {2004}),\ \bibfield  {title} {\enquote {\bibinfo {title} {{I}mage
  {S}tates in {M}etal {C}lusters},}\ }\href@noop {} {\bibfield  {journal}
  {\bibinfo  {journal} {Phys.\ Rev.\ A}\ }\textbf {\bibinfo {volume} {70}},\
  \bibinfo {pages} {063201}}\BibitemShut {NoStop}%
\bibitem [{\citenamefont {Rinke}\ \emph {et~al.}(2009)\citenamefont {Rinke},
  \citenamefont {Janotti}, \citenamefont {Scheffler},\ and\ \citenamefont {{Van
  de Walle}}}]{Rinke/etal:2009}%
  \BibitemOpen
  \bibfield  {author} {\bibinfo {author} {\bibnamefont {Rinke}, \bibfnamefont
  {P}}, \bibinfo {author} {\bibfnamefont {A.}~\bibnamefont {Janotti}}, \bibinfo
  {author} {\bibfnamefont {M.}~\bibnamefont {Scheffler}}, \ and\ \bibinfo
  {author} {\bibfnamefont {C.~G.}\ \bibnamefont {{Van de Walle}}}} (\bibinfo
  {year} {2009}),\ \bibfield  {title} {\enquote {\bibinfo {title} {{D}efect
  {F}ormation {E}nergies without the {B}and-{G}ap {P}roblem: {C}ombining
  {D}ensity-{F}unctional {T}heory and the ${G}{W}$ {A}pproach for the {S}ilicon
  {S}elf-{I}nterstitial},}\ }\href@noop {} {\bibfield  {journal} {\bibinfo
  {journal} {Phys.\ Rev.\ Lett.}\ }\textbf {\bibinfo {volume} {102}},\ \bibinfo
  {pages} {026402}}\BibitemShut {NoStop}%
\bibitem [{\citenamefont {Rinke}\ \emph {et~al.}(2005)\citenamefont {Rinke},
  \citenamefont {Qteish}, \citenamefont {Neugebauer}, \citenamefont
  {Freysoldt},\ and\ \citenamefont {Scheffler}}]{Rinke/etal:2005}%
  \BibitemOpen
  \bibfield  {author} {\bibinfo {author} {\bibnamefont {Rinke}, \bibfnamefont
  {P}}, \bibinfo {author} {\bibfnamefont {A.}~\bibnamefont {Qteish}}, \bibinfo
  {author} {\bibfnamefont {J.}~\bibnamefont {Neugebauer}}, \bibinfo {author}
  {\bibfnamefont {C.}~\bibnamefont {Freysoldt}}, \ and\ \bibinfo {author}
  {\bibfnamefont {M.}~\bibnamefont {Scheffler}}} (\bibinfo {year} {2005}),\
  \bibfield  {title} {\enquote {\bibinfo {title} {{C}ombining \textit{GW}
  calculations with exact-exchange density-functional theory: {A}n analysis of
  valence-band photoemission for compound semiconductors},}\ }\href@noop {}
  {\bibfield  {journal} {\bibinfo  {journal} {New J. Phys.}\ }\textbf {\bibinfo
  {volume} {7}},\ \bibinfo {pages} {126}}\BibitemShut {NoStop}%
\bibitem [{\citenamefont {Rinke}\ \emph
  {et~al.}(2008{\natexlab{a}})\citenamefont {Rinke}, \citenamefont {Qteish},
  \citenamefont {Neugebauer},\ and\ \citenamefont {Scheffler}}]{Rinke/pssb}%
  \BibitemOpen
  \bibfield  {author} {\bibinfo {author} {\bibnamefont {Rinke}, \bibfnamefont
  {P}}, \bibinfo {author} {\bibfnamefont {A.}~\bibnamefont {Qteish}}, \bibinfo
  {author} {\bibfnamefont {J.}~\bibnamefont {Neugebauer}}, \ and\ \bibinfo
  {author} {\bibfnamefont {M.}~\bibnamefont {Scheffler}}} (\bibinfo {year}
  {2008}{\natexlab{a}}),\ \bibfield  {title} {\enquote {\bibinfo {title}
  {{E}xciting prospects for solids: {E}xact-exchange based functionals meet
  quasiparticle energy calculations},}\ }\href@noop {} {\bibfield  {journal}
  {\bibinfo  {journal} {Phys. Status Solidi B}\ }\textbf {\bibinfo {volume}
  {245}}~(\bibinfo {number} {5}),\ \bibinfo {pages} {929--945}}\BibitemShut
  {NoStop}%
\bibitem [{\citenamefont {Rinke}\ \emph {et~al.}(2006)\citenamefont {Rinke},
  \citenamefont {Qteish}, \citenamefont {Winkelnkemper}, \citenamefont
  {Bimberg}, \citenamefont {Neugebauer},\ and\ \citenamefont
  {Scheffler}}]{Rinke/etal:2006}%
  \BibitemOpen
  \bibfield  {author} {\bibinfo {author} {\bibnamefont {Rinke}, \bibfnamefont
  {P}}, \bibinfo {author} {\bibfnamefont {A.}~\bibnamefont {Qteish}}, \bibinfo
  {author} {\bibfnamefont {M.}~\bibnamefont {Winkelnkemper}}, \bibinfo {author}
  {\bibfnamefont {D.}~\bibnamefont {Bimberg}}, \bibinfo {author} {\bibfnamefont
  {J.}~\bibnamefont {Neugebauer}}, \ and\ \bibinfo {author} {\bibfnamefont
  {M.}~\bibnamefont {Scheffler}}} (\bibinfo {year} {2006}),\ \bibfield  {title}
  {\enquote {\bibinfo {title} {{B}and gap and band parameters of {I}n{N} and
  {G}a{N} from quasiparticle energy calculations based on exact-exchange
  density-functional theory},}\ }\href@noop {} {\bibfield  {journal} {\bibinfo
  {journal} {Appl.\ Phys.\ Lett.}\ }\textbf {\bibinfo {volume} {89}},\ \bibinfo
  {pages} {161919}}\BibitemShut {NoStop}%
\bibitem [{\citenamefont {Rinke}\ \emph {et~al.}(2012)\citenamefont {Rinke},
  \citenamefont {Schleife}, \citenamefont {Kioupakis}, \citenamefont {Janotti},
  \citenamefont {R\"odl}, \citenamefont {Bechstedt}, \citenamefont
  {Scheffler},\ and\ \citenamefont {de~Walle}}]{Rinke/etal:2012}%
  \BibitemOpen
  \bibfield  {author} {\bibinfo {author} {\bibnamefont {Rinke}, \bibfnamefont
  {P}}, \bibinfo {author} {\bibfnamefont {A.}~\bibnamefont {Schleife}},
  \bibinfo {author} {\bibfnamefont {E.}~\bibnamefont {Kioupakis}}, \bibinfo
  {author} {\bibfnamefont {A.}~\bibnamefont {Janotti}}, \bibinfo {author}
  {\bibfnamefont {C.}~\bibnamefont {R\"odl}}, \bibinfo {author} {\bibfnamefont
  {F.}~\bibnamefont {Bechstedt}}, \bibinfo {author} {\bibfnamefont
  {M.}~\bibnamefont {Scheffler}}, \ and\ \bibinfo {author} {\bibfnamefont
  {C.~G.~Van}\ \bibnamefont {de~Walle}}} (\bibinfo {year} {2012}),\ \bibfield
  {title} {\enquote {\bibinfo {title} {{F}irst-{P}rinciples {O}ptical {S}pectra
  for ${F}$ {C}enters in {M}g{O}},}\ }\href@noop {} {\bibfield  {journal}
  {\bibinfo  {journal} {Phys. Rev. Lett.}\ }\textbf {\bibinfo {volume} {108}},\
  \bibinfo {pages} {126404}}\BibitemShut {NoStop}%
\bibitem [{\citenamefont {Rinke}\ \emph
  {et~al.}(2008{\natexlab{b}})\citenamefont {Rinke}, \citenamefont
  {Winkelnkemper}, \citenamefont {Qteish}, \citenamefont {Bimberg},
  \citenamefont {Neugebauer},\ and\ \citenamefont
  {Scheffler}}]{Rinke/etal:2008}%
  \BibitemOpen
  \bibfield  {author} {\bibinfo {author} {\bibnamefont {Rinke}, \bibfnamefont
  {P}}, \bibinfo {author} {\bibfnamefont {M.}~\bibnamefont {Winkelnkemper}},
  \bibinfo {author} {\bibfnamefont {A.}~\bibnamefont {Qteish}}, \bibinfo
  {author} {\bibfnamefont {D.}~\bibnamefont {Bimberg}}, \bibinfo {author}
  {\bibfnamefont {J.}~\bibnamefont {Neugebauer}}, \ and\ \bibinfo {author}
  {\bibfnamefont {M.}~\bibnamefont {Scheffler}}} (\bibinfo {year}
  {2008}{\natexlab{b}}),\ \bibfield  {title} {\enquote {\bibinfo {title}
  {{C}onsistent set of band parameters for the group-{I}{I}{I} nitrides
  {A}l{N}, {G}a{N}, and {I}n{N}},}\ }\href@noop {} {\bibfield  {journal}
  {\bibinfo  {journal} {Phys. Rev. B}\ }\textbf {\bibinfo {volume} {77}},\
  \bibinfo {pages} {075202}}\BibitemShut {NoStop}%
\bibitem [{\citenamefont {Robert}\ \emph {et~al.}(2016)\citenamefont {Robert},
  \citenamefont {Picard}, \citenamefont {Lagarde}, \citenamefont {Wang},
  \citenamefont {Echeverry}, \citenamefont {Cadiz}, \citenamefont {Renucci},
  \citenamefont {H\"ogele}, \citenamefont {Amand}, \citenamefont {Marie},
  \citenamefont {Gerber},\ and\ \citenamefont {Urbaszek}}]{robert_prb_94}%
  \BibitemOpen
  \bibfield  {author} {\bibinfo {author} {\bibnamefont {Robert}, \bibfnamefont
  {C}}, \bibinfo {author} {\bibfnamefont {R.}~\bibnamefont {Picard}}, \bibinfo
  {author} {\bibfnamefont {D.}~\bibnamefont {Lagarde}}, \bibinfo {author}
  {\bibfnamefont {G.}~\bibnamefont {Wang}}, \bibinfo {author} {\bibfnamefont
  {J.~P.}\ \bibnamefont {Echeverry}}, \bibinfo {author} {\bibfnamefont
  {F.}~\bibnamefont {Cadiz}}, \bibinfo {author} {\bibfnamefont
  {P.}~\bibnamefont {Renucci}}, \bibinfo {author} {\bibfnamefont
  {A.}~\bibnamefont {H\"ogele}}, \bibinfo {author} {\bibfnamefont
  {T.}~\bibnamefont {Amand}}, \bibinfo {author} {\bibfnamefont
  {X.}~\bibnamefont {Marie}}, \bibinfo {author} {\bibfnamefont {I.~C.}\
  \bibnamefont {Gerber}}, \ and\ \bibinfo {author} {\bibfnamefont
  {B.}~\bibnamefont {Urbaszek}}} (\bibinfo {year} {2016}),\ \bibfield  {title}
  {\enquote {\bibinfo {title} {{E}xcitonic properties of semiconducting
  monolayer and bilayer $\mathrm{MoT}{\mathrm{e}}_{2}$},}\ }\href {\doibase
  10.1103/PhysRevB.94.155425} {\bibfield  {journal} {\bibinfo  {journal} {Phys.
  Rev. B}\ }\textbf {\bibinfo {volume} {94}},\ \bibinfo {pages}
  {155425}}\BibitemShut {NoStop}%
\bibitem [{\citenamefont {R\"odl}\ \emph {et~al.}(2008)\citenamefont {R\"odl},
  \citenamefont {Fuchs}, \citenamefont {Furthm\"uller},\ and\ \citenamefont
  {Bechstedt}}]{Rodl/etal:2008}%
  \BibitemOpen
  \bibfield  {author} {\bibinfo {author} {\bibnamefont {R\"odl}, \bibfnamefont
  {C}}, \bibinfo {author} {\bibfnamefont {F.}~\bibnamefont {Fuchs}}, \bibinfo
  {author} {\bibfnamefont {J.}~\bibnamefont {Furthm\"uller}}, \ and\ \bibinfo
  {author} {\bibfnamefont {F.}~\bibnamefont {Bechstedt}}} (\bibinfo {year}
  {2008}),\ \bibfield  {title} {\enquote {\bibinfo {title} {{A}b initio theory
  of excitons and optical properties for spin-polarized systems: {A}pplication
  to antiferromagnetic {M}n{O}},}\ }\href@noop {} {\bibfield  {journal}
  {\bibinfo  {journal} {Phys. Rev. B}\ }\textbf {\bibinfo {volume} {77}},\
  \bibinfo {pages} {184408}}\BibitemShut {NoStop}%
\bibitem [{\citenamefont {R{\"o}dl}\ \emph {et~al.}(2009)\citenamefont
  {R{\"o}dl}, \citenamefont {Furthm{\"u}ller},\ and\ \citenamefont
  {Bechstedt}}]{Roedl/etal:2009}%
  \BibitemOpen
  \bibfield  {author} {\bibinfo {author} {\bibnamefont {R{\"o}dl},
  \bibfnamefont {C}}, \bibinfo {author} {\bibfnamefont {J.}~\bibnamefont
  {Furthm{\"u}ller}}, \ and\ \bibinfo {author} {\bibfnamefont {F.}~\bibnamefont
  {Bechstedt}}} (\bibinfo {year} {2009}),\ \bibfield  {title} {\enquote
  {\bibinfo {title} {{Q}uasiparticle band structures of the antiferromagnetic
  transition-metal oxides {M}n{O}, {F}e{O}, {C}o{O}, and {N}i{O}},}\
  }\href@noop {} {\bibfield  {journal} {\bibinfo  {journal} {Phys.\ Rev.\ B}\
  }\textbf {\bibinfo {volume} {79}},\ \bibinfo {pages} {235114}}\BibitemShut
  {NoStop}%
\bibitem [{\citenamefont {R\"odl}\ \emph {et~al.}(2017)\citenamefont {R\"odl},
  \citenamefont {Ruotsalainen}, \citenamefont {Sottile}, \citenamefont
  {Honkanen}, \citenamefont {Ablett}, \citenamefont {Rueff}, \citenamefont
  {Sirotti}, \citenamefont {Verbeni}, \citenamefont {Al-Zein}, \citenamefont
  {Reining},\ and\ \citenamefont {Huotari}}]{Roedl/etal:2017}%
  \BibitemOpen
  \bibfield  {author} {\bibinfo {author} {\bibnamefont {R\"odl}, \bibfnamefont
  {C}}, \bibinfo {author} {\bibfnamefont {K.~O.}\ \bibnamefont {Ruotsalainen}},
  \bibinfo {author} {\bibfnamefont {F.}~\bibnamefont {Sottile}}, \bibinfo
  {author} {\bibfnamefont {A.-P.}\ \bibnamefont {Honkanen}}, \bibinfo {author}
  {\bibfnamefont {J.~M.}\ \bibnamefont {Ablett}}, \bibinfo {author}
  {\bibfnamefont {J.-P.}\ \bibnamefont {Rueff}}, \bibinfo {author}
  {\bibfnamefont {F.}~\bibnamefont {Sirotti}}, \bibinfo {author} {\bibfnamefont
  {R.}~\bibnamefont {Verbeni}}, \bibinfo {author} {\bibfnamefont
  {A.}~\bibnamefont {Al-Zein}}, \bibinfo {author} {\bibfnamefont
  {L.}~\bibnamefont {Reining}}, \ and\ \bibinfo {author} {\bibfnamefont
  {S.}~\bibnamefont {Huotari}}} (\bibinfo {year} {2017}),\ \bibfield  {title}
  {\enquote {\bibinfo {title} {{L}ow-energy electronic excitations and band-gap
  renormalization in {C}u{O}},}\ }\href@noop {} {\bibfield  {journal} {\bibinfo
   {journal} {Phys. Rev. B}\ }\textbf {\bibinfo {volume} {95}},\ \bibinfo
  {pages} {195142}}\BibitemShut {NoStop}%
\bibitem [{\citenamefont {R\"odl}\ \emph {et~al.}(2015)\citenamefont {R\"odl},
  \citenamefont {Sottile},\ and\ \citenamefont
  {Reining}}]{Roedl/Sottile/Reining:2015}%
  \BibitemOpen
  \bibfield  {author} {\bibinfo {author} {\bibnamefont {R\"odl}, \bibfnamefont
  {C}}, \bibinfo {author} {\bibfnamefont {F.}~\bibnamefont {Sottile}}, \ and\
  \bibinfo {author} {\bibfnamefont {L.}~\bibnamefont {Reining}}} (\bibinfo
  {year} {2015}),\ \bibfield  {title} {\enquote {\bibinfo {title}
  {{Q}uasiparticle excitations in the photoemission spectrum of {C}u{O} from
  first principles: {A} ${G}{W}$ study},}\ }\href {\doibase
  10.1103/PhysRevB.91.045102} {\bibfield  {journal} {\bibinfo  {journal} {Phys.
  Rev. B}\ }\textbf {\bibinfo {volume} {91}},\ \bibinfo {pages}
  {045102}}\BibitemShut {NoStop}%
\bibitem [{\citenamefont {Rohlfing}(2001)}]{Rohlfing:2001}%
  \BibitemOpen
  \bibfield  {author} {\bibinfo {author} {\bibnamefont {Rohlfing},
  \bibfnamefont {M}}} (\bibinfo {year} {2001}),\ \bibfield  {title} {\enquote
  {\bibinfo {title} {{Q}uasiparticle spectrum and optical excitations of
  semiconductor surfaces.}}\ }\href@noop {} {\bibfield  {journal} {\bibinfo
  {journal} {Appl. Phys. A}\ }\textbf {\bibinfo {volume} {72}}~(\bibinfo
  {number} {4}),\ \bibinfo {pages} {413}}\BibitemShut {NoStop}%
\bibitem [{\citenamefont {Rohlfing}\ \emph
  {et~al.}(1995{\natexlab{a}})\citenamefont {Rohlfing}, \citenamefont
  {Kr\"uger},\ and\ \citenamefont {Pollmann}}]{Rohlfing/Krueger/Pollmann:1995}%
  \BibitemOpen
  \bibfield  {author} {\bibinfo {author} {\bibnamefont {Rohlfing},
  \bibfnamefont {M}}, \bibinfo {author} {\bibfnamefont {P.}~\bibnamefont
  {Kr\"uger}}, \ and\ \bibinfo {author} {\bibfnamefont {J.}~\bibnamefont
  {Pollmann}}} (\bibinfo {year} {1995}{\natexlab{a}}),\ \bibfield  {title}
  {\enquote {\bibinfo {title} {{E}fficient scheme for ${G}{W}$ quasiparticle
  band-structure calculations with applications to bulk {S}i and to the
  {S}i(001)-(2x1) surface},}\ }\href {\doibase 10.1103/PhysRevB.52.1905}
  {\bibfield  {journal} {\bibinfo  {journal} {Phys. Rev. B}\ }\textbf {\bibinfo
  {volume} {52}}~(\bibinfo {number} {3}),\ \bibinfo {pages}
  {1905--1917}}\BibitemShut {NoStop}%
\bibitem [{\citenamefont {Rohlfing}\ \emph
  {et~al.}(1995{\natexlab{b}})\citenamefont {Rohlfing}, \citenamefont
  {Kr\"uger},\ and\ \citenamefont
  {Pollmann}}]{Rohlfing/Krueger/Pollmann:1995_CdS}%
  \BibitemOpen
  \bibfield  {author} {\bibinfo {author} {\bibnamefont {Rohlfing},
  \bibfnamefont {M}}, \bibinfo {author} {\bibfnamefont {P.}~\bibnamefont
  {Kr\"uger}}, \ and\ \bibinfo {author} {\bibfnamefont {J.}~\bibnamefont
  {Pollmann}}} (\bibinfo {year} {1995}{\natexlab{b}}),\ \bibfield  {title}
  {\enquote {\bibinfo {title} {{Q}uasiparticle {B}and {S}tructure of
  {C}d{S}},}\ }\href@noop {} {\bibfield  {journal} {\bibinfo  {journal} {Phys.
  Rev. Lett.}\ }\textbf {\bibinfo {volume} {75}},\ \bibinfo {pages}
  {3489--3492}}\BibitemShut {NoStop}%
\bibitem [{\citenamefont {Rohlfing}\ \emph {et~al.}(1996)\citenamefont
  {Rohlfing}, \citenamefont {Kr\"uger},\ and\ \citenamefont
  {Pollmann}}]{Rohlfing/Krueger/Pollmann:1996}%
  \BibitemOpen
  \bibfield  {author} {\bibinfo {author} {\bibnamefont {Rohlfing},
  \bibfnamefont {M}}, \bibinfo {author} {\bibfnamefont {P.}~\bibnamefont
  {Kr\"uger}}, \ and\ \bibinfo {author} {\bibfnamefont {J.}~\bibnamefont
  {Pollmann}}} (\bibinfo {year} {1996}),\ \bibfield  {title} {\enquote
  {\bibinfo {title} {{Q}uasiparticle band structures of clean, hydrogen-, and
  sulfur-terminated {G}e(001) surfaces},}\ }\href@noop {} {\bibfield  {journal}
  {\bibinfo  {journal} {Phys. Rev. B}\ }\textbf {\bibinfo {volume} {54}},\
  \bibinfo {pages} {13759--13766}}\BibitemShut {NoStop}%
\bibitem [{\citenamefont {Rohlfing}\ \emph
  {et~al.}(1997{\natexlab{a}})\citenamefont {Rohlfing}, \citenamefont
  {Kr\"uger},\ and\ \citenamefont
  {Pollmann}}]{Rohlfing/Krueger/Pollmann:1997_semicore}%
  \BibitemOpen
  \bibfield  {author} {\bibinfo {author} {\bibnamefont {Rohlfing},
  \bibfnamefont {M}}, \bibinfo {author} {\bibfnamefont {P.}~\bibnamefont
  {Kr\"uger}}, \ and\ \bibinfo {author} {\bibfnamefont {J.}~\bibnamefont
  {Pollmann}}} (\bibinfo {year} {1997}{\natexlab{a}}),\ \bibfield  {title}
  {\enquote {\bibinfo {title} {{Q}uasiparticle calculations of semicore states
  in {S}i, {G}e, and {C}d{S}},}\ }\href@noop {} {\bibfield  {journal} {\bibinfo
   {journal} {Phys. Rev. B}\ }\textbf {\bibinfo {volume} {56}},\ \bibinfo
  {pages} {R7065--R7068}}\BibitemShut {NoStop}%
\bibitem [{\citenamefont {Rohlfing}\ \emph
  {et~al.}(1997{\natexlab{b}})\citenamefont {Rohlfing}, \citenamefont
  {Kr\"uger},\ and\ \citenamefont {Pollmann}}]{Rohlfing/Krueger/Pollmann:1997}%
  \BibitemOpen
  \bibfield  {author} {\bibinfo {author} {\bibnamefont {Rohlfing},
  \bibfnamefont {M}}, \bibinfo {author} {\bibfnamefont {P.}~\bibnamefont
  {Kr\"uger}}, \ and\ \bibinfo {author} {\bibfnamefont {J.}~\bibnamefont
  {Pollmann}}} (\bibinfo {year} {1997}{\natexlab{b}}),\ \bibfield  {title}
  {\enquote {\bibinfo {title} {{Q}uasiparticle calculations of surface
  core-level shifts},}\ }\href@noop {} {\bibfield  {journal} {\bibinfo
  {journal} {Phys. Rev. B}\ }\textbf {\bibinfo {volume} {56}},\ \bibinfo
  {pages} {2191--2197}}\BibitemShut {NoStop}%
\bibitem [{\citenamefont {Rohlfing}\ \emph {et~al.}(1998)\citenamefont
  {Rohlfing}, \citenamefont {Kr\"uger},\ and\ \citenamefont
  {Pollmann}}]{Rohlfing/Krueger/Pollmann:1998}%
  \BibitemOpen
  \bibfield  {author} {\bibinfo {author} {\bibnamefont {Rohlfing},
  \bibfnamefont {M}}, \bibinfo {author} {\bibfnamefont {P.}~\bibnamefont
  {Kr\"uger}}, \ and\ \bibinfo {author} {\bibfnamefont {J.}~\bibnamefont
  {Pollmann}}} (\bibinfo {year} {1998}),\ \bibfield  {title} {\enquote
  {\bibinfo {title} {{R}ole of semicore $d$ electrons in quasiparticle
  band-structure calculations},}\ }\href {\doibase 10.1103/PhysRevB.57.6485}
  {\bibfield  {journal} {\bibinfo  {journal} {Phys. Rev. B}\ }\textbf {\bibinfo
  {volume} {57}},\ \bibinfo {pages} {6485--6492}}\BibitemShut {NoStop}%
\bibitem [{\citenamefont {Rohlfing}\ and\ \citenamefont
  {Louie}(1998)}]{Rohlfing/etal:1998}%
  \BibitemOpen
  \bibfield  {author} {\bibinfo {author} {\bibnamefont {Rohlfing},
  \bibfnamefont {M}}, \ and\ \bibinfo {author} {\bibfnamefont {S.~G.}\
  \bibnamefont {Louie}}} (\bibinfo {year} {1998}),\ \bibfield  {title}
  {\enquote {\bibinfo {title} {{E}lectron-{H}ole {E}xcitations in
  {S}emiconductors and {I}nsulators},}\ }\href@noop {} {\bibfield  {journal}
  {\bibinfo  {journal} {Phys. Rev. Lett.}\ }\textbf {\bibinfo {volume} {81}},\
  \bibinfo {pages} {2312--2315}}\BibitemShut {NoStop}%
\bibitem [{\citenamefont {Rohlfing}\ and\ \citenamefont
  {Louie}(1999)}]{Rohlfing/Louie:1999}%
  \BibitemOpen
  \bibfield  {author} {\bibinfo {author} {\bibnamefont {Rohlfing},
  \bibfnamefont {M}}, \ and\ \bibinfo {author} {\bibfnamefont {S.~G.}\
  \bibnamefont {Louie}}} (\bibinfo {year} {1999}),\ \bibfield  {title}
  {\enquote {\bibinfo {title} {{E}xcitons and {O}ptical {S}pectrum of the
  $\mathrm{Si}(111)\ensuremath{-}(2\ifmmode\times\else\texttimes\fi{}1)$
  {S}urface},}\ }\href@noop {} {\bibfield  {journal} {\bibinfo  {journal}
  {Phys. Rev. Lett.}\ }\textbf {\bibinfo {volume} {83}},\ \bibinfo {pages}
  {856--859}}\BibitemShut {NoStop}%
\bibitem [{\citenamefont {Rohlfing}\ and\ \citenamefont
  {Louie}(2000)}]{Rohlfing/etal:2000}%
  \BibitemOpen
  \bibfield  {author} {\bibinfo {author} {\bibnamefont {Rohlfing},
  \bibfnamefont {M}}, \ and\ \bibinfo {author} {\bibfnamefont {S.~G.}\
  \bibnamefont {Louie}}} (\bibinfo {year} {2000}),\ \bibfield  {title}
  {\enquote {\bibinfo {title} {{E}lectron-hole excitations and optical spectra
  from first principles},}\ }\href@noop {} {\bibfield  {journal} {\bibinfo
  {journal} {Phys. Rev. B}\ }\textbf {\bibinfo {volume} {62}},\ \bibinfo
  {pages} {4927--4944}}\BibitemShut {NoStop}%
\bibitem [{\citenamefont {Rohlfing}\ \emph {et~al.}(2003)\citenamefont
  {Rohlfing}, \citenamefont {Wang}, \citenamefont {Kr\"uger},\ and\
  \citenamefont {Pollmann}}]{Rohlfing/Wang/Krueger/Pollmann:2003}%
  \BibitemOpen
  \bibfield  {author} {\bibinfo {author} {\bibnamefont {Rohlfing},
  \bibfnamefont {M}}, \bibinfo {author} {\bibfnamefont {N.-P.}\ \bibnamefont
  {Wang}}, \bibinfo {author} {\bibfnamefont {P.}~\bibnamefont {Kr\"uger}}, \
  and\ \bibinfo {author} {\bibfnamefont {J.}~\bibnamefont {Pollmann}}}
  (\bibinfo {year} {2003}),\ \bibfield  {title} {\enquote {\bibinfo {title}
  {{I}mage {S}tates and {E}xcitons at {I}nsulator {S}urfaces with {N}egative
  {E}lectron {A}ffinity},}\ }\href@noop {} {\bibfield  {journal} {\bibinfo
  {journal} {Phys.\ Rev.\ Lett.}\ }\textbf {\bibinfo {volume} {91}},\ \bibinfo
  {pages} {256802}}\BibitemShut {NoStop}%
\bibitem [{\citenamefont {Rojas}\ \emph {et~al.}(1995)\citenamefont {Rojas},
  \citenamefont {Godby},\ and\ \citenamefont {Needs}}]{Rojas/Godby/Needs:1995}%
  \BibitemOpen
  \bibfield  {author} {\bibinfo {author} {\bibnamefont {Rojas}, \bibfnamefont
  {H~N}}, \bibinfo {author} {\bibfnamefont {R.~W.}\ \bibnamefont {Godby}}, \
  and\ \bibinfo {author} {\bibfnamefont {R.~J.}\ \bibnamefont {Needs}}}
  (\bibinfo {year} {1995}),\ \bibfield  {title} {\enquote {\bibinfo {title}
  {{S}pace-{T}ime {M}ethod for {\it {a}b {i}nitio} {C}alculations of
  {S}elf-{E}nergies and {D}ielectric {R}esponse {F}unctions of {S}olids},}\
  }\href@noop {} {\bibfield  {journal} {\bibinfo  {journal} {Phys. Rev. Lett.}\
  }\textbf {\bibinfo {volume} {74}},\ \bibinfo {pages} {1827}}\BibitemShut
  {NoStop}%
\bibitem [{\citenamefont {Romaniello}\ \emph {et~al.}(2012)\citenamefont
  {Romaniello}, \citenamefont {Bechstedt},\ and\ \citenamefont
  {Reining}}]{Romaniello/etal:2012}%
  \BibitemOpen
  \bibfield  {author} {\bibinfo {author} {\bibnamefont {Romaniello},
  \bibfnamefont {P}}, \bibinfo {author} {\bibfnamefont {F.}~\bibnamefont
  {Bechstedt}}, \ and\ \bibinfo {author} {\bibfnamefont {L.}~\bibnamefont
  {Reining}}} (\bibinfo {year} {2012}),\ \bibfield  {title} {\enquote {\bibinfo
  {title} {{B}eyond the ${G}{W}$ approximation: {C}ombining correlation
  channels},}\ }\href {\doibase 10.1103/PhysRevB.85.155131} {\bibfield
  {journal} {\bibinfo  {journal} {Phys. Rev. B}\ }\textbf {\bibinfo {volume}
  {85}},\ \bibinfo {pages} {155131}}\BibitemShut {NoStop}%
\bibitem [{\citenamefont {Romaniello}\ \emph {et~al.}(2009)\citenamefont
  {Romaniello}, \citenamefont {Guyot},\ and\ \citenamefont
  {Reining}}]{Romaniello/Guyot/Reining:2009}%
  \BibitemOpen
  \bibfield  {author} {\bibinfo {author} {\bibnamefont {Romaniello},
  \bibfnamefont {P}}, \bibinfo {author} {\bibfnamefont {S.}~\bibnamefont
  {Guyot}}, \ and\ \bibinfo {author} {\bibfnamefont {L.}~\bibnamefont
  {Reining}}} (\bibinfo {year} {2009}),\ \bibfield  {title} {\enquote {\bibinfo
  {title} {{T}he self-energy beyond ${G}{W}$: {L}ocal and nonlocal vertex
  corrections},}\ }\href {\doibase 10.1063/1.3249965} {\bibfield  {journal}
  {\bibinfo  {journal} {J. Chem. Phys.}\ }\textbf {\bibinfo {volume}
  {131}}~(\bibinfo {number} {15}),\ \bibinfo {eid} {154111}}\BibitemShut
  {NoStop}%
\bibitem [{\citenamefont {Rossi}\ and\ \citenamefont
  {Werner}(2015)}]{Rossi_2015}%
  \BibitemOpen
  \bibfield  {author} {\bibinfo {author} {\bibnamefont {Rossi}, \bibfnamefont
  {R}}, \ and\ \bibinfo {author} {\bibfnamefont {F.}~\bibnamefont {Werner}}}
  (\bibinfo {year} {2015}),\ \bibfield  {title} {\enquote {\bibinfo {title}
  {{S}keleton series and multivaluedness of the self-energy functional in zero
  space-time dimensions},}\ }\href {\doibase 10.1088/1751-8113/48/48/485202}
  {\bibfield  {journal} {\bibinfo  {journal} {J. Phys. A: Math. Theor.}\
  }\textbf {\bibinfo {volume} {48}}~(\bibinfo {number} {48}),\ \bibinfo {pages}
  {485202}}\BibitemShut {NoStop}%
\bibitem [{\citenamefont {Rostgaard}(2009)}]{Rostgaard2009}%
  \BibitemOpen
  \bibfield  {author} {\bibinfo {author} {\bibnamefont {Rostgaard},
  \bibfnamefont {C}}} (\bibinfo {year} {2009}),\ \bibfield  {title} {\enquote
  {\bibinfo {title} {{T}he {P}rojector {A}ugmented-wave {M}ethod},}\
  }\href@noop {} {\bibinfo  {journal} {arXiv:0910.1921}\ }\BibitemShut
  {NoStop}%
\bibitem [{\citenamefont {Rostgaard}\ \emph {et~al.}(2010)\citenamefont
  {Rostgaard}, \citenamefont {Jacobsen},\ and\ \citenamefont
  {Thygesen}}]{Rostgaard/Jacobsen/Thygesen:2010}%
  \BibitemOpen
\bibfield  {journal} {  }\bibfield  {author} {\bibinfo {author} {\bibnamefont
  {Rostgaard}, \bibfnamefont {C}}, \bibinfo {author} {\bibfnamefont {K.~W.}\
  \bibnamefont {Jacobsen}}, \ and\ \bibinfo {author} {\bibfnamefont {K.~S.}\
  \bibnamefont {Thygesen}}} (\bibinfo {year} {2010}),\ \bibfield  {title}
  {\enquote {\bibinfo {title} {{F}ully self-consistent ${G}{W}$ calculations
  for molecules},}\ }\href@noop {} {\bibfield  {journal} {\bibinfo  {journal}
  {Phys. Rev. B}\ }\textbf {\bibinfo {volume} {81}},\ \bibinfo {pages}
  {085103}}\BibitemShut {NoStop}%
\bibitem [{\citenamefont {Rozzi}\ \emph {et~al.}(2006)\citenamefont {Rozzi},
  \citenamefont {Varsano}, \citenamefont {Marini}, \citenamefont {Gross},\ and\
  \citenamefont {Rubio}}]{Rozzi/etal:2006}%
  \BibitemOpen
  \bibfield  {author} {\bibinfo {author} {\bibnamefont {Rozzi}, \bibfnamefont
  {C~A}}, \bibinfo {author} {\bibfnamefont {D.}~\bibnamefont {Varsano}},
  \bibinfo {author} {\bibfnamefont {A.}~\bibnamefont {Marini}}, \bibinfo
  {author} {\bibfnamefont {E.~K.~U.}\ \bibnamefont {Gross}}, \ and\ \bibinfo
  {author} {\bibfnamefont {A.}~\bibnamefont {Rubio}}} (\bibinfo {year}
  {2006}),\ \bibfield  {title} {\enquote {\bibinfo {title} {{E}xact {C}oulomb
  cutoff technique for supercell calculations},}\ }\href@noop {} {\bibfield
  {journal} {\bibinfo  {journal} {Phys. Rev. B}\ }\textbf {\bibinfo {volume}
  {73}},\ \bibinfo {pages} {205119}}\BibitemShut {NoStop}%
\bibitem [{\citenamefont {Rudenko}\ \emph {et~al.}(2015)\citenamefont
  {Rudenko}, \citenamefont {Yuan},\ and\ \citenamefont
  {Katsnelson}}]{rudenko_prb_92}%
  \BibitemOpen
  \bibfield  {author} {\bibinfo {author} {\bibnamefont {Rudenko}, \bibfnamefont
  {A~N}}, \bibinfo {author} {\bibfnamefont {S.}~\bibnamefont {Yuan}}, \ and\
  \bibinfo {author} {\bibfnamefont {M.~I.}\ \bibnamefont {Katsnelson}}}
  (\bibinfo {year} {2015}),\ \bibfield  {title} {\enquote {\bibinfo {title}
  {{T}oward a realistic description of multilayer black phosphorus: {F}rom
  ${G}{W}$ approximation to large-scale tight-binding simulations},}\
  }\href@noop {} {\bibfield  {journal} {\bibinfo  {journal} {Phys. Rev. B}\
  }\textbf {\bibinfo {volume} {92}},\ \bibinfo {pages} {085419}}\BibitemShut
  {NoStop}%
\bibitem [{\citenamefont {Ruzsinszky}\ \emph {et~al.}(2006)\citenamefont
  {Ruzsinszky}, \citenamefont {Perdew}, \citenamefont {Csonka}, \citenamefont
  {Vydrov},\ and\ \citenamefont {Scuseria}}]{Ruzsinszky2006}%
  \BibitemOpen
  \bibfield  {author} {\bibinfo {author} {\bibnamefont {Ruzsinszky},
  \bibfnamefont {A}}, \bibinfo {author} {\bibfnamefont {J.~P.}\ \bibnamefont
  {Perdew}}, \bibinfo {author} {\bibfnamefont {G.~I.}\ \bibnamefont {Csonka}},
  \bibinfo {author} {\bibfnamefont {O.~A.}\ \bibnamefont {Vydrov}}, \ and\
  \bibinfo {author} {\bibfnamefont {G.~E.}\ \bibnamefont {Scuseria}}} (\bibinfo
  {year} {2006}),\ \bibfield  {title} {\enquote {\bibinfo {title} {{S}purious
  fractional charge on dissociated atoms: {P}ervasive and resilient
  self-interaction error of common density functionals},}\ }\href@noop {}
  {\bibfield  {journal} {\bibinfo  {journal} {J. Chem. Phys.}\ }\textbf
  {\bibinfo {volume} {125}}~(\bibinfo {number} {19}),\ \bibinfo {pages}
  {194112}}\BibitemShut {NoStop}%
\bibitem [{\citenamefont {Sakuma}\ \emph {et~al.}(2011)\citenamefont {Sakuma},
  \citenamefont {Friedrich}, \citenamefont {Miyake}, \citenamefont {Bl\"ugel},\
  and\ \citenamefont {Aryasetiawan}}]{Sakuma/etal:2011}%
  \BibitemOpen
  \bibfield  {author} {\bibinfo {author} {\bibnamefont {Sakuma}, \bibfnamefont
  {R}}, \bibinfo {author} {\bibfnamefont {C.}~\bibnamefont {Friedrich}},
  \bibinfo {author} {\bibfnamefont {T.}~\bibnamefont {Miyake}}, \bibinfo
  {author} {\bibfnamefont {S.}~\bibnamefont {Bl\"ugel}}, \ and\ \bibinfo
  {author} {\bibfnamefont {F.}~\bibnamefont {Aryasetiawan}}} (\bibinfo {year}
  {2011}),\ \bibfield  {title} {\enquote {\bibinfo {title} {${G}{W}$
  calculations including spin-orbit coupling: {A}pplication to {H}g
  chalcogenides},}\ }\href {\doibase 10.1103/PhysRevB.84.085144} {\bibfield
  {journal} {\bibinfo  {journal} {Phys. Rev. B}\ }\textbf {\bibinfo {volume}
  {84}},\ \bibinfo {pages} {085144}}\BibitemShut {NoStop}%
\bibitem [{\citenamefont {Sakuma}\ \emph {et~al.}(2012)\citenamefont {Sakuma},
  \citenamefont {Miyake},\ and\ \citenamefont
  {Aryasetiawan}}]{Sakuma/etal:2012}%
  \BibitemOpen
  \bibfield  {author} {\bibinfo {author} {\bibnamefont {Sakuma}, \bibfnamefont
  {R}}, \bibinfo {author} {\bibfnamefont {T.}~\bibnamefont {Miyake}}, \ and\
  \bibinfo {author} {\bibfnamefont {F.}~\bibnamefont {Aryasetiawan}}} (\bibinfo
  {year} {2012}),\ \bibfield  {title} {\enquote {\bibinfo {title}
  {{S}elf-energy and spectral function of {C}e within the $\mathit{GW}$
  approximation},}\ }\href@noop {} {\bibfield  {journal} {\bibinfo  {journal}
  {Phys. Rev. B}\ }\textbf {\bibinfo {volume} {86}},\ \bibinfo {pages}
  {245126}}\BibitemShut {NoStop}%
\bibitem [{\citenamefont {Sakuma}\ \emph {et~al.}(2013)\citenamefont {Sakuma},
  \citenamefont {Werner},\ and\ \citenamefont
  {Aryasetiawan}}]{Sakuma/Werner/Aryasetiawan:2013}%
  \BibitemOpen
  \bibfield  {author} {\bibinfo {author} {\bibnamefont {Sakuma}, \bibfnamefont
  {R}}, \bibinfo {author} {\bibfnamefont {P.}~\bibnamefont {Werner}}, \ and\
  \bibinfo {author} {\bibfnamefont {F.}~\bibnamefont {Aryasetiawan}}} (\bibinfo
  {year} {2013}),\ \bibfield  {title} {\enquote {\bibinfo {title} {{E}lectronic
  structure of {S}r{V}{O}$_3$ within ${G}{W}$+{D}{M}{F}{T}},}\ }\href@noop {}
  {\bibfield  {journal} {\bibinfo  {journal} {Phys. Rev. B}\ }\textbf {\bibinfo
  {volume} {88}},\ \bibinfo {pages} {235110}}\BibitemShut {NoStop}%
\bibitem [{\citenamefont {Salpeter}\ and\ \citenamefont
  {Bethe}(1951)}]{Salpeter/etal:1951}%
  \BibitemOpen
  \bibfield  {author} {\bibinfo {author} {\bibnamefont {Salpeter},
  \bibfnamefont {E~E}}, \ and\ \bibinfo {author} {\bibfnamefont {H.~A.}\
  \bibnamefont {Bethe}}} (\bibinfo {year} {1951}),\ \bibfield  {title}
  {\enquote {\bibinfo {title} {{A} {R}elativistic {E}quation for
  {B}ound-{S}tate {P}roblems},}\ }\href@noop {} {\bibfield  {journal} {\bibinfo
   {journal} {Phys. Rev.}\ }\textbf {\bibinfo {volume} {84}},\ \bibinfo {pages}
  {1232--1242}}\BibitemShut {NoStop}%
\bibitem [{\citenamefont {Sato}\ \emph {et~al.}(1981)\citenamefont {Sato},
  \citenamefont {Seki},\ and\ \citenamefont {Inokuchi}}]{Sato1981}%
  \BibitemOpen
  \bibfield  {author} {\bibinfo {author} {\bibnamefont {Sato}, \bibfnamefont
  {N}}, \bibinfo {author} {\bibfnamefont {K.}~\bibnamefont {Seki}}, \ and\
  \bibinfo {author} {\bibfnamefont {H.}~\bibnamefont {Inokuchi}}} (\bibinfo
  {year} {1981}),\ \bibfield  {title} {\enquote {\bibinfo {title}
  {{P}olarization energies of organic solids determined by ultraviolet
  photoelectron spectroscopy},}\ }\href {\doibase 10.1039/F29817701621}
  {\bibfield  {journal} {\bibinfo  {journal} {J. Chem. Soc.{,} Faraday Trans.
  2}\ }\textbf {\bibinfo {volume} {77}},\ \bibinfo {pages}
  {1621--1633}}\BibitemShut {NoStop}%
\bibitem [{\citenamefont {Scherpelz}\ \emph {et~al.}(2016)\citenamefont
  {Scherpelz}, \citenamefont {Govoni}, \citenamefont {Hamada},\ and\
  \citenamefont {Galli}}]{Scherpelz/etal:2016}%
  \BibitemOpen
  \bibfield  {author} {\bibinfo {author} {\bibnamefont {Scherpelz},
  \bibfnamefont {P}}, \bibinfo {author} {\bibfnamefont {M.}~\bibnamefont
  {Govoni}}, \bibinfo {author} {\bibfnamefont {I.}~\bibnamefont {Hamada}}, \
  and\ \bibinfo {author} {\bibfnamefont {G.}~\bibnamefont {Galli}}} (\bibinfo
  {year} {2016}),\ \bibfield  {title} {\enquote {\bibinfo {title}
  {{I}mplementation and {V}alidation of {F}ully {R}elativistic ${G}{W}$
  {C}alculations: {S}pin–{O}rbit {C}oupling in {M}olecules, {N}anocrystals,
  and {S}olids},}\ }\href@noop {} {\bibfield  {journal} {\bibinfo  {journal}
  {J. Chem. Theory Comput.}\ }\textbf {\bibinfo {volume} {12}}~(\bibinfo
  {number} {8}),\ \bibinfo {pages} {3523--3544}}\BibitemShut {NoStop}%
\bibitem [{\citenamefont {van Schilfgaarde}\ \emph {et~al.}(2006)\citenamefont
  {van Schilfgaarde}, \citenamefont {Kotani},\ and\ \citenamefont
  {Faleev}}]{Schilfgaarde/Kotani/Faleev:2006}%
  \BibitemOpen
  \bibfield  {author} {\bibinfo {author} {\bibnamefont {van Schilfgaarde},
  \bibfnamefont {M}}, \bibinfo {author} {\bibfnamefont {T.}~\bibnamefont
  {Kotani}}, \ and\ \bibinfo {author} {\bibfnamefont {S.}~\bibnamefont
  {Faleev}}} (\bibinfo {year} {2006}),\ \bibfield  {title} {\enquote {\bibinfo
  {title} {{Q}uasiparticle {S}elf-consistent ${G}{W}$ theory},}\ }\href@noop {}
  {\bibfield  {journal} {\bibinfo  {journal} {Phys. Rev. Lett.}\ }\textbf
  {\bibinfo {volume} {96}},\ \bibinfo {pages} {226402}}\BibitemShut {NoStop}%
\bibitem [{\citenamefont {Schindlmayr}(1997)}]{Schindlmayr:1997}%
  \BibitemOpen
  \bibfield  {author} {\bibinfo {author} {\bibnamefont {Schindlmayr},
  \bibfnamefont {A}}} (\bibinfo {year} {1997}),\ \bibfield  {title} {\enquote
  {\bibinfo {title} {{V}iolation of particle number conservation in the
  ${G}{W}$ approximation},}\ }\href@noop {} {\bibfield  {journal} {\bibinfo
  {journal} {Phys.\ Rev.\ B}\ }\textbf {\bibinfo {volume} {56}},\ \bibinfo
  {pages} {3528}}\BibitemShut {NoStop}%
\bibitem [{\citenamefont {Schleife}\ \emph {et~al.}(2009)\citenamefont
  {Schleife}, \citenamefont {Fuchs}, \citenamefont {R\"odl}, \citenamefont
  {Furthm\"uller},\ and\ \citenamefont {Bechstedt}}]{Schleife/etal:2009}%
  \BibitemOpen
  \bibfield  {author} {\bibinfo {author} {\bibnamefont {Schleife},
  \bibfnamefont {A}}, \bibinfo {author} {\bibfnamefont {F.}~\bibnamefont
  {Fuchs}}, \bibinfo {author} {\bibfnamefont {C.}~\bibnamefont {R\"odl}},
  \bibinfo {author} {\bibfnamefont {J.}~\bibnamefont {Furthm\"uller}}, \ and\
  \bibinfo {author} {\bibfnamefont {F.}~\bibnamefont {Bechstedt}}} (\bibinfo
  {year} {2009}),\ \bibfield  {title} {\enquote {\bibinfo {title}
  {{B}and-structure and optical-transition parameters of wurtzite {M}g{O},
  {Z}n{O}, and {C}d{O} from quasiparticle calculations},}\ }\href@noop {}
  {\bibfield  {journal} {\bibinfo  {journal} {Phys. Status Solidi B}\ }\textbf
  {\bibinfo {volume} {246}}~(\bibinfo {number} {9}),\ \bibinfo {pages}
  {2150--2153}}\BibitemShut {NoStop}%
\bibitem [{\citenamefont {Schleife}\ \emph {et~al.}(2018)\citenamefont
  {Schleife}, \citenamefont {Neumann}, \citenamefont {Esser}, \citenamefont
  {Galazka}, \citenamefont {Gottwald}, \citenamefont {Nixdorf}, \citenamefont
  {Goldhahn},\ and\ \citenamefont {Feneberg}}]{Schleife/etal:2018}%
  \BibitemOpen
  \bibfield  {author} {\bibinfo {author} {\bibnamefont {Schleife},
  \bibfnamefont {A}}, \bibinfo {author} {\bibfnamefont {M.~D.}\ \bibnamefont
  {Neumann}}, \bibinfo {author} {\bibfnamefont {N.}~\bibnamefont {Esser}},
  \bibinfo {author} {\bibfnamefont {Z.}~\bibnamefont {Galazka}}, \bibinfo
  {author} {\bibfnamefont {A.}~\bibnamefont {Gottwald}}, \bibinfo {author}
  {\bibfnamefont {J.}~\bibnamefont {Nixdorf}}, \bibinfo {author} {\bibfnamefont
  {R.}~\bibnamefont {Goldhahn}}, \ and\ \bibinfo {author} {\bibfnamefont
  {M.}~\bibnamefont {Feneberg}}} (\bibinfo {year} {2018}),\ \bibfield  {title}
  {\enquote {\bibinfo {title} {{O}ptical properties of {I}n$_2${O}$_3$ from
  experiment and first-principles theory: influence of lattice screening},}\
  }\href {http://stacks.iop.org/1367-2630/20/i=5/a=053016} {\bibfield
  {journal} {\bibinfo  {journal} {New J. Phys.}\ }\textbf {\bibinfo {volume}
  {20}}~(\bibinfo {number} {5}),\ \bibinfo {pages} {053016}}\BibitemShut
  {NoStop}%
\bibitem [{\citenamefont {Schleife}\ \emph {et~al.}(2011)\citenamefont
  {Schleife}, \citenamefont {Varley}, \citenamefont {Fuchs}, \citenamefont
  {R\"odl}, \citenamefont {Bechstedt}, \citenamefont {Rinke}, \citenamefont
  {Janotti},\ and\ \citenamefont {de~Walle}}]{Schleife/etal:2011}%
  \BibitemOpen
  \bibfield  {author} {\bibinfo {author} {\bibnamefont {Schleife},
  \bibfnamefont {A}}, \bibinfo {author} {\bibfnamefont {J.~B.}\ \bibnamefont
  {Varley}}, \bibinfo {author} {\bibfnamefont {F.}~\bibnamefont {Fuchs}},
  \bibinfo {author} {\bibfnamefont {C.}~\bibnamefont {R\"odl}}, \bibinfo
  {author} {\bibfnamefont {F.}~\bibnamefont {Bechstedt}}, \bibinfo {author}
  {\bibfnamefont {P.}~\bibnamefont {Rinke}}, \bibinfo {author} {\bibfnamefont
  {A.}~\bibnamefont {Janotti}}, \ and\ \bibinfo {author} {\bibfnamefont
  {C.~G.~Van}\ \bibnamefont {de~Walle}}} (\bibinfo {year} {2011}),\ \bibfield
  {title} {\enquote {\bibinfo {title} {{T}in dioxide from first principles:
  {Q}uasiparticle electronic states and optical properties},}\ }\href@noop {}
  {\bibfield  {journal} {\bibinfo  {journal} {Phys. Rev. B}\ }\textbf {\bibinfo
  {volume} {83}},\ \bibinfo {pages} {035116}}\BibitemShut {NoStop}%
\bibitem [{\citenamefont {Schlipf}\ \emph {et~al.}(2019)\citenamefont
  {Schlipf}, \citenamefont {Lambert}, \citenamefont {Zibouche},\ and\
  \citenamefont {Giustino}}]{Schlipf2019}%
  \BibitemOpen
  \bibfield  {author} {\bibinfo {author} {\bibnamefont {Schlipf}, \bibfnamefont
  {M}}, \bibinfo {author} {\bibfnamefont {H.}~\bibnamefont {Lambert}}, \bibinfo
  {author} {\bibfnamefont {N.}~\bibnamefont {Zibouche}}, \ and\ \bibinfo
  {author} {\bibfnamefont {F.}~\bibnamefont {Giustino}}} (\bibinfo {year}
  {2019}),\ \bibfield  {title} {\enquote {\bibinfo {title}
  {{S}ternheimer${G}{W}$: a program for calculating ${G}{W}$ quasiparticle band
  structures and spectral functions without unoccupied states},}\ }\href@noop
  {} {\bibinfo  {journal} {arXiv:1812.03717}\ }\BibitemShut {NoStop}%
\bibitem [{\citenamefont {Schmidt}\ \emph {et~al.}(2017)\citenamefont
  {Schmidt}, \citenamefont {Patrick},\ and\ \citenamefont
  {Thygesen}}]{Thygesen/etal:2017}%
  \BibitemOpen
\bibfield  {journal} {  }\bibfield  {author} {\bibinfo {author} {\bibnamefont
  {Schmidt}, \bibfnamefont {P~S}}, \bibinfo {author} {\bibfnamefont {C.~E.}\
  \bibnamefont {Patrick}}, \ and\ \bibinfo {author} {\bibfnamefont {K.~S.}\
  \bibnamefont {Thygesen}}} (\bibinfo {year} {2017}),\ \bibfield  {title}
  {\enquote {\bibinfo {title} {{S}imple vertex correction improves ${G}{W}$
  band energies of bulk and two-dimensional crystals},}\ }\href@noop {}
  {\bibfield  {journal} {\bibinfo  {journal} {Phys.\ Rev.\ B}\ }\textbf
  {\bibinfo {volume} {96}},\ \bibinfo {pages} {205206}}\BibitemShut {NoStop}%
\bibitem [{\citenamefont {Schmidt}\ \emph {et~al.}(2000)\citenamefont
  {Schmidt}, \citenamefont {Esser}, \citenamefont {Frisch}, \citenamefont
  {Vogt}, \citenamefont {Bernholc}, \citenamefont {Bechstedt}, \citenamefont
  {Zorn}, \citenamefont {Hannappel}, \citenamefont {Visbeck}, \citenamefont
  {Willig},\ and\ \citenamefont {Richter}}]{Schmidt/etal:2000}%
  \BibitemOpen
  \bibfield  {author} {\bibinfo {author} {\bibnamefont {Schmidt}, \bibfnamefont
  {W~G}}, \bibinfo {author} {\bibfnamefont {N.}~\bibnamefont {Esser}}, \bibinfo
  {author} {\bibfnamefont {A.~M.}\ \bibnamefont {Frisch}}, \bibinfo {author}
  {\bibfnamefont {P.}~\bibnamefont {Vogt}}, \bibinfo {author} {\bibfnamefont
  {J.}~\bibnamefont {Bernholc}}, \bibinfo {author} {\bibfnamefont
  {F.}~\bibnamefont {Bechstedt}}, \bibinfo {author} {\bibfnamefont
  {M.}~\bibnamefont {Zorn}}, \bibinfo {author} {\bibfnamefont {T.}~\bibnamefont
  {Hannappel}}, \bibinfo {author} {\bibfnamefont {S.}~\bibnamefont {Visbeck}},
  \bibinfo {author} {\bibfnamefont {F.}~\bibnamefont {Willig}}, \ and\ \bibinfo
  {author} {\bibfnamefont {W.}~\bibnamefont {Richter}}} (\bibinfo {year}
  {2000}),\ \bibfield  {title} {\enquote {\bibinfo {title} {{U}nderstanding
  reflectance anisotropy: {S}urface-state signatures and bulk-related features
  in the optical spectrum of
  $\mathrm{InP}(001)(2\ifmmode\times\else\texttimes\fi{}4)$},}\ }\href@noop {}
  {\bibfield  {journal} {\bibinfo  {journal} {Phys. Rev. B}\ }\textbf {\bibinfo
  {volume} {61}},\ \bibinfo {pages} {R16335--R16338}}\BibitemShut {NoStop}%
\bibitem [{\citenamefont {Schmidt}\ \emph {et~al.}(1999)\citenamefont
  {Schmidt}, \citenamefont {Fattebert}, \citenamefont {Bernholc},\ and\
  \citenamefont {Bechstedt}}]{Schmidt/etal:1999}%
  \BibitemOpen
  \bibfield  {author} {\bibinfo {author} {\bibnamefont {Schmidt}, \bibfnamefont
  {W~G}}, \bibinfo {author} {\bibfnamefont {J.~L.}\ \bibnamefont {Fattebert}},
  \bibinfo {author} {\bibfnamefont {J.}~\bibnamefont {Bernholc}}, \ and\
  \bibinfo {author} {\bibfnamefont {F.}~\bibnamefont {Bechstedt}}} (\bibinfo
  {year} {1999}),\ \bibfield  {title} {\enquote {\bibinfo {title}
  {{S}elf-{E}energy {E}ffects in the {O}ptical {A}nisotropy of {G}a{P}(001)},}\
  }\href@noop {} {\bibfield  {journal} {\bibinfo  {journal} {Surf. Rev. Lett.}\
  }\textbf {\bibinfo {volume} {06}}~(\bibinfo {number} {06}),\ \bibinfo {pages}
  {1159--1165}}\BibitemShut {NoStop}%
\bibitem [{\citenamefont {Sch\"one}\ and\ \citenamefont
  {Eguiluz}(1998)}]{Schoene/Eguiluz:1998}%
  \BibitemOpen
  \bibfield  {author} {\bibinfo {author} {\bibnamefont {Sch\"one},
  \bibfnamefont {W-D}}, \ and\ \bibinfo {author} {\bibfnamefont {A.~G.}\
  \bibnamefont {Eguiluz}}} (\bibinfo {year} {1998}),\ \bibfield  {title}
  {\enquote {\bibinfo {title} {{S}elf-{C}onsistent {C}alculations of
  {Q}uasiparticle {S}tates in {M}etals and {S}emiconductors},}\ }\href@noop {}
  {\bibfield  {journal} {\bibinfo  {journal} {Phys. Rev. Lett.}\ }\textbf
  {\bibinfo {volume} {81}},\ \bibinfo {pages} {1662}}\BibitemShut {NoStop}%
\bibitem [{\citenamefont {Schwinger}(1951)}]{Schwinger:1951_1}%
  \BibitemOpen
  \bibfield  {author} {\bibinfo {author} {\bibnamefont {Schwinger},
  \bibfnamefont {J}}} (\bibinfo {year} {1951}),\ \bibfield  {title} {\enquote
  {\bibinfo {title} {{O}n the {G}reen's functions of quantized fields. {I}},}\
  }\href {\doibase 10.1073/pnas.37.7.452} {\bibfield  {journal} {\bibinfo
  {journal} {Proc. Natl. Acad. Sci. U.S.A.}\ }\textbf {\bibinfo {volume}
  {37}}~(\bibinfo {number} {7}),\ \bibinfo {pages} {452--455}}\BibitemShut
  {NoStop}%
\bibitem [{\citenamefont {van Setten}\ \emph {et~al.}(2015)\citenamefont {van
  Setten}, \citenamefont {Caruso}, \citenamefont {Sharifzadeh}, \citenamefont
  {Ren}, \citenamefont {Scheffler}, \citenamefont {Liu}, \citenamefont
  {Lischner}, \citenamefont {Lin}, \citenamefont {Deslippe}, \citenamefont
  {Louie}, \citenamefont {Yang}, \citenamefont {Weigend}, \citenamefont
  {Neaton}, \citenamefont {Evers},\ and\ \citenamefont {Rinke}}]{Setten2015}%
  \BibitemOpen
  \bibfield  {author} {\bibinfo {author} {\bibnamefont {van Setten},
  \bibfnamefont {M~J}}, \bibinfo {author} {\bibfnamefont {F.}~\bibnamefont
  {Caruso}}, \bibinfo {author} {\bibfnamefont {S.}~\bibnamefont {Sharifzadeh}},
  \bibinfo {author} {\bibfnamefont {X.}~\bibnamefont {Ren}}, \bibinfo {author}
  {\bibfnamefont {M.}~\bibnamefont {Scheffler}}, \bibinfo {author}
  {\bibfnamefont {F.}~\bibnamefont {Liu}}, \bibinfo {author} {\bibfnamefont
  {J.}~\bibnamefont {Lischner}}, \bibinfo {author} {\bibfnamefont
  {L.}~\bibnamefont {Lin}}, \bibinfo {author} {\bibfnamefont {J.~R.}\
  \bibnamefont {Deslippe}}, \bibinfo {author} {\bibfnamefont {S.~G.}\
  \bibnamefont {Louie}}, \bibinfo {author} {\bibfnamefont {C.}~\bibnamefont
  {Yang}}, \bibinfo {author} {\bibfnamefont {F.}~\bibnamefont {Weigend}},
  \bibinfo {author} {\bibfnamefont {J.~B.}\ \bibnamefont {Neaton}}, \bibinfo
  {author} {\bibfnamefont {F.}~\bibnamefont {Evers}}, \ and\ \bibinfo {author}
  {\bibfnamefont {P.}~\bibnamefont {Rinke}}} (\bibinfo {year} {2015}),\
  \bibfield  {title} {\enquote {\bibinfo {title} {${G}{W}$100: {B}enchmarking
  ${G}_0{W}_0$ for {M}olecular {S}ystems},}\ }\href {\doibase
  10.1021/acs.jctc.5b00453} {\bibfield  {journal} {\bibinfo  {journal} {J.
  Chem. Theory Comput.}\ }\textbf {\bibinfo {volume} {11}}~(\bibinfo {number}
  {12}),\ \bibinfo {pages} {5665--5687}}\BibitemShut {NoStop}%
\bibitem [{\citenamefont {van Setten}\ \emph {et~al.}(2018)\citenamefont {van
  Setten}, \citenamefont {Giantomassi}, \citenamefont {Bousquet}, \citenamefont
  {Verstraete}, \citenamefont {Hamann}, \citenamefont {X.},\ and\ \citenamefont
  {Rignanese}}]{VanSetten/etal:2018}%
  \BibitemOpen
  \bibfield  {author} {\bibinfo {author} {\bibnamefont {van Setten},
  \bibfnamefont {M~J}}, \bibinfo {author} {\bibfnamefont {M.}~\bibnamefont
  {Giantomassi}}, \bibinfo {author} {\bibfnamefont {E.}~\bibnamefont
  {Bousquet}}, \bibinfo {author} {\bibfnamefont {M.}~\bibnamefont
  {Verstraete}}, \bibinfo {author} {\bibfnamefont {D.}~\bibnamefont {Hamann}},
  \bibinfo {author} {\bibnamefont {X.}}, \ and\ \bibinfo {author}
  {\bibfnamefont {G.-M.}\ \bibnamefont {Rignanese}}} (\bibinfo {year} {2018}),\
  \bibfield  {title} {\enquote {\bibinfo {title} {{T}he {P}seudo{D}ojo:
  {T}raining and grading a 85 element optimized norm-conserving pseudopotential
  table},}\ }\href@noop {} {\bibfield  {journal} {\bibinfo  {journal} {Comput.
  Phys. Commun.}\ }\textbf {\bibinfo {volume} {226}},\ \bibinfo {pages} {39 --
  54}}\BibitemShut {NoStop}%
\bibitem [{\citenamefont {van Setten}\ \emph {et~al.}(2017)\citenamefont {van
  Setten}, \citenamefont {Giantomassi}, \citenamefont {X.}, \citenamefont
  {Rignanese},\ and\ \citenamefont {Hautier}}]{Setten2017}%
  \BibitemOpen
  \bibfield  {author} {\bibinfo {author} {\bibnamefont {van Setten},
  \bibfnamefont {M~J}}, \bibinfo {author} {\bibfnamefont {M.}~\bibnamefont
  {Giantomassi}}, \bibinfo {author} {\bibnamefont {X.}}, \bibinfo {author}
  {\bibfnamefont {G.-M.}\ \bibnamefont {Rignanese}}, \ and\ \bibinfo {author}
  {\bibfnamefont {G.}~\bibnamefont {Hautier}}} (\bibinfo {year} {2017}),\
  \bibfield  {title} {\enquote {\bibinfo {title} {{A}utomation methodologies
  and large-scale validation for ${G}{W}$: {T}owards high-throughput ${G}{W}$
  calculations},}\ }\href {\doibase 10.1103/PhysRevB.96.155207} {\bibfield
  {journal} {\bibinfo  {journal} {Phys. Rev. B}\ }\textbf {\bibinfo {volume}
  {96}},\ \bibinfo {pages} {155207}}\BibitemShut {NoStop}%
\bibitem [{\citenamefont {van Setten}\ \emph {et~al.}(2013)\citenamefont {van
  Setten}, \citenamefont {Weigend},\ and\ \citenamefont
  {Evers}}]{vanSetten/etal:2013}%
  \BibitemOpen
  \bibfield  {author} {\bibinfo {author} {\bibnamefont {van Setten},
  \bibfnamefont {M~J}}, \bibinfo {author} {\bibfnamefont {F.}~\bibnamefont
  {Weigend}}, \ and\ \bibinfo {author} {\bibfnamefont {F.}~\bibnamefont
  {Evers}}} (\bibinfo {year} {2013}),\ \bibfield  {title} {\enquote {\bibinfo
  {title} {{T}he ${G}{W}$-{M}ethod for {Q}uantum {C}hemistry {A}pplications:
  {T}heory and {I}mplementation},}\ }\href@noop {} {\bibfield  {journal}
  {\bibinfo  {journal} {J. Chem. Theory Comput.}\ }\textbf {\bibinfo {volume}
  {9}}~(\bibinfo {number} {1}),\ \bibinfo {pages} {232--246}}\BibitemShut
  {NoStop}%
\bibitem [{\citenamefont {Shaltaf}\ \emph {et~al.}(2008)\citenamefont
  {Shaltaf}, \citenamefont {Rignanese}, \citenamefont {X.}, \citenamefont
  {Giustino},\ and\ \citenamefont {Pasquarello}}]{Shaltaf/etal:2008}%
  \BibitemOpen
  \bibfield  {author} {\bibinfo {author} {\bibnamefont {Shaltaf}, \bibfnamefont
  {R}}, \bibinfo {author} {\bibfnamefont {G.-M.}\ \bibnamefont {Rignanese}},
  \bibinfo {author} {\bibnamefont {X.}}, \bibinfo {author} {\bibfnamefont
  {F.}~\bibnamefont {Giustino}}, \ and\ \bibinfo {author} {\bibfnamefont
  {A.}~\bibnamefont {Pasquarello}}} (\bibinfo {year} {2008}),\ \bibfield
  {title} {\enquote {\bibinfo {title} {{B}and {O}ffsets at the
  {S}i/{S}i{O}$_{2}$ {I}nterface from {M}any-{B}ody {P}erturbation {T}heory},}\
  }\href {\doibase 10.1103/PhysRevLett.100.186401} {\bibfield  {journal}
  {\bibinfo  {journal} {Phys. Rev. Lett.}\ }\textbf {\bibinfo {volume}
  {100}}~(\bibinfo {number} {18}),\ \bibinfo {pages} {186401}}\BibitemShut
  {NoStop}%
\bibitem [{\citenamefont {Shang}\ \emph {et~al.}(2018)\citenamefont {Shang},
  \citenamefont {Raimbault}, \citenamefont {Rinke}, \citenamefont {Scheffler},
  \citenamefont {Rossi},\ and\ \citenamefont {Carbogno}}]{Shang/etal:2018}%
  \BibitemOpen
  \bibfield  {author} {\bibinfo {author} {\bibnamefont {Shang}, \bibfnamefont
  {H}}, \bibinfo {author} {\bibfnamefont {N.}~\bibnamefont {Raimbault}},
  \bibinfo {author} {\bibfnamefont {P.}~\bibnamefont {Rinke}}, \bibinfo
  {author} {\bibfnamefont {M.}~\bibnamefont {Scheffler}}, \bibinfo {author}
  {\bibfnamefont {M.}~\bibnamefont {Rossi}}, \ and\ \bibinfo {author}
  {\bibfnamefont {C.}~\bibnamefont {Carbogno}}} (\bibinfo {year} {2018}),\
  \bibfield  {title} {\enquote {\bibinfo {title} {{A}ll-electron, real-space
  perturbation theory for homogeneous electric fields: theory, implementation,
  and application within {D}{F}{T}},}\ }\href@noop {} {\bibfield  {journal}
  {\bibinfo  {journal} {New J. Phys.}\ }\textbf {\bibinfo {volume}
  {20}}~(\bibinfo {number} {7}),\ \bibinfo {pages} {073040}}\BibitemShut
  {NoStop}%
\bibitem [{\citenamefont {Sharifzadeh}\ \emph {et~al.}(2012)\citenamefont
  {Sharifzadeh}, \citenamefont {Biller}, \citenamefont {Kronik},\ and\
  \citenamefont {Neaton}}]{Sharifzadeh2012}%
  \BibitemOpen
  \bibfield  {author} {\bibinfo {author} {\bibnamefont {Sharifzadeh},
  \bibfnamefont {S}}, \bibinfo {author} {\bibfnamefont {A.}~\bibnamefont
  {Biller}}, \bibinfo {author} {\bibfnamefont {L.}~\bibnamefont {Kronik}}, \
  and\ \bibinfo {author} {\bibfnamefont {J.~B.}\ \bibnamefont {Neaton}}}
  (\bibinfo {year} {2012}),\ \bibfield  {title} {\enquote {\bibinfo {title}
  {{Q}uasiparticle and optical spectroscopy of the organic semiconductors
  pentacene and {P}{T}{C}{D}{A} from first principles},}\ }\href {\doibase
  10.1103/PhysRevB.85.125307} {\bibfield  {journal} {\bibinfo  {journal} {Phys.
  Rev. B}\ }\textbf {\bibinfo {volume} {85}},\ \bibinfo {pages}
  {125307}}\BibitemShut {NoStop}%
\bibitem [{\citenamefont {Sharifzadeh}\ \emph {et~al.}(2015)\citenamefont
  {Sharifzadeh}, \citenamefont {Wong}, \citenamefont {Wu}, \citenamefont
  {Cotts}, \citenamefont {Kronik}, \citenamefont {Ginsberg},\ and\
  \citenamefont {Neaton}}]{Sharifzadeh2015}%
  \BibitemOpen
  \bibfield  {author} {\bibinfo {author} {\bibnamefont {Sharifzadeh},
  \bibfnamefont {S}}, \bibinfo {author} {\bibfnamefont {C.~Y.}\ \bibnamefont
  {Wong}}, \bibinfo {author} {\bibfnamefont {H.}~\bibnamefont {Wu}}, \bibinfo
  {author} {\bibfnamefont {B.~L.}\ \bibnamefont {Cotts}}, \bibinfo {author}
  {\bibfnamefont {L.}~\bibnamefont {Kronik}}, \bibinfo {author} {\bibfnamefont
  {N.~S.}\ \bibnamefont {Ginsberg}}, \ and\ \bibinfo {author} {\bibfnamefont
  {J.~B.}\ \bibnamefont {Neaton}}} (\bibinfo {year} {2015}),\ \bibfield
  {title} {\enquote {\bibinfo {title} {{R}elating the {P}hysical {S}tructure
  and {O}ptoelectronic {F}unction of {C}rystalline {T}{I}{P}{S}-{P}entacene},}\
  }\href {\doibase 10.1002/adfm.201403005} {\bibfield  {journal} {\bibinfo
  {journal} {Adv. Funct. Mater.}\ }\textbf {\bibinfo {volume} {25}}~(\bibinfo
  {number} {13}),\ \bibinfo {pages} {2038--2046}}\BibitemShut {NoStop}%
\bibitem [{\citenamefont {Shi}\ \emph {et~al.}(2013)\citenamefont {Shi},
  \citenamefont {Pan}, \citenamefont {Zhang},\ and\ \citenamefont
  {Yakobson}}]{Shi/etal:2013}%
  \BibitemOpen
  \bibfield  {author} {\bibinfo {author} {\bibnamefont {Shi}, \bibfnamefont
  {H}}, \bibinfo {author} {\bibfnamefont {H.}~\bibnamefont {Pan}}, \bibinfo
  {author} {\bibfnamefont {Y.-W.}\ \bibnamefont {Zhang}}, \ and\ \bibinfo
  {author} {\bibfnamefont {B.~I.}\ \bibnamefont {Yakobson}}} (\bibinfo {year}
  {2013}),\ \bibfield  {title} {\enquote {\bibinfo {title} {{Q}uasiparticle
  band structures and optical properties of strained monolayer {M}o{S}${}_{2}$
  and {W}{S}${}_{2}$},}\ }\href@noop {} {\bibfield  {journal} {\bibinfo
  {journal} {Phys. Rev. B}\ }\textbf {\bibinfo {volume} {87}},\ \bibinfo
  {pages} {155304}}\BibitemShut {NoStop}%
\bibitem [{\citenamefont {Shih}\ \emph {et~al.}(2010)\citenamefont {Shih},
  \citenamefont {Xue}, \citenamefont {Zhang}, \citenamefont {Cohen},\ and\
  \citenamefont {Louie}}]{Shih2010}%
  \BibitemOpen
  \bibfield  {author} {\bibinfo {author} {\bibnamefont {Shih}, \bibfnamefont
  {B-C}}, \bibinfo {author} {\bibfnamefont {Y.}~\bibnamefont {Xue}}, \bibinfo
  {author} {\bibfnamefont {P.}~\bibnamefont {Zhang}}, \bibinfo {author}
  {\bibfnamefont {M.~L.}\ \bibnamefont {Cohen}}, \ and\ \bibinfo {author}
  {\bibfnamefont {S.~G.}\ \bibnamefont {Louie}}} (\bibinfo {year} {2010}),\
  \bibfield  {title} {\enquote {\bibinfo {title} {{Q}uasiparticle {B}and {G}ap
  of {Z}n{O}: {H}igh {A}ccuracy from the {C}onventional ${G}_{0}{W}_{0}$
  {A}pproach},}\ }\href {\doibase 10.1103/PhysRevLett.105.146401} {\bibfield
  {journal} {\bibinfo  {journal} {Phys. Rev. Lett.}\ }\textbf {\bibinfo
  {volume} {105}},\ \bibinfo {pages} {146401}}\BibitemShut {NoStop}%
\bibitem [{\citenamefont {Shirley}(1996)}]{Shirley:1996}%
  \BibitemOpen
  \bibfield  {author} {\bibinfo {author} {\bibnamefont {Shirley}, \bibfnamefont
  {E~L}}} (\bibinfo {year} {1996}),\ \bibfield  {title} {\enquote {\bibinfo
  {title} {{S}elf-consistent \textit{GW} and higher-order calculations of
  electron states in metals},}\ }\href {\doibase 10.1103/PhysRevB.54.7758}
  {\bibfield  {journal} {\bibinfo  {journal} {Phys. Rev. B}\ }\textbf {\bibinfo
  {volume} {54}},\ \bibinfo {pages} {7758--7764}}\BibitemShut {NoStop}%
\bibitem [{\citenamefont {Shirley}\ and\ \citenamefont
  {Louie}(1993)}]{Shirley1993}%
  \BibitemOpen
  \bibfield  {author} {\bibinfo {author} {\bibnamefont {Shirley}, \bibfnamefont
  {E~L}}, \ and\ \bibinfo {author} {\bibfnamefont {S.~G.}\ \bibnamefont
  {Louie}}} (\bibinfo {year} {1993}),\ \bibfield  {title} {\enquote {\bibinfo
  {title} {{E}lectron excitations in solid {C}$_{60}$: {E}nergy gap, band
  dispersions, and effects of orientational disorder},}\ }\href {\doibase
  10.1103/PhysRevLett.71.133} {\bibfield  {journal} {\bibinfo  {journal} {Phys.
  Rev. Lett.}\ }\textbf {\bibinfo {volume} {71}},\ \bibinfo {pages}
  {133--136}}\BibitemShut {NoStop}%
\bibitem [{\citenamefont {Shirley}\ and\ \citenamefont
  {Martin}(1993{\natexlab{a}})}]{Shirley/Martin:1993}%
  \BibitemOpen
  \bibfield  {author} {\bibinfo {author} {\bibnamefont {Shirley}, \bibfnamefont
  {E~L}}, \ and\ \bibinfo {author} {\bibfnamefont {R.~M.}\ \bibnamefont
  {Martin}}} (\bibinfo {year} {1993}{\natexlab{a}}),\ \bibfield  {title}
  {\enquote {\bibinfo {title} {{M}any-body core-valence partitioning},}\
  }\href@noop {} {\bibfield  {journal} {\bibinfo  {journal} {Phys. Rev. B}\
  }\textbf {\bibinfo {volume} {47}},\ \bibinfo {pages} {15413}}\BibitemShut
  {NoStop}%
\bibitem [{\citenamefont {Shirley}\ and\ \citenamefont
  {Martin}(1993{\natexlab{b}})}]{Shirley/Martin:1993_2}%
  \BibitemOpen
  \bibfield  {author} {\bibinfo {author} {\bibnamefont {Shirley}, \bibfnamefont
  {E~L}}, \ and\ \bibinfo {author} {\bibfnamefont {R.~M.}\ \bibnamefont
  {Martin}}} (\bibinfo {year} {1993}{\natexlab{b}}),\ \bibfield  {title}
  {\enquote {\bibinfo {title} {\textit{GW} quasiparticle calculations in
  atoms},}\ }\href@noop {} {\bibfield  {journal} {\bibinfo  {journal} {Phys.
  Rev. B}\ }\textbf {\bibinfo {volume} {47}},\ \bibinfo {pages}
  {15404--15412}}\BibitemShut {NoStop}%
\bibitem [{\citenamefont {Shirley}\ \emph {et~al.}(1990)\citenamefont
  {Shirley}, \citenamefont {Martin}, \citenamefont {Bachelet},\ and\
  \citenamefont {Ceperley}}]{Shirley/Martin/Bachelet/Ceperley:1990}%
  \BibitemOpen
  \bibfield  {author} {\bibinfo {author} {\bibnamefont {Shirley}, \bibfnamefont
  {E~L}}, \bibinfo {author} {\bibfnamefont {R.~M.}\ \bibnamefont {Martin}},
  \bibinfo {author} {\bibfnamefont {G.~B.}\ \bibnamefont {Bachelet}}, \ and\
  \bibinfo {author} {\bibfnamefont {D.~M.}\ \bibnamefont {Ceperley}}} (\bibinfo
  {year} {1990}),\ \bibfield  {title} {\enquote {\bibinfo {title} {{R}ole of
  forms of exchange and correlation used in generating pseudopotentials},}\
  }\href@noop {} {\bibfield  {journal} {\bibinfo  {journal} {Phys.\ Rev.\ B}\
  }\textbf {\bibinfo {volume} {42}},\ \bibinfo {pages} {5057}}\BibitemShut
  {NoStop}%
\bibitem [{\citenamefont {Shirley}\ \emph {et~al.}(1997)\citenamefont
  {Shirley}, \citenamefont {Zhu},\ and\ \citenamefont
  {Louie}}]{Shirley/Zhu/Louie:1997}%
  \BibitemOpen
  \bibfield  {author} {\bibinfo {author} {\bibnamefont {Shirley}, \bibfnamefont
  {E~L}}, \bibinfo {author} {\bibfnamefont {X.}~\bibnamefont {Zhu}}, \ and\
  \bibinfo {author} {\bibfnamefont {S.~G.}\ \bibnamefont {Louie}}} (\bibinfo
  {year} {1997}),\ \bibfield  {title} {\enquote {\bibinfo {title} {{C}ore
  polarization in solids: {F}ormulation and application to semiconductors},}\
  }\href@noop {} {\bibfield  {journal} {\bibinfo  {journal} {Phys.\ Rev.\ B}\
  }\textbf {\bibinfo {volume} {56}},\ \bibinfo {pages} {6648}}\BibitemShut
  {NoStop}%
\bibitem [{\citenamefont {Shishkin}\ and\ \citenamefont
  {Kresse}(2006{\natexlab{a}})}]{Shishkin06}%
  \BibitemOpen
  \bibfield  {author} {\bibinfo {author} {\bibnamefont {Shishkin},
  \bibfnamefont {M}}, \ and\ \bibinfo {author} {\bibfnamefont {G.}~\bibnamefont
  {Kresse}}} (\bibinfo {year} {2006}{\natexlab{a}}),\ \bibfield  {title}
  {\enquote {\bibinfo {title} {{I}mplementation and performance of the
  frequency-dependent ${G}{W}$ method within the {P}{A}{W} framework},}\
  }\href@noop {} {\bibfield  {journal} {\bibinfo  {journal} {Phys. Rev. B}\
  }\textbf {\bibinfo {volume} {74}},\ \bibinfo {pages} {035101}}\BibitemShut
  {NoStop}%
\bibitem [{\citenamefont {Shishkin}\ and\ \citenamefont
  {Kresse}(2006{\natexlab{b}})}]{Shishkin/Kresse:2006}%
  \BibitemOpen
  \bibfield  {author} {\bibinfo {author} {\bibnamefont {Shishkin},
  \bibfnamefont {M}}, \ and\ \bibinfo {author} {\bibfnamefont {G.}~\bibnamefont
  {Kresse}}} (\bibinfo {year} {2006}{\natexlab{b}}),\ \bibfield  {title}
  {\enquote {\bibinfo {title} {{S}elf-consistent ${G}{W}$ calculations for
  semiconductors and insulators},}\ }\href@noop {} {\bibfield  {journal}
  {\bibinfo  {journal} {Phys. Rev. B}\ }\textbf {\bibinfo {volume} {75}},\
  \bibinfo {pages} {235102}}\BibitemShut {NoStop}%
\bibitem [{\citenamefont {Siegbahn}\ \emph {et~al.}(1969)\citenamefont
  {Siegbahn}, \citenamefont {Nordling}, \citenamefont {Johansson},
  \citenamefont {Hedman}, \citenamefont {Hed\'{e}n}, \citenamefont {Hamrin},
  \citenamefont {Gelius}, \citenamefont {Bergmark}, \citenamefont {Werme},
  \citenamefont {Manne},\ and\ \citenamefont {Baer}}]{Siegbahn1969all}%
  \BibitemOpen
  \bibfield  {author} {\bibinfo {author} {\bibnamefont {Siegbahn},
  \bibfnamefont {K}}, \bibinfo {author} {\bibfnamefont {C.}~\bibnamefont
  {Nordling}}, \bibinfo {author} {\bibfnamefont {G.}~\bibnamefont {Johansson}},
  \bibinfo {author} {\bibfnamefont {J.}~\bibnamefont {Hedman}}, \bibinfo
  {author} {\bibfnamefont {P.~F.}\ \bibnamefont {Hed\'{e}n}}, \bibinfo {author}
  {\bibfnamefont {K.}~\bibnamefont {Hamrin}}, \bibinfo {author} {\bibfnamefont
  {U.}~\bibnamefont {Gelius}}, \bibinfo {author} {\bibfnamefont
  {T.}~\bibnamefont {Bergmark}}, \bibinfo {author} {\bibfnamefont {L.~O.}\
  \bibnamefont {Werme}}, \bibinfo {author} {\bibfnamefont {R.}~\bibnamefont
  {Manne}}, \ and\ \bibinfo {author} {\bibfnamefont {Y.}~\bibnamefont {Baer}}}
  (\bibinfo {year} {1969}),\ \href@noop {} {\emph {\bibinfo {title}
  {{E}{S}{C}{A} applied to free molecules}}}\ (\bibinfo  {publisher}
  {North-Holland Publishing Company Amsterdam-London})\BibitemShut {NoStop}%
\bibitem [{\citenamefont {Siegel}\ \emph {et~al.}(2011)\citenamefont {Siegel},
  \citenamefont {Park}, \citenamefont {Hwang}, \citenamefont {Deslippe},
  \citenamefont {Fedorov}, \citenamefont {Louie},\ and\ \citenamefont
  {Lanzara}}]{siegel_pnas_108}%
  \BibitemOpen
  \bibfield  {author} {\bibinfo {author} {\bibnamefont {Siegel}, \bibfnamefont
  {D~A}}, \bibinfo {author} {\bibfnamefont {C.-H.}\ \bibnamefont {Park}},
  \bibinfo {author} {\bibfnamefont {C.}~\bibnamefont {Hwang}}, \bibinfo
  {author} {\bibfnamefont {J.}~\bibnamefont {Deslippe}}, \bibinfo {author}
  {\bibfnamefont {A.~V.}\ \bibnamefont {Fedorov}}, \bibinfo {author}
  {\bibfnamefont {S.~G.}\ \bibnamefont {Louie}}, \ and\ \bibinfo {author}
  {\bibfnamefont {A.}~\bibnamefont {Lanzara}}} (\bibinfo {year} {2011}),\
  \bibfield  {title} {\enquote {\bibinfo {title} {{M}any-body interactions in
  quasi-freestanding graphene},}\ }\href {\doibase 10.1073/pnas.1100242108}
  {\bibfield  {journal} {\bibinfo  {journal} {Proc. Natl. Acad. Sci. USA}\
  }\textbf {\bibinfo {volume} {108}}~(\bibinfo {number} {28}),\ \bibinfo
  {pages} {11365--11369}}\BibitemShut {NoStop}%
\bibitem [{\citenamefont {Singh}(1991)}]{Singh1991}%
  \BibitemOpen
  \bibfield  {author} {\bibinfo {author} {\bibnamefont {Singh}, \bibfnamefont
  {D}}} (\bibinfo {year} {1991}),\ \bibfield  {title} {\enquote {\bibinfo
  {title} {{G}round-state properties of lanthanum: {T}reatment of extended-core
  states},}\ }\href {\doibase 10.1103/PhysRevB.43.6388} {\bibfield  {journal}
  {\bibinfo  {journal} {Phys. Rev. B}\ }\textbf {\bibinfo {volume} {43}},\
  \bibinfo {pages} {6388--6392}}\BibitemShut {NoStop}%
\bibitem [{\citenamefont {Skl{\'e}nard}\ \emph {et~al.}(2018)\citenamefont
  {Skl{\'e}nard}, \citenamefont {Dragoni}, \citenamefont {Triozon},\ and\
  \citenamefont {Olevano}}]{Benoit/etal:2018}%
  \BibitemOpen
  \bibfield  {author} {\bibinfo {author} {\bibnamefont {Skl{\'e}nard},
  \bibfnamefont {B}}, \bibinfo {author} {\bibfnamefont {A.}~\bibnamefont
  {Dragoni}}, \bibinfo {author} {\bibfnamefont {F.}~\bibnamefont {Triozon}}, \
  and\ \bibinfo {author} {\bibfnamefont {V.}~\bibnamefont {Olevano}}} (\bibinfo
  {year} {2018}),\ \bibfield  {title} {\enquote {\bibinfo {title} {{O}ptical vs
  electronic gap of hafnia by ab initio {B}ethe-{S}alpeter equation},}\
  }\href@noop {} {\bibfield  {journal} {\bibinfo  {journal} {Appl. Phys.
  Lett.}\ }\textbf {\bibinfo {volume} {113}}~(\bibinfo {number} {17}),\
  \bibinfo {pages} {172903}}\BibitemShut {NoStop}%
\bibitem [{\citenamefont {Slater}(1937)}]{Slater1937}%
  \BibitemOpen
  \bibfield  {author} {\bibinfo {author} {\bibnamefont {Slater}, \bibfnamefont
  {J~C}}} (\bibinfo {year} {1937}),\ \bibfield  {title} {\enquote {\bibinfo
  {title} {{W}ave {F}unctions in a {P}eriodic {P}otential},}\ }\href {\doibase
  10.1103/PhysRev.51.846} {\bibfield  {journal} {\bibinfo  {journal} {Phys.
  Rev.}\ }\textbf {\bibinfo {volume} {51}},\ \bibinfo {pages}
  {846--851}}\BibitemShut {NoStop}%
\bibitem [{\citenamefont {Smith}(1988)}]{Smith:1988}%
  \BibitemOpen
  \bibfield  {author} {\bibinfo {author} {\bibnamefont {Smith}, \bibfnamefont
  {N~V}}} (\bibinfo {year} {1988}),\ \bibfield  {title} {\enquote {\bibinfo
  {title} {{I}nverse {P}hotoemission},}\ }\href@noop {} {\bibfield  {journal}
  {\bibinfo  {journal} {Rep.\ Prog.\ Phys.}\ }\textbf {\bibinfo {volume}
  {51}},\ \bibinfo {pages} {1227}}\BibitemShut {NoStop}%
\bibitem [{\citenamefont {Smondyrev}(1986)}]{Smondyrev:1986}%
  \BibitemOpen
  \bibfield  {author} {\bibinfo {author} {\bibnamefont {Smondyrev},
  \bibfnamefont {M~A}}} (\bibinfo {year} {1986}),\ \bibfield  {title}
  {{\selectlanguage {English}\enquote {\bibinfo {title} {{D}iagrams in the
  polaron model},}\ }}\href {\doibase 10.1007/BF01017794} {\bibfield  {journal}
  {\bibinfo  {journal} {Theor. Math. Phys.}\ }\textbf {\bibinfo {volume}
  {68}}~(\bibinfo {number} {1}),\ \bibinfo {pages} {653--664}}\BibitemShut
  {NoStop}%
\bibitem [{\citenamefont {Soininen}\ \emph {et~al.}(2003)\citenamefont
  {Soininen}, \citenamefont {Rehr},\ and\ \citenamefont
  {Shirley}}]{Soininen2003}%
  \BibitemOpen
  \bibfield  {author} {\bibinfo {author} {\bibnamefont {Soininen},
  \bibfnamefont {J~A}}, \bibinfo {author} {\bibfnamefont {J.~J.}\ \bibnamefont
  {Rehr}}, \ and\ \bibinfo {author} {\bibfnamefont {E.~L.}\ \bibnamefont
  {Shirley}}} (\bibinfo {year} {2003}),\ \bibfield  {title} {\enquote {\bibinfo
  {title} {{E}lectron self-energy calculation using a general multi-pole
  approximation},}\ }\href {\doibase 10.1088/0953-8984/15/17/312} {\bibfield
  {journal} {\bibinfo  {journal} {J. Phys.: Condens. Matter}\ }\textbf
  {\bibinfo {volume} {15}}~(\bibinfo {number} {17}),\ \bibinfo {pages}
  {2573--2586}}\BibitemShut {NoStop}%
\bibitem [{\citenamefont {Soininen}\ \emph {et~al.}(2005)\citenamefont
  {Soininen}, \citenamefont {Rehr},\ and\ \citenamefont
  {Shirley}}]{Soininen2005}%
  \BibitemOpen
  \bibfield  {author} {\bibinfo {author} {\bibnamefont {Soininen},
  \bibfnamefont {J~A}}, \bibinfo {author} {\bibfnamefont {J.~J.}\ \bibnamefont
  {Rehr}}, \ and\ \bibinfo {author} {\bibfnamefont {E.~L.}\ \bibnamefont
  {Shirley}}} (\bibinfo {year} {2005}),\ \bibfield  {title} {\enquote {\bibinfo
  {title} {{M}ultipole {R}epresentation of the {D}ielectric {M}atrix},}\ }\href
  {\doibase 10.1238/physica.topical.115a00243} {\bibfield  {journal} {\bibinfo
  {journal} {Phys. Scr.}\ }\textbf {\bibinfo {volume} {T115}},\ \bibinfo
  {pages} {243}}\BibitemShut {NoStop}%
\bibitem [{\citenamefont {Sottile}\ \emph {et~al.}(2003)\citenamefont
  {Sottile}, \citenamefont {Olevano},\ and\ \citenamefont
  {Reining}}]{Sottile/etal:2003}%
  \BibitemOpen
  \bibfield  {author} {\bibinfo {author} {\bibnamefont {Sottile}, \bibfnamefont
  {F}}, \bibinfo {author} {\bibfnamefont {V.}~\bibnamefont {Olevano}}, \ and\
  \bibinfo {author} {\bibfnamefont {L.}~\bibnamefont {Reining}}} (\bibinfo
  {year} {2003}),\ \bibfield  {title} {\enquote {\bibinfo {title}
  {{P}arameter-{F}ree {C}alculation of {R}esponse {F}unctions in
  {T}ime-{D}ependent {D}ensity-{F}unctional {T}heory},}\ }\href@noop {}
  {\bibfield  {journal} {\bibinfo  {journal} {Phys. Rev. Lett.}\ }\textbf
  {\bibinfo {volume} {91}},\ \bibinfo {pages} {056402}}\BibitemShut {NoStop}%
\bibitem [{\citenamefont {Spataru}\ \emph
  {et~al.}(2004{\natexlab{a}})\citenamefont {Spataru}, \citenamefont
  {Benedict},\ and\ \citenamefont {Louie}}]{Spataru/Benedict/Louie:2004}%
  \BibitemOpen
  \bibfield  {author} {\bibinfo {author} {\bibnamefont {Spataru}, \bibfnamefont
  {C~D}}, \bibinfo {author} {\bibfnamefont {L.~X.}\ \bibnamefont {Benedict}}, \
  and\ \bibinfo {author} {\bibfnamefont {S.~G.}\ \bibnamefont {Louie}}}
  (\bibinfo {year} {2004}{\natexlab{a}}),\ \bibfield  {title} {\enquote
  {\bibinfo {title} {{A}b initio calculation of band-gap renormalization in
  highly excited {G}a{A}s},}\ }\href@noop {} {\bibfield  {journal} {\bibinfo
  {journal} {Phys. Rev. B}\ }\textbf {\bibinfo {volume} {69}},\ \bibinfo
  {pages} {205204}}\BibitemShut {NoStop}%
\bibitem [{\citenamefont {Spataru}\ \emph {et~al.}(2009)\citenamefont
  {Spataru}, \citenamefont {Hybertsen}, \citenamefont {Louie},\ and\
  \citenamefont {Millis}}]{Spataru/Hybertsen/Louie:2009}%
  \BibitemOpen
  \bibfield  {author} {\bibinfo {author} {\bibnamefont {Spataru}, \bibfnamefont
  {C~D}}, \bibinfo {author} {\bibfnamefont {M.~S.}\ \bibnamefont {Hybertsen}},
  \bibinfo {author} {\bibfnamefont {S.~G.}\ \bibnamefont {Louie}}, \ and\
  \bibinfo {author} {\bibfnamefont {A.~J.}\ \bibnamefont {Millis}}} (\bibinfo
  {year} {2009}),\ \bibfield  {title} {\enquote {\bibinfo {title} {${G}{W}$
  approach to {A}nderson model out of equilibrium: {C}oulomb blockade and false
  hysteresis in the ${I}\text{\ensuremath{-}}{V}$ characteristics},}\
  }\href@noop {} {\bibfield  {journal} {\bibinfo  {journal} {Phys. Rev. B}\
  }\textbf {\bibinfo {volume} {79}},\ \bibinfo {pages} {155110}}\BibitemShut
  {NoStop}%
\bibitem [{\citenamefont {Spataru}\ \emph
  {et~al.}(2004{\natexlab{b}})\citenamefont {Spataru}, \citenamefont
  {Ismail-Beigi}, \citenamefont {Benedict},\ and\ \citenamefont
  {Louie}}]{Spataru/etal:2004}%
  \BibitemOpen
  \bibfield  {author} {\bibinfo {author} {\bibnamefont {Spataru}, \bibfnamefont
  {C~D}}, \bibinfo {author} {\bibfnamefont {S.}~\bibnamefont {Ismail-Beigi}},
  \bibinfo {author} {\bibfnamefont {L.~X.}\ \bibnamefont {Benedict}}, \ and\
  \bibinfo {author} {\bibfnamefont {S.~G.}\ \bibnamefont {Louie}}} (\bibinfo
  {year} {2004}{\natexlab{b}}),\ \bibfield  {title} {\enquote {\bibinfo {title}
  {{Q}uasiparticle energies, excitonic effects and optical absorption spectra
  of small-diameter single-walled carbon nanotubes},}\ }\href@noop {}
  {\bibfield  {journal} {\bibinfo  {journal} {Appl. Phys. A}\ }\textbf
  {\bibinfo {volume} {78}},\ \bibinfo {pages} {1129}}\BibitemShut {NoStop}%
\bibitem [{\citenamefont {Springer}\ \emph {et~al.}(1998)\citenamefont
  {Springer}, \citenamefont {Aryasetiawan},\ and\ \citenamefont
  {Karlsson}}]{Springer/etal:1998}%
  \BibitemOpen
  \bibfield  {author} {\bibinfo {author} {\bibnamefont {Springer},
  \bibfnamefont {M}}, \bibinfo {author} {\bibfnamefont {F.}~\bibnamefont
  {Aryasetiawan}}, \ and\ \bibinfo {author} {\bibfnamefont {K.}~\bibnamefont
  {Karlsson}}} (\bibinfo {year} {1998}),\ \bibfield  {title} {\enquote
  {\bibinfo {title} {{F}irst-{P}rinciples $\mathit{T}$-{M}atrix {T}heory with
  {A}pplication to the 6 e{V} {S}atellite in {N}i},}\ }\href@noop {} {\bibfield
   {journal} {\bibinfo  {journal} {Phys. Rev. Lett.}\ }\textbf {\bibinfo
  {volume} {80}},\ \bibinfo {pages} {2389--2392}}\BibitemShut {NoStop}%
\bibitem [{\citenamefont {Stan}\ \emph {et~al.}(2006)\citenamefont {Stan},
  \citenamefont {Dahlen},\ and\ \citenamefont {van Leeuwen}}]{Stan/etal:2006}%
  \BibitemOpen
  \bibfield  {author} {\bibinfo {author} {\bibnamefont {Stan}, \bibfnamefont
  {A}}, \bibinfo {author} {\bibfnamefont {N.~E.}\ \bibnamefont {Dahlen}}, \
  and\ \bibinfo {author} {\bibfnamefont {R.}~\bibnamefont {van Leeuwen}}}
  (\bibinfo {year} {2006}),\ \bibfield  {title} {\enquote {\bibinfo {title}
  {{F}ully self-consistent ${G}{W}$ calculations for atoms and molecules},}\
  }\href@noop {} {\bibfield  {journal} {\bibinfo  {journal} {Europhys. Lett.}\
  }\textbf {\bibinfo {volume} {76}},\ \bibinfo {pages} {298}}\BibitemShut
  {NoStop}%
\bibitem [{\citenamefont {Stan}\ \emph {et~al.}(2009)\citenamefont {Stan},
  \citenamefont {Dahlen},\ and\ \citenamefont {van Leeuwen}}]{Stan/etal:2009}%
  \BibitemOpen
  \bibfield  {author} {\bibinfo {author} {\bibnamefont {Stan}, \bibfnamefont
  {A}}, \bibinfo {author} {\bibfnamefont {N.~E.}\ \bibnamefont {Dahlen}}, \
  and\ \bibinfo {author} {\bibfnamefont {R.}~\bibnamefont {van Leeuwen}}}
  (\bibinfo {year} {2009}),\ \bibfield  {title} {\enquote {\bibinfo {title}
  {{L}evels of self-consistency in the ${G}{W}$ approximation},}\ }\href@noop
  {} {\bibfield  {journal} {\bibinfo  {journal} {J. Chem. Phys.}\ }\textbf
  {\bibinfo {volume} {130}}~(\bibinfo {number} {11}),\ \bibinfo {eid}
  {114105}}\BibitemShut {NoStop}%
\bibitem [{\citenamefont {Stan}\ \emph {et~al.}(2015)\citenamefont {Stan},
  \citenamefont {Romaniello}, \citenamefont {Rigamonti}, \citenamefont
  {Reining},\ and\ \citenamefont {Berger}}]{Stan_2015}%
  \BibitemOpen
  \bibfield  {author} {\bibinfo {author} {\bibnamefont {Stan}, \bibfnamefont
  {A}}, \bibinfo {author} {\bibfnamefont {P.}~\bibnamefont {Romaniello}},
  \bibinfo {author} {\bibfnamefont {S.}~\bibnamefont {Rigamonti}}, \bibinfo
  {author} {\bibfnamefont {L.}~\bibnamefont {Reining}}, \ and\ \bibinfo
  {author} {\bibfnamefont {J.~A.}\ \bibnamefont {Berger}}} (\bibinfo {year}
  {2015}),\ \bibfield  {title} {\enquote {\bibinfo {title} {{U}nphysical and
  physical solutions in many-body theories: from weak to strong correlation},}\
  }\href {\doibase 10.1088/1367-2630/17/9/093045} {\bibfield  {journal}
  {\bibinfo  {journal} {New J. Phys.}\ }\textbf {\bibinfo {volume}
  {17}}~(\bibinfo {number} {9}),\ \bibinfo {pages} {093045}}\BibitemShut
  {NoStop}%
\bibitem [{\citenamefont {Stankovski}\ \emph {et~al.}(2011)\citenamefont
  {Stankovski}, \citenamefont {Antonius}, \citenamefont {Waroquiers},
  \citenamefont {Miglio}, \citenamefont {Dixit}, \citenamefont {Sankaran},
  \citenamefont {Giantomassi}, \citenamefont {X.}, \citenamefont {C\^ot\'e},\
  and\ \citenamefont {Rignanese}}]{Stankovski/etal:2011}%
  \BibitemOpen
  \bibfield  {author} {\bibinfo {author} {\bibnamefont {Stankovski},
  \bibfnamefont {M}}, \bibinfo {author} {\bibfnamefont {G.}~\bibnamefont
  {Antonius}}, \bibinfo {author} {\bibfnamefont {D.}~\bibnamefont
  {Waroquiers}}, \bibinfo {author} {\bibfnamefont {A.}~\bibnamefont {Miglio}},
  \bibinfo {author} {\bibfnamefont {H.}~\bibnamefont {Dixit}}, \bibinfo
  {author} {\bibfnamefont {K.}~\bibnamefont {Sankaran}}, \bibinfo {author}
  {\bibfnamefont {M.}~\bibnamefont {Giantomassi}}, \bibinfo {author}
  {\bibnamefont {X.}}, \bibinfo {author} {\bibfnamefont {M.}~\bibnamefont
  {C\^ot\'e}}, \ and\ \bibinfo {author} {\bibfnamefont {G.-M.}\ \bibnamefont
  {Rignanese}}} (\bibinfo {year} {2011}),\ \bibfield  {title} {\enquote
  {\bibinfo {title} {${G}_{0}{W}_{0}$ band gap of {Z}n{O}: {E}ffects of
  plasmon-pole models},}\ }\href@noop {} {\bibfield  {journal} {\bibinfo
  {journal} {Phys. Rev. B}\ }\textbf {\bibinfo {volume} {84}},\ \bibinfo
  {pages} {241201}}\BibitemShut {NoStop}%
\bibitem [{\citenamefont {Stefanucci}\ and\ \citenamefont
  {Leeuwen}(2013)}]{MBPT_book_Stefanucci/Leeuwen:2013}%
  \BibitemOpen
  \bibfield  {author} {\bibinfo {author} {\bibnamefont {Stefanucci},
  \bibfnamefont {G}}, \ and\ \bibinfo {author} {\bibfnamefont {R.~V.}\
  \bibnamefont {Leeuwen}}} (\bibinfo {year} {2013}),\ \href@noop {} {\emph
  {\bibinfo {title} {{N}onequilibrium {M}any-{B}ody {T}heory of {Q}uantum
  {S}ystems}}}\ (\bibinfo  {publisher} {Cambridge University
  Press})\BibitemShut {NoStop}%
\bibitem [{\citenamefont {Steinkasserer}\ \emph {et~al.}(2016)\citenamefont
  {Steinkasserer}, \citenamefont {Suhr},\ and\ \citenamefont
  {Paulus}}]{steinkasserer_prb_94}%
  \BibitemOpen
  \bibfield  {author} {\bibinfo {author} {\bibnamefont {Steinkasserer},
  \bibfnamefont {L~E~Marsoner}}, \bibinfo {author} {\bibfnamefont
  {S.}~\bibnamefont {Suhr}}, \ and\ \bibinfo {author} {\bibfnamefont
  {B.}~\bibnamefont {Paulus}}} (\bibinfo {year} {2016}),\ \bibfield  {title}
  {\enquote {\bibinfo {title} {{B}and-gap control in phosphorene/{B}{N}
  structures from first-principles calculations},}\ }\href {\doibase
  10.1103/PhysRevB.94.125444} {\bibfield  {journal} {\bibinfo  {journal} {Phys.
  Rev. B}\ }\textbf {\bibinfo {volume} {94}},\ \bibinfo {pages}
  {125444}}\BibitemShut {NoStop}%
\bibitem [{\citenamefont {Sternheimer}(1954)}]{Sternheimer1954}%
  \BibitemOpen
  \bibfield  {author} {\bibinfo {author} {\bibnamefont {Sternheimer},
  \bibfnamefont {R~M}}} (\bibinfo {year} {1954}),\ \bibfield  {title} {\enquote
  {\bibinfo {title} {{E}lectronic {P}olarizabilities of {I}ons from the
  {H}artree-{F}ock {W}ave {F}unctions},}\ }\href {\doibase
  10.1103/PhysRev.96.951} {\bibfield  {journal} {\bibinfo  {journal} {Phys.
  Rev.}\ }\textbf {\bibinfo {volume} {96}},\ \bibinfo {pages}
  {951--968}}\BibitemShut {NoStop}%
\bibitem [{\citenamefont {Strange}\ \emph {et~al.}(2011)\citenamefont
  {Strange}, \citenamefont {Rostgaard}, \citenamefont {H\"akkinen},\ and\
  \citenamefont {Thygesen}}]{Strange/etal:2011}%
  \BibitemOpen
  \bibfield  {author} {\bibinfo {author} {\bibnamefont {Strange}, \bibfnamefont
  {M}}, \bibinfo {author} {\bibfnamefont {C.}~\bibnamefont {Rostgaard}},
  \bibinfo {author} {\bibfnamefont {H.}~\bibnamefont {H\"akkinen}}, \ and\
  \bibinfo {author} {\bibfnamefont {K.~S.}\ \bibnamefont {Thygesen}}} (\bibinfo
  {year} {2011}),\ \bibfield  {title} {\enquote {\bibinfo {title}
  {{S}elf-consistent ${G}{W}$ calculations of electronic transport in thiol-
  and amine-linked molecular junctions},}\ }\href@noop {} {\bibfield  {journal}
  {\bibinfo  {journal} {Phys. Rev. B}\ }\textbf {\bibinfo {volume} {83}},\
  \bibinfo {pages} {115108}}\BibitemShut {NoStop}%
\bibitem [{\citenamefont {Strange}\ and\ \citenamefont
  {Thygesen}(2011)}]{Strange/Thygesen:2011}%
  \BibitemOpen
  \bibfield  {author} {\bibinfo {author} {\bibnamefont {Strange}, \bibfnamefont
  {M}}, \ and\ \bibinfo {author} {\bibfnamefont {K.~S.}\ \bibnamefont
  {Thygesen}}} (\bibinfo {year} {2011}),\ \bibfield  {title} {\enquote
  {\bibinfo {title} {{T}owards quantitative accuracy in first-principles
  transport calculations: {T}he ${G}{W}$ method applied to alkane/gold
  junctions},}\ }\href@noop {} {\bibfield  {journal} {\bibinfo  {journal}
  {Beilstein J. Nanotechnol.}\ }\textbf {\bibinfo {volume} {2}},\ \bibinfo
  {pages} {746--754}}\BibitemShut {NoStop}%
\bibitem [{\citenamefont {Strange}\ and\ \citenamefont
  {Thygesen}(2012)}]{Strange/Thygesen:2012}%
  \BibitemOpen
  \bibfield  {author} {\bibinfo {author} {\bibnamefont {Strange}, \bibfnamefont
  {M}}, \ and\ \bibinfo {author} {\bibfnamefont {K.~S.}\ \bibnamefont
  {Thygesen}}} (\bibinfo {year} {2012}),\ \bibfield  {title} {\enquote
  {\bibinfo {title} {{I}mage-charge-induced localization of molecular orbitals
  at metal-molecule interfaces: {S}elf-consistent ${G}{W}$ calculations},}\
  }\href@noop {} {\bibfield  {journal} {\bibinfo  {journal} {Phys. Rev. B}\
  }\textbf {\bibinfo {volume} {86}},\ \bibinfo {pages} {195121}}\BibitemShut
  {NoStop}%
\bibitem [{\citenamefont
  {Streitenberger}(1984{\natexlab{a}})}]{streitenberger_pssb_125}%
  \BibitemOpen
  \bibfield  {author} {\bibinfo {author} {\bibnamefont {Streitenberger},
  \bibfnamefont {P}}} (\bibinfo {year} {1984}{\natexlab{a}}),\ \bibfield
  {title} {\enquote {\bibinfo {title} {{E}ffective {E}xchange and {C}orrelation
  {I}nteraction and {D}ielectric {S}creening in the {E}lectron {G}as. {A}
  {F}unctional {A}pproach to the {L}ocal-{F}ield {C}orrection},}\ }\href
  {\doibase 10.1002/pssb.2221250229} {\bibfield  {journal} {\bibinfo  {journal}
  {Phys. Status Solidi B}\ }\textbf {\bibinfo {volume} {125}}~(\bibinfo
  {number} {2}),\ \bibinfo {pages} {681--692}}\BibitemShut {NoStop}%
\bibitem [{\citenamefont
  {Streitenberger}(1984{\natexlab{b}})}]{streitenberger_pla_106}%
  \BibitemOpen
  \bibfield  {author} {\bibinfo {author} {\bibnamefont {Streitenberger},
  \bibfnamefont {P}}} (\bibinfo {year} {1984}{\natexlab{b}}),\ \bibfield
  {title} {\enquote {\bibinfo {title} {{F}unctional approach to the local-field
  effects in interacting electron systems},}\ }\href {\doibase
  https://doi.org/10.1016/0375-9601(84)90493-6} {\bibfield  {journal} {\bibinfo
   {journal} {Phys. Lett. A}\ }\textbf {\bibinfo {volume} {106}}~(\bibinfo
  {number} {1}),\ \bibinfo {pages} {57--60}}\BibitemShut {NoStop}%
\bibitem [{\citenamefont {Strinati}(1982)}]{Strinati:1982}%
  \BibitemOpen
  \bibfield  {author} {\bibinfo {author} {\bibnamefont {Strinati},
  \bibfnamefont {G}}} (\bibinfo {year} {1982}),\ \bibfield  {title} {\enquote
  {\bibinfo {title} {{D}ynamical {S}hift and {B}roadening of {C}ore {E}xcitons
  in {S}emiconductors},}\ }\href@noop {} {\bibfield  {journal} {\bibinfo
  {journal} {Phys. Rev. Lett.}\ }\textbf {\bibinfo {volume} {49}},\ \bibinfo
  {pages} {1519--1522}}\BibitemShut {NoStop}%
\bibitem [{\citenamefont {Strinati}(1984)}]{Strinati:1984}%
  \BibitemOpen
  \bibfield  {author} {\bibinfo {author} {\bibnamefont {Strinati},
  \bibfnamefont {G}}} (\bibinfo {year} {1984}),\ \bibfield  {title} {\enquote
  {\bibinfo {title} {{E}ffects of dynamical screening on resonances at
  inner-shell thresholds in semiconductors},}\ }\href@noop {} {\bibfield
  {journal} {\bibinfo  {journal} {Phys. Rev. B}\ }\textbf {\bibinfo {volume}
  {29}},\ \bibinfo {pages} {5718--5726}}\BibitemShut {NoStop}%
\bibitem [{\citenamefont {Strinati}(1988)}]{Strinati:1988}%
  \BibitemOpen
  \bibfield  {author} {\bibinfo {author} {\bibnamefont {Strinati},
  \bibfnamefont {G}}} (\bibinfo {year} {1988}),\ \bibfield  {title} {\enquote
  {\bibinfo {title} {{A}pplication of the {G}reen's function method to the
  study of the optical properties of semiconductors},}\ }\href
  {http://dx.doi.org/10.1007/BF02725962} {\bibfield  {journal} {\bibinfo
  {journal} {Riv. Nuovo Cimento (1978-1999)}\ }\textbf {\bibinfo {volume}
  {11}},\ \bibinfo {pages} {1--86}}\BibitemShut {NoStop}%
\bibitem [{\citenamefont {Strinati}\ \emph {et~al.}(1980)\citenamefont
  {Strinati}, \citenamefont {Mattausch},\ and\ \citenamefont
  {Hanke}}]{Strinati/etal:1980}%
  \BibitemOpen
  \bibfield  {author} {\bibinfo {author} {\bibnamefont {Strinati},
  \bibfnamefont {G}}, \bibinfo {author} {\bibfnamefont {H.~J.}\ \bibnamefont
  {Mattausch}}, \ and\ \bibinfo {author} {\bibfnamefont {W.}~\bibnamefont
  {Hanke}}} (\bibinfo {year} {1980}),\ \bibfield  {title} {\enquote {\bibinfo
  {title} {{D}ynamical {C}orrelation {E}ffects on the {Q}uasiparticle {B}loch
  {S}tates of a {C}ovalent {C}rystal},}\ }\href@noop {} {\bibfield  {journal}
  {\bibinfo  {journal} {Phys. Rev. Lett.}\ }\textbf {\bibinfo {volume} {45}},\
  \bibinfo {pages} {290--294}}\BibitemShut {NoStop}%
\bibitem [{\citenamefont {Strinati}\ \emph {et~al.}(1982)\citenamefont
  {Strinati}, \citenamefont {Mattausch},\ and\ \citenamefont
  {Hanke}}]{Strinati/etal:1982}%
  \BibitemOpen
  \bibfield  {author} {\bibinfo {author} {\bibnamefont {Strinati},
  \bibfnamefont {G}}, \bibinfo {author} {\bibfnamefont {H.~J.}\ \bibnamefont
  {Mattausch}}, \ and\ \bibinfo {author} {\bibfnamefont {W.}~\bibnamefont
  {Hanke}}} (\bibinfo {year} {1982}),\ \bibfield  {title} {\enquote {\bibinfo
  {title} {{D}ynamical aspects of correlation corrections in a covalent
  crystal},}\ }\href@noop {} {\bibfield  {journal} {\bibinfo  {journal} {Phys.
  Rev. B}\ }\textbf {\bibinfo {volume} {25}},\ \bibinfo {pages}
  {2867--2888}}\BibitemShut {NoStop}%
\bibitem [{\citenamefont {Surh}\ \emph {et~al.}(1988)\citenamefont {Surh},
  \citenamefont {Northrup},\ and\ \citenamefont {Louie}}]{Surh/etal:1988}%
  \BibitemOpen
  \bibfield  {author} {\bibinfo {author} {\bibnamefont {Surh}, \bibfnamefont
  {M~P}}, \bibinfo {author} {\bibfnamefont {J.~E.}\ \bibnamefont {Northrup}}, \
  and\ \bibinfo {author} {\bibfnamefont {S.~G.}\ \bibnamefont {Louie}}}
  (\bibinfo {year} {1988}),\ \bibfield  {title} {\enquote {\bibinfo {title}
  {{O}ccupied quasiparticle bandwidth of potassium},}\ }\href@noop {}
  {\bibfield  {journal} {\bibinfo  {journal} {Phys. Rev. B}\ }\textbf {\bibinfo
  {volume} {38}},\ \bibinfo {pages} {5976--5980}}\BibitemShut {NoStop}%
\bibitem [{\citenamefont {Svane}\ \emph {et~al.}(2010)\citenamefont {Svane},
  \citenamefont {Christensen}, \citenamefont {Gorczyca}, \citenamefont {van
  Schilfgaarde}, \citenamefont {Chantis},\ and\ \citenamefont
  {Kotani}}]{Svane/etal:2010}%
  \BibitemOpen
  \bibfield  {author} {\bibinfo {author} {\bibnamefont {Svane}, \bibfnamefont
  {A}}, \bibinfo {author} {\bibfnamefont {N.~E.}\ \bibnamefont {Christensen}},
  \bibinfo {author} {\bibfnamefont {I.}~\bibnamefont {Gorczyca}}, \bibinfo
  {author} {\bibfnamefont {M.}~\bibnamefont {van Schilfgaarde}}, \bibinfo
  {author} {\bibfnamefont {A.~N.}\ \bibnamefont {Chantis}}, \ and\ \bibinfo
  {author} {\bibfnamefont {T.}~\bibnamefont {Kotani}}} (\bibinfo {year}
  {2010}),\ \bibfield  {title} {\enquote {\bibinfo {title} {{Q}uasiparticle
  self-consistent ${G}{W}$ theory of {I}{I}{I}-{V} nitride semiconductors:
  {B}ands, gap bowing, and effective masses},}\ }\href@noop {} {\bibfield
  {journal} {\bibinfo  {journal} {Phys. Rev. B}\ }\textbf {\bibinfo {volume}
  {82}},\ \bibinfo {pages} {115102}}\BibitemShut {NoStop}%
\bibitem [{\citenamefont {Szabo}\ and\ \citenamefont
  {Ostlund}(1989)}]{Szabo/Ostlund:1989}%
  \BibitemOpen
  \bibfield  {author} {\bibinfo {author} {\bibnamefont {Szabo}, \bibfnamefont
  {A}}, \ and\ \bibinfo {author} {\bibfnamefont {N.~S.}\ \bibnamefont
  {Ostlund}}} (\bibinfo {year} {1989}),\ \href@noop {} {\emph {\bibinfo {title}
  {\textit{Modern {Q}uantum {C}hemistry: {I}ntroduction to {A}dvanced
  {E}lectronic {S}tructure {T}heory}}}}\ (\bibinfo  {publisher} {McGraw-Hill},\
  \bibinfo {address} {New York})\BibitemShut {NoStop}%
\bibitem [{\citenamefont {Talirz}\ \emph {et~al.}(2017)\citenamefont {Talirz},
  \citenamefont {S\"ode}, \citenamefont {Dumslaff}, \citenamefont {Wang},
  \citenamefont {Sanchez-Valencia}, \citenamefont {Liu}, \citenamefont
  {Shinde}, \citenamefont {Pignedoli}, \citenamefont {Liang}, \citenamefont
  {Meunier}, \citenamefont {Plumb}, \citenamefont {Shi}, \citenamefont {Feng},
  \citenamefont {Narita}, \citenamefont {M\"ullen}, \citenamefont {Fasel},\
  and\ \citenamefont {Ruffieux}}]{Talirz2017}%
  \BibitemOpen
  \bibfield  {author} {\bibinfo {author} {\bibnamefont {Talirz}, \bibfnamefont
  {L}}, \bibinfo {author} {\bibfnamefont {H.}~\bibnamefont {S\"ode}}, \bibinfo
  {author} {\bibfnamefont {T.}~\bibnamefont {Dumslaff}}, \bibinfo {author}
  {\bibfnamefont {S.}~\bibnamefont {Wang}}, \bibinfo {author} {\bibfnamefont
  {J.~R.}\ \bibnamefont {Sanchez-Valencia}}, \bibinfo {author} {\bibfnamefont
  {J.}~\bibnamefont {Liu}}, \bibinfo {author} {\bibfnamefont {P.}~\bibnamefont
  {Shinde}}, \bibinfo {author} {\bibfnamefont {C.~A.}\ \bibnamefont
  {Pignedoli}}, \bibinfo {author} {\bibfnamefont {L.}~\bibnamefont {Liang}},
  \bibinfo {author} {\bibfnamefont {V.}~\bibnamefont {Meunier}}, \bibinfo
  {author} {\bibfnamefont {N.~C.}\ \bibnamefont {Plumb}}, \bibinfo {author}
  {\bibfnamefont {M.}~\bibnamefont {Shi}}, \bibinfo {author} {\bibfnamefont
  {X.}~\bibnamefont {Feng}}, \bibinfo {author} {\bibfnamefont {A.}~\bibnamefont
  {Narita}}, \bibinfo {author} {\bibfnamefont {K.}~\bibnamefont {M\"ullen}},
  \bibinfo {author} {\bibfnamefont {R.}~\bibnamefont {Fasel}}, \ and\ \bibinfo
  {author} {\bibfnamefont {P.}~\bibnamefont {Ruffieux}}} (\bibinfo {year}
  {2017}),\ \bibfield  {title} {\enquote {\bibinfo {title} {{O}n-{S}urface
  {S}ynthesis and {C}haracterization of 9-{A}tom {W}ide {A}rmchair {G}raphene
  {N}anoribbons},}\ }\href@noop {} {\bibfield  {journal} {\bibinfo  {journal}
  {ACS Nano}\ }\textbf {\bibinfo {volume} {11}}~(\bibinfo {number} {2}),\
  \bibinfo {pages} {1380--1388}}\BibitemShut {NoStop}%
\bibitem [{\citenamefont {Tamblyn}\ \emph {et~al.}(2011)\citenamefont
  {Tamblyn}, \citenamefont {Darancet}, \citenamefont {Quek}, \citenamefont
  {Bonev},\ and\ \citenamefont {Neaton}}]{Tamblyn/etal:2011}%
  \BibitemOpen
  \bibfield  {author} {\bibinfo {author} {\bibnamefont {Tamblyn}, \bibfnamefont
  {I}}, \bibinfo {author} {\bibfnamefont {P.}~\bibnamefont {Darancet}},
  \bibinfo {author} {\bibfnamefont {S.~Y.}\ \bibnamefont {Quek}}, \bibinfo
  {author} {\bibfnamefont {S.~A.}\ \bibnamefont {Bonev}}, \ and\ \bibinfo
  {author} {\bibfnamefont {J.~B.}\ \bibnamefont {Neaton}}} (\bibinfo {year}
  {2011}),\ \bibfield  {title} {\enquote {\bibinfo {title} {{E}lectronic energy
  level alignment at metal-molecule interfaces with a ${G}{W}$ approach},}\
  }\href {\doibase 10.1103/PhysRevB.84.201402} {\bibfield  {journal} {\bibinfo
  {journal} {Phys. Rev. B}\ }\textbf {\bibinfo {volume} {84}},\ \bibinfo
  {pages} {201402}}\BibitemShut {NoStop}%
\bibitem [{\citenamefont {Tandetzky}\ \emph {et~al.}(2015)\citenamefont
  {Tandetzky}, \citenamefont {Dewhurst}, \citenamefont {Sharma},\ and\
  \citenamefont {Gross}}]{tandetzky_prb_92}%
  \BibitemOpen
  \bibfield  {author} {\bibinfo {author} {\bibnamefont {Tandetzky},
  \bibfnamefont {F}}, \bibinfo {author} {\bibfnamefont {J.~K.}\ \bibnamefont
  {Dewhurst}}, \bibinfo {author} {\bibfnamefont {S.}~\bibnamefont {Sharma}}, \
  and\ \bibinfo {author} {\bibfnamefont {E.~K.~U.}\ \bibnamefont {Gross}}}
  (\bibinfo {year} {2015}),\ \bibfield  {title} {\enquote {\bibinfo {title}
  {{M}ultiplicity of solutions to ${G}{W}$-type approximations},}\ }\href
  {\doibase 10.1103/PhysRevB.92.115125} {\bibfield  {journal} {\bibinfo
  {journal} {Phys. Rev. B}\ }\textbf {\bibinfo {volume} {92}},\ \bibinfo
  {pages} {115125}}\BibitemShut {NoStop}%
\bibitem [{\citenamefont {Tarantino}\ \emph {et~al.}(2017)\citenamefont
  {Tarantino}, \citenamefont {Romaniello}, \citenamefont {Berger},\ and\
  \citenamefont {Reining}}]{tarantino_prb_96}%
  \BibitemOpen
  \bibfield  {author} {\bibinfo {author} {\bibnamefont {Tarantino},
  \bibfnamefont {W}}, \bibinfo {author} {\bibfnamefont {P.}~\bibnamefont
  {Romaniello}}, \bibinfo {author} {\bibfnamefont {J.~A.}\ \bibnamefont
  {Berger}}, \ and\ \bibinfo {author} {\bibfnamefont {L.}~\bibnamefont
  {Reining}}} (\bibinfo {year} {2017}),\ \bibfield  {title} {\enquote {\bibinfo
  {title} {{S}elf-consistent {D}yson equation and self-energy functionals: {A}n
  analysis and illustration on the example of the {H}ubbard atom},}\ }\href
  {\doibase 10.1103/PhysRevB.96.045124} {\bibfield  {journal} {\bibinfo
  {journal} {Phys. Rev. B}\ }\textbf {\bibinfo {volume} {96}},\ \bibinfo
  {pages} {045124}}\BibitemShut {NoStop}%
\bibitem [{\citenamefont {Thoss}\ and\ \citenamefont
  {Evers}(2018)}]{Thoss/Evers:2018}%
  \BibitemOpen
  \bibfield  {author} {\bibinfo {author} {\bibnamefont {Thoss}, \bibfnamefont
  {M}}, \ and\ \bibinfo {author} {\bibfnamefont {F.}~\bibnamefont {Evers}}}
  (\bibinfo {year} {2018}),\ \bibfield  {title} {\enquote {\bibinfo {title}
  {{P}erspective: {T}heory of quantum transport in molecular junctions},}\
  }\href@noop {} {\bibfield  {journal} {\bibinfo  {journal} {J. Chem. Phys.}\
  }\textbf {\bibinfo {volume} {148}}~(\bibinfo {number} {3}),\ \bibinfo {pages}
  {030901}}\BibitemShut {NoStop}%
\bibitem [{\citenamefont {Thygesen}(2017)}]{thygesen_2d_4}%
  \BibitemOpen
  \bibfield  {author} {\bibinfo {author} {\bibnamefont {Thygesen},
  \bibfnamefont {K~S}}} (\bibinfo {year} {2017}),\ \bibfield  {title} {\enquote
  {\bibinfo {title} {{C}alculating excitons, plasmons, and quasiparticles in
  2{D} materials and van der {W}aals heterostructures},}\ }\href {\doibase
  10.1088/2053-1583/aa6432} {\bibfield  {journal} {\bibinfo  {journal} {2D
  Mater.}\ }\textbf {\bibinfo {volume} {4}}~(\bibinfo {number} {2}),\ \bibinfo
  {pages} {022004}}\BibitemShut {NoStop}%
\bibitem [{\citenamefont {Thygesen}\ and\ \citenamefont
  {Rubio}(2008)}]{Thygesen/Rubio:2008}%
  \BibitemOpen
  \bibfield  {author} {\bibinfo {author} {\bibnamefont {Thygesen},
  \bibfnamefont {K~S}}, \ and\ \bibinfo {author} {\bibfnamefont
  {A.}~\bibnamefont {Rubio}}} (\bibinfo {year} {2008}),\ \bibfield  {title}
  {\enquote {\bibinfo {title} {{C}onserving ${G}{W}$ scheme for nonequilibrium
  quantum transport in molecular contacts},}\ }\href@noop {} {\bibfield
  {journal} {\bibinfo  {journal} {Phys. Rev. B}\ }\textbf {\bibinfo {volume}
  {77}},\ \bibinfo {pages} {115333}}\BibitemShut {NoStop}%
\bibitem [{\citenamefont {Thygesen}\ and\ \citenamefont
  {Rubio}(2009)}]{Thygesen/Rubio:2009}%
  \BibitemOpen
  \bibfield  {author} {\bibinfo {author} {\bibnamefont {Thygesen},
  \bibfnamefont {K~S}}, \ and\ \bibinfo {author} {\bibfnamefont
  {A.}~\bibnamefont {Rubio}}} (\bibinfo {year} {2009}),\ \bibfield  {title}
  {\enquote {\bibinfo {title} {{R}enormalization of {M}olecular {Q}uasiparticle
  {L}evels at {M}etal-{M}olecule {I}nterfaces: {T}rends across {B}inding
  {R}egimes},}\ }\href@noop {} {\bibfield  {journal} {\bibinfo  {journal}
  {Phys. Rev. Lett.}\ }\textbf {\bibinfo {volume} {102}},\ \bibinfo {pages}
  {046802}}\BibitemShut {NoStop}%
\bibitem [{\citenamefont {Tiago}\ and\ \citenamefont
  {Chelikowsky}(2006)}]{Tiago/Chelikowsky:2006}%
  \BibitemOpen
  \bibfield  {author} {\bibinfo {author} {\bibnamefont {Tiago}, \bibfnamefont
  {M~L}}, \ and\ \bibinfo {author} {\bibfnamefont {J.~R.}\ \bibnamefont
  {Chelikowsky}}} (\bibinfo {year} {2006}),\ \bibfield  {title} {\enquote
  {\bibinfo {title} {{O}ptical excitations in organic molecules, clusters, and
  defects studied by first-principles {G}reen's function methods},}\
  }\href@noop {} {\bibfield  {journal} {\bibinfo  {journal} {Phys. Rev. B}\
  }\textbf {\bibinfo {volume} {73}},\ \bibinfo {pages} {205334}}\BibitemShut
  {NoStop}%
\bibitem [{\citenamefont {Tiago}\ \emph
  {et~al.}(2003{\natexlab{a}})\citenamefont {Tiago}, \citenamefont
  {Ismail-Beigi},\ and\ \citenamefont {Louie}}]{Tiago/Ismail-Beigi/Louie:2003}%
  \BibitemOpen
  \bibfield  {author} {\bibinfo {author} {\bibnamefont {Tiago}, \bibfnamefont
  {M~L}}, \bibinfo {author} {\bibfnamefont {S.}~\bibnamefont {Ismail-Beigi}}, \
  and\ \bibinfo {author} {\bibfnamefont {S.~G.}\ \bibnamefont {Louie}}}
  (\bibinfo {year} {2003}{\natexlab{a}}),\ \bibfield  {title} {\enquote
  {\bibinfo {title} {{E}ffect of semicore orbitals on the electronic band gaps
  of {S}i, {G}e and {G}a{A}s within the ${G}{W}$ approximation},}\ }\href@noop
  {} {\bibfield  {journal} {\bibinfo  {journal} {Phys. Rev. B}\ }\textbf
  {\bibinfo {volume} {69}},\ \bibinfo {pages} {125212}}\BibitemShut {NoStop}%
\bibitem [{\citenamefont {Tiago}\ \emph
  {et~al.}(2003{\natexlab{b}})\citenamefont {Tiago}, \citenamefont {Northrup},\
  and\ \citenamefont {Louie}}]{Tiago2003}%
  \BibitemOpen
  \bibfield  {author} {\bibinfo {author} {\bibnamefont {Tiago}, \bibfnamefont
  {M~L}}, \bibinfo {author} {\bibfnamefont {J.~E.}\ \bibnamefont {Northrup}}, \
  and\ \bibinfo {author} {\bibfnamefont {S.~G.}\ \bibnamefont {Louie}}}
  (\bibinfo {year} {2003}{\natexlab{b}}),\ \bibfield  {title} {\enquote
  {\bibinfo {title} {{A}b initio calculation of the electronic and optical
  properties of solid pentacene},}\ }\href {\doibase
  10.1103/PhysRevB.67.115212} {\bibfield  {journal} {\bibinfo  {journal} {Phys.
  Rev. B}\ }\textbf {\bibinfo {volume} {67}},\ \bibinfo {pages}
  {115212}}\BibitemShut {NoStop}%
\bibitem [{\citenamefont {Tkatchenko}\ and\ \citenamefont
  {Scheffler}(2009)}]{Tkatchenko2009}%
  \BibitemOpen
  \bibfield  {author} {\bibinfo {author} {\bibnamefont {Tkatchenko},
  \bibfnamefont {A}}, \ and\ \bibinfo {author} {\bibfnamefont {M.}~\bibnamefont
  {Scheffler}}} (\bibinfo {year} {2009}),\ \bibfield  {title} {\enquote
  {\bibinfo {title} {{A}ccurate {M}olecular {V}an {D}er {W}aals {I}nteractions
  from {G}round-{S}tate {E}lectron {D}ensity and {F}ree-{A}tom {R}eference
  {D}ata},}\ }\href@noop {} {\bibfield  {journal} {\bibinfo  {journal} {Phys.
  Rev. Lett.}\ }\textbf {\bibinfo {volume} {102}},\ \bibinfo {pages}
  {073005}}\BibitemShut {NoStop}%
\bibitem [{\citenamefont {Tomczak}\ \emph {et~al.}(2012)\citenamefont
  {Tomczak}, \citenamefont {Casula}, \citenamefont {Miyake}, \citenamefont
  {Aryasetiawan},\ and\ \citenamefont {Biermann}}]{Tomczak/etal:2012b}%
  \BibitemOpen
  \bibfield  {author} {\bibinfo {author} {\bibnamefont {Tomczak}, \bibfnamefont
  {J~M}}, \bibinfo {author} {\bibfnamefont {M.}~\bibnamefont {Casula}},
  \bibinfo {author} {\bibfnamefont {T.}~\bibnamefont {Miyake}}, \bibinfo
  {author} {\bibfnamefont {F.}~\bibnamefont {Aryasetiawan}}, \ and\ \bibinfo
  {author} {\bibfnamefont {S.}~\bibnamefont {Biermann}}} (\bibinfo {year}
  {2012}),\ \bibfield  {title} {\enquote {\bibinfo {title} {{C}ombined ${G}{W}$
  and dynamical mean-field theory: {D}ynamical screening effects in transition
  metal oxides},}\ }\href@noop {} {\bibfield  {journal} {\bibinfo  {journal}
  {EPL}\ }\textbf {\bibinfo {volume} {100}}~(\bibinfo {number} {6}),\ \bibinfo
  {pages} {67001}}\BibitemShut {NoStop}%
\bibitem [{\citenamefont {Tomczak}\ \emph {et~al.}(2014)\citenamefont
  {Tomczak}, \citenamefont {Casula}, \citenamefont {Miyake},\ and\
  \citenamefont {Biermann}}]{Tomczak/etal:2014}%
  \BibitemOpen
  \bibfield  {author} {\bibinfo {author} {\bibnamefont {Tomczak}, \bibfnamefont
  {J~M}}, \bibinfo {author} {\bibfnamefont {M.}~\bibnamefont {Casula}},
  \bibinfo {author} {\bibfnamefont {T.}~\bibnamefont {Miyake}}, \ and\ \bibinfo
  {author} {\bibfnamefont {S.}~\bibnamefont {Biermann}}} (\bibinfo {year}
  {2014}),\ \bibfield  {title} {\enquote {\bibinfo {title} {{A}symmetry in band
  widening and quasiparticle lifetimes in {S}r{V}{O}$_{3}$: {C}ompetition
  between screened exchange and local correlations from combined ${G}{W}$ and
  dynamical mean-field theory ${G}{W} + \mathrm{D{M}{F}T}$},}\ }\href@noop {}
  {\bibfield  {journal} {\bibinfo  {journal} {Phys. Rev. B}\ }\textbf {\bibinfo
  {volume} {90}},\ \bibinfo {pages} {165138}}\BibitemShut {NoStop}%
\bibitem [{\citenamefont {Tomczak}\ \emph {et~al.}(2017)\citenamefont
  {Tomczak}, \citenamefont {Liu}, \citenamefont {Toschi}, \citenamefont
  {Kresse},\ and\ \citenamefont {Held}}]{Tomczak/etal:2017}%
  \BibitemOpen
  \bibfield  {author} {\bibinfo {author} {\bibnamefont {Tomczak}, \bibfnamefont
  {J~M}}, \bibinfo {author} {\bibfnamefont {P.}~\bibnamefont {Liu}}, \bibinfo
  {author} {\bibfnamefont {A.}~\bibnamefont {Toschi}}, \bibinfo {author}
  {\bibfnamefont {G.}~\bibnamefont {Kresse}}, \ and\ \bibinfo {author}
  {\bibfnamefont {K.}~\bibnamefont {Held}}} (\bibinfo {year} {2017}),\
  \bibfield  {title} {\enquote {\bibinfo {title} {{M}erging ${G}{W}$ with
  {D}{M}{F}{T} and non-local correlations beyond},}\ }\href@noop {} {\bibfield
  {journal} {\bibinfo  {journal} {Eur. Phys. J. Special Topics}\ }\textbf
  {\bibinfo {volume} {226}}~(\bibinfo {number} {11}),\ \bibinfo {pages}
  {2565--2590}}\BibitemShut {NoStop}%
\bibitem [{\citenamefont {Tran}\ \emph {et~al.}(2014)\citenamefont {Tran},
  \citenamefont {Soklaski}, \citenamefont {Liang},\ and\ \citenamefont
  {Yang}}]{tran_prb_89}%
  \BibitemOpen
  \bibfield  {author} {\bibinfo {author} {\bibnamefont {Tran}, \bibfnamefont
  {V}}, \bibinfo {author} {\bibfnamefont {R.}~\bibnamefont {Soklaski}},
  \bibinfo {author} {\bibfnamefont {Y.}~\bibnamefont {Liang}}, \ and\ \bibinfo
  {author} {\bibfnamefont {L.}~\bibnamefont {Yang}}} (\bibinfo {year} {2014}),\
  \bibfield  {title} {\enquote {\bibinfo {title} {{L}ayer-controlled band gap
  and anisotropic excitons in few-layer black phosphorus},}\ }\href {\doibase
  10.1103/PhysRevB.89.235319} {\bibfield  {journal} {\bibinfo  {journal} {Phys.
  Rev. B}\ }\textbf {\bibinfo {volume} {89}},\ \bibinfo {pages}
  {235319}}\BibitemShut {NoStop}%
\bibitem [{\citenamefont {Trevisanutto}\ \emph {et~al.}(2008)\citenamefont
  {Trevisanutto}, \citenamefont {Giorgetti}, \citenamefont {Reining},
  \citenamefont {Ladisa},\ and\ \citenamefont
  {Olevano}}]{trevisanutto_prl_101}%
  \BibitemOpen
  \bibfield  {author} {\bibinfo {author} {\bibnamefont {Trevisanutto},
  \bibfnamefont {P~E}}, \bibinfo {author} {\bibfnamefont {C.}~\bibnamefont
  {Giorgetti}}, \bibinfo {author} {\bibfnamefont {L.}~\bibnamefont {Reining}},
  \bibinfo {author} {\bibfnamefont {M.}~\bibnamefont {Ladisa}}, \ and\ \bibinfo
  {author} {\bibfnamefont {V.}~\bibnamefont {Olevano}}} (\bibinfo {year}
  {2008}),\ \bibfield  {title} {\enquote {\bibinfo {title} {{A}b {I}nitio
  ${G}{W}$ {M}any-{B}ody {E}ffects in {G}raphene},}\ }\href {\doibase
  10.1103/PhysRevLett.101.226405} {\bibfield  {journal} {\bibinfo  {journal}
  {Phys. Rev. Lett.}\ }\textbf {\bibinfo {volume} {101}},\ \bibinfo {pages}
  {226405}}\BibitemShut {NoStop}%
\bibitem [{\citenamefont {Troullier}\ and\ \citenamefont
  {Martins}(1991)}]{Troullier/Martins:1991}%
  \BibitemOpen
  \bibfield  {author} {\bibinfo {author} {\bibnamefont {Troullier},
  \bibfnamefont {N}}, \ and\ \bibinfo {author} {\bibfnamefont {J.~L.}\
  \bibnamefont {Martins}}} (\bibinfo {year} {1991}),\ \bibfield  {title}
  {\enquote {\bibinfo {title} {{E}fficient pseudopotentials for plane-wave
  calculations},}\ }\href@noop {} {\bibfield  {journal} {\bibinfo  {journal}
  {Phys.\ Rev.\ B}\ }\textbf {\bibinfo {volume} {43}},\ \bibinfo {pages}
  {1993}}\BibitemShut {NoStop}%
\bibitem [{\citenamefont {Tse}\ \emph {et~al.}(2008)\citenamefont {Tse},
  \citenamefont {Klug}, \citenamefont {Yao},\ and\ \citenamefont
  {Desgreniers}}]{Tse/etal:2008}%
  \BibitemOpen
  \bibfield  {author} {\bibinfo {author} {\bibnamefont {Tse}, \bibfnamefont
  {J~S}}, \bibinfo {author} {\bibfnamefont {D.~D.}\ \bibnamefont {Klug}},
  \bibinfo {author} {\bibfnamefont {Y.}~\bibnamefont {Yao}}, \ and\ \bibinfo
  {author} {\bibfnamefont {S.}~\bibnamefont {Desgreniers}}} (\bibinfo {year}
  {2008}),\ \bibfield  {title} {\enquote {\bibinfo {title} {{E}lectronic
  structure of $\ensuremath{\epsilon}$-oxygen at high pressure: ${G}{W}$
  calculations},}\ }\href@noop {} {\bibfield  {journal} {\bibinfo  {journal}
  {Phys. Rev. B}\ }\textbf {\bibinfo {volume} {78}},\ \bibinfo {pages}
  {132101}}\BibitemShut {NoStop}%
\bibitem [{\citenamefont {Ugeda}\ \emph {et~al.}(2014)\citenamefont {Ugeda},
  \citenamefont {Bradley}, \citenamefont {Shi}, \citenamefont {da~Jornada},
  \citenamefont {Zhang}, \citenamefont {Qiu}, \citenamefont {Ruan},
  \citenamefont {Mo}, \citenamefont {Hussain}, \citenamefont {Shen},
  \citenamefont {Wang}, \citenamefont {Louie},\ and\ \citenamefont
  {Crommie}}]{Ugeda/etal:2014}%
  \BibitemOpen
  \bibfield  {author} {\bibinfo {author} {\bibnamefont {Ugeda}, \bibfnamefont
  {M~M}}, \bibinfo {author} {\bibfnamefont {A.~J.}\ \bibnamefont {Bradley}},
  \bibinfo {author} {\bibfnamefont {S.-F.}\ \bibnamefont {Shi}}, \bibinfo
  {author} {\bibfnamefont {F.~H.}\ \bibnamefont {da~Jornada}}, \bibinfo
  {author} {\bibfnamefont {Y.}~\bibnamefont {Zhang}}, \bibinfo {author}
  {\bibfnamefont {D.~Y.}\ \bibnamefont {Qiu}}, \bibinfo {author} {\bibfnamefont
  {W.}~\bibnamefont {Ruan}}, \bibinfo {author} {\bibfnamefont {S.-K.}\
  \bibnamefont {Mo}}, \bibinfo {author} {\bibfnamefont {Z.}~\bibnamefont
  {Hussain}}, \bibinfo {author} {\bibfnamefont {Z.-X.}\ \bibnamefont {Shen}},
  \bibinfo {author} {\bibfnamefont {F.}~\bibnamefont {Wang}}, \bibinfo {author}
  {\bibfnamefont {S.~G.}\ \bibnamefont {Louie}}, \ and\ \bibinfo {author}
  {\bibfnamefont {M.~F.}\ \bibnamefont {Crommie}}} (\bibinfo {year} {2014}),\
  \bibfield  {title} {\enquote {\bibinfo {title} {{G}iant bandgap
  renormalization and excitonic effects in a monolayer transition metal
  dichalcogenide semiconductor},}\ }\href@noop {} {\bibfield  {journal}
  {\bibinfo  {journal} {Nat. Mat.}\ }\textbf {\bibinfo {volume} {13}},\
  \bibinfo {pages} {1091}}\BibitemShut {NoStop}%
\bibitem [{\citenamefont {Umari}\ \emph {et~al.}(2014)\citenamefont {Umari},
  \citenamefont {Mosconi},\ and\ \citenamefont
  {Angelis}}]{Umari/Mosconi/DeAngelis:2014}%
  \BibitemOpen
  \bibfield  {author} {\bibinfo {author} {\bibnamefont {Umari}, \bibfnamefont
  {P}}, \bibinfo {author} {\bibfnamefont {E.}~\bibnamefont {Mosconi}}, \ and\
  \bibinfo {author} {\bibfnamefont {F.~De}\ \bibnamefont {Angelis}}} (\bibinfo
  {year} {2014}),\ \bibfield  {title} {\enquote {\bibinfo {title}
  {{R}elativistic ${G}{W}$ calculations on {C}{H}$_3${N}{H}$_3${P}b{I}$_3$ and
  {C}{H}$_3${N}{H}$_3${S}n{I}$_3$ {P}erovskites for {S}olar {C}ell
  {A}pplications},}\ }\href@noop {} {\bibfield  {journal} {\bibinfo  {journal}
  {Sci. Rep.}\ }\textbf {\bibinfo {volume} {4}},\ \bibinfo {pages}
  {4467}}\BibitemShut {NoStop}%
\bibitem [{\citenamefont {Umari}\ \emph {et~al.}(2009)\citenamefont {Umari},
  \citenamefont {Stenuit},\ and\ \citenamefont
  {Baroni}}]{Umari/Stenuit/Baroni:2009}%
  \BibitemOpen
  \bibfield  {author} {\bibinfo {author} {\bibnamefont {Umari}, \bibfnamefont
  {P}}, \bibinfo {author} {\bibfnamefont {G.}~\bibnamefont {Stenuit}}, \ and\
  \bibinfo {author} {\bibfnamefont {S.}~\bibnamefont {Baroni}}} (\bibinfo
  {year} {2009}),\ \bibfield  {title} {\enquote {\bibinfo {title} {{O}ptimal
  representation of the polarization propagator for large-scale ${G}{W}$
  calculations},}\ }\href {\doibase 10.1103/PhysRevB.79.201104} {\bibfield
  {journal} {\bibinfo  {journal} {Phys. Rev. B}\ }\textbf {\bibinfo {volume}
  {79}}~(\bibinfo {number} {20}),\ \bibinfo {pages} {201104}}\BibitemShut
  {NoStop}%
\bibitem [{\citenamefont {Umari}\ \emph {et~al.}(2010)\citenamefont {Umari},
  \citenamefont {Stenuit},\ and\ \citenamefont
  {Baroni}}]{Umari/Stenuit/Baroni:2010}%
  \BibitemOpen
  \bibfield  {author} {\bibinfo {author} {\bibnamefont {Umari}, \bibfnamefont
  {P}}, \bibinfo {author} {\bibfnamefont {G.}~\bibnamefont {Stenuit}}, \ and\
  \bibinfo {author} {\bibfnamefont {S.}~\bibnamefont {Baroni}}} (\bibinfo
  {year} {2010}),\ \bibfield  {title} {\enquote {\bibinfo {title} {${G}{W}$
  quasiparticle spectra from occupied states only},}\ }\href {\doibase
  10.1103/PhysRevB.81.115104} {\bibfield  {journal} {\bibinfo  {journal} {Phys.
  Rev. B}\ }\textbf {\bibinfo {volume} {81}}~(\bibinfo {number} {11}),\
  \bibinfo {pages} {115104}}\BibitemShut {NoStop}%
\bibitem [{\citenamefont {Vahtras}\ \emph {et~al.}(1993)\citenamefont
  {Vahtras}, \citenamefont {Alml\"{o}f},\ and\ \citenamefont
  {Feyereisen}}]{Vahtras1993}%
  \BibitemOpen
  \bibfield  {author} {\bibinfo {author} {\bibnamefont {Vahtras}, \bibfnamefont
  {O}}, \bibinfo {author} {\bibfnamefont {J.}~\bibnamefont {Alml\"{o}f}}, \
  and\ \bibinfo {author} {\bibfnamefont {M.~W.}\ \bibnamefont {Feyereisen}}}
  (\bibinfo {year} {1993}),\ \bibfield  {title} {\enquote {\bibinfo {title}
  {{I}ntegral approximations for {L}{C}{A}{O}-{S}{C}{F} calculations},}\ }\href
  {\doibase http://dx.doi.org/10.1016/0009-2614(93)89151-7} {\bibfield
  {journal} {\bibinfo  {journal} {Chem. Phys. Lett.}\ }\textbf {\bibinfo
  {volume} {213}}~(\bibinfo {number} {5–6}),\ \bibinfo {pages}
  {514--518}}\BibitemShut {NoStop}%
\bibitem [{\citenamefont {V\'{e}ril}\ \emph {et~al.}(2018)\citenamefont
  {V\'{e}ril}, \citenamefont {Romaniello}, \citenamefont {Berger},\ and\
  \citenamefont {Loos}}]{veril_jctc_14}%
  \BibitemOpen
  \bibfield  {author} {\bibinfo {author} {\bibnamefont {V\'{e}ril},
  \bibfnamefont {M}}, \bibinfo {author} {\bibfnamefont {P.}~\bibnamefont
  {Romaniello}}, \bibinfo {author} {\bibfnamefont {J.~A.}\ \bibnamefont
  {Berger}}, \ and\ \bibinfo {author} {\bibfnamefont {P.-F.}\ \bibnamefont
  {Loos}}} (\bibinfo {year} {2018}),\ \bibfield  {title} {\enquote {\bibinfo
  {title} {{U}nphysical {D}iscontinuities in ${G}{W}$ {M}ethods},}\ }\href@noop
  {} {\bibfield  {journal} {\bibinfo  {journal} {J. Chem. Theory Comput.}\
  }\textbf {\bibinfo {volume} {14}}~(\bibinfo {number} {10}),\ \bibinfo {pages}
  {5220--5228}}\BibitemShut {NoStop}%
\bibitem [{\citenamefont {Vidberg}\ and\ \citenamefont
  {Serene}(1977)}]{Vidberg1977}%
  \BibitemOpen
  \bibfield  {author} {\bibinfo {author} {\bibnamefont {Vidberg}, \bibfnamefont
  {H~J}}, \ and\ \bibinfo {author} {\bibfnamefont {J.~W.}\ \bibnamefont
  {Serene}}} (\bibinfo {year} {1977}),\ \bibfield  {title} {\enquote {\bibinfo
  {title} {{S}olving the {E}liashberg equations by means of {N}-point
  {P}ad{\'e} approximants},}\ }\href {\doibase 10.1007/BF00655090} {\bibfield
  {journal} {\bibinfo  {journal} {J. Low Temp. Phys.}\ }\textbf {\bibinfo
  {volume} {29}}~(\bibinfo {number} {3}),\ \bibinfo {pages}
  {179--192}}\BibitemShut {NoStop}%
\bibitem [{\citenamefont {Vl\v{c}ek}\ \emph {et~al.}(2017)\citenamefont
  {Vl\v{c}ek}, \citenamefont {Rabani}, \citenamefont {Neuhauser},\ and\
  \citenamefont {Baer}}]{Vlcek2017}%
  \BibitemOpen
  \bibfield  {author} {\bibinfo {author} {\bibnamefont {Vl\v{c}ek},
  \bibfnamefont {V}}, \bibinfo {author} {\bibfnamefont {E.}~\bibnamefont
  {Rabani}}, \bibinfo {author} {\bibfnamefont {D.}~\bibnamefont {Neuhauser}}, \
  and\ \bibinfo {author} {\bibfnamefont {R.}~\bibnamefont {Baer}}} (\bibinfo
  {year} {2017}),\ \bibfield  {title} {\enquote {\bibinfo {title} {{S}tochastic
  ${G}{W}$ {C}alculations for {M}olecules},}\ }\href@noop {} {\bibfield
  {journal} {\bibinfo  {journal} {J. Chem. Theory Comput.}\ }\textbf {\bibinfo
  {volume} {13}}~(\bibinfo {number} {10}),\ \bibinfo {pages}
  {4997--5003}}\BibitemShut {NoStop}%
\bibitem [{\citenamefont {Vu\v{c}i\v{c}evi\'{c}}\ \emph
  {et~al.}(2018)\citenamefont {Vu\v{c}i\v{c}evi\'{c}}, \citenamefont
  {Wentzell}, \citenamefont {Ferrero},\ and\ \citenamefont
  {Parcollet}}]{parcollet_prb_97}%
  \BibitemOpen
  \bibfield  {author} {\bibinfo {author} {\bibnamefont {Vu\v{c}i\v{c}evi\'{c}},
  \bibfnamefont {J}}, \bibinfo {author} {\bibfnamefont {N.}~\bibnamefont
  {Wentzell}}, \bibinfo {author} {\bibfnamefont {M.}~\bibnamefont {Ferrero}}, \
  and\ \bibinfo {author} {\bibfnamefont {O.}~\bibnamefont {Parcollet}}}
  (\bibinfo {year} {2018}),\ \bibfield  {title} {\enquote {\bibinfo {title}
  {{P}ractical consequences of the {L}uttinger-{W}ard functional
  multivaluedness for cluster {D}{M}{F}{T} methods},}\ }\href {\doibase
  10.1103/PhysRevB.97.125141} {\bibfield  {journal} {\bibinfo  {journal} {Phys.
  Rev. B}\ }\textbf {\bibinfo {volume} {97}},\ \bibinfo {pages}
  {125141}}\BibitemShut {NoStop}%
\bibitem [{\citenamefont {Wallace}(1947)}]{wallace_pr_71}%
  \BibitemOpen
  \bibfield  {author} {\bibinfo {author} {\bibnamefont {Wallace}, \bibfnamefont
  {P~R}}} (\bibinfo {year} {1947}),\ \bibfield  {title} {\enquote {\bibinfo
  {title} {{T}he {B}and {T}heory of {G}raphite},}\ }\href {\doibase
  10.1103/PhysRev.71.622} {\bibfield  {journal} {\bibinfo  {journal} {Phys.
  Rev.}\ }\textbf {\bibinfo {volume} {71}},\ \bibinfo {pages}
  {622--634}}\BibitemShut {NoStop}%
\bibitem [{\citenamefont {Wang}\ \emph {et~al.}(2003)\citenamefont {Wang},
  \citenamefont {Rohlfing}, \citenamefont {Kr\"uger},\ and\ \citenamefont
  {Pollmann}}]{Wang/etal:2003}%
  \BibitemOpen
  \bibfield  {author} {\bibinfo {author} {\bibnamefont {Wang}, \bibfnamefont
  {N-P}}, \bibinfo {author} {\bibfnamefont {M.}~\bibnamefont {Rohlfing}},
  \bibinfo {author} {\bibfnamefont {P.}~\bibnamefont {Kr\"uger}}, \ and\
  \bibinfo {author} {\bibfnamefont {J.}~\bibnamefont {Pollmann}}} (\bibinfo
  {year} {2003}),\ \bibfield  {title} {\enquote {\bibinfo {title}
  {{Q}uasiparticle band structure and optical spectrum of {L}i{F}(001)},}\
  }\href@noop {} {\bibfield  {journal} {\bibinfo  {journal} {Phys. Rev. B}\
  }\textbf {\bibinfo {volume} {67}},\ \bibinfo {pages} {115111}}\BibitemShut
  {NoStop}%
\bibitem [{\citenamefont {Wang}\ \emph {et~al.}(2015)\citenamefont {Wang},
  \citenamefont {Xu},\ and\ \citenamefont {Pi}}]{wang_cpb_24}%
  \BibitemOpen
  \bibfield  {author} {\bibinfo {author} {\bibnamefont {Wang}, \bibfnamefont
  {R}}, \bibinfo {author} {\bibfnamefont {M.-S.}\ \bibnamefont {Xu}}, \ and\
  \bibinfo {author} {\bibfnamefont {X.-D.}\ \bibnamefont {Pi}}} (\bibinfo
  {year} {2015}),\ \bibfield  {title} {\enquote {\bibinfo {title} {{C}hemical
  modification of silicene},}\ }\href {\doibase 10.1088/1674-1056/24/8/086807}
  {\bibfield  {journal} {\bibinfo  {journal} {Chin. Phys. B}\ }\textbf
  {\bibinfo {volume} {24}}~(\bibinfo {number} {8}),\ \bibinfo {pages}
  {086807}}\BibitemShut {NoStop}%
\bibitem [{\citenamefont {Wang}\ \emph {et~al.}(2016)\citenamefont {Wang},
  \citenamefont {Talirz}, \citenamefont {Pignedoli}, \citenamefont {Feng},
  \citenamefont {M\"ullen}, \citenamefont {Fasel},\ and\ \citenamefont
  {Ruffieux}}]{Wang2016}%
  \BibitemOpen
  \bibfield  {author} {\bibinfo {author} {\bibnamefont {Wang}, \bibfnamefont
  {S}}, \bibinfo {author} {\bibfnamefont {L.}~\bibnamefont {Talirz}}, \bibinfo
  {author} {\bibfnamefont {C.~A.}\ \bibnamefont {Pignedoli}}, \bibinfo {author}
  {\bibfnamefont {X.}~\bibnamefont {Feng}}, \bibinfo {author} {\bibfnamefont
  {K.}~\bibnamefont {M\"ullen}}, \bibinfo {author} {\bibfnamefont
  {R.}~\bibnamefont {Fasel}}, \ and\ \bibinfo {author} {\bibfnamefont
  {P.}~\bibnamefont {Ruffieux}}} (\bibinfo {year} {2016}),\ \bibfield  {title}
  {\enquote {\bibinfo {title} {{G}iant edge state splitting at atomically
  precise graphene zigzag edges},}\ }\href
  {http://dx.doi.org/10.1038/ncomms11507} {\bibfield  {journal} {\bibinfo
  {journal} {Nat. Commun.}\ }\textbf {\bibinfo {volume} {7}},\ \bibinfo {pages}
  {11507--}}\BibitemShut {NoStop}%
\bibitem [{\citenamefont {Wang}\ and\ \citenamefont
  {Wu}(2017)}]{wang_pccp_2017}%
  \BibitemOpen
  \bibfield  {author} {\bibinfo {author} {\bibnamefont {Wang}, \bibfnamefont
  {X}}, \ and\ \bibinfo {author} {\bibfnamefont {Z.}~\bibnamefont {Wu}}}
  (\bibinfo {year} {2017}),\ \bibfield  {title} {\enquote {\bibinfo {title}
  {{I}ntrinsic magnetism and spontaneous band gap opening in bilayer silicene
  and germanene},}\ }\href {\doibase 10.1039/C6CP07184H} {\bibfield  {journal}
  {\bibinfo  {journal} {Phys. Chem. Chem. Phys.}\ }\textbf {\bibinfo {volume}
  {19}},\ \bibinfo {pages} {2148--2152}}\BibitemShut {NoStop}%
\bibitem [{\citenamefont {Wang}\ and\ \citenamefont
  {Shi}(2010)}]{wang_ssc_150}%
  \BibitemOpen
  \bibfield  {author} {\bibinfo {author} {\bibnamefont {Wang}, \bibfnamefont
  {Y}}, \ and\ \bibinfo {author} {\bibfnamefont {S.}~\bibnamefont {Shi}}}
  (\bibinfo {year} {2010}),\ \bibfield  {title} {\enquote {\bibinfo {title}
  {{S}tructural and electronic properties of monolayer hydrogenated honeycomb
  {I}{I}{I}-{V} sheets from first-principles},}\ }\href {\doibase
  https://doi.org/10.1016/j.ssc.2010.05.031} {\bibfield  {journal} {\bibinfo
  {journal} {Solid State Commun.}\ }\textbf {\bibinfo {volume} {150}}~(\bibinfo
  {number} {31}),\ \bibinfo {pages} {1473--1478}}\BibitemShut {NoStop}%
\bibitem [{\citenamefont {Weber}\ \emph {et~al.}(2007)\citenamefont {Weber},
  \citenamefont {Janotti}, \citenamefont {Rinke},\ and\ \citenamefont {{Van de
  Walle}}}]{Weber/Janotti/Rinke/VandeWalle:2007}%
  \BibitemOpen
  \bibfield  {author} {\bibinfo {author} {\bibnamefont {Weber}, \bibfnamefont
  {J}}, \bibinfo {author} {\bibfnamefont {A.}~\bibnamefont {Janotti}}, \bibinfo
  {author} {\bibfnamefont {P.}~\bibnamefont {Rinke}}, \ and\ \bibinfo {author}
  {\bibfnamefont {C.~G.}\ \bibnamefont {{Van de Walle}}}} (\bibinfo {year}
  {2007}),\ \bibfield  {title} {\enquote {\bibinfo {title} {{D}angling-bond
  defects and hydrogen passivation in germanium},}\ }\href@noop {} {\bibfield
  {journal} {\bibinfo  {journal} {Appl.\ Phys.\ Lett.}\ }\textbf {\bibinfo
  {volume} {91}},\ \bibinfo {pages} {142101}}\BibitemShut {NoStop}%
\bibitem [{\citenamefont {Wei}\ \emph {et~al.}(2013)\citenamefont {Wei},
  \citenamefont {Dai}, \citenamefont {Huang},\ and\ \citenamefont
  {Jacob}}]{wei_pccp_15}%
  \BibitemOpen
  \bibfield  {author} {\bibinfo {author} {\bibnamefont {Wei}, \bibfnamefont
  {W}}, \bibinfo {author} {\bibfnamefont {Y.}~\bibnamefont {Dai}}, \bibinfo
  {author} {\bibfnamefont {B.}~\bibnamefont {Huang}}, \ and\ \bibinfo {author}
  {\bibfnamefont {T.}~\bibnamefont {Jacob}}} (\bibinfo {year} {2013}),\
  \bibfield  {title} {\enquote {\bibinfo {title} {{M}any-body effects in
  silicene{,} silicane{,} germanene and germanane},}\ }\href {\doibase
  10.1039/C3CP51078F} {\bibfield  {journal} {\bibinfo  {journal} {Phys. Chem.
  Chem. Phys.}\ }\textbf {\bibinfo {volume} {15}},\ \bibinfo {pages}
  {8789--8794}}\BibitemShut {NoStop}%
\bibitem [{\citenamefont {Wei}\ and\ \citenamefont
  {Jacob}(2013{\natexlab{a}})}]{wei_prb_87}%
  \BibitemOpen
  \bibfield  {author} {\bibinfo {author} {\bibnamefont {Wei}, \bibfnamefont
  {W}}, \ and\ \bibinfo {author} {\bibfnamefont {T.}~\bibnamefont {Jacob}}}
  (\bibinfo {year} {2013}{\natexlab{a}}),\ \bibfield  {title} {\enquote
  {\bibinfo {title} {{E}lectronic and optical properties of fluorinated
  graphene: {A} many-body perturbation theory study},}\ }\href {\doibase
  10.1103/PhysRevB.87.115431} {\bibfield  {journal} {\bibinfo  {journal} {Phys.
  Rev. B}\ }\textbf {\bibinfo {volume} {87}},\ \bibinfo {pages}
  {115431}}\BibitemShut {NoStop}%
\bibitem [{\citenamefont {Wei}\ and\ \citenamefont
  {Jacob}(2013{\natexlab{b}})}]{wei_prb_88}%
  \BibitemOpen
  \bibfield  {author} {\bibinfo {author} {\bibnamefont {Wei}, \bibfnamefont
  {W}}, \ and\ \bibinfo {author} {\bibfnamefont {T.}~\bibnamefont {Jacob}}}
  (\bibinfo {year} {2013}{\natexlab{b}}),\ \bibfield  {title} {\enquote
  {\bibinfo {title} {{S}trong many-body effects in silicene-based
  structures},}\ }\href {\doibase 10.1103/PhysRevB.88.045203} {\bibfield
  {journal} {\bibinfo  {journal} {Phys. Rev. B}\ }\textbf {\bibinfo {volume}
  {88}},\ \bibinfo {pages} {045203}}\BibitemShut {NoStop}%
\bibitem [{\citenamefont {Weigend}\ and\ \citenamefont
  {Ahlrichs}(2005)}]{Weigend2005}%
  \BibitemOpen
  \bibfield  {author} {\bibinfo {author} {\bibnamefont {Weigend}, \bibfnamefont
  {F}}, \ and\ \bibinfo {author} {\bibfnamefont {R.}~\bibnamefont {Ahlrichs}}}
  (\bibinfo {year} {2005}),\ \bibfield  {title} {\enquote {\bibinfo {title}
  {{B}alanced basis sets of split valence{,} triple zeta valence and quadruple
  zeta valence quality for {H} to {R}n: {D}esign and assessment of accuracy},}\
  }\href {\doibase 10.1039/B508541A} {\bibfield  {journal} {\bibinfo  {journal}
  {Phys. Chem. Chem. Phys.}\ }\textbf {\bibinfo {volume} {7}},\ \bibinfo
  {pages} {3297--3305}}\BibitemShut {NoStop}%
\bibitem [{\citenamefont {Weinelt}\ \emph {et~al.}(2004)\citenamefont
  {Weinelt}, \citenamefont {Kutschera}, \citenamefont {Fauster},\ and\
  \citenamefont {Rohlfing}}]{Weinelt/etal:2004}%
  \BibitemOpen
  \bibfield  {author} {\bibinfo {author} {\bibnamefont {Weinelt}, \bibfnamefont
  {M}}, \bibinfo {author} {\bibfnamefont {M.}~\bibnamefont {Kutschera}},
  \bibinfo {author} {\bibfnamefont {T.}~\bibnamefont {Fauster}}, \ and\
  \bibinfo {author} {\bibfnamefont {M.}~\bibnamefont {Rohlfing}}} (\bibinfo
  {year} {2004}),\ \bibfield  {title} {\enquote {\bibinfo {title} {{D}ynamics
  of {E}xciton {F}ormation at the {S}i(100)
  $c(4\ifmmode\times\else\texttimes\fi{}2)$ {S}urface},}\ }\href@noop {}
  {\bibfield  {journal} {\bibinfo  {journal} {Phys. Rev. Lett.}\ }\textbf
  {\bibinfo {volume} {92}},\ \bibinfo {pages} {126801}}\BibitemShut {NoStop}%
\bibitem [{\citenamefont {White}\ \emph {et~al.}(1997)\citenamefont {White},
  \citenamefont {Godby}, \citenamefont {Rieger},\ and\ \citenamefont
  {Needs}}]{White/Godby/Rieger/Needs:1997}%
  \BibitemOpen
  \bibfield  {author} {\bibinfo {author} {\bibnamefont {White}, \bibfnamefont
  {I~D}}, \bibinfo {author} {\bibfnamefont {R.~W.}\ \bibnamefont {Godby}},
  \bibinfo {author} {\bibfnamefont {M.~M.}\ \bibnamefont {Rieger}}, \ and\
  \bibinfo {author} {\bibfnamefont {R.~J.}\ \bibnamefont {Needs}}} (\bibinfo
  {year} {1997}),\ \bibfield  {title} {\enquote {\bibinfo {title} {{D}ynamic
  {I}mage {P}otential at an {A}l(111) {S}urface},}\ }\href@noop {} {\bibfield
  {journal} {\bibinfo  {journal} {Phys.\ Rev.\ Lett.}\ }\textbf {\bibinfo
  {volume} {80}},\ \bibinfo {pages} {4265}}\BibitemShut {NoStop}%
\bibitem [{\citenamefont {Wigner}\ and\ \citenamefont
  {Huntington}(1935)}]{Wigner/etal:1935}%
  \BibitemOpen
  \bibfield  {author} {\bibinfo {author} {\bibnamefont {Wigner}, \bibfnamefont
  {E}}, \ and\ \bibinfo {author} {\bibfnamefont {H.~B.}\ \bibnamefont
  {Huntington}}} (\bibinfo {year} {1935}),\ \bibfield  {title} {\enquote
  {\bibinfo {title} {{O}n the {P}ossibility of a {M}etallic {M}odification of
  {H}ydrogen},}\ }\href@noop {} {\bibfield  {journal} {\bibinfo  {journal} {J.
  Chem. Phys.}\ }\textbf {\bibinfo {volume} {3}}~(\bibinfo {number} {12}),\
  \bibinfo {pages} {764--770}}\BibitemShut {NoStop}%
\bibitem [{\citenamefont {Wilhelm}\ \emph {et~al.}(2016)\citenamefont
  {Wilhelm}, \citenamefont {Ben},\ and\ \citenamefont {Hutter}}]{Wilhelm2016}%
  \BibitemOpen
  \bibfield  {author} {\bibinfo {author} {\bibnamefont {Wilhelm}, \bibfnamefont
  {J}}, \bibinfo {author} {\bibfnamefont {M.~Del}\ \bibnamefont {Ben}}, \ and\
  \bibinfo {author} {\bibfnamefont {J.}~\bibnamefont {Hutter}}} (\bibinfo
  {year} {2016}),\ \bibfield  {title} {\enquote {\bibinfo {title} {${G}{W}$ in
  the {G}aussian and {P}lane {W}aves {S}cheme with {A}pplication to {L}inear
  {A}cenes},}\ }\href@noop {} {\bibfield  {journal} {\bibinfo  {journal} {J.
  Chem. Theory Comput.}\ }\textbf {\bibinfo {volume} {12}}~(\bibinfo {number}
  {8}),\ \bibinfo {pages} {3623--3635}}\BibitemShut {NoStop}%
\bibitem [{\citenamefont {Wilhelm}\ \emph {et~al.}(2018)\citenamefont
  {Wilhelm}, \citenamefont {Golze}, \citenamefont {Talirz}, \citenamefont
  {Hutter},\ and\ \citenamefont {Pignedoli}}]{Wilhelm2018}%
  \BibitemOpen
  \bibfield  {author} {\bibinfo {author} {\bibnamefont {Wilhelm}, \bibfnamefont
  {J}}, \bibinfo {author} {\bibfnamefont {D.}~\bibnamefont {Golze}}, \bibinfo
  {author} {\bibfnamefont {L.}~\bibnamefont {Talirz}}, \bibinfo {author}
  {\bibfnamefont {J.}~\bibnamefont {Hutter}}, \ and\ \bibinfo {author}
  {\bibfnamefont {C.~A.}\ \bibnamefont {Pignedoli}}} (\bibinfo {year} {2018}),\
  \bibfield  {title} {\enquote {\bibinfo {title} {{T}oward ${G}{W}$
  {C}alculations on {T}housands of {A}toms},}\ }\href@noop {} {\bibfield
  {journal} {\bibinfo  {journal} {J. Phys. Chem. Lett.}\ }\textbf {\bibinfo
  {volume} {9}}~(\bibinfo {number} {2}),\ \bibinfo {pages}
  {306--312}}\BibitemShut {NoStop}%
\bibitem [{\citenamefont {Wilhelm}\ and\ \citenamefont
  {Hutter}(2017)}]{Wilhelm2017}%
  \BibitemOpen
  \bibfield  {author} {\bibinfo {author} {\bibnamefont {Wilhelm}, \bibfnamefont
  {J}}, \ and\ \bibinfo {author} {\bibfnamefont {J.}~\bibnamefont {Hutter}}}
  (\bibinfo {year} {2017}),\ \bibfield  {title} {\enquote {\bibinfo {title}
  {{P}eriodic ${G}{W}$ calculations in the {G}aussian and plane-waves
  scheme},}\ }\href {\doibase 10.1103/PhysRevB.95.235123} {\bibfield  {journal}
  {\bibinfo  {journal} {Phys. Rev. B}\ }\textbf {\bibinfo {volume} {95}},\
  \bibinfo {pages} {235123}}\BibitemShut {NoStop}%
\bibitem [{\citenamefont {Wilson}\ \emph {et~al.}(1996)\citenamefont {Wilson},
  \citenamefont {Mourik},\ and\ \citenamefont {Dunning}}]{Wilson1996}%
  \BibitemOpen
  \bibfield  {author} {\bibinfo {author} {\bibnamefont {Wilson}, \bibfnamefont
  {A~K}}, \bibinfo {author} {\bibfnamefont {T.~V.}\ \bibnamefont {Mourik}}, \
  and\ \bibinfo {author} {\bibfnamefont {T.~H.}\ \bibnamefont {Dunning}}}
  (\bibinfo {year} {1996}),\ \bibfield  {title} {\enquote {\bibinfo {title}
  {{G}aussian basis sets for use in correlated molecular calculations. {V}{I}.
  {S}extuple zeta correlation consistent basis sets for boron through neon},}\
  }\href {\doibase https://doi.org/10.1016/S0166-1280(96)80048-0} {\bibfield
  {journal} {\bibinfo  {journal} {J. Mol. Struct. (Theochem)}\ }\textbf
  {\bibinfo {volume} {388}},\ \bibinfo {pages} {339 -- 349}}\BibitemShut
  {NoStop}%
\bibitem [{\citenamefont {Wilson}\ \emph {et~al.}(2008)\citenamefont {Wilson},
  \citenamefont {Gygi},\ and\ \citenamefont {Galli}}]{Wilson2008}%
  \BibitemOpen
  \bibfield  {author} {\bibinfo {author} {\bibnamefont {Wilson}, \bibfnamefont
  {H~F}}, \bibinfo {author} {\bibfnamefont {F.}~\bibnamefont {Gygi}}, \ and\
  \bibinfo {author} {\bibfnamefont {G.}~\bibnamefont {Galli}}} (\bibinfo {year}
  {2008}),\ \bibfield  {title} {\enquote {\bibinfo {title} {{E}fficient
  iterative method for calculations of dielectric matrices},}\ }\href {\doibase
  10.1103/PhysRevB.78.113303} {\bibfield  {journal} {\bibinfo  {journal} {Phys.
  Rev. B}\ }\textbf {\bibinfo {volume} {78}},\ \bibinfo {pages}
  {113303}}\BibitemShut {NoStop}%
\bibitem [{\citenamefont {Wilson}\ \emph {et~al.}(2009)\citenamefont {Wilson},
  \citenamefont {Lu}, \citenamefont {Gygi},\ and\ \citenamefont
  {Galli}}]{Wilson2009}%
  \BibitemOpen
  \bibfield  {author} {\bibinfo {author} {\bibnamefont {Wilson}, \bibfnamefont
  {H~F}}, \bibinfo {author} {\bibfnamefont {D.}~\bibnamefont {Lu}}, \bibinfo
  {author} {\bibfnamefont {F.}~\bibnamefont {Gygi}}, \ and\ \bibinfo {author}
  {\bibfnamefont {G.}~\bibnamefont {Galli}}} (\bibinfo {year} {2009}),\
  \bibfield  {title} {\enquote {\bibinfo {title} {{I}terative calculations of
  dielectric eigenvalue spectra},}\ }\href {\doibase
  10.1103/PhysRevB.79.245106} {\bibfield  {journal} {\bibinfo  {journal} {Phys.
  Rev. B}\ }\textbf {\bibinfo {volume} {79}},\ \bibinfo {pages}
  {245106}}\BibitemShut {NoStop}%
\bibitem [{\citenamefont {Wimmer}\ \emph {et~al.}(1981)\citenamefont {Wimmer},
  \citenamefont {Krakauer}, \citenamefont {Weinert},\ and\ \citenamefont
  {Freeman}}]{Wimmer1981}%
  \BibitemOpen
  \bibfield  {author} {\bibinfo {author} {\bibnamefont {Wimmer}, \bibfnamefont
  {E}}, \bibinfo {author} {\bibfnamefont {H.}~\bibnamefont {Krakauer}},
  \bibinfo {author} {\bibfnamefont {M.}~\bibnamefont {Weinert}}, \ and\
  \bibinfo {author} {\bibfnamefont {A.~J.}\ \bibnamefont {Freeman}}} (\bibinfo
  {year} {1981}),\ \bibfield  {title} {\enquote {\bibinfo {title}
  {{F}ull-potential self-consistent linearized-augmented-plane-wave method for
  calculating the electronic structure of molecules and surfaces: {O}$_{2}$
  molecule},}\ }\href {\doibase 10.1103/PhysRevB.24.864} {\bibfield  {journal}
  {\bibinfo  {journal} {Phys. Rev. B}\ }\textbf {\bibinfo {volume} {24}},\
  \bibinfo {pages} {864--875}}\BibitemShut {NoStop}%
\bibitem [{\citenamefont {Winther}\ and\ \citenamefont
  {Thygesen}(2017)}]{winther_2dm_4}%
  \BibitemOpen
  \bibfield  {author} {\bibinfo {author} {\bibnamefont {Winther}, \bibfnamefont
  {K~T}}, \ and\ \bibinfo {author} {\bibfnamefont {K.~S.}\ \bibnamefont
  {Thygesen}}} (\bibinfo {year} {2017}),\ \bibfield  {title} {\enquote
  {\bibinfo {title} {{B}and structure engineering in van der {W}aals
  heterostructures via dielectric screening: the {G}{$\Delta$}{W} method},}\
  }\href@noop {} {\bibfield  {journal} {\bibinfo  {journal} {2D Mater.}\
  }\textbf {\bibinfo {volume} {4}}~(\bibinfo {number} {2}),\ \bibinfo {pages}
  {025059}}\BibitemShut {NoStop}%
\bibitem [{\citenamefont {Wippermann}\ \emph {et~al.}(2014)\citenamefont
  {Wippermann}, \citenamefont {V\"or\"os}, \citenamefont {Gali}, \citenamefont
  {Gygi}, \citenamefont {Zimanyi},\ and\ \citenamefont
  {Galli}}]{Wippermann/etal:2014}%
  \BibitemOpen
  \bibfield  {author} {\bibinfo {author} {\bibnamefont {Wippermann},
  \bibfnamefont {S}}, \bibinfo {author} {\bibfnamefont {M.}~\bibnamefont
  {V\"or\"os}}, \bibinfo {author} {\bibfnamefont {A.}~\bibnamefont {Gali}},
  \bibinfo {author} {\bibfnamefont {F.}~\bibnamefont {Gygi}}, \bibinfo {author}
  {\bibfnamefont {G.~T.}\ \bibnamefont {Zimanyi}}, \ and\ \bibinfo {author}
  {\bibfnamefont {G.}~\bibnamefont {Galli}}} (\bibinfo {year} {2014}),\
  \bibfield  {title} {\enquote {\bibinfo {title} {{S}olar {N}anocomposites with
  {C}omplementary {C}harge {E}xtraction {P}athways for {E}lectrons and {H}oles:
  {S}i {E}mbedded in {Z}n{S}},}\ }\href@noop {} {\bibfield  {journal} {\bibinfo
   {journal} {Phys. Rev. Lett.}\ }\textbf {\bibinfo {volume} {112}},\ \bibinfo
  {pages} {106801}}\BibitemShut {NoStop}%
\bibitem [{\citenamefont {Wirtz}\ \emph {et~al.}(2005)\citenamefont {Wirtz},
  \citenamefont {Marini},\ and\ \citenamefont {Rubio}}]{wirtz_aip_786}%
  \BibitemOpen
  \bibfield  {author} {\bibinfo {author} {\bibnamefont {Wirtz}, \bibfnamefont
  {L}}, \bibinfo {author} {\bibfnamefont {A.}~\bibnamefont {Marini}}, \ and\
  \bibinfo {author} {\bibfnamefont {A.}~\bibnamefont {Rubio}}} (\bibinfo {year}
  {2005}),\ \bibfield  {title} {\enquote {\bibinfo {title} {{O}ptical
  {A}bsorption of hexagonal {B}oron {N}itride and {B}{N} nanotubes},}\ }\href
  {\doibase 10.1063/1.2103894} {\bibfield  {journal} {\bibinfo  {journal} {AIP
  Conf. Proc.}\ }\textbf {\bibinfo {volume} {786}}~(\bibinfo {number} {1}),\
  \bibinfo {pages} {391--395}}\BibitemShut {NoStop}%
\bibitem [{\citenamefont {Wiser}(1963)}]{Wiser:1963}%
  \BibitemOpen
  \bibfield  {author} {\bibinfo {author} {\bibnamefont {Wiser}, \bibfnamefont
  {N}}} (\bibinfo {year} {1963}),\ \bibfield  {title} {\enquote {\bibinfo
  {title} {{D}ielectric constant with local field effect included},}\
  }\href@noop {} {\bibfield  {journal} {\bibinfo  {journal} {Phys. Rev.}\
  }\textbf {\bibinfo {volume} {129}},\ \bibinfo {pages} {62}}\BibitemShut
  {NoStop}%
\bibitem [{\citenamefont {Yan}\ \emph {et~al.}(2015)\citenamefont {Yan},
  \citenamefont {Gao}, \citenamefont {Stein},\ and\ \citenamefont
  {Coard}}]{yan_prb_91}%
  \BibitemOpen
  \bibfield  {author} {\bibinfo {author} {\bibnamefont {Yan}, \bibfnamefont
  {J-A}}, \bibinfo {author} {\bibfnamefont {S.-P.}\ \bibnamefont {Gao}},
  \bibinfo {author} {\bibfnamefont {R.}~\bibnamefont {Stein}}, \ and\ \bibinfo
  {author} {\bibfnamefont {G.}~\bibnamefont {Coard}}} (\bibinfo {year}
  {2015}),\ \bibfield  {title} {\enquote {\bibinfo {title} {{T}uning the
  electronic structure of silicene and germanene by biaxial strain and electric
  field},}\ }\href {\doibase 10.1103/PhysRevB.91.245403} {\bibfield  {journal}
  {\bibinfo  {journal} {Phys. Rev. B}\ }\textbf {\bibinfo {volume} {91}},\
  \bibinfo {pages} {245403}}\BibitemShut {NoStop}%
\bibitem [{\citenamefont {Yan}\ \emph {et~al.}(2011)\citenamefont {Yan},
  \citenamefont {Rinke}, \citenamefont {Winkelnkemper}, \citenamefont {Qteish},
  \citenamefont {Bimberg}, \citenamefont {Scheffler},\ and\ \citenamefont {{Van
  de Walle}}}]{Yan/etal:2011}%
  \BibitemOpen
  \bibfield  {author} {\bibinfo {author} {\bibnamefont {Yan}, \bibfnamefont
  {Q}}, \bibinfo {author} {\bibfnamefont {P.}~\bibnamefont {Rinke}}, \bibinfo
  {author} {\bibfnamefont {M.}~\bibnamefont {Winkelnkemper}}, \bibinfo {author}
  {\bibfnamefont {A.}~\bibnamefont {Qteish}}, \bibinfo {author} {\bibfnamefont
  {D.}~\bibnamefont {Bimberg}}, \bibinfo {author} {\bibfnamefont
  {M.}~\bibnamefont {Scheffler}}, \ and\ \bibinfo {author} {\bibfnamefont
  {C.~G.}\ \bibnamefont {{Van de Walle}}}} (\bibinfo {year} {2011}),\ \bibfield
   {title} {\enquote {\bibinfo {title} {{B}and parameters and strain effects in
  {Z}n{O} and group-{I}{I}{I} nitrides},}\ }\href@noop {} {\bibfield  {journal}
  {\bibinfo  {journal} {Semicond. Sci. Technol.}\ }\textbf {\bibinfo {volume}
  {26}},\ \bibinfo {pages} {014037}}\BibitemShut {NoStop}%
\bibitem [{\citenamefont {Yan}\ \emph {et~al.}(2012)\citenamefont {Yan},
  \citenamefont {Rinke}, \citenamefont {Winkelnkemper}, \citenamefont {Qteish},
  \citenamefont {Bimberg}, \citenamefont {Scheffler},\ and\ \citenamefont {{Van
  de Walle}}}]{Yan/etal:2012}%
  \BibitemOpen
  \bibfield  {author} {\bibinfo {author} {\bibnamefont {Yan}, \bibfnamefont
  {Q}}, \bibinfo {author} {\bibfnamefont {P.}~\bibnamefont {Rinke}}, \bibinfo
  {author} {\bibfnamefont {M.}~\bibnamefont {Winkelnkemper}}, \bibinfo {author}
  {\bibfnamefont {A.}~\bibnamefont {Qteish}}, \bibinfo {author} {\bibfnamefont
  {D.}~\bibnamefont {Bimberg}}, \bibinfo {author} {\bibfnamefont
  {M.}~\bibnamefont {Scheffler}}, \ and\ \bibinfo {author} {\bibfnamefont
  {C.~G.}\ \bibnamefont {{Van de Walle}}}} (\bibinfo {year} {2012}),\ \bibfield
   {title} {\enquote {\bibinfo {title} {{S}train effects and band parameters in
  {M}g{O}, {Z}n{O}, and {C}d{O}},}\ }\href@noop {} {\bibfield  {journal}
  {\bibinfo  {journal} {Appl. Phys. Lett.}\ }\textbf {\bibinfo {volume}
  {101}}~(\bibinfo {number} {15}),\ \bibinfo {eid} {152105}}\BibitemShut
  {NoStop}%
\bibitem [{\citenamefont {Yanagisawa}\ and\ \citenamefont
  {Hamada}(2017)}]{Yanagisawa2017}%
  \BibitemOpen
  \bibfield  {author} {\bibinfo {author} {\bibnamefont {Yanagisawa},
  \bibfnamefont {S}}, \ and\ \bibinfo {author} {\bibfnamefont {I.}~\bibnamefont
  {Hamada}}} (\bibinfo {year} {2017}),\ \bibfield  {title} {\enquote {\bibinfo
  {title} {{D}etermination of geometric and electronic structures of organic
  crystals from first principles: {R}ole of the molecular configuration on the
  electronic structure},}\ }\href {\doibase 10.1063/1.4974844} {\bibfield
  {journal} {\bibinfo  {journal} {J. Appl. Phys.}\ }\textbf {\bibinfo {volume}
  {121}}~(\bibinfo {number} {4}),\ \bibinfo {pages} {045501}}\BibitemShut
  {NoStop}%
\bibitem [{\citenamefont {Yanagisawa}\ \emph {et~al.}(2013)\citenamefont
  {Yanagisawa}, \citenamefont {Morikawa},\ and\ \citenamefont
  {Schindlmayr}}]{Yanagisawa/etal:2013}%
  \BibitemOpen
  \bibfield  {author} {\bibinfo {author} {\bibnamefont {Yanagisawa},
  \bibfnamefont {S}}, \bibinfo {author} {\bibfnamefont {Y.}~\bibnamefont
  {Morikawa}}, \ and\ \bibinfo {author} {\bibfnamefont {A.}~\bibnamefont
  {Schindlmayr}}} (\bibinfo {year} {2013}),\ \bibfield  {title} {\enquote
  {\bibinfo {title} {{H}{O}{M}{O} band dispersion of crystalline rubrene:
  {E}ffects of self-energy corrections within the ${G}{W}$ approximation},}\
  }\href@noop {} {\bibfield  {journal} {\bibinfo  {journal} {Phys. Rev. B}\
  }\textbf {\bibinfo {volume} {88}},\ \bibinfo {pages} {115438}}\BibitemShut
  {NoStop}%
\bibitem [{\citenamefont {Yanagisawa}\ \emph {et~al.}(2014)\citenamefont
  {Yanagisawa}, \citenamefont {Morikawa},\ and\ \citenamefont
  {Schindlmayr}}]{Yanagisawa2014}%
  \BibitemOpen
  \bibfield  {author} {\bibinfo {author} {\bibnamefont {Yanagisawa},
  \bibfnamefont {S}}, \bibinfo {author} {\bibfnamefont {Y.}~\bibnamefont
  {Morikawa}}, \ and\ \bibinfo {author} {\bibfnamefont {A.}~\bibnamefont
  {Schindlmayr}}} (\bibinfo {year} {2014}),\ \bibfield  {title} {\enquote
  {\bibinfo {title} {{T}heoretical investigation of the band structure of
  picene single crystals within the ${G}{W}$ approximation},}\ }\href {\doibase
  10.7567/jjap.53.05fy02} {\bibfield  {journal} {\bibinfo  {journal} {Jpn. J.
  Appl. Phys.}\ }\textbf {\bibinfo {volume} {53}}~(\bibinfo {number} {5S1}),\
  \bibinfo {pages} {05FY02}}\BibitemShut {NoStop}%
\bibitem [{\citenamefont {Yang}(2017)}]{Yong:2017}%
  \BibitemOpen
  \bibfield  {author} {\bibinfo {author} {\bibnamefont {Yang}, \bibfnamefont
  {Y}}} (\bibinfo {year} {2017}),\ \bibfield  {title} {\enquote {\bibinfo
  {title} {{U}nexpected robustness of the band gaps of {T}i{O}$_2$ under high
  pressures},}\ }\href {http://stacks.iop.org/2399-6528/1/i=5/a=055014}
  {\bibfield  {journal} {\bibinfo  {journal} {J. Phys. Commun.}\ }\textbf
  {\bibinfo {volume} {1}}~(\bibinfo {number} {5}),\ \bibinfo {pages}
  {055014}}\BibitemShut {NoStop}%
\bibitem [{\citenamefont {Yi}\ \emph {et~al.}(2010)\citenamefont {Yi},
  \citenamefont {Ma}, \citenamefont {Rohlfing}, \citenamefont {Silkin},\ and\
  \citenamefont {Chulkov}}]{Yi/etal:2010}%
  \BibitemOpen
  \bibfield  {author} {\bibinfo {author} {\bibnamefont {Yi}, \bibfnamefont
  {Z}}, \bibinfo {author} {\bibfnamefont {Y.}~\bibnamefont {Ma}}, \bibinfo
  {author} {\bibfnamefont {M.}~\bibnamefont {Rohlfing}}, \bibinfo {author}
  {\bibfnamefont {V.~M.}\ \bibnamefont {Silkin}}, \ and\ \bibinfo {author}
  {\bibfnamefont {E.~V.}\ \bibnamefont {Chulkov}}} (\bibinfo {year} {2010}),\
  \bibfield  {title} {\enquote {\bibinfo {title} {{Q}uasiparticle band
  structures and lifetimes in noble metals using {G}aussian orbital basis
  sets},}\ }\href@noop {} {\bibfield  {journal} {\bibinfo  {journal} {Phys.
  Rev. B}\ }\textbf {\bibinfo {volume} {81}},\ \bibinfo {pages}
  {125125}}\BibitemShut {NoStop}%
\bibitem [{\citenamefont {Zgid}\ \emph {et~al.}(2012)\citenamefont {Zgid},
  \citenamefont {Gull},\ and\ \citenamefont {Chan}}]{zgid/etal:2012}%
  \BibitemOpen
  \bibfield  {author} {\bibinfo {author} {\bibnamefont {Zgid}, \bibfnamefont
  {D}}, \bibinfo {author} {\bibfnamefont {E.}~\bibnamefont {Gull}}, \ and\
  \bibinfo {author} {\bibfnamefont {G.~K.-L.}\ \bibnamefont {Chan}}} (\bibinfo
  {year} {2012}),\ \bibfield  {title} {\enquote {\bibinfo {title} {{T}runcated
  configuration interaction expansions as solvers for correlated quantum
  impurity models and dynamical mean-field theory},}\ }\href@noop {} {\bibfield
   {journal} {\bibinfo  {journal} {Phys. Rev. B}\ }\textbf {\bibinfo {volume}
  {86}},\ \bibinfo {pages} {165128}}\BibitemShut {NoStop}%
\bibitem [{\citenamefont {Zhang}\ \emph {et~al.}(2016)\citenamefont {Zhang},
  \citenamefont {Gong}, \citenamefont {Nie}, \citenamefont {Min}, \citenamefont
  {Liang}, \citenamefont {Oh}, \citenamefont {Zhang}, \citenamefont {Wang},
  \citenamefont {Hong}, \citenamefont {Colombo}, \citenamefont {Wallace},\ and\
  \citenamefont {Cho}}]{zhang_2dm_4}%
  \BibitemOpen
  \bibfield  {author} {\bibinfo {author} {\bibnamefont {Zhang}, \bibfnamefont
  {C}}, \bibinfo {author} {\bibfnamefont {C.}~\bibnamefont {Gong}}, \bibinfo
  {author} {\bibfnamefont {Y.}~\bibnamefont {Nie}}, \bibinfo {author}
  {\bibfnamefont {K.-A.}\ \bibnamefont {Min}}, \bibinfo {author} {\bibfnamefont
  {C.}~\bibnamefont {Liang}}, \bibinfo {author} {\bibfnamefont {Y.~J.}\
  \bibnamefont {Oh}}, \bibinfo {author} {\bibfnamefont {H.}~\bibnamefont
  {Zhang}}, \bibinfo {author} {\bibfnamefont {W.}~\bibnamefont {Wang}},
  \bibinfo {author} {\bibfnamefont {S.}~\bibnamefont {Hong}}, \bibinfo {author}
  {\bibfnamefont {L.}~\bibnamefont {Colombo}}, \bibinfo {author} {\bibfnamefont
  {R.~M.}\ \bibnamefont {Wallace}}, \ and\ \bibinfo {author} {\bibfnamefont
  {K.}~\bibnamefont {Cho}}} (\bibinfo {year} {2016}),\ \bibfield  {title}
  {\enquote {\bibinfo {title} {{S}ystematic study of electronic structure and
  band alignment of monolayer transition metal dichalcogenides in {V}an der
  {W}aals heterostructures},}\ }\href {\doibase 10.1088/2053-1583/4/1/015026}
  {\bibfield  {journal} {\bibinfo  {journal} {2D Mater.}\ }\textbf {\bibinfo
  {volume} {4}}~(\bibinfo {number} {1}),\ \bibinfo {pages}
  {015026}}\BibitemShut {NoStop}%
\bibitem [{\citenamefont {Zhang}\ \emph {et~al.}(2017)\citenamefont {Zhang},
  \citenamefont {Su},\ and\ \citenamefont {Yang}}]{Zhang/Su/Yang:2017}%
  \BibitemOpen
  \bibfield  {author} {\bibinfo {author} {\bibnamefont {Zhang}, \bibfnamefont
  {D}}, \bibinfo {author} {\bibfnamefont {N.~Q.}\ \bibnamefont {Su}}, \ and\
  \bibinfo {author} {\bibfnamefont {W.}~\bibnamefont {Yang}}} (\bibinfo {year}
  {2017}),\ \bibfield  {title} {\enquote {\bibinfo {title} {{A}ccurate
  {Q}uasiparticle {S}pectra from the {T}-{M}atrix {S}elf-{E}nergy and the
  {P}article-{P}article {R}andom {P}hase {A}pproximation},}\ }\href@noop {}
  {\bibfield  {journal} {\bibinfo  {journal} {J. Phys. Chem. Lett.}\ }\textbf
  {\bibinfo {volume} {8}}~(\bibinfo {number} {14}),\ \bibinfo {pages}
  {3223--3227}}\BibitemShut {NoStop}%
\bibitem [{\citenamefont {Zhou}\ \emph {et~al.}(2018)\citenamefont {Zhou},
  \citenamefont {Gatti}, \citenamefont {Kas}, \citenamefont {Rehr},\ and\
  \citenamefont {Reining}}]{Zhou/etal:2018}%
  \BibitemOpen
  \bibfield  {author} {\bibinfo {author} {\bibnamefont {Zhou}, \bibfnamefont
  {J~S}}, \bibinfo {author} {\bibfnamefont {M.}~\bibnamefont {Gatti}}, \bibinfo
  {author} {\bibfnamefont {J.~J.}\ \bibnamefont {Kas}}, \bibinfo {author}
  {\bibfnamefont {J.~J.}\ \bibnamefont {Rehr}}, \ and\ \bibinfo {author}
  {\bibfnamefont {L.}~\bibnamefont {Reining}}} (\bibinfo {year} {2018}),\
  \bibfield  {title} {\enquote {\bibinfo {title} {{C}umulant {G}reen's function
  calculations of plasmon satellites in bulk sodium: {I}nfluence of screening
  and the crystal environment},}\ }\href@noop {} {\bibfield  {journal}
  {\bibinfo  {journal} {Phys. Rev. B}\ }\textbf {\bibinfo {volume} {97}},\
  \bibinfo {pages} {035137}}\BibitemShut {NoStop}%
\bibitem [{\citenamefont {Zhou}\ \emph {et~al.}(2015)\citenamefont {Zhou},
  \citenamefont {Kas}, \citenamefont {Sponza}, \citenamefont {Reshetnyak},
  \citenamefont {Guzzo}, \citenamefont {Giorgetti}, \citenamefont {Gatti},
  \citenamefont {Sottile}, \citenamefont {Rehr},\ and\ \citenamefont
  {Reining}}]{Zhou/etal:2015}%
  \BibitemOpen
  \bibfield  {author} {\bibinfo {author} {\bibnamefont {Zhou}, \bibfnamefont
  {J~S}}, \bibinfo {author} {\bibfnamefont {J.~J.}\ \bibnamefont {Kas}},
  \bibinfo {author} {\bibfnamefont {L.}~\bibnamefont {Sponza}}, \bibinfo
  {author} {\bibfnamefont {I.}~\bibnamefont {Reshetnyak}}, \bibinfo {author}
  {\bibfnamefont {M.}~\bibnamefont {Guzzo}}, \bibinfo {author} {\bibfnamefont
  {C.}~\bibnamefont {Giorgetti}}, \bibinfo {author} {\bibfnamefont
  {M.}~\bibnamefont {Gatti}}, \bibinfo {author} {\bibfnamefont
  {F.}~\bibnamefont {Sottile}}, \bibinfo {author} {\bibfnamefont {J.~J.}\
  \bibnamefont {Rehr}}, \ and\ \bibinfo {author} {\bibfnamefont
  {L.}~\bibnamefont {Reining}}} (\bibinfo {year} {2015}),\ \bibfield  {title}
  {\enquote {\bibinfo {title} {{D}ynamical effects in electron spectroscopy},}\
  }\href@noop {} {\bibfield  {journal} {\bibinfo  {journal} {J. Chem. Phys.}\
  }\textbf {\bibinfo {volume} {143}}~(\bibinfo {number} {18}),\ \bibinfo
  {pages} {184109}}\BibitemShut {NoStop}%
\bibitem [{\citenamefont {Zhukov}\ \emph {et~al.}(2004)\citenamefont {Zhukov},
  \citenamefont {Chulkov},\ and\ \citenamefont
  {Echenique}}]{Zhukov/Chulkov/Echenique:2004}%
  \BibitemOpen
  \bibfield  {author} {\bibinfo {author} {\bibnamefont {Zhukov}, \bibfnamefont
  {V~P}}, \bibinfo {author} {\bibfnamefont {E.~V.}\ \bibnamefont {Chulkov}}, \
  and\ \bibinfo {author} {\bibfnamefont {P.~M.}\ \bibnamefont {Echenique}}}
  (\bibinfo {year} {2004}),\ \bibfield  {title} {\enquote {\bibinfo {title}
  {{L}ifetimes of {E}xcited {E}lectrons {I}n {F}e {A}nd {N}i:
  {F}irst-{P}rinciples ${G}{W}$ and the ${T}$-{M}atrix {T}heory},}\ }\href
  {\doibase 10.1103/PhysRevLett.93.096401} {\bibfield  {journal} {\bibinfo
  {journal} {Phys. Rev. Lett.}\ }\textbf {\bibinfo {volume} {93}},\ \bibinfo
  {pages} {096401}}\BibitemShut {NoStop}%
\bibitem [{\citenamefont {Zhukov}\ \emph {et~al.}(2005)\citenamefont {Zhukov},
  \citenamefont {Chulkov},\ and\ \citenamefont {Echenique}}]{Zhukov/etal:2005}%
  \BibitemOpen
  \bibfield  {author} {\bibinfo {author} {\bibnamefont {Zhukov}, \bibfnamefont
  {V~P}}, \bibinfo {author} {\bibfnamefont {E.~V.}\ \bibnamefont {Chulkov}}, \
  and\ \bibinfo {author} {\bibfnamefont {P.~M.}\ \bibnamefont {Echenique}}}
  (\bibinfo {year} {2005}),\ \bibfield  {title} {\enquote {\bibinfo {title}
  {$\mathit{GW}+{T}$ theory of excited electron lifetimes in metals},}\
  }\href@noop {} {\bibfield  {journal} {\bibinfo  {journal} {Phys. Rev. B}\
  }\textbf {\bibinfo {volume} {72}},\ \bibinfo {pages} {155109}}\BibitemShut
  {NoStop}%
\bibitem [{\citenamefont {Zhukov}\ \emph {et~al.}(2006)\citenamefont {Zhukov},
  \citenamefont {Chulkov},\ and\ \citenamefont {Echenique}}]{Zhukov/etal:2006}%
  \BibitemOpen
  \bibfield  {author} {\bibinfo {author} {\bibnamefont {Zhukov}, \bibfnamefont
  {V~P}}, \bibinfo {author} {\bibfnamefont {E.~V.}\ \bibnamefont {Chulkov}}, \
  and\ \bibinfo {author} {\bibfnamefont {P.~M.}\ \bibnamefont {Echenique}}}
  (\bibinfo {year} {2006}),\ \bibfield  {title} {\enquote {\bibinfo {title}
  {{L}ifetimes and inelastic mean free path of low-energy excited electrons in
  {F}e, {N}i, {P}t, and {A}u: {A}b initio $\mathrm{GW}+\mathrm{T}$
  calculations},}\ }\href@noop {} {\bibfield  {journal} {\bibinfo  {journal}
  {Phys. Rev. B}\ }\textbf {\bibinfo {volume} {73}},\ \bibinfo {pages}
  {125105}}\BibitemShut {NoStop}%
\bibitem [{\citenamefont {Zoppi}\ \emph {et~al.}(2011)\citenamefont {Zoppi},
  \citenamefont {Martin-Samos},\ and\ \citenamefont {Baldridge}}]{Zoppi2011}%
  \BibitemOpen
  \bibfield  {author} {\bibinfo {author} {\bibnamefont {Zoppi}, \bibfnamefont
  {L}}, \bibinfo {author} {\bibfnamefont {L.}~\bibnamefont {Martin-Samos}}, \
  and\ \bibinfo {author} {\bibfnamefont {K.~K.}\ \bibnamefont {Baldridge}}}
  (\bibinfo {year} {2011}),\ \bibfield  {title} {\enquote {\bibinfo {title}
  {{E}ffect of {M}olecular {P}acking on {C}orannulene-{B}ased {M}aterials
  {E}lectroluminescence},}\ }\href {\doibase 10.1021/ja2040688} {\bibfield
  {journal} {\bibinfo  {journal} {J. Am. Chem. Soc.}\ }\textbf {\bibinfo
  {volume} {133}}~(\bibinfo {number} {35}),\ \bibinfo {pages}
  {14002--14009}}\BibitemShut {NoStop}%
\end{thebibliography}%

\end{document}